%% file: thesis_main.tex
\documentclass [MS] {uclathes}

\usepackage{booktabs}

\usepackage[margin=1.0in]{geometry}
\usepackage{calc}
\usepackage{amsmath,amsthm,amssymb,mathrsfs}
\numberwithin{equation}{chapter}
\usepackage{chngcntr}
\counterwithin{figure}{chapter}
\usepackage{mathtools}
\usepackage{physics}
\usepackage{textgreek}
\usepackage{array}
\usepackage{graphicx}
\usepackage{svg}
\usepackage{float}
\usepackage{rotating}  
\usepackage{dcolumn}
\usepackage{tablefootnote}
\usepackage{cancel}

\usepackage{tikz}
\usepackage{pgfplots}
\pgfplotsset{compat=1.18}
\usepackage[american]{circuitikz}
\usetikzlibrary{circuits}
\usetikzlibrary{arrows.meta}
\usetikzlibrary{shapes,arrows}
\tikzset{every picture/.style={line width=1pt}}
\ctikzset{bipoles/thickness=1}
\makeatletter
\pgfcircdeclarebipolescaled{instruments}
{
    \anchor{text}{\pgfextracty{\pgf@circ@res@up}{\northeast}
        \pgfpoint{-.5\wd\pgfnodeparttextbox}{
            \dimexpr.5\dp\pgfnodeparttextbox+.5\ht\pgfnodeparttextbox+\pgf@circ@res@up\relax
        }
    }
}
{\ctikzvalof{bipoles/oscope/height}}
{josephson}
{\ctikzvalof{bipoles/oscope/height}}
{\ctikzvalof{bipoles/oscope/width}}
{
    \pgf@circ@setlinewidth{bipoles}{\pgfstartlinewidth}
    \pgfextracty{\pgf@circ@res@up}{\northeast}
    \pgfextractx{\pgf@circ@res@right}{\northeast}
    \pgfextractx{\pgf@circ@res@left}{\southwest}
    \pgfextracty{\pgf@circ@res@down}{\southwest}
    \pgfmathsetlength{\pgf@circ@res@step}{0.25*\pgf@circ@res@up}
    \pgfscope
        \pgfpathrectanglecorners{\pgfpoint{\pgf@circ@res@left}{\pgf@circ@res@down}}{\pgfpoint{\pgf@circ@res@right}{\pgf@circ@res@up}}
        \pgf@circ@draworfill
    \endpgfscope
    \pgfscope
      \pgfpathmoveto{\pgfpoint{\pgf@circ@res@left}{\pgf@circ@res@up}}%
      \pgfpathlineto{\pgfpoint{\pgf@circ@res@right}{\pgf@circ@res@down}}%
      \pgfpathmoveto{\pgfpoint{\pgf@circ@res@right}{\pgf@circ@res@up}}%
      \pgfpathlineto{\pgfpoint{\pgf@circ@res@left}{\pgf@circ@res@down}}%
      \pgfusepath{draw}
    \endpgfscope
}
\def\pgf@circ@josephson@path#1{\pgf@circ@bipole@path{josephson}{#1}}
\tikzset{josephson/.style = {\circuitikzbasekey, /tikz/to path=\pgf@circ@josephson@path, l=#1}}
\makeatletter
\pgfcircdeclarebipolescaled{instruments}
{
    \anchor{text}{\pgfextracty{\pgf@circ@res@up}{\northeast}
        \pgfpoint{-.5\wd\pgfnodeparttextbox}{
            \dimexpr.5\dp\pgfnodeparttextbox+.5\ht\pgfnodeparttextbox+\pgf@circ@res@up\relax
        }
    }
}
{\ctikzvalof{bipoles/oscope/height}}
{JJ}
{\ctikzvalof{bipoles/oscope/height}}
{\ctikzvalof{bipoles/oscope/width}}
{
    \pgf@circ@setlinewidth{bipoles}{\pgfstartlinewidth}
    \pgfextracty{\pgf@circ@res@up}{\northeast}
    \pgfextractx{\pgf@circ@res@right}{\northeast}
    \pgfextractx{\pgf@circ@res@left}{\southwest}
    \pgfextracty{\pgf@circ@res@down}{\southwest}
    \pgfmathsetlength{\pgf@circ@res@step}{0.25*\pgf@circ@res@up}
    \pgfscope
      \pgfpathmoveto{\pgfpoint{\pgf@circ@res@left}{\pgf@circ@res@up}}%
      \pgfpathlineto{\pgfpoint{\pgf@circ@res@right}{\pgf@circ@res@down}}%
      \pgfpathmoveto{\pgfpoint{\pgf@circ@res@right}{\pgf@circ@res@up}}%
      \pgfpathlineto{\pgfpoint{\pgf@circ@res@left}{\pgf@circ@res@down}}%
      \pgfpathmoveto{\pgfpoint{\pgf@circ@res@left}{0}}%
      \pgfpathlineto{\pgfpoint{\pgf@circ@res@right}{0}}%
      \pgfusepath{draw}
    \endpgfscope
}
\def\pgf@circ@JJ@path#1{\pgf@circ@bipole@path{JJ}{#1}}
\tikzset{JJ/.style = {\circuitikzbasekey, /tikz/to path=\pgf@circ@JJ@path, l=#1}}

\usepackage[font=normal,labelfont=bf, textfont=it,margin=5pt]{caption}
\counterwithout{footnote}{chapter}

\DeclareMathAlphabet{\mathscrbf}{OMS}{mdugm}{b}{n} 

\DeclareFontFamily{U}{BOONDOX-calo}{\skewchar\font=45 }
\DeclareFontShape{U}{BOONDOX-calo}{m}{n}{
  <-> s*[1.05] BOONDOX-r-calo}{}
\DeclareFontShape{U}{BOONDOX-calo}{b}{n}{
  <-> s*[1.05] BOONDOX-b-calo}{}
\DeclareMathAlphabet{\mathcalboondox}{U}{BOONDOX-calo}{m}{n}
\SetMathAlphabet{\mathcalboondox}{bold}{U}{BOONDOX-calo}{b}{n}
\DeclareMathAlphabet{\mathbcalboondox}{U}{BOONDOX-calo}{b}{n}

\newcommand{\ci}{\mathrm{i}}
\newcommand{\cj}{\mathrm{j}}

\usepackage{scalerel,stackengine}
\newcommand\Bigs[1]{\scalerel*[5.5pt]{\Big#1}{%
  \ensurestackMath{\addstackgap[1.5pt]{\big#1}}}}
\newcommand\Bigsl[1]{\mathopen{\Bigs{#1}}}
\newcommand\Bigsr[1]{\mathclose{\Bigs{#1}}}

\usepackage{mdframed}
\theoremstyle{definition}
\newmdtheoremenv{exercise}{Exercise}[chapter]

\usepackage{hyperref}
\hypersetup{
    colorlinks,
    citecolor=black,
    filecolor=black,
    linkcolor=black,
    urlcolor=black
}

\def\dsp{\def\baselinestretch{2}\large\normalsize}
\dsp

\title          {Stochastic Model of Qudit Measurement \\ for Superconducting Quantum Information Processing}
\author         {Kangdi Yu}
\department     {Electrical and Computer Engineering}
\degreeyear     {2023}

\chair          {Chee Wei Wong}
\member         {Louis-Serge Bouchard}
\member         {Wesley C. Campbell}
\member         {Bahman Gharesi Fard}

\dedication     {To Qi Song and Min Yu}

\acknowledgments {This material is based upon work supported by the National Science Foundation Graduate Research Fellowship Program under Grant Number DGE-2034835. Any opinions, findings, and conclusions or recommendations expressed in this material are those of the author(s) and do not necessarily reflect the views of the National Science Foundation. 

My heartfelt appreciation extends to those who have supported and guided me throughout this research journey. The insightful discussions with Prof. Chee Wei Wong and Prof. Louis Bouchard enriched various aspects of my work, offering new perspectives and understanding. I am grateful to Prof. Wesley Campbell for his critical review of my manuscript. Prof. Bahman Gharesifard's course in stochastic calculus and control laid the foundational knowledge vital for this research. The superconducting cavity and transmon chips are kindly provided by Jonathan L. DuBois, Yaniv J. Rosen, and Sean R. O'Kelley from Lawrence Livermore National Laboratory. Jin Ho Kang's management of the dilution fridge in our research group was a crucial element in our experimental success. Lastly, my gratitude extends to Prof. Ming Wu, who introduced me to the realm of physical and wave electronics during my undergraduate studies, equipping me with a solid background essential for my quantum research.}

\abstract       {The field of superconducting quantum computing, based on Josephson junctions, has recently seen remarkable strides in scaling the number of logical qubits. In particular, the fidelities of one- and two-qubit gates are close to the breakeven point with the novel error mitigation and correction methods. Parallel to these advances is the effort to expand the Hilbert space within a single device by employing high-dimensional qubits, otherwise known as qudits. Research has demonstrated the possibility of driving higher-order transitions in a transmon or designing innovative multimode superconducting circuits, termed multimons. These advances can significantly expand the computational basis while simplifying the interconnects in a large-scale quantum processor. This thesis provides a detailed introduction to the superconducting qudit and demonstrates a comprehensive analysis of decoherence in an artificial atom with more than two levels using Lindblad master equations and stochastic master equations (SMEs). After extending the theory of the design, control, and readout of a conventional superconducting qubit to that of a qudit, the thesis focuses on modeling the dispersive measurement of a transmon qutrit in an open quantum system using quadrature detections. Under the Markov assumption, master equations with different levels of abstraction are proposed and solved; in addition, both the ensemble-averaged and the quantum-jump approach of decoherence analysis are presented and compared analytically and numerically. The thesis ends with a series of experimental results on a transmon-type qutrit, verifying the validity of the stochastic model.}

\begin {document}
\makeintropages


\input{include_chapters/chapter_introduction}
\input{include_chapters/chapter_quantum_light}

\input{include_chapters/chapter_light_matter_interaction}

\input{include_chapters/chapter_qubit}
\input{include_chapters/chapter_open_systems}

\appendix

\input{include_appendices/appendix_Fourier}
\input{include_appendices/appendix_QHO_Field}
\input{include_appendices/appendix_LC_TL}
\input{include_appendices/appendix_Semiclassical_Rabi}
\input{include_appendices/appendix_Displacement_op}
\input{include_appendices/appendix_Dephasing}

\bibliographystyle{uclathes}
\bibliography{references}


\end{document}

%% file: include_chapters/chapter_introduction.tex
\chapter{Introduction}
\section{Quantum Information Processing and the Necessity of High-Dimensional Computational Space}
Advances in quantum processors using superconducting Josephson junctions have showcased their advantages over traditional supercomputers in certain tasks \cite{s41586_019_1666_5, nature23879}. These processors, though lacking fault tolerance, have already proven their worth as quantum simulators for multi-particle physical systems in chemistry \cite{doi:10.1126/science.abb9811, PhysRevX.10.021060}. However, their current implementation relies on fixed sets of one- and two-qubit gates, similar to CMOS VLSI design methodology. While this approach allows for easy scalability in CMOS technology, it presents challenges in the quantum realm due to the large size of qubit cavities and error propagation. To overcome these limitations, higher-dimensional artificial atoms (also known as qudits) have brought much attention since they enable a significant increase in computational space \cite{ doi:10.1126/science.1173440, PhysRevLett.115.137002, PhysRevApplied.7.054025, s41534_021_00388_0, pnas.2221736120} within the same footprint.

For a concrete comparison, consider the Toffoli gate (also known as the controlled-controlled-not or CCNOT gate), which plays a vital role in various quantum algorithms and forms a universal gate set when combined with the Hadamard gate. However, constructing a Toffoli gate using only one- and two-qubit gate sets requires 16 gates \cite{nielsen_chuang_2010}. This inefficiency is, of course, a result of the limited utilization of higher energy levels in an artificial atom. In contrast, one can construct the same Toffoli gate by utilizing only 2 two-qubit gates but with 1 three-level system (also known as a qutrit), leading to significant resource savings as we scale up towards universal quantum computing.

This thesis is one step toward the understanding of quantum computation using superconducting qudit. Of fundamental importance in the current framework of superconducting quantum computation are the transmon qubits \cite{PhysRevA.76.042319, s41567-020-0806-z, s41467-021-22030-5}. Though being used as a qubit widely, a transmon is an anharmonic oscillator with unequally spaced energy levels that enable the control and detection of higher-order excitations. Hence, the superconducting platform can be rapidly modified to include the usage of qudits. The goal of this thesis is to establish a mathematical framework for the state detection of a superconducting qudit by extending some of the known results about measuring a transmon-type qubit.

\section{Superconducting Quantum Computation}
As a part of the introductory chapter, we briefly go over the experimental setup for superconducting quantum information processing. Shown in Figure \ref{fig:experiment_setup} is the schematic of a typical setup for controlling and measuring superconducting transmon qudits hosted by a readout resonator. In general, the full experiment consists of two parts -- the room-temperature electronics and the dilution fridge. 

The dilution fridge (DF) operates at cryogenic temperatures below 20 mK at the mixing-chamber stage. Its purpose is to minimize thermal excitation in the quantum processor and create a pristine environment for executing quantum algorithms. Microwave control and readout signals from the room-temperature classical electronics enter the DF and undergo approximately 60 dB attenuation to suppress thermal photons before reaching the resonator and the qudit. The readout resonator can be configured for transmission- or reflection-mode readout. In either case, the readout signal coming out of the resonator propagates through multiple stages of amplification and filtering and is ultimately digitized by the analog-to-digital converter (ADC) at room temperature.

To generate microwave signals, the output of an arbitrary waveform generator (AWG) is mixed with a local oscillator (LO). The AWG output acts as a ``gate'' signal, enabling precise control of the pulse duration in the mixed signal. Additionally, the AWG, typically operated with a sampling rate above 1 GHz, provides versatile control over the microwave pulse envelope. On the receiver side, the readout signal coming out of the DF is demodulated by the same LO used for modulating the input to the DF. This synchronized modulation-demodulation scheme establishes a phase relationship between the input and output readout signals, facilitating information retrieval regarding the phase change induced by the qubit-resonator interaction.

\begin{figure}
    \centering
    \begin{tikzpicture}[scale=0.75]
    \ctikzset{tripoles/mos style/arrows};
    \begin{scope}[shift={(-7.5,3)}]
        \draw 
            (0,0) ellipse (0.6 and 1.2);
        \draw 
            (4.5,-1.2) arc(-90:90:0.6 and 1.2);
        \draw
            (0,1.2) to (4.5,1.2)
            (0,-1.2) to (4.5,-1.2)
            (2.25, 1.5) node[]{$Z_0$};
        \draw[dashed]
            (0,0.2) to (4.5,0.2)
            (0,-0.2) to (4.5,-0.2);
        \draw 
            (0,0) ellipse (0.1 and 0.2)
            (4.5,0) ellipse (0.1 and 0.2);
        \draw 
            (2.25,-1.2) node[/tikz/circuitikz/bipoles/length=30pt,sground]{};
        \draw 
            (0,0) to [/tikz/circuitikz/bipoles/length=30pt, R, l^=$R_{\text{s}}$] (-3,0) 
            (-3,0) to [/tikz/circuitikz/bipoles/length=40pt, sV, l_=$V_{\text{s}}$](-3,-3) node[/tikz/circuitikz/bipoles/length=30pt,sground]{};
        \draw 
            (4.5,0) to (5.5,0); 
    \end{scope}
        \draw
            (0,3) to (3,3) 
            (3,0) to (0,0) 
            (0,3) to [/tikz/circuitikz/bipoles/length=40pt, L, l_=$L_{\text{r}}$] (0,0)
            (3,3) to (3,2) to [/tikz/circuitikz/bipoles/length=40pt, C, l_=$C_{\text{r}}$] (3,1) to (3,0)
            (1.5, 0) node[/tikz/circuitikz/bipoles/length=30pt,sground]{}
            (1.5, -1.2) node[]{{\Large${\displaystyle \substack{\text{Readout} \\[1mm] \text{Resonator}}}$}};
        \draw
            (5,3) to (8,3) 
            (5,0) to (8,0) 
            (5,3) to [/tikz/circuitikz/bipoles/length=30pt, JJ, l_=$L_{\text{J}}$] (5,0)
            (8,3) to (8,2) to [/tikz/circuitikz/bipoles/length=40pt, C, l_=$C_{\text{q}}$] (8,1) to (8,0)
            (6.5, 0) node[/tikz/circuitikz/bipoles/length=30pt,sground]{}
            (6.5, -1.2) node[]{${\text{Qubit/Qudit}}$};
        \draw
            (3,3) to [/tikz/circuitikz/bipoles/length=40pt, C, l^=$C_{g}$] (5,3);
        \draw
            (-2,3) to [/tikz/circuitikz/bipoles/length=40pt, C, l^=$C_{\kappa}$] (0,3);
        \draw 
            (1.5, 3) to [/tikz/circuitikz/bipoles/length=20pt, C, l^=$C_{\text{br}}$] (2.5,5.85)
            (6.5, 3) to [/tikz/circuitikz/bipoles/length=20pt, C, l_=$C_{\text{bq}}$] (5.5,5.85);
        \draw
            (4,6.5) ellipse (2 and 1)
            (4,6.5) node[]{$\text{Bath}$};
    \end{tikzpicture}
    \caption{A summary of the superconducting quantum computing.}
    \label{fig:General_model_superconducting_quantum_computing}
\end{figure}
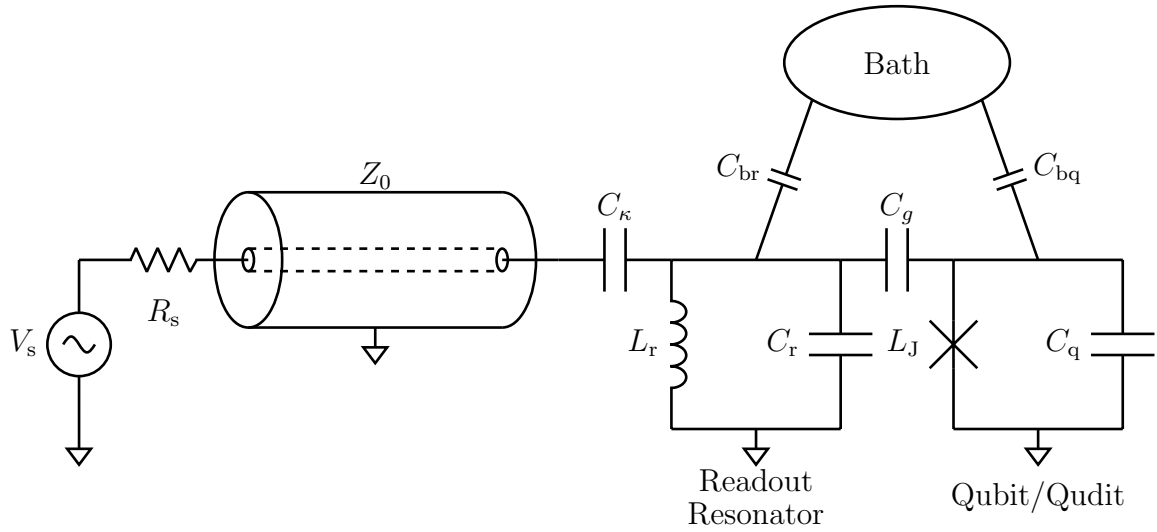

In a more abstract representation, quantum control and measurement can be depicted as a circuit diagram, as shown in Figure \ref{fig:General_model_superconducting_quantum_computing}. The coaxial cables connecting the DF and the room-temperature electronics are modeled as transmission lines. The figure illustrates a readout in the reflection mode, but additional transmission line sections can be included for transmission-mode measurements. The couplings between the qudit and the resonator and between the resonator and the cable are represented by capacitors $C_g$ and $C_{\kappa}$, respectively. However, it is important to note that the quantum system is inevitably affected by its surrounding environment, leading to the decoherence of quantum states. These effects are depicted by the coupling of the qudit and the resonator to the bath. Each aspect of the circuit diagram will be covered in the main chapters of the thesis.

\begin{figure}
    \centering
    \begin{circuitikz}[scale=0.65]
    \ctikzset{tripoles/mos style/arrows};
    \draw[line width=0.8mm] 
        (6,0) to (12,0) to (12,18) to (0,18) to (0,0) to (6,0);
    \draw[dashed]
        (0,9) to (12,9) 
        (0,14) to (12,14); 
    \draw (12,0.5) node[left]{${\text{16 mK}}$};
    \draw (12,9.5) node[left]{${\text{800 mK}}$};
    \draw (12,14.5) node[left]{${\text{4 K}}$};
    \draw (6,0) node[below]{${\textbf{dilution refrigerator}}$};
    
    \draw 
        (3.5,19) to (3.5,22) to [/tikz/circuitikz/bipoles/length=30pt, amp] (6,22) to (8,22) to (8,23);
    \draw 
        (0.8,20) to (0.8, 19) to (2.2, 19) to (2.2, 20.4) to (0.8, 20.4) to (0.8,20)
        (1.5, 19) to (1.5, 19.8)
        (1.5, 19.8) to (2, 20.4) to (2,21) 
        (1.5, 19.8) to (1, 20.4) to (1,21) to (-3,21); 
    \begin{scope}[shift={(11, 7.4)}]
        \draw 
            (0.8,20) to (0.8, 19) to (2.2, 19) to (2.2, 20.4) to (0.8, 20.4) to (0.8,20)
            (1.5, 20.4) to (1.5, 19.6)
            (1.5, 19.6) to (2, 19) to (2, 17.6) to (1, 17.6)
            (1.5, 19.6) to (1, 19) to (1,18.3);
        \draw
            (1.5,20.4) to (1.5,22.1) node[above]{$\text{LO}_{\text{r}}$};
    \end{scope}
    \begin{scope}[shift={(-4, 4)}]
        \draw 
            (0,20) node[mixer,scale=0.7] (mixQ) {}
            (mixQ.w) to (-2,20) to (-2,22) to [/tikz/circuitikz/bipoles/length=30pt, lowpass2] (-2,24) to [/tikz/circuitikz/bipoles/length=20pt, twoportsplit] (-2,25.5) node[above]{$I_{\text{q}}$}
            (mixQ.e) to (2,20) to (2,22) to [/tikz/circuitikz/bipoles/length=30pt, lowpass2] (2,24) to [/tikz/circuitikz/bipoles/length=20pt, twoportsplit] (2,25.5) node[above]{$Q_{\text{q}}$}
            (mixQ.n) to (0,25.5)node[above]{$\text{LO}_{\text{q}}$}
            (mixQ.s) to [/tikz/circuitikz/bipoles/length=30pt, amp](0,17) to (1,17);
    \end{scope}
    \begin{scope}[shift={(2, 4)}]
        \draw 
            (0,20) node[mixer,scale=0.7] (mixQ) {}
            (mixQ.w) to (-2,20) to (-2,22) to [/tikz/circuitikz/bipoles/length=30pt, lowpass2] (-2,24) to [/tikz/circuitikz/bipoles/length=20pt, twoportsplit] (-2,25.5) node[above]{$I_{\text{r,in}}$}
            (mixQ.e) to (2,20) to (2,22) to [/tikz/circuitikz/bipoles/length=30pt, lowpass2] (2,24) to [/tikz/circuitikz/bipoles/length=20pt, twoportsplit] (2,25.5) node[above]{$Q_{\text{r,in}}$}
            (mixQ.s) to [/tikz/circuitikz/bipoles/length=20pt, twoportsplit](0,17)
            (mixQ.n) to (0,21.5) to (1,21.5) to[/tikz/circuitikz/bipoles/crossing/size=0.4, crossing] (3,21.5) to [/tikz/circuitikz/bipoles/crossing/size=0.4, crossing] (5,21.5) to (6,21.5) to [/tikz/circuitikz/bipoles/crossing/size=0.4, crossing] (10,21.5) to (10,22);
    \end{scope}
    \begin{scope}[shift={(8, 4)}]
        \draw 
            (0,20) node[mixer,scale=0.7] (mixQ) {}
            (mixQ.w) to (-2,20) to (-2,22) to [/tikz/circuitikz/bipoles/length=30pt, amp] (-2,24) to [/tikz/circuitikz/bipoles/length=30pt, lowpass2] (-2,25.5) node[above]{$I_{\text{r,out}}$}
            (mixQ.e) to (2,20) to (2,22) to [/tikz/circuitikz/bipoles/length=30pt, amp] (2,24) to [/tikz/circuitikz/bipoles/length=30pt, lowpass2] (2,25.5) node[above]{$Q_{\text{r,out}}$}
            (mixQ.s) to (0,19)
            (mixQ.n) to (0,21) to [/tikz/circuitikz/bipoles/crossing/size=0.4, crossing] (4,21);
    \end{scope}
    \begin{scope}[shift={(0.5, 0.5)}]
        \draw[line width=0.8mm]  
            (2.5,0) to (5,0) to (5,3) to (0,3) to (0,0) to (2.5,0)
            (5,1.85) node[right]{${\text{resonator}}$}
            (5,1.15) node[right]{${+ \text{ qudits}}$};
        \draw 
            (0.6,1.5) to (0.6,1) to (1,0.6) to (4,0.6) to (4.4,1) 
            to (4.4,2) to (4,2.4) to (1,2.4) to (0.6,2) to (0.6,1.5)
            ;
        \draw 
            (2, 1) to (2, 1.2) to (3,1.2) to (3, 0.9) to (2, 0.9) to (2, 1) 
            (2, 2) to (2, 1.8) to (3,1.8) to (3, 2.1) to (2, 2.1) to (2, 2) 
            (2.5, 1.2) to [/tikz/circuitikz/bipoles/length=10pt, josephson] (2.5, 1.8);
        \ctikzset{european, bipoles/inductors/core distance=4pt}
        \draw 
            (2, 2.4) to (2,3.5) to (1,3.5) to [/tikz/circuitikz/bipoles/length=25pt, lowpass2] (1, 5) to [/tikz/circuitikz/bipoles/length=20pt, L] (1,6.5)
            to [/tikz/circuitikz/bipoles/length=20pt, twoportsplit] (1,8) to (1,8.5);
        \draw 
            (3, 2.4) to (3,2.7)
            (3,4) node[circulator, xscale=-1] (circulator1) {}
            (circulator1.down) to (3,2.7)
            (2.25, 3.8) to (2.1,3.8) to (2.1,4.2) to (2.25, 4.2)
            (3,5.75) node[circulator, xscale=-1] (circulator2) {}
            (circulator2.down) to (circulator1.up)
            (circulator2.up) to [/tikz/circuitikz/bipoles/length=20pt, L] (3,8) to (3,8.5)
            (circulator2.left) to (5, 5.75) to (5, 8.5);
    \end{scope}
    \begin{scope}[shift={(0.5, 0.5)}]
        \ctikzset{american, bipoles/inductors/core distance=4pt}
        \draw
            (1, 8.5) to (1, 11.5) to [/tikz/circuitikz/bipoles/length=20pt, twoportsplit] (1, 13) to (1, 13.5);
        \draw
            (3, 8.5) to (3, 13.5);
        \draw [line width=0.8mm] 
            (4.5,10) to (4.5,9.7) to (7.2,9.7) to (7.2, 11.3) to (4.5,11.3) to (4.5,10);
        \draw 
            (5, 8.5) to (5, 10.5) to [/tikz/circuitikz/bipoles/length=30pt, amp, invert] (7,10.5) to (8,10.5)
            (5.85,11.3) node[above]{${\text{JPA}}$};
        \draw
            (7.5,10) to (7.5,9.7) to (9,9.7) to (9, 12.5) to (7.5,12.5) to (7.5,10)
            (8,10.5) to [/tikz/circuitikz/bipoles/length=25pt, C] (9,10.5) to (10,10.5) to (10, 15.5) to [/tikz/circuitikz/bipoles/length=20pt, twoportsplit] (10, 17.5) to (10,20) node[above]{${I_{\text{pump}}}$}
            (8,10.5) to [/tikz/circuitikz/bipoles/length=25pt, L](8,12.5) to (8,20) node[above]{${I_{\text{bias}}}$};
    \end{scope}
    \begin{scope}[shift={(0.5, 0.5)}]
        \draw
            (1, 13.5) to (1, 15.5) to [/tikz/circuitikz/bipoles/length=20pt, twoportsplit] (1, 17) to (1, 17.5) to [/tikz/circuitikz/bipoles/length=25pt, C] (1, 18.5);
        \draw
            (3, 13.5) to [/tikz/circuitikz/bipoles/length=25pt, lowpass2] (3, 15.5)
            to [/tikz/circuitikz/bipoles/length=30pt, amp](3,17.5) to [/tikz/circuitikz/bipoles/length=25pt, C] (3, 18.5)
            (3.5, 16) node[right]{${\text{HEMT}}$}
            (3.3,16.5) to (6, 16.5) to (6, 20) node[above]{${V_{\text{bias}}}$};
    \end{scope}
    
    \begin{scope}[shift={(-8, 4)}]
        \draw 
            (-1,10) to [/tikz/circuitikz/bipoles/length=20pt, amp] (1,10)
            (1.5,10) node[right]{${\text{amplifier}}$};
        \draw 
            (-1,8.5) to [/tikz/circuitikz/bipoles/length=20pt, twoportsplit] (1,8.5)
            (1.5,8.5) node[right]{${\text{attenuator}}$};
        \draw 
            (-1,7) to [/tikz/circuitikz/bipoles/length=20pt, L] (1,7)
            (1.5,7) node[right]{${\text{choke}}$};
        \ctikzset{european, bipoles/inductors/core distance=4pt}
        \draw 
            (0,5.5) ellipse (0.5 and 0.5);
        \draw[-{Latex[length=1.5mm]}] 
            (0.25,5.5) arc (0:315:0.25 and 0.25);
        \draw 
            (0,6) to (0,6.25)
            (-0.5,5.5) to (-1, 5.5)
            (0.5,5.5) to (1, 5.5);;
        \draw (1.5,5.5) node[right]{${\text{circulator}}$};
        \draw 
            (-1,4) to [/tikz/circuitikz/bipoles/length=20pt, C] (1,4)
            (1.5,4) node[right]{${\text{DC block}}$};
        \draw 
            (-1,2.5) to [/tikz/circuitikz/bipoles/length=20pt, L] (1,2.5)
            (1.5,2.5) node[right]{${\text{eccosorb filter}}$};
        \draw 
            (-1,1) to [/tikz/circuitikz/bipoles/length=20pt, lowpass2] (1,1)
            (1.5,1) node[right]{${\text{low-pass filter}}$};
        \draw 
            (0,-0.5) node[mixer,scale=0.6] (mix_annotated) {}
            (mix_annotated.e) to (1,-0.5)
            (mix_annotated.w) to (-1,-0.5)
            (mix_annotated.n) to (0,0.25)
            (mix_annotated.s) to (0,-1.25)
            (1.5,-0.5) node[right]{${\text{mixer}}$};
        \draw 
            (-0.5, -2) to (-0.5, -1.6) to (0.5,-1.6) to (0.5, -2.4) to (-0.5,-2.4) to (-0.5,-2);
        \draw
            (-1,-2) to (0.1, -2)
            (0.1, -2) to (0.5,-1.75) to (1,-1.75)
            (0.1, -2) to (0.5,-2.25) to (1,-2.25)
            (1.5,-2) node[right]{${\text{splitter (combiner)}}$};
    \end{scope}
    \end{circuitikz}
    \caption{A simplified experimental setup}
    \label{fig:experiment_setup}
\end{figure}
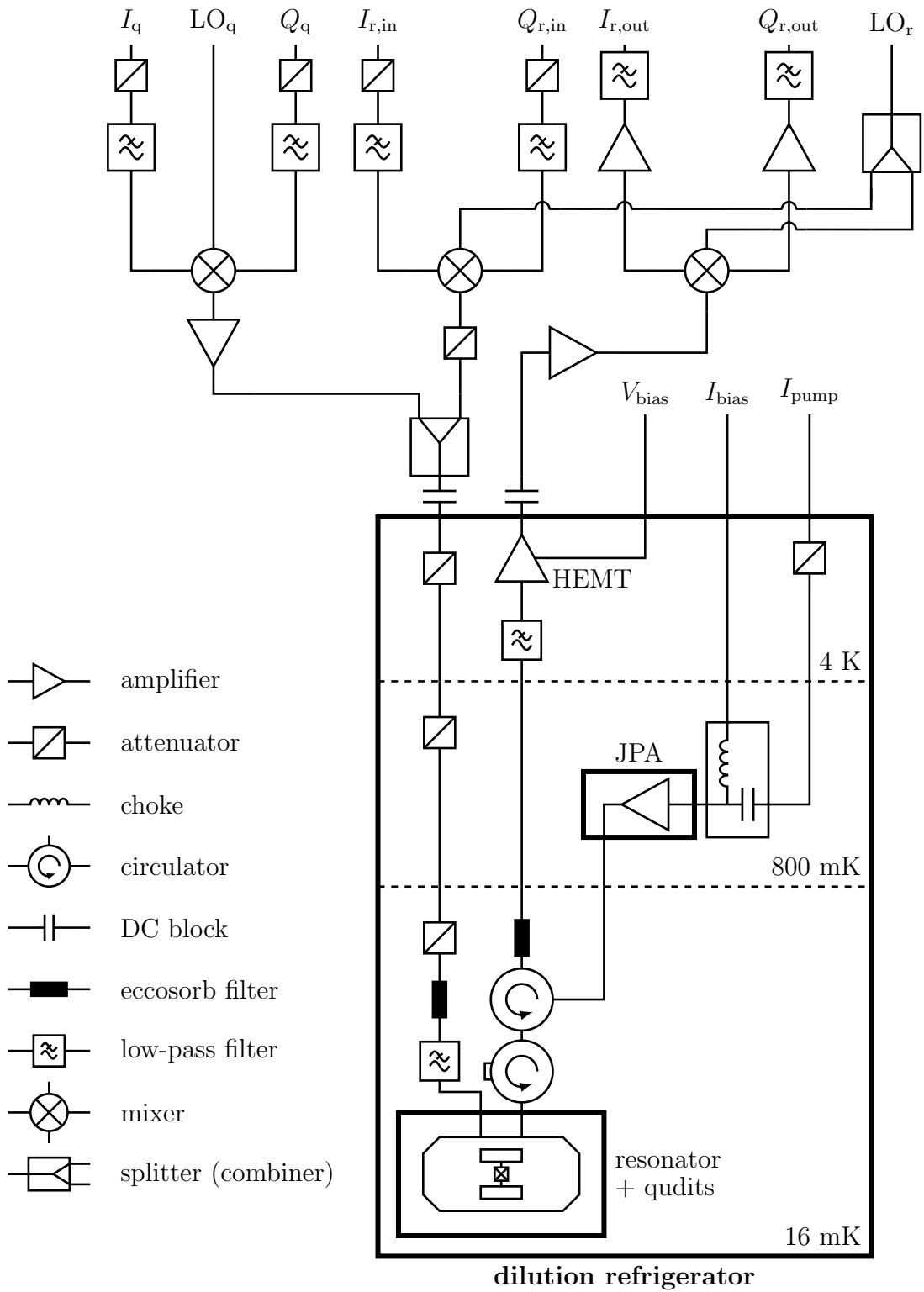

\section{Overview of the Thesis}
The thesis is organized as follows. Chapter 2 offers a concise overview of field quantization, highlighting various properties of a quantized field. These properties play a crucial role in the interaction with the quantum processor. In Chapter 3, we explore the interaction between qudits and the electromagnetic field comprehensively. We introduce quantum gates and delve into the concept of dispersive coupling, which facilitates the readout of qudit states. Chapter 4 focuses on the realization and modeling of superconducting qudits. We examine the multi-level nature of a transmon qudit and provide an analysis of a general Josephson network, also referred to as a multimon. Moving into the realm of open quantum systems, Chapter 5 showcases practical methods for controlling and detecting quantum systems. We introduce the quantum Langevin equations and master equations, enabling us to quantitatively evaluate the effectiveness of control and readout. Lastly, the thesis concludes with a discussion of the stochastic master equation. This equation serves as a valuable tool for modeling the probabilistic nature of dispersive readout, offering insights into the quantification of measurement outcomes.

%% file: include_chapters/chapter_quantum_light.tex
\chapter{The Quantum Description of Light}

\newcounter{qcounter}

\section{The Classical Description of Light-Matter Interaction}

\subsection{Equations of Motion}
In classical electrodynamics, the four \textit{Maxwell's equations} (with the additional boundary conditions) provide a full description of the dynamics of the electric and magnetic fields, $\mathbf{E}(\mathbf{r},t)$ and $\mathbf{B}(\mathbf{r},t)$, given that the sources $\rho(\mathbf{r},t)$ and $\mathbf{J}(\mathbf{r},t)$ are also specified:
\begin{align}\label{eq:maxwell_1}
    &\nabla \cdot     
            \mathbf{E}(\mathbf{r},t) 
        = \frac{\rho(\mathbf{r},t)}{\epsilon_0},
\\[2mm] \label{eq:maxwell_2}
    &\nabla \cdot 
            \mathbf{B}(\mathbf{r},t) 
        = 0,
\\[2mm] \label{eq:maxwell_3}
    &\nabla \times 
            \mathbf{E}(\mathbf{r},t) 
        = - \frac{\partial}{\partial t}     
            \mathbf{B}(\mathbf{r},t),
\\[2mm] \label{eq:maxwell_4}
    &\nabla \times     
            \mathbf{B}(\mathbf{r},t) 
        = \frac{1}{\epsilon_0 c^2}
                \mathbf{J}(\mathbf{r},t) 
            + \frac{1}{c^2} \frac{\partial }{\partial t} 
                \mathbf{E}(\mathbf{r},t).
\end{align}
From Eq.(\ref{eq:maxwell_1}) and (\ref{eq:maxwell_4}), we can derive the continuity equation
\begin{equation}
    \frac{\partial}{\partial t} \rho(\mathbf{r},t)
    + \nabla \cdot \mathbf{J}(\mathbf{r},t) = 0,
\end{equation}
which is a manifestation of the conservation of charges.

The dynamics can also be formulated in terms of the vector potential $\mathbf{A}(\mathbf{r},t)$ and scalar potential $U(\mathbf{r},t)$ defined via
\begin{align}
    \mathbf{B}(\mathbf{r},t) 
    &= \nabla \times \mathbf{A}(\mathbf{r},t),
\nonumber \\
    \mathbf{E}(\mathbf{r},t) 
    &= - \frac{\partial}{\partial t} \mathbf{A}(\mathbf{r},t) 
        - \nabla U(\mathbf{r},t).
\end{align}
Consequently, Maxwell's equations can be rewritten compactly as
\begin{align} \label{eq:scalar_potential_pde}
    \Delta U(\mathbf{r},t)  
    &= - \frac{\rho(\mathbf{r},t)}{\epsilon_0}
        - \nabla \cdot \frac{\partial}{\partial t}
        \mathbf{A}(\mathbf{r},t),
\\ \label{eq:vector_potential_pde}
    \square \mathbf{A}(\mathbf{r},t)  
    &= \frac{1}{\epsilon_0 c^2} 
            \mathbf{J}(\mathbf{r},t)  
        - \nabla
            \left[ 
                \nabla \cdot \mathbf{A}(\mathbf{r},t)  
                + \frac{1}{c^2} \frac{\partial}{\partial t} U(\mathbf{r},t) 
            \right],
\end{align}
where $\square = (1/c^2) \partial^2_t - \Delta$ is the d'Alembert operator. However, it turns out the potentials are not uniquely defined. Consider any scalar function $F(\mathbf{r},t)$ (called the gauge function), $\mathbf{E}$ and $\mathbf{B}$ are invariant under the following transformation 
\begin{equation}
    \mathbf{A}'(\mathbf{r},t) 
    = \mathbf{A}(\mathbf{r},t) 
        + \nabla F(\mathbf{r},t),
\end{equation}
\begin{equation}
    U'(\mathbf{r},t)
    = U(\mathbf{r},t)
        - \frac{\partial}{\partial t} F(\mathbf{r},t).
\end{equation}
This means that the vector and scalar potentials are unique up to some choice of the gauge, which leads to the question of whether vector and scalar potentials are the fundamental description of the field or merely some mathematical tools for calculating the electric and magnetic fields. Nevertheless, thanks to the discovery of the Aharanov-Bohm effect, we now know that the potentials are more fundamental than the fields.

So far, we have ignored the dynamics of the charged particles (i.e., the position $\mathbf{r}_{\alpha}(t)$ and velocity $\mathbf{v}_{\alpha}(t) = \dot{\mathbf{r}}_{\alpha}(t)$ for particle $\alpha$), which obviously affect the fields via Maxwell's equations\footnote{Mathematically, we can model the charge and current distribution by 
\begin{equation}
    \rho(\mathbf{r},t)
    = \sum_{\alpha} 
        q_{\alpha} 
        \delta^{(3)}
        [\mathbf{r} - \mathbf{r}_{\alpha}(t)],
\end{equation}
\begin{equation}
    \mathbf{J}(\mathbf{r},t)
    = \sum_{\alpha} 
        q_{\alpha}\mathbf{v}_{\alpha}(t) 
        \delta^{(3)}
        [\mathbf{r} - \mathbf{r}_{\alpha}(t)].
\end{equation}
It's then clear that $\rho$ and $\mathbf{J}$ are functions of the particle's degrees of freedom only. This point will be important when we separate the longitudinal and transverse components of the field.}. However, the interaction is not one-way and the ``backaction'' of the field on the sources is governed by the Newton-Lorentz equations
\begin{align}
    m_{\alpha} 
        \frac{\mathrm{d}}{\mathrm{d} t} 
        \mathbf{v}_{\alpha}(t)
    = q_{\alpha} 
        \Big[
            \mathbf{E}(\mathbf{r}_{\alpha}(t),t)
            + \mathbf{v}_{\alpha}(t) 
                \times 
                \mathbf{B}(\mathbf{r}_{\alpha}(t),t)
        \Big]
\end{align}
for each particle $\alpha$ with mass $m_{\alpha}$ and charge $q_{\alpha}$\footnote{The parameter $\mathbf{r}$ in Maxwell's equations is not a dynamical variable like $\mathbf{r}_{\alpha}(t)$. In the Newton-Lorentz equations, $\mathbf{E}$ and $\mathbf{B}$ are evaluated at $\mathbf{r}=\mathbf{r}_{\alpha}(t)$ to specify a local force acting on the particles. We should think of the fields as a collection of infinitely many dynamical variables indexed by a continuous label $\mathbf{r}$, just like $\mathbf{r}_{\alpha}$ is indexed by $\alpha$.}. Maxwell's equations combined with the Newton-Lorentz equations give the classical theory of light-matter interaction. However, not all the variables in the current description are independent, and the goal of the next subsection is to perform a cleverly chosen reduction of Maxwell's equations to filter out the truly independent degrees of freedom.

\subsection{Transverse and Longitudinal Fields}
\begin{table}
    \scriptsize{
    \centering
    \setlength\tabcolsep{8pt}
    \renewcommand{\arraystretch}{0.8}
    \begin{tabular}{c|c c}
    \toprule
        & Real space & $k$-space
    \\  \midrule \midrule
    \\
        Observables
        &$\displaystyle \mathbf{E}, \mathbf{B}, U, \mathbf{A}, \rho, \mathbf{J}$ 
        & $\displaystyle \mathscrbf{E}, \mathscrbf{B}, \mathscr{U}, \mathscrbf{A}, \varrho, \mathscrbf{J}$ 
    \\ \\ \hline \\
        $\substack{
                \displaystyle \text{Conjugate} 
            \\[2mm] 
                \displaystyle \text{symmetry}
        }$
        & e.g., $\mathbf{E}_{\perp}(\mathbf{r},t)$ is real
        & e.g., $\mathscrbf{E}_{\perp}^*(-\mathbf{k},t) = \mathscrbf{E}_{\perp}(\mathbf{k},t)$
    \\ \\ \hline \\
        $\substack{
                \displaystyle \text{Maxwell's} 
            \\[2mm] 
                \displaystyle \text{equations}
            \\[2mm] 
                \displaystyle \text{(fields)}
        }$
        &$\substack{
                \displaystyle 
                \nabla \cdot     
                \mathbf{E}(\mathbf{r},t) 
                = \frac{\rho(\mathbf{r},t)}{\epsilon_0}
            \\[4mm] 
                \displaystyle 
                \nabla \cdot 
                    \mathbf{B}(\mathbf{r},t) 
                = 0
            \\[4mm] 
                \displaystyle 
                \nabla \times 
                    \mathbf{E}(\mathbf{r},t) 
                = - \frac{\partial}{\partial t}     
                    \mathbf{B}(\mathbf{r},t)
            \\[4mm] 
                \displaystyle 
                \nabla \times     
                    \mathbf{B}(\mathbf{r},t) 
                = \frac{1}{\epsilon_0 c^2}
                        \mathbf{J}(\mathbf{r},t) 
                    + \frac{1}{c^2} \frac{\partial }{\partial t} 
                        \mathbf{E}(\mathbf{r},t)
            }$
        &$\substack{
                \displaystyle 
                \ci\mathbf{k} 
                    \cdot 
                    \mathscrbf{E}(\mathbf{k},t) 
                = \frac{\varrho (\mathbf{k},t) }{\epsilon_0}
            \\[4mm] 
                \displaystyle 
                \ci\mathbf{k} 
                    \cdot 
                    \mathscrbf{B}(\mathbf{k},t) 
                = 0 
            \\[4mm] 
                \displaystyle 
                \ci\mathbf{k} 
                    \times 
                    \mathscrbf{E}(\mathbf{k},t) 
                = - \frac{\partial}{\partial t} 
                        \mathscrbf{B} (\mathbf{k},t)
            \\[4mm] 
                \displaystyle 
                \ci\mathbf{k} 
                    \times 
                    \mathscrbf{B}(\mathbf{k},t) 
                = \frac{1}{\epsilon_0 c^2} \mathscrbf{J}(\mathbf{k},t)
                    + \frac{1}{c^2}
                        \frac{\partial}{\partial t}
                        \mathscrbf{E}(\mathbf{k},t)
        }$
    \\ \\ \hline \\
        $\substack{
                \displaystyle \text{Continuity} 
            \\[2mm] 
                \displaystyle \text{equation}
        }$
        &$\displaystyle \frac{\partial}{\partial t} \rho(\mathbf{r},t)
            + \nabla \cdot \mathbf{J}(\mathbf{r},t) = 0$
        & $\displaystyle \frac{\partial}{\partial t} \varrho(\mathbf{k},t)
            + \ci \mathbf{k} \cdot \mathscrbf{J}(\mathbf{k},t) = 0$
    \\ \\ \hline \\
        $\substack{
                \displaystyle \text{Maxwell's} 
            \\[2mm] 
                \displaystyle \text{equations}
            \\[2mm] 
                \displaystyle \text{(potentials)}
        }$
        &$\substack{
                \displaystyle 
                    \mathbf{B}(\mathbf{r},t) 
                    = \nabla \times \mathbf{A}(\mathbf{r},t)
            \\[4mm] 
                \displaystyle 
                \mathbf{E}(\mathbf{r},t) 
                = - \frac{\partial}{\partial t} 
                        \mathbf{A}(\mathbf{r},t) 
                    - \nabla U(\mathbf{r},t)
            \\[4mm] 
                \displaystyle 
                \Delta U(\mathbf{r},t)  
                = - \frac{\rho(\mathbf{r},t)}{\epsilon_0}
                    - \nabla \cdot \frac{\partial}{\partial t}
                    \mathbf{A}(\mathbf{r},t)
            \\[4mm] 
                \displaystyle 
                \square \mathbf{A}(\mathbf{r},t) 
                        \ \ \ \ \ \ \ \ \ \ \ \ \ \ \ \ \ 
                        \ \ \ \ \ \ \ \ \ \ \ \ \ \ \ \ \
                        \ \ \ \ \ \ \ \ \ \ \ \ \ \
            \\[1.5mm]  
                \displaystyle
                = \frac{1}{\epsilon_0 c^2} 
                \mathbf{J}(\mathbf{r},t)  
                - \nabla
                    \left[ 
                        \nabla \cdot \mathbf{A}(\mathbf{r},t)  
                        + \frac{1}{c^2} \frac{\partial}{\partial t} U(\mathbf{r},t) 
                    \right]
            }$
        &$\substack{
                \displaystyle 
                \mathscrbf{B}(\mathbf{k},t) 
                = \ci\mathbf{k} 
                    \times \mathscrbf{A}(\mathbf{k},t)
            \\[4mm] 
                \displaystyle 
                \mathscrbf{E}(\mathbf{k},t) 
                = - \frac{\partial}{\partial t} 
                    \mathscrbf{A}(\mathbf{k},t)
                    - \ci\mathbf{k} \mathscr{U}(\mathbf{k},t)
            \\[4mm] 
                \displaystyle 
                k^2 \mathscr{U}(\mathbf{k},t)  
                = \frac{\varrho(\mathbf{k},t)}{\epsilon_0} 
                    + \ci\mathbf{k} 
                        \cdot 
                        \frac{\partial}{\partial t}
                        \mathscrbf{A}(\mathbf{k},t)
            \\[2mm] 
                \displaystyle 
                \frac{1}{c^2} 
                \frac{\partial^2}{\partial t^2} \mathscrbf{A}(\mathbf{k},t)  
                    + k^2 \mathscrbf{A}(\mathbf{k},t)  
                \ \ \ \ \ \ \ \ \ \ \ \ \ \ \ \ \ 
                \ \ \ \ \ \ \ \ \ \ \ \ \
            \\
                \displaystyle = \frac{1}{\epsilon_0 c^2} 
                    \mathscrbf{J}(\mathbf{k},t)  
                - \ci\mathbf{k} 
                    \left[ 
                        \ci\mathbf{k} \cdot \mathscrbf{A}(\mathbf{k},t)  
                        + \frac{1}{c^2} 
                            \frac{\partial}{\partial t}\mathscr{U}(\mathbf{k},t) 
                    \right]
        }$
    \\ \\ \hline \\
        $\substack{
                \displaystyle \text{Gauge} 
            \\[2mm] 
                \displaystyle \text{invariance}
        }$
        & $\substack{
                \displaystyle 
                \mathbf{A}'(\mathbf{r},t) 
                = \mathbf{A}(\mathbf{r},t) 
                    + \nabla F(\mathbf{r},t)
            \\[4mm]
                \displaystyle 
                U'(\mathbf{r},t)
                = U(\mathbf{r},t)
                    - \frac{\partial}{\partial t} F(\mathbf{r},t)
        }$
        & $\substack{
                \displaystyle 
                \mathscrbf{A}'(\mathbf{k},t) 
                = \mathscrbf{A}(\mathbf{k},t) 
                    + \ci\mathbf{k} \mathscr{F}(\mathbf{k},t)
            \\[4mm]
                \displaystyle 
                \mathscr{U}'(\mathbf{k},t)
                = \mathscr{U}(\mathbf{k},t)
                    - \frac{\partial}{\partial t} 
                        \mathscr{F}(\mathbf{k},t)
        }$
    \\ \\ \bottomrule
    \end{tabular}
    }
    \caption{A summary of the equations in classical electrodynamics.}
    \label{table:classical_em_eqns}
\end{table}

It's a well-known fact that electromagnetic waves can travel in the absence of any source or medium. Mathematically, this means Eq.(\ref{eq:maxwell_1})-(\ref{eq:maxwell_4}) or Eq.(\ref{eq:scalar_potential_pde})-(\ref{eq:vector_potential_pde}) admit non-trivial solutions even when the sources are set to zero, which suggests that the electromagnetic fields can be divided into two parts -- (1) fields that are ``attached'' to the sources and (2) fields that propagate in space.

A rigorous derivation of the above conjecture relies on Helmholtz theorem, the fundamental theorem of vector calculus, which states that any vector field $\mathbf{F}(\mathbf{r})$ can be decomposed into a curl-free term $\mathbf{F}_{||}(\mathbf{r})$ and a divergence-free term $\mathbf{F}_{\perp}(\mathbf{r})$ through the formula
\begin{equation}
    \mathbf{F}(\mathbf{r}) 
    = - \underbrace{\nabla \left[ \frac{1}{4 \pi}\nabla \cdot 
        \int_D \mathrm{d}^3 r' \, \frac{\mathbf{F}(\mathbf{r}')}{|\mathbf{r}-\mathbf{r}'|} \right]}_{\text{curl-free } \mathbf{F}_{||}}
        + \underbrace{\nabla \times \left[\frac{1}{4 \pi}\nabla \times
        \int_D \mathrm{d}^3 r' \, \frac{\mathbf{F}(\mathbf{r}')}{|\mathbf{r}-\mathbf{r}'|} \right]}_{\text{divergence-free } \mathbf{F}_{\perp}}.
\end{equation}
In the spatial Fourier domain (also known as $k$-space, where the variable is the wavevector $\mathbf{k}$), this statement is equivalent to the fact that a field can be written as a sum of the transverse and longitudinal components\footnote{We use a script letter to denote the three-dimensional spatial Fourier transform of a vector field
\begin{equation}
        \mathscrbf{F}(\mathbf{k},t) 
        = \frac{1}{(2 \pi)^{3/2}} 
            \int \mathrm{d}^3 r \, 
            \mathbf{F}(\mathbf{r},t) 
            e^{-\ci\mathbf{k} \cdot \mathbf{r}},
    \end{equation}
    \begin{equation}
        \mathbf{F}(\mathbf{r},t) 
        = \frac{1}{(2 \pi)^{3/2}} 
            \int \mathrm{d}^3 k \, 
            \mathscrbf{F}(\mathbf{k},t) 
            e^{\ci\mathbf{k} \cdot \mathbf{r}}.
    \end{equation}
For a list of useful identities related to the Fourier transforms, see Appendix \ref{appendix:fourier_properties}.}
\begin{equation}
    \mathscrbf{F} (\mathbf{k})
    = \mathscrbf{F}_{||}(\mathbf{k}) 
        +  \mathscrbf{F}_{\perp}(\mathbf{k}),
    \end{equation}
where $\mathbf{k} \times \mathscrbf{F}_{||}(\mathbf{k}) = \mathbf{0}$ and $\mathbf{k} \cdot \mathscrbf{F}_{\perp}(\mathbf{k})= 0$ for all $\mathbf{k}$. In other words, $\mathscrbf{F}_{||}$, the Fourier transform of $\mathbf{F}_{||}$, is always parallel to $\mathbf{k}$ while $\mathscrbf{F}_{\perp}$, the Fourier transform of $\mathbf{F}_{\perp}$, is perpendicular to $\mathbf{k}$. 

Table \ref{table:classical_em_eqns} summarizes the important equations in electrodynamics and their counterparts in $k$-space. By going into $k$-space, spatial differentiation reduces into multiplication with $\ci \mathbf{k}$, making the analysis of the longitudinal and transverse fields much easier than that in real space. 
As a result of such simplification, we can show (see Exercise \ref{exercise:E_perp}) that the electric field can be separated into a curl-free/longitudinal ``Coulomb field''\footnote{This is not exactly the Coulomb field in electrostatics since $\rho$ is time-dependent. But the expression is otherwise the same so we will call it a Coulomb field.} \cite{cohen1989photons}
\begin{align}\label{eq:long_E_field}
        \mathbf{E}_{||}(\mathbf{r},t)
        = \frac{1}{4 \pi \epsilon_0} 
            \int \mathrm{d}^3 r \, 
                \rho(\mathbf{r}', t) 
                \frac{\mathbf{r} - \mathbf{r}'}{|\mathbf{r}-\mathbf{r}'|^3}
        = \frac{1}{4 \pi \epsilon_0} 
            \sum_{\alpha} 
                q_{\alpha}
                \frac{\mathbf{r} - \mathbf{r}_{\alpha}(t)}{|\mathbf{r}-\mathbf{r}_{\alpha}(t)|^3},
    \end{align}
which only depends on the particle variables $\mathbf{r}_{\alpha}$, and a divergence-free/transverse field $\mathbf{E}_{\perp}(\mathbf{r},t)$ whose initial conditions must be separately specified to solve Maxwell's equations. That is, $\mathbf{E}_{||}$ should not be treated as separate dynamical variables. Moreover, the magnetic field is always transverse due to Eq.(\ref{eq:maxwell_2}) (i.e., $\mathbf{k} \cdot \mathscrbf{B} = 0$); hence, besides $\mathbf{r}_{\alpha}$ and $\mathbf{v}_{\alpha}$, we should look for independent degrees of freedom only from the transverse electromagnetic field, i.e., the field that detaches from the source and radiates.

\subsection{Normal-Mode Expansion}
To extract the true degrees of freedom of the radiating field, we first define the so-called \textbf{normal modes} \cite{cohen1989photons}
\begin{equation}
    \label{eq:normal_mode_a_k}
    \mathbf{a}(\mathbf{k},t) 
    = -\frac{\ci}{2\mathcal{N}(k)}
    \left[
        \mathscrbf{E}_{\perp}(\mathbf{k},t)
        - c \, \hat{\mathbf{e}}_{\mathbf{k}} \times \mathscrbf{B}(\mathbf{k},t)
    \right],
\end{equation}
\noindent where $\mathcal{N}(k)$ is some normalization factor to be chosen\footnote{We denote the magnitude of $\mathbf{k}$ by $k=\abs{\mathbf{k}}$ and the unit vector along $\mathbf{k}$ by $\hat{\mathbf{e}}_{\mathbf{k}} = \mathbf{k}/k$.}. Then, it can be shown, by using Maxwell's equations in $k$-space, that
\begin{equation} \label{eq:normal_mode_diff_eqn}
    \frac{\partial}{\partial t} \mathbf{a}(\mathbf{k},t)
    = - \ci c k \, \mathbf{a}(\mathbf{k},t)
        + \frac{\ci}{2 \mathcal{N}(k) \epsilon_0}   
            \mathscrbf{J}_{\perp}(\mathbf{k},t).
\end{equation}
We call $\mathbf{a}(\mathbf{k},t)$ the normal modes since the solution to Eq.(\ref{eq:normal_mode_diff_eqn}) in the absence of any source $\mathscrbf{J}_{\perp}$ traces a circle in the complex plane with angular frequency $\omega(k) = c k$; that is, 
\begin{equation}\label{eq:time_evo_a_k}
    \mathbf{a}(\mathbf{k},t) 
    = \mathbf{a}(\mathbf{k}) e^{-\ci\omega(k)t},
\end{equation}
which reminds us of the time evolution of a mechanical oscillator plotted in the phase space (i.e., the position-momentum space). To be more precise, Eq.(\ref{eq:time_evo_a_k}) actually represents two oscillators since $\mathbf{a}$ is a vector orthogonal to $\mathbf{k}$; in other words, for each $\mathbf{k}$, we have 
\begin{equation}
    \mathbf{a}(\mathbf{k},t) 
    = \sum_{\lambda = 1}^{2} \hat{\mathbf{e}}_{\lambda}(\mathbf{k}) a_{\lambda}(\mathbf{k}) e^{-\ci\omega(k)t},
\end{equation}
where $\hat{\mathbf{e}}_{\lambda}(\mathbf{k})$ with $\lambda = 1,2$ can be any two unit basis vectors orthogonal to $\mathbf{k}$. The extra label $\lambda$ used to index the two directions is called the polarization of the field.

Consequently, we can rewrite the total electromagnetic energy in terms of the normal modes,
\begin{align} \label{eq:em_total_energy}
    \mathcal{H}_{\text{EM}}(t) 
    &= \frac{\epsilon_0}{2} 
        \int \mathrm{d}^3 r \, \Big[\abs{\mathbf{E}(\mathbf{r},t)}^2
        + c^2 \abs{\mathbf{B}(\mathbf{r},t)}^2
        \Big]
\nonumber \\
    &= \frac{\epsilon_0}{2} 
                \int \mathrm{d}^3 k \,
                    \abs{\mathscrbf{E}_{||}(\mathbf{k},t)}^2
        + \frac{\epsilon_0}{2} 
            \int \mathrm{d}^3 k \,
                \left[
                    \abs{\mathscrbf{E}_{\perp}(\mathbf{k},t)}^2
                    + c^2 \abs{\mathscrbf{B}(\mathbf{k},t)}^2
                \right]
\nonumber \\
    &= \mathcal{H}_{||}(t) + \mathcal{H}_{\perp}(t) ,
\end{align}
where
\begin{gather} \label{eq:em_longitudinal_energy}
    \mathcal{H}_{||}(t)
    = \frac{1}{8 \pi \epsilon_0} 
        \int \mathrm{d}^3 r 
        \int \mathrm{d}^3 r'
            \frac{
                \rho(\mathbf{r},t)
                \rho(\mathbf{r}',t)
            }{|\mathbf{r}-\mathbf{r}'|},
\\ \label{eq:em_transverse_energy}
    \mathcal{H}_{\perp}(t) 
    = \epsilon_0 
        \int 
            \mathrm{d}^3 k \,
            \sum_{\lambda = 1}^{2}
                |\mathcal{N}(k)|^2
                \Big[
                    a_{\lambda}^{*}(\mathbf{k},t)
                    a_{\lambda}(\mathbf{k},t)
                    + 
                    a_{\lambda}(\mathbf{k},t)
                    a_{\lambda}^{*}(\mathbf{k},t)
                \Big] .
\end{gather}
On one hand, $\mathcal{H}_{||}(t)$ belongs to the longitudinal field (i.e., the Coulomb potential) and is a function of the particle's degrees of freedom. For example, when we solve the energy levels of a hydrogen atom, the Coulomb potential can be treated as a part of the atom's Hamiltonian.
On the other hand, $\mathcal{H}_{\perp}(t)$ consists of infinitely many harmonic oscillators summed over the wavevectors $\mathbf{k}$ and polarization $\lambda$. In particular, if we choose\footnote{Despite the appearance of $\hbar$ in $\mathcal{N}$, there is nothing quantum-mechanical at this point. However, we choose $\mathcal{N}$ such that the normal mode $a_{\lambda}(\mathbf{k})$ will eventually become the annihilation operator for the oscillator labeled by $(\mathbf{k}, \lambda)$. As a result of imposing the canonical commutation relations, $aa^*$ in the integrand will be turned into $a^*a$ with an extra $1/2$, matching the Hamiltonian of a quantum harmonic oscillator.} $\mathcal{N}(k) = \sqrt{\hbar \omega(k) /2\epsilon_0}$, the transverse energy becomes
\begin{equation}
    \mathcal{H}_{\perp}(t) 
    = \int \mathrm{d}^3 k \,
        \sum_{\lambda = 1}^{2}
            \frac{1}{2} 
            \hbar \omega(k) 
            \Big[
                a_{\lambda}^{*}(\mathbf{k},t)
                a_{\lambda}(\mathbf{k},t)
                + 
                a_{\lambda}(\mathbf{k},t)
                a_{\lambda}^{*}(\mathbf{k},t)
            \Big].
\end{equation}

Moreover, the transverse vector potential and the electric field can be treated as the ``position'' and ``momentum'' of each mode since they can be written as the real and imaginary parts of the normal mode, respectively,
\begin{align} \label{eq:vector_potential_polarized_k_space}
    \mathscr{A}_{\perp, \lambda}(\mathbf{k},t) 
    &= \sqrt{\frac{\hbar}{2\epsilon_0 \omega(k)}}
        \, 
        \Big[
            a_{\lambda}(\mathbf{k},t) 
            + a_{\lambda}^{*}(-\mathbf{k},t) 
        \Big],
\\ \label{eq:electric_field_polarized_k_space}
    \mathscr{E}_{\perp,\lambda}(\mathbf{k},t) 
    &= \ci 
        \sqrt{\frac{\hbar \omega(k)}{2\epsilon_0}}
        \,
        \Big[
            a_{\lambda}(\mathbf{k},t) 
            - a_{\lambda}^{*}(-\mathbf{k},t) 
        \Big].
\end{align}
Taking the inverse Fourier transform of Eq.(\ref{eq:vector_potential_polarized_k_space}) and (\ref{eq:electric_field_polarized_k_space}) and summing over the two polarization for each $\mathbf{k}$, one obtains
\begin{align}
    \mathbf{A}_{\perp}(\mathbf{r},t) 
    &= \int 
        \mathrm{d}^3k \,
        \sum_{\lambda = 1}^{2}
            \frac{\mathscr{E}_0(k)}{\omega(k)}
            \, 
            \hat{\mathbf{e}}_{\lambda}(\mathbf{k})
            \Big[
                a_{\lambda}(\mathbf{k},t) 
                    e^{\ci\mathbf{k} \cdot \mathbf{r}}
                + a^*_{\lambda}(\mathbf{k},t)         
                    e^{-\ci\mathbf{k} \cdot \mathbf{r}}
            \Big],
\nonumber \\
    \mathbf{E}_{\perp}(\mathbf{r},t) 
    &= \int 
        \mathrm{d}^3k \,
        \sum_{\lambda = 1}^{2}
            \ci
            \mathscr{E}_0(k) 
            \hat{\mathbf{e}}_{\lambda} (\mathbf{k})
            \Big[
                a_{\lambda}(\mathbf{k},t) 
                    e^{\ci\mathbf{k} \cdot \mathbf{r}}
                - a^*_{\lambda}(\mathbf{k},t)     
                    e^{- \ci\mathbf{k} \cdot \mathbf{r}}
            \Big],
\end{align}
where $\mathscr{E}_0(k) = \sqrt{\hbar\omega(k) / 2\epsilon_0 (2\pi)^3}$ is the zero-point fluctuation of the electric field (for a mode with wavenumber $k$). The significance of $\mathscr{E}_0(k)$ will be explored later.

At this point, we can heuristically argue that the transverse electromagnetic fields are equivalent to many \textit{independent} harmonic oscillators indexed by $\mathbf{k}$ and $\lambda$ and we can simply apply the theory of the quantum harmonic oscillator (QHO) to each of them.

\begin{exercise}\label{exercise:E_perp} Use Gauss's law to show that the transverse component of the electric field is
\begin{equation}\label{eq:longitudinal_electric}
    \mathscrbf{E}_{||} (\mathbf{k},t)
        = \hat{\mathbf{e}}_{\mathbf{k}} 
            \Big[
                \hat{\mathbf{e}}_{\mathbf{k}} \cdot \mathscrbf{E}(\mathbf{k},t)
            \Big]
        = -\frac{\ci\mathbf{k}}{k^2}\frac{\varrho(\mathbf{k},t)}{\epsilon_0}
\end{equation}
and thus its inverse Fourier transform is given by Eq.(\ref{eq:long_E_field}). In addition, use Eq.(\ref{eq:longitudinal_electric}) to derive the longitudinal energy $\mathcal{H}_{||}$, i.e., Eq.(\ref{eq:em_longitudinal_energy}).
\end{exercise}

\begin{exercise}\label{exercise:normal_modes} Show that 
\begin{align} \label{eq:transverse_electric_normal_mode}
    \mathscr{E}_{\perp}(\mathbf{k},t) 
    &= \ci\mathcal{N}(k) 
        \big[
            \mathbf{a}(\mathbf{k},t) 
            - \mathbf{a}^{*}(-\mathbf{k},t) 
        \big],
\\ \label{eq:magnetic_normal_mode}
    \mathscr{B}(\mathbf{k},t) 
    &= \frac{\ci\mathcal{N}(k)}{c}
        \left[
            \hat{\mathbf{e}}_{\mathbf{k}} \times 
                \mathbf{a}(\mathbf{k},t)
            + \hat{\mathbf{e}}_{\mathbf{k}} \times
                \mathbf{a}^{*}(-\mathbf{k},t) 
        \right].
\end{align}
(Hint: $\mathscr{E}_{\perp}^*(-\mathbf{k},t) = \mathscr{E}_{\perp}(\mathbf{k},t)$ and $\mathscr{B}^*(-\mathbf{k},t) = \mathscr{B}(\mathbf{k},t)$.) Then, use Eq.(\ref{eq:transverse_electric_normal_mode}) and (\ref{eq:magnetic_normal_mode}) to derive the transverse energy, i.e., Eq.(\ref{eq:em_transverse_energy}).
\end{exercise}

\subsection{Lagrangian and Hamiltonian Formulations}
Formally, the process of quantization should start with the definition of a Lagrangian, which, in the case of classical electrodynamics, is given by
\begin{align}
    \mathcal{L}
    &= \sum_{\alpha} 
            \frac{1}{2} m_{\alpha}
            |\dot{\mathbf{r}}_{\alpha}|^2
        + \int 
            \mathrm{d}^3 r 
            \left[
                \frac{\epsilon_0}{2}  
                    \abs{\mathbf{E}(\mathbf{r})}^2
                - \frac{1}{2 \mu_0} 
                    \abs{\mathbf{B}(\mathbf{r})}^2
                + \mathbf{J}(\mathbf{r}) 
                        \cdot \mathbf{A}(\mathbf{r})
                - \rho(\mathbf{r}) U(\mathbf{r})
            \right]
\nonumber \\
    &= \sum_{\alpha} 
            \frac{1}{2} m_{\alpha}
            |\dot{\mathbf{r}}_{\alpha}|^2
        + \frac{\epsilon_0}{2} 
            \int 
                \mathrm{d}^3 r 
                \Big[
                    \abs{\mathbf{E}(\mathbf{r})}^2
                    - c^2 \abs{\mathbf{B}(\mathbf{r})}^2
                \Big]
        + \sum_{\alpha} 
            \Big[
                q_{\alpha} \dot{\mathbf{r}}_{\alpha} 
                    \cdot \mathbf{A}(\mathbf{r}_{\alpha})
                - q_{\alpha} U(\mathbf{r}_{\alpha})
            \Big].
\end{align}
One can readily derive all the equations of motion (i.e., Maxwell's equations and Newton-Lorentz equations) by simply writing down all the Euler-Lagrange equations for this Lagrangian\footnote{The dynamical variables of the Lagrangian are $\mathbf{r}_{\alpha}$, $\mathbf{A}(\mathbf{r})$, and $U(\mathbf{r})$ and their associated velocities. The electric and magnetic fields are derived quantities of $\mathbf{A}(\mathbf{r})$ and $U(\mathbf{r})$: 
\begin{align}
    \mathbf{E}(\mathbf{r})
    &\doteq - \dot{\mathbf{A}}(\mathbf{r}) - \nabla U (\mathbf{r}),
\\
    \mathbf{B}(\mathbf{r})
    &\doteq \nabla \times \mathbf{A}(\mathbf{r}).
\end{align}
They do not participate in the definition of Lagrangian density directly.}. 

With the Legendre transformation \cite{evans_optimal_control, gelfand_cov}, the Hamiltonian is found to be \cite{cohen1989photons}
\begin{equation}\label{eq:full_classical_hamiltonian}
    \mathcal{H} 
    = \sum_{\alpha}
        \frac{1}{2m_{\alpha}}
        \big|
            \mathbf{p}_{\alpha}
            - q_{\alpha} \mathbf{A}(\mathbf{r}_{\alpha})
        \big|^2 
        + \mathcal{H}_{||} 
        + \mathcal{H}_{\perp},
\end{equation}
where $\mathbf{p}_{\alpha}$ are the momenta conjugate to $\mathbf{r}_{\alpha}$, and $\mathcal{H}_{||}$ and $\mathcal{H}_{\perp}$, the longitudinal and transverse energy, are given by Eq.(\ref{eq:em_longitudinal_energy}) and (\ref{eq:em_transverse_energy}). The analysis in the previous subsection gives a heuristic justification of the Coulomb and radiating terms of the fields; nevertheless, one can show rigorously \cite{sakurai1967advancedQM} that the reduced set of degrees of freedom identified in the last subsection indeed is correct based on the Lagrangian and Hamiltonian formalism. 

However, unlike the heuristic argument based on the QHO, here we can say more about the combined system of particles and fields -- namely, we can provide a formal description of the \textit{interaction between particles and fields}. In particular, all classical atom-light interaction can now be understood as a consequence of the coupling between $\mathbf{p}_{\alpha}$ and $\mathbf{A}$ at position $\mathbf{r}_{\alpha}$ in the quadratic term $\left| \mathbf{p}_{\alpha} - q_{\alpha} \mathbf{A}(\mathbf{r}_{\alpha}) \right|^2$. We are now ready to quantize the harmonic oscillators and study Eq.(\ref{eq:full_classical_hamiltonian}) in the quantum-mechanical setting.

\begin{exercise} Define the Lagrangian density of the field to be
    \begin{equation}
        \mathscr{L}
            \Bigsl (
                \mathbf{r}, 
                \mathbf{r}_{\alpha}, 
                \dot{\mathbf{r}}_{\alpha}, 
                \mathbf{A}, 
                \dot{\mathbf{A}}, 
                \partial_{i} \mathbf{A}, 
                U, 
                \partial_{i} U
            \Bigsr ) 
        = \frac{\epsilon_0}{2} 
            \left[  
                \abs{\mathbf{E}}^2
                - c^2 \abs{\mathbf{B}}^2
            \right]
            + \mathbf{J} \cdot \mathbf{A}
                    - \rho U,
    \end{equation}
    where the electric and magnetic fields and the charge and current densities are treated as derived quantities of the generalized coordinates:
    \begin{gather}
        \rho(\mathbf{r},\mathbf{r}_{\alpha})
        = \sum_{\alpha} 
            q_{\alpha} 
            \delta^{(3)}
            (\mathbf{r} - \mathbf{r}_{\alpha}),
    \\
        \mathbf{J}(
            \mathbf{r}, 
            \mathbf{r}_{\alpha}, 
            \dot{\mathbf{r}}_{\alpha} 
            )
        = \sum_{\alpha} 
            q_{\alpha} 
            \dot{\mathbf{r}}_{\alpha} 
            \delta^{(3)}
            (\mathbf{r} - \mathbf{r}_{\alpha}),
    \\
        E_i(\mathbf{A}, 
                \dot{\mathbf{A}}, 
                \partial_{i} \mathbf{A}, 
                U, 
                \partial_{i} U)
        \doteq - \dot{A}_{i} - \partial_i U  ,
        \ \ \text{ for } \ \ i = x,y,z,
    \\
        B_i
            (\mathbf{A}, 
                \dot{\mathbf{A}}, 
                \partial_{i} \mathbf{A}, 
                U, 
                \partial_{i} U)
        \doteq 
            \sum_{jk} \epsilon_{ijk} \partial_j A_k,
        \ \ \text{ for } \ \ i = x,y,z.
    \end{gather}
    (Note that $\mathbf{r}$ is not a dynamical variable but an index for integration.) Show that the full Lagrangian can be rewritten as
    \begin{multline}
        \mathcal{L}
            \Bigsl (
                \mathbf{r}_{\alpha}, 
                \dot{\mathbf{r}}_{\alpha}, \mathbf{A}(\mathbf{r}),
                \dot{\mathbf{A}}(\mathbf{r}),
                U(\mathbf{r})
            \Bigsr )
    \\
        = \sum_{\alpha} 
                \frac{1}{2} 
                m_{\alpha}
                |\dot{\mathbf{r}}_{\alpha}|^2
            + \int 
                \mathrm{d}^3 r \,
                \mathscr{L}
                    \Bigsl (
                        \mathbf{r},
                        \mathbf{r}_{\alpha}, 
                        \dot{\mathbf{r}}_{\alpha}, 
                        \mathbf{A}(\mathbf{r}), 
                        \dot{\mathbf{A}}(\mathbf{r}), 
                        \partial_{i} \mathbf{A}(\mathbf{r}), 
                        U(\mathbf{r}), 
                        \partial_{i} U(\mathbf{r})
                    \Bigsr ),
    \end{multline}
    and the momentum conjugate to $\mathbf{A}(\mathbf{r})$ is the electric field
    \begin{equation}
        \nabla_{\dot{\mathbf{A}}} \mathscr{L} 
        = - \epsilon_0 \mathbf{E}.
    \end{equation}
    Moreover, note that $\mathscr{L}$ is independent of $\dot{U}$, which is another hint that $U$, the longitudinal part of the field, is not an independent degree of freedom.
\end{exercise}

\section{Quantization of the Electromagnetic Field}
\subsection{Born–Von Karman Periodic Boundary Conditions}
To avoid the computation of integrals over $\mathbf{k}$, it is common to constrain the system to a periodic box of size $V=L^3$. By requiring, for instance,
\begin{equation}
    \mathbf{E}_{\perp}(\mathbf{r},t) = \mathbf{E}_{\perp}(\mathbf{r} + \hat{\mathbf{e}}_x L,t),
\end{equation}
the wavevector can only take discretized values along the $\hat{\mathbf{x}}$ direction now:
\begin{equation}
    k_x = \frac{2\pi n_x}{L}
    \ \ \ \ 
    \forall \, n_x \in \mathbf{Z}.
\end{equation}
Applying the same periodic boundary condition to the three orthogonal directions, we can transform any integral over $\mathbf{k}$ to a sum over the lattice
\begin{equation}
    \mathbf{k} 
    = (k_x,k_y,k_z) 
    = \left(
            \frac{2\pi n_x}{L}, 
            \frac{2\pi n_y}{L}, 
            \frac{2\pi n_z}{L}
        \right)
    \ \ \ \ 
    \forall \, 
        n_x, n_y, n_z 
        \in \mathbf{Z}.
\end{equation}
To account for the volume occupied by each lattice point in the reciprocal space, we need a scale factor so that the sum reproduces the integral as $V$ goes to infinity:
\begin{equation}\label{eq:integral_to_sum}
    \left(
        \frac{2\pi}{L}
    \right)^3
    \sum_{\mathbf{k}}
        \left( \cdot \right)
    \ \ \longleftrightarrow \ \ 
    \int 
        \mathrm{d}^3 k 
        \left( \cdot \right)
\end{equation}
Note that we can always let $V$ go to infinity to reproduce the unbounded space, so the artificial boundary conditions won't change the physics.

Apply Eq.(\ref{eq:integral_to_sum}) to the transverse electric field, we obtain
\begin{equation}
    \mathbf{E}_{\perp}(\mathbf{r},t) 
    = \left(
            \frac{2\pi}{L}
        \right)^3
        \sum_{\mathbf{k}}
        \sum_{\lambda = 1}^{2}
            \ci
            \sqrt{\frac{\hbar \omega(k)}{2 \epsilon_0 (2\pi)^3}} 
            \, \hat{\mathbf{e}}_{\lambda}(\mathbf{k})
            \Big[
                a_{\lambda}(\mathbf{k},t) 
                    e^{\ci\mathbf{k} \cdot \mathbf{r}}
                - a^*_{\lambda}(\mathbf{k},t)     
                    e^{- \ci\mathbf{k} \cdot \mathbf{r}}
            \Big].
\end{equation}
In addition, we redefine the discrete normal mode and electric field zero-point fluctuation to absorb the extra constants:
\begin{equation}
    \mathscr{E}_{0, k} 
    = \sqrt{\frac{\hbar \omega_{k}}{2 \epsilon_0 V}} 
    \ \ \text{ and } \ \
    a_{\mathbf{k},\lambda}(t)
    = \left(\frac{2\pi}{L}\right)^{3/2} 
        a_{\lambda}(\mathbf{k},t),
\end{equation}
where, for clarity, the dependence on discrete variables is indicated by subscripts. In summary, 
    \begin{align}
        \mathbf{E}_{\perp}(\mathbf{r},t) 
        &= \sum_{\mathbf{k},\lambda}
                \ci \mathscr{E}_{0, k} \hat{\mathbf{e}}_{\mathbf{k},\lambda}
                \Big[
                    a_{\mathbf{k},\lambda}(t)
                        e^{\ci\mathbf{k} \cdot \mathbf{r}}
                    - a_{\mathbf{k},\lambda}^{*}(t)     
                        e^{- \ci\mathbf{k} \cdot \mathbf{r}}
                \Big],
    \\
        \mathbf{B}(\mathbf{r},t) 
        &= \sum_{\mathbf{k},\lambda}
                \frac{\ci\mathscr{E}_{0, k}}{c}
                \,
                \hat{\mathbf{e}}_{\mathbf{k}} 
                \times
                \hat{\mathbf{e}}_{\mathbf{k},\lambda}
                \Big[
                    a_{\mathbf{k},\lambda}(t)
                        e^{\ci\mathbf{k} \cdot \mathbf{r}}
                    - a_{\mathbf{k},\lambda}^{*}(t)   
                        e^{-\ci\mathbf{k} \cdot \mathbf{r}}
                \Big],
    \\
        \mathbf{A}_{\perp}(\mathbf{r},t) 
        &= \sum_{\mathbf{k},\lambda}
                \frac{\mathscr{E}_{0, k}}{\omega_{k}}
                \,
                \hat{\mathbf{e}}_{\mathbf{k},\lambda}
                \Big[
                    a_{\mathbf{k},\lambda}(t) 
                        e^{\ci\mathbf{k} \cdot \mathbf{r}}
                    + a_{\mathbf{k},\lambda}^{*}(t)
                        e^{-\ci\mathbf{k} \cdot \mathbf{r}}
                \Big],
    \\
        \mathcal{H}_{\perp}(t)
        &= \sum_{\mathbf{k},\lambda}
            \frac{1}{2} \hbar \omega_{k} 
            \Big[
                a_{\mathbf{k},\lambda}^{*}(t) 
                    a_{\mathbf{k},\lambda}(t) 
                + 
                a_{\mathbf{k},\lambda}(t) 
                    a_{\mathbf{k},\lambda}^{*}(t) 
            \Big].
    \end{align} 
\subsection{Field Quantization}
Mathematically speaking, quantizing the electromagnetic field boils down to a mapping of the terminologies from the familiar QHO to each mode of oscillation in the field. In other words, understanding the QHO is the key to appreciating a quantized field.

To begin with, recall that the ladder operator $\hat{a}$ (or $\hat{a}^{\dagger}$) lowers (or raises) the Fock states (the eigenstates of the number operator $\hat{N} = \hat{a}^{\dagger} \hat{a}$) of a QHO. If we assign each excitation of the ladder the meaning of a photon, then $a_{\mathbf{k}, \lambda}(t)$ and $a_{\mathbf{k}, \lambda}^{*}(t)$, once promoted to quantum operators, can be used to create and annihilate a photon in the mode $(\mathbf{k}, \lambda)$, respectively. Hence, we promote each mode amplitude $a_{\mathbf{k}, \lambda}$ to an \textbf{annihilation operator} and replace its complex conjugate with the adjoint of the annihilation operator, i.e., the \textbf{creation operator}:
\begin{gather}
    a_{\mathbf{k},\lambda}
    \ \ \longrightarrow \ \ 
    \hat{a}_{\mathbf{k},\lambda}
    \\
    a^{*}_{\mathbf{k},\lambda}
    \ \ \longrightarrow \ \ 
    \hat{a}_{\mathbf{k},\lambda}^{\dagger}
\end{gather}
Since we have a sum of QHOs, the commutation relations must be modified to reflect all the indices. We define the commutation relations between the annihilation and creation operators to be
\begin{equation}
    \Big[
        \hat{a}_{\mathbf{k},\lambda},
        \hat{a}_{\mathbf{k}',\lambda'}^{\dagger}
    \Big] 
    = \delta_{\mathbf{k} \mathbf{k}'} 
        \delta_{\lambda \lambda'} \hat{1},
\end{equation}
where $\hat{1}$ is the identity operator (usually omitted for simplicity). The commutation relation postulated above implies that operators acting on a single mode follow the usual commutation relation of a QHO whereas operators associated with different modes commute with one another\footnote{
    If we did not use the periodic boundary conditions, the commutator would have been
    \begin{equation}
        \Big[
            \hat{a}_{\lambda}(\mathbf{k}),
            \hat{a}_{\lambda'}^{\dagger}(\mathbf{k}')
        \Big] 
        = \delta^{(3)}(\mathbf{k} - \mathbf{k}') 
            \delta_{\lambda \lambda'} \hat{1}.\\[-5mm]
    \end{equation}
}.

Note that we are in the Schr\"odinger picture where the operators are constant in time. Adding the time evolution $e^{-\ci\omega t}$ to the mode amplitude in the classical treatment is similar to moving to the Heisenberg picture (or the interaction picture). To see this in action, consider the free space first (i.e., no sources and no longitudinal field). 
Since
\begin{equation}\label{eq:commutator_H_a}
    \Big[
            \hat{H}_{\perp}, 
            \hat{a}_{\mathbf{k},\lambda}
        \Big]
    = \sum_{\mathbf{k},\lambda}
        \hbar \omega_{k} 
        \Big[
            \hat{a}_{\mathbf{k},\lambda}^{\dagger} 
                \hat{a}_{\mathbf{k},\lambda}, 
            \hat{a}_{\mathbf{k},\lambda}
        \Big]
    = - \hbar \omega_{k} 
        \hat{a}_{\mathbf{k},\lambda},
\end{equation}
by using the Baker–Campbell–Hausdorff formula, the annihilation operator in the Heisenberg picture is indeed given by 
\begin{align}
    \hat{a}_{\mathbf{k},\lambda}(t)
    &= \hat{U}^{\dagger}(t) 
        \hat{a}_{\mathbf{k},\lambda} 
        \hat{U}(t)
    = \exp(\frac{\ci \hat{H}_{\perp} t}{\hbar})     
        \hat{a}_{\mathbf{k},\lambda}
        \exp(\frac{- \ci \hat{H}_{\perp} t}{\hbar})
        \hat{a}_{\mathbf{k},\lambda}(t)
\nonumber \\[-1mm]
    &= \hat{a}_{\mathbf{k},\lambda}
        + \left(\frac{\ci t}{\hbar} \right)  
            \! \Big[
                \hat{H}_{\perp},
                \hat{a}_{\mathbf{k},\lambda}
            \Big]
        + \left(\frac{\ci t}{\hbar} \right)^2 
            \! 
            \bigg[
                \hat{H}_{\perp}, 
                \Big[
                    \hat{H}_{\perp},
                    \hat{a}_{\mathbf{k},\lambda}
                \Big]
            \bigg]
        + \cdots
\nonumber \\
    &= \hat{a}_{\mathbf{k},\lambda}
        + \left(\frac{\ci t}{\hbar} \right)  
            \! (-\hbar \omega_{k}) 
            \hat{a}_{\mathbf{k},\lambda}
        + \left(\frac{\ci t}{\hbar} \right)^2 
            \! (-\hbar \omega_{k})^2
            \hat{a}_{\mathbf{k},\lambda}
        + \cdots
\nonumber \\ \label{eq:annihilation_op_no_source}
    &= \hat{a}_{\mathbf{k},\lambda} e^{-\ci \omega_{k} t}.
\end{align}
However, Eq.(\ref{eq:annihilation_op_no_source}) is only true in the absence of any source\footnote{If we move to the interaction picture defined using $\hat{U}_0 ={\exp} \big(\!-\ci \hat{H}_{\perp} t /\hbar \, \big)$ and treat the other part of the full Hamiltonian as the interaction, the annihilation operator will only acquire a phase factor, which demonstrates the advantage of using the interaction picture.}. Both in the classical and quantum cases, the time evolution of the mode amplitude will be more complicated if the field is coupled to a source\footnote{The classical field driven by the current source is described by Eq.(\ref{eq:normal_mode_diff_eqn}), and it's straightforward to show, for the quantized field, that
\begin{equation}
    \dot{\hat{a}}_{\mathbf{k},\lambda}(t)
    = -\ci \omega_{k} \hat{a}_{\mathbf{k},\lambda}(t) 
        + \frac{\ci}{\sqrt{2\epsilon_0 \hbar \omega_{\mathbf{k},\lambda}}} 
            \hat{\mathscr{J}}_{\perp,\mathbf{k},\lambda}(t),
\end{equation}
provided we defined the symmetrized current density operator
\begin{equation}
    \hat{\mathbf{J}} (\mathbf{r}, t)
    = \frac{1}{2} 
        \sum_{\alpha} 
            q_{\alpha}
            \left\{
                \dot{\hat{\mathbf{r}}}_{\alpha}(t)
                \delta^{(3)}[\mathbf{r} - \hat{\mathbf{r}}_{\alpha}(t)]
                +
                \delta^{(3)}[\mathbf{r} - \hat{\mathbf{r}}_{\alpha}(t)] \dot{\hat{\mathbf{r}}}_{\alpha}(t)
            \right\}
\end{equation}
with the particle velocity $\dot{\hat{\mathbf{r}}}_{\alpha}(t) = \hat{\mathbf{p}}_{\alpha} (t) - q_{\alpha} \hat{\mathbf{A}}(\hat{\mathbf{r}}_{\alpha})$. Also note that the Fourier transform of the transverse current density, i.e.,
\begin{equation}
    \hat{\mathscr{J}}_{\perp, \mathbf{k}, \lambda} (t)
    = \frac{1}{\sqrt{V}}
        \int_V \mathrm{d}^3 r \, 
            e^{-\ci \mathbf{k} \cdot \mathbf{r}}
            \, 
            \hat{\mathbf{e}}_{\mathbf{k},\lambda}
            \cdot \hat{\mathbf{J}}(\mathbf{r},t),
\end{equation}
has been modified due to the periodic boundary condition.}.

Since $\hat{a}_{\mathbf{k}, \lambda}$ are now operators, the physical observables defined using the normal modes are also operators now. For example, the transverse electric field and vector potential are promoted to
\begin{align} \label{eq:electric_field_quantum_operator}
    \hat{\mathbf{E}}_{\perp}(\mathbf{r}) 
    &= \sum_{\mathbf{k},\lambda}
        \ci \mathscr{E}_{0, k} 
        \hat{\mathbf{e}}_{\mathbf{k},\lambda}
        \Big(
            \hat{a}_{\mathbf{k},\lambda} 
                e^{\ci\mathbf{k} \cdot \mathbf{r}}
            - \hat{a}_{\mathbf{k},\lambda}^{\dagger}   
                e^{- \ci\mathbf{k} \cdot \mathbf{r}}
        \Big),
\\ \label{eq:vector_potential_quantum_operator}
    \hat{\mathbf{A}}_{\perp}(\mathbf{r}) 
    &= \sum_{\mathbf{k},\lambda}
        \frac{\mathscr{E}_{0, k}}{\omega_{k}}
        \,
        \hat{\mathbf{e}}_{\mathbf{k},\lambda}
        \Big(
            \hat{a}_{\mathbf{k},\lambda} 
                e^{\ci\mathbf{k} \cdot \mathbf{r}}
            + \hat{a}^{\dagger}_{\mathbf{k},\lambda} 
                e^{-\ci\mathbf{k} \cdot \mathbf{r}}
        \Big).
\end{align}
They are clearly Hermitian, satisfying the postulates of quantum mechanics. In addition, the transverse energy now becomes
\begin{equation}
    \hat{H}_{\perp}
    = \sum_{\mathbf{k},\lambda} 
    \frac{1}{2} \hbar \omega_{k} 
        \Big(
            \hat{a}_{\mathbf{k},\lambda}^{\dagger} 
            \hat{a}_{\mathbf{k},\lambda}
            + 
            \hat{a}_{\mathbf{k},\lambda} 
            \hat{a}_{\mathbf{k},\lambda}^{\dagger} 
        \Big),
\end{equation}
which is also Hermitian as expected. Moreover, by applying the commutation relations of $\hat{a}_{\mathbf{k},\lambda}$, we can express the Hamiltonian in terms of the number operators $\hat{N}_{\mathbf{k},\lambda} = \hat{a}_{\mathbf{k},\lambda}^{\dagger}\hat{a}_{\mathbf{k},\lambda}$ just like the Hamiltonian of a single quantum harmonic oscillator):
\begin{equation}
    \hat{H}_{\perp}
    = \sum_{\mathbf{k},\lambda} 
        \hbar \omega_{k} 
        \bigg(
            \hat{a}_{\mathbf{k},\lambda}^{\dagger} 
                \hat{a}_{\mathbf{k},\lambda}
            + \frac{1}{2}
        \bigg)
    = \sum_{\mathbf{k},\lambda} 
        \hbar \omega_{k}  
        \bigg(
            \hat{N}_{\mathbf{k},\lambda}
            + \frac{1}{2}
        \bigg).
\end{equation}
Note that if we didn't keep the order of multiplication in the classical Hamiltonian, we would have lost the ground state energy, $\hbar \omega_k / 2$, in the usual QHO Hamiltonian.

\subsection{Photons and Number States}
For a QHO labelled by $(\mathbf{k}, \lambda)$, the Fock state $\ket{n}_{\mathbf{k}, \lambda}$ corresponds to an electromagnetic mode $(\mathbf{k}, \lambda)$ with $n$ photons; hence, $\{\ket{n}_{\mathbf{k}, \lambda}\}_{n=0}^{\infty}$ are called the photon \textbf{number states} in the mode $(\mathbf{k}, \lambda)$. 
With no excitation (i.e., zero photons) in the mode, the QHO has a ground-state energy
\begin{equation}
    E_{\mathbf{k},\lambda,0} = \frac{1}{2} \hbar \omega_k,
\end{equation}
known as the \textbf{zero-point energy}. For each photon added to the same mode, the QHO gains a quantum of energy of $\hbar \omega_k$, i.e., the energy difference between $\ket{n}_{\mathbf{k},\lambda}$ and $\ket{0}_{\mathbf{k},\lambda}$ is
\begin{equation}
    E_{\mathbf{k},\lambda, n} - E_{\mathbf{k},\lambda,0}
    = n \hbar \omega_k.
\end{equation}
In other words, a photon is simply the unit of the energy quanta in a mode.

In addition, the action of the creation and annihilation operators on the number states follows directly from the theory of QHO, i.e.,
\begin{gather} \label{eq:formula_annhilation}
    \hat{a}_{\mathbf{k},\lambda} \ket{n}_{\mathbf{k},\lambda}
    = \sqrt{n} \ket{n-1}_{\mathbf{k},\lambda},
\\ \label{eq:formula_creation}
    \hat{a}_{\mathbf{k},\lambda}^{\dagger} 
        \ket{n}_{\mathbf{k},\lambda}
    = \sqrt{n + 1} 
        \ket{n + 1}_{\mathbf{k},\lambda}.
\end{gather}
Note that these operators are mode-dependent; when embedding into a full Fock state of all the modes, we will use a tensor product to describe the number of photons in each mode. In particular, suppose the set $\{ n_{\mathbf{k}',\lambda'} \}_{\mathbf{k}',\lambda'}$ lists the number of photons in each mode, then we say that the electromagnetic field is in the Fock state
\begin{equation}
    \ket{\{n_{\mathbf{k},\lambda}\}}
    = \bigotimes_{\mathbf{k}',\lambda'}     
            \ket{n_{\mathbf{k}',\lambda'}}_{\mathbf{k}',\lambda'}.
\end{equation}
Moreoever, $\hat{a}_{\mathbf{k},\lambda}$ and $\hat{a}_{\mathbf{k},\lambda}^{\dagger}$ create and annihilate, respectively, a photon in one mode at a time
\begin{gather}\label{eq:a_acts_on_n}
    \hat{a}_{\mathbf{k},\lambda} \ket{\{n_{\mathbf{k},\lambda}\}}
    = \sqrt{n_{\mathbf{k},\lambda}}   
        \ket{n_{\mathbf{k},\lambda} - 1}_{\mathbf{k},\lambda}     
        \bigotimes_{(\mathbf{k}',\lambda') \neq     (\mathbf{k},\lambda)}     
            \ket{n_{\mathbf{k}',\lambda'}}_{\mathbf{k}',\lambda'},
\\ \label{eq:a_dagger_acts_on_n}
    \hat{a}_{\mathbf{k},\lambda}^{\dagger} \ket{\{n_{\mathbf{k},\lambda}\}}
    = \sqrt{n_{\mathbf{k},\lambda} + 1}   
        \ket{n_{\mathbf{k},\lambda} + 1}_{\mathbf{k},\lambda}     
        \bigotimes_{(\mathbf{k}',\lambda') \neq     (\mathbf{k},\lambda)}     
            \ket{n_{\mathbf{k}',\lambda'}}_{\mathbf{k}',\lambda'};
\end{gather}
that is, the operators for a particular mode only talk to the sub-states in that mode. Consequently, if $n_{\mathbf{k},\lambda} = 0$, $\hat{a}_{\mathbf{k},\lambda}$ will annihilate the entire Fock state regardless of the number of photons in the other modes.

The only Fock state that has zero photons in all the modes is called the \textbf{vacuum state}, denoted by $\ket{0}$; in other words,
\begin{equation}
    \ket{0}
    = \bigotimes_{\mathbf{k},\lambda} \ket{0}_{\mathbf{k},\lambda}.
\end{equation}
Any single-photon state can then be expressed as
\begin{equation}
    \hat{a}_{\mathbf{k},\lambda}^{\dagger} \ket{0}
    = \ket{1}_{\mathbf{k},\lambda}     
        \bigotimes_{(\mathbf{k}',\lambda') \neq     (\mathbf{k},\lambda)}     
            \ket{0}_{\mathbf{k}',\lambda'}.
\end{equation}
By induction, we can build all the Fock states from the vacuum state by using the creation operators. By normalizing the action of the creation operators using Eq.(\ref{eq:a_dagger_acts_on_n}), we can write
\begin{equation} \label{eq:build_number_state_general}
    \ket{\{n_{\mathbf{k},\lambda}\}}
    = \bigotimes_{\mathbf{k},\lambda} 
        \frac{1}{\sqrt{n_{\mathbf{k},\lambda} !}} 
        \Big(  
            \hat{a}_{\mathbf{k},\lambda}^{\dagger} 
        \Big)^{\displaystyle n_{\mathbf{k},\lambda}} \ket{0}.
\end{equation}
Like the other elementary particles, photons living in the same mode are treated as identical particles, requiring extra steps to be symmetrized in quantum mechanics. However, note that the normalization factor $1/\sqrt{n_{\mathbf{k},\lambda}!}$ in Eq.(\ref{eq:build_number_state_general}) is exactly the one used to symmetrize a state of multiple bosons (i.e., there are $n_{\mathbf{k},\lambda}!$ ways of ordering $n_{\mathbf{k},\lambda}$ bosons in a mode). Moreover, since the creation operators of two different modes commute, how we order the creation operators in Eq.(\ref{eq:build_number_state_general}) has no effect on the overall sign of the state; for example, 
\begin{equation}
    \hat{a}_1^{\dagger} \hat{a}_2^{\dagger} \ket{0} = \hat{a}_2^{\dagger}\hat{a}_1^{\dagger}
    \ket{0} = \ket{1_1,1_2}.
\end{equation}
In conclusion, the field theory we developed above well respects the bosonic nature of photons.
        
\subsection{Coherent States}
Classically, we are familiar with the traveling wave of the form 
\begin{equation}\label{eq:classical_electric_field}
    \mathbf{E}(\mathbf{r},t) 
    = - \hat{\mathbf{e}}_{\mathbf{k}, \lambda} E_0 \sin(\mathbf{k} \cdot \mathbf{r} - \omega_k t + \phi),
\end{equation}
which bears no resemblance to a photon, i.e., a particle. Additionally, we know that quantum theory gives rise to the duality between particle and wave, so how can particles like photons behave like a macroscopic wave in the classical limit?

It turns out that the eigenstates of each annihilation operator $\hat{a}_{\mathbf{k}, \lambda}$ serve as the most ``classical'' states in a quantum world. Before even writing down the eigenstate, it should be emphasized that $\hat{a}_{\mathbf{k}, \lambda}$ is not Hermitian, so its eigenvalues could be complex and its eigenstates do not form an orthonormal basis of the Fock space (though they can be shown to be complete). We call the eigenstates of $\hat{a}_{\mathbf{k}, \lambda}$ the \text{coherent states}, labelled by $\ket{\alpha}$, satisfying the eigenvalue equation
    \begin{equation}
        \hat{a}_{\mathbf{k}, \lambda} \ket{\alpha} 
        = \alpha \ket{\alpha}.
    \end{equation}
Again, $\alpha$ is, in general, a complex scalar, whose value can be thought of as the normal mode amplitude $a_{\mathbf{k}, \lambda}(t)$ used before the quantization. 

Whenever we talk about the coherent state, we usually are referring to a single-mode coherent state; hence, let's focus on a single mode of electromagnetic field and use $\ket{n}$ to denote the number states in mode $(\mathbf{k},\lambda)$. Now, we claim that
\begin{equation}
    \ket{\alpha} 
    = e^{-\abs{\alpha}^2/2} \sum_{n=0}^{\infty} \frac{\alpha^n}{\sqrt{n!}} \ket{n}
\end{equation}
is an eigenstate of $\hat{a}_{\mathbf{k}, \lambda}$ for any complex number $\alpha$. Verifying this is straightforward:
\begin{align}
    \hat{a}_{\mathbf{k}, \lambda} \ket{\alpha} 
    &= e^{-\abs{\alpha}^2/2} \sum_{n=0}^{\infty} \frac{\alpha^n}{\sqrt{n!}} \sqrt{n}\ket{n-1}
\nonumber \\
    &= \alpha e^{-\abs{\alpha}^2/2} \sum_{n=1}^{\infty} \frac{\alpha^{n-1}}{\sqrt{(n-1)!}} \ket{n-1}
    = \alpha \ket{\alpha}
\end{align}
One property of the coherent state $\ket{\alpha}$ can be derived immediately: The probabilities of measuring the number states are distributed according to a Poisson distribution with a ``rate'' parameter $\abs{\alpha}^2$. Hence, the expectation value of the number operator (i.e., the mean photon number in a mode) is $\Bigsl\langle \hat{N}_{\mathbf{k},\lambda} \Bigsr\rangle = \abs{\alpha}^2$ and so is the variance of the number operator.

To get a feeling of the coherent state, let us examine its time evolution. Recall that the number states of a QHO are stationary states, following the trivial time evolution
\begin{equation}
    \hat{U}(t) \ket{n} 
    = \ket{n} e^{-\ci \left( n \omega_k + 1/2 \right) t}.
\end{equation}
The evolution of the coherent state is thus a superposition of that of all the number states. To be more precise, suppose $\ket{\alpha(0)}$ is an eigenstate of $\hat{a}_{\mathbf{k}, \lambda}$ with eigenvalue $\alpha(0)$, then
\begin{align}
    \hat{U}(t) \ket{\alpha(0)} 
    &= e^{-\abs{\alpha(0)}^2/2} \sum_{n=0}^{\infty} \frac{\alpha(0)^n e^{-\ci \left( n\omega_k + 1/2 \right) t} }{\sqrt{n!}} \ket{n}
\nonumber \\
    &= e^{-\ci t /2} e^{-\abs{\alpha(0)}^2/2} 
        \sum_{n=0}^{\infty} 
            \frac{
                \left[\alpha(0) e^{- \ci \omega_k t} \right]^n
            }{
                \sqrt{n!}
            } 
            \ket{n}
\nonumber \\
    &= e^{-\ci t /2} \ket{\alpha(0) e^{- \ci \omega_k t}} ,
\end{align}
where $\ket{\alpha(0) e^{- \ci \omega_k t}}$ is another coherent state with eigenvalue $\alpha(0) e^{- \ci \omega_k t}$. In other words, up to an extra rotation due to the zero-point energy\footnote{When $\omega_k$ is large, the extra rotation due to the constant zero-point energy is negligible.}, the evolution of the coherent state (in the absence of any sources) can be visualized as a rotation in the complex plane,
\begin{equation}
    \ket{\alpha(0)}
    \ \ \longrightarrow \ \ 
    \ket{\alpha(t)}
    = \ket{\alpha(0) e^{- \ci \omega_k t}},
\end{equation}
exactly the same as that of a classical normal mode. In addition, we can also say that a coherent state is a quantum state whose eigenvalue, as a function of time, follows the Euler-Lagrange equation; hence, it is the ``most'' classical state we can construct in a quantized setting.
    
We can take one step further to justify the ``classicalness'' of the coherent state. In particular, we can check the expectation values of the field operators in a coherent state. For example, let $\alpha(0) = \sqrt{n} e^{\ci \phi}$, i.e., the mean photon number is $n$. Then, the mean electric field is given by
\begin{align}
    \bra{\alpha(t)} 
            \hat{\mathbf{E}}_{\perp} (\mathbf{r}) 
        \ket{\alpha(t)}
    &= \ci 
        \sum_{\mathbf{k}',\lambda'} 
            \mathscr{E}_{0,k'}
            \hat{\mathbf{e}}_{\mathbf{k}', \lambda'}
            \Big[
                \bra{\alpha(t)} 
                    \hat{a}_{\mathbf{k}', \lambda'}
                    \ket{\alpha(t)} 
                    e^{\ci \mathbf{k}' \cdot \mathbf{r}} 
                - \bra{\alpha(t)} 
                    \hat{a}_{\mathbf{k}', \lambda'}^{\dagger}
                    \ket{\alpha(t)} 
                    e^{-\ci \mathbf{k}'\cdot \mathbf{r}} 
            \Big]
\nonumber \\[-1mm]
    &= \ci \hat{\mathbf{e}}_{\mathbf{k},\lambda} 
        \mathscr{E}_{0,k}
        \Big[
            \sqrt{n} e^{\ci(\mathbf{k} \cdot \mathbf{r} - \omega_k t + \phi)}
            - \sqrt{n} e^{-\ci(\mathbf{k} \cdot \mathbf{r} - \omega_k t + \phi)}
        \Big]
\nonumber \\[1mm]
    &= - \hat{\mathbf{e}}_{\mathbf{k},\lambda} 
        E_{0} 
        \sin(\mathbf{k} \cdot \mathbf{r} - \omega_k t + \phi),
\end{align}
where, to match exactly with Eq.(\ref{eq:classical_electric_field}), we have set $E_{0} = 2 \sqrt{n} \mathscr{E}_{0}$. Again, we reproduce a classical harmonic field in terms of the expectation value.

We have demonstrated, from two perspectives, that a coherent state links the quantum field with a classical field. However, we cannot say a coherent state is a classical object because it still induces a finite amount of uncertainty governed by the Poisson statistics. This also implies that the magnitude of the field observables (e.g., the mean photon number) increases linearly in $n$ whereas the uncertainty (e.g., standard derivation of the photon number) grows only according to $\sqrt{n}$. Thus, for large $n$ (i.e., for macroscopic fields), the uncertainty is effectively unobservable.

\subsection{A Mode Driven by a Classical Source}
To conclude this section, let us bring into the picture a drive. We will study the matter-field interaction more in-depth later; for now, imagine a single-mode quantum field driven by a classical source of \textit{the same} frequency. 

To be concrete, consider a classical, mechanically-driven mass-spring system. By including an on-resonance force $F_0 \sin (\omega t - \phi)$,  Hamilton's equations become
\begin{align}
    \dot{x} 
    &= \frac{\partial \mathcal{H}}{\partial p} = \frac{p}{m},
\\[1mm]
    \dot{p} 
    &= - \frac{\partial \mathcal{H}}{\partial x} = -kx - F_0 \sin(\omega t - \phi).
\end{align}
Equivalently, the Hamiltonian for such a system now contains an interaction term
\begin{equation}
    \mathcal{H} 
    = \frac{p^2}{2m} 
        + \frac{1}{2}k x^2
        + F_0 \sin(\omega t - \phi) x.
\end{equation}
For an electromagnetic mode, we expect the coupling to appear in the same way mathematically (the precise value of the coupling strength will be discussed in the later chapters); that is, in the quantized case, we should have 
\begin{equation} \label{eq:semiclassically_drive_QHO}
    \hat{H} 
    = \hbar \omega
            \bigg(
                \hat{a}^{\dagger} \hat{a} 
                + \frac{1}{2} 
            \bigg)
        + \hbar \Omega_{\text{d}} 
            \sin(\omega t - \phi) 
            \Big(
                \hat{a} 
                + \hat{a}^{\dagger}
            \Big).
\end{equation}
The position $x$ is replaced by a generalized coordinate (e.g., the vector potential at $\mathbf{r} = \mathbf{0}$), and all the constants are lumped into the energy parameter $\hbar\Omega_{\text{d}}$ (i.e., $\Omega_{\text{d}}$ has the unit of angular frequency).

To solve the time evolution, we will go to the interaction picture. By defining 
\begin{equation}
    \hat{H}_0 
    = \hbar \omega
        \bigg(
            \hat{a}^{\dagger} \hat{a} 
            + \frac{1}{2} 
        \bigg),
\end{equation}
\begin{equation}
    \hat{H}_{\text{int}}(t)
    = \hbar \Omega_{\text{d}} 
        \sin(\omega t - \phi) 
        \Big(
            \hat{a} 
            + \hat{a}^{\dagger}
        \Big),
\end{equation}
and the unitary operator $\hat{U}_0(t) = \exp \Bigsl( -\ci \hat{H}_0 t/\hbar \Bigsr)$, we obtain
    \begin{align}
        \hat{\Tilde{H}}_{\text{int}} (t)
        &= \hat{U}_0^{\dagger}(t) \hat{H}_{\text{int}}(t) \hat{U}_0(t)
    \nonumber \\
        &= \hbar \Omega_{\text{d}} 
            \sin(\omega t) 
            \Big(
                \hat{a} e^{-\ci \omega t} 
                + \hat{a}^{\dagger} e^{\ci \omega t}
            \Big) 
    \nonumber \\ \label{eq:classical_drive_interaction_picture}
        &= \frac{\ci \hbar\Omega_{\text{d}}}{2} 
            \Big( 
                e^{\ci \phi} \hat{a}^{\dagger}
                - e^{-\ci \phi} \hat{a} 
                + e^{-\ci \phi} 
                    e^{\ci 2\omega t} \hat{a}^{\dagger}
                - e^{\ci \phi} 
                    e^{- \ci 2\omega t} \hat{a}
            \Big).
    \end{align}
To simplify the interaction Hamiltonian further, we argue that any terms with a fast oscillation at frequency $2\omega$ will be averaged to 0 as long as $\omega \gg \Omega_{\text{d}}$, i.e.,
\begin{equation}
    \int_0^{t} \mathrm{d}\tau 
        e^{\pm \ci 2\omega \tau} 
    \ \ \substack{\omega \gg \Omega_{\text{d}} \\ \displaystyle \longrightarrow \\ {}} \ \
    0.
\end{equation}
As we will see later, there is always a characteristic time (i.e., the inverse of the Rabi frequency) for describing the change of the quantum state in the interaction picture. Thus, the wave function will not be able to respond to an oscillating term whose period is much shorter than the characteristic timescale. Consequently, we drop the last two terms in Eq.(\ref{eq:classical_drive_interaction_picture}), making the interaction Hamiltonian time-independent:
\begin{equation}
    \hat{\Tilde{H}}_{\text{int}}
    \approx 
        \frac{\ci \hbar \Omega_{\text{d}}}{2} 
        \Big( 
            e^{\ci \phi} \hat{a}^{\dagger}
            - e^{-\ci \phi} \hat{a} 
        \Big)
    = \ci \hbar 
        \Big(
            \alpha \hat{a}^{\dagger} 
            - \alpha^* \hat{a}
        \Big),
\end{equation}
with $\alpha = \Omega_{\text{d}}e^{\ci \phi} / 2$. The approximation we just made is known as the \textbf{rotating wave approximation (RWA)} and will show up frequently when talking about quantum control.

Under the RWA, the time evolution operator in the interaction picture is given by the unitary operator
\begin{equation}\label{eq:displacement_operator_driven_oscillator}
    \hat{\Tilde{U}}(t=1) 
    = \exp(-\frac{\ci \hat{\Tilde{H}}_{\text{int}}}{\hbar})
    = e^{\alpha \hat{a}^{\dagger} - \alpha^* \hat{a}}.
\end{equation}
An operator of the form of Eq.(\ref{eq:displacement_operator_driven_oscillator}) is, in general, called a \textbf{displacement operator} and denoted by
\begin{equation}
    \hat{D}_{\mathbf{k}, \lambda}(\alpha) 
    = \exp(\alpha \hat{a}_{\mathbf{k}, \lambda}^{\dagger} - \alpha^* \hat{a}_{\mathbf{k}, \lambda})
\end{equation}
when multiple modes are considered. When acting on a vacuum state, the displacement operator simply displaces the vacuum to a coherent state with amplitude $\alpha$ \cite{nature08005, PhysRevA.92.040303}. This fact follows from the property that
\begin{equation}
    \hat{D}^{\dagger}(\alpha)
    \hat{a}
    \hat{D}(\alpha)
    = \hat{a} + \alpha
\end{equation}
\begin{equation}
    \hat{D}^{\dagger}(\alpha)
    \hat{a}^{\dagger}
    \hat{D}(\alpha)
    = \hat{a} + \alpha^*
\end{equation}
for a single-mode displacement operator, which can be proven by using the Baker–Campbell-Hausdorff formula. Intuitively, when we drive a resonator with a classical signal at the resonant frequency, we expect to see oscillations with an amplitude that grows linearly in time. The coherent state produced by the displacement operator is precisely the ``classical'' resonance behavior. In addition, since the displacement operator is unitary, one can imagine a reverse process where a coherent state loses its energy and reduces to a vacuum state.

\section{Examples of (Isolated) Electromagnetic Systems}
Equipped with the general theory of electrodynamics introduced above, let us land ourselves on some concrete examples of electromagnetic systems. As mentioned in the introduction chapter, in the typical superconducting quantum computation, a qubit modeled by a nonlinear LC circuit is placed inside or next to a readout resonator to communicate with the room-temperature electronics through transmission lines. Therefore, we will examine each component, i.e., the LC circuit, transmission line, and cavity, one by one and also use them to verify the generality of the theory derived in the last two sections.
\subsection{Zero-Dimensional System: an LC Circuit}
\begin{figure}[b]
    \centering
    \begin{tikzpicture}[scale=1, transform shape]
    \ctikzset{tripoles/mos style/arrows};
        \draw
            (0,3) to (4,3) 
            (4,0) to (0,0) 
            (0,3) to [/tikz/circuitikz/bipoles/length=40pt, L, l_=$L$] (0,0)
            (4,3) to [short, i>^={${I(t)}$}](4,2) to [/tikz/circuitikz/bipoles/length=40pt, C, l_=$C$] (4,1) to (4,0);
        \draw
            (2,2.6) node[]{$+$} 
            (2,1.5) node[]{$V(t)$}
            (2,0.4) node[]{$-$}
            (2,3.4) node[]{$\Phi(t)$}
            (4.6,1.85) node[]{$+Q(t)$} 
            (4.6,1.15) node[]{$-Q(t)$};
    \end{tikzpicture}
    \caption{An LC circuit.}
    \label{fig:LC_circuit_isolated}
\end{figure}
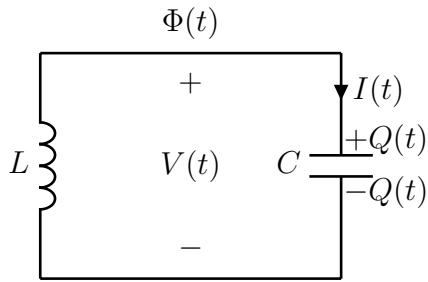
The simplest example of an electromagnetic system is a lumped circuit consisting of an inductor and a capacitor as shown in Figure \ref{fig:LC_circuit_isolated}. We ignore any losses for now and treat the LC circuit as an isolated system\footnote{We assume somehow the circuit has some non-trivial excitation at $t=0$ but is then completely isolated from the environment.}. The energy stored on the capacitor and inductor is given by
\begin{equation}
    T_{\text{cap}}(t) 
    = \frac{1}{2} C V(t)^2 
    = \frac{1}{2} C \dot{\Phi}(t)^2,
\end{equation}
\begin{equation}
    U_{\text{ind}}(t) 
    = \frac{1}{2} L I(t)^2
    = \frac{1}{2L} \Phi(t)^2,
\end{equation}
respectively. Identify the node flux $\Phi$ and the voltage drop $\dot{\Phi}$ as the generalized position and velocity of the system \cite{Vool_2017_into_cQED}; treat $T_{\text{cap}}$ as the kinetic energy and $U_{\text{ind}}$ as the potential energy, then the Lagrangian is given by
\begin{equation}
    \mathcal{L}_{\text{LC}} (\Phi,\dot{\Phi})
    = T_{\text{cap}} - U_{\text{ind}}
    = \frac{1}{2} C \dot{\Phi}^2 - \frac{1}{2L} \Phi^2
\end{equation}
and the conjugate momentum
\begin{equation}\label{eq:LC_conjugate_momentum}
    \frac{\partial \mathcal{L}_{\text{LC}}}{\partial \dot{\Phi}}
    = C \dot{\Phi} = Q
\end{equation}
is simply the charge on the capacitor. Inverting Eq.(\ref{eq:LC_conjugate_momentum}) gives $\dot{\Phi} = Q/C$; thus, the Hamiltonian is given by
\begin{equation}
    \mathcal{H}_{\text{LC}}(\Phi, Q) 
    = Q \dot{\Phi} - \mathcal{L}_{\text{LC}}
    =  \frac{1}{2C}  Q^2 + \frac{1}{2L} \Phi^2.
\end{equation}

We can also approach the problem by defining the normal mode from the conjugate variables
\begin{equation} \label{eq:annihilation_op_LC_definition}
    a
    = \frac{1}{\sqrt{2 \hbar Z_{\text{r}0}}}
        \Big(
            \Phi 
            + \ci Z_{\text{r}0} Q
        \Big),
\end{equation}
where $Z_{\text{r}0} = \sqrt{L/C}$ is the characteristic impedance of the LC oscillator. Then, the Hamiltonian, as expected, can be rewritten as
\begin{equation}
    \mathcal{H}_{\text{LC}} (a,a^*) 
    = \frac{1}{2} \hbar \omega_{\text{r}} 
        \Big(
            a^* a
            + a a^*
        \Big),
\end{equation}
where $\omega_{\text{r}} = 1/\sqrt{LC}$ is the resonant frequency of the oscillator. However, note that $\mathcal{H}_{\text{LC}}$ represents a \textit{single} oscillator instead of infinitely many because there is no notion of space in the lumped-element model; for this reason, we can say that an LC oscillator is a zero-dimensional electromagnetic problem.

To quantize the oscillator, we promote the classical observables to their quantum operators and impose the canonical commutation relation \cite{Vool_2017_into_cQED, doi:10.1063/1.5089550}
\begin{equation}
    \Big[ \hat{\Phi}, \hat{Q} \Big] 
    = \ci \hbar \hat{1}
    \ \ \text{ or } \ \ 
    \Big[ \hat{a}, \hat{a}^{\dagger} \Big] 
    = \hat{1}.
\end{equation}
From this point, everything follows from the theory of the QHO. In addition, it's also useful to express all observables in terms of the annihilation and creation operators; for example, we have
\begin{gather}
    \hat{H}_{\text{LC}} 
    = \hbar \omega_{\text{r}} 
        \bigg(
            \hat{a}^{\dagger} \hat{a} 
            + \frac{1}{2}
        \bigg),
\\ \label{eq:res_charge_operator}
    \hat{Q} 
    = -\ci \sqrt{\frac{\hbar}{2 Z_{\text{r}0}}} 
        \Big(
            \hat{a} 
            - \hat{a}^{\dagger}
        \Big),
\\
    \hat{\Phi} 
    = \sqrt{\frac{\hbar Z_{\text{r}0}}{2}} 
        \Big(
            \hat{a} 
            + \hat{a}^{\dagger}
        \Big).
\end{gather}
In fact, we are not restricted to the conjugate variables used in defining the Hamiltonian; the operators for voltage and current can be generalized in the same way:
\begin{gather} \label{eq:res_voltage_operator}
    \hat{V} 
    = \frac{\hat{Q}}{C}
    = - \ci \sqrt{\frac{\hbar\omega_{\text{r}}}{2C}} 
        \Big(
            \hat{a} 
            - \hat{a}^{\dagger}
        \Big),
\\
    \hat{I} 
    = \frac{\hat{\Phi}}{L}
    = \sqrt{\frac{\hbar\omega_{\text{r}}}{2L}} 
        \Big(
            \hat{a} 
            + \hat{a}^{\dagger}
        \Big).
\end{gather}

Within the field of circuit quantum electrodynamics (cQED), it is custom to normalize the charge $Q$ on the capacitor by two units of electron charge (i.e., $2e$) and the node flux $\Phi$ by the reduced magnetic flux quantum $\phi_0 = \Phi_0/2\pi = \hbar/2e$ \cite{RevModPhys.73.357, annurev-conmatphys-031119-050605, PRXQuantum.2.040202}. The reason for this particular normalization will be apparent when we discuss the theory of superconductivity; for now, this is merely a matter of notation. We thus define the (classical) reduced charge and flux to be, respectively,
\begin{equation}
    n = \frac{Q}{2e}
\ \ \text{ and } \ \ 
    \varphi = \frac{\Phi}{\phi_0}.
\end{equation}
With the newly defined variables, the Hamiltonian becomes
\begin{equation}
    \mathcal{H}_{\text{LC}} (\varphi, n)
    = 4 E_C n^2 + \frac{1}{2} E_L \varphi^2,
\end{equation}
where the two energy constants\footnote{One might ask why don't we absorb the factor $1/2$ into the definition of $E_L$. The reason is that the Josephson junction to be discussed in Chapter 4 is associated with the Josephson energy $E_{\text{J}} = \phi_0^2/L_{\text{J}}$, where $L_{\text{J}}$ is the effective inductance of the junction. Defining the inductive energy constant to have the same form as the Josephson energy will make the mapping to the anharmonic oscillator easier later.} are related to the capacitance and inductance used in the circuit by
\begin{equation} \label{eq:energy_linear_inductor}
    E_C = \frac{e^2}{2C}
\ \ \text{ and } \ \ 
    E_L = \frac{\phi_0^2}{L}.
\end{equation}
Moreover, due to the additional normalization factors, the quantum operators $\hat{n}$ and $\hat{\varphi}$ satisfy the commutation relation
\begin{equation}
    \Bigsl [ \hat{\varphi}, \hat{n} \Bigsr]
    = \ci.
\end{equation}
A more systemic way of purposing the commutation relation would be to first find the Poisson bracket $\{\varphi, n\}$ with respect to the conjugate variables, $\Phi$ and $Q$, and then define $\Bigsl [ \hat{\varphi}, \hat{n} \Bigsr] = \ci \hbar \{\varphi, n\}$. For a summary of the results related to the LC circuit, see Table \ref{table:normal_mode_expansion}.
\subsection{One-Dimensional System: an Infinite Transmission Line}
In practice, microwave signals used to control and read out the state of the qubit or cavity travel on the coaxial cables, which can be modeled as a one-dimensional electromagnetic system. Clearly, a real transmission line has a finite length and is loaded on both ends; however, we will, for now,  consider an ideal line of infinite length along the $x$-axis. In the later chapter, we will connect the LC circuit to a semi-infinite line (i.e., a line lives on the positive $x$-axis) at $x=0$.

\begin{figure}[t]
    \centering
    \begin{circuitikz}[scale=0.9, transform shape]
    \draw 
        (-1,-0.5) to [/tikz/circuitikz/bipoles/length=30pt, R, l = ${\Delta x R_{l}}$] (1,-0.5) 
        (1,-0.5) to [L,l = ${\Delta x L_{l}}$] (3,-0.5) 
        (3,-0.5) to (3,-1) 
        (3,-1) to (2.5,-1) to [/tikz/circuitikz/bipoles/length=30pt, R] (2.5,-3) to (3,-3)
        (2.5, -2) node[label = {[font =\fontsize{10}{0}\selectfont] 180: ${\Delta x G_{l}}$}]{}
        (3,-1) to (3.5,-1) to [C] (3.5,-3) to (3,-3)
        (3.7, -2) node[label = {[font =\fontsize{10}{0}\selectfont] 0: ${\Delta x C_{l}}$}]{}
        (3,-3) to (3,-3.5) node[/tikz/circuitikz/bipoles/length=30pt,sground]{};
    \draw 
        (3,-0.5) to [/tikz/circuitikz/bipoles/length=30pt, R, l = ${\Delta x R_{l}}$] (5,-0.5)
        (5,-0.5) to [L,l = ${\Delta x L_{l}}$] (7,-0.5) to (8,-0.5)
        (7,-0.5) to (7,-1) 
        (7,-1)  to (6.5,-1) to [/tikz/circuitikz/bipoles/length=30pt, R] (6.5,-3) to (7,-3) 
        (6.5, -2) node[label = {[font =\fontsize{10}{0}\selectfont] 180: ${\Delta x G_{l}}$}]{}
        (7,-1) to (7.5,-1) to [C](7.5,-3) to (7,-3)
        (7.7, -2) node[label = {[font =\fontsize{10}{0}\selectfont] 0: ${\Delta x C_{l}}$}]{}
        (7,-3) to (7,-3.5) node[/tikz/circuitikz/bipoles/length=30pt,sground]{};
    \draw 
        (-1, -0.5) node[label = {left:$\cdots$}]{}
        (8, -0.5) node[label = {right:$\cdots$}]{};
    \end{circuitikz}
    $$\textbf{(a)}$$
    \begin{circuitikz}[scale=0.8, transform shape]
    \begin{scope}[shift={(-11,-2)}]
    \draw 
        (0,0) ellipse (0.6 and 1.2);
    \draw 
        (4.5,-1.2) arc(-90:90:0.6 and 1.2);
    \draw
        (0,1.2) to (4.5,1.2)
        (0,-1.2) to (4.5,-1.2);
    \draw[dashed]
        (0,0.2) to (4.5,0.2)
        (0,-0.2) to (4.5,-0.2);
    \draw 
        (0,0) ellipse (0.1 and 0.2)
        (4.5,0) ellipse (0.1 and 0.2);
    \draw 
        (2.25,-1.2) node[/tikz/circuitikz/bipoles/length=30pt,sground]{};
    \draw 
        (0,0) to (-1.5,0)
        (4.5,0) to (6,0);
    \draw
        (-1.5,-0.3) node[]{$+$} 
        (-1.5,-0.9) node[]{$V(x_k,t)$}
        (-1.5,-1.5) node[]{$-$}
        (6,-0.3) node[]{$+$} 
        (6.3,-0.9) node[]{${V(x_k+\Delta x,t)}$}
        (6,-1.5) node[]{$-$};
    \draw[-{Latex[length=2mm]}] 
        (1.25,0) to (3.25,0) ;
    \draw 
        (2.25,0.5) node[]{${I(x_k,t)}$};
    \end{scope}
    \draw 
        (-2,-0.5) to (-1,-0.5) to [short, i>_={${I(x_k,t)}$}, L,l = ${\Delta x L_{l}}$] (3,-0.5) 
        (3,-0.5) to (4.8, -0.5) 
        (3,-0.5) to (3,-1) 
        (3,-1) to [C] (3,-3)
        (2.7, -2) node[label = {[font =\fontsize{10}{0}\selectfont] 180: ${\Delta x C_{l}}$}]{}
        (3,-3) to (3,-3.5) node[/tikz/circuitikz/bipoles/length=30pt,sground]{};
    \draw 
        (-2,-0.1) node[]{$\Phi(x_k, t)$}
        (3,-0.1) node[]{${\Phi(x_k + \Delta x, t)}$}
        (-2,-0.9) node[]{$+$} 
        (-2,-2) node[]{$V(x_k,t)$}
        (-2,-3.3) node[]{$-$}
        (4.8,-0.9) node[]{$+$} 
        (4.8,-2) node[]{${V(x_k+\Delta x,t)}$}
        (4.8,-3.3) node[]{$-$};
    \draw 
        (-2, -0.5) node[label = {left:$\cdots$}]{}
        (4.8, -0.5) node[label = {right:$\cdots$}]{};
    \end{circuitikz}\\
    $\textbf{(b)}$
    \caption{The LC-model of an infinite transmission line. \textbf{a.} The general circuit-level representation of a lossy line. $R_l$, $G_l$, $C_l$, and $L_l$ are the resistance, conductance, capacitance, and inductance per unit length, respectively \textbf{b.} The LC model of a lossless line of infinitesimal length $\Delta x$.}
    \label{fig:lossy_transmission_line_model}
\end{figure}
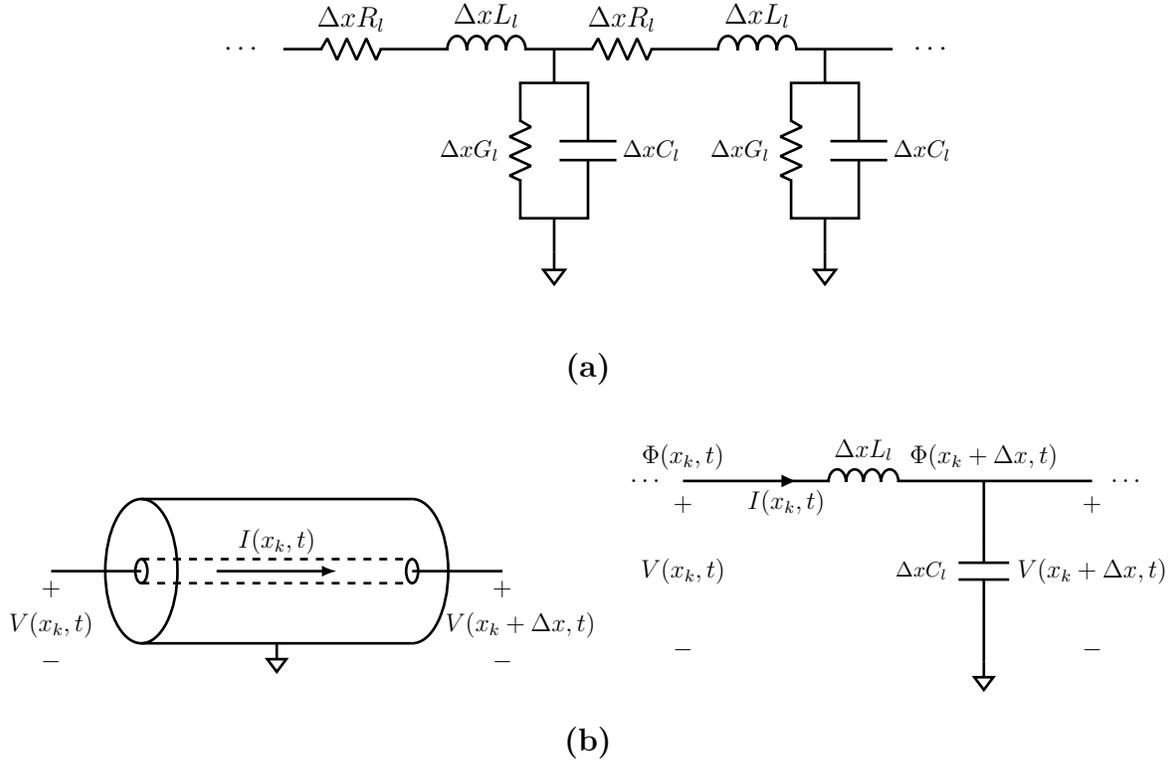

Recall that, in general, a transmission line can be modeled by a concatenation of LC circuits of infinitesimal size as drawn in Figure \ref{fig:lossy_transmission_line_model}(a). In addition, dielectric and conduction losses per unit length can be modeled, respectively, by adding a conductance in parallel with the capacitance and a resistance in series with the inductance. Since quantum time evolution for an isolated system should always be unitary, resistors cannot appear in a quantum circuit directly; thus, we will first ignore the losses and discuss other ways of modeling dissipation in later chapters when we introduce the idea of open quantum systems.

Thus, the model of a transmission line per small length $\Delta x$ reduces to Figure \ref{fig:lossy_transmission_line_model}(b) and we can start by writing down the capacitive and inductive energy, $T_{\text{cap}}$ and $U_{\text{ind}}$, stored on the line, generalizing the idea used in analyzing the LC circuit (see Appendix \ref{appendix:LC_to_TL} for the classical theory of the transmission line). Let $\Phi(x,t)$ be the node flux at position $x$ in the LC model shown in Figure \ref{fig:lossy_transmission_line_model}. By summing the energy stored on each capacitor/inductor and setting $\Delta x \rightarrow 0$, we obtain \cite{cQED_lectures_Girvin}
\begin{align}
    T_{\text{cap}}(t) 
    &= \int_{-\infty}^{\infty}
        \mathrm{d} x \, \frac{C_l}{2} \left(\frac{\partial \Phi }{\partial t}\right)^2,
\\
    U_{\text{ind}}(t) 
    &= \int_{-\infty}^{\infty}
            \mathrm{d} x \, 
            \frac{1}{2L_l} 
            \left(
                \frac{\partial \Phi }{\partial x}
            \right)^2.
\end{align}
By choosing the node flux (as a function of the position $x$) to be a continuum of degrees of freedom of the infinite line, the Lagrangian $\mathcal{L}_{\text{TL}}$ and its density $\mathscr{L}_{\text{TL}}$ are found to be
\begin{align}
    \mathcal{L}_{\text{TL}}(\Phi(x), \dot{\Phi}(x))
    &= \int_{-\infty}^{\infty}
        \mathrm{d} x \, 
        \mathscr{L}_{\text{TL}}(\Phi(x), \dot{\Phi}(x), \partial_x \Phi(x))
\nonumber \\
    &= \int_{-\infty}^{\infty}
        \mathrm{d} x 
            \left[
                \frac{C_l}{2} 
                    \dot{\Phi}(x)^2
                - \frac{1}{2 L_l}
                    (\partial_x \Phi(x))^2
            \right],
\end{align}
yielding the conjugate momentum
\begin{equation} \label{eq:conjugate_variable_t_line}
    \Pi(x) 
    = \frac{\partial \mathscr{L}_{\text{TL}}}{\partial \dot{\Phi}} 
        \Bigg|_{\dot{\Phi} = \dot{\Phi}(x)} 
    = C_l \dot{\Phi}(x) 
    = C_l V(x)
\end{equation}
for each $x$. Note that there are infinitely many coordinates, $\Phi(x)$, indexed by the position $x$ on the real line; hence, what we are doing is actually the classical field theory. 

Despite the continuous nature of the line, its Hamiltonian can still be derived from the Legendre transform. However, to reveal the normal modes supported on the line, we shall perform a (one-dimensional) spatial Fourier transform to all the variables \cite{Clerk_2010}, which eventually gives the Hamiltonian in $k$-space (see Appendix \ref{appendix:LC_to_TL} for the details):
\begin{equation}
    \mathcal{H}_{\text{TL}}
        \Bigsl(
            \tilde{\Phi}(k), \tilde{\Pi}(k)
        \Bigsr)
    = \int_{-\infty}^{\infty}
        \mathrm{d} k 
        \left[
            \frac{1}{2C_l} 
                \Bigsl| \tilde{\Pi}(k) \Bigsr|^2
            + \frac{\omega_k^2 C_l}{2} 
                \Bigsl| \tilde{\Phi}(k) \Bigsr|^2
        \right]
\end{equation}
Consequently, by defining the normal modes to be
\begin{equation}
    a(k) 
    = \sqrt{\frac{\omega_k C_l}{2\hbar}} 
            \, \tilde{\Phi}(k)
        + \frac{\ci}{\sqrt{2 \hbar \omega_k C_l}} 
            \, \tilde{\Pi}(k),
\end{equation}
for each wavenumber $k \in \mathbf{R}$, we obtain a Hamiltonian in the familiar form
\begin{equation}
    \mathcal{H}_{\text{TL}} 
        \Bigsl(
            a(k), a^*(k)
        \Bigsr)
    = \int_{-\infty}^{\infty} 
        \mathrm{d} k \, 
        \frac{1}{2} 
        \hbar \omega_k 
        \Big[
            a^*(k) a(k) 
            + a(k) a^*(k)
        \Big].
\end{equation}
    
Moreover, although not necessary, we can also impose the periodic boundary conditions over a length $L$ to discretize $k$-space and reduce the integral to a sum
\begin{equation}
    \mathcal{H}_{\text{TL}} 
        \Bigsl(
            a_k, a_k^*
        \Bigsr)
    = \sum_{k=-\infty}^{\infty}
        \frac{1}{2} 
        \hbar \omega_k 
        \Big(
            a_k^* a_k
            + a_k a_k^*
        \Big)
\end{equation}
by re-scaling the normal mode to be $a_k = \sqrt{2 \pi / L} \, a(k)$. We have thus demonstrated that an infinite transmission line is simply a one-dimensional free space and can be quantized in exactly the same way. We will come back to this model later when discussing the coupling of a resonator with a line.

\subsection{Three-Dimensional System: a Microwave Cavity}\label{section:microwave_cavity}
\begin{figure}
    \centering
    \includegraphics[scale=0.28]{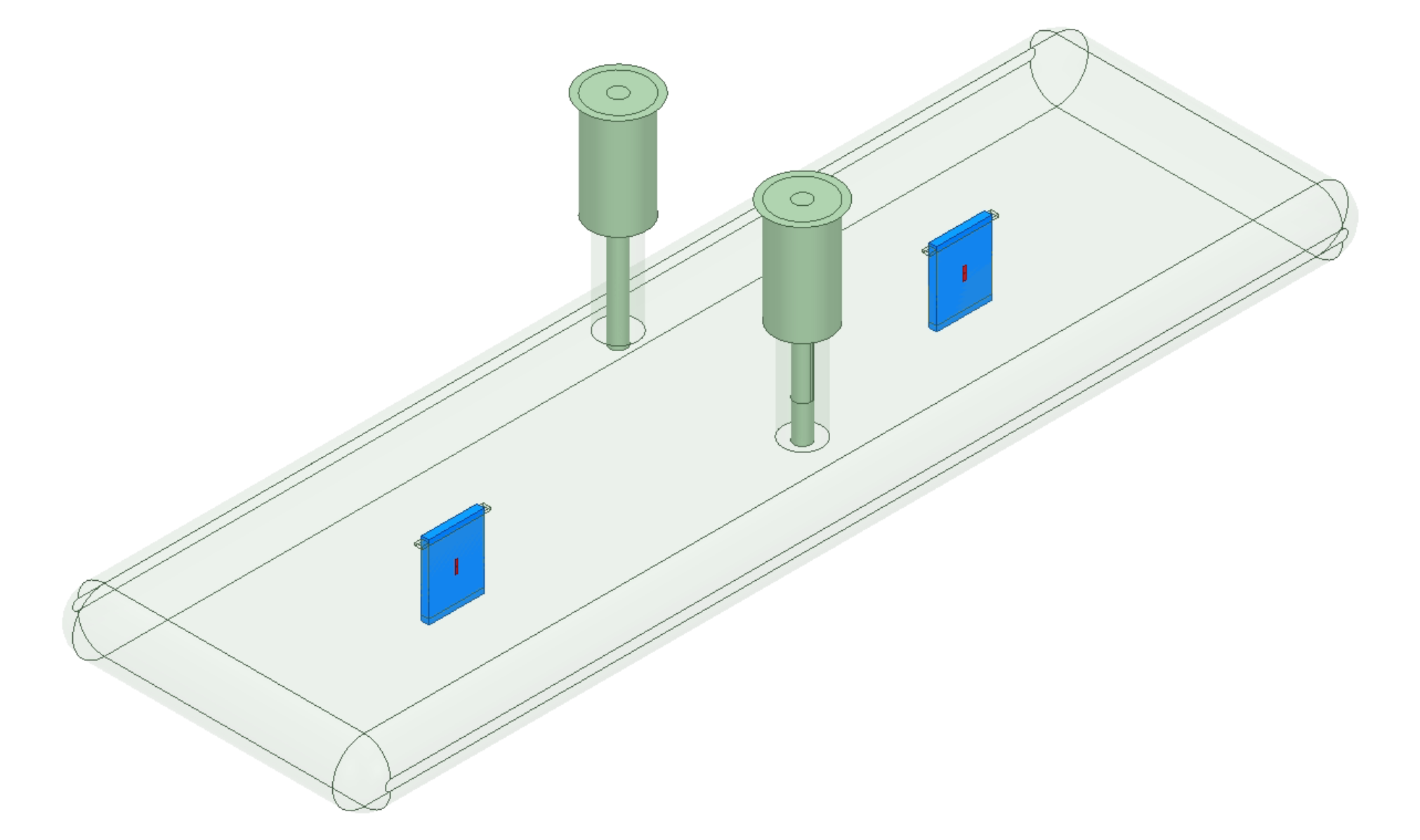}
    \caption{A microwave cavity with two qubit chips installed inside. The two cylindrical structures on the cavity are the SMA ports used to send and receive signals into and out of the cavity.}
    \label{fig:cavity_example}
\end{figure}
Finally, we apply the general framework to microwave cavities, which will be used to host the qudits in superconducting quantum computing. Compared to the two previous examples and the free space formulation, a 3D cavity is different in the sense that it is subject to a non-trivial boundary condition\footnote{Although we did use the periodic boundary conditions to discretize the continuous mode of the free space, we can always remove the boundary conditions.}. An ideal (i.e., lossless) cavity is an enclosed volume, denoted by $V$, with a perfect-electric-conductor (PEC) boundary condition 
\begin{equation}
    \hat{\mathbf{e}}_{\text{n}}(\mathbf{r})
        \times 
        \mathbf{E}(\mathbf{r},t) 
    = \mathbf{0}
    \ \ \text{ for } \ \ 
    \mathbf{r} 
        \in \partial V,
\end{equation}
where $\partial V$ is the boundary surface of $V$ and $\hat{\mathbf{e}}_{\text{n}}(\mathbf{r})$ is the unit normal vector of $\partial V$ at $\mathbf{r}$. 

It turns out that we can mode-expand the electromagnetic field inside the cavity \cite{RevModPhys.18.441, steck_quantum_optics, yariv1989quantum} just like we can decompose an arbitrary waveform on a guitar string into the orthonormal Fourier components. In particular, it's possible to find an orthonormal set (of infinite size) of vector fields $\{\mathbf{f}_n(\mathbf{r})\}_{n=1}^{\infty}$ that solve the vector Helmholtz equation (i.e., the temporal Fourier transform of the wave equation) 
\begin{equation}
    \Delta \mathbf{f}_n(\mathbf{r}) 
        + k_n^2 \mathbf{f}_n(\mathbf{r}) 
    = \mathbf{0}
\end{equation}
subject to the PEC boundary condition. The orthonormality is defined within $V$ to be
\begin{equation}
    \int_{V} 
        \mathrm{d}^3 r \, 
        \mathbf{f}_m(\mathbf{r}) 
        \cdot 
        \mathbf{f}_n(\mathbf{r})
    = \delta_{mn}.
\end{equation}

The vector fields $\{\mathbf{f}_n(\mathbf{r})\}_{n=1}^{\infty}$ can be solved numerically (e.g., the Eigenmode solver in HFSS). Mathematically, once we obtain $\{\mathbf{f}_n(\mathbf{r})\}_{n=1}^{\infty}$, the electric and magnetic fields, the vector potential, and the Hamiltonian of the cavity are given by
\begin{gather}
    \mathbf{E}(\mathbf{r},t) 
    = \sum_{n}
            \ci \mathscr{E}_{0, n} \mathbf{f}_n(\mathbf{r})
            \Big[
                a_{n}(t)
                - a_{n}^{*}(t)     
            \Big] ,
\\
    \mathbf{B}(\mathbf{r},t) 
    = \sum_{n}
            \frac{\mathscr{E}_{0, n}}{c}
            \,
            \frac{\nabla \times
            \mathbf{f}_n(\mathbf{r})}{k_n}
            \Big[
                a_{n}(t)
                + a_{n}^{*}(t)     
            \Big] ,
\\
    \mathbf{A}(\mathbf{r},t) 
    = \sum_{n}
            \frac{\mathscr{E}_{0, n}}{\omega_{n}}
            \,
            \mathbf{f}_n(\mathbf{r})
            \Big[
                a_{n}(t)
                + a_{n}^{*}(t)     
            \Big],
\\ 
    \mathcal{H}_{\text{C}}(t)
    = \sum_{n}
        \frac{1}{2} \hbar \omega_{n} 
        \Big[
            a_{n}^{*}(t) 
            a_{n}(t) 
            + 
            a_{n}(t) 
            a_{n}^{*}(t) 
        \Big]
\end{gather} 
for a set of coefficient functions $\{a_n(t)\}$ that solve
\begin{equation}
    \dot{a}_n(t) 
    = - \ci \omega_n a_n(t)
\end{equation}
with some given initial conditions. Just like in the free-space formulation, for each mode in a cavity, we have a linear dispersion relation $\omega_n = c k_n$ and the zero-point fluctuation (averaged over the cavity volume)
\begin{equation}
    \frac{\mathscr{E}_{0, n}}{\sqrt{V}}     
    = \sqrt{\frac{\hbar \omega_{n}}{2 \epsilon_0 V}}.
\end{equation}
It's clear that we are just generalizing the free space mode expansion by replacing a plane wave basis $\{e^{\ci \mathbf{k} \cdot \mathbf{r}} / \sqrt{V}\}_{\mathbf{k}}$ with a more general set of basis functions $\{\mathbf{f}_n(\mathbf{r})\}_n$. However, due to the boundary condition, the mode spectrum is discretized in the true sense, allowing us to address each mode separately using a narrow-band microwave source. Table \ref{table:normal_mode_expansion} compares the free space and cavity mode expansion.

\subsection{A Half-Wavelength Transmission-Line Resonator}
To reduce the footprint of the design, it is of practical interest to replace a 3D cavity with a planar cavity. In particular, we can readily make a resonator by imposing boundary conditions on the two sides of a finite transmission line of length $l$ (between $x=0$ and $l$). In practice, most electromagnetic systems can be coupled capacitively since electric dipole interaction is at least an order larger than the magnetic dipole (if spins are ignored, see Exercise \ref{ex:3_1}); hence, one usually deploys a half-wavelength ($\lambda/2$) transmission-line resonator where the two ends are open (i.e., the current at the two ends vanishes).

We have already seen the consequence of adding boundary conditions in the three-dimensional case. Imposing open-circuit conditions on the two ends of a transmission line will also discretize the mode spectrum, resulting in wavenumbers $k_m = m \pi /l$ (or $\omega_m = v_{\text{p}} k_m = k_m / \sqrt{L_l C_l}$) for $m=0,1,2,...$ As a result, the flux $\Phi(x,t)$ on the line can be expanded in terms of the standing-wave normal modes 
\begin{equation}
    \Phi_m(t) 
    = \sqrt{\frac{\hbar Z_{\mathrm{r0},m}}{2}}
        \Big[ a_m(t) + a_m^*(t) \Big]
\end{equation}
for $m = 0,1,2,...$, where $Z_{\mathrm{r0},m} = 2/(l C_l \omega_m)$ is the impedance\footnote{The capacitance and inductance of the effective LC circuit are $C_m = l C_l/2$ and $L_m = 1/ (C_m \omega_m^2)$, respectively.} of the effective LC model of the $m$th mode \cite{pozar1990microwave}. The mode amplitudes satisfy the familiar equation $\dot{a}_m(t) = - \ci \omega_m a_m(t)$ and each of them is associated with a standing wave pattern $u_m(x) = \cos(k_m x)$. Linearly combining the normal modes gives the total flux 
\begin{equation}
    \Phi(x,t) 
    = \sum_{m=0}^{\infty} 
        u_m(x)
        \Phi_m(t) 
    = \sum_{m=0}^{\infty} 
        \sqrt{\frac{\hbar Z_{\mathrm{r0},m}}{2}} 
        \cos(k_m x) 
        \Big[a_m(t) + a_m^*(t) \Big].
\end{equation}
Similarly, the conjugate variable (charge per unit length) is found to be 
\begin{equation}
    \Pi(x,t)
    = - \ci \sqrt{\frac{\hbar}{2Z_{\mathrm{r0},m}}} 
        \cos(k_m x) 
        \Big[a_m(t) - a_m^*(t) \Big],
\end{equation}
and the voltage developed along the line is given by
\begin{equation}
    V(x,t) = \frac{\partial}{\partial t} \Phi(x,t) 
    = - \ci \sum_{m=0}^{\infty} 
        \sqrt{\frac{\hbar \omega_{m}}{2 C_m}} 
        \cos(k_m x) 
        \Big[a_m(t) - a_m^*(t) \Big].
\end{equation}
When the voltage is evaluated near the two ends, $\cos(k_m x) \approx \pm 1$ and the expression for $V(x,t)$ reduces to Eq.(\ref{eq:res_voltage_operator}) up to a minus sign. The possible minus sign changes the phase of the coupling coefficients, which will be introduced in the next chapter. 

\begin{sidewaystable}
    \scriptsize
    \centering
    \setlength\tabcolsep{8pt}
    \renewcommand{\arraystretch}{0.4}
\begin{tabular}{c | c c c}
\toprule
    $\substack{\displaystyle \text{EM System} \\[1mm] \displaystyle \text{(Dimensions)}}$ &  boundary condition  &  Spatial orthonormal basis & Quantum Observables 
\\  \midrule \midrule
\\
    LC circuit (0) &  N/A  &  N/A 
    & $\substack{
            \displaystyle \hat{\Phi} 
            = \sqrt{\frac{\hbar Z_{\text{r}0}}{2}} 
                \Big(
                    \hat{a} 
                    + \hat{a}^{\dagger}
                \Big)
            \ \text{ or } \
            \hat{\varphi}
            = \left(\frac{8E_C}{E_L}\right)^{\!\! 1/4}
            \frac{\hat{a} + \hat{a}^{\dagger}}{\sqrt{2}}
        \\[1mm]
            \displaystyle \hat{Q} 
            = -\ci \sqrt{\frac{\hbar}{2 Z_{\text{r}0}}} 
                \Big(
                    \hat{a} 
                    - \hat{a}^{\dagger}
                \Big)
            \ \text{ or } \
            \hat{n}
            = - \ci \left(\frac{E_L}{8E_C}\right)^{\!\! 1/4} \frac{\hat{a} - \hat{a}^{\dagger}}{\sqrt{2}}
        \\[1mm] 
            \displaystyle \hat{H}_{\text{LC}} 
            = \hbar \omega_{\text{r}} 
                \bigg(
                    \hat{a}^{\dagger} \hat{a} 
                    + \frac{1}{2}
                \bigg)
    }$  
\\ \\ \hline \\
    $\substack{\displaystyle \text{Transmission} \\[2mm] \displaystyle \text{line (1)}}$ &  N/A  
    & $\substack{
            \displaystyle 
            \frac{1}{\sqrt{2\pi}} e^{\ci k x} \text{ for } k \in \mathbf{R} 
        \\[1mm] 
            \displaystyle 
            \int_{-\infty}^{\infty} 
                \mathrm{d} x \,
                \Big(
                    \frac{1}{\sqrt{2\pi}} e^{-\ci k x} 
                \Big)
                \Big(
                    \frac{1}{\sqrt{2\pi}} e^{\ci k' x}
                \Big)
        = \delta(k-k')
    }$ 
    & $\substack{
            \displaystyle
            \hat{\Phi}(x) 
            = \int_{-\infty}^{\infty}  
                \mathrm{d} k  
                \sqrt{\frac{\hbar}{4\pi \omega_k C_l}}
                \left[
                    \hat{a}(k) e^{\ci kx}
                    + \hat{a}^{\dagger}(k) e^{- \ci kx}
                \right]
        \\[1mm] \displaystyle
            \hat{\Pi}(x) 
            = - \ci \int_{-\infty}^{\infty}  
                \mathrm{d} k  
                \sqrt{\frac{\hbar \omega_k C_l}{4\pi}}
                \left[
                    \hat{a}(k) e^{\ci kx} 
                    - \hat{a}^{\dagger}(k) e^{- \ci kx} 
                \right]
        \\[1mm]
            \displaystyle \hat{H}_{\text{TL}}
            = \int_{-\infty}^{\infty} 
                \mathrm{d} k \, 
                \hbar \omega_k 
                \left[
                    \hat{a}^{\dagger}(k) \hat{a}(k) 
                    + \frac{1}{2}
                \right]
        }$ 
\\ \\ \hline \\
    Cavity (3) 
    & $\substack{
        \displaystyle \hat{\mathbf{e}}_{\text{n}} \times \mathbf{E}(\mathbf{r},t) = \mathbf{0} \\[2mm] 
        \displaystyle \text{ for } \mathbf{r} \in \partial V
    }$  
    & $\substack{
            \displaystyle 
            \mathbf{f}_n(\mathbf{r}) \text{ for }  n \in \mathbf{Z}_{+} 
        \\[1mm] \displaystyle 
            \text{such that }
            \Delta \mathbf{f}_n(\mathbf{r}) + k_n^2 \mathbf{f}_n(\mathbf{r}) = \mathbf{0}
        \\[1mm] \displaystyle 
            \text{subject to the boundary condition}
        \\[1mm]
            \displaystyle 
            \int_{V} 
                \mathrm{d}^3 r \, 
                \mathbf{f}_m(\mathbf{r}) 
                \cdot 
                \mathbf{f}_n(\mathbf{r})
            = \delta_{mn} 
    }$
    & $\substack{
            \displaystyle \hat{\mathbf{E}}(\mathbf{r}) 
            = \sum_{n}
                    \ci \mathscr{E}_{0, n} \mathbf{f}_n(\mathbf{r})
                    \left(
                        \hat{a}_{n}
                        - \hat{a}_{n}^{\dagger}         
                    \right) 
        \\[1mm]
            \displaystyle \hat{\mathbf{B}}(\mathbf{r}) 
            = \sum_{n}
                    \frac{\mathscr{E}_{0, n}}{c}
                    \,
                    \frac{\nabla \times
                    \mathbf{f}_n(\mathbf{r})}{k_n}
                    \left(
                        \hat{a}_{n}
                        + \hat{a}_{n}^{\dagger}      
                    \right)
        \\[1mm]
            \displaystyle \hat{\mathbf{A}}(\mathbf{r}) 
            = \sum_{n}
                    \frac{\mathscr{E}_{0, n}}{\omega_{n}}
                    \,
                    \mathbf{f}_n(\mathbf{r})
                    \left(
                        \hat{a}_{n}
                        + \hat{a}_{n}^{\dagger}     
                    \right)
        \\[1mm] 
            \displaystyle \hat{H}_{\text{C}}
            = \sum_{n}
                \hbar \omega_{n} 
                \left(
                    \hat{a}_{n}^{\dagger}
                    \hat{a}_{n}
                    + \frac{1}{2}
                \right)
    }$
\\ \\ \hline \\
    Free Space (3) &  $\substack{\displaystyle \mathbf{E}_{\perp}(\mathbf{r},t) = \mathbf{E}_{\perp}(\mathbf{r} + \hat{\mathbf{e}}_i L,t) \\[2mm] \displaystyle \text{ for } \ \ i = x,y,z}$ 
    & $\substack{
            \displaystyle \frac{\hat{\mathbf{e}}_{\mathbf{k},\lambda}}{\sqrt{V}} e^{\ci \mathrm{k} \cdot \mathbf{r}} 
            \text{ for } \mathbf{k} 
            = \left(
                    \frac{2\pi n_x}{L}, 
                    \frac{2\pi n_y}{L}, 
                    \frac{2\pi n_z}{L}
                \right) 
        \\[1mm] 
            \displaystyle \text{ and } n_x, n_y, n_z \in \mathbf{Z} 
        \\[1mm]
            \displaystyle 
            \int_V 
                \mathrm{d}^3 r
                \left(
                    \frac{\hat{\mathbf{e}}_{\mathbf{k},\lambda}}{\sqrt{V}} e^{-\ci \mathrm{k} \cdot \mathbf{r}} 
                \right)
                \cdot
                \left(
                    \frac{\hat{\mathbf{e}}_{\mathbf{k}',\lambda'}}{\sqrt{V}} e^{\ci \mathrm{k}' \cdot \mathbf{r}} 
                \right) 
            = \delta_{\mathbf{k}\mathbf{k}'} \delta_{\lambda \lambda'}
    }$ 
    & $\substack{
            \displaystyle \hat{\mathbf{E}}_{\perp}(\mathbf{r}) 
            = \sum_{\mathbf{k},\lambda}
                \ci \mathscr{E}_{0, k} 
                \hat{\mathbf{e}}_{\mathbf{k},\lambda}
                \Big(
                    \hat{a}_{\mathbf{k},\lambda} 
                        e^{\ci\mathbf{k} \cdot \mathbf{r}}
                    - \hat{a}_{\mathbf{k},\lambda}^{\dagger}   
                        e^{- \ci\mathbf{k} \cdot \mathbf{r}}
                \Big)
        \\[1mm]
            \displaystyle \hat{\mathbf{B}}(\mathbf{r}) 
            = \sum_{\mathbf{k},\lambda}
                \frac{\ci\mathscr{E}_{0, k}}{c}
                \,
                \hat{\mathbf{e}}_{\mathbf{k}} 
                \times
                \hat{\mathbf{e}}_{\mathbf{k},\lambda}
                \Big(
                    \hat{a}_{\mathbf{k},\lambda} 
                        e^{\ci\mathbf{k} \cdot \mathbf{r}}
                    - \hat{a}_{\mathbf{k},\lambda}^{\dagger}   
                        e^{- \ci\mathbf{k} \cdot \mathbf{r}}
                \Big)
        \\[1mm]
            \displaystyle \hat{\mathbf{A}}_{\perp}(\mathbf{r}) 
            = \sum_{\mathbf{k},\lambda}
                \frac{\mathscr{E}_{0, k}}{\omega_{k}}
                \,
                \hat{\mathbf{e}}_{\mathbf{k},\lambda}
                \Big(
                    \hat{a}_{\mathbf{k},\lambda} 
                        e^{\ci\mathbf{k} \cdot \mathbf{r}}
                    + \hat{a}^{\dagger}_{\mathbf{k},\lambda} 
                        e^{-\ci\mathbf{k} \cdot \mathbf{r}}
                \Big)
        \\[1mm] 
            \displaystyle \hat{H}_{\perp}
            = \sum_{\mathbf{k},\lambda}
                \hbar \omega_{k} 
                \left(
                    \hat{a}_{\mathbf{k},\lambda}^{\dagger}
                    \hat{a}_{\mathbf{k},\lambda} 
                    + \frac{1}{2}
                \right)
    }$
\\ \\ \bottomrule
    \end{tabular}
    \caption{Normal mode expansion of isolated (i.e., sourceless) electromagnetic systems.}
    \label{table:normal_mode_expansion}
\end{sidewaystable}

\section{Uncertainty and Noise of a Quantized Field}
\subsection{Commutation Relation and Heisenberg Uncertainty}
Recall that given two Hermitian operators $\hat{A}$ and $\hat{B}$, the Heisenberg uncertainty, derived from the Cauchy-Schwarz inequality, is given by
\begin{equation} \label{eq:heisenberg_uncertainty}
    \Bigsl( \Delta \hat{A} \Bigsr)^2 
        \Bigsl( \Delta \hat{B} \Bigsr)^2 
    \geq \left(
            \frac{1}{2\ci}
                \Bigsl \langle 
                    \Bigsl[ \hat{A}, \hat{B} \Bigsr]
                \Bigsr \rangle
        \right)^2
        + \left(
                \frac{1}{2}
                    \Bigsl \langle 
                        \Bigsl\{ 
                            \hat{A}, \hat{B} 
                        \Bigsr \}
                    \Bigsr \rangle
                - \Bigsl\langle 
                        \hat{A} 
                    \Bigsr\rangle  
                    \Bigsl\langle 
                        \hat{B} 
                    \Bigsr\rangle 
            \right)^2,
\end{equation}
where the uncertainty of an operator (with respect to some state) is defined to be
\begin{equation}
     \Delta \hat{A} 
     = \Bigsl\langle 
                        \hat{A}^2 
                    \Bigsr\rangle  
        - \Bigsl\langle 
                        \hat{A} 
                    \Bigsr\rangle^2.
\end{equation}

It can be shown that the statistical correlation between quantum operators cannot be encoded by a classical joint probability distribution unless the operators commute \cite{nelson_brownian_motion_1967}. Along the same logic, Eq.(\ref{eq:heisenberg_uncertainty}) can be treated as a quantum generalization of the following inequality in classical probability theory:
\begin{equation}
    \sigma_X \sigma_Y \geq \mathrm{Cov}(X,Y).
\end{equation}
Specifically, since operators, in general, do not commute, a classical product of two observables requires symmetrization when promoted to a product of two operators, i.e.,
\begin{equation}
    AB = BA
    \ \ \longrightarrow \ \ 
    \frac{1}{2}
        \left(
            \hat{A}\hat{B}
            + \hat{B}\hat{A}
        \right).
\end{equation}
Subsequently, the quantum covariance of two operators can be defined via the mapping
\begin{equation}
    \mathrm{Cov}(A,B) 
    = \mathbb{E}(AB) 
        - \mathbb{E}(A)\mathbb{E}(B)
    \ \ \longrightarrow \ \  
    \frac{1}{2}
            \Bigsl \langle 
                \Bigsl\{ 
                    \hat{A}, \hat{B} 
                \Bigsr \}
            \Bigsr \rangle
        - \Bigsl\langle 
                \hat{A} 
            \Bigsr\rangle  
            \Bigsl\langle 
                \hat{B} 
            \Bigsr\rangle,
\end{equation}
which appears as the second term in Eq.(\ref{eq:heisenberg_uncertainty}). However, the first term in Eq.(\ref{eq:heisenberg_uncertainty}) does not have a classical counterpart and is usually the most interesting part of the uncertainty principle.

Furthermore, the idea of symmetrization is meaningful for non-Hermitian operators as well. For example, the Hamiltonian of a QHO is essentially a symmetrized variance of the annihilation operator. As a result, we get an extra $1/2$ when taking the expectation of the Hamiltonian (normalized by $\hbar \omega$)
\begin{equation} \label{eq:energy_fluctuation}
    \frac{1}{2} 
        \bra{\Psi} 
            \Big\{
                \hat{a}, \hat{a}^{\dagger}
            \Big\}
        \ket{\Psi} 
    = \frac{1}{2}
        \bra{\Psi} 
        \left(
            \hat{a}^{\dagger} \hat{a} 
            + \hat{a} \hat{a}^{\dagger}
        \right)
        \ket{\Psi} 
    = \bra{\Psi} \hat{N} \ket{\Psi} 
        + \frac{1}{2}.
\end{equation}
Following this observation, we will now discuss the zero-point fluctuation associated with a field observable.
\subsection{The Zero-Point Fluctuation}
The state we will be focusing on is the vacuum state $\ket{0}$ of the electromagnetic field since it is perhaps the simplest state of the field but still reveals the essential result of quantum fluctuation. 

By using Eq.(\ref{eq:formula_annhilation}) and (\ref{eq:formula_creation}), we find the average electric field of $\ket{0}$ to be
    \begin{equation}
        \bra{0}\hat{\mathbf{E}}_{\perp}(\mathbf{r})\ket{0}
        = \sum_{\mathbf{k},\lambda}
            \ci\mathscr{E}_{0, k} \hat{\mathbf{e}}_{\mathbf{k},\lambda}
            \left[
                \bra{0}\hat{a}_{\mathbf{k},\lambda}\ket{0}
                    e^{\ci\mathbf{k} \cdot \mathbf{r}}
                - \bra{0}\hat{a}_{\mathbf{k},\lambda}^{\dagger}\ket{0}     
                    e^{- \ci\mathbf{k} \cdot \mathbf{r}}
            \right]
        = \mathbf{0},
    \end{equation}
as suggested by the fact that $\ket{0}$ has no photon in any mode. Note that, at this point, we have not used the fact that the creation and annihilation operators in the same mode do not commute.

The next quantity to examine is, of course, the second moment of the field; since the field is a vectorial quantity, we are talking about the dot product of the field with itself. Since the annihilation and creation operators of different modes do not talk to one another, we can further simplify the dot product to
\begin{equation}
    \bra{0} 
            \hat{\mathbf{E}}_{\perp}(\mathbf{r}) 
            \cdot  
            \hat{\mathbf{E}}_{\perp}(\mathbf{r}) 
        \ket{0}
    = \sum_{\mathbf{k},\lambda} 
        \bra{0} \hat{\mathbf{E}}_{\perp,\mathbf{k},\lambda}(\mathbf{r}) \cdot  \hat{\mathbf{E}}_{\perp,\mathbf{k},\lambda}(\mathbf{r})\ket{0},
\end{equation}
where
\begin{equation}
    \hat{\mathbf{E}}_{\perp,\mathbf{k},\lambda}
    = \ci \mathscr{E}_{0, k} 
        \hat{\mathbf{e}}_{\mathbf{k},\lambda}
        \Big(
            \hat{a}_{\mathbf{k},\lambda} 
                e^{\ci\mathbf{k} \cdot \mathbf{r}}
            - \hat{a}_{\mathbf{k},\lambda}^{\dagger}   
                e^{- \ci\mathbf{k} \cdot \mathbf{r}}
        \Big)
\end{equation}
is the transverse electric field of mode $(\mathbf{k},\lambda)$; in other words, the second moment of the entire field is the sum of the second moment of each mode. For a single mode, we have
\begin{align}
    \bra{0} \hat{\mathbf{E}}_{\perp,\mathbf{k},\lambda}(\mathbf{r}) \cdot  \hat{\mathbf{E}}_{\perp,\mathbf{k},\lambda}(\mathbf{r})\ket{0}
    &= \mathscr{E}_{0, k}^2
        \bra{0} 
            \left( 
                \hat{a}_{\mathbf{k},\lambda}^{\dagger} \hat{a}_{\mathbf{k},\lambda} + \hat{a}_{\mathbf{k},\lambda} \hat{a}_{\mathbf{k},\lambda}^{\dagger}
            \right)
        \ket{0}
\nonumber \\ \label{eq:single_mode_uncertainty}
    &= \mathscr{E}_{0, k}^2
        \left( 
            \bra{0} \hat{N}_{\mathbf{k},\lambda} \ket{0} 
            + 1
        \right)
    = \mathscr{E}_{0, k}^2.
\end{align}
Eq.(\ref{eq:energy_fluctuation}) and (\ref{eq:single_mode_uncertainty}) have the same spirit, i.e., they both have an extra constant due to the fact that $\hat{a}\hat{a}^{\dagger} = \hat{a}^{\dagger} \hat{a} + 1$. Moreover, we arrive at an important fact about a quantized field: The uncertainty in a single-mode electric field, i.e.,
\begin{equation}
    \Delta     
        \Bigsl|
            \hat{\mathbf{E}}_{\perp,\mathbf{k},\lambda}
        \Bigsr| 
    \doteq \sqrt{
        \bra{0}
        \Bigsl|
            \hat{\mathbf{E}}_{\perp,\mathbf{k},\lambda}
        \Bigsr|^2
        \ket{0}
        }
    = \mathscr{E}_{0,k}
    = \sqrt{\frac{\hbar \omega_{k}}{2\epsilon_0 V}}
\end{equation}
is nonzero even for a state with no photon. This nonzero uncertainty of the observable in the ground state is known as the \textbf{zero-point fluctuation (ZPF)}. 

Following the same computation, we see that any Hermitian operator of the form
\begin{equation}
    \hat{O} = O_0 \left( e^{\ci \phi} \hat{a} + e^{-\ci \phi}\hat{a}^{\dagger} \right)
    \ \ \text{ or } \ \ 
    \ci O_0 \left( e^{\ci \phi} \hat{a} - e^{-\ci \phi}\hat{a}^{\dagger} \right)
\end{equation}
with a real non-negative number $O_0$ has a ZPF equal to $O_0$. For example, the ZPF of the vector potential and magnetic field are given, respectively, by
\begin{equation}
    \Delta 
        \Bigsl|
            \hat{\mathbf{A}}_{\perp, \mathbf{k}, \lambda}
        \Bigsr|
    = \frac{\mathscr{E}_{0, k}}{\omega_k} 
    = \sqrt{\frac{\hbar }{2 \epsilon_0 \omega_k V}}
    \ \ \text{ and } \ \ 
    \Delta 
        \Bigsl|
            \hat{\mathbf{B}}_{\mathbf{k}, \lambda}
        \Bigsr|
    = \frac{\mathscr{E}_{0, k}}{c} 
    = \sqrt{\frac{\hbar \omega_{k}}{2 \epsilon_0 c^2 V}}.
\end{equation}

\subsection{Uncertainty in the Phase Plane}
Since the electromagnetic field is a collection of QHOs, the uncertainty principle related to the conjugate variables of the field is exactly the same as that of a QHO. Hence, our job is to restate the position-momentum uncertainty in the language of quantum electrodynamics.

Since the vector potential and the electric field are conjugate variables generalizing the idea of position and momentum, the uncertainty relation should be exactly the same up to a rescaling based on the ZPF of the two quadratures. Hence, let us define the normalized quadrature operators
\begin{equation}
    \hat{X}_{\mathbf{k},\lambda} = \frac{\hat{a}_{\mathbf{k},\lambda}+\hat{a}_{\mathbf{k},\lambda}^{\dagger}}{2}
\ \ \text{ and } \ \ 
    \hat{P}_{\mathbf{k},\lambda} = \frac{\hat{a}_{\mathbf{k},\lambda}-\hat{a}_{\mathbf{k},\lambda}^{\dagger}}{2\ci}
\end{equation}
for a single mode of the electromagnetic field. Then, for any coherent state of the mode (including the vacuum state), one can easily show that each quadrature operator has an uncertainty of $1/2$, resulting in the relation
\begin{equation}
    \Delta \hat{X}_{\mathbf{k},\lambda} \Delta \hat{P}_{\mathbf{k},\lambda} = \frac{1}{4},
\end{equation}
which is the minimum uncertainty allowed by Eq.(\ref{eq:heisenberg_uncertainty}) since $\Bigsl[ \hat{X}_{\mathbf{k},\lambda}, \hat{P}_{\mathbf{k},\lambda} \Bigsr] = \ci /2$. Later, we will show that the quantum measurement of the qudits is heavily affected by this uncertainty.

\subsection{Quantum-Limited Amplification}
A related consequence of the bosonic commutation relations shows up when we try to amplify a signal. Classically, an ideal amplifier with an infinite bandwidth is a device that simply multiplies the input power by some gain $G>1$ (ignore the trivial case of a unit-gain buffer). However, as we will prove shortly, an ideal amplifier under the law of quantum mechanics cannot amplify a signal without introducing any noise.

Before working out the math, we first need to introduce the traveling-wave annihilation and creation operators, $\hat{a}^{\rightleftarrows}$ and $\hat{a}^{\rightleftarrows \dagger}$, which is closely related to the annihilation and creation operators introduced before. From classical microwave theory, we know that any signal on a transmission line can always be decomposed into the left- and right-traveling parts. We can generalize this idea easily for a quantized transmission line; consequently, we obtain a different type of annihilation and creation operators. We will introduce the concept properly in Chapter 5; for now, we only need the bosonic commutation relations
\begin{equation}
    \Big[
        \hat{a}^{\rightarrow}, \hat{a}^{\rightarrow \dagger}
    \Big] = \hat{1},
\ \ \ \ 
    \Big[
        \hat{a}^{\leftarrow}, \hat{a}^{\leftarrow \dagger}
    \Big] = \hat{1}.
\end{equation}
We have also, for simplicity, constrained ourselves to a narrow-bandwidth signal on the transmission line; nevertheless, the main point is that the traveling-wave annihilation and creation operators carry the usual bosonic commutation relations \cite{ROY2016740}.

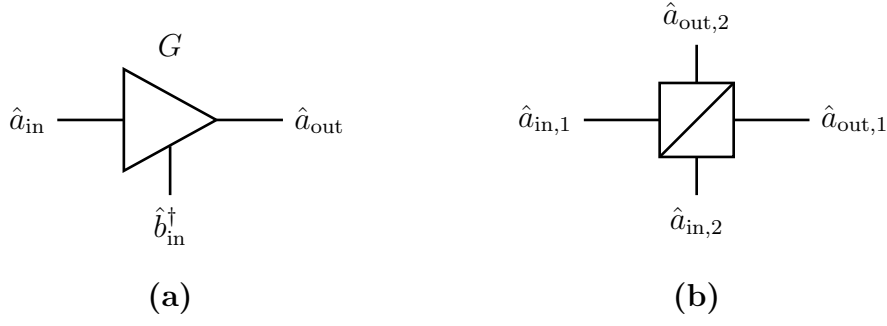
\begin{figure}
    \centering
    \begin{circuitikz}
    \draw
        (-0.5,0) to [/tikz/circuitikz/bipoles/length=50pt,amp](2.5,0)
        (1,1) node[]{$G$}
        (1,-0.35) to (1,-1);
    \draw
        (-0.5,0) node[left]{${\hat{a}_{\text{in}}}$}
        (2.5,0) node[right]{${\hat{a}_{\text{out}}}$}
        (1,-1) node[below]{${\hat{b}_{\text{in}}^{\dagger}}$};
    \draw
        (1,-2) node[below]{$$\textbf{(a)}$$};
    \begin{scope}[shift={(10,0)}]
    \draw
        (-3.5,0) to [/tikz/circuitikz/bipoles/length=40pt, twoportsplit] (-0.5,0)
        (-2,0.5) to (-2,1)
        (-2,-0.5) to (-2,-1);
    \draw
        (-3.5,0) node[left]{${\hat{a}_{\text{in,1}}}$}
        (-2,-1) node[below]{${\hat{a}_{\text{in,2}}}$}
        (-0.5,0) node[right]{${\hat{a}_{\text{out,1}}}$}
        (-2,1) node[above]{${\hat{a}_{\text{out,2}}}$};
    \draw
        (-2,-2) node[below]{$$\textbf{(b)}$$};
    \end{scope}
    \end{circuitikz}
    \caption{Diagram of \textbf{a.} a generic amplifier and \textbf{b.} a beam splitter. The energy supplier of the amplifier is not shown; for a parametric amplifier, we need both a DC bias and an RF pump.}
    \label{fig:generic_amplifier_and_splitter}
\end{figure}

We can now study the input and output of an amplifier formally as signals on two transmission lines using $\hat{a}_{\text{in}}$ and $\hat{a}_{\text{out}}$. Suppose we naively set the output to be a scaled version of the input, i.e, $\hat{a}_{\text{out}} = \sqrt{G} \, \hat{a}_{\text{in}}$ (remember that $\hat{a}_{\text{in}}^{\dagger} \hat{a}_{\text{in}}$ is proportional to the power), then we would have
\begin{equation}
    \Big[
            \hat{a}_{\text{out}}, 
            \hat{a}_{\text{out}}^{\dagger}
        \Big]
    = \Big[ 
            \sqrt{G} \, \hat{a}_{\text{in}}, 
            \sqrt{G} \, \hat{a}_{\text{in}}^{\dagger}
        \Big] 
    = G \hat{1} \neq \hat{1},
\end{equation}
which violates the commutation relation we just introduced as long as $G \neq 1$. Hence, to have a non-trivial gain, we must modify the input-output relation to 
\begin{equation}
    \hat{a}_{\text{out}}
    = \sqrt{G} \, \hat{a}_{\text{in}}
                + \hat{F},
\end{equation}
where $\hat{F}$ is some generic operator related to the amplifier and is assumed to be independent of the input signal, i.e., $\Bigsl[\hat{a}_{\text{in}}, \hat{F}\Bigsr] = 0$\footnote{This assumption results in non-degenerate amplification since the signal and idler modes are different. If, however, we assume $\Bigsl[\hat{a}_{\text{in}}, \hat{F}\Bigsr] \neq 0$, then we will discover that $\hat{F} \propto \hat{a}_{\text{in}}^{\dagger}$, resulting in the so-called degenerate amplification. A degenerate amplifier does not have a quantum limit (i.e., it does not add the extra $1/2$ photons); instead, it redistributes the noise in the two quadratures via squeezing.}. Consequently, we should have
\begin{equation}
    \hat{1} = \Big[
            \hat{a}_{\text{out}}, 
            \hat{a}_{\text{out}}^{\dagger}
        \Big]
    = \Big[ 
            \sqrt{G} \, \hat{a}_{\text{in}}
                + \hat{F}, 
            \sqrt{G} \, \hat{a}_{\text{in}}^{\dagger}
                + \hat{F}^{\dagger}
        \Big]
    = G \hat{1} + \Big[\hat{F}, \hat{F}^{\dagger}\Big],
\end{equation}
implying that $\Bigsl[\hat{F}, \hat{F}^{\dagger}\Bigsr] = (1-G) \hat{1}$. Since $1-G < 0$, we can define a new operator $\hat{b}_{\text{in}} = \hat{F}^{\dagger}/\sqrt{G-1}$ such that $\hat{b}_{\text{in}}$ also satisfies the bosonic commutation relation
\begin{equation} \label{eq:b_in_commutator}
    \Big[
            \hat{b}_{\text{in}}, 
            \hat{b}_{\text{in}}^{\dagger}
        \Big]
    = \hat{1},
\end{equation}
i.e., $\hat{b}_{\text{in}}$ can be treated as another annihilation operator (with a frequency different from that of $\hat{a}_{\text{in}}$ since $\hat{a}_{\text{in}}$ and $\hat{b}_{\text{in}}$ commute).
In conclusion, to amplify a signal by $\sqrt{G}$ in amplitude, it's required that we introduce another mode, known as the \textbf{idler mode} with a traveling-wave annihilation operator $\hat{b}_{\text{in}}$ as shown in Figure \ref{eq:b_in_commutator}(a). The output mode consequently contains both the signal and idler modes
\begin{equation} \label{eq:amp_input_output_relation}
    \hat{a}_{\text{out}} 
    = \sqrt{G} \, \hat{a}_{\text{in}} 
        + \sqrt{G-1} \, \hat{b}_{\text{in}}^{\dagger}
    = \sqrt{G} 
        \left(
            \hat{a}_{\text{in}} 
            + \sqrt{1-\frac{1}{G}} \, \hat{b}_{\text{in}}^{\dagger}
        \right);
\end{equation}
in particular, the idler mode shows up as its creation operator instead of the annihilation operator, so the average output photon number is not simply the sum of the signal and idler photon numbers due to the bosonic commutation relation defined in Eq.(\ref{eq:b_in_commutator}):
\begin{align}
    \hat{a}_{\text{out}}^{\dagger} \hat{a}_{\text{out}} 
    &= G \hat{a}_{\text{in}}^{\dagger} \hat{a}_{\text{in}} 
        + (G-1) \hat{b}_{\text{in}} \hat{b}_{\text{in}}^{\dagger}
\nonumber \\ \label{eq:quantum_noise_in_amp}
    &= G \hat{a}_{\text{in}}^{\dagger} \hat{a}_{\text{in}} 
        + (G-1) 
            \hat{b}_{\text{in}}^{\dagger}
            \hat{b}_{\text{in}} 
        + (G-1) 
\end{align}
That is, even if the idler mode is in a vacuum state, i.e., $\Bigsl \langle \hat{b}_{\text{in}}^{\dagger} \hat{b}_{\text{in}} \Bigsr \rangle = 0$, an additive term $G-1$ still shows up as unwanted quantum noise \cite{IEEEMicoMaga9134828}. We emphasize that the extra $(G-1)$ is unrelated to the concept of thermal noise whose appearance only affects $\Bigsl \langle \hat{a}_{\text{in}}^{\dagger} \hat{a}_{\text{in}} \Bigsr \rangle$ and $\Bigsl \langle \hat{b}_{\text{in}}^{\dagger} \hat{b}_{\text{in}} \Bigsr \rangle$. Therefore, an ideal amplifier subject to the law of quantum mechanics, also known as a \textbf{quantum-limited amplifier}, is one that only adds the minimal unavoidable quantum noise to the output. Using the symmetrized variance introduced before, the noise of the output field can be formally quantified to be
\begin{align}
        \left \langle 
            \frac{1}{2} 
            \Bigsl \{
                \hat{a}_{\text{out}},
                \hat{a}_{\text{out}}^{\dagger} 
            \Bigsr \}
        \right \rangle
    &= G \left[
        \left \langle 
            \frac{1}{2} 
            \Bigsl \{
                \hat{a}_{\text{in}},
                \hat{a}_{\text{in}}^{\dagger} 
            \Bigsr \}
        \right \rangle
        + \frac{G - 1}{G}
            \left(
                \left \langle 
                        \hat{b}_{\text{in}}^{\dagger} 
                        \hat{b}_{\text{in}}
                    \right \rangle
                + \frac{1}{2}
            \right)
        \right]
\nonumber \\ \label{eq:quantum_limited_amp}
    &\approx 
        G \left(\left \langle 
            \frac{1}{2} 
            \Bigsl \{
                \hat{a}_{\text{in}},
                \hat{a}_{\text{in}}^{\dagger} 
            \Bigsr \}
        \right \rangle
        + \frac{1}{2}
        \right)
\end{align}
when $\Bigsl \langle \hat{b}_{\text{in}}^{\dagger} \hat{b}_{\text{in}} \Bigsr \rangle = 0$ and $G \gg 1$.
On one hand, for an input signal with a macroscopic amplitude, the first term of Eq.(\ref{eq:quantum_limited_amp}) dominates the quantum noise, which is a constant; thus, we can essentially ignore the quantum noise and reproduce the classical input-output relation. On the other hand, for a small signal photon number, effectively half a photon is introduced in the quantum-limited case. 

\subsection{Attenuation and Quantum Efficiency}
Another frequently encountered concept in an actual experiment is the attenuation of the electromagnetic field. Attenuation could happen as the signal travels along the transmission line for a long distance or might be intentionally achieved by using an attenuator. Classically, the loss of signal can be modeled easily by some decay parameter; for example, for a lossy transmission line, the decay of a single-frequency right-traveling wave is given by
\begin{equation}
    V^{\rightarrow}(x,t) 
    = V^{\rightarrow}_0 
            e^{-\alpha x} 
            \cos (\beta x - \omega t),
\end{equation}
where the propagation constant $\gamma$ derived from the transmission line model is separated into the decay constant $\alpha > 0$ and the wavenumber $\beta > 0$ \cite{pozar1990microwave}:
\begin{equation}
    \gamma(\omega) 
    = \sqrt{(R_l + \ci \omega L_l)(G_l + \ci \omega C_l)}
    = \alpha(\omega) + \ci \beta(\omega).
\end{equation}

However, to preserve a unitary time evolution in quantum mechanics, something else must also be injected into the attenuated signal, similar to how quantum-limited amplification works. The model we use is that of a beam splitter, which is, in general, a four-port device with two possible inputs and two possible outputs. Due to energy conservation, a classical beam splitter is described by a unitary matrix
\begin{equation}
    U_{\text{BS}}
    = \begin{pmatrix}
            \cos \alpha & \sin \alpha \\[1mm]
            \sin \alpha & -\cos \alpha
        \end{pmatrix}
    = \begin{pmatrix}
            \sqrt{\eta} & \sqrt{1-\eta} \\[1mm]
            \sqrt{1-\eta} & -\sqrt{\eta}
        \end{pmatrix};
\end{equation}
that is, the output electric fields of the two ports are computed by a simple matrix multiplication
\begin{equation} \label{eq:beam_splitter_eqn_classical}
    \begin{pmatrix}
            E_{\text{out},1} \\[1mm]
            E_{\text{out},2}
        \end{pmatrix}
    = U_{\text{BS}} 
        \begin{pmatrix}
            E_{\text{in},1} \\[1mm]
            E_{\text{in},2}
        \end{pmatrix}
    = \begin{pmatrix}
            \sqrt{\eta} \, E_{\text{in},1} + \sqrt{1-\eta} \, E_{\text{in},2} \\[1mm]
            \sqrt{1-\eta} \, E_{\text{in},1} -\sqrt{\eta} \, E_{\text{in},2}
        \end{pmatrix}.
\end{equation}
Hence, for a classical beam-splitter, if $E_{\text{in,2}} = 0$, $E_{\text{out,1}} = \sqrt{\eta} \, E_{\text{in},1}$, which models a power attenuation of $1-\eta$ (since power is proportional to $|E|^2$.) Note that we are just using the physics of a beam splitter as a way to describe loss with a unitary transformation. In the optical domain, a beam splitter is just a partially transmissive and partially reflective interface while in the microwave domain, a ``splitter'' can be realized, for example, by a directional coupler. Despite the different physical realizations, the mathematical relation between the inputs and outputs is captured by Eq.(\ref{eq:beam_splitter_eqn_classical}).

Given that the input-output transformation is already unitary in the classical definition, we propose to apply it directly to quantum operators as shown in Figure \ref{fig:generic_amplifier_and_splitter}(b); that is,
\begin{equation}
    \begin{pmatrix}
            \hat{a}_{\text{out},1} \\[1mm]
            \hat{a}_{\text{out},2}
        \end{pmatrix}
    = \hat{U}_{\text{BS}} 
        \begin{pmatrix}
            \hat{a}_{\text{in},1} \\[1mm]
            \hat{a}_{\text{in},2}
        \end{pmatrix}
    = \begin{pmatrix}
            \sqrt{\eta} \, \hat{a}_{\text{in},1} + \sqrt{1-\eta} \, \hat{a}_{\text{in},2} \\[1mm]
            \sqrt{1-\eta} \, \hat{a}_{\text{in},1} -\sqrt{\eta} \, \hat{a}_{\text{in},2}
        \end{pmatrix}.
\end{equation}
In addition, we assume that the annihilation/creation operators of the two input fields are independent and satisfy the commutation relations
\begin{equation}
    \Bigsl [ 
            \hat{a}_{\text{out},i}
            , \hat{a}^{\dagger}_{\text{out},j} 
        \Bigsr] 
    = \delta_{ij},
\end{equation}
\begin{equation}
    \Bigsl [ 
            \hat{a}_{\text{out},i}, 
            \hat{a}_{\text{out},j} 
        \Bigsr ] 
    = 0.
\end{equation}
There is rich physics born out of these simple assumptions; nevertheless, currently, we focus on using the beam splitter only as an attenuator.

To define an attenuator, we clearly only need one input and one output out of the four ports of a beam splitter; however, unlike the classical case where the second input can be turned off completely (i.e., $E_{\text{in}, 2} = 0$), a quantum-mechanical attenuator is described by the operator equation
\begin{equation} \label{eq:attenuation_eqn_quantum}
    \hat{a}_{\text{out},1} 
    = \sqrt{\eta} \, \hat{a}_{\text{in},1} + \sqrt{1-\eta} \, \hat{a}_{\text{in},2}
\end{equation}
and, thus, is subject to various commutation relations and quantum fluctuation from both ports. By comparing Eq.(\ref{eq:amp_input_output_relation}) and (\ref{eq:attenuation_eqn_quantum}), we observe a common theme in the discussion of quantum amplification and attenuation; that is, a signal cannot be modified without introducing an idler mode. To verify that Eq.(\ref{eq:attenuation_eqn_quantum}) indeed describes an attenuator, we compute the average photon flux at the output (which is proportional to the average output power)
\begin{align}
    \bar{n}_{\text{out},1} 
    &= \Bigsl\langle
            \hat{a}_{\text{out},1}^{\dagger} 
            \hat{a}_{\text{out},1}
        \Bigsr\rangle
\nonumber \\
    &= \eta \, 
            \Bigsl\langle 
                \hat{a}_{\text{in},1}^{\dagger} 
                \hat{a}_{\text{in},1}
            \Bigsr\rangle
        + (1 - \eta) \, 
            \Bigsl\langle 
                \hat{a}_{\text{in},2}^{\dagger} \hat{a}_{\text{in},2}
            \Bigsr\rangle
\nonumber \\
    &= \eta 
            \, \bar{n}_{\text{in},1} 
        + (1 - \eta) 
            \, \bar{n}_{\text{in},2}.
\end{align}
Hence, if the second input is in the zero-photon state, i.e., any input state of the form $\ket{\psi_{\text{in,1}}} \otimes \ket{0_{\text{in,2}}}$, we reproduce the behavior of an attenuator.

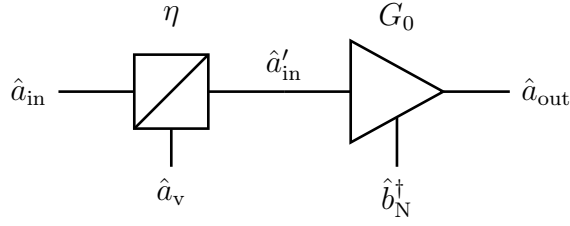
\begin{figure}
    \centering
    \begin{circuitikz}
    \draw
        (-0.5,0) to [/tikz/circuitikz/bipoles/length=50pt,amp](2.5,0)
        (1,1) node[]{$G_0$}
        (1,-0.35) to (1,-1);
    \draw 
        (-3.5,0) to [/tikz/circuitikz/bipoles/length=40pt, twoportsplit] (-0.5,0)
        (-2,1) node[]{${\eta}$}
        (-2,-0.5) to (-2,-1);
    \draw
        (-3.5,0) node[left]{${\hat{a}_{\text{in}}}$}
        (-2,-1) node[below]{${\hat{a}_{\text{v}}}$}
        (-0.5,0) node[above]{${\hat{a}_{\text{in}}'}$}
        (2.5,0) node[right]{${\hat{a}_{\text{out}}}$}
        (1,-1) node[below]{${\hat{b}_{\text{N}}^{\dagger}}$};
    \end{circuitikz}
    \caption{An amplifier with cable losses in front of the input port.}
    \label{fig:noisy_generic_amplifier}
\end{figure}

As an example, let us add the attenuation of cabling to the amplifier input; in other words, we will connect an attenuator to the input port of the amplifier as shown in Figure \ref{fig:noisy_generic_amplifier}. Then, the output is given by
\begin{align}
    \hat{a}_{\text{out}}
    &= \sqrt{G_0} \,
            \hat{a}_{\text{in}}'
        + \sqrt{G_0 - 1} \, 
            \hat{b}_{\text{N}}^{\dagger}
\nonumber \\
    &= \sqrt{G_0} 
        \Big(
            \sqrt{\eta} \, \hat{a}_{\text{in}} 
            + \sqrt{1 - \eta} \,  \hat{a}_{\text{v}}
        \Big)
        + \sqrt{G_0 - 1} \, \hat{b}_{\text{N}}^{\dagger}
\nonumber \\
    &= \sqrt{\eta G_0} \, \hat{a}_{\text{in}} 
        + \sqrt{(1 - \eta) G_0} \,  \hat{a}_{\text{v}}
        + \sqrt{G_0 - 1} \, \hat{b}_{\text{N}}^{\dagger}
\end{align}
with an average photon number
\begin{align}
    \bar{n}_{\text{out}} 
    &= \Bigsl\langle
            \hat{a}_{\text{out}}^{\dagger} 
            \hat{a}_{\text{out}}
        \Bigsr\rangle
    = \Bigsl\langle
                \hat{a}_{\text{in}}^{\dagger} 
                \hat{a}_{\text{in}}
            \Bigsr\rangle
        + \Bigsl\langle
                \hat{a}_{\text{v}}^{\dagger} 
                \hat{a}_{\text{v}}
            \Bigsr\rangle
        + \Bigsl\langle
                \hat{b}_{\text{N}}
                \hat{b}_{\text{N}}^{\dagger} 
            \Bigsr\rangle
\nonumber \\
   &= \eta G_0 \bar{n}_{\text{in}} 
        + (G_0 - 1) (\bar{n}_{\text{N}} + 1).
\end{align}
Similar to the characterization of a classical noisy amplifier, we can define the noise figure to be the ratio of the total output noise power to the amplified input noise power, i.e.,
\begin{align} \label{eq:noise_figure_single}
    F 
    = \frac{\bar{n}_{\text{out}}}{G \bar{n}_{\text{in}}}
    = 1 + \frac{G_0 - 1}{\eta G_0} (\bar{n}_{\text{N}} + 1).
\end{align}
(An alternative definition can be given based on the symmetrized noise, but the result does not change much.) However, the classical noise figure is defined with respect to a temperature, whereas Eq.(\ref{eq:noise_figure_single}) does not have any reference to the temperature of the system. Hence, Eq.(\ref{eq:noise_figure_single}) works when the thermal noise is negligible and the quantum noise dominates. 

Such a model can be easily generalized to a concatenation of amplifiers and cables; for instance, the output mode of a two-stage amplifier with lossy cabling can be calculated to be \cite{RevModPhys.93.025005}
\begin{align}
    \hat{a}_{\text{out}}
    &= \sqrt{G_{0,2}}
        \Bigg\{ 
            \sqrt{\eta_2} 
                \bigg[
                    \sqrt{G_{0,1}} 
                    \Big(
                        \sqrt{\eta_1} \,    
                            \hat{a}_{\text{in}} 
                        + \sqrt{1 - \eta_1} \,  
                            \hat{a}_{\text{v,1}}
                    \Big)
                    + \sqrt{G_{0,1} - 1} \, 
                        \hat{b}_{\text{N,1}}^{\dagger}
                \bigg]
\nonumber \\
    & \ \ \ \ \ \ \ \ \ \ \ \ \ \ \ \ \ \ \ \ \ \ \ \ \ \ \ \ \ \ \ \ \ \ \ \ \ \ \ \ \ \ \ \ \ \ \ \ \ \ \ \ \ \ \ \ \ \ \ \ \ \ \
            + \sqrt{1 - \eta_2} \,
                \hat{a}_{\text{v,2}}
        \Bigg\}
        + \sqrt{G_{0,2} - 1} \, \hat{b}_{\text{N,2}}^{\dagger}
\nonumber \\
    &= \sqrt{\eta_1 G_{0,1} \eta_2 G_{0,2}} \, \hat{a}_{\text{in}} 
        + \sqrt{(1-\eta_1) G_{0,1} \eta_2 G_{0,2}} \, \hat{a}_{\text{v,1}}
        + \sqrt{(1-\eta_2) G_{0,2}} \, \hat{a}_{\text{v,2}}
\nonumber \\
    & \ \ \ \ \ \ \ \ \ \ \ \ \ \ \ \ \ \ \ \ \ \ \ \ \ \ \ \ \ \ \ \ \ \ \ \ \ \ \ \ \ \ \ \ \ \ \ \ \ \ \ \ 
        + \sqrt{(G_{0,1} - 1) \eta_2 G_{0,2}} \, \hat{b}_{\text{N,1}}^{\dagger}
        + \sqrt{G_{0,2} - 1} \, \hat{b}_{\text{N,2}}^{\dagger},
\end{align}
resulting in the output photon number
\begin{align}
    \bar{n}_{\text{out}} 
    &= \Bigsl\langle
            \hat{a}_{\text{out}}^{\dagger} 
            \hat{a}_{\text{out}}
        \Bigsr\rangle
\nonumber\\ \label{eq:two_stage_output_photons}
    &= \eta_1 G_{0,1} \eta_2 G_{0,2} 
        \bar{n}_{\text{in}} 
        + (G_{0,1} - 1) \eta_2 G_{0,2}
            (\bar{n}_{\text{N,1}} + 1)
        + (G_{0,2} - 1)
            (\bar{n}_{\text{N,2}} + 1)
\\
    &= G \bar{n}_{\text{in}} 
        + (G - 1)
            \left[
                \frac{(G_{0,1} - 1) \eta_2 G_{0,2}}{G - 1} (\bar{n}_{\text{N,1}} + 1)
                + \frac{G_{0,2} - 1}{G - 1}
                 (\bar{n}_{\text{N,2}} + 1)
            \right],
\end{align}
where $G = \eta_1 G_1 \eta_2 G_2 $ is the total gain. By using Eq.(\ref{eq:two_stage_output_photons}) and defining, based on Eq.(\ref{eq:noise_figure_single}), the single-stage noise figure
\begin{equation}
    F_i 
    = 1 + \frac{G_{0,i} - 1}{\eta_i G_{0,i}} (\bar{n}_{\text{N},i} + 1)
    \ \ \text{ for } \ \ 
    i = 1,2,
\end{equation}
the total noise figure can be expressed as 
\begin{align}
    F 
    = \frac{
        \bar{n}_{\text{out}} 
    }{
        G \bar{n}_{\text{in}}
    }
    = 1 
        + \frac{G_{0,1} - 1}{G_1}
            (\bar{n}_{\text{N,1}} + 1)
        + \frac{G_{0,2} - 1}{G_1 G_2}
            (\bar{n}_{\text{N,2}} + 1)
    = F_1 + \frac{F_2 - 1}{G_1},
\end{align}
which clearly satisfies the Friis formula from classical microwave theory
\begin{equation}
    F = F_1 
        + \sum_{i=2}^{M}
            \frac{F_{i} - 1}{\prod_{j=1}^{i-1} G_j}.
\end{equation}

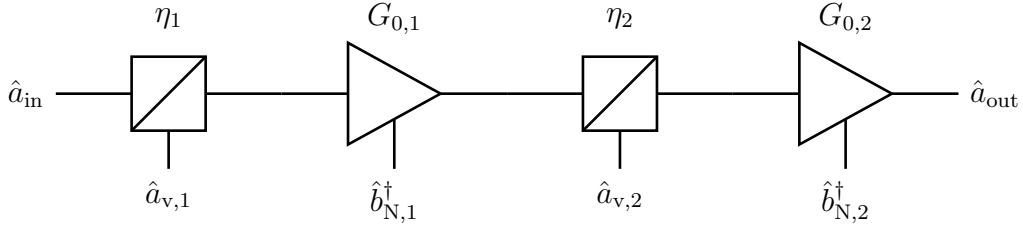
\begin{figure}
    \centering
    \begin{circuitikz}
    \draw 
        (-9.5,0) to [/tikz/circuitikz/bipoles/length=40pt, twoportsplit] (-6.5,0)
        (-8,1) node[]{${\eta_1}$}
        (-8,-0.5) to (-8,-1);
    \draw
        (-6.5,0) to [/tikz/circuitikz/bipoles/length=50pt,amp](-3.5,0)
        (-5,1) node[]{${G_{0,1}}$}
        (-5,-0.35) to (-5,-1);
    \draw 
        (-3.5,0) to [/tikz/circuitikz/bipoles/length=40pt, twoportsplit] (-0.5,0)
        (-2,1) node[]{${\eta_2}$}
        (-2,-0.5) to (-2,-1);
    \draw
        (-0.5,0) to [/tikz/circuitikz/bipoles/length=50pt,amp](2.5,0)
        (1,1) node[]{${G_{0,2}}$}
        (1,-0.35) to (1,-1);
    \draw
        (-9.5,0) node[left]{${\hat{a}_{\text{in}}}$}
        (-8,-1) node[below]{${\hat{a}_{\text{v,1}}}$}
        (-2,-1) node[below]{${\hat{a}_{\text{v,2}}}$}
        (2.5,0) node[right]{${\hat{a}_{\text{out}}}$}
        (-5,-1) node[below]{${\hat{b}_{\text{N,1}}^{\dagger}}$}
        (1,-1) node[below]{${\hat{b}_{\text{N,2}}^{\dagger}}$};
    \end{circuitikz}
    \caption{Cascading two non-ideal amplifiers.}
    \label{fig:effective noise_calculation}
\end{figure}

\subsection{The Classical Limit}
After introducing the idea of a coherent state, it is perhaps not surprising that we can approximate the quantized field with a classical normal mode when the field amplitude is sufficiently large. Recall that in the Heisenberg picture, the annihilation operator (in the absence of any sources) essentially follows the Euler-Lagrange equation; thus, 
\begin{equation}
    \hat{a}(t) \ket{\alpha(0)} 
    = e^{-\ci \omega t} \hat{a} \ket{\alpha(0)}
    = \alpha(0) e^{-\ci \omega t} \ket{\alpha(0)},
\end{equation}
Taking the classical limit, thus, amounts to assuming that we always have a coherent state and then replacing the annihilation operator with its (time-dependent) eigenvalue. In addition, the bosonic commutation relation and the quantum noise come into the picture since the coherent states are not eigenstates of the creation operator; thus,
\begin{equation}
    \hat{a}^{\dagger}(t) \ket{\alpha(0)}
    = e^{\ci \omega t} \hat{a}^{\dagger} \ket{\alpha(0)}
    \approx \alpha^*(0) e^{\ci \omega t} \ket{\alpha(0)}
\end{equation}
is only an approximation. Nevertheless, as we have seen before in several contexts, the extra constant resulting from switching the order of $\hat{a}$ and $\hat{a}^{\dagger}$ is negligible for large field amplitude.

Consequently, any operator defined in terms of $\hat{a}$ and $\hat{a}^{\dagger}$ can also be mapped to a classical expression. For example, a single-mode electric field in the classical limit is given by
\begin{align}
    \mathbf{E}_{\perp}(\mathbf{r},t)
    &\approx 
        \ci \mathscr{E}_{0, k} 
        \hat{\mathbf{e}}_{\mathbf{k},\lambda}
        \left[
            \alpha(0) e^{-\ci \omega t}
                e^{\ci\mathbf{k} \cdot \mathbf{r}}
            - \alpha^{*}(0) e^{\ci \omega t} 
                e^{- \ci\mathbf{k} \cdot \mathbf{r}}
        \right]
\nonumber \\ \label{eq:classical_limit_electric field}
    &= - \hat{\mathbf{e}}_{\mathbf{k},\lambda} E_0 
        \sin(\mathbf{k} \cdot \mathbf{r} - \omega t + \phi)
\end{align}
where $\alpha(0) = \sqrt{n} e^{\ci \phi}$ and 
\begin{equation} \label{eq:connecting_classical_and_quantum_field}
    E_0 = 2\mathscr{E}_{0, k} \sqrt{n}.
\end{equation}

%% file: include_chapters/chapter_light_matter_interaction.tex
\chapter{Control of Quantum Systems}
\section{Light-Matter Interaction with a Quantized Field}
\subsection{The Electrical Dipole Approximation}
The theory developed so far assumes the absence of any energy source and sink. To use the electromagnetic field as a medium for information processing, we bring the particles back to the picture and use the field as a means of controlling or reading out the quantum states of the particles. 

As shown before, the full Hamiltonian that contains the particles, fields, and their interaction (ignoring the spin of the particles) is given by
\cite{cohen1992atom}
\begin{align}
    \hat{H}
    &= \sum_{\alpha} 
        \hat{H}_{\alpha}
        + \hat{H}_{||} 
        + \hat{H}_{\perp} 
\nonumber \\ \label{eq:full_particle_field_Hamiltonian}
    &= \sum_{\alpha} 
        \frac{1}{2m_{\alpha}}
        |\hat{\mathbf{p}}_{\alpha} - q_{\alpha} \hat{\mathbf{A}}(\hat{\mathbf{r}}_{\alpha})|^2
    + \hat{H}_{||} 
    + \sum_{\mathbf{k},\lambda} 
        \hbar \omega_{k} 
            \left(
                \hat{a}_{\mathbf{k},\lambda}^{\dagger} 
                \hat{a}_{\mathbf{k},\lambda}
                + \frac{1}{2}
            \right).
\end{align}
If there was no interaction between the matter and fields, we would have a separable Hamiltonian
\begin{equation}
    \hat{H}_0 
    = \underbrace{\sum_{\alpha} 
            \frac{|\hat{\mathbf{p}}_{\alpha}|^2}{2m_{\alpha}}
        + \hat{H}_{||} }_{\hat{H}_{\text{p}}}
        + \underbrace{\sum_{\mathbf{k},\lambda} 
            \hbar \omega_{k} 
                \left(
                    \hat{a}_{\mathbf{k},\lambda}^{\dagger} 
                    \hat{a}_{\mathbf{k},\lambda}
                    + \frac{1}{2}
                \right)}_{\hat{H}_{\perp}},
\end{equation}
where the particle Hamiltonian $\hat{H}_{\text{p}}$ and the transverse-field Hamiltonian $\hat{H}_{\perp}$ can be solved independently. Consequently, the technique of perturbation theory and the interaction picture can be adopted if the interaction Hamiltonian $\hat{H}_{\text{int}}$ is additive to $\hat{H}_0$, i.e.,
\begin{equation}
    \hat{H} 
    = \hat{H}_0 + \hat{H}_{\text{int}}
    = \hat{H}_{\text{p}} + \hat{H}_{\perp} + \hat{H}_{\text{int}}.
\end{equation}
Looking at Eq.(\ref{eq:full_particle_field_Hamiltonian}), one might expand the quadratic term and define (in the Coulomb gauge where $\hat{\mathbf{p}}_{\alpha} \cdot \hat{\mathbf{A}}(\mathbf{r}_{\alpha}) \propto \nabla_{\mathbf{r}_{\alpha}} \cdot \hat{\mathbf{A}}(\mathbf{r}_{\alpha}) = 0$)
\begin{equation} \label{eq:interaction_hamiltonian_expanded}
    \hat{H}_{\text{int}} 
    = \underbrace{
            \sum_{\alpha} 
                \frac{q_{\alpha}}{m_{\alpha}}
                \hat{\mathbf{A}} (\hat{\mathbf{r}}_{\alpha}) 
                \cdot 
                \hat{\mathbf{p}}_{\alpha} 
        }_{\hat{H}_{\text{int},1} }
        + \underbrace{
            \sum_{\alpha} 
                \frac{q_{\alpha}^2}{2 m_{\alpha}} 
                \hat{\mathbf{A}}^2 (\hat{\mathbf{r}}_{\alpha})
        }_{\hat{H}_{\text{int},2} }.
\end{equation}

A cleaner way of analyzing Eq.(\ref{eq:full_particle_field_Hamiltonian}) exists with the introduction of a (time-independent) unitary transformation
\begin{equation}
    \hat{T} = \exp\left[-\frac{\ci}{\hbar} \sum_{\alpha} \hat{\mathbf{r}}_{\alpha}\cdot q_{\alpha} \hat{\mathbf{A}}(\mathbf{0}) \right] 
    = \exp\left[-\frac{\ci}{\hbar} \hat{\mathbf{d}} \cdot \hat{\mathbf{A}}(\mathbf{0}) \right] ,
\end{equation}
where $\hat{\mathbf{d}} = \sum_{\alpha} q_{\alpha} \hat{\mathbf{r}}_{\alpha}$ is the electric dipole operator related to the charge distribution. Recall that the momentum translation operator for one particle is of the form $\exp \!\big(-\ci \mathbf{p} \cdot \hat{\mathbf{r}} / \hbar \big)$; thus, $\hat{T}$ has the effect of shifting each momentum $\hat{\mathbf{p}}_{\alpha}$ by $q_{\alpha} \hat{\mathbf{A}}(\mathbf{0})$, removing the coupling in the quadratic term of Eq.(\ref{eq:full_particle_field_Hamiltonian}). 

But, there is still one caveat -- in Eq.(\ref{eq:full_particle_field_Hamiltonian}), we have $q_{\alpha} \hat{\mathbf{A}}(\hat{\mathbf{r}}_{\alpha})$ instead of $q_{\alpha} \hat{\mathbf{A}}(\mathbf{0})$. Hence, we further make the assumption that the region where the particles can move is much smaller than the wavelength of the field, then the field can be treated as a constant from the perspective of the particles; that is, we will make the \textbf{long-wavelength approximation}
\begin{equation}
    \hat{\mathbf{A}}(\hat{\mathbf{r}}_{\alpha})
    \approx \hat{\mathbf{A}}(\mathbf{0})
    \ \ \text{ and } \ \ 
    \hat{\mathbf{E}}_{\perp}(\hat{\mathbf{r}}_{\alpha})
    \approx \hat{\mathbf{E}}_{\perp}(\mathbf{0}),
\end{equation}
where have redefined the origin so that the particles locate around $\mathbf{r} = \mathbf{0}$. Typically, the size of a superconducting qubit is on the order of $100 \, \mu\text{m}$, whereas the wavelength of the microwave field is about $5 \,\text{cm}$; therefore, the long-wavelength approximation is valid for our application.

By making the approximation and applying the transformation $\hat{T}$, we obtain
\begin{equation}
    \hat{H}' 
    = \hat{T} \hat{H} \hat{T}^{\dagger}
    = \underbrace{
            \sum_{\alpha} 
                \frac{|\hat{\mathbf{p}}_{\alpha}|^2}{2m_{\alpha}}
            + \hat{H}_{||} 
            + \hat{H}_{\text{dipole}}
        }_{\hat{H}_{\text{p}}}
        + \underbrace{
            \sum_{\mathbf{k},\lambda} 
                \hbar \omega_{k} 
                \left(
                    \hat{a}_{\mathbf{k},\lambda}^{\dagger} 
                    \hat{a}_{\mathbf{k},\lambda}
                    + \frac{1}{2}
                \right)
        }_{\hat{H}_{\perp}}
        \underbrace{
            - \, \hat{\mathbf{d}} \cdot \hat{\mathbf{E}}_{\perp} (\mathbf{0})
        }_{\hat{H}_{\text{int}}}.
\end{equation}
As expected, $q_{\alpha} \hat{\mathbf{A}}(\hat{\mathbf{r}}_{\alpha})$ is eliminated by $\hat{T}$. Nevertheless, $\hat{T}$ also interacts with the field\footnote{In fact, after some manipulation, we can rewrite $\hat{T}$ as a product of displacement operators $\bigotimes_{\mathbf{k},\lambda}\hat{D}_{\mathbf{k},\lambda}(\hat{\Lambda}_{\mathbf{k},\lambda})$ with
\begin{equation}
    \hat{\Lambda}_{\mathbf{k},\lambda} = \frac{-\ci}{\sqrt{2\epsilon_0 \hbar \omega_k V}} \hat{\mathbf{d}} \cdot \hat{\mathbf{e}}_{\mathbf{k},\lambda}.
\end{equation}}; the net effect is the creation of an electric dipole self-energy $\hat{H}_{\text{dipole}}$ and the electric dipole interaction $- \, \hat{\mathbf{d}} \cdot \hat{\mathbf{E}}_{\perp} (\mathbf{0})$. Since $\hat{H}_{\text{dipole}}$ only depends on the operators associated with the particles, it can be grouped into $\hat{H}_{\text{p}}$; as a result, we arrive at a simple interaction Hamiltonian in the transformed frame
\begin{equation}
    \hat{H}_{\text{int}} = - \, \hat{\mathbf{d}} \cdot \hat{\mathbf{E}}_{\perp} (\mathbf{0}).
\end{equation}
More importantly, due to the long-wavelength approximation, the interaction only depends on the field evaluated at $\mathbf{r} = \mathbf{0}$.

\begin{exercise}\label{ex:3_1}
Show that $\hat{H}_{\text{int},2}$ in Eq.(\ref{eq:interaction_hamiltonian_expanded}) is negligible compared to $\hat{H}_{\text{int},1}$ when the photon momentum $\hbar |\mathbf{k}|$ of the field is much smaller than the particle momentum $\hat{\mathbf{p}}$. Next, expand $e^{\ci \mathbf{k} \cdot \mathbf{r}} = 1 + \ci \mathbf{k} \cdot \mathbf{r} + (\ci \mathbf{k} \cdot \mathbf{r})^2 + \cdots$ in $\hat{\mathbf{A}}$ (see Eq.(\ref{eq:vector_potential_quantum_operator})) to show that, under the RWA, 
\begin{equation}
    \hat{H}_{\text{int},1} 
    =  - \, \hat{\mathbf{d}} 
                \cdot 
                \hat{\mathbf{E}}_{\perp}(\mathbf{0})
        - \hat{\boldsymbol{\mu}} \cdot \hat{\mathbf{B}}(\mathbf{0}) 
        - \frac{1}{6}
            \sum_{i,j} 
                \hat{Q}_{ij} 
                \frac{\partial \hat{E}_{\perp,j}}{\partial r_i}(\mathbf{0})
        + \cdots,
\end{equation}
where
\begin{gather}
    \hat{\mathbf{d}} 
    = \sum_{\alpha} q_{\alpha} \hat{\mathbf{r}}_{\alpha},
\\
    \hat{\boldsymbol{\mu}}
    = \sum_{\alpha} 
        \frac{{q}_{\alpha}}{2m_{\alpha}} 
        \hat{\mathbf{r}}_{\alpha} 
        \times 
        \hat{\mathbf{p}}_{\alpha},
\\
    \hat{Q}_{ij}
    = \sum_{\alpha}
        q_{\alpha}
            \left(
                3 \hat{r}_{\alpha, i} \hat{r}_{\alpha, j}
                - \delta_{ij} |\hat{\mathbf{r}}_{\alpha}|^2
            \right)
    \ \ \text{ for } \ \ 
    i,j = x,y,z
\end{gather}
are the electric dipole, magnetic dipole, and electric quadrupole operators, respectively. In other words, the quantum-mechanical interaction generalizes the classical multipole interaction. (Also, note that $\hat{\mathbf{A}}_{\perp} = \hat{\mathbf{A}}$ in the Coulomb gauge.)
\end{exercise}

\subsection{Two-Level Systems}
\begin{figure}
    \centering
    \begin{circuitikz}[scale=0.7]
    \draw 
        (0,0) ellipse (4 and 4);
    \draw[dashed]
        (0,0) ellipse (4 and 1.3);
    \draw[-{Latex[length=2mm]}] 
        (0,0) to (-2.1, -1.7) node[below]{$x$};
    \draw[-{Latex[length=2mm]}] 
        (0,0) to (4.8, 0) node[below]{$y$};
    \draw[-{Latex[length=2mm]}] 
        (0,0) to (0,4.8) node[right]{$z$};
    \draw[line width=0.8mm, -{Latex[length=3.2mm]}] 
        (0,0) to (1.4,1.7) node[right]{${\mathbf{r}}$};
    \draw[dashed]
        (1.4, 1.7) to (1.4, -0.92)
        (0,0) to (1.4, -0.92);
    \draw 
        (-0.3,-0.27) arc (255:290:1.2 and 0.4)
        (0,-0.5) node[]{$\phi$};
    \draw 
        (0,0.6) arc (90:50:0.4 and 1.2)
        (0.2,0.8) node[]{$\theta$};
    \end{circuitikz}
    \caption{Representing a two-level system as a unit vector on the Bloch sphere.}
    \label{fig:bloch_sphere}
\end{figure}
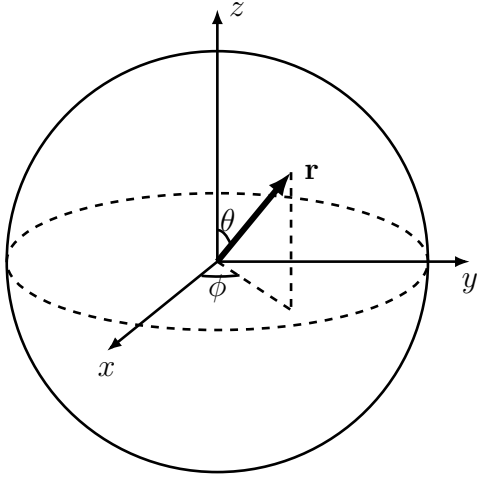

At this point, we restrain from discussing how to realize a qubit or qudit; instead, we will assume an ideal two-level system and try to understand the way we can control or infer the state of the two-level system. For the qudit case, it turns out that the techniques for manipulating a qubit still apply there, so we will first focus on a two-level system, whose states can be generally expressed as
\begin{equation}
    \ket{\Psi} 
    = \alpha \ket{0} + \beta \ket{1}
\end{equation}
for two orthonormal basis vectors $\{\ket{0}, \ket{1}\}$. For the state to be normalized, we must always have $\abs{\alpha}^2 + \abs{\beta}^2 = 1$. Moreover, it's not hard to show that a qubit state can also be written, up to a global phase, as
\begin{equation} \label{eq:general_state_Bloch_vector}
     \ket{\Psi} 
    = \cos \frac{\theta}{2} \ket{0} + e^{\ci \phi}\sin \frac{\theta}{2}\ket{1},
\end{equation}
where $\theta \in [0, \pi]$ and $\phi \in [0, 2\pi)$. As a result, we can encode any qubit state as a unit vector $\mathbf{r} = (r=1, \theta, \phi)$ on a unit sphere, also known as the Bloch sphere, as shown in Figure \ref{fig:bloch_sphere}. For example, $\ket{0}$ and $\ket{1}$ can be mapped to the north and south poles, respectively, whereas the state $\ket{+} \doteq (\ket{0} + \ket{1})/\sqrt{2}$ is a unit vector pointing in the $x$-direction.

To study the dynamics of the state, we also need the observables. Recall that the Pauli matrices (including the $2 \times 2$ identity matrix), i.e.,
\begin{equation}
    \text{\small $   
        \hat{\sigma}_0 
        = \hat{1} 
        = \begin{pmatrix}
                1 & 0 \\
                0 & 1
            \end{pmatrix},
        \ \ 
        \hat{\sigma}_1 
        = \hat{\sigma}_x 
        = \begin{pmatrix}
                0 & 1 \\
                1 & 0
            \end{pmatrix},
        \ \ 
        \hat{\sigma}_2
        = \hat{\sigma}_y 
        = \begin{pmatrix}
                0 & -\ci \\
                \ci & 0
            \end{pmatrix},
        \ \     
        \hat{\sigma}_3 
        = \hat{\sigma}_z 
        = \begin{pmatrix}
                1 & 0 \\
                0 & -1
            \end{pmatrix},
    $}
\end{equation}
form a complete basis for the space of all Hermitian matrices on a two-dimensional Hilbert space $\mathscr{H} = \mathbf{C}^2$. To be concrete, we will define ${\ket{0}}$ and ${\ket{1}}$ to be the (orthonormal) eigenvectors of $\hat{\sigma}_z$ with eigenvalues $1$ and $-1$, respectively; that is,
\begin{equation}
    \ket{0} 
    = \begin{pmatrix}
            1 \\
            0 
        \end{pmatrix}
    \ \ \text{ and } \ \ 
    \ket{1} 
    = \begin{pmatrix}
            0 \\
            1 
        \end{pmatrix}.
\end{equation}
For short, we will refer to the basis above, $\beta_{z} \doteq \{ \ket{0}, \ket{1} \}$, as the $z$-basis and the eigenbasis of $\hat{\sigma}_x$ and $\hat{\sigma}_y$ as the $x$-basis and $y$-basis, respectively. By the spectral theorem, we can now write $\hat{\sigma}_z = \ket{0} \! \bra{0} - \ket{1} \! \bra{1}$ and express the Hamiltonian of a generic two-level system using the Pauli matrices
\begin{equation}
    \hat{H}_{\text{q}} 
    = - \frac{1}{2} \hbar \omega_{\text{q}} \hat{\sigma}_z
    = - \frac{1}{2} \hbar \omega_{\text{q}} \ket{0} \! \bra{0}
        + \frac{1}{2} \hbar \omega_{\text{q}} \ket{1} \! \bra{1}
\end{equation}
such that $\ket{0}$ and $\ket{1}$ are also the energy eigenstates of the system with energy $\mp \hbar \omega_{\text{q}} / 2$; in other words, the two-level system has an energy spacing of $\hbar \omega_{\text{q}}$, so $\omega_{\text{q}}$ is called the qubit frequency. Since we can always redefine the reference energy and rotate the orthonormal basis, our choice of $\hat{H}_{\text{q}}$ is completely general. 

Note also that, by setting $\ket{0}$ and $\ket{1}$ to be eigenvectors of $\hat{\sigma}_z$, we have explicitly picked a representation of the two-level system; hence, keep in mind that all the matrices and vectors are defined here with respective to the $z$-basis. Suppose we express everything instead in the $x$-basis, i.e.,
\begin{equation}
    \beta_x 
    =
    \left \{ \ket{+} \doteq  \frac{1}{\sqrt{2}} (\ket{0} + \ket{1}), \ \ 
    \ket{-} \doteq  \frac{1}{\sqrt{2}} (\ket{0} - \ket{1})
    \right \},
\end{equation}
then the matrix representation of the Pauli matrices changes to
\begin{equation}
        [ \hat{\sigma}_x ]_{\beta_x } 
        = \begin{pmatrix}
                1 & 0 \\
                0 & -1
            \end{pmatrix},
        \ \ 
        [ \hat{\sigma}_y ]_{\beta_x } 
        = \begin{pmatrix}
                0 & \ci \\
                - \ci & 0
            \end{pmatrix},
        \ \     
        [ \hat{\sigma}_z ]_{\beta_x } 
        = \begin{pmatrix}
                0 & 1 \\
                1 & 0
            \end{pmatrix}.
    \end{equation}

Additionally, it's instructive to compare a two-level system to a QHO which has infinitely many energy levels. We define, for a qubit, the lowing and raising operators as
\begin{align}
    \hat{\sigma}_{-} 
    &= \ket{0} \! \bra{1}
    = \begin{pmatrix}
            0 & 1 \\
            0 & 0
        \end{pmatrix},
\\
    \hat{\sigma}_{+} 
    &= \ket{1} \! \bra{0}
    = \begin{pmatrix}
            0 & 0 \\
            1 & 0
        \end{pmatrix},
\end{align}
respectively. The action of the two operators on the qubit energy eigenstates is analogous to that of the annihilation and creation operations on the number states of a QHO; for example, 
\begin{align}
    \hat{\sigma}_{-} \ket{1} 
    &= \ket{0},
\\
    \hat{\sigma}_{-} \ket{0} 
    &= 0,
\\
    \hat{\sigma}_{+} \ket{0} 
    &= \ket{1},
\\
    \hat{\sigma}_{+} \ket{1} 
    &= 0.
\end{align}
Unlike the QHO where the creation operator can in principle keep adding excitations to a mode, a qubit has its excitation capped at $\ket{1}$.
Moreover, we clearly have $\hat{\sigma}_{+} = \hat{\sigma}_{-}^{\dagger} $ and 
\begin{align}
    \hat{\sigma}_{x} 
    &= \hat{\sigma}_{-} + \hat{\sigma}_{+},
\\
    \hat{\sigma}_{y} 
    &= - \ci \hat{\sigma}_{-} + \ci \hat{\sigma}_{+}.
\end{align}
We see that $\hat{\sigma}_x$ and $\hat{\sigma}_y$ can be interpreted as conjugate quantities like the position and momentum of a QHO. This analogy can be further explained by comparing the matrix representation of $\hat{a}$ and $\hat{a}^{\dagger}$ in the number basis, i.e.,
\begin{equation}
    \hat{a}
    = \begin{pmatrix}
            0 & 1 & 0 & 0 & \cdots \\[-1mm]
            0 & 0 & \sqrt{2} & 0 & \cdots \\[-1mm]
            0 & 0 & 0 & \sqrt{3} &  \cdots \\[-1mm]
            0 & 0 & 0 & 0 & \cdots \\[-1mm]
            \vdots & \vdots & \vdots & \vdots & \ddots \\
        \end{pmatrix},
\end{equation}
\begin{equation}
    \hat{a}^{\dagger}
    = \begin{pmatrix}
            0 & 0 & 0 & 0 & \cdots \\[-1mm]
            1 & 0 & 0 & 0 & \cdots \\[-1mm]
            0 & \sqrt{2} & 0 & 0 &  \cdots \\[-1mm]
            0 & 0 & \sqrt{3} & 0 & \cdots \\[-1mm]
            \vdots & \vdots & \vdots & \vdots & \ddots \\
        \end{pmatrix},
\end{equation}
with that of $\hat{\sigma}_{-}$ and $\hat{\sigma}_{+}$ (in the $z$-basis). We notice that $\hat{\sigma}_{-}$ and $\hat{\sigma}_{+}$ can be reproduced by restricting $\hat{a}$ and $\hat{a}^{\dagger}$ to the two-dimensional subspace spanned by the number states $\ket{0}$ and $\ket{1}$ of the QHO. Along the same line of argument, we can define the ``number operator'' for a qubit as 
\begin{equation}
    \hat{N}_{\text{q}} 
    = \hat{\sigma}_{+} \hat{\sigma}_{-}
    = \begin{pmatrix}
            0 & 0 \\
            0 & 1
        \end{pmatrix}
\end{equation}
so that the expectation of $\hat{N}_{\text{q}} $ gives the number of excitation in the qubit
\begin{equation}
    \bra{0} \hat{N}_{\text{q}} \ket{0} = 0
    \ \ \text{ and } \ \ 
    \bra{1} \hat{N}_{\text{q}} \ket{1} = 1.
\end{equation}

\subsection{The Jaynes-Cummings Hamiltonian}
By replacing the particle part of the Hamiltonian with the two-level model, the full Hamiltonian now takes the form
\begin{equation}
    \hat{H}
    = - \frac{1}{2} 
            \hbar \omega_{\text{q}} \hat{\sigma}_z
            + \sum_{\mathbf{k},\lambda} 
                \hbar \omega_{k} 
                \left(
                    \hat{a}_{\mathbf{k},\lambda}^{\dagger} 
                    \hat{a}_{\mathbf{k},\lambda}
                    + \frac{1}{2}
                \right)
            - \, \hat{\mathbf{d}} 
                \cdot 
                \hat{\mathbf{E}}_{\perp} (\mathbf{0}).
\end{equation}
As mentioned before, in principle, we can use perturbation theory to study the effect of $\hat{H}_{\text{int}}$ by taking $\hat{H}_{\text{p}} + \hat{H}_{\perp}$ as the unperturbed Hamiltonian. For example, we can obtain Fermi's golden rule for the rate of spontaneous emission of the particles via this approach. However, to perform quantum computation, we generally want a coherent control of the system, i.e., we don't want to talk about the state of a qubit only in the average sense. Therefore, our agenda now is to derive a closed-form solution to the qubit time evolution subject to the electric dipole interaction.

In practice, only a single mode of the electric field is used to interact with the qubit (e.g., the fundamental mode in a CPW resonator or a microwave cavity); therefore, let's restrict our attention to a single-mode electric field
\begin{equation}
    \hat{\mathbf{E}}_{\perp}(\mathbf{0})
    = \hat{\mathbf{e}}_x \ci \mathscr{E}_0 
        \Big(
            \hat{a} - \hat{a}^{\dagger}
        \Big) 
\end{equation}
with polarization\footnote{Note that we can always rotate the coordinates so that the polarization points in $\hat{\mathbf{e}}_x$. This is possible only because of the long wavelength approximation; otherwise, the electric field could point in different directions depending on the position $\mathbf{r}$, which will create higher-order multipole interaction.} $\hat{\mathbf{e}}_x$ at $\mathbf{r} = \mathbf{0}$ and frequency $\omega_k = \omega_{\text{r}}$. 

\textbf{The Dipole Operator:} We assume the wavefunction of the energy eigenstates have either even or odd parity (think about the lowest two energy eigenstates of a harmonic oscillator or of a particle in a box); thus, the diagonal matrix elements of the dipole operator must vanish. Moreover, note that any operator restricted to a two-dimensional manifold can be written as a linear combination of the Pauli matrices (including the identity matrix). Hence, we have
\begin{align}
    \hat{\mathbf{d}} 
    &= \sum_{i = {x,y,z}} 
            \hat{\mathbf{e}}_i 
            \Big(
                d_{i,0} \hat{1} 
                + d_{i,1} \hat{\sigma}_x 
                + d_{i,2} \hat{\sigma}_y 
                + d_{i,3} \hat{\sigma}_z
            \Big) 
    & (\text{\scriptsize 2-level system})
\nonumber\\
    &= \sum_{i = {x,y,z}} 
            \hat{\mathbf{e}}_i 
            \Big(
                d_{i,1} \hat{\sigma}_x 
                + d_{i,2} \hat{\sigma}_y
            \Big) 
    & (\text{\scriptsize parity selection rule})
\nonumber\\
    &= \hat{\mathbf{e}}_x 
        \Big(
            d_{x,1} \hat{\sigma}_x
            + d_{x,2} \hat{\sigma}_y
        \Big) 
    & (\text{\scriptsize assume $\hat{\mathbf{E}}_{\perp} (\mathbf{0})$ points in $\hat{\mathbf{e}}_x$})
\nonumber\\[2mm]
    &= \hat{\mathbf{e}}_x d_0 
        \Big(
           e^{\ci \phi_{\text{di}}} \hat{\sigma}_{+} 
           + e^{-\ci \phi_{\text{di}}} \hat{\sigma}_{-} 
        \Big),
\end{align}
where $d_0 = \sqrt{|d_{x,1}|^2 + |d_{x,2}|^2}$ is a real scalar representing the magnitude of the dipole and $\phi_{\text{di}} = \tan^{-1}(d_{x,2} / d_{x,1})$. In the following chapter, we will make the expression concrete; for now, $d_0$ and $\phi_{\text{di}}$ are just some numbers.
    
\begin{figure}
    \centering
    \includegraphics[scale=0.33]{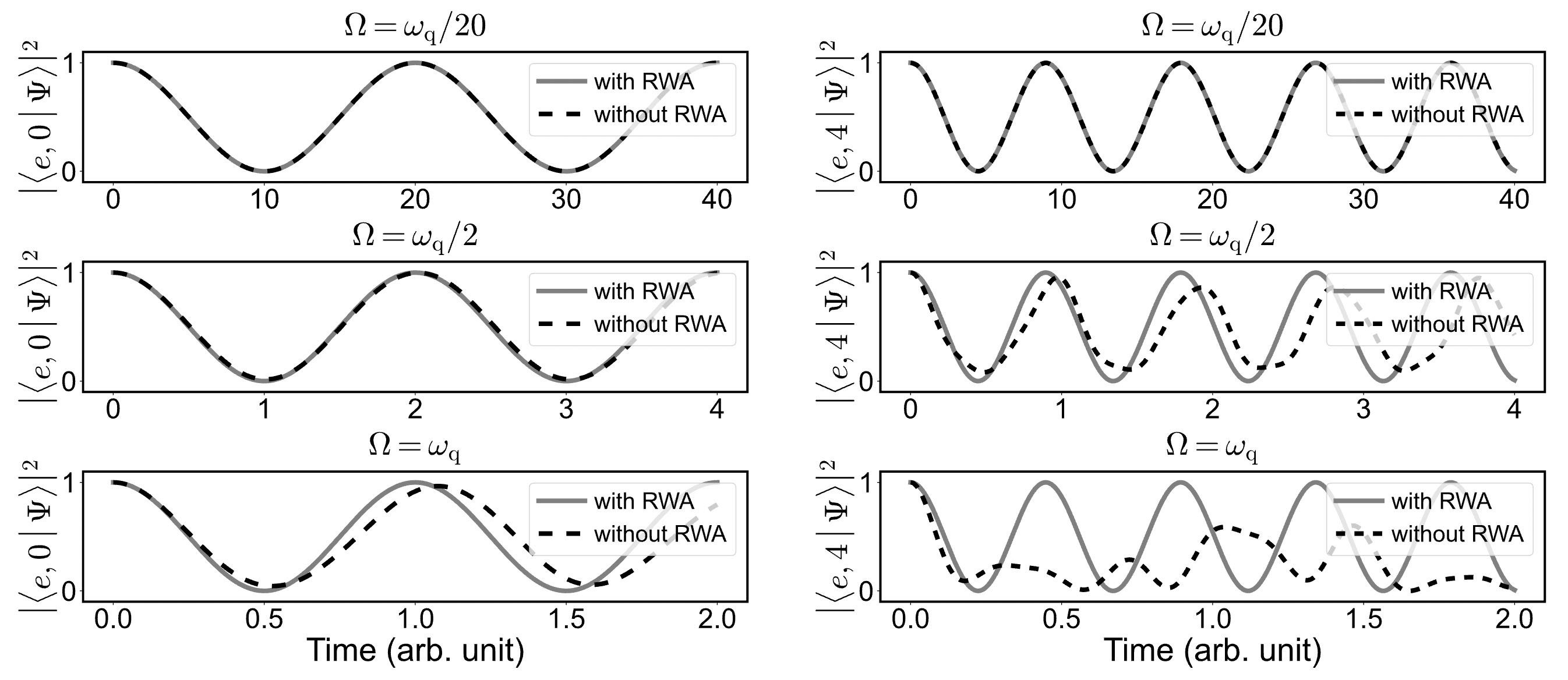}
    \caption{Simulation of a two-level system coupled to a resonator with (i.e., Eq.(\ref{eq:interaction_hamiltonian_after_RWA_for_comparison})) or without (i.e., Eq.(\ref{eq:interaction_hamiltonian_before_RWA})) the RWA. In the simulation, $\omega_{\mathrm{r}} = \omega_{\mathrm{q}}$, and the Hilbert space is truncated at $n=100$ for the resonator. Left column: $\ket{\Psi(0)} = \ket{e,0}$. Since there is at most one photon in the system, the RWA produces the correct time evolution even when $\Omega$ is close to $\omega_{\mathrm{q}}$ and $\omega_{\mathrm{r}}$. Right column: $\ket{\Psi(0)} = \ket{e,4}$. As the number of photons in the system increases, the RWA is only valid if $\Omega \ll \omega_{\mathrm{q}}$ and $\omega_{\mathrm{r}}$.}
    \label{fig:RWA_verification}
\end{figure}

With all the assumptions and simplifications, we obtain the coupling Hamiltonian
\begin{align}
    \hat{H}_{\text{int}} 
    = - \, \hat{\mathbf{d}} \cdot \hat{\mathbf{E}}_{\perp}(\mathbf{0})
    &= - \Big[ 
            d_0 
            \Big(
               e^{\ci \phi_{\text{di}}} \hat{\sigma}_{+} 
               + e^{-\ci \phi_{\text{di}}} \hat{\sigma}_{-} 
            \Big)
        \Big]
        \Big[
            \ci \mathscr{E}_0 
            \Big(
                \hat{a} - \hat{a}^{\dagger}
            \Big)
        \Big]
\nonumber \\ \label{eq:interaction_hamiltonian_before_RWA}
    &= - \ci (d_0 \mathscr{E}_0) 
        \Big( 
            e^{\ci \phi_{\text{di}}} 
                \hat{a} 
                \hat{\sigma}_{+} 
            - e^{-\ci \phi_{\text{di}}} 
                \hat{a}^{\dagger} 
                \hat{\sigma}_{-}
            + e^{-\ci \phi_{\text{di}}}
                \hat{a} 
                \hat{\sigma}_{-}
            - e^{\ci \phi_{\text{di}}} 
                \hat{a}^{\dagger} 
                \hat{\sigma}_{+}
        \Big)
\\
    &\approx - \ci (d_0 \mathscr{E}_0) 
        \Big( 
            e^{\ci \phi_{\text{di}}} 
                \hat{a} 
                \hat{\sigma}_{+} 
            - e^{-\ci \phi_{\text{di}}} 
                \hat{a}^{\dagger} 
                \hat{\sigma}_{-}
        \Big)
\nonumber \\ \label{eq:interaction_hamiltonian_after_RWA_for_comparison}
    &= - \hbar 
        \Big (
            g \hat{a}
                \hat{\sigma}_{+}
            + g^* \hat{a}^{\dagger} 
                \hat{\sigma}_{-}
        \Big),
\end{align}
where\footnote{For simplicity, we sometimes assume $g$ is real, which allows us to drop the phase factor $e^{\ci \phi_g}$. Usually, such an assumption will not change the qualitative behavior of the interaction as we will see soon; however, it turns out that the phase does matter when building quantum gates that rotate the qubit along different axes on the Bloch sphere, so I decide to make it as general as possible here.}
\begin{equation}
    g = \frac{\ci e^{\ci \phi_{\text{di}}} d_0 
        \mathscr{E}_0}{\hbar}
    = \frac{\Omega}{2} 
        e^{\ci \phi_g}
\end{equation}
is called the \textbf{coupling coefficient} and $\Omega = 2\abs{g} = 2|d_0 \mathscr{E}_0|/\hbar$ is the \textbf{vacuum Rabi frequency} whose meaning will be explored in the next two sub-sections. Importantly, we have made again the RWA in the above derivation to reduce the four coupling terms down to two. We applied the RWA to a classical field coupled to a QHO before; however, in the case where both the qubit and the electromagnetic field are quantized, we can argue for the RWA intuitively based on energy conservation. $\hat{a}$ and $\hat{a}^{\dagger}$ annihilates and creates a photon with frequency $\omega_{\text{r}}$, respectively, while $\hat{\sigma}_{+}$ and $\hat{\sigma}_{-}$ excites and de-excites the qubit, respectively. If the frequency of the drive field is about the same as the transition frequency of the qubit (which is usually the case as we will see soon), then the four terms in Eq.(\ref{eq:interaction_hamiltonian_before_RWA}) have the following interpretation: (1) $\hat{a} \hat{\sigma}_{+}$ excites the qubit by absorbing a photon; (2) $\hat{a}^{\dagger} \hat{\sigma}_{-}$ de-excites the qubit by emitting a photon; (3) $\hat{a} \hat{\sigma}_{-}$ de-excites the qubit and absorbs a photon; (4) $\hat{a}^{\dagger} \hat{\sigma}_{+}$ excites the qubit and emits a photon.
Clearly, the last two processes do not satisfy energy conservation and can be ignored in most of the applications we are interested in. We can also solve the time evolution numerically to check the validity of the RWA. As shown in Figure \ref{fig:RWA_verification}, the RWA produces the correct time evolution as long as 
\begin{equation}
    \sqrt{\text{photon number}} \times \Omega 
    \ \ll \ 
    \omega_{\text{q}} \text{ and } \omega_{\text{r}},
\end{equation}
also known as the weak coupling regime. Consequently, we arrive at the Jaynes-Cummings model of qubit-oscillator interaction
\begin{equation} \label{eq:JC_Hamiltonian_general}
    \hat{H}_{\text{JC}} 
    = - \frac{1}{2} 
            \hbar \omega_{\text{q}} 
            \hat{\sigma}_z
        + \hbar \omega_{\text{r}} \! 
            \left(
                \hat{a}^{\dagger}\hat{a} 
                + \frac{1}{2}
            \right)
        - \hbar
            \Big( 
                g \hat{a} \hat{\sigma}_{+}
                + g^* \hat{a}^{\dagger} \hat{\sigma}_{-}
            \Big).
\end{equation}

As a note on the terminology, for cQED application, we refer to the single-mode field as a resonator mode, a cavity mode, or a drive field depending on the context. For discussing the general features of the Jaynes-Cummings Hamiltonian, I will use ``resonator'' for concreteness; however, keep in mind that the same mathematics can be applied to completely different situations. For instance, we use a readout resonator whose frequency is far away from the qubit frequency to measure the qubit state, while we apply drive fields that typically have the same frequency as that of the qubit. This is also the reason why we should study the JC Hamiltonian with full generality as opposed to some special cases (such as $\omega_{\text{q}} = \omega_{\text{r}}$ or $\Bigsl \langle \hat{a}^{\dagger} \hat{a} \Bigsr \rangle \gg 1$).
\subsection{Diagonalizing the JC Hamiltonian}
The good news is that the Jaynes-Cummings Hamiltonian admits a closed-form solution. First, we need to pick a basis for the matrix representation. The natural choice is the set of all product states of the qubit states and photon number states. For clarity, we use $\ket{g} = \ket{0}$ and $\ket{e} = \ket{1}$ to denote the ground and excited states of the qubit, respectively, and $\{\ket{n}\}_{n \in \mathbf{Z}_{+}}$ to represent the Fock states of a resonator mode. The product states are thus given by 
\begin{align}
    & \ket{g,0}, 
\nonumber \\
    & \ket{g,1}, \ket{e,0},
\nonumber \\
    & \ket{g,2}, \ket{e,1},
\nonumber \\
    & \ket{g,3}, \ket{e,2},
\nonumber \\
    & \cdots 
\nonumber
\end{align}

Using the basis states listed above, the JC Hamiltonian has the matrix representation
\begin{equation} \label{eq:JC_matrix_representation}
    \hat{H}_{\text{JC}}
    = \begin{pmatrix}
        \hat{H}_0 & \hat{0} & \hat{0} & \cdots & \hat{0} & \cdots  \\
        \hat{0} & \hat{H}_1 & \hat{0} & \cdots & \hat{0} &  \cdots  \\
        \hat{0} & \hat{0} & \hat{H}_2 & \cdots & \hat{0} & \cdots  \\
        \vdots & \vdots & \vdots & \ddots & \vdots & \vdots  \\
        \hat{0} & \hat{0} & \hat{0} & \cdots & \hat{H}_n & \cdots  \\
        \vdots & \vdots & \vdots & \cdots & \vdots &  \ddots \\
    \end{pmatrix}
\end{equation}
if we define
\begin{equation}
    \hat{H}_0 = \bra{g,0} \hat{H}_{\text{JC}} \ket{g,0}
\end{equation}
and 
\begin{align}
    \hat{H}_n 
    &= \begin{pmatrix}
        \bra{g,n} \hat{H}_{\text{JC}} \ket{g,n}
        & \bra{g,n} \hat{H}_{\text{JC}} \ket{e,n-1} \\
        \bra{e,n-1} \hat{H}_{\text{JC}} \ket{g,n}
        & \bra{e,n-1} \hat{H}_{\text{JC}} \ket{e,n-1}
    \end{pmatrix}
\nonumber \\
    &= \begin{pmatrix}
        n\hbar\omega_{\text{r}} - \frac{1}{2} \hbar \Delta 
        & -\sqrt{n} \, \hbar g^* \\
        -\sqrt{n} \, \hbar g 
        & n \hbar \omega_{\text{r}} + \frac{1}{2} \hbar \Delta
    \end{pmatrix}
    = \hbar \begin{pmatrix}
        n \omega_{\text{r}} - \frac{1}{2} \Delta 
        & - \sqrt{n} |g| e^{- \ci \phi_g}\\
        -  \sqrt{n}  |g| e^{\ci \phi_g}
        & n \omega_{\text{r}} + \frac{1}{2} \Delta
    \end{pmatrix}
\end{align}
for $n = 1,2,...$ We have introduced a new variable $\Delta = \omega_{\text{q}} - \omega_{\text{r}}$, known as the detuning between the qubit and the resonator. According to the argument made for the RWA, we intuitively expect $\Delta$ to be small for the qubit and resonator to exchange photons of the same frequency. As shown below, the qubit and resonator can still interact even if $\Delta$ is large, but, of course, the interaction will not be as strong as in the case where $\Delta = 0$.

As indicated by Eq.(\ref{eq:JC_matrix_representation}), the matrix representation is block-diagonal in the chosen basis, which means that we can analyze the Hamiltonian block by block, reducing an infinite-dimensional problem into infinitely many finite-dimensional problems. In addition, note that $\hat{H}_0$ is a $1 \times 1$ sub-block, implying that $\ket{g,0}$ is one eigenstate of the JC Hamiltonian and is completely isolated from any other states (unless we augment $\hat{H}_{\text{JC}}$ with an external drive). Other than $\hat{H}_0$, all the other sub-blocks are $2 \times 2$, and the interpretation is intuitive: The qubit can be excited from $\ket{g}$ to $\ket{e}$ by absorbing a photon in the resonator mode or be de-excited from $\ket{e}$ to $\ket{g}$ by emitting a photon into the resonator mode; hence, $\ket{g,n}$ and $\ket{e,n-1}$ share a two-dimensional subspace and are isolated from other two-dimensional subspaces with a different photon number $n$.

\begin{figure}
    \centering
    \begin{circuitikz}[scale=0.9, transform shape]
    \draw 
        (0,0) ellipse (4 and 4);
    \draw[dashed]
        (0,0) ellipse (4 and 1.3);
    \draw[-{Latex[length=2mm]}] 
        (0,0) to (-2.1, -1.7) node[below]{$x$};
    \draw[-{Latex[length=2mm]}] 
        (0,0) to (4.8, 0) node[below]{$y$};
    \draw[-{Latex[length=2mm]}] 
        (0,0) to (0,4.8) node[right]{$z$};
    \draw[line width=0.8mm, -{Latex[length=3.2mm]}] 
        (0,0) to (1.4,-0.4);
    \draw
        (1.26,-0.26) node[right]{${\hat{\mathbf{e}}}_{\Omega_n'}$};
    \draw[dashed]
        (1.4,-0.4) to (1.4, -0.92)
        (0,0) to (1.4, -0.92);
    \draw 
        (-0.25,-0.22) arc (255:290:0.9 and 0.3)
        (0,-0.45) node[]{${\phi_g}$};
    \draw 
        (0,0.5) arc (90:40:0.6 and 1.8)
        (0.5,0.5) node[]{${\theta_{\text{m}}}$};
    \draw 
        (-0.2,-0.9) node[right]{${\Omega_n/\Omega_n'}$}
        (1.35,-0.78) node[right]{${\Delta/\Omega_n'}$};
    \end{circuitikz}
    \caption{Bloch vector representation of the dressed state $\ket{g,-}$. The mixing angle $\theta_{\mathrm{m}}$ controls the $z$ coordinate while the Rabi frequency $\Omega_n$ and the phase $\phi_g$ determine the $x$ and $y$ coordinates.}
    \label{fig:generalized_rabi_visualization}
\end{figure}
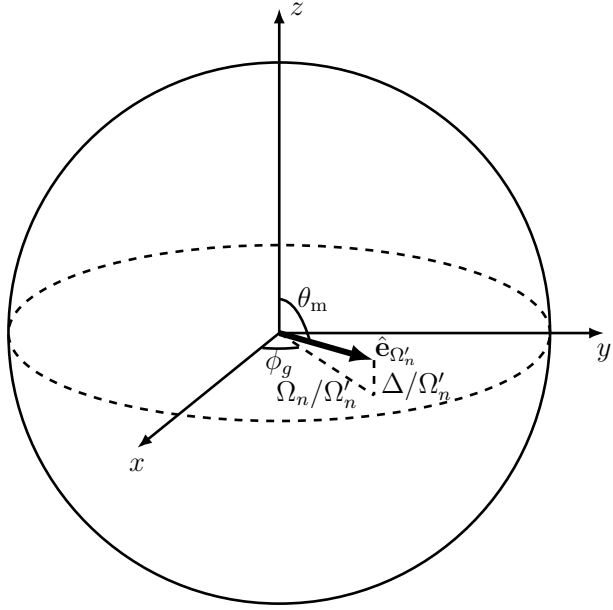

Since all the $2 \times 2$ sub-blocks along the diagonal are parametrized by $n$, the only task left is to diagonalize one $2 \times 2$ matrix as a function of $n$. With some algebra, one conclude that $\hat{H}_n$ has eigenvalues
\begin{equation} \label{eq:eigenvalue_JC_hamiltonian}
    E_{n,\pm} 
    = n \hbar \omega_{\text{r}}
        \pm 
            \frac{1}{2} 
            \hbar 
            \sqrt{(2\sqrt{n}|g|)^2 + \Delta^2}
    = n \hbar \omega_{\text{r}}
        \pm 
            \frac{1}{2} 
            \hbar 
            \sqrt{(\sqrt{n}\Omega)^2 + \Delta^2},
\end{equation} 
corresponding, respectively, to eigenvectors
\begin{equation} \label{eq:dressed_state_n_plus}
    \ket{n,+} 
    = e^{-\ci \phi_g} \sin \frac{\theta_n}{2} \ket{g,n} 
    - \cos \frac{\theta_n}{2} \ket{e,n-1},
\end{equation}
\begin{equation} \label{eq:dressed_state_n_minus}
    \ket{n,-} 
    = \cos \frac{\theta_n}{2} \ket{g,n} 
    + e^{\ci \phi_g} \sin \frac{\theta_n}{2} \ket{e,n-1},
\end{equation}
where we have defined the mixing angle
\begin{equation}
    \theta_{\text{m}, n} 
    = \tan^{-1}(2\sqrt{n}|g| /\Delta)
    = \tan^{-1}(\sqrt{n}\Omega /\Delta).
\end{equation}
We call the eigenstates of the JC Hamiltonian the \textbf{dressed states} of two subsystems (i.e., the qubit and resonator). To understand these eigenstates better, let us consider two limiting cases before discussing the general interpretation:
\begin{list}
{\text{(\roman{qcounter})}~}
{
\usecounter{qcounter}
\setlength\labelwidth{2em}
\setlength\listparindent{1.5em}
}
    \item $\Delta \gg \sqrt{n}\Omega$: When the qubit and the resonator mode are sufficiently detuned, the eigenenergies are simply
    \begin{equation}
        E_{n,+} 
        \approx \hbar \omega_{\text{r}} \bigg(n + \frac{1}{2}\bigg) 
            - \frac{1}{2} \hbar \omega_{\text{q}},
    \end{equation}
    \begin{equation}
        E_{n,-} 
        \approx \hbar \omega_{\text{r}} \bigg(n - 1 + \frac{1}{2}\bigg) 
            + \frac{1}{2} \hbar \omega_{\text{q}},
    \end{equation}
    corresponding to energies of $\ket{g,n}$ and $\ket{e,n-1}$, respectively\footnote{I have assumed $\Delta < 0$ in the calculation since the frequency of the readout resonator ($\sim $ 7 GHz) is higher than that of the qubit ($\sim $ 5 GHz) in a typical cQED experiment. If $\Delta >0$, then
    \begin{equation}
        E_{n,+} 
        \approx \hbar \omega_{\text{r}} \bigg(n - 1 + \frac{1}{2}\bigg) 
            + \frac{1}{2} \hbar \omega_{\text{q}},
    \end{equation}
    \begin{equation}
        E_{n,-} 
        \approx \hbar \omega_{\text{r}} \bigg(n + \frac{1}{2}\bigg) 
            - \frac{1}{2} \hbar \omega_{\text{q}}
    \end{equation}
    since $E_{n,+} \geq E_{n,-}$ by definition.
    }. Indeed, the dressed states are approximately given by the product states since the mixing angle $\theta_{\text{m}, n}  \approx 0$ for large $\Delta$. In other words, the qubit and resonator are decoupled in the large-detuning limit.

    \item $|\Delta| \ll \sqrt{n} \Omega$: When the qubit is in resonance with the resonator mode, the energy of the original product states, $\ket{g,n}$ and $\ket{e,n-1}$, are both $n \hbar \omega_{\text{r}} \approx n \hbar \omega_{\text{q}}$. However, the energies of the dressed states,
    \begin{equation}
        E_{n,\pm} 
    \approx n \hbar \omega_{\text{r}}
        \pm 
            \hbar 
            \sqrt{n}|g|
    = n \hbar \omega_{\text{r}}
        \pm 
            \frac{1}{2} 
            \hbar 
            \sqrt{n} \Omega,
    \end{equation}
    split around $n \hbar \omega_{\text{r}}$ with a total separation given by the vacuum Rabi frequency $\Omega$ scaled by $\sqrt{n}$. Thus, a larger photon number will induce a stronger splitting. However, note that our result is still subject to the RWA, which means the splitting $\sqrt{n} \Omega$ cannot be of the same order as $\omega_{\text{r}}$ (as demonstrated in Figure \ref{fig:RWA_verification}); in other words, the ordering $E_{g,0}, E_{1,-}, E_{1,+}, ..., E_{n,-}, E_{n,+}, ...$ must hold under the RWA, which is sometimes referred to as the JC ladder.

    The dressed states are now in equal superpositions of the product states; assuming $\Delta > 0$, we have $\theta_{\text{m},n} \approx \pi/2$ and
    \begin{equation} 
        \ket{n,+} 
        = \frac{e^{-\ci \phi_g}}{\sqrt{2}} 
                \ket{g,n} 
            - \frac{1}{\sqrt{2}} 
                \ket{e,n-1},
    \end{equation}
    \begin{equation}
        \ket{n,-} 
        = \frac{1}{\sqrt{2}} 
                \ket{g,n} 
            + \frac{e^{\ci \phi_g}}{2}
                \ket{e,n-1}.
    \end{equation}

    \item (general case) First, we introduce more terminologies that will make more sense after introducing the Rabi oscillations. We call $\Omega_n = \sqrt{n} \Omega$ the Rabi frequency (with $n$ photons) and we've already introduced the special case when $n = 1$, i.e., the vacuum Rabi frequency $\Omega$. In addition, we name the quantity
    \begin{equation}
        \Omega_n' = \sqrt{\big(\sqrt{n}\Omega\big)^2 + \Delta^2}
    \end{equation}
    the generalized Rabi frequency (with $n$ photons). For small $\Delta$, we naturally have $\Omega_n' \approx \Omega_n$. 
    
    Geometrically, we can view $\sqrt{n} \Omega$ and $\Delta$ as the length of the two legs of a right triangle and $\Omega_n'$ as the length of the hypotenuse. Clearly, the mixing angle also fits into this picture. In fact, we can plot this right triangle in the Bloch sphere if we treat $\ket{g,n}$ and $\ket{e,n-1}$ as $\ket{0}$ and $\ket{1}$, respectively. Comparing Eq.(\ref{eq:dressed_state_n_minus}) with Eq.(\ref{eq:general_state_Bloch_vector}), we realize that $\ket{n,-}$ is already in the spherical-coordinate representation of the Bloch vector. The Bloch vector
    \begin{equation}
        \hat{\mathbf{e}}_{\Omega'_n} 
        = (r=1, \theta = \theta_{\text{m},n}, \phi = \phi_g)
    \end{equation}
    for $\ket{n,-}$ is plotted in Figure \ref{fig:generalized_rabi_visualization}; the size of the right triangle is scaled down by $\Omega_n'$ since the Bloch sphere is of unit radius. It's also easy to show that $\ket{n,+}$ points in the opposite direction such that $\ket{n,+}$ and $\ket{n,-}$ always form a line going through the center of the sphere.
\end{list}

\subsection{Rabi Oscillations}
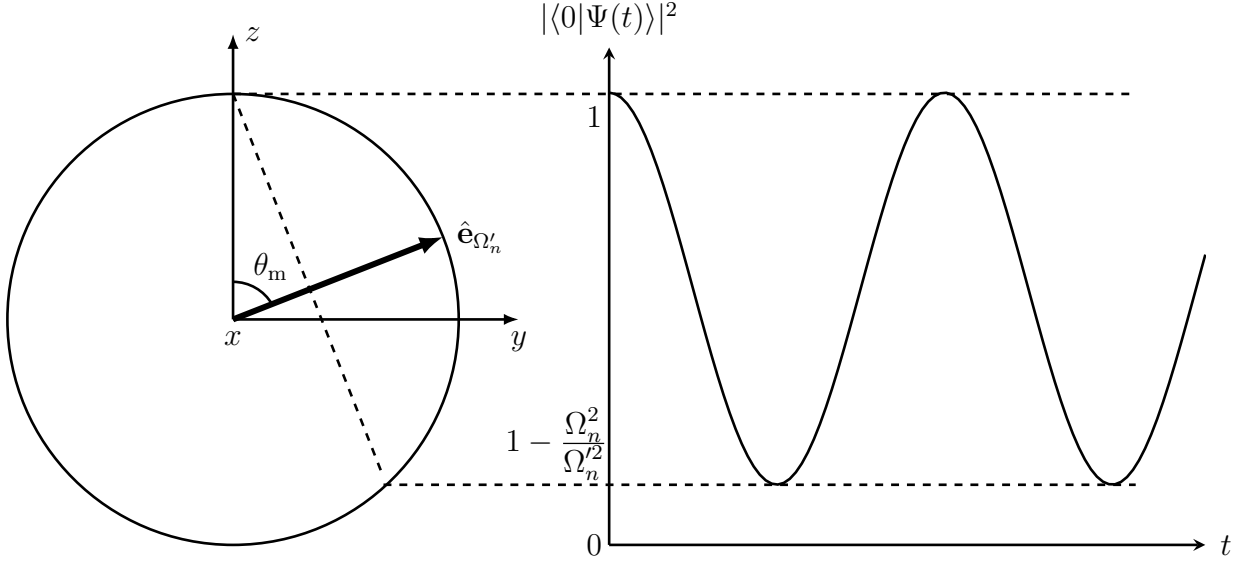
\begin{figure}
    \centering
    \begin{circuitikz}[scale=1, transform shape]
    \draw 
        (0,0) ellipse (3 and 3);
    \draw[-{Latex[length=2mm]}] 
        (0,0) to (3.8, 0) node[below]{$y$};
    \draw[-{Latex[length=2mm]}] 
        (0,0) to (0,3.8) node[right]{$z$};
    \draw
        (0,0) node[below]{$x$};
    \draw[line width=0.8mm, -{Latex[length=3.2mm]}] 
        (0,0) to (2.8, 1.1) node[right]{${\hat{\mathbf{e}}}_{\Omega_n'}$};
    \draw 
        (0,0.5) arc (90:30:0.6 and 0.6)
        (0.5,0.7) node[]{${\theta_{\text{m}}}$};
    \draw[dashed]
        (0,3) to (2,-2.1);
    \draw[dashed]
        (0, 3) to (12, 3)
        (2,-2.2) to (12, -2.2);
    \draw   
        (13.2,-3) node[]{$t$}
        (5,4) node[]{${|\!\bra{0}\ket{\Psi(t)}\!|^2}$}
        (4.8,-3) node[]{$0$}
        (4.8,2.7) node[]{$1$}
        (4.3,-1.65) node[]{${1-\frac{\displaystyle\Omega_n^2}{\displaystyle\Omega_n'^{2}}}$};
    \begin{scope}[shift={(5,-3)}]
    \begin{axis}[
        height=8.2cm,
        axis lines = center,
        xtick={0},
        ytick={0},
        samples=100,
        domain=0:3.2,
        xmin=0,
        ymin=0,
        ymax=1.1,
        ]
        \addplot []
            ({\x}, {2.6 / 6 * cos(200 * \x) + (1 - 2.6 / 6)});
    \end{axis}
    \end{scope}
    \end{circuitikz}
    \caption{Relationship between the axis of rotation and the population contrast.}
    \label{fig:population_contrast}
\end{figure}

Once the JC Hamiltonian is diagonalized, we have all the information to understand the dynamics of the qubit when coupled to a resonator since any state can be written as the linear combination of the two dressed states. In particular, suppose the system is initially $\ket{\Psi(0)} = \ket{g,n}$, then the phenomenon of Rabi flopping will occur as shown in Figure \ref{fig:population_contrast}. The state evolution can be solved exactly (up to a global phase $e^{-\ci n \omega_{\text{r}} t}$ which is ignored here)
\begin{equation}
    \ket{\Psi(t)}
    = \left[
                \cos\left( \frac{1}{2} \Omega_n' t \right) 
                + \ci \frac{\Delta}{\Omega_n'} \sin \left( \frac{1}{2}\Omega_n' t \right) 
            \right] 
            \ket{g,n}
    + \ci e^{\ci \phi_{g}} 
            \frac{\sqrt{n} \Omega}{\Omega_n'} \sin\left( \frac{1}{2}\Omega_n' t \right) \ket{e,n-1} ,
\end{equation}
which gives the change in probabilities as functions of time
\begin{equation}
    p_g(t) = \abs{\bra{g}\ket{\Psi(t)}}^2
    = 1- \frac{\Omega^2}{\Omega'^2}
        \sin^2 \left( \frac{1}{2} \Omega' t \right) ,
\end{equation}
\begin{equation}
    p_e(t) = \abs{\bra{e}\ket{\Psi(t)}}^2
    = \frac{\Omega^2}{\Omega'^2}  
        \sin^2 \left( \frac{1}{2}\Omega' t \right) .
\end{equation}
The probability oscillates at the generalized Rabi frequency if the detuning is nonzero. At resonance (i.e., $\Delta = 0$), the generalized Rabi frequency reduces to the ideal Rabi frequency $\Omega_n = \sqrt{n} \Omega$. If the field has only 1 photon initially, the population would oscillate at the vacuum Rabi frequency $\Omega$.

Graphically, any non-stationary state will rotate along the axis defined by the Bloch vector of $\ket{n,-}$. Hence, to flip the qubit state (i.e., move from $\ket{g,n}$ to $\ket{e,n-1}$), the dress states must lie exactly on the $xy$-plane of the Bloch sphere, implying a resonance between the qubit and the field. A pulse (i.e., the duration for which the dipole interaction is turned on) that moves the qubit from the ground to the excited state is known as a $\pi$-pulse; assuming $\Delta = 0$, a $\pi$-pulse would correspond to a pulse length of $T_{\pi} = \pi/\Omega$. Similarly, a pulse that creates an equal superposition state (i.e., the Bloch vector points along the $xy$-plane) from the ground state is called a $\pi/2$-pulse; the pulse duration is $T_{\pi} = \pi/2\Omega$ when $\Delta=0$. 

Moreover, there is a continuum of possible axes of rotation on the $xy$-plane of the Bloch sphere that could result in 100\% flopping; these axes of rotation are parametrized by the phase $\phi_g$. Although the time evolution of the probability would look the same for all $\phi_g$, the actual probability amplitudes are different due to the phase factor $e^{\ci \phi_g}$. In fact, starting from the ground state of the qubit, $\pi/2$-pulses with $\phi_g = 0$ and $\phi_g = \pi/2$ create Bloch vectors that point in the $+y$ and $-x$ directions,
respectively; hence, the pulses with different phases do represent physically distinct gates in quantum computation.

\subsection{The Generalized JC Model for a Qudit}
Most systems in reality are not two-level systems, not to mention the enormous computational space provided by a multi-level system. Thus, it's also critical to extend the simple result of a qubit to that of a qudit. In general, the Hamiltonian of a $D$-level system can be expressed as
\begin{equation}
    \hat{H}_{\text{q}} = \sum_{j=0}^{D-1} \hbar \omega_{j}\ket{j}\!\bra{j},
\end{equation}
where $E_j = \hbar \omega_j$ is the energy of the eigenstate $\ket{j}$ for $j = 0,...,D-1$. The interaction between the qudit and a single-mode resonator field is, again, described by the dipole Hamiltonian \cite{RevModPhys.93.025005}
\begin{align}
    \hat{H}
    &= \underbrace{
            \sum_{j=0}^{D-1} 
                \hbar \omega_{j} \ket{j}\!\bra{j}
            + \hbar \omega_{\text{r}} 
                \left(
                    \hat{a}^{\dagger}\hat{a} 
                    + \frac{1}{2}
                \right)
        }_{\hat{H}_0}
        \underbrace{
            - \, \hat{\mathbf{d}} 
                \cdot 
                \hat{\mathbf{E}}_{\perp} (\mathbf{0})
        }_{\hat{H}_{\text{int}}}
\nonumber \\
    &= \sum_{j=0}^{D-1} \hbar \omega_{j} \ket{j}\!\bra{j}
        + \hbar \omega_{\text{r}} 
            \left(
                \hat{a}^{\dagger}\hat{a} 
                + \frac{1}{2}
            \right)
        - \sum_{j, k =0}^{D-1} 
            \hbar
            \Big( 
                g_{jk} \hat{a} \ket{j} \! \bra{k}
                + g_{jk}^* \hat{a}^{\dagger} \ket{k} \! \bra{j}
            \Big),
\end{align}
where
\begin{equation}
    g_{jk} 
    = \frac{\ci \mathscr{E}_0 \bra{j} \hat{\mathbf{d}} \cdot \hat{\mathbf{e}}_x \ket{k}}{\hbar}
\end{equation}
are the coupling coefficients. If, in addition, the qudit is constructed by truncating a perturbed version of the QHO (e.g., the first $D$ levels of a weakly-anharmonic oscillator), then $g_{jk} \approx 0$ for $|j-k| \neq 1$ since $g_{jk}$ is proportional to the matrix element of the ``position'' operator of the perturbed QHO. Consequently, the Hamiltonian of the coupled system, after applying the RWA, reduces to the so-called \textbf{generalized Jaynes-Cummings Hamiltonian}:
\begin{equation}
    \hat{H}_{\text{GJC}}
    = \sum_{j=0}^{D-1} \hbar \omega_{j} \ket{j}\!\bra{j}
        + \hbar \omega_{\text{r}} 
            \left(
                \hat{a}^{\dagger}\hat{a} 
                + \frac{1}{2}
            \right)
        - \sum_{j=0}^{D-2} 
            \hbar
            \Big( 
                g_{j+1,j} \hat{a} \ket{j+1} \! \bra{j}
                + g_{j+1,j}^* \hat{a}^{\dagger} \ket{j} \! \bra{j+1}
            \Big).
\end{equation}
From the perturbed QHO model, we can further assume that
\begin{equation}
    g_{j+1,j} \approx \sqrt{j+1} g_{10},
\end{equation}
just like how the matrix elements of the annihilation operator of a QHO scale. Combine all the approximations, we arrive at
\begin{equation}
    \hat{H}_{\text{GJC}}
    = \sum_{j=0}^{D-1} \hbar \omega_{j} \ket{j}\!\bra{j}
        + \hbar \omega_{\text{r}}
            \left(
                \hat{a}^{\dagger}\hat{a} 
                + \frac{1}{2}
            \right)
        - \sum_{j=0}^{D-2} 
            \hbar
            \sqrt{j+1}
            \Big( 
                g_{10} \hat{a} \ket{j+1} \! \bra{j}
                + g_{10}^* \hat{a}^{\dagger} \ket{j} \! \bra{j+1}
            \Big).
\end{equation}

Unfortunately, we cannot derive a closed-form solution for the Rabi flopping between two levels in a qudit; nevertheless, under certain assumptions, we can keep the two levels we want to control and ignore the other energy levels. To see this, we first assume that all the (single-photon) transition frequencies within the qudit are distinct\footnote{This is why we want to use an anharmonic oscillator to describe the qudit. If the qudit itself is an ideal QHO, then all the neighboring transitions will have the same frequency. Consequently, we cannot address individual transitions separately.}. Next, if we move to the interaction picture, the interaction Hamiltonian takes the form
\begin{equation}
    - \sum_{j=0}^{D-2} 
            \hbar
            \Big( 
                g_{j+1,j} e^{\ci (\omega_{j} - \omega_k - \omega_{\text{r}})t} \hat{a} \ket{j+1} \! \bra{j}
                + g_{j+1,j}^* e^{- \ci (\omega_{j} - \omega_k - \omega_{\text{r}})t} \hat{a}^{\dagger} \ket{j} \! \bra{j+1}
            \Big).
\end{equation}
Suppose the resonator frequency is tuned to one of the transitions, then $e^{\ci (\omega_{j} - \omega_k - \omega_{\text{r}})t}$ will oscillate for all but one transition. Hence, by making another RWA, we can apply the conclusions drawn in the two-level case to a general qudit. However, one may ask to what extent will the other transitions average to zero. Clearly, if the transition happens much faster than the detuning between the field and all the other transitions, then there will be leakage into the other energy levels since there is not enough time for $e^{\ci (\omega_{j} - \omega_k - \omega_{\text{r}})t}$ to average to zero. In reality, leakage is always a main concern since the gate fidelity drops if we excite some unwanted transitions.

\section{Dispersive Coupling}
In the last section, we found a general solution to the time evolution induced by the JC Hamiltonian. In principle, all physics contained in the JC Hamiltonian can be readily analyzed using the general solution; however, in the limit where the qubit and resonator frequencies are far detuned from each other, there is a more insightful way to understand the qubit-resonator interaction. More precisely, we restrict ourselves to the case where
$|g| \ll |\Delta| \ll \omega_\text{q} \text{ and } \omega_{\text{r}}$, i.e., the detuning is still much smaller than the qubit and resonator frequencies for the RWA to still hold. In the cQED community, we say the qubit is dispersively coupled to the resonator \cite{s41567-020-0806-z}.

\subsection{Schrieffer–Wolff Transformation}
Before studying the JC Hamiltonian, let us consider a generic problem
\begin{equation}
    \hat{H} = \hat{H}_0 + \lambda \hat{V},
\end{equation}
where $\hat{V}$ is a perturbation added to some initial Hamiltonian $\hat{H}_0$ and $\lambda$ is a unitless real number, called the adiabaticity parameter, that controls the strength of the perturbation. The magnitude of $\hat{V}$ (e.g., in terms of its eigenvalues) can be of the same order as that of $\hat{H}_0$, but $\lambda$ is small by assumption. In addition, we assume that $\bra{k}\hat{V}\ket{k} = 0$ for any energy eigenstate $\ket{k}$ of $\hat{H}_0$, which is always possible by absorbing the diagonal entries of $\hat{V}$ into $\hat{H}_0$.

In the current frame, $\hat{H}$ is affine in $\lambda$; our goal is to adiabatically eliminate the first-order perturbation for small $\lambda$ by introducing a unitary transformation
\begin{equation}
    \hat{H}'
    = e^{- \lambda \hat{S}} \hat{H} e^{\lambda \hat{S}}
    \approx \hat{H}_0 + \lambda^2 \hat{V}',
\end{equation}
where $\hat{S}$ is an anti-Hermitian operator to be determined and $\hat{V}'$ is the effective perturbation in the transformed frame independent of $\lambda$. Since the unitary transformation does not alter the energy eigenvalues, it allows us to study the dependence of the energy spectrum of $\hat{H}$ on $\lambda$ when the perturbation is weak. Note that the same $\lambda$ is used in defining the transformation, which will be the key to eliminating the first-order interaction.

Using the Baker–Campbell–Hausdorff formula, we expand $\hat{H}'$ to the second order in $\lambda$:
\begin{align}
    \hat{H}'
    = e^{- \lambda \hat{S}} \hat{H} e^{\lambda \hat{S}}
    &= \hat{H} 
        + \lambda \Bigsl[\hat{H}, \hat{S} \Bigsr]
        + \frac{\lambda^2}{2} 
            \Big[ 
                \Bigsl[\hat{H}, \hat{S} \Bigsr],
                \hat{S}
            \Big]
        + \cdots
\nonumber\\
    &= \hat{H}_0 
        + \lambda \hat{V}  
        + \lambda \Bigsl[\hat{H}_0, \hat{S} \Bigsr] 
        + \lambda^2 \Bigsl[\hat{V}, \hat{S} \Bigsr]
        + \frac{\lambda^2}{2} 
            \Big[ 
                \Bigsl[\hat{H}_0, \hat{S} \Bigsr],
                \hat{S}
            \Big]
        + \cdots
\end{align}
The terms linear in $\lambda$ will vanish if
\begin{equation} \label{eq:Schrieffer_Wolff_conditions}
    \hat{V} + \Bigsl[\hat{H}_0, \hat{S} \Bigsr] = \hat{0},
\end{equation}
in which case we have
\begin{equation} \label{eq:transformed_hamiltonian_2nd_order}
    \hat{H}' 
    \approx \hat{H}_0 
        + \lambda^2\Bigsl[\hat{V}, \hat{S} \Bigsr]
        + \frac{\lambda^2}{2} 
            \Big[ 
                \Bigsl[\hat{H}_0, \hat{S} \Bigsr],
                \hat{S}
            \Big]
    = \hat{H}_0 
        + \frac{\lambda^2}{2} \Bigsl[\hat{V}, \hat{S} \Bigsr]
\end{equation}
to the second order in $\lambda$. Hence, our problem reduces to solving Eq.(\ref{eq:Schrieffer_Wolff_conditions}). Moreover, the leading correction to this approximation is third order in $\lambda$ and first order in $\hat{H}_0$ and $\hat{V}$; for example, if the energy scale related to $\hat{H}_0$ and $\hat{V}$ is several GHz and $\lambda \sim 0.05$, then the error associated with the transformation would be on the order of 100 kHz.

Most of the time, we will actually absorb $\lambda$ into the definition of $\hat{V}$, i.e., 
\begin{equation}
    \hat{H} = \hat{H}_0 + \hat{V}
    \ \ \text{ and } \ \ 
    \hat{H}' = e^{-\hat{S}} \hat{H} e^{\hat{S}},
\end{equation}
as long as we can identify a small adiabatic parameter in $\hat{V}$. Under this convention, we still require that $\hat{V} + \Bigsl[\hat{H}_0, \hat{S} \Bigsr] = \hat{0}$, but the transformed Hamiltonian to second-order is now given by
\begin{equation} \label{eq:transformed_hamiltonian_2nd_order_no_lambda}
    \hat{H}' 
    \approx \hat{H}_0 
        + \frac{1}{2} \Bigsl[\hat{V}, \hat{S} \Bigsr].
\end{equation}

It should be noted that the transformation we introduced, also known as the \textbf{Schrieffer–Wolff transformation}, is nothing more than an operator representation of the traditional second-order time-independent perturbation theory. We can eliminate the first-order perturbation simply because we have lumped the possible first-order energy shift $\bra{k}\hat{V}\ket{k} = 0$ into $\hat{H}_0$. Nevertheless, this formalism provides an alternative interpretation of the perturbation theory as an intentionally picked unitary transformation.

\subsection{JC Hamiltonian in the Dispersive Limit}
Now, we apply the Schrieffer–Wolff transformation to the JC Hamiltonian under the assumption that $|g| \ll |\Delta|$. Let the unperturbed Hamiltonian be
\begin{equation}
    \hat{H}_0 
    = - \frac{1}{2} 
            \hbar \omega_{\text{q}} 
            \hat{\sigma}_z
        + \hbar \omega_{\text{r}} \! 
            \left(
                \hat{a}^{\dagger}\hat{a} 
                + \frac{1}{2}
            \right),
\end{equation}
describing the situation where the qubit and the resonator are decoupled; the perturbation is then
\begin{equation} \label{eq:SW_transform_JC_interaction}
    \hat{V}
    = \frac{\Delta}{|g|} \hat{H}_{\text{int}}
    = - \hbar \Delta
        \Big( 
            e^{\ci \phi_g} \hat{a} \hat{\sigma}_{+}
            + e^{-\ci \phi_g} \hat{a}^{\dagger} \hat{\sigma}_{-}
        \Big)
\end{equation}
with $\lambda = |g|/\Delta$, which is small by our assumption. In addition, the diagonal matrix elements of $\hat{V}$ in the energy eigenbasis of $\hat{H}_0$ are zero, so the technique is indeed applicable. However, it should be noted that the magnitude of the matrix elements of $\hat{H}_{\text{int}}$ is unbounded since we can always move up along the QHO ladder. Another way to see this is to recall that the Rabi frequency linearly increases with $\sqrt{n}$. Hence, there naturally exists a critical photon number $n_{\text{crit}}$ such that the strength of the interaction is no longer suppressed by $\lambda$ and cannot be treated as a perturbation. 

Since $\hat{H}_{\text{JC}}$ is block-diagonal in the eigenbasis of $\hat{H}_0$, it's clear that $\hat{V}$ must also be block-diagonal in the same basis; thus, $\hat{S}$ can be decomposed into blocks along the diagonal 
\begin{equation}
    \hat{S}
    = \begin{pmatrix}
        \hat{S}_0 & \hat{0} & \hat{0} & \cdots & \hat{0} & \cdots  \\[-1mm]
        \hat{0} & \hat{S}_1 & \hat{0} & \cdots & \hat{0} &  \cdots  \\[-1mm]
        \hat{0} & \hat{0} & \hat{S}_2 & \cdots & \hat{0} & \cdots  \\[-1mm]
        \vdots & \vdots & \vdots & \ddots & \vdots & \vdots  \\[-1mm]
        \hat{0} & \hat{0} & \hat{0} & \cdots & \hat{S}_n & \cdots  \\[-1mm]
        \vdots & \vdots & \vdots & \cdots & \vdots &  \ddots \\
    \end{pmatrix}
\end{equation}
and Eq.(\ref{eq:Schrieffer_Wolff_conditions}) is equivalent to the following set of operator equations
\begin{equation}
    \hat{S}_0 = 0,
\end{equation}
\begin{equation} \label{eq:matrix_equation_for_S_n}
    \hbar \Delta
        \begin{pmatrix}
            0 
            & - \sqrt{n} e^{- \ci \phi_g}\\
            -  \sqrt{n} e^{\ci \phi_g}
            & 0
        \end{pmatrix}
    + \Bigg[
        \hbar 
        \begin{pmatrix}
            n \omega_{\text{r}} - \frac{1}{2} \Delta 
            & 0\\
            0
            & n \omega_{\text{r}} + \frac{1}{2} \Delta
        \end{pmatrix}
        , \hat{S}_n \Bigg] 
        = \hat{0}
    \ \ \ \ \text{for } n = 1,2,...
\end{equation}
By expanding Eq.(\ref{eq:matrix_equation_for_S_n}), we can show that
\begin{equation}
    \hat{S}_n 
    = \sqrt{n} e^{\ci \phi_g} \hat{\sigma}_{+} 
        - \sqrt{n} e^{-\ci \phi_g}\hat{\sigma}_{-}
\end{equation}
and, thus,
\begin{equation}
    \hat{S} 
    = e^{\ci \phi_g} \hat{a}\hat{\sigma}_{+} 
        - e^{-\ci \phi_g} \hat{a}^{\dagger} \hat{\sigma}_{-},
\end{equation}
\begin{equation}\label{eq:SW_transform_qubit_resonator_dispersive_coupling}
    \hat{T} = e^{\lambda \hat{S}}
    = \exp(\frac{g}{\Delta} \hat{a}\hat{\sigma}_{+} - \frac{g^*}{\Delta} \hat{a}^{\dagger} \hat{\sigma}_{-}).
\end{equation}
Moreover, according to Eq.(\ref{eq:transformed_hamiltonian_2nd_order}), the Hamiltonian in transformed frame is given by \cite{PhysRevA.75.032329}
\begin{equation}\\ \label{eq:dispersive_shift_derivation}
    \hat{H}^{\text{disp}}_{\text{JC}}
    \approx \hat{H}_0 
        + \frac{\lambda^2}{2} \Bigsl[\hat{V}, \hat{S} \Bigsr]
    = - \frac{1}{2} 
            \hbar 
            \bigg(
                \omega_{\text{q}} 
                + \frac{|g|^2}{\Delta}
            \bigg)
            \hat{\sigma}_z
        + \hbar \omega_{\text{r}} 
                \hat{a}^{\dagger}\hat{a} 
        - \frac{\hbar |g|^2}{\Delta} 
            \hat{\sigma}_z 
            \hat{a}^{\dagger}\hat{a},
\end{equation}
where we have ignored a constant energy $\hbar\omega_{\text{r}} / 2 + \hbar|g|^2/2\Delta $. On one hand, when $\lambda = |g|/\Delta \rightarrow 0$, we do retrieve $\hat{H}_0$ as expected. On the other hand, for small but finite $\lambda$, two modifications appear: 1) The qubit acquires a frequency shift
\begin{equation}
    \chi = \frac{|g|^2}{\Delta},
\end{equation}
known as the \textbf{Lamb shift}, which is independent of the resonator state. 2) The qubit-resonator interaction in the transformed frame, $\hat{\sigma}_z \hat{a}^{\dagger}\hat{a}$, commutes with $\hat{H}_0$; such an interaction is said to be longitudinal as opposed to the transverse coupling using $\hat{\sigma}_{+}$ and $\hat{\sigma}_{-}$. We discuss some of the consequences of the dispersive interaction now.

\subsection{Dispersive Shift}
Eq.(\ref{eq:dispersive_shift_derivation}) can be understood from two perspectives based on how we group the three terms. If we lump the last term with the resonator Hamiltonian, i.e., 
\begin{equation}
    \hat{H}^{\text{disp}}_{\text{JC}}
    = - \frac{1}{2} 
            \hbar 
            \big(
                \omega_{\text{q}} 
                + \chi
            \big)
            \hat{\sigma}_z
        + \hbar 
            \big( 
                \omega_{\text{r}} 
                - \chi
                    \hat{\sigma}_z 
            \big) 
            \hat{a}^{\dagger}\hat{a},
\end{equation}
then we see a qubit-state-dependent shift of the resonator frequency. Suppose the qubit is in $\ket{g}$, then $\langle \hat{\sigma}_z \rangle = 1$ and the resonator effectively has a frequency $\omega_{\text{r}} - \chi$. However, if the qubit is excited to $\ket{e}$, the frequency of the resonator shifts in the opposite direction.

We could have derived this frequency shift by directly expanding the eigenvalue of $\ket{n,+} \approx \ket{g,n}$ and $\ket{n+1,-} \approx \ket{e,n}$ in Eq.(\ref{eq:eigenvalue_JC_hamiltonian}) to the second order in $|g|/\Delta$ (with $\Delta < 0$ as used before):
\begin{gather}
    E_{n,+} 
    = n \hbar \omega_{\text{r}}
        + 
            \frac{1}{2} 
            \hbar 
            \sqrt{4n|g|^2 + \Delta^2}
    \approx 
        \hbar 
            \bigg[ 
                (
                    \omega_{\text{r}} 
                    - \chi
                ) n
            + \frac{\omega_{\text{r}}}{2}
            \bigg] 
            - \frac{1}{2} \hbar \omega_{\text{q}} ,
\\
    E_{n+1,-} 
    = n \hbar \omega_{\text{r}}
        - 
            \frac{1}{2} 
            \hbar 
            \sqrt{4(n+1)|g|^2 + \Delta^2}
    \approx 
        \hbar 
            \bigg[ 
                (
                    \omega_{\text{r}} 
                    + \chi
                ) n
                + \frac{\omega_{\text{r}}}{2}
            \bigg] 
            + \frac{1}{2} \hbar \omega_{\text{q}}
            + \hbar \chi.
\end{gather}
By treating $n$ as $\hat{a}^{\dagger}\hat{a}$, we see the same dispersive shift conditioned on the qubit states being in $\ket{g}$ or $\ket{e}$; additionally, the constant qubit frequency shift $\chi$ appears in the expansion.

However, note that $\langle \hat{\sigma}_z \rangle$, in general, can take any value between -1 and 1, so the frequency shift of the resonator can fall between $-|\chi|$ and $|\chi|$. In Chapter 5, we will use this dispersive shift of the resonator as a way to infer the qubit or qudit state.

\subsection{AC Stark Shift}
If we lump the third term in Eq.(\ref{eq:dispersive_shift_derivation}) into the qubit Hamiltonian, then we see a photon-number-induced shift of the qubit frequency instead
\begin{equation}
    \hat{H}^{\text{disp}}_{\text{JC}}
    = - \frac{1}{2} 
            \hbar 
            \Big(
                \omega_{\text{q}} 
                + \chi
                + 2 \chi \hat{a}^{\dagger}\hat{a}
            \Big)
            \hat{\sigma}_z
        + \hbar 
            \omega_{\text{r}} 
            \hat{a}^{\dagger}\hat{a}.
\end{equation}
Experimentally, this can be observed by measuring the change of the qubit frequency while sweeping the number of photons (i.e., power of the cavity readout signal) in the cavity. Alternatively, one can explicitly add a coherent pump field near the qubit frequency to observe the pump-power-dependent shift of the qubit frequency. In the latter application, it's more intuitive to model a coherent drive as a classical electromagnetic field as will be discussed in Section \ref{sec:light_matter_semiclassical}; nevertheless, both the fully-quantized and semiclassical approaches result in the same AC Stark effect.

Note that the dispersive coupling and AC Stark shift are just two different interpretations of the same Hamiltonian. In general, it's more appropriate to treat such a phenomenon as a two-photon scattering process. In fact, later we use the Schrieffer–Wolff transformation to explicitly derive a two-photon Rabi flopping. In the AC Stark effect, the initial and final states are the same (i.e., a photon is absorbed and then emitted to create a scattering event) while for a two-photon Rabi flopping, the initial and final states are different (i.e., two photons of the same frequency are absorbed and then emitted to create an oscillation).

\subsection{Dispersive Coupling for a Qudit}
Next, we want to generalize the results above to a multi-level system \cite{RevModPhys.93.025005}. We still consider a single mode of a resonator, i.e., $\hat{\mathbf{E}}_{\perp}(\mathbf{0}) = \hat{\mathbf{e}}_x \ci \mathscr{E}_0 \Bigsl (\hat{a} - \hat{a}^{\dagger} \Bigsr)$ coupled to a qudit, which is described by 
\begin{align}
    \hat{H}
    = \sum_{j=0}^{D-1} \hbar \omega_{j} \ket{j}\!\bra{j}
        + \hbar \omega_{\text{r}} 
            \left(
                \hat{a}^{\dagger}\hat{a} 
                + \frac{1}{2}
            \right)
        - \sum_{j, k=0}^{D-1} 
            \hbar
            \Big( 
                g_{jk} \hat{a} \ket{j} \! \bra{k}
                + g_{jk}^* \hat{a}^{\dagger} \ket{k} \! \bra{j}
            \Big).
\end{align}
To be completely general, we will not use the generalized JC model for the derivation (in other words, the results we are going to obtain will be valid for any kind of qudit systems.)

To apply the Schrieffer–Wolff transformation, we first identify
\begin{equation}
    \hat{H}_0 
    = \sum_{j=0}^{D-1} \hbar \omega_{j} \ket{j}\!\bra{j}
        + \hbar \omega_{\text{r}} 
            \left(
                \hat{a}^{\dagger}\hat{a} 
                + \frac{1}{2}
            \right)
\end{equation}
and 
\begin{equation}
    \hat{V} 
    = - \sum_{j, k=0}^{D-1} 
            \hbar
            \Big( 
                g_{jk} \hat{a} \ket{j} \! \bra{k}
                + g_{jk}^* \hat{a}^{\dagger} \ket{k} \! \bra{j}
            \Big)
\end{equation}
as the unperturbed Hamiltonian and perturbative interaction, respectively. Here, we will use the convention where the adiabatic parameter is absorbed into $\hat{S}$. One can verify easily that
\begin{equation}
    \hat{S} 
    = \sum_{j,k=0}^{D-1} 
        \frac{1}{\omega_j - \omega_k - \omega_{\text{r}}}
        \Big( 
            g_{jk} \hat{a} \ket{j} \! \bra{k}
            - g_{jk}^* \hat{a}^{\dagger} \ket{k} \! \bra{j}
        \Big)
\end{equation}
is one possible solution to $\hat{V} + \Bigsl[\hat{H}_0, \hat{S} \Bigsr] = \hat{0}$. Hence, what is left is to compute the transformed Hamiltonian $\hat{H}^{\text{disp}}$ using Eq.(\ref{eq:transformed_hamiltonian_2nd_order_no_lambda}).

In the calculation of $\hat{H}^{\text{disp}}$, we encounter products of the form $\hat{a}^{\dagger} \hat{a} \ket{j} \!\bra{k}$, $\hat{a} \hat{a}^{\dagger} \ket{j} \!\bra{k}$, $\hat{a}^2 \ket{j} \!\bra{k}$, and $\hat{a}^{\dagger 2} \ket{j} \!\bra{k}$. In the interaction picture, 
\begin{align}
    \hat{a}^{\dagger} \hat{a} \ket{j} \!\bra{k} 
    &\ \ \longrightarrow \ \ 
    \hat{a}^{\dagger} \hat{a} \ket{j} \!\bra{k} e^{\ci(\omega_j - \omega_k) t},
\\
    \hat{a} \hat{a}^{\dagger} \ket{j} \!\bra{k} 
    &\ \ \longrightarrow \ \ 
    \hat{a} \hat{a}^{\dagger} \ket{j} \!\bra{k} e^{\ci(\omega_j - \omega_k) t},
\\
    \hat{a}^2 \ket{j} \!\bra{k}
    &\ \ \longrightarrow \ \ 
    \hat{a}^2 \ket{j} \!\bra{k} e^{\ci(\omega_j - \omega_k - 2 \omega_{\text{r}})t},
\\
    \hat{a}^{\dagger 2} \ket{j} \!\bra{k}
    &\ \ \longrightarrow \ \ 
    \hat{a}^{\dagger 2} \ket{j} \!\bra{k} e^{\ci(\omega_j - \omega_k + 2 \omega_{\text{r}})t}.
\end{align}
Suppose the qudit energy levels are nondegenerate and no two-photon resonance conditions are met (i.e., $|\omega_j - \omega_k - 2 \omega_{\text{r}}| \gg 0$), we can apply the RWA to omit the fast oscillating terms. Consequently, 
\begin{equation} \label{eq:derivation_qudit_SW_transformation}
    \text{\small $
    \frac{1}{2} \Big[ \hat{V}, \hat{S} \Big]
    \approx \sum_{j,k=0}^{D-1} 
            \left(
                \frac{\hbar |g_{jk}|^2 }{\omega_j - \omega_k - \omega_{\text{r}}}
                - \frac{\hbar |g_{jk}|^2 }{\omega_k - \omega_j - \omega_{\text{r}}}
            \right)
            \hat{a}^{\dagger} \hat{a}
            \ket{j} \! \bra{j}
        + \sum_{j,k=0}^{D-1}
            \frac{\hbar |g_{jk}|^2 }{\omega_j - \omega_k - \omega_{\text{r}}} 
            \ket{j} \! \bra{j},
    $}
\end{equation}
from which we can define
\begin{equation} \label{eq:general_chi_shifts}
    \chi_{jk} 
    = \frac{|g_{jk}|^2 }{\omega_j - \omega_k - \omega_{\text{r}}},
\end{equation}
\begin{equation} \label{eq:general_lambda_shifts}
    \Lambda_j
    = \sum_{k=0}^{D-1} \chi_{jk}
    = \sum_{k=0}^{D-1}
        \frac{|g_{jk}|^2 }{\omega_j - \omega_k - \omega_{\text{r}}} .
\end{equation}
The Hamiltonian in the dispersive regime is thus given by
\begin{equation}\label{eq:dispersive_Hamiltonian_qudit_derivation}
    \text{\small $
    \hat{H}^{\text{disp}} 
    \approx 
        \hat{H}_0 
        + \frac{1}{2} \Bigsl[\hat{V}, \hat{S} \Bigsr]
    \approx 
        \sum_{j=0}^{D-1} \hbar (\omega_{j} + \Lambda_j) 
            \ket{j}\!\bra{j}
        + \hbar \omega_{\text{r}} \hat{a}^{\dagger} \hat{a}
        + \sum_{j=0}^{D-1} 
            \left[
                \sum_{k=0}^{D-1} 
                    \hbar (\chi_{jk} - \chi_{kj}) 
            \right] \!
            \hat{a}^{\dagger}
            \hat{a} 
            \ket{j} \! \bra{j}.
    $}
\end{equation}
Just like the qubit case, there are two features in the transformed Hamiltonian: 1) As a consequence of the zero-point fluctuation, the qudit is augmented by the Lamb shifts $\Lambda_j$. To put it in another way, one realizes, during the derivation of Eq.(\ref{eq:derivation_qudit_SW_transformation}), that each $\Lambda_j$ appears because of the bosonic commutation relation $\big[\hat{a},\hat{a}^{\dagger}\big] = \hat{1}$; therefore, Lamb shifts reflect the quantum nature of the electromagnetic field. 2) The qudit-resonator interaction is longitudinal in the sense that $\ket{j}\!\bra{j}$ appears instead of $\ket{j}\!\bra{k}$ in the last term of Eq.(\ref{eq:dispersive_Hamiltonian_qudit_derivation}). 

The two interpretations of the dispersive coupling discussed before also apply in the qudit case: The $j$th qudit energy level feels the AC Stark shift 
\begin{equation} \label{eq:qudit_ac_stark_shift_quantized}
    \left \langle 
        \left[
            \sum_{k} 
                \hbar (\chi_{jk} - \chi_{kj}) 
        \right]
        \hat{a}^{\dagger}
        \hat{a} 
    \right \rangle
\end{equation}
from the resonator field while the resonator sees a qudit-state-dependent frequency shift
\begin{equation}
    \left \langle 
        \sum_{j} 
            \left[
                \sum_{k} 
                    \hbar (\chi_{jk} - \chi_{kj}) 
            \right]
            \ket{j} \! \bra{j}
    \right \rangle.
\end{equation}


\subsection{Two-Photon Transitions}
In a multi-level quantum system, some of the transitions are forbidden in the sense that the matrix elements of the interaction connecting the initial and final states are negligibly small. For example, if the qudit is modeled by an anharmonic oscillator, the coupling between the ground and second excited states is practically null; thus, we want to explore a way to drive the ``forbidden'' transition directly. Intuitively, if the first-order dipole transition is not allowed, we might construct a second-order interaction that jumps from one state to the other via an intermediate state. If the transitions between the initial and intermediate states and between the intermediate and final states are allowed, then we expect the net effect to be a transition from the initial state to the final state. However, note that we do not want two consecutive pulses targeting two different transitions; to save hardware resources, we would like to reach a forbidden transition using a single frequency (maybe with a practically achievable slow varying envelope to suppress the bandwidth of the pulse due to the Fourier uncertainty in a fast pulse).

In fact, we already have all the ingredients to reveal such a transition. In the previous derivation, we have assumed that there do not exist two qudit levels $j_1$ and $j_2$ such that $\omega_{j_2} - \omega_{j_1} - 2 \omega_{\text{r}} \approx 0$, which implicitly suppresses the second-order transitions in the Hamiltonian. However, if we reserve the possibility that some transitions might be on-resonance with two photons, we would obtain the extra terms
\begin{align}
        &- \frac{1}{2}
            \sum_{j_1, j_2} 
            \sum_{k} 
            \left(
                \frac{\hbar g_{j_2 k} g_{k j_1}}{\omega_k - \omega_{j_1} - \omega_{\text{r}}}
                - \frac{\hbar g_{j_2 k} g_{k j_1}}{\omega_{j_2} - \omega_k - \omega_{\text{r}}}
            \right)
        \hat{a}^{2} \ket{j_2} \! \bra{j_1}
\nonumber \\
    & \ \ \    \ \ \ \ 
        - \frac{1}{2}
            \sum_{j_1, j_2} 
            \sum_{k} 
            \left(
                \frac{\hbar g_{j_2 k}^* g_{k j_1}^*}{\omega_k - \omega_{j_1} - \omega_{\text{r}}}
                - \frac{\hbar g_{j_2 k}^* g_{k j_1}^*}{\omega_{j_2} - \omega_k - \omega_{\text{r}}}
            \right)
            \hat{a}^{\dagger 2} \ket{j_1} \! \bra{j_2}
\end{align}
in $\Bigsl[ \hat{V}, \hat{S} \Bigsr]/2$.
Furthermore, since $\omega_{j_2} - \omega_{j_1} \approx 2 \omega_{\text{r}}$ near resonance, we have $\omega_k - \omega_{j_1} - \omega_{\text{r}} \approx - (\omega_{j_2} - \omega_k - \omega_{\text{r}})$ and 
\begin{align}
    \frac{1}{2} \Big[ \hat{V}, \hat{S} \Big]
    &\approx 
        \sum_{j=0}^{D-1}  \hbar \Lambda_j
            \ket{j}\!\bra{j}
        + \sum_{j=0}^{D-1}  
            \left[
                \sum_{k} 
                    \hbar (\chi_{jk} - \chi_{kj}) 
            \right]
            \hat{a}^{\dagger}
            \hat{a} 
            \ket{j} \! \bra{j}
\nonumber\\
    &\ \ \ \ - \sum_{j_1, j_2=0}^{D-1}   \sum_{k=0}^{D-1} 
        \left(
            \frac{\hbar g_{j_2 k} g_{k j_1}}{\omega_k - \omega_{j_1} - \omega_{\text{r}}}
                \hat{a}^{2} \ket{j_2} \! \bra{j_1}
            + \frac{\hbar g_{j_2 k}^* g_{k j_1}^*}{\omega_k - \omega_{j_1} - \omega_{\text{r}}}
                \hat{a}^{\dagger 2} \ket{j_1} \! \bra{j_2}
        \right)
\nonumber \\
    &= \sum_{j=0}^{D-1}  \hbar \Lambda_j
            \ket{j}\!\bra{j}
        + \sum_{j=0}^{D-1}  
            \left[
                \sum_{k=0}^{D-1}  
                    \hbar (\chi_{jk} - \chi_{kj}) 
            \right]
            \hat{a}^{\dagger}
            \hat{a} 
            \ket{j} \! \bra{j}
\nonumber\\ \label{eq:fully_quantized_two_photon_transition_derivation}
    &\ \ \ \
        - \sum_{j_1, j_2=0}^{D-1} 
        \frac{1}{2} 
        \left(
            g^{(2)}_{j_2, j_1}
                \hat{a}^{2} \ket{j_2} \! \bra{j_1}
            + g^{(2)*}_{j_2, j_1}
                \hat{a}^{\dagger 2} \ket{j_1} \! \bra{j_2}
        \right),
\end{align}
where
\begin{equation}
    g^{(2)}_{j_2, j_1} 
    = \sum_{k=0}^{D-1} 
        \frac{2 \hbar g_{j_2 k} g_{k j_1}}{\omega_k - \omega_{j_1} - \omega_{\text{r}}}.
\end{equation}
are the coupling coefficients of the two-photon transition. 

Eq.(\ref{eq:fully_quantized_two_photon_transition_derivation}) provides the full description of coherent two-photon processes in the quantized-field picture. However, in practice, driving a two-photon transition requires a much higher power than driving a one-photon transition (about $20 \text{ dB}$ larger); hence, it's much easier to treat the drive as a classical drive (i.e., a coherent state), which we will discuss now.

\section{Light-Matter Interaction with a Classical Field}\label{sec:light_matter_semiclassical}
When the strength of the electromagnetic field is macroscopically large, we can adopt the classical limit introduced before. This new point of view is particularly useful for describing the qubit control pulses since they are usually microwave fields generated by room-temperature electronics; in other words, control signals are well described by coherent states with a large amplitude $\alpha$. In contrast, in order for the dispersive coupling to be valid\footnote{Recall that the dispersive regime is derived by assuming the coupling is perturbative. When the mean photon number is large, $\hat{a}$ and $\hat{a}^{\dagger}$ in Eq.(\ref{eq:SW_transform_JC_interaction}) will give a large matrix element, thus violating the perturbative assumption.}, the cavity readout signals are around the single-photon level and are usually described with a quantized field to account for the quantum fluctuation.

The Hamiltonian for a qubit interacting with a classical electric field is given by
\begin{align} \label{eq:hamiltonian_semi_classical_interaction}
    \hat{H}
    = - \frac{1}{2} \hbar \omega_{\text{q}} \hat{\sigma}_z
        - \, \hat{\mathbf{d}} \cdot \mathbf{E}(\mathbf{r}, t)
    \approx - \frac{1}{2} \hbar \omega_{\text{q}} \hat{\sigma}_z
        - \, \hat{\mathbf{d}} \cdot \mathbf{E}(\mathbf{0}, t),
\end{align}
where we apply the long-wavelength approximation again so that the field is position-independent. The electric field is assumed to be a single harmonic function
\begin{equation}
    \mathbf{E}(\mathbf{0}, t)
    = \mathbf{E}(\mathbf{0})
        \sin(\omega_{\text{d}} t - \phi)
    = \hat{\mathbf{e}}_x \frac{\ci E_0}{2}
        \left(
            e^{- \ci \omega_{\text{d}} t + \ci \phi} 
            - e^{\ci \omega_{\text{d}} t - \ci \phi} 
        \right)
\end{equation}
as introduced in Eq.(\ref{eq:classical_limit_electric field}) with drive frequency $\omega_{\text{d}}$. In the semi-classical analysis, we no longer include the Hamiltonian of the field, so the Hilbert space is only two-dimensional.

The interaction Hamiltonian in the semi-classical picture is given by
\begin{align}
    \hat{H}_{\text{int}}(t)
    &= - \frac{\ci d_0 E_0}{2}  \left(
            e^{- \ci \omega_{\text{d}} t + \ci \phi} 
            - e^{\ci \omega_{\text{d}} t - \ci \phi} 
        \right) \!
        \left(
            e^{\ci \phi_{\text{di}}} \hat{\sigma}_{+} 
            + e^{-\ci \phi_{\text{di}}} \hat{\sigma}_{-} 
        \right)
\nonumber \\
    &= - \frac{\ci}{2} \hbar \Omega 
            \left(
                e^{-\ci \omega_{\text{d}} t + \ci \phi}e^{\ci \phi_{\text{di}}} \hat{\sigma}_{+}
                - e^{\ci \omega_{\text{d}} t - \ci \phi}e^{-\ci \phi_{\text{di}}} \hat{\sigma}_{-} 
                + e^{-\ci \omega_{\text{d}} t + \ci \phi}e^{-\ci \phi_{\text{di}}} \hat{\sigma}_{-} 
                - e^{\ci \omega_{\text{d}} t - \ci \phi}e^{\ci \phi_{\text{di}}} \hat{\sigma}_{+}
            \right),
\end{align}
where $\Omega = d_0 E_0 / \hbar$ will turn out to be the Rabi frequency in the semi-classical picture. Moreover, recall that the JC Hamiltonian, before making the RWA, takes the form
\begin{align}
    \hat{H} 
    &= - \frac{1}{2} \hbar \omega_{\text{q}} \hat{\sigma}_z
        + \hbar \omega_{\text{d}} \! 
            \left(
                \hat{a}^{\dagger} \hat{a} 
                + \frac{1}{2}
            \right)
\nonumber \\
    & \ \ \ \  \ \ \ \  \ \ \ \  \ \ \ \ 
        - \ci (d_0 \mathscr{E}_0) 
        \Big( 
            e^{\ci \phi_{\text{di}}} 
                \hat{a} 
                \hat{\sigma}_{+} 
            - e^{-\ci \phi_{\text{di}}} 
                \hat{a}^{\dagger} 
                \hat{\sigma}_{-}
            + e^{-\ci \phi_{\text{di}}}
                \hat{a} 
                \hat{\sigma}_{-}
            - e^{\ci \phi_{\text{di}}} 
                \hat{a}^{\dagger} 
                \hat{\sigma}_{+}
        \Big).
\end{align}
Hence, we can intuitively connect the semi-classical and fully quantized pictures by making the following mapping
\begin{equation}
    \frac{E_0}{2} e^{-\ci \omega_{\text{d}} t + \ci \phi}  
    \ \ \longleftrightarrow \ \ 
    \mathscr{E}_0 \hat{a}.
\end{equation}
That is, writing down the semi-classical interaction Hamiltonian is almost the same as going into the interaction picture of the quantized field (i.e., $\hat{a} \ \rightarrow \ \hat{a} e^{-\ci \omega_{\text{d}} t} \sim \alpha e^{-\ci \omega_{\text{d}} t}$). Of course, what we have ignored by going into the classical description is the quantum uncertainty associated with the coherent state.

With Eq.(\ref{eq:hamiltonian_semi_classical_interaction}), the solution to the Schr\"{o}dinger equation is derived in detail in Appendix \ref{appendix:semiclassical_matter_light}. In the next sub-section, we summarize the results from the appendix. 

\subsection{Rabi Oscillations}
To begin with, the derivation uses the RWA again so that
\begin{align} 
    \hat{H}
    &\approx - \frac{1}{2} \hbar \omega_{\text{q}} \hat{\sigma}_z
        - \frac{\ci}{2} \hbar \Omega 
            \left(
                e^{-\ci \omega_{\text{d}} t + \ci \phi} 
                    e^{\ci \phi_{\text{di}}} 
                    \hat{\sigma}_{+}
                - e^{\ci \omega_{\text{d}} t - \ci \phi} 
                    e^{-\ci \phi_{\text{di}}} 
                    \hat{\sigma}_{-} 
            \right)
\nonumber \\ \label{eq:RWA_in_semi_classical_analysis}
    &= - \frac{1}{2} \hbar \omega_{\text{q}} \hat{\sigma}_z
        - \frac{1}{2} \hbar \Omega 
            \left( 
                e^{-\ci \omega_{\text{d}} t + \ci \phi_g} \hat{\sigma}_{+}
                + e^{\ci \omega_{\text{d}} t - \ci \phi_g} \hat{\sigma}_{-}
            \right),
\end{align}
where $\phi_g = \phi + \phi_{\text{di}} + \pi/2$. The validity of the approximation can be more clearly shown by going into the interaction picture with $\hat{U}_0(t) = \exp(\ci \omega_{\text{q}} t \hat{\sigma}_z/2)$;  the interaction Hamiltonian in the interaction picture is
\begin{align}
    \hat{\Tilde{H}}_{\text{int}}(t) 
    &= \hat{U}_0^{\dagger}(t)
        \hat{H}_{\text{int}}(t) 
        \hat{U}_0(t)
\nonumber \\
    &= - \frac{\ci}{2} \hbar \Omega
        \Big[ 
            e^{\ci(\phi + \phi_{\text{di}})} \hat{\sigma}_{+}
            - e^{-\ci(\phi + \phi_{\text{di}})} \hat{\sigma}_{-}
            + e^{-2\ci \omega_{\text{d}} t} e^{\ci \phi_g} \hat{\sigma}_{-}
            - e^{2\ci \omega_{\text{d}} t} e^{-\ci \phi_g} \hat{\sigma}_{+}
        \Big]
\nonumber \\
    &\approx - \frac{1}{2} \hbar \Omega
        \Big( 
            e^{\ci \phi_g} \hat{\sigma}_{+}
            + e^{-\ci \phi_g} \hat{\sigma}_{-}
        \Big),
\end{align}
where we see that the two terms neglected in Eq.(\ref{eq:RWA_in_semi_classical_analysis}) are terms that oscillates with angular frequency $2 \omega_{\text{d}}$. Hence, if the state of the qubit varies on a timescale that is much slower than $2\pi / \omega_{\text{d}}$, the RWA will stand.

Under the RWA, the Schr\"{o}dinger equation can be solved exactly. In general, the qubit state, which lives in a two-dimensional space, can be expressed as
\begin{equation}\label{eq:general_2level_state}
    \ket{\Psi(t)} = a(t) \ket{g} + b(t) \ket{e},\\[-1mm]
\end{equation}
where $a(0)$ and $b(0)$ are specified quantities. As shown in Appendix \ref{appendix:semiclassical_matter_light}, the probability amplitude associated with $\ket{g}$ and $\ket{e}$ are given, respectively, as
\begin{equation}
    a(t) 
    = a(0) e^{\ci \omega_{\text{d}} t /2} 
            \left[
                \cos \! \left( \frac{1}{2} \Omega' t \right) 
                + \ci \frac{\Delta}{\Omega'} \sin \left( \frac{1}{2}\Omega' t \right) \!
            \right]
        + \ci b(0) e^{\ci \omega_{\text{d}} t /2 -\ci \phi_{g}} 
                \frac{\Omega}{\Omega'} \sin \! \left( \frac{1}{2}\Omega' t \right),
\end{equation}
\begin{equation}
    b(t)
    = \ci a(0) e^{-\ci \omega_{\text{d}} t/2 + \ci \phi_{g}} 
            \frac{\Omega}{\Omega'} \sin \! \left( \frac{1}{2}\Omega' t \right) 
        + b(0) e^{-\ci \omega_{\text{d}} t/2} 
            \left[
                \cos \! \left( \frac{1}{2}\Omega' t \right) 
                - \ci \frac{\Delta}{\Omega'} \sin \left( \frac{1}{2}\Omega' t \right) \!
            \right], \\[2mm]
\end{equation}
where $\Delta = \omega_{\text{q}} - \omega_{\text{d}}$ is the detuning and $\Omega' = \sqrt{\Omega^2 + \Delta^2}$ is the generalized Rabi frequency defined analogously to that in the fully-quantized analysis. Now, by setting $a(0) = 1$ and $b(t) = 0$, we obtain the probabilities of measuring $\ket{g}$ and $\ket{e}$, respectively,
\begin{equation}
    p_g(t)
    = \abs{a(t)}^2
    = 1- \frac{\Omega^2}{\Omega^2 + \Delta^2}
        \sin^2 \left( \frac{1}{2}\sqrt{\Omega^2 + \Delta^2} \,  t \right),
\end{equation}
\begin{equation}
    p_e(t) 
    = \abs{b(t)}^2
    = \frac{\Omega^2}{\Omega^2 + \Delta^2}  
        \sin^2 \left( \frac{1}{2}\sqrt{\Omega^2 + \Delta^2} \,  t \right),
\end{equation}
which clearly entails the Rabi flopping.

Unlike the fully quantized case, we have a single Rabi frequency defined in the semi-classical picture since a photon is not a well-defined object in a classical field. Nevertheless, remember that $\Omega$ is a function of the electric field amplitude $E_0$, which, according to Eq.(\ref{eq:connecting_classical_and_quantum_field}), is proportional to $\sqrt{n}$ in the classical limit, where $n$ is the mean photon number of a coherent field. In other words, if we use Eq.(\ref{eq:connecting_classical_and_quantum_field}), i.e., $E_0(n) = 2\mathscr{E}_{0, k} \sqrt{n}$, then
\begin{equation} \label{eq:classical_quantum_rabi_frequency_comparison}
    \Omega(n) 
    = \frac{d_0 E_0(n)}{\hbar} 
    = \frac{2 d_0 \mathscr{E}_{0, k} \sqrt{n}}{\hbar} 
    = \Omega_n
    = 2\sqrt{n} g
\end{equation}
such that the classical and quantized fields produce a consistent Rabi frequency.

Furthermore, similar to the result with the quantized field, there is a $\phi_g$-dependence in the Rabi flopping. However, in the semi-classical case, it's clear how we can control this phase. Since $\phi_g$ can be modified by $\phi$, the phase of the classical field, we can change the axis of rotation by simply asking the room-temperature electronics to output the drive at a different phase.

\subsection{AC Stark Shift and Two-Photon Transitions with a Classical Drive}
We have established several equivalences between the fully quantized and semi-classical approaches; in particular, the semiclassical Hamiltonian is very similar to the two-dimensional subblocks of the JC Hamiltonian. Therefore, we also expect to reveal the AC Stark shifts and two-photon transitions using the Schrieffer–Wolff transformation. However, the classical drive is the same as a coherent state only up to the removal of the quantum fluctuation, implying the fact that we will not see Lamb shifts in the semi-classical calculation. 

We now apply the Schrieffer–Wolff transformation to the qudit Hamiltonian (of course, the qubit case is also included) plus a classical drive: 
\begin{align}
    \hat{H}
    &= \sum_{j=0}^{D-1} 
            \hbar \omega_{j} \ket{j}\!\bra{j}
        - \, \hat{\mathbf{d}} 
            \cdot 
            \mathbf{E}(\mathbf{0},t)
\nonumber \\
    &= \sum_{j=0}^{D-1}  \hbar \omega_{j} \ket{j}\!\bra{j}
        - \sum_{j, k=0}^{D-1} 
            \hbar
            \left( 
                \frac{\Omega_{jk}}{2} 
                    e^{-\ci \omega_{\text{d}} t} 
                    \ket{j} \! \bra{k}
                + \frac{\Omega_{jk}^*}{2} 
                    e^{\ci \omega_{\text{d}} t}  
                    \ket{k} \! \bra{j}
            \right),
\end{align}
where we have defined the complex Rabi frequencies\footnote{By using Eq.(\ref{eq:classical_quantum_rabi_frequency_comparison}), we can map the complex Rabi frequencies to the coupling coefficients in the fully-quantized picture: $\Omega_{jk} = 2\sqrt{n} g_{jk}$, where $n$ is the mean photon number of the classical drive.}
\begin{equation}
    \Omega_{jk}
    = \frac{\ci \bra{j}\hat{\mathbf{d}} \ket{k} \cdot \hat{\mathbf{e}}_{x} E_0 e^{\ci \phi}}{\hbar},
\end{equation}
To simplify the calculation, we first move to a frame rotating at frequency $\omega_{\text{d}}$ by using the unitary transformation
\begin{equation}
    \hat{R}(t)
    = \exp [ 
            - \ci \sum_{j=0}^{D-1} j \omega_{\text{d}} t \ket{j} \! \bra{j}
        ].
\end{equation}
Recall that given any unitary operator $\hat{U}(t)$ (which in general is time-dependent) and the state $\ket{\Psi(t)}$ that solves the usual Schr\"{o}dinger equation with Hamiltonian $\hat{H}$, the transformed state $\ket{\Psi'(t)} = \hat{U}^{\dagger}(t) \ket{\Psi(t)}$ satisfies a new Schr\"{o}dinger equation
\begin{equation}
    \ci \hbar \frac{\mathrm{d}}{\mathrm{d}t} \ket{\Psi'(t)} = \hat{H}'(t) \ket{\Psi'(t)}
\end{equation}
with the transformed Hamiltonian
\begin{equation}
    \hat{H}'(t) 
    = \hat{U}^{\dagger}(t) \hat{H} \hat{U}(t) 
        - \ci \hbar \hat{U}^{\dagger}(t) \frac{\mathrm{d}}{\mathrm{d} t} \hat{U}(t).
\end{equation}
Let $\hat{U}(t) = \hat{R}(t)$, the effective Hamiltonian in the rotating frame is given by
\begin{align}
    \hat{H}_{\text{rot}} 
    &= \hat{R}^{\dagger}(t) \hat{H} \hat{R}(t)
        - \ci \hbar \hat{R}^{\dagger}(t) \frac{\mathrm{d}}{\mathrm{d} t} \hat{R}(t)
\nonumber \\
    &= \sum_{j=0}^{D-1}  
            \hbar (\omega_{j} - j \omega_{\text{d}}) 
            \ket{j} \!\bra{j}
        - \sum_{j, k=0}^{D-1} 
            \hbar
            \left[ 
                \frac{\Omega_{jk}}{2} 
                    e^{\ci (j - k - 1) \omega_{\text{d}} t} 
                    \ket{j} \! \bra{k}
                + \frac{\Omega_{jk}^*}{2} 
                    e^{- \ci (j - k - 1)\omega_{\text{d}} t}  
                    \ket{k} \! \bra{j}
            \right].
\end{align}
In practice, the qudit is modeled as an anharmonic oscillator; hence, we will assume $\Omega_{jk} \approx 0$ if $|j-k|\neq 0$, resulting in
\begin{equation}
    \hat{H}_{\text{rot}} 
    \approx 
    \sum_{j=0}^{D-1} 
            \hbar (\omega_{j} - j \omega_{\text{d}}) 
            \ket{j} \!\bra{j}
        - \sum_{j=0}^{D-2} 
            \hbar
            \left( 
                \frac{\Omega_{j+1,j}}{2} 
                    \ket{j+1} \! \bra{j}
                + \frac{\Omega_{j+1,j}^*}{2}  
                    \ket{j} \! \bra{j+1}
            \right).
\end{equation}

\begin{figure}[t]
    \centering
    \includegraphics[scale=0.4]{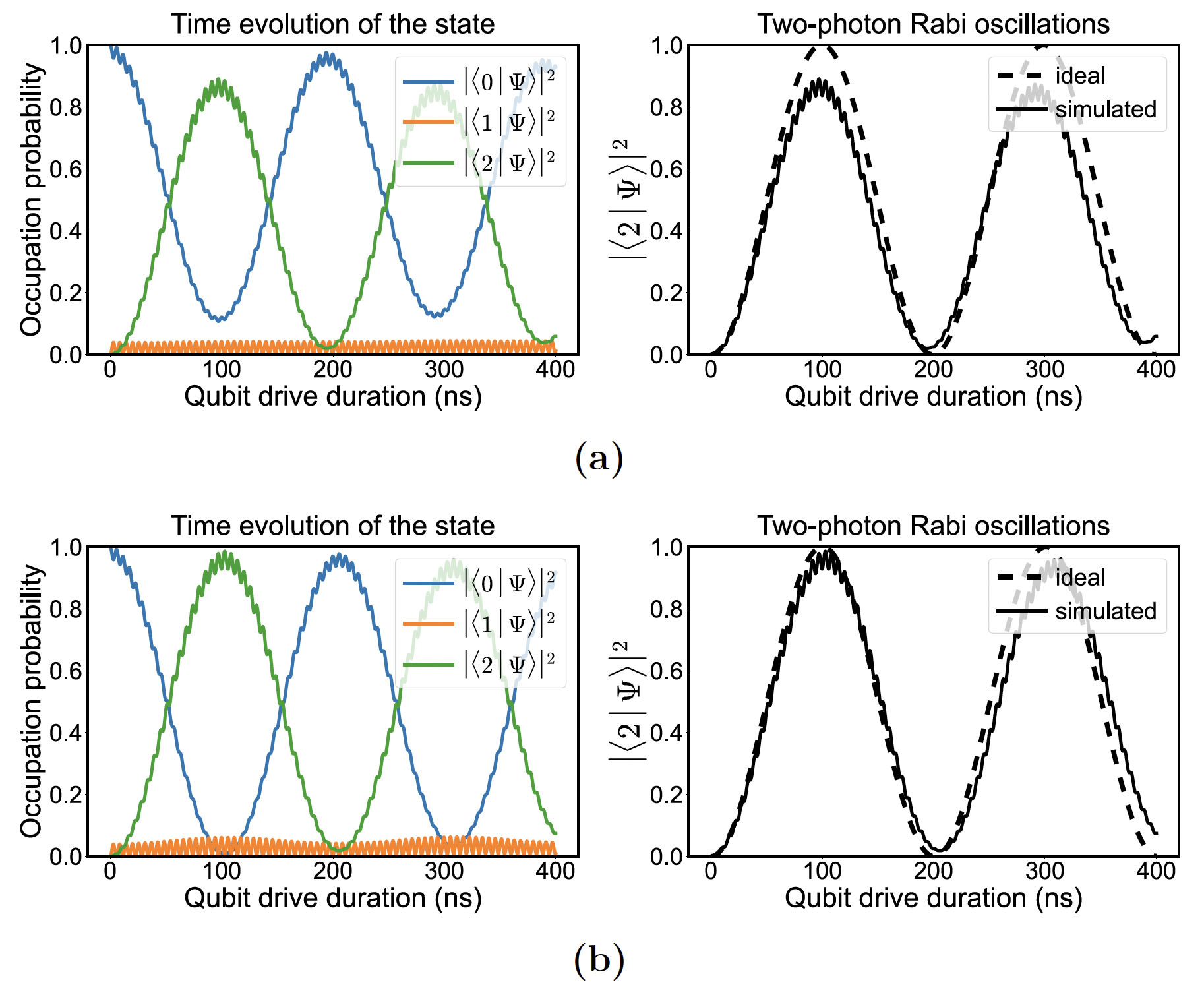} 
    \caption{Master equation simulations of the two-photon Rabi flopping with a $25$-kHz decay rate and a $100$-kHz pure dephasing rate added to each energy level. \textbf{a.} The (rectangular) control pulse has a drive frequency $\omega_{\mathrm{d}} = (\omega_{2} - \omega_{0})/2$. The idea Rabi frequency is calculated from Eq.(\ref{eq:two_photon_rabi_frequency_semiclassical}) \textbf{b.} The drive frequency takes into account the AC Stark shifts. Since the AC Stark shifts are themselves drive-frequency-dependent, the frequency shifts are calculated using fixed-point iterations.}
    \label{fig:two_photon_nonideal}
\end{figure}

Next, by setting
\begin{equation}
    \hat{H}_0 
    = \sum_{j=0}^{D-1} 
            \hbar (\omega_{j} - j \omega_{\text{d}}) 
            \ket{j} \!\bra{j}
\end{equation}
and
\begin{equation}
    \hat{V} 
    = - \sum_{j=0}^{D-2} 
            \hbar
            \left( 
                \frac{\Omega_{j+1,j}}{2} 
                    \ket{j+1} \! \bra{j}
                + \frac{\Omega_{j+1,j}^*}{2}  
                    \ket{j} \! \bra{j+1}
            \right),
\end{equation}
one can show that
\begin{equation}
    \hat{S} 
    = \sum_{j=0}^{D-2} 
        \frac{1}{\omega_{j+1} - \omega_j - \omega_{\text{d}}}
        \left( 
            \frac{\Omega_{j+1,j}}{2} 
                \ket{j+1} \! \bra{j}
            -\frac{\Omega_{j+1,j}^*}{2}  
                \ket{j} \! \bra{j+1}
        \right)
\end{equation}
solves $\hat{V} + \Bigsl[\hat{H}_0, \hat{S} \Bigsr] = \hat{0}$ as required by the Schrieffer–Wolff transformation. By applying the appropriate RWA while allowing the possibility of two-photon transitions, we arrive at
\begin{align}
    \hat{H}_{\text{rot}}' 
    &\approx \hat{H}_0 + \frac{1}{2} \Big[ \hat{V}, \hat{S} \Big]
\nonumber \\
    &\approx \sum_{j=0}^{D-1} 
            \hbar (\omega_{j} - j \omega_{\text{d}} + \chi_j) 
            \ket{j} \!\bra{j}
        - \sum_{j=0}^{D-3} \frac{\hbar }{2}
            \left(
                \Omega^{(2)}_{j+2,j} 
                \ket{j+2} \! \bra{j}
                + \Omega^{(2)*}_{j+2,j} 
                \ket{j} \! \bra{j+2}
            \right),
\end{align}
where 
\begin{equation}\label{eq:qudit_ac_stark_shift_semiclassical}
    \chi_j
    = \frac{|\Omega_{j,j-1}|^2}{4(\omega_{j} - \omega_{j-1} - \omega_{\text{d}})} 
        - \frac{|\Omega_{j+1,j}|^2 }{4(\omega_{j+1} - \omega_j - \omega_{\text{d}})}
\end{equation}
are the AC Stark shifts (since they depend on the strength of the drive) and 
\begin{equation} \label{eq:two_photon_rabi_frequency_semiclassical}
    \Omega^{(2)}_{j+2,j} 
    = \frac{\Omega_{j+2, j+1} \Omega_{j+1, j}}{2(\omega_{j+1} - \omega_{j} - \omega_{\text{d}})}
\end{equation}
are the (complex) Rabi frequencies of the two-photon transitions. Indeed, we do not see any Lamb shift; this discrepancy between the semi-classical and fully quantized picture can be neglected since the AC Stark shifts due to the classical drive (i.e., with a macroscopic photon number) are much larger than the photon-number-independent Lamb shift. Nevertheless, in both formulations, we have AC Stark shifts to the original energy levels, suggesting that we should not drive the two-photon transition exactly at $\omega_{\text{d}} = (\omega_{j_2} - \omega_{j_1})/2$ since the qudit is simultaneously subject to the AC Stark shifts from the same field. Two master equation simulations of the two-photon transitions are shown in Figure \ref{fig:two_photon_nonideal}; Figure \ref{fig:two_photon_nonideal}(a) shows the change of populations in a qutrit when subject to a drive at exactly $\omega_{\text{d}} = (\omega_{2} - \omega_{0})/2$ while Figure \ref{fig:two_photon_nonideal} has a drive frequency that includes the shifts in Eq.(\ref{eq:qudit_ac_stark_shift_semiclassical}), proving that the qutrit indeed has shifted energy levels.

\begin{figure}
    \centering
    \includegraphics[scale=0.34]{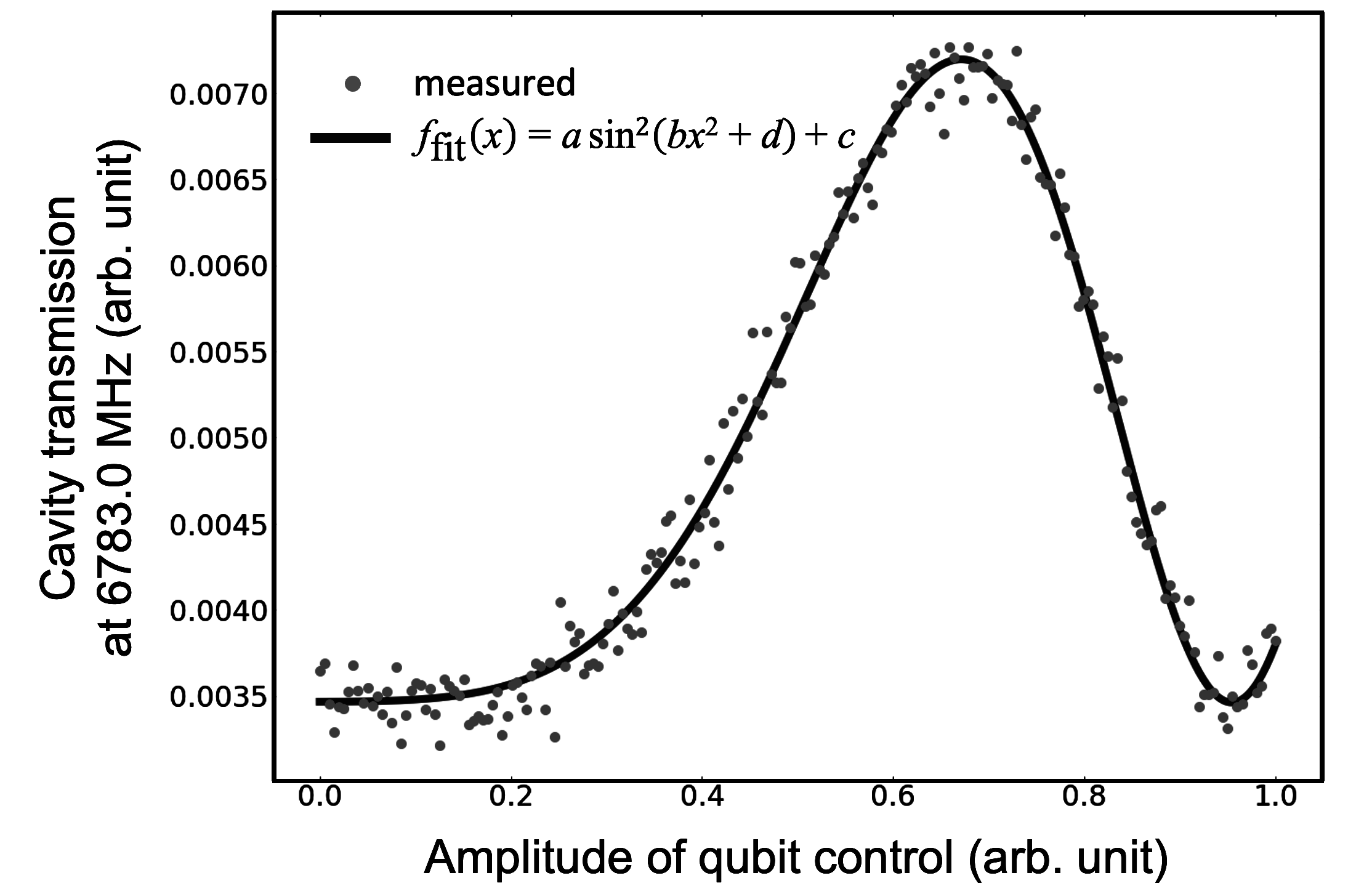}
    \caption{Power dependence of the two-photon Rabi flopping inferred from the averaged cavity transmission. Note that the power of the drive is proportional to the drive amplitude squared.}
    \label{fig:power_dependence_two_photon_rabi}
\end{figure}

As mentioned before, a two-photon transition requires more power than a single-photon transition since the coupling strength, as shown in Eq.(\ref{eq:two_photon_rabi_frequency_semiclassical}), is suppressed by the detuning between the drive frequency $\omega_{\text{d}}$ and the virtual transition frequency $\omega_{j+1} - \omega_{j}$. In addition, the Rabi frequency is proportional to the power of the field instead of the amplitude since 
\begin{equation}
    \Omega^{(2)}_{j+2,j} 
    \propto \Omega_{j+2, j+1} \Omega_{j+1, j}
    \propto E_0^2.
\end{equation}
This power dependence can be easily observed in experiments as shown in Figure \ref{fig:power_dependence_two_photon_rabi}. 

Furthermore, since an AC Stark shift is also a two-photon process, it has the same power dependence and usually requires substantial signal power to create any visible effect. For example, we might use the AC Stark shift to add an extra phase $e^{-\ci \chi_j T}$ to the eigenstates of the qudits by bringing a pump drive near the qudit frequencies for a pulse duration $T$. However, such a way of creating a phase gate is not practical in superconducting quantum computation because the high-photon-number pump will drastically increase the base temperature of the dilution fridge. Instead, people often use DC magnetic fields combined with SQUIDs (see the next chapter) to tune the qudit frequencies.

\section{Single-Qubit Gates and Optimization}
\subsection{Time Evolution Operators for a General Single-Qubit Gate}
In the last three sections, we have studied various ways of building single-qubit gates, and it's not hard to see that we can construct any single-qubit gates by manipulating the JC or the semi-classical Hamiltonian. However, in practice, we often prefer to abstract away the underlying physics and simply use matrix algebra to describe the action of a sequence of gates.

In both the fully quantized and semi-classical cases, we can lump all the single-qubit gates into a single matrix. For example, given that the JC Hamiltonian is block-diagonalizable, it is possible to obtain the time evolution operator restricted to a specific two-dimensional subspace by simply exponentiating the Hamiltonian $\hat{H}_n$. To find the matrix exponential of $\hat{H}_n$, one first observe that
\begin{align}
    \hat{H}_n 
    &= \hbar \begin{pmatrix}
        n \omega_{\text{r}} - \frac{1}{2} \Delta 
        & - \sqrt{n} |g| e^{- \ci \phi_g}\\
        -  \sqrt{n}  |g| e^{\ci \phi_g}
        & n \omega_{\text{r}} + \frac{1}{2} \Delta
    \end{pmatrix}
\nonumber \\
    &= n\omega_{\text{r}} \hat{1}
        - \frac{\Omega_n \cos\phi_g}{2} \hat{\sigma}_x
        - \frac{\Omega_n \sin\phi_g}{2} \hat{\sigma}_y
        - \frac{\Delta}{2} \hat{\sigma}_z
    = n\omega_{\text{r}} \hat{1}
        - \frac{\Omega'_n}{2} 
            \hat{\mathbf{e}}_{\Omega'_n} 
            \cdot
            \hat{\boldsymbol{\sigma}}.
\end{align}
Thus, by using the identity
\begin{equation}
    \exp(\ci \alpha \hat{\mathbf{e}} \cdot \hat{\boldsymbol{\sigma}})
    = \hat{1} \cos \alpha 
        + \ci \hat{\mathbf{e}} \cdot \hat{\boldsymbol{\sigma}} \sin \alpha 
\end{equation}
for any real number $\alpha$ and unit vector $\hat{\mathbf{e}}$, we obtain the time evolution operator
\begin{align}
    \hat{U}_n(t)
    &= \exp \!
        \left(
            - \frac{\ci}{\hbar} \hat{H}_n t
        \right)
    = \exp \!
        \left[
            - \ci
            \left(
                n\omega_{\text{r}} \hat{1}
                - \frac{\Omega'_n}{2} 
                    \hat{\mathbf{e}}_{\Omega'_n} 
                    \cdot
                    \hat{\boldsymbol{\sigma}}
            \right)
        \right]
\nonumber \\[2mm]
    &= e^{-\ci n\omega_{\text{r}} t} 
    \left[
        \cos\left( \frac{1}{2}\Omega'_n t \right) \hat{1}
        + \ci \sin\left( \frac{1}{2}\Omega'_n t \right) 
            \left(
                \frac{\sqrt{n} \Omega \cos \phi_{g}}{\Omega'_n} \hat{\sigma}_x
                + \frac{\sqrt{n} \Omega \sin \phi_{g}}{\Omega'_n} \hat{\sigma}_y
                + \frac{\Delta}{\Omega'_n} \hat{\sigma}_z
            \right)
    \right]
\nonumber\\[2mm] \label{eq:single_qubit_gate_unitary_op}
    &= e^{-\ci n\omega_{\text{r}} t}
    \begin{pmatrix} \displaystyle
        \cos\left( \frac{1}{2} \Omega'_n t \right) 
        + \ci \frac{\Delta}{\Omega'_n} \sin \left( \frac{1}{2}\Omega'_n t \right) 
        & \displaystyle \ci \frac{\sqrt{n} \Omega}{\Omega'_n} e^{-\ci \phi_{g}} \sin\left( \frac{1}{2}\Omega'_n t \right) \\[4mm]
        \displaystyle \ci \frac{\sqrt{n}\Omega}{\Omega'_n} e^{\ci \phi_{g}} \sin\left( \frac{1}{2}\Omega'_n t \right) 
        & \displaystyle \cos \left( \frac{1}{2}\Omega'_n t \right) 
        - \ci \frac{\Delta}{\Omega'_n} \sin \left( \frac{1}{2}\Omega'_n t \right) 
    \end{pmatrix}.
\end{align}
Eq.(\ref{eq:single_qubit_gate_unitary_op}) can be applied to all regimes mentioned before under the RWA and is thus general for describing all the possible gates enabled by a weak coupling to the field. It can be easily shown that $\hat{U}(t)$, with carefully chosen parameters, can realize any single-qubit gate (put in another way, we can write any unitary operators up to some global phase in terms of $\hat{U}(t)$ by designing $\Delta$, $\Omega$ and $\phi_g$). For example, setting $\Delta=0$ gives the ideal Rabi oscillations while making $\sqrt{n}\Omega / \Delta \ll 1$ generates the AC Stark shifts. Although practically not very useful, it's also possible to write down a general time evolution operator without referring to the photon number basis \cite{STENHOLM19731}:
\begin{align}
    \hat{U}_{\text{JC}}(t) 
    &= \ket{g} \! \bra{g} \otimes
        e^{-\ci \omega_{\text{r}} \hat{N} t} 
        \text{\footnotesize $
        \left[
                \cos 
                \left(
                    \frac{t}{2}
                    \sqrt{\Omega^2 \hat{N} + \Delta^2} 
                \right)
            + \ci \frac{\Delta}{\sqrt{\Omega^2 \hat{N} + \Delta^2}}
                \sin 
                \left( 
                    \frac{t}{2} \sqrt{\Omega^2 \hat{N} + \Delta^2}
                \right) 
        \right]
        $}
\nonumber \\[2mm] 
    &\ \ \ \ 
        +  \ket{g} \! \bra{e} \otimes e^{- \ci \omega_{\text{r}} \hat{N} t} 
        \text{\footnotesize $
        \left[
            \ci \frac{\Omega}{\sqrt{\Omega^2 \hat{N} + \Delta^2}} e^{- \ci \phi_{g}} \sin \left( \frac{t}{2} \sqrt{\Omega^2 \hat{N} + \Delta^2} \right) 
        \right] 
        $}
        \hat{a}^{\dagger}
\nonumber \\[2mm]
    &\ \ \ \ 
        + \ket{e} \! \bra{g} \otimes e^{-\ci \omega_{\text{r}} \hat{N} t} 
        \, \, \hat{a} 
        \text{\footnotesize $
        \left[
            \ci \frac{\Omega}{\sqrt{\Omega^2 \hat{N} + \Delta^2}} e^{\ci \phi_{g}} \sin \left(\frac{t}{2} \sqrt{\Omega^2\hat{N} + \Delta^2} \right) 
        \right]
        $}
\nonumber \\[2mm] 
    &\ \ \ \ 
        + \ket{e} \! \bra{e} 
            \otimes 
            e^{
                -\ci \omega_{\text{r}} 
                \big(
                    \hat{N} + 1 
                \big)  
                t
            } 
\nonumber \\[2mm] 
    &\ \ \ \  \ \ \ \ 
        \text{\footnotesize $\left[
            \cos 
                \left( 
                    \frac{t}{2}
                    \sqrt{\Omega^2 
                            \Bigsl(
                                \hat{N} \! + \! 1 
                            \Bigsr) 
                        + \Delta^2}
                \right) 
            - \ci 
                \frac{\Delta}{
                    \sqrt{\Omega^2 
                        \Bigsl(
                                \hat{N} \! + \! 1 
                            \Bigsr) 
                        + \Delta^2
                    }
                }
                \sin 
                \left( 
                    \frac{t}{2} 
                    \sqrt{\Omega^2 
                        \Bigsl(
                                \hat{N} \! + \! 1 
                            \Bigsr) 
                        + \Delta^2
                    }
                \right) 
        \right].
        $}
\nonumber
\end{align}
\normalsize
\begin{equation}
    {}
\end{equation}
Note the extra $\hat{a}$ and $\hat{a}^{\dagger}$ in the second and third terms, respectively, in order to make the math work out.\footnote{Also note that the singularity is the avoided because $\sin(x)/x$ is well-defined at $x=0$.} 

\begin{figure}[ht]
    \centering
    \includegraphics[scale=0.25]{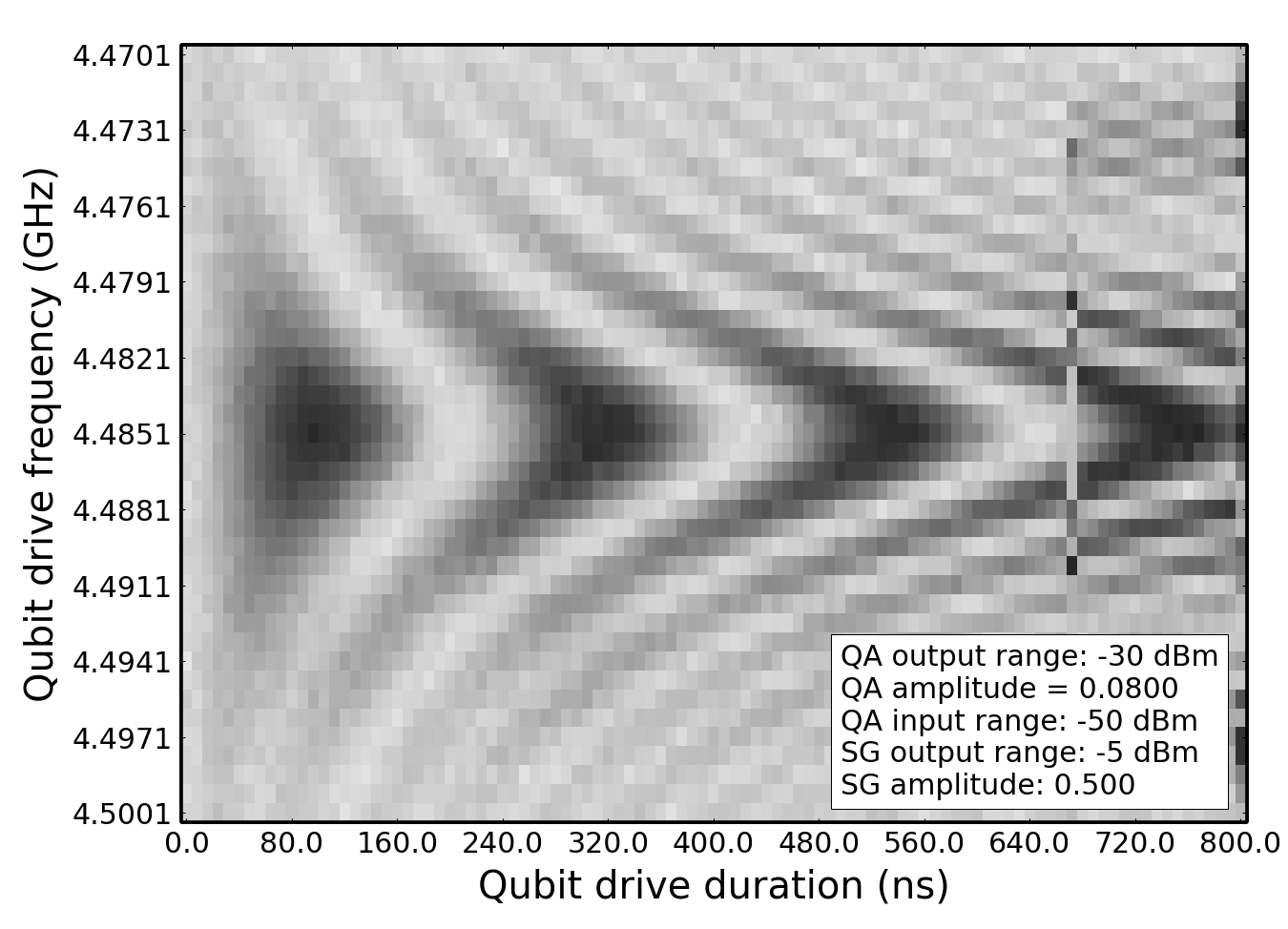}
    \caption{Rabi flopping with different detunings. The horizontal axis represents the duration of the Rabi drive, while the vertical axis shows a sweep of the drive frequency near the actual qubit frequency. The color of the plot encodes the population of the ground state, with white representing $p_g = 1$ and black representing $p_g = 0$. At the qubit frequency, the Rabi flopping has the slowest oscillation frequency and the strongest population contrast; as we increase the detuning, the population contrast decreases while the oscillation frequency increases. (QA: quantum analyzer for state readout. SG: signal generator for qubit control.)}
    \label{fig:chevron_plot}
\end{figure}

More often used in practice is the time evolution operator from semi-classical analysis, which is given by
\begin{equation} \label{eq:semiclassical_single_qubit_gate}
    \hat{U}(t)\!
    = \begin{pmatrix} 
        e^{\ci \omega_{\text{d}} t /2} \cos\!\left( \frac{1}{2} \Omega' t \right) 
        + \ci e^{\ci \omega_{\text{d}} t /2} \frac{\Delta}{\Omega'} \sin\! \left( \frac{1}{2}\Omega' t \right) 
        &  \ci e^{\ci \omega_{\text{d}} t /2 -\ci \phi_{g}} \frac{\Omega}{\Omega'} \sin\! \left( \frac{1}{2}\Omega' t \right) \\[4mm]
         \ci e^{-\ci \omega_{\text{d}} t /2 + \ci \phi_{g}} \frac{\Omega}{\Omega'} \sin\!\left( \frac{1}{2}\Omega' t \right) 
        &  e^{-\ci \omega_{\text{d}} t /2} \cos\! \left( \frac{1}{2}\Omega' t \right) 
        - \ci e^{-\ci \omega_{\text{d}} t /2} \frac{\Delta}{\Omega'} \sin\! \left( \frac{1}{2}\Omega' t \right) 
    \end{pmatrix}.
\end{equation}
The Rabi frequency $\Omega$ is proportional to the amplitude of the microwave signal coming out from the signal generator; hence, to design a single-qubit gate, one often first sweeps the microwave power to pin down a reasonable Rabi frequency by measuring how fast the qubit oscillates between $\ket{g}$ and $\ket{e}$ (this oscillation frequency must be the generalized Rabi frequency). To find the resonance condition (i.e., $\Delta=0$), one then fixes the signal power and sweeps the drive frequency to find the minimum generalized Rabi frequency because $\Omega' = \sqrt{\Delta^2 +\Omega^2}$ is minimized when $\Delta = 0$. Figure \ref{fig:chevron_plot} shows the experimentally measured Rabi oscillations (as a function of time) for a list of the detuning around the qubit transition frequency ($4.4851 \text{ GHz}$ in the case).

\subsection{Area Theorem}
So far we have always assumed that the signal generator is suddenly turned on and off to define a pulse, which leads to the Rabi oscillations with a well-defined sinusoidal behavior. In this ideal case, a $\pi$-pulse is defined by setting the pulse length to be $T_{\pi} = \pi/\Omega$. However, it's straightforward to show, by matrix exponentiation of the semiclassical Hamiltonian, that we will still have Rabi oscillations even if the Rabi frequency is time-varying (i.e., the electric field has a time-varying envelope besides the carrier signal at $\omega_{\text{d}}$). This result is particularly useful since a real signal must start and end with zero amplitude. For a $\pi$-pulse, what becomes important is thus the ``area'' swept by the generalized Rabi frequency, i.e.,
\begin{equation}
    \pi = \int_{0}^{T_{\pi}} \Omega'(t) \, \mathrm{d} t.
\end{equation}
Clearly, if $\Omega'$ is constant, the area theorem reduces to the usual definition of a $\pi$-pulse. A similar generalization applies to other pulses, such as the $\pi/2$-pulse.

\subsection{Derivative Removal by Adiabatic Gate (DRAG)}
The time evolution operators we just obtained are useful guidance for designing a single-qubit gate; however, one often needs to optimize the pulses in practice because a physical qubit is not an ideal two-level system. Consequently, for a multi-level system, short rectangular pulses (i.e., pulses that have no rise and fall time) will contain a wide range of frequencies, thus leading to unwanted excitations\footnote{For an ideal two-level system, the Fourier theorem is not an issue since the Hilbert space is strictly two-dimensional; leakage is not possible by definition.}. In addition, to use higher energy eigenstates for quantum computation, we also need to understand and mitigate leakage to nearby levels when exciting the intended transition.

The simplest solution to avoid the wide spectrum of the rectangular pulse is to add a finite rise and fall time to the pulse. For example, we can replace the constant pulse with a Gaussian pulse (with the carrier frequency being the drive frequency $\omega_{\text{d}}$). By the Fourier theorem, the spectrum of the Gaussian pulse will also be a Gaussian centered at $\omega_{\text{d}}$; the wider the pulse is in the time domain, the narrower its spectrum will be. However, this clearly does not rescue us completely from the unwanted transition since we cannot make the spectrum infinitely peaked at $\omega_{\text{d}}$ in practice. To further remove the leakage, we introduce two quadratures of the drive signals, i.e.,
\begin{equation}
    \mathbf{E}(t) 
    = \hat{\mathbf{e}}_{x} 
        \Big[ 
            E_1(t) \cos(\omega_{\text{d}} t)
            + E_2(t) \sin(\omega_{\text{d}} t)
        \Big],
\end{equation}
where $E_1(t)$ and $E_2(t)$ are slow-varying envelopes of the two quadrature fields. As discussed before, the phase of the drive signal determines the axis of rotation; the hope is that adding this extra degree of freedom can help us counter the leakage to higher energy levels in a real qubit. It should be emphasized that the technique we are going to derive, known as \textbf{Derivative Removal by Adiabatic Gate (DRAG)} \cite{PhysRevA.83.012308, PhysRevLett.103.110501}, only applies to the transition between the ground and the first excited state, which is what we care about if we only use a multi-level system as a qubit.

Given the two drive fields, the Hamiltonian of the driven qudit is
\begin{equation}
    \hat{H}(t)
    = \sum_{j = 0}^{D-1} 
            \hbar \omega_j \ket{j} \! \bra{j}
        + \sum_{j = 1}^{D-1}
            \mathbf{E}(t) 
            \cdot 
            \Big(
                \mathbf{d}_{j,j-1} \ket{j} \! \bra{j-1}
                + \mathbf{d}_{j,j-1}^* \ket{j-1} \! \bra{j}
            \Big),
\end{equation}
where we have assumed the anharmonic model of the qudit and defined $\mathbf{d}_{j,j-1} = \bra{j} \hat{\mathbf{d}} \ket{j-1}$. To remove the fast oscillating field, we move to a frame that rotates at $\omega_{\text{d}}$ by using
\begin{equation}
    \hat{R}(t) 
    = \exp[
            - \ci
            \sum_{j=0}^{D-1}
                j \omega_{\text{d}} t  \ket{j} \! \bra{j}
        ],
\end{equation}
which results in the transformed Hamiltonian
\begin{align}
    \hat{H}_{\text{rot}} (t)
    &= \hat{R}^{\dagger}(t) \hat{H}(t) \hat{R}(t) - \ci \hbar \hat{R}^{\dagger}(t) \frac{\mathrm{d}}{\mathrm{d} t} \hat{R}(t)
\nonumber \\
    &\approx \hbar \sum_{j = 0}^{D-1} 
        \left[
            (\omega_j - j \omega_{\text{d}})
                \ket{j} \! \bra{j}
            - \frac{\Omega_{x,j, j-1}(t)}{2} 
                \hat{\sigma}_{x, j, j-1}
            - \frac{\Omega_{y,j, j-1}(t)}{2} 
                \hat{\sigma}_{y, j, j-1}
        \right],
\end{align}
where, for simplicity, we have assumed that $\mathbf{d}_{j,j-1}$ are real-valued\footnote{One can always compensate the phase in the dipole by an extra phase added to the two quadratures of the drive field.} so that
\begin{equation}
    \Omega_{x, j, j-1}(t)
    = \frac{\mathbf{d}_{j,j-1} \cdot \hat{\mathbf{e}}_{x} E_1(t)}{\hbar},
    \ \ \ \ 
    \Omega_{y, j, j-1}(t)
    = \frac{\mathbf{d}_{j,j-1} \cdot \hat{\mathbf{e}}_{x} E_2(t)}{\hbar}
\end{equation}
are real\footnote{Note that the subscripts $x$ and $y$ are unrelated to the real space coordinates. Instead, they represent the rotations along the $x$- and $y$-axes of the Bloch sphere.} and 
\begin{equation}
    \hat{\sigma}_{x, j, j-1}
    = \ket{j} \! \bra{j-1} + \ket{j-1} \! \bra{j},
    \ \ \ \ 
    \hat{\sigma}_{y, j, j-1}
    = \ci \ket{j} \! \bra{j-1} - \ci \ket{j-1} \! \bra{j}.
\end{equation}
Since a low-fidelity gate is due to leakage to the most nearby level, i.e., the second excited state, we will restrict the Hamiltonian to only three energy levels ($\ket{0}$, $\ket{1}$, and $\ket{2}$\footnote{Since we do not have photon number states in the semi-classical derivation, the use of $j=0,1,2$ as the qudit levels should not cause any confusion.}); in other words, our goal is to prevent the excitation of $\ket{2}$ in the following Hamiltonian of a driven qutrit
\begin{align}
    \hat{H}_{\text{rot}} (t)
    &= \hbar \Delta \ket{1} \! \bra{1}
        +  \hbar (2\Delta + \alpha) \ket{2} \! \bra{2}
\nonumber \\
    & \ \ \ \ 
        - \frac{\hbar \Omega_x(t)}{2} \hat{\sigma}_{x,10} 
        - \frac{\hbar \Omega_y(t)}{2} \hat{\sigma}_{y,10}
        - \frac{\hbar \lambda \Omega_x(t)}{2} \hat{\sigma}_{x,21} 
        - \frac{\hbar \lambda \Omega_y(t)}{2} \hat{\sigma}_{y,21},
\end{align}
where we have set $\omega_0 = 0$, $\Omega_{x,10} = \Omega_x$, and $\Omega_{y,10} = \Omega_y$ and have defined $\Delta = \omega_1 - \omega_{\text{d}}$, $\alpha = \omega_2 - 2\omega_1$, and $\lambda = \Omega_{x,21}/\Omega_{x,10} = \Omega_{y,21}/\Omega_{y,10} = |\mathbf{d}_{21}|/|\mathbf{d}_{10}|$. The parameter $\alpha$ is known as the anharmonicity since, for an anharmonic oscillator, $\alpha$ describes the deviation of the second energy level from the ideal level of a QHO. In addition, $\lambda \approx \sqrt{2}$ for a weakly anharmonic oscillator.

Mathematically, our objective is the same as finding two envelope functions such that $\hat{H}_{\text{rot}}$ can be block-diagonalized into a two-dimensional subspace spanned by $\ket{0}$ and $\ket{1}$ and another one-dimensional space spanned by $\ket{2}$. The block-diagonalization can be achieved by first introducing the following unitary transformation, 
\begin{equation}
    \hat{T}(t) 
    = \exp[ 
            - \frac{\ci \Omega_x(t)}{2 \alpha} 
            \Big(
                \hat{\sigma}_{y,10} 
                + \lambda \hat{\sigma}_{y,21}
            \Big)
        ].
\end{equation}
Note that if we assume the pulse starts and ends with zero amplitude (i.e., $\Omega_x(t=0) = \Omega_x(t=T_{\text{p}}) = 0$, where $T_{\text{p}}$ is the pulse duration), then $\hat{T}(0) = \hat{T}(T_{\text{p}}) = \hat{1}$ and
\begin{equation}
    \ket{\Psi'(0 \text{ or } T_{\text{p}})} = \hat{T}^{\dagger}(0\text{ or } T_{\text{p}}) \ket{\Psi(0\text{ or } T_{\text{p}})} = \ket{\Psi(0\text{ or } T_{\text{p}})}.
\end{equation}
In other words, if we can reach the target state in the transformed frame, so can we in the rest frame. To compute the transformed Hamiltonian, we, again, use the formula $\hat{H}_{\text{rot}}'(t) = \hat{T}^{\dagger} \hat{H}_{\text{rot}} \hat{T} - \ci \hbar \hat{T}^{\dagger} \dot{\hat{T}}$. The first term in the formula, to the first order in $\Omega_x/\alpha$, is given by
\begin{align}
    \hat{T}^{\dagger} \hat{H}_{\text{rot}} \hat{T}
    &\approx 
        \hbar 
        \begin{pmatrix}
            \displaystyle \frac{\Omega_x^2(\Delta - 2\alpha)}{4 \alpha^2} & \displaystyle -\frac{\Omega_x}{2} + \frac{\Delta \Omega_x}{2\alpha} + \frac{\ci \Omega_y}{2} & \displaystyle \frac{\lambda \Omega_x^2}{8\alpha} \\[4mm]
            \displaystyle - \frac{\Omega_x}{2} + \frac{\Delta \Omega_x}{2\alpha} - \frac{\ci \Omega_y}{2} & \displaystyle \Delta - \frac{(\lambda^2 - 2) \Omega_x^2}{4\alpha} & \displaystyle \frac{\lambda \Delta \Omega_x}{2\alpha} + \frac{\ci \lambda \Omega_y}{2} \\[4mm]
            \displaystyle \frac{\lambda \Omega_x^2}{8\alpha} & \displaystyle \frac{\lambda \Delta \Omega_x}{2\alpha} - \frac{\ci \lambda \Omega_y}{2} & \displaystyle \alpha + 2 \Delta + \frac{\lambda^2 \Omega_x^2}{4 \alpha}
        \end{pmatrix}
\nonumber \\[1mm]
    &\approx 
        \hbar 
            \left[
                \Delta - \frac{(\lambda^2 - 4) \Omega_x^2}{4\alpha}
            \right] 
            \ket{1} \! \bra{1}
        + \hbar 
            \left[
                (2 \Delta + \alpha) 
                + \frac{(\lambda^2 + 2) \Omega_x^2}{4\alpha}
            \right] 
            \ket{2} \! \bra{2}
\nonumber \\
    & \ \ \ \   \ \ \ \   \ \ \ \   \ \ \ \   \ \ \ \   \ \ \ \   \ \ \ \   \ \ \ \ 
        - \frac{\hbar \Omega_x}{2} 
            \hat{\sigma}_{x,10} 
        + \frac{\hbar \lambda \Omega_x^2}{8 \alpha} 
            \hat{\sigma}_{x,20}
        - \frac{\hbar \Omega_y}{2} 
            \Big( 
                \hat{\sigma}_{y,10} 
                + \lambda \hat{\sigma}_{y,21} 
            \Big),
\end{align}
where we have shifted the reference energy to the ground-state energy in the second line. Consequently, the transformed Hamiltonian becomes
\begin{align}
    \hat{H}_{\text{rot}}'(t) 
    &= \hat{T}^{\dagger} \hat{H}_{\text{rot}} \hat{T}
        - \ci \hbar \hat{T}^{\dagger} \dot{\hat{T}}
\nonumber \\
    &\approx
        \hbar 
            \left[
                \Delta - \frac{(\lambda^2 - 4) \Omega_x^2(t)}{4\alpha}
            \right] 
            \ket{1} \! \bra{1}
        + \hbar 
            \left[
                (2 \Delta + \alpha) 
                + \frac{(\lambda^2 + 2) \Omega_x^2(t)}{4\alpha}
            \right] 
            \ket{2} \! \bra{2}
\nonumber \\
    & \ \ \ \   \ \ \ \ 
        - \frac{\hbar \Omega_x(t)}{2} 
            \hat{\sigma}_{x,10} 
        + \frac{\hbar \lambda \Omega_x^2(t)}{8 \alpha} 
            \hat{\sigma}_{x,20}
        - \hbar 
            \left[ 
                \frac{\Omega_y(t)}{2} + \frac{\dot{\Omega}_x(t)}{2\alpha}
            \right]
            \Big( 
                \hat{\sigma}_{y,10} 
                + \lambda \hat{\sigma}_{y,21} 
            \Big).
\end{align}

For a real superconducting qubit design, $\alpha > 200 \text{ MHz}$ and $\Omega \approx 10 \text{ MHz}$; thus, the coupling strength $ \lambda \Omega_x^2/ 8 \alpha$ between $\ket{0}$ and $\ket{2}$ is much smaller than the coupling between $\ket{0}$ and $\ket{1}$ and between $\ket{1}$ and $\ket{2}$. What's not negligible is the coupling between $\ket{1}$ and $\ket{2}$; however, we can eliminate the term involving $\hat{\sigma}_{y,21}$ by designing $\Omega_y(t)$ to satisfy
\begin{equation}
    \frac{\Omega_y(t)}{2} 
        + \frac{\dot{\Omega}_x(t)}{2\alpha} 
    = 0.
\end{equation}
As a result, the projection operator $\ket{2}\!\bra{2}$ in $\hat{H}_{\text{rot}}'$ no longer talks to the qubit subspace; thus, leakage into $\ket{2}$ is turned off. In addition, the AC Stark shift of $\ket{1}$ can be removed if\\[2mm]
\begin{equation}
    \Delta - \frac{(\lambda^2 - 4) \Omega_x^2(t)}{4\alpha} = 0;
\end{equation}
however, in practice, it's difficult to vary the detuning $\Delta$ (i.e., the carrier frequency $\omega_{\text{d}}$), so one might add the average of the time-varying detuning as a constant offset to the drive frequency. The typical DRAG sequence used in a real experiment consists of the old Gaussian pulse with some width $\sigma_{\text{p}}$ as the in-phase signal (i.e., $\Omega_x(t)$) and its derivative as the quadrature signal (i.e., $\Omega_y(t)$); that is
\begin{equation}
    \Omega_x(t) = \frac{1}{\sqrt{2\pi \sigma_{\text{p}}}} e^{-\frac{(t - T_{\text{p}}/2)^2}{2\sigma_{\text{p}}^2}}
    \ \ \text{ and } \ \ 
    \Omega_y(t) = \frac{\dot{\Omega}_x(t)}{\alpha}.
\end{equation}
Note that the quadrature signal is proportional to the rate of change of the in-phase signal.
Intuitively, when $\sigma_{\text{p}}$ is small, the spectrum of the in-phase signal is wider, thus, requiring more corrections from the quadrature signal. 

\begin{figure}
    \centering
    \includegraphics[scale=0.35]{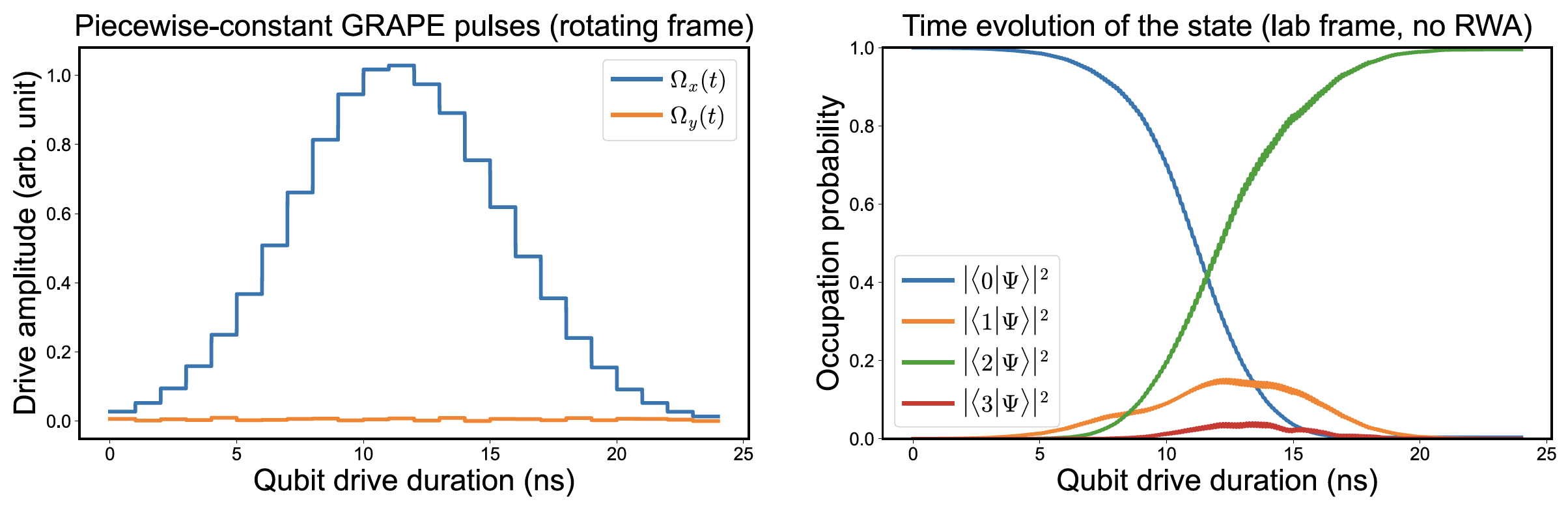}
    \caption{A two-photon transition designed by GRAPE. Left: The envelope of the control signals optimized by GRAPE. The actual control drives are modulated at a carrier frequency around the two-photon resonance. Right: The simulated state evolution using the optimized pulses. Indeed, the qudit reaches the target state with high fidelity by utilizing virtual states near $\ket{1}$ and $\ket{3}$.}
    \label{fig:GRAPE_two_photon}
\end{figure}

For optimization of the transitions between higher energy levels, we need more general approaches, usually numerical optimization based on optimal control theory \cite{PhysRevA.101.022321, s41534_020_00346_2}. For example, one can use \textbf{Gradient Ascent Pulse Engineering (GRAPE)} \cite{KHANEJA2005296} to optimize an arbitrary trajectory with given initial and target states by numerically maximizing the fidelity of the gate. An example of a two-photon transition optimized by GRAPE is shown in Figure \ref{fig:GRAPE_two_photon}.

\section{Two-Qubit Gates}
We end this chapter by providing several ways of realizing two-qubit gates. Since a complete gate set can be constructed by several single-qubit gates plus a two-qubit gate, it's crucial that we understand the physics behind building a two-qubit gate.
\subsection{Virtual Excitation of a Resonator}
We first consider two qubits with frequency $\omega_1$ and $\omega_2$ coupled to a common resonator with frequency $\omega_{\text{r}}$. We define $\Delta_1 = \omega_1 - \omega_{\text{r}}$ and $\Delta_2 = \omega_1 - \omega_{\text{r}}$ to be the detuning between the resonator and the two qubits, respectively; in addition, it is assumed that the resonator frequency is far away from the qubit frequencies. The Hamiltonian describing the composite system is nothing else but two copies of the JC Hamiltonian:
\begin{align}
    \hat{H}_{\text{JC},2} 
    = - \frac{1}{2} 
            \hbar \omega_{1} 
            \hat{\sigma}_{z,1}
        &- \frac{1}{2} 
            \hbar \omega_{2} 
            \hat{\sigma}_{z,2}
        + \hbar \omega_{\text{r}} \! 
            \left(
                \hat{a}^{\dagger}\hat{a} 
                + \frac{1}{2}
            \right)
\nonumber \\
        &- \hbar
            \Big( 
                g_1 \hat{a} \hat{\sigma}_{+,1}
                + g_1^* \hat{a}^{\dagger} \hat{\sigma}_{-,1}
            \Big)
        - \hbar
            \Big( 
                g_2 \hat{a} \hat{\sigma}_{+,2}
                + g_2^* \hat{a}^{\dagger} \hat{\sigma}_{-,2}
            \Big),
\end{align}
also known as the Tavis-Cummings model.
In the next chapter, we will discuss the physical origin of the qubit-resonator coupling; for now, we simply use $g_j$ to denote the coupling coefficients between the qubit $j$ and the resonator for $j=1,2$.

Under the condition\footnote{As pointed out before, the condition $|g_1|, |g_2| \ll |\Delta_1|, |\Delta_2|$ is not enough to ensure that the interaction is perturbative; in addition, we need to assume that the resonator is not excited to a high-photon-number state. In practice, since the system is cooled down to millikelvin temperature, the excitation on the resonator is negligible.} $|g_1|, |g_2| \ll |\Delta_1|, |\Delta_2|$, we can use the Schrieffer–Wolff transformation to adiabatically eliminate the qubit-resonator coupling. As usual, we define 
\begin{equation}
    \hat{H}_0 
    = - \frac{1}{2} 
            \hbar \omega_{1} 
            \hat{\sigma}_{z,1}
        - \frac{1}{2} 
            \hbar \omega_{2} 
            \hat{\sigma}_{z,2}
        + \hbar \omega_{\text{r}} \! 
            \left(
                \hat{a}^{\dagger}\hat{a} 
                + \frac{1}{2}
            \right)
\end{equation}
and 
\begin{equation}
    \hat{V} 
    = - \hbar
            \Big( 
                g_1 \hat{a} \hat{\sigma}_{+,1}
                + g_1^* \hat{a}^{\dagger} \hat{\sigma}_{-,1}
            \Big)
        - \hbar
            \Big( 
                g_2 \hat{a} \hat{\sigma}_{+,2}
                + g_2^* \hat{a}^{\dagger} \hat{\sigma}_{-,2}
            \Big),
\end{equation}
which leads to the choice
\begin{equation} \label{eq:S_for_two_qubit_res_SW_transform}
    \hat{S}
    = \left(
                \frac{g_1}{\Delta_1} 
                    \hat{a}\hat{\sigma}_{+,1} 
                - \frac{g_1^*}{\Delta_1} 
                    \hat{a}^{\dagger} \hat{\sigma}_{-,1}
            \right)
            + \left(
                \frac{g_2}{\Delta_2} 
                    \hat{a}\hat{\sigma}_{+,2} 
                - \frac{g_2^*}{\Delta_2} 
                    \hat{a}^{\dagger} \hat{\sigma}_{-,2}
            \right)
\end{equation}
and the transformed Hamiltonian
\begin{align}
    \hat{H}'
    &\approx \hat{H}_0 
        + \frac{1}{2} \Bigsl[\hat{V}, \hat{S} \Bigsr]
\nonumber \\
    &= - \frac{1}{2} 
            \hbar 
            \bigg(
                \omega_{1} 
                + \frac{|g_1|^2}{\Delta_1}
            \bigg)
            \hat{\sigma}_{z,1}
        - \frac{1}{2} 
            \hbar 
            \bigg(
                \omega_{1} 
                + \frac{|g_2|^2}{\Delta_2}
            \bigg)
            \hat{\sigma}_{z,1}
\nonumber \\
    & \ \ \    \ \ \ \ 
        + \hbar \omega_{\text{r}} 
                \hat{a}^{\dagger} \hat{a} 
        - \left( 
                \frac{\hbar |g_1|^2}{\Delta_1} 
                    \hat{\sigma}_{z,1} 
                + \frac{\hbar |g_2|^2}{\Delta_2} 
                    \hat{\sigma}_{z,2} 
            \right)
            \hat{a}^{\dagger}\hat{a}
\nonumber \\
    & \ \ \    \ \ \ \ 
        + \hbar \left[
            \frac{g_1 g_2^* (\Delta_1 + \Delta_2)}{2 \Delta_1 \Delta_2} 
                \hat{\sigma}_{+,1} \hat{\sigma}_{-,2}
            + \frac{g_1^* g_2 (\Delta_1 + \Delta_2)}{2 \Delta_1 \Delta_2}
                \hat{\sigma}_{-,1} \hat{\sigma}_{+,2}
        \right],
\end{align}
where we have ignored a constant energy $\hbar \omega_{\text{r}} / 2 + \hbar|g_1|^2/2 \Delta_1 + \hbar|g_2|^2/2 \Delta_2$. Like the single-qubit case, we can define the Lamb shift
\begin{equation}
    \chi_j = \frac{|g_j|^2}{\Delta_j}
\end{equation}
and the effective qubit frequency $\tilde{\omega}_j = \omega_j + \chi_j$ for $j = 1, 2$. In addition, we will define 
\begin{equation}\label{eq:two_qubit_gate_J_12}
    J_{12} 
    = - \frac{g_1 g_2^* (\Delta_1 + \Delta_2)}{2 \Delta_1 \Delta_2} 
    = |J_{12}| e^{\ci \phi_{J_{12}}}
\end{equation}
as the qubit-qubit coupling strength. Consequently, when $\hat{a}^{\dagger}\hat{a} = 0$, i.e., the resonator is unexcited, the coupling between the two qubits is explicitly derived:
\begin{equation}
    \hat{H}^{'}
    = - \frac{1}{2} 
            \hbar \Tilde{\omega}_{1} 
            \hat{\sigma}_{z,1}
        - \frac{1}{2} 
            \hbar \Tilde{\omega}_{2} 
            \hat{\sigma}_{z,2}
        - \hbar \Big(
            J_{12} \hat{\sigma}_{+,1} \hat{\sigma}_{-,2}
            + J_{12}^* \hat{\sigma}_{-,1} \hat{\sigma}_{+,2}
        \Big),
\end{equation}
with the resonator Hamiltonian decoupled and thus omitted. (If $\hat{a}^{\dagger}\hat{a} > 0$, each qubit will experience an AC Stark shift.)

One can then show that $J_{12} \hat{\sigma}_{+,1} \hat{\sigma}_{-,2} + J_{12}^* \hat{\sigma}_{-,1} \hat{\sigma}_{+,2}$ leads to an oscillation between the two qubits at a generalized Rabi frequency $\Omega_{12}' = \sqrt{(2|J_{12}|)^2 + (\Tilde{\omega}_1-\Tilde{\omega}_2)^2}$, similar to the qubit-resonator Rabi oscillations. 
For instance, if we start with the state $\ket{\Psi(0)} = \ket{1}_1 \otimes \ket{0}_2$ (i.e., the first qubit is excited and the second qubit is in the ground state), the state evolution is given by
\begin{align}
    \ket{\Psi(t)}
    &= \left[
                \cos\left( \frac{1}{2} \Omega_{12}' t \right) 
                - \ci \frac{\Tilde{\omega}_1-\Tilde{\omega}_2}{\Omega_n'} \sin \left( \frac{1}{2}\Omega_{12}' t \right) 
            \right] 
            \ket{1}_1 \otimes \ket{0}_2
\nonumber \\
    & \ \ \ \ \ \ \ \ 
    - \ci e^{\ci \phi_{J_{12}}} 
            \frac{2J_{12}}{\Omega_{12}'} \sin\left( \frac{1}{2}\Omega_{12}' t \right) \ket{0}_1 \otimes \ket{1}_2,
\end{align}
which also resembles the result derived from the JC Hamiltonian. Thus, to have a perfect Rabi flopping, we need the two qubits to have exactly the same frequency, which can be achieved by tuning the frequency of one of the qubits with a DC magnetic field (see next chapter). One important application of this type of two-qubit gate is to create Bell states. For instance, under the resonance condition $\Tilde{\omega}_1 = \Tilde{\omega}_2$, by starting with $\ket{\Psi(0)} = \ket{1}_1 \otimes \ket{0}_2$ and waiting for $T_{\text{B}} = \pi / 2\Omega'_{12}$, we end up with the entangled state 
\begin{equation}
    \ket{\Psi(T_{\text{B}})} 
    = \frac{\ket{1}_1 \otimes \ket{0}_2 - \ci e^{\ci \phi_{J_{12}}} \ket{0}_1 \otimes \ket{1}_2}{\sqrt{2}},
\end{equation}
which is one maximally-entangled state.

Moreover, to have a reasonable gate time, the detuning $\Delta_1$ and $\Delta_2$ cannot be too large according to Eq.(\ref{eq:two_qubit_gate_J_12}); this also gives us another way to turn on and off the coupling: To turn on the gate, move the resonator close to the qubit frequency such that $J_{12}$ is sufficiently large but $\Delta_1$ and $\Delta_2$ are still much larger than $g_1$ and $g_2$ in magnitude. To turn off the coupling, simply move the resonator frequency farther away so that $J_{12}$ is practically zero.

\subsection{Tunable Coupling}
In reality, qubits are physically placed near each other, resulting in unwanted static crosstalk. We can model this coupling by the dipole-dipole coupling 
\begin{equation} \label{eq:dipole_dipole_interaction_two_qubit}
    \hat{H}_{\text{dd}}
    = \frac{\hat{\mathbf{d}}_1 \cdot \hat{\mathbf{d}}_2}{4\pi \epsilon_0 |\hat{\mathbf{r}}_1 - \hat{\mathbf{r}}_2|^3 }
    \approx - \hbar 
        \Big(
            g_{12} \hat{\sigma}_{+,1} \hat{\sigma}_{-,2}
            + g_{12}^* \hat{\sigma}_{-,1} \hat{\sigma}_{+,2}
        \Big),
\end{equation}
where, as usual, we have used the RWA to eliminate $\hat{\sigma}_{-,1} \hat{\sigma}_{-,2}$ and $\hat{\sigma}_{+,1} \hat{\sigma}_{+,2}$. This native coupling results in unwanted entanglement between the qubits even when the two-qubit gates are turned off (i.e., by moving the resonator frequency far away to suppress $J_{12}$). We now provide a way to turn off the static coupling using resonator-mediated coupling.

First, note that the previous Schrieffer–Wolff transformation is no longer exact since $\hat{V} + \Bigsl[\hat{H}_0, \hat{S} \Bigsr] \neq \hat{0}$ when $\hat{H}_{\text{dd}}$ is added to $\hat{H}_0$. Nevertheless, if we restrict our calculation to the subspace where the resonator is unexcited (i.e., the resonator is in the vacuum state), the transformation is still valid, resulting in
\begin{align}
    \hat{H}'
    &= - \frac{1}{2} 
            \hbar \Tilde{\omega}_{1} 
            \hat{\sigma}_{z,1}
        - \frac{1}{2} 
            \hbar \Tilde{\omega}_{2} 
            \hat{\sigma}_{z,2}
        - \hbar \Big(
            J_{12} \hat{\sigma}_{+,1} \hat{\sigma}_{-,2}
            + J_{12}^* \hat{\sigma}_{-,1} \hat{\sigma}_{+,2}
        \Big)
\nonumber \\
    & \ \ \ \  \ \ \ \  \ \ \ \  \ \ \ \  \ \ \ \  \ \ \ \  \ \ \ \  \ \ \ \  \ \ \ \  
        - \hbar \Big(
            g_{12} \hat{\sigma}_{+,1} \hat{\sigma}_{-,2}
            + g_{12}^* \hat{\sigma}_{-,1} \hat{\sigma}_{+,2}
        \Big).
\end{align}
In other words, we have the resonator-mediated coupling plus the direct coupling between the two qubits. Then, if $J_{12}$ and $g_{12}$ have opposite signs, it's possible to completely turn off the coupling between the qubits outside the operation of two-qubit gates by tweaking the resonator frequency such that $J_{12} = g_{12}$ \cite{PhysRevApplied.10.054062}. In the next chapter, we will show that $J_{12}$ and $g_{12}$ indeed have the desired signs to make the cancellation possible.

When designing a real qubit, one should always keep in mind the multi-level nature of the qubit. The higher energy levels can also induce unwanted coupling between the qubits other than the static transverse coupling in Eq.(\ref{eq:dipole_dipole_interaction_two_qubit}). For example, one can show that two anharmonic oscillators will also experience a static $ZZ$ interaction (i.e., terms involving $\hat{\sigma}_{z,1} \hat{\sigma}_{z,1}$). Like the idea of the tunable coupling, one can design multiple paths (e.g., more tunable resonator-mediated couplings) between the qubits such that the probability amplitudes of various transitions cancel \cite{PhysRevLett.127.130501}.

\subsection{Cross Resonance}
The previous two-qubit interaction is known as a transverse coupling since the operators involved are $\hat{\sigma}_{+}$ and $\hat{\sigma}_{-}$. We now briefly discuss another type of two-qubit gate known as the $ZX$-gate, which involves the term $\hat{\sigma}_{z,1} \hat{\sigma}_{x,2}$.

The starting point is a system with some transverse qubit-qubit coupling. It does not matter whether this coupling is established using a resonator or a direct interaction; what we need is just some coupling coefficient $J_{12}$ between the two qubits. Next, we send a classical drive to the first qubit, however, \textit{at the (Lamb-shifted) transition frequency of the second qubit (i.e., $\omega_{\text{d}} = \Tilde{\omega}_2$)}, hence, resulting in the Hamiltonian
\begin{align}
    \hat{H} 
    &= - \frac{1}{2} \hbar \Tilde{\omega}_{1} 
            \hat{\sigma}_{z,1}
        - \frac{1}{2} \hbar \Tilde{\omega}_{2} 
            \hat{\sigma}_{z,2}
        - \hbar \Big( 
                J_{12} \hat{\sigma}_{+,1} \hat{\sigma}_{-,2}
                + J_{12}^* \hat{\sigma}_{-,1} \hat{\sigma}_{+,2}
            \Big)
\nonumber \\
    & \ \ \ \  \ \ \ \  \ \ \ \  \ \ \ \  \ \ \ \  \ \ \ \  \ \ \ \  \ \ \ \  \ \ \ \  
        - \hbar \Big( 
                g_{1\text{d}} e^{-\ci \Tilde{\omega}_{2} t} \hat{\sigma}_{+,1}
                + g_{1\text{d}}^* e^{\ci \Tilde{\omega}_{2} t} \hat{\sigma}_{-,1}
            \Big),
\end{align}
where $g_{1\text{d}}$ is the coupling coefficient due to the classical drive (i.e., it is half of the Rabi frequency associated with the drive).
To remove the time dependence of the control drive, we move to a rotating frame with frequency $\Tilde{\omega}_2$ by using
\begin{equation}
    \hat{R}(t)
    = \exp [ 
            \ci \Tilde{\omega}_2 t
            \Bigsl(
                \hat{\sigma}_{z,1} 
                + \hat{\sigma}_{z,2}
            \Bigsr) / 2
        ].
\end{equation}
After some algebra, we arrive at
\begin{align}
    \hat{H}_{\text{rot}}
    &= \hat{R}^{\dagger} \hat{H} \hat{R} - \ci \hbar \hat{R}^{\dagger} \dot{\hat{R}}
\nonumber \\
    &= - \frac{1}{2} \hbar \Tilde{\Delta}_{12}
            \hat{\sigma}_{z,1}
        - \hbar \Big( 
                J_{12} \hat{\sigma}_{+,1} \hat{\sigma}_{-,2}
                + J_{12}^* \hat{\sigma}_{-,1} \hat{\sigma}_{+,2}
            \Big)
        - \hbar \Big( 
                g_{1\text{d}} \hat{\sigma}_{+,1}
                + g_{1\text{d}}^* \hat{\sigma}_{-,1}
            \Big)
\nonumber \\
    &= \hbar \begin{pmatrix}
            - \Tilde{\Delta}_{12} / 2 & 0 & -g_{1\text{d}}^* & 0 \\[-0.75mm]
            0 & - \Tilde{\Delta}_{12} / 2 & -J_{12}^* & -g_{1\text{d}}^* \\[-0.75mm]
            -g_{1\text{d}} & -J_{12} & \Tilde{\Delta}_{12} / 2 & 0 \\[-0.75mm]
            0 & -g_{1\text{d}} & 0 & \Tilde{\Delta}_{12} / 2 
        \end{pmatrix},
\end{align}
where $\Tilde{\Delta}_{12} = \Tilde{\omega}_1 - \Tilde{\omega}_2$.

Unlike the previous case where we set $\tilde{\Delta}_{12} = 0$ for maximal entanglement, to observe the phenomenon of \textbf{cross resonance (CR)} \cite{PhysRevA.101.052308}, we intentionally detune the qubit frequencies so that $\tilde{\Delta}_{12} \gg g_{1\text{d}}$. For instance, if $g_{1\text{d}} \sim 5 \text{ MHz}$, then $\tilde{\Delta}_{12} \sim 100 \text{ MHz}$ would be appropriate. Under this assumption, we can apply the Schrieffer–Wolff transformation to eliminate the drive on the first qubit: Let
\begin{equation}
    \hat{H}_0 
    = - \frac{1}{2} \hbar \Tilde{\Delta}_{12}
            \hat{\sigma}_{z,1}
    = \hbar \begin{pmatrix}
            - \Tilde{\Delta}_{12} / 2 & 0 & 0 & 0 \\[-0.75mm]
            0 & - \Tilde{\Delta}_{12} / 2 & 0 & 0 \\[-0.75mm]
            0 & 0 & \Tilde{\Delta}_{12} / 2 & 0 \\[-0.75mm]
            0 & 0 & 0 & \Tilde{\Delta}_{12} / 2 
        \end{pmatrix}
\end{equation}
and 
\begin{align}
    \lambda \hat{V}
    &= - \hbar \Big( 
                J_{12} \hat{\sigma}_{+,1} \hat{\sigma}_{-,2}
                + J_{12}^* \hat{\sigma}_{-,1} \hat{\sigma}_{+,2}
            \Big)
        - \hbar \Big( 
                g_{1\text{d}} \hat{\sigma}_{+,1}
                + g_{1\text{d}}^* \hat{\sigma}_{-,1}
            \Big)
\nonumber \\
    &= \underbrace{\left(\frac{|g_{1\text{d}}|}{\Tilde{\Delta}_{12}} \right)}_{\lambda} 
        \underbrace{ 
            \frac{\hbar \Tilde{\Delta}_{12}}{|g_{1\text{d}}|}
            \begin{pmatrix}
                0 & 0 & -g_{1\text{d}}^* & 0 \\[-0.75mm]
                0 & 0 & -J_{12}^* & -g_{1\text{d}}^* \\[-0.75mm]
                -g_{1\text{d}} & -J_{12} & 0 & 0 \\[-0.75mm]
                0 & -g_{1\text{d}} & 0 & 0
            \end{pmatrix}}_{\hat{V}},
\end{align}
where we have explicitly written out the adiabatic parameter $\lambda$ to show the validity of the transformation. One then finds that
\begin{equation}
    \hat{S} 
    = \frac{1}{|g_{1\text{d}}|}
        \begin{pmatrix}
            0 & 0 & -g_{1\text{d}}^* & 0 \\[-0.75mm]
            0 & 0 & -J_{12}^* & -g_{1\text{d}}^* \\[-0.75mm]
            g_{1\text{d}} & J_{12} & 0 & 0 \\[-0.75mm]
            0 & g_{1\text{d}} & 0 & 0
        \end{pmatrix}
\end{equation}
can be used to eliminate the interaction to the first order. The transformed Hamiltonian is, thus, given by
\begin{align}
    \hat{H}_{\text{rot}}^{\text{CR}}
    &\approx \hat{H}_0 
        + \frac{\lambda^2}{2} \Bigsl[\hat{V}, \hat{S} \Bigsr]
\nonumber \\
    &= \hbar 
        \begin{pmatrix}
            \displaystyle - \frac{\Tilde{\Delta}_{12}}{2} - \frac{|g_{1\text{d}}|^2}{\Tilde{\Delta}_{12}} & \displaystyle - \frac{g_{1\text{d}}^* J_{12}}{\Delta} & 0 & 0 \\[2.75mm]
            \displaystyle - \frac{g_{1\text{d}} J_{12}^*}{\Tilde{\Delta}_{12}} & \displaystyle - \frac{\Tilde{\Delta}_{12}}{2} - \frac{|g_{1\text{d}}|^2 + |J_{12}|^2}{\Tilde{\Delta}_{12}} & 0 & 0 \\[2.75mm]
            0 & 0 & \displaystyle \frac{\Tilde{\Delta}_{12}}{2} + \frac{|g_{1\text{d}}|^2 + |J_{12}|^2}{\Tilde{\Delta}_{12}} & \displaystyle \frac{g_{1\text{d}}^* J_{12}}{\Tilde{\Delta}_{12}} \\[2.75mm]
            0 & 0 & \displaystyle \frac{g_{1\text{d}} J_{12}^*}{\Tilde{\Delta}_{12}} & \displaystyle \frac{\Tilde{\Delta}_{12}}{2} + \frac{|g_{1\text{d}}|^2}{\Tilde{\Delta}_{12}}
        \end{pmatrix}.
\end{align}
In terms of the basic two-qubit gates,
\begin{align}
    \hat{H}_{\text{rot}}^{\text{CR}}
    &= - \frac{1}{2} \hbar
            \left(
                \Tilde{\Delta}_{12}
                + \frac{|J_{12}|^2}{\Tilde{\Delta}_{12}}
                + \frac{|g_{1\text{d}}|^2}{\Tilde{\Delta}_{12}}
            \right)
            \hat{\sigma}_{z,1} \hat{1}_{2}
        - \frac{1}{2} \hbar
            \left(
                - \frac{|J_{12}|^2}{\Tilde{\Delta}_{12}}
            \right)
            \hat{1}_{1} \hat{\sigma}_{z,2}
\nonumber \\ \label{eq:CR_gate_derivation}
    & \ \ \    \ \ \ \  \ \ \ \  \ \ \ \  \ \ \ \  \ \ \ \  \ \ \ \  \ \ \ \  \ \ \ \  \ \ \ \  \ \ \ \  \ \ \ \ 
        - \hbar \left(
                \frac{g_{1\text{d}} J_{12}^*}{\Tilde{\Delta}_{12}}
                    \hat{\sigma}_{z,1} \hat{\sigma}_{+,2} 
                + \frac{g_{1\text{d}}^* J_{12}}{\Tilde{\Delta}_{12}}
                    \hat{\sigma}_{z,1} \hat{\sigma}_{-,2} 
            \right).
\end{align}
Consequently, if $J_{12}$ and $g_{1\text{d}}$ are real numbers, the last term in Eq.(\ref{eq:CR_gate_derivation}) is the $ZX$-gate (also known as the CR gate) we are looking for.

In practice, however, a CR gate is always accompanied by unwanted interactions. This nonideality is also implied by Eq.(\ref{eq:CR_gate_derivation}): 1) Both qubits acquire some Lamb shift that is drive-power-independent (remember that we are in the rotating frame, so the frequencies are shifted by $\tilde{\omega}_2$). 2) the first qubit also feels an AC Stark shift due to the control drive it receives. In reality, due to the multi-level nature of the qubits, there will be other undesired terms appearing in the Hamiltonian; thus, more complicated error mitigation techniques are required to achieve an ideal CR gate.



%% file: include_chapters/chapter_qubit.tex
\chapter{Superconducting Circuits, Nonlinearity, and Qudits}

We now turn to the problem of designing superconducting qubits and qudits that reflect the properties we have discussed in the last chapter. We start by introducing superconductivity and then move to one important application of the superconducting phenomena, namely the Josephson tunneling junction. As hinted at in the last chapter, all the transition frequencies in a multi-level system must be distinct in order to be addressed separately. In particular, an anharmonic oscillator will have unevenly spaced energy levels, leading to practical realizations of qubits/qudits. We will see how the Josephson junction, when shunted by a capacitor, can lead to a weakly anharmonic oscillator and thus to the well-known transmon qubit.
\section{Introduction to Superconductivity}
\subsection{Particle-Wave Duality}
This beginning section serves as a short introduction to superconductivity. Throughout this non-rigorous discussion, we emphasize the analogy of the theory of superconductivity to QED introduced in the last two chapters.

Recall that we have been relying on the Lagrangian 
\begin{align} \label{eq:old_particle_field_lagrangian}
    &\mathcal{L}
        \Bigsl(
            \mathbf{r}_{\alpha}, 
            \dot{\mathbf{r}}_{\alpha}, 
            \mathbf{A}(\mathbf{r}), 
            \dot{\mathbf{A}}(\mathbf{r}), 
            U(\mathbf{r})
        \Bigsr)
\nonumber\\
    &= \sum_{\alpha} 
            \frac{1}{2} m_{\alpha}
            |\dot{\mathbf{r}}_{\alpha}|^2
        + \frac{\epsilon_0}{2} \int \mathrm{d}^3 r \,
            \Big[
                \abs{\mathbf{E}(\mathbf{r})}^2
                - c^2 \abs{\mathbf{B}(\mathbf{r})}^2
            \Big]
        + \sum_{\alpha} 
            \Big[
                q_{\alpha} \dot{\mathbf{r}}_{\alpha} \cdot \mathbf{A}(\mathbf{r}_{\alpha})
                - q_{\alpha} U(\mathbf{r}_{\alpha})
            \Big].
\end{align}
to describe the matter-light interaction, which is a mixture of particle and field theories. However, just like we can interpret photons as excitations of the electromagnetic field, we might as well construct a field for the other particles and treat the particles as some quanta of the field. One simple, non-relativistic (i.e., respect Galilean invariance) classical field one can imagine is known as the \textbf{classical Schr\"{o}dinger field}, whose Lagrangian is given by \cite{lancaster2014qft_for_amateur, murayama_QFT_notes2}
\begin{align} \label{eq:larangian_schrodinger_field}
    \mathcal{L}
        \Bigsl(
            \psi(\mathbf{r}), \dot{\psi}(\mathbf{r})
        \Bigsr)
    = \int \mathrm{d}^3 r 
        \left[
            \ci \hbar 
                \psi^{*}
                \dot{\psi}
            - \frac{\hbar^2}{2m} 
                \nabla_{\mathbf{r}}
                    \psi^*
                \nabla_{\mathbf{r}}
                    \psi
            + \mu \psi^{*}
                \psi
            - \frac{\lambda}{2}
                \psi^*\psi^*\psi\psi
        \right]
\end{align}
and corresponds to the Euler-Lagrange equation 
\begin{equation} \label{eq:E_L_schrodinger_field}
    \ci \hbar \dot{\psi}
    = -\frac{\hbar^2}{2m}  
        \nabla_{\mathbf{r}}^2 \psi
        - \mu \psi
        + \lambda \psi^* \psi \psi.
\end{equation}

Although Eq.(\ref{eq:E_L_schrodinger_field}) looks just like the beloved Schr\"{o}dinger equation, it is in general nonlinear due to the last term. In fact, $\psi(\mathbf{r})$ should not be thought of as the wave function, but as a quantity similar to the normal mode $a_{\mathbf{k},\lambda}$ in classical electrodynamics. Hence, after quantizing the Schr\"{o}dinger field, $\psi$ is promoted to the annihilation operator $\hat{\psi}(\mathbf{r})$, whose job is, not surprisingly, to annihilate particles of the Schr\"{o}dinger field at position $\mathbf{r}$. Recall that we have terms like $\hbar\omega\hat{a}^{\dagger} \hat{a}$ in QED, which is the energy required to put $\langle \hat{a}^{\dagger} \hat{a}\rangle$ photons into a single mode; comparing to the QED case, $-\mu \psi^* \psi$ and $\lambda \psi^* \psi^* \psi \psi / 2 $ in Eq.(\ref{eq:larangian_schrodinger_field}), thus, are the energy associated with having one or two bosonic particles at this same position $\mathbf{r}$. The two-particle potential should be interpreted as the energy cost to place two bosons close to one another, i.e., a weak repulsion is added explicitly.

Moreover, it can be shown that the quantized Schr\"{o}dinger field theory implies the old Schr\"{o}dinger equation when the number of particles is fixed and the weak repulsion is ignored\footnote{As mentioned in the case of QED, the particle number is not conserved in a quantum field theory due to the existence of the creation and annihilation operators. This is still true in the Schr\"{o}dinger field theory, making it more powerful than a Schr\"{o}dinger equation with a fixed particle number.}; thus, we can go ahead and replace the charged-particle part of the Lagrangian in Eq.(\ref{eq:old_particle_field_lagrangian}) by the Schr\"{o}dinger field Lagrangian. In order to respect the gauge invariance of the electromagnetic field, we need to change the normal derivatives by their covariant counterparts
\begin{align}
    - \ci \hbar \nabla 
    \ \ & \longrightarrow \ \ 
    - \ci \hbar \nabla 
        -q \mathbf{A}(\mathbf{r},t),
\\
    \ci \hbar \partial_t
    \ \ & \longrightarrow \ \ 
    \ci \hbar \partial_t 
        - qU(\mathbf{r},t),
\end{align}
resulting in the following Lagrangian
\begin{align} \label{eq:gauge_invariant_L_level1_em_coupled}
    &\mathcal{L}
        \Bigsl (
            \psi(\mathbf{r}),
            \dot{\psi}(\mathbf{r}),
            \mathbf{A}(\mathbf{r}),
            \dot{\mathbf{A}}(\mathbf{r}),
            U(\mathbf{r}),
        \Bigsr )
\nonumber \\
    &= \int \mathrm{d}^3 r \, 
        \bigg\{
            \ci \hbar \psi^* \dot{\psi} 
            - \frac{1}{2m}
                \Bigsl |
                    \big(
                        \! - \ci \hbar \nabla 
                        - q \mathbf{A}
                    \big) \psi
                \Bigsr |^2
\nonumber \\[-1mm]
    & \ \ \ \ \ \ \ \ \ \ \ \ \ \ \ \ \ \ \ \ \ \ \
            - (qU - \mu) \psi^* \psi
            - \frac{\lambda}{2}
                \psi^*\psi^*\psi\psi
        +  \frac{\epsilon_0}{2}
            \left(
                \abs{\mathbf{E}}^2
                - c^2 \abs{\mathbf{B}}^2
            \right)
        \bigg\}.
\end{align}

Eq.(\ref{eq:gauge_invariant_L_level1_em_coupled}) describes the coupling of a field of charged bosons to the electromagnetic field, which can be thought as a direct generalization of Eq.(\ref{eq:old_particle_field_lagrangian}); in fact, the general structure of Eq.(\ref{eq:old_particle_field_lagrangian}) is kept in Eq.(\ref{eq:gauge_invariant_L_level1_em_coupled}). In addition, as an exercise, the reader can verify that the Euler-Lagrange equation associated with Eq.(\ref{eq:gauge_invariant_L_level1_em_coupled}) can be reduced to the following set of equations: 
\begin{gather} \label{eq:E_L_eqn_superfluidity}
    \big (
            \ci \hbar \partial_t - qU
        \big) \psi
    = \frac{1}{2m}
            \big|\! 
                -\ci\hbar \nabla_{\mathbf{r}} 
                - q \mathbf{A} 
            \big|^2 \psi
        - \mu \psi
        + \lambda \psi^* \psi \psi,
\\ \label{eq:E_L_eqn_scalar_potential_pde}
    \Delta U(\mathbf{r},t)  
    = - \frac{\rho(\mathbf{r},t)}{\epsilon_0}
        - \nabla \cdot \frac{\partial}{\partial t}
        \mathbf{A}(\mathbf{r},t),
\\ \label{eq:E_L_eqn_vector_potential_pde}
    \square \mathbf{A}(\mathbf{r},t)  
    = \frac{1}{\epsilon_0 c^2} 
            \mathbf{J}(\mathbf{r},t)  
        - \nabla
            \left[ 
                \nabla \cdot \mathbf{A}(\mathbf{r},t)  
                + \frac{1}{c^2} \frac{\partial}{\partial t} U(\mathbf{r},t) 
            \right],
\\ \label{eq:field_theory_charge_density_general}
    \rho(\mathbf{r},t) 
    = q \psi^* \psi,
\\[1mm] \label{eq:field_theory_current_density_general}
    \mathbf{J}(\mathbf{r},t)
    = \frac{q}{2m} 
        \Big[ 
            - \ci \hbar 
                \psi^*
                \nabla
                \psi
            + \ci \hbar 
                \nabla
                \psi^*
                \psi
            - 2q \mathbf{A}(\mathbf{r},t) \psi^* \psi
        \Big].
\end{gather}
Eq.(\ref{eq:E_L_eqn_superfluidity}) is clearly a generalization of Eq.(\ref{eq:E_L_schrodinger_field}), which now includes the coupling $\big|\! -\ci\hbar \nabla_{\mathbf{r}} - q \mathbf{A} \big|^2 \psi$. Eq.(\ref{eq:E_L_eqn_scalar_potential_pde}) and (\ref{eq:E_L_eqn_vector_potential_pde}) are nothing but Maxwell's equations in terms of the potentials. Lastly, Eq.(\ref{eq:field_theory_charge_density_general}) and (\ref{eq:field_theory_current_density_general}) replace the old definition of charge and current densities; in particular, without the charge $q$, the expressions of $\rho$ and $\mathbf{J}$ resemble that of the probability density and probability current density associated with a wavefunction $\Psi(x,t)$ (nevertheless, we emphasize again that $\psi$ should be interpreted as the annihilation operator, not a quantum state.).

At this point, one may wonder why we have focused on deriving the classical equations of a field instead of quantizing the field and looking at the operator $\hat{\psi}$ like what we did for $\hat{a}$. The reason is that a quantum coherent state can be well approximated by the solution to the Euler-Lagrange equation as we have discussed extensively in the last chapter. It turns out the phenomenon of superconductivity can be explained by using the coherent states, i.e., the Bose–Einstein condensate, of the field; therefore, it is sufficient, at least for the purpose of this paper, to only look at the classical equations.

\subsection{A Simplified Model of Superconductivity}
The field theory we derived above works for bosons (since they can stack on one another with a small energy cost $\lambda$) and can be adopted to explain the phenomenon of superfluidity; however, superconductivity is clearly related to the motion of electrons, which are fermion. Nevertheless, it turns out that in a superconducting material, two electrons can pair together by some phonon-assisted process, resulting in a quasi-particle, known as a \textbf{Cooper pair} \cite{PhysRev.104.1189}, with an integer spin and, thus, behaves like a boson. Without going into the deep theory of superconductivity, we will now use Eq.(\ref{eq:E_L_eqn_superfluidity})-(\ref{eq:field_theory_current_density_general}) as an elementary model to study the physics of superconductivity \cite{lancaster2014qft_for_amateur}.

As motivated in the last sub-section, we will solve the classical equations for a coherent ground state of the Schr\"{o}dinger field of the Cooper pairs. In particular, a static solution can be solved by setting the time derivatives to $0$, yielding the phenomenological \textbf{Ginzburg–Landau equations} \cite{RevModPhys.72.969}
\begin{gather} \label{eq:ginzburg_landau_equation}
    \frac{1}{2m^*}
            \big|\! -\ci\hbar \nabla_{\mathbf{r}} 
                + 2e \mathbf{A} 
            \big|^2 \psi
        - \mu \psi
        + \lambda \psi^* \psi \psi
    = 0,
\\
    \nabla \times \mathbf{B} 
    = \mu_0 \mathbf{J},
\\ \label{eq:charge_density_superconductor}
    \rho 
    = -2e \psi^*\psi
\\ \label{eq:current_density_superconductor}
    \mathbf{J}
    = -\frac{e}{m^*} 
        \Big[ 
            - \ci \hbar 
                \psi^*
                \nabla
                \psi
            + \ci \hbar 
                \nabla
                \psi^*
                \psi
            + 4e \mathbf{A} \psi^* \psi
        \Big].
\end{gather}
Note that we have set $q = -2e$ since a Cooper pair is made of two electrons; $m^*$ is the effective mass of a Cooper pair in the lattice of the superconductor. We have also assumed the absence of an external voltage (i.e., $U = 0$) at this point. 

Consequently, it's straightforward to verify that in the absence of any electromagnetic field (i.e., $\mathbf{A} = 0$ and $U=0$),
\begin{equation}\label{eq:coherent_state_superconductor}
    \psi 
    = \sqrt{\frac{\mu}{\lambda}} 
        e^{\ci \phi}
    = \sqrt{n} 
        e^{\ci \phi}
\end{equation}
is one solution to Eq.(\ref{eq:ginzburg_landau_equation}) with some constant phase $\phi$ across the entire superconductor if $n = \mu/\lambda > 0$; that is, all Cooper pairs collectively behave like a classic field with a constant phase everywhere. Due to the pairing of electrons, it turns out that $n$ can be positive for a superconductor below some critical temperature $T_{\text{c}}$. Moreover, from Eq.(\ref{eq:charge_density_superconductor}), it's clear that $|\psi|^2 = n$ represents the number density of the Cooper pairs in the superconductor, similar to how $|\alpha|^2 = N$ is the mean photon number in a coherent state $\ket{\alpha}$ of the electromagnetic field; indeed, in a more rigorous derivation of the superconductivity, Eq.(\ref{eq:charge_density_superconductor}) is interpreted as the coherent ground state (i.e., the vacuum state). However, unlike the vacuum state of an electromagnetic field, the ground state of a superconductor has nonzero particle density due to symmetry breaking.

When the vector potential is nonzero, the solution in general can be solved numerically. In particular, for a superconducting loop, it can be shown that Eq.(\ref{eq:coherent_state_superconductor}) is a good approximation if the loop has a large enough radius. 

\subsection{Flux Quantization}
Another interesting consequence one can deduce from the Euler-Lagrange equation is related to the magnetic flux across a superconducting loop. Since current cannot flow inside a perfect conductor, Eq.(\ref{eq:current_density_superconductor}) must vanishes (inside the superconducting), i.e., 
\begin{equation}
    \ci \hbar 
            \psi^*
            \nabla
            \psi
        - \ci \hbar 
            \nabla
            \psi^*
            \psi
    = 4e \mathbf{A} \psi^* \psi .
\end{equation}
If we substitute $\psi = \sqrt{n(\mathbf{r})} e^{\ci \phi (\mathbf{r})}$ as an ansatz for the field, we obtain $- 2 \hbar n \nabla \phi = 4e \mathbf{A} n$, which, upon a closed line integration along the superconducting loop $C$ gives
\begin{equation}
    2 \pi m 
    = \oint_C 
        \nabla \phi \cdot 
        \mathrm{d} \mathbf{l} 
    = \frac{2e}{\hbar} 
        \oint_C 
            \mathbf{A} 
            \cdot \mathrm{d} \mathbf{l} 
    = \frac{2e}{\hbar} \Phi_{\mathbf{B}} 
    \ \ \text{ for } \ \ m \in \mathbf{Z}.
\end{equation}
We have used the fact that the phase $\phi$ must be single-valued modulo $2\pi$. By defining the flux quantum to be $\Phi_0 = 2e / h$, we obtain the condition of \textbf{flux quantization}
\begin{equation}
    \Phi_{\mathbf{B}} 
    = m \Phi_0 
    = m \frac{h}{2e};
\end{equation}
in other words, the total magnetic flux across the surface enclosed by a superconducting loop must be an integer multiple of the constant $\Phi_0$. For future reference, we also define a related quantity $\phi_0 = \Phi_0/2\pi = \hbar/2e$, called the reduced flux quantum.

The above quantization condition is derived for a bulk superconducting loop only. If the superconducting loop is interrupted by weak links (e.g., Josephson junctions or a section of superconducting nanowire), we cannot set the current to be zero near those weak links. In general, we would obtain
\begin{equation}
    2 \pi m = \frac{\Phi_{\mathbf{B}}}{\phi_0} + \sum_{j} \delta_j,
\end{equation}
where $\delta_j$ represents the superconducting phase accumulated when the line integration goes through the $j$th weak link. Intuitively speaking, $\delta_j$ is related to the change in the momenta of the Cooper pairs when traveling across a weak link; consequently, a voltage is developed across the weak link similar to the voltage generated by a linear inductor when a current is suddenly applied. In general, the voltage across the weak link is the time derivative of $\delta_j$. We will derive this relation explicitly for a Josephson junction.

The flux quantization, similar to the Aharonov–Bohm effect, results from the gauge invariance of the electromagnetic field. Intuitively, the $2\pi$ multiples can be thought of as a consequence of evaluating a closed loop integral around some singularities of the vector potential. The effective singularities are then counted in units of $\Phi_{0}$. Moreover, based on the Aharonov–Bohm effect, we can also interpret the flux quantization as an interference between the two branches of a loop.

\section{The Josephson Junction}
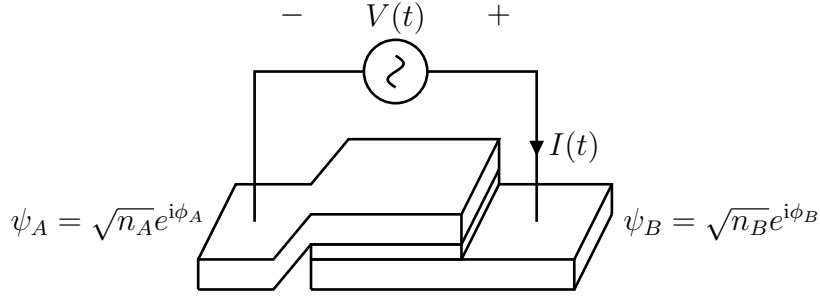
\begin{figure}
    \centering
    \begin{circuitikz}[scale=1, american]
    \ctikzset{tripoles/mos style/arrows};
        \draw
            (-1,0) to (1,0) to (1,0.2) to (-1,0.2) to (-1,0);
        \draw
            (-1,-0.4) to (2.5,-0.4) to (2.5,0) to (-1,0) to (-1,-0.4);
        \draw
            (-2.5, -0.4) to (-1.5,-0.4) to (-1,0.2) to (1,0.2) to (1,0.6) to (-1,0.6) to (-1.5,0) to (-2.5, 0) to (-2.5, -0.4);
        \draw 
            (1,0.6) to (1.5,1.6) to (-0.5,1.6) to (-1,1) to
            (-2, 1) to (-2.5, 0);
        \draw 
            (1.5,1.6) to (1.5,1)
            (1.5,1.2) to (1,0.2)
            (1.5,1) to (1,0);
        \draw
            (1.5,1) to (3,1) to (3, 0.6)
            (3,1) to (2.5,0)
            (3,0.6) to (2.5,-0.4);
        \draw 
            (2, 0.5) to [short, i<_={${I(t)}$}](2,2.5) to [/tikz/circuitikz/bipoles/length=40pt, sV, l_={${V(t)}$}](-1.75, 2.5) to (-1.75,0.5)
            (1.5,3.25) node[]{$+$} 
            (-1.25,3.25) node[]{$-$};
        \draw 
            (-2.25, 0.5) node[left]{${\psi_A = \sqrt{n_{A}} e^{\ci \phi_{A}}}$}
            (3, 0.5) node[right]{${\psi_{B} = \sqrt{n_{B}} e^{\ci \phi_{B}}}$};
    \end{circuitikz}
    \caption{The Josephson junction.}
    \label{fig:physical_josephson_junction}
\end{figure}

As mentioned at the beginning of the chapter, to encode information on an artificial atom, we need to be able to address each transition without accidentally exciting other transitions; one way to construct an artificial atom with unequally spaced energy levels is to add some nonlinearity to a harmonic oscillator. It turns out that, with superconductivity, we can build a non-dissipative nonlinear element, known as the \textbf{Josephson junction (JJ)} \cite{JOSEPHSON1962251, RevModPhys.73.357}, by placing two superconductors in close proximity to allow quantum tunneling. In practice, a JJ is made by depositing one superconductor (A) on top of another superconductor (B) but separated by a nanometer-thin oxide layer as shown in Figure \ref{fig:physical_josephson_junction}. As one of the most popular sources of nonlinearity in superconducting circuits, JJs not only enable a wide range of qubit and qudit designs \cite{nature19718, science.1081045, PhysRevB.68.224518, PhysRevA.76.042319, science.1175552, PhysRevLett.107.240501, PhysRevA.87.052306, PhysRevA.98.052318, PhysRevX.9.041041} but also provide means to achieve parametric effects such as three- and four-wave conversion \cite{nphys2035, PhysRevLett.120.200501, NComms.s41467.023.41104.0} and amplification \cite{nature09035, ApplPhysLett.1.4984142, PhysRevApplied.7.054020}.

\subsection{Dynamics of the Josephson Junction}
In the static case, a global order parameter $\psi$ describes a single piece of superconductor; to analyze the dynamics of a JJ, we can attribute each superconducting section with one order parameter. That is, we still approximately each superconductor with their coherent ground states $\psi_A = \sqrt{n_{A}} e^{\ci \phi_{A}}$ and $\psi_B = \sqrt{n_{A}} e^{\ci \phi_{A}}$; however, we will now go back to Eq.(\ref{eq:E_L_eqn_superfluidity}) (i.e., the Euler-Lagrange equation) to obtain two time-dependent equations for each coherent state. In addition, we assume charges are balanced on both sides by an externally connected battery as shown in Figure \ref{fig:physical_josephson_junction}.

To account for the possible tunneling effect, we heuristically add a coupling term in each differential equation. If no magnetic field is applied (i.e., $\mathbf{A} = \mathbf{0}$), we obtain the following coupled differential equations\footnote{Due to the choice of ansatze, $- \mu \psi + \lambda \psi^* \psi \psi = 0$ in the Euler-Lagrange equation. In addition, our ansatze are position-independent inside each piece of superconductor so $\nabla_{\mathrm{r}}$ is also ignored.}
\begin{equation}
    \ci \hbar \, \frac{\partial}{\partial t} \!
        \begin{pmatrix}
            \displaystyle 
                \sqrt{n_{A}}
                e^{\ci \phi_{A}} \\ 
            \displaystyle 
                \sqrt{n_{B}} 
                e^{\ci \phi_{B}}
        \end{pmatrix}
    = \begin{pmatrix} 
            eV & K \\
            K & -eV 
        \end{pmatrix}
        \begin{pmatrix}
            \displaystyle 
                \sqrt{n_{A}} 
                e^{\ci \phi_{A}} \\
            \displaystyle 
                \sqrt {n_{B}} 
                e^{\ci \phi_{B}}
        \end{pmatrix},
\end{equation}
where $K$ models the coupling strength across the junction and $V$ is the applied field across the junction (thus each side experiences a potential drop $U = \pm (2e \cdot V)/2 = \pm eV$).
The solution\footnote{Remember that $n_{A} = n_{B} = n_0$ will be held constant by the external battery (see \cite{feynman_vol3} for more elaboration on this point).}, which only depends on the relative phase\footnote{Technically, we should define phase difference as
\begin{equation}
    \delta 
    = \phi_B - \phi_A 
        + \frac{2 \pi}{\Phi_0} 
            \int_A^B  
            \mathbf{A}(\mathbf{r}) \cdot 
            \mathrm{d} \mathbf{l}
\end{equation}
when $\mathbf{A} \neq \mathbf{0}$. This ensures that the phase difference is gauge-invariant, which leads to the appearance of magnetic flux in the SQUID equations.} $\delta = \phi_{B} - \phi_{A}$, describes a nonlinear circuit element with the current-voltage ($I$-$V$) relations
\begin{equation} \label{eq:JJ_current}
    I(t) 
    = I_{\text{c}} 
        \sin \delta(t) ,
\end{equation}
\begin{equation} \label{eq:JJ_voltage}
    V(t)
    = \frac{\hbar}{2e} 
        \dot{\delta}(t)
    = \phi_0 
        \dot{\delta}(t)
\end{equation}
with the critical current
\begin{equation} \label{eq:current_current_definition}
    I_{\text{c}} 
    \propto \frac{2K n_0 A}{\hbar}.
\end{equation}

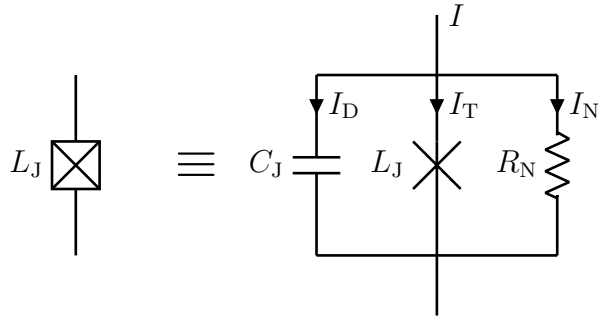
\begin{figure}[b]
    \centering
    \begin{tikzpicture}[scale=0.8]
    \begin{scope}[shift={(-4,0)}]
        \draw 
            (0,3) to [/tikz/circuitikz/bipoles/length=30pt, josephson, l_=$L_{\text{J}}$] (0,0);
        \draw 
            (2,1.5) node[]{{\LARGE${\equiv}$}};
    \end{scope}
    \ctikzset{tripoles/mos style/arrows};
        \draw
            (0,3) to (4,3) 
            (4,0) to (0,0) 
            (0,3) to [short, i>^={${I_{\text{D}}}$}] (0,2) to [/tikz/circuitikz/bipoles/length=30pt, C, l_=$C_{\text{J}}$] (0,1) to (0,0)
            (2,3) to [short, i>^={${I_{\text{T}}}$}] (2,2) to [/tikz/circuitikz/bipoles/length=30pt, JJ, l_=${L_{\text{J}}}$] (2,1) to (2,0)
            (4,3) to [short, i>^={${I_{\text{N}}}$}](4,2) to [/tikz/circuitikz/bipoles/length=30pt, R, l_=$R_{\text{N}}$] (4,1) to (4,0);
        \draw
            (2,3) to [short, -](2,4) node[right]{$I$}
            (2,0) to [short, -](2,-1);
    \end{tikzpicture}
    \caption{The RCSJ model of the Josephson junction.}
    \label{fig:RCSJ_model}
\end{figure}

From the $I$-$V$ relations, it is clear that the current $I$ cannot exceed the critical current $I_{\text{c}}$, which depends on the coupling strength $K$, the number of Cooper pairs $n_A = n_B = n_0$, and the area of the junction (not shown in Eq.(\ref{eq:current_current_definition})). When $I > I_{\text{c}}$, the current is due to the normal current going through an effective resistor of the junction or the displacement current due to the junction capacitance associated with a physical JJ as shown in Figure \ref{fig:RCSJ_model}.

\subsection{Josephson Inductance and Energy}
In general, the loop voltage induced by a time-varying magnetic field is equal to the time derivative of the net flux flowing across the loop, i.e., $V = \dot{\Phi}$. Recall that we defined the reduced flux to be $\varphi = \Phi/\phi_0$, which leads to 
\begin{equation}
    V(t) = \phi_0 \dot{\varphi}(t).
\end{equation}
Similarly, Eq.(\ref{eq:JJ_voltage}) relates the voltage across a JJ with $\dot{\delta}$, hence, suggesting that we treat the phase difference $\delta$ as the reduced ``flux'' developed across the JJ. However, since $\mathbf{A} = \mathbf{0}$, the inductance we derived is unrelated to the magnetic energy but is due to the kinetic energy of the Cooper pairs in the superconductor (since they don't see any resistance and have the inertial to keep moving).

Classically, the self-inductance $L$ of a one-port device is defined via
\begin{equation}
    L 
    = \frac{V}{\dot{I}}.
\end{equation}
In other words, the inductance characterizes the voltage response of the device against a \textit{change} of the current in time.
Following the same logic, we can define an effective inductance for the JJ to be
\begin{equation}
    L(\delta)
    = \frac{V}{\dot{I}}
    = \frac{\phi_0}{I_{\text{c}} \cos \delta}
    = \frac{L_{\text{J}}}{\sqrt{1-(I/I_{\text{c}})^2}},
\end{equation}
with a newly defined constant
\begin{equation}
    L_{\text{J}} 
    = \frac{\phi_0}{I_c},
\end{equation}
critical when designing a qubit. Importantly, $L(\delta)$ is nonlinear in the sense that it depends on the magnitude of the current $I$ following through the junction in comparison to the critical current $I_{\text{c}}$. When $I = 0$, $L = L_{\text{J}}$, which is why $L_{\text{J}}$ is also called the \textbf{zero-bias Josephson inductance} or simply the Josephson inductance. Again, the definition breaks down when $I > I_{\text{c}}$.

We know that only a linear inductor or capacitor stores energy that is quadratic in the current or voltage. Since $L(\delta)$ is nonlinear, we expect it to give a different energy landscape; to wit, we compute the electromagnetic energy stored in a JJ by integrating the power with respect to time
\begin{align}
    U_{\text{J}} (t)
    &= \int_{-\infty}^{t}
            \mathrm{d} t' 
            \, 
            V(t') I(t') 
\nonumber\\
    &= \int_{-\infty}^{t} 
            \mathrm{d} t' 
            \, 
            \phi_0 \dot{\delta}(t') 
            I_{\text{c}} \sin \delta(t')
    = \phi_0 I_{\text{c}} 
        \int_{\delta(-\infty)}^{\delta(t)} 
            \mathrm{d} \delta'
            \, 
            \sin \delta' 
\nonumber\\
    &= \phi_0 I_{\text{c}}  
        \Big[
            \cos \delta(-\infty) - \cos \delta(t)
        \Big].
\end{align}
By assuming $\delta(-\infty) = 0$ (energy reference can be set arbitrarily) and define $E_{\text{J}} = \phi_0 I_{\text{c}} = \phi_0^2 / L_{\text{J}}$ to be the \textbf{Josephson energy}, we obtain
\begin{equation} \label{eq:Taylor_expansion_JJ_energy}
    U_{\text{J}} (\delta)
    = E_{\text{J}} (1-\cos \delta)
    = E_{\text{J}}
        \left(
            \frac{\delta^2}{2} - \frac{\delta^4}{24} + \cdots
        \right),
\end{equation}
which is quadratic in $\delta$ for small $\delta$ but is weakly anharmonic as $\delta$ moves away from the origin.

Recall that the energy stored on a linear inductor $L$ (see Eq.(\ref{eq:energy_linear_inductor})) is given by
\begin{equation}
    U_{L} 
    = \frac{1}{2} E_{L} \varphi^2.
\end{equation}
Thus, $U_{\text{J}}$ agrees with $U_L$ to the third-order if we identify the Josephson energy $E_{\text{J}}$ as the linear inductive energy $E_{L}$ and interpret $\delta$ as the reduced flux $\varphi$ for the nonlinear inductor. Furthermore, shunting the JJ with a large capacitor gives rise to a nonlinear LC circuit with the Hamiltonian
\begin{equation}
    \mathcal{H}
    = T_{\text{cap}} + U_{\text{J}}
    = 4 E_C n^2 
        + \frac{1}{2} E_{\text{J}} \delta^2 
        - \frac{1}{24} E_{\text{J}} \delta^4 
        + \cdots .
\end{equation}
Since a linear resonator describing by
\begin{equation}
    \mathcal{H}_{\text{lin}} 
    = T_C + U_L
    = 4E_C n^2 + \frac{1}{2} E_L \varphi^2
\end{equation}
has a resonant frequency of $\omega = \hbar^{-1}\sqrt{8E_C E_{L}}$, we can define a similar frequency $\omega_{\text{J}}$, called the \textbf{Josephson plasma frequency}, by simply replacing $E_{L}$ by $E_{\text{J}}$:
\begin{equation}\label{eq:josephson_plasma_frequency}
    \omega_{\text{J}} 
    = \frac{\sqrt{8E_C E_{\text{J}}}}{\hbar}
\end{equation}
However, note that $\omega_{\text{J}}$ is defined only for the linear part of $\mathcal{H}$. Once we quantize the anharmonic oscillator, the originally equally spaced energy levels of the QHO will experience different shifts due to the nonlinear part of $\mathcal{H}$, thus, allowing us to address each transition separately.

\subsection{DC-SQUID}
\begin{figure}
    \centering
    \begin{circuitikz}[scale=1,european voltages]
    \ctikzset{tripoles/mos style/arrows};
        \draw
            (0,3) to (5,3) 
            (5,0) to (0,0) 
            (0,3) to  [/tikz/circuitikz/bipoles/length=20pt, josephson, l^=${L_{\text{J1}}}$, v=${\delta_1 = \delta - \varphi_{\text{ext}}}$] (0,0)
            (5,3) to  [/tikz/circuitikz/bipoles/length=30pt, josephson, l_=${L_{\text{J2}}}$, v^=${\delta_2 = \delta + \varphi_{\text{ext}}}$] (5,0);
        \draw
            (2.5,1.8) node[]{\large$\Phi_{\text{ext}}$}
            (2.5,1.2) node[]{\large${\left(\varphi_{\text{ext}} = \frac{\pi \Phi_{\text{ext}}}{\Phi_0}\right)}$};
    \end{circuitikz}
    \caption{An (asymmetric) DC-SQUID threaded by an external magnetic flux.}
    \label{fig:DC_SQUID}
\end{figure}
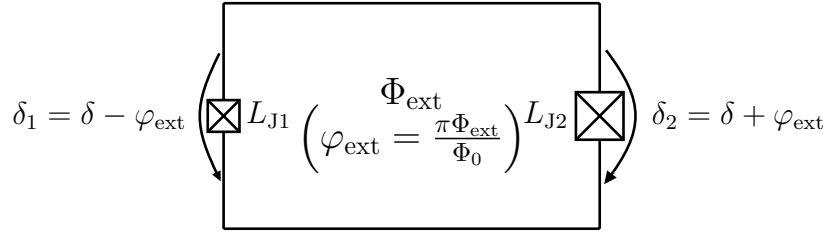

One property of the order parameter is the quantization of magnetic flux in a superconducting loop. As pointed out in the discussion of the JJ, the superconducting phase $\delta$ of the order parameter is conceptually indistinguishable from the reduced flux $\varphi$ due to a real magnetic field. In other words, the sum of the superconducting phase and the external flux around a superconducting loop must be subject to the flux quantization condition. Now, let's start with the simplest nontrivial superconducting loop - the parallel combination of two JJs as shown in Figure \ref{fig:DC_SQUID}; this configuration is known as a \textbf{superconducting quantum interference device (SQUID)} \cite{RevModPhys.72.969, doi:10.1063/1.5089550} since flux quantization can be thought of as an interference between the two paths (i.e., the two JJs).

Applying the flux quantization condition yields the expression
\begin{equation} \label{eq:flux_quantization_SQUID}
    \delta_1 - \delta_2 + \frac{\Phi_{\text{ext}}}{\phi_0} = 2 n \pi,
\end{equation}
where we have defined the parities of the two phases so that they are measured from the same node (see Figure \ref{fig:DC_SQUID}). Since we can adjust the external flux $\Phi_{\text{ext}}$ to achieve any multiple of $2\pi$, we will set the right-hand side of Eq.(\ref{eq:flux_quantization_SQUID}) to zero without loss of generality (in other words, $\Phi_{\text{ext}}/\phi_0$ is already the value modulo $2\pi$). In addition, we define half of the reduced external flux by
\begin{equation}
    \varphi_{\text{ext}} 
    = \frac{\Phi_{\text{ext}}}{2\phi_0} 
    = \frac{\pi \Phi_{\text{ext}}}{\Phi_0},
\end{equation}
resulting in
\begin{equation}
    (\delta_1 + \varphi_{\text{ext}}) - (\delta_2 - \varphi_{\text{ext}}) = 0.
\end{equation}
Therefore, the external flux leaves an imprint on the phases of the two JJs, thus creating a mismatch of the superconducting currents flowing through the two junctions. 

To better study the interference, we define the average phase across the two junctions
\begin{equation}
    \delta = \frac{\delta_1 + \delta_2}{2};
\end{equation}
then, the flux quantization can be expressed as
\begin{equation}
    \delta_1 
    = \delta - \varphi_{\text{ext}}
\end{equation}
\begin{equation}
    \delta_2 
    = \delta + \varphi_{\text{ext}},
\end{equation}

\begin{figure}
    \centering
    \includegraphics[scale=0.35]{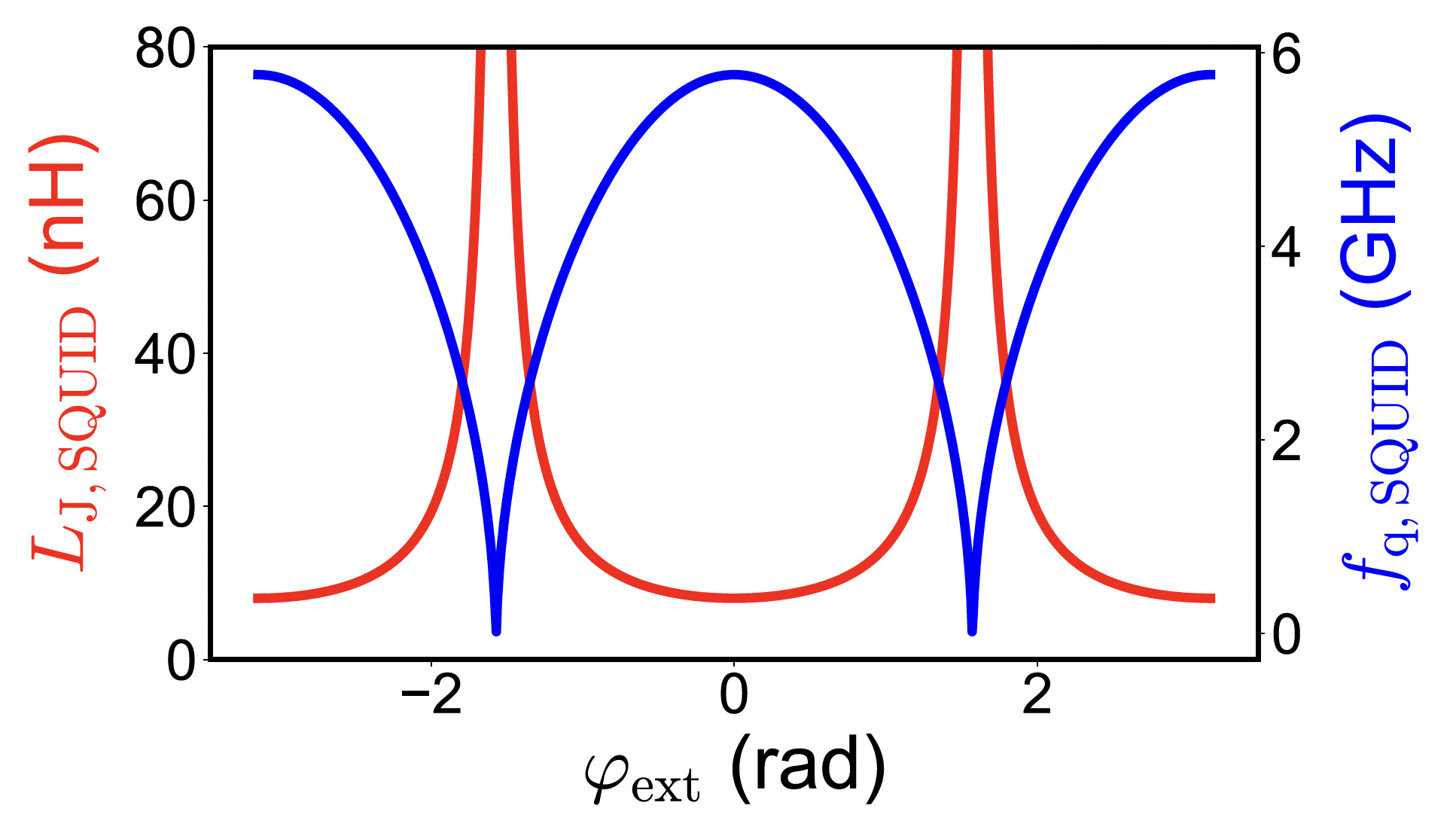}
    \caption{The effective Josephson inductance $L_{\mathrm{J,SQUID}} = \phi_0 / I_{\mathrm{c,SQUID}}$ of a symmetric SQUID and the corresponding transmon qubit frequency $f_{\mathrm{q,SQUID}} = (\sqrt{8E_C E_{\mathrm{J}}} - E_C) / h$ (see Section \ref{section:qubit_analysis}) plotted as functions of the external flux $\varphi_{\mathrm{ext}} = \pi \Phi_{\mathrm{ext}}/\Phi_0$.}
    \label{fig:SQUID_LJ_freq}
\end{figure}

\textbf{Syemmtric SQUID}: For simplicity, let's first consider the case of two identical junctions, i.e., the critical currents, $I_{\text{c}}$, of the two JJs are the same. Then, the total current flowing through the parallel circuit is given by
\begin{align}
    I_{\text{SQUID}}(\delta; \varphi_{\text{ext}})
    &= I_{\text{c}} \sin(\delta - \varphi_{\text{ext}})
    + I_{\text{c}} \sin(\delta + \varphi_{\text{ext}})
\nonumber \\
    &= 2I_{\text{c}} \cos \varphi_{\text{ext}}
    \sin (\delta)
    = 2I_{\text{c}} \abs{\cos \varphi_{\text{ext}}}
    \sin (\delta + \delta_0),
\end{align}
where $\delta_0$, treated as the initial condition of $\delta(t)$, accounts for the sign of $\cos \varphi_{\text{ext}}$ and is singled out to make $2I_{\text{c}} \abs{\cos \varphi_{\text{ext}}}$ a positive number. Comparing with the single junction current, we realize that a SQUID can be treated as a single JJ\footnote{It's left as an exercise to show that the voltage equation in the Josephson $I$-$V$ relations is also satisfied trivially.} with an effective critical current of
\begin{equation} \label{eq:SQUID_effective_critical_current}
    I_{\text{c,SQUID}} 
    = 2I_{\text{c}} \abs{\cos \varphi_{\text{ext}}}.
\end{equation}
Moreover, the critical current and thus the Josephson inductance (i.e., $L_{\text{J,SQUID}} = \phi_0 / I_{\text{c,SQUID}} $) can be tuned by an external flux. Combined with the anharmonic oscillator introduced in the previous subsection, we now have a qubit with tunable transition frequencies (see Figure \ref{fig:SQUID_LJ_freq}).

\begin{figure}
    \centering
    \includegraphics[scale=0.39]{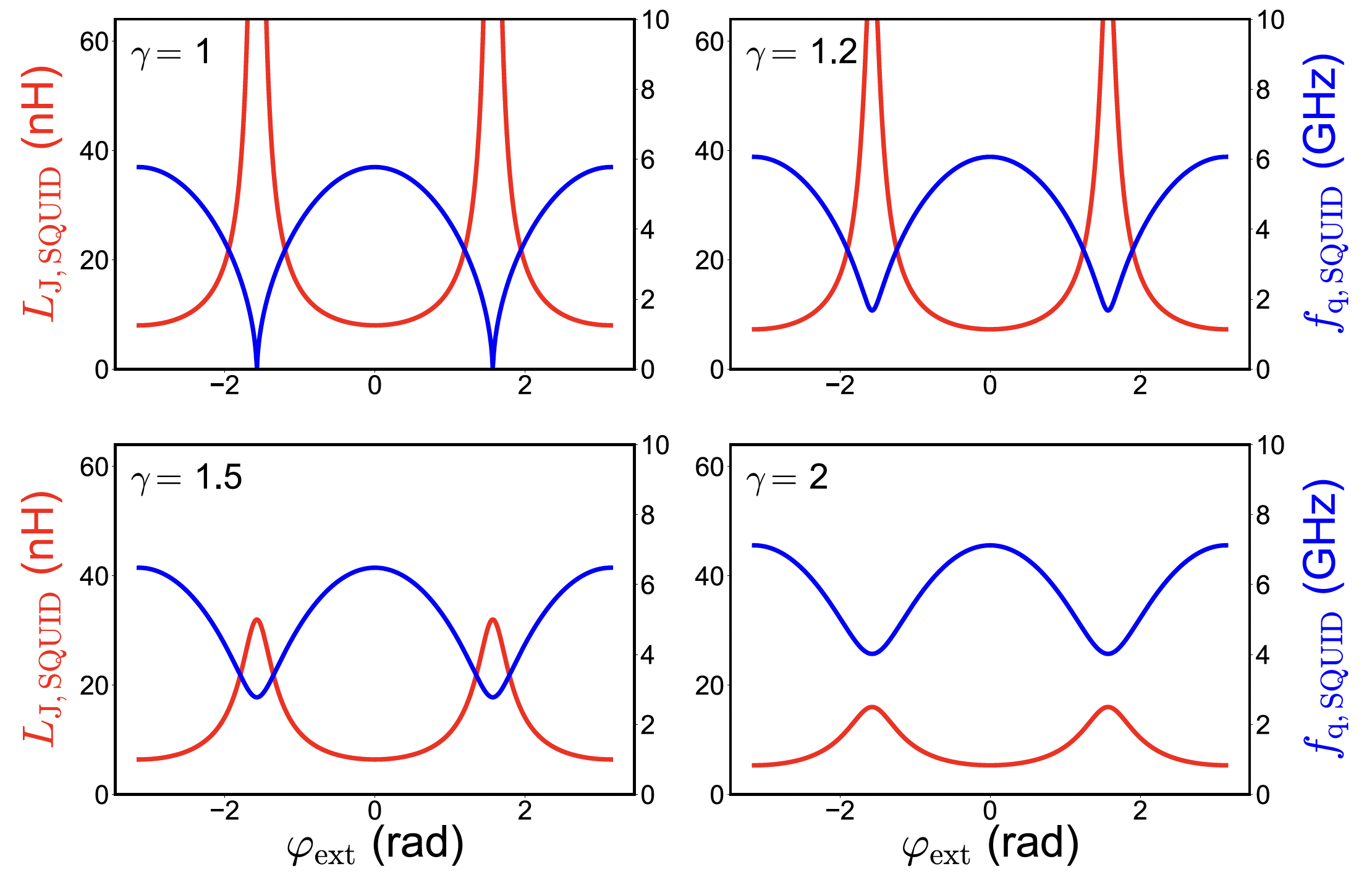}
    \caption{The effective Josephson inductance and the corresponding transmon qubit frequency as a function of the external flux for $\gamma = I_{\mathrm{c2}} / I_{\mathrm{c1}} = 1, 1.2, 1.5, 2$.}
    \label{fig:SQUID_LJ_freq_gamma}
\end{figure}

\textbf{Asymmetric SQUID}: If we differentiate the critical current in Eq.(\ref{eq:SQUID_effective_critical_current}) with respect to $\varphi_{\text{ext}}$, we see that the slope near $\varphi_{\text{ext}} = \pm \pi/2$ can be large; moreover, as shown in Figure \ref{fig:SQUID_LJ_freq}, the corresponding qubit frequency (i.e., Eq.(\ref{eq:josephson_plasma_frequency}) is also not well-defined since the effective critical current is zero. In practice, the large slope translates to the strong sensitivity of the qubit properties to the noise on the external flux line, creating an unwanted channel for qubit decoherence. To make the SQUID insensitive to the flux noise, we introduce the asymmetric SQUID, where the two JJs no longer have the same critical current. In this case, the total current is given by
\begin{align}
    I_{\text{SQUID}}(\delta; \varphi_{\text{ext}})
    &= I_{\text{c1}} \sin(\delta - \varphi_{\text{ext}})
    + I_{\text{c2}} \sin(\delta + \varphi_{\text{ext}})
\nonumber \\
    &= (I_{\text{c1}} + I_{\text{c2}}) 
        \sqrt{
            \cos^2 \varphi_{\text{ext}}
            + d^2 \sin^2 \varphi_{\text{ext}}
        }
        \, \sin(\delta + \delta_0),
\end{align}
where we have introduced
\begin{equation}
    \gamma = I_{\text{c2}}/I_{\text{c1}}
\end{equation}
and 
\begin{equation}
    d 
    = \frac{\gamma - 1}{\gamma + 1}
    = \frac{I_{\text{c2}}-I_{\text{c1}}}{I_{\text{c2}}+I_{\text{c1}}}
\end{equation}
to characterize the level of asymmetry in the two branches. Again, $\delta_0$ can be absorbed into the initial condition of $\delta$. Hence, the effective critical current is now
\begin{equation}
\label{eq:asymmetrix_SQUID_effective_critical_current}
    I_{\text{c,SQUID}} 
    = (I_{\text{c1}} + I_{\text{c2}}) 
        \sqrt{
            \cos^2 \varphi_{\text{ext}}
            + d^2 \sin^2 \varphi_{\text{ext}}
        }.
\end{equation}
As shown in Figure \ref{fig:SQUID_LJ_freq_gamma}, by increasing the parameter $\gamma$, the sensitivity of effective Josephson inductance to $\varphi_{\text{ext}}$ drops considerably. Consequently, the qubit frequency is also protected from flux noise.

\section{Introductory Superconducting Circuits Theory}
With the nonlinear circuit elements introduced, we are now ready to study the superconducting qubits and also their coupling to the readout resonators and the control lines (i.e., coupling to the external drive). However, unlike the classical circuits which are analyzed using simple Kirchhoff's current and voltage laws, circuit quantum electrodynamics (cQED) are generally studied by writing down the Lagrangian and Hamiltonian of the electrical network because knowing the generalized coordinates and momenta of a system allows us to apply canonical quantization. 

For a weakly anharmonic system (i.e., placing JJs in an otherwise linear circuitry), we usually start with the linearized system in order to define the independent normal modes. Subsequently, the mode amplitudes are promoted to the annihilation operators using the procedure of canonical quantization. Finally, the nonlinear terms are added as perturbations to the system and all are expressed using the annihilation and creation operators.

\subsection{Circuit-Level Dipole Interaction}
\begin{figure}
    \centering
    \begin{tikzpicture}[scale=1, transform shape]
    \ctikzset{tripoles/mos style/arrows};
    \begin{scope}[shift={(-8,0)}]
        \draw 
            (0,3) to [/tikz/circuitikz/bipoles/length=30pt, C, l^=$C_{g}$] (-2,3) to [/tikz/circuitikz/bipoles/length=40pt, sV, l_=${V_{\text{d}}}$] (-2,0) node[/tikz/circuitikz/bipoles/length=30pt,sground]{}
            (-2.7,2.6) node[]{$+$} 
            (-2.7,0.4) node[]{$-$};
        \draw
            (0,3) to (4,3) 
            (4,0) to (0,0) 
            (2,0) node[/tikz/circuitikz/bipoles/length=30pt,sground]{}
            (0,3) to [/tikz/circuitikz/bipoles/length=30pt, josephson, l=${L_{\text{J}}, C_{\text{J}}}$] (0,0)
            (4,3) to (4,2) to [/tikz/circuitikz/bipoles/length=40pt, C, l_=$C_{\text{S}}$] (4,1) to (4,0);
        \draw 
            (1,-1) node[]{${\textbf{(a)}}$};
    \end{scope}
        \draw 
            (0,3) to [/tikz/circuitikz/bipoles/length=30pt, C, l^=$C_{g}$] (-2,3) to [/tikz/circuitikz/bipoles/length=40pt, sV, l_=${V_{\text{d}}}$] (-2,0) node[/tikz/circuitikz/bipoles/length=30pt,sground]{}
            (-2.7,2.6) node[]{$+$} 
            (-2.7,0.4) node[]{$-$};
        \draw
            (0,3) to (4,3) 
            (4,0) to (0,0) 
            (2,0) node[/tikz/circuitikz/bipoles/length=30pt,sground]{}
            (0,3) to [/tikz/circuitikz/bipoles/length=40pt, L, l_=$L$] (0,0)
            (4,3) to (4,2) to [/tikz/circuitikz/bipoles/length=40pt, C, l_=$C_0$] (4,1) to (4,0);
        \draw
            (2,2.6) node[]{$+$} 
            (2,1.5) node[]{${\dot{\Phi}(t)}$}
            (2,0.4) node[]{$-$}
            (2,3) node[above]{$\Phi(t)$}
            (4.7,1.85) node[]{$+Q_0(t)$}
            (-0.2,3) node[above]{$+Q_g(t)$};
        \draw 
            (1,-1) node[]{${\textbf{(b)}}$};
    \end{tikzpicture}
    \caption{LC circuits driven by a classical source $V_{\text{d}}$. \textbf{a.} The LC circuit is constructed from a JJ shunted by a capacitor $C_{\text{S}}$. This topology is also known as the Cooper-pair box in the early history of superconducting quantum computing. \textbf{b.} The linearized circuit derived from \textbf{a}. The node flux $\Phi$ is treated as the generalized coordinate.}
    \label{fig:classically_driven_LC_circuit}
\end{figure}
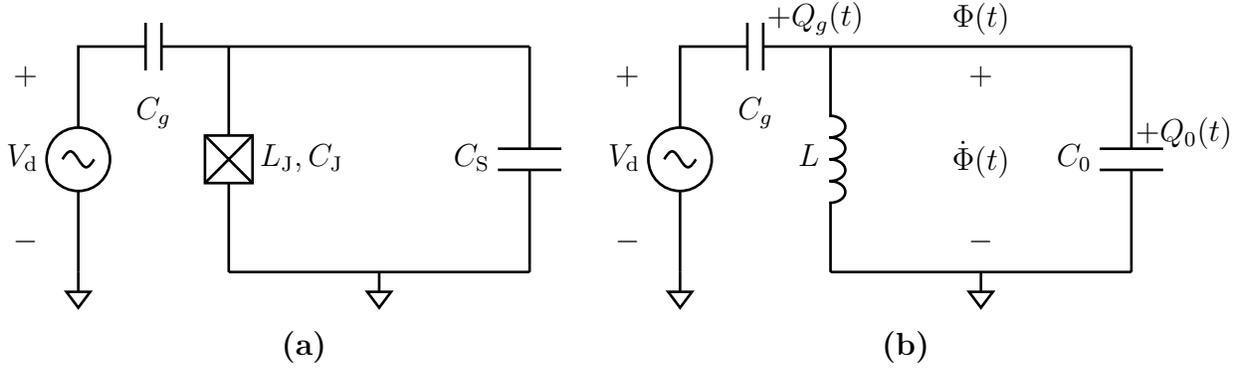

As discussed before, the prototype of a superconducting qubit consists of a JJ in parallel with a capacitor\footnote{Note that a physical JJ always has a parasitic capacitance $C_{\text{S}}$ between the two electrodes of the junction, so it's natural to include a capacitor in parallel with the junction even if no capacitor is added explicitly.}, a topology also known as the \textbf{Cooper-pair box}. In order to drive the qubit, we need some kind of coupling to the qubit; for an electrical circuit, coupling to the external environment is most naturally expressed as a capacitor as shown in Figure \ref{fig:classically_driven_LC_circuit}(a). Inductive coupling is also possible; nevertheless, it's always easier to make capacitive coupling by bringing the qubit close to a conductive pad in a miniaturized design. As outlined at the beginning of the section, we will first analyze the linearized circuit shown in Figure \ref{fig:classically_driven_LC_circuit}(b). According to the Taylor expansion in Eq.(\ref{eq:Taylor_expansion_JJ_energy}), the linearized JJ is nothing but a linear inductance of value $L = L_{\text{J}}$. In addition, we will lump the capacitance in parallel to the linear inductance into $C_0 = C_{\text{J}} + C_{\text{S}}$.
Hence, our objective is to understand an LC oscillator coupled to a voltage source via the capacitance $C_g$.

To find the generalized coordinates and momenta for canonical quantization, we start from the Lagrangian, which is given by the difference between the capacitive and inductive energy. Unlike the isolated LC oscillator solved in Chapter 2, we expect the coupling to the external source to change the expression of the conjugate momentum. By basic circuit analysis, one finds that
\begin{align}
    \mathcal{L}
        \Bigsl( 
            \Phi, \dot{\Phi}
        \Bigsr)
    &= \frac{1}{2} C_{g} (\dot{\Phi} - V_{\text{d}})^2
        + \frac{1}{2} C_0 \dot{\Phi}^2
        - \frac{1}{2L} \Phi^2
\nonumber \\
    &= \frac{1}{2} (C_0 + C_g) \dot{\Phi}^2
        - C_{g} \dot{\Phi} V_{\text{d}}
        - \frac{1}{2L} \Phi^2 
        \ \ 
        \text{ (ignore the constant $C_{g} V_{\text{d}}^2/2$)},
\end{align}
which gives the conjugate momentum
\begin{align}
    Q \doteq 
    \frac{\partial \mathcal{L}}{\partial \dot{\Phi}}
    &= (C_0 + C_g) \dot{\Phi} 
        - C_{g} V_{\text{d}}
\nonumber\\
    &= C_g (\dot{\Phi} - V_{\text{d}}) + C \dot{\Phi} = Q_0 + Q_g.
\end{align}
Instead of being the charge on $C_0$, the conjugate momentum is now the charge on $C = C_0 + C_g$ at the node where $\Phi$ is defined; we can also argue this modification via the standard circuit theory by nulling the voltage source and calculating the drive capacitance seen by the inductor.

Consequently, the Hamiltonian is
\begin{equation} \label{eq:hamiltonian_driven_LC}
    \mathcal{H}(\Phi, Q)
    = Q \dot{\Phi} - \mathcal{L}
    = \frac{1}{2C}
        (Q + C_{g} V_{\text{d}})^2 + \frac{1}{2L} \Phi^2,
\end{equation}
which, by adopting the notation $n = Q/2e$ and $\varphi = \Phi/\phi_0$ and defining $E_C = e^2/2C = e^2/2(C_0 + C_g)$ and $E_L = \phi_0^2/L$, becomes
\begin{equation} \label{eq:hamiltonian_canonical_coupling_model}
    \mathcal{H}(\varphi, n)  
    = 4 E_{C} (n - n_{\text{g}})^2 
        + \frac{1}{2} E_{L} \varphi^2.
\end{equation}

Compared to the case of an isolated LC, we now have a coupling term similar to the kinetic momentum $|\mathbf{p} - q\mathbf{A}|^2/2m$ with the vector potential replaced by an external parameter
\begin{equation}
    n_{\text{g}} 
    = -\frac{ C_{g} V_{\text{d}}}{2e},
\end{equation}
commonly called the gate charge. Hence, Eq.(\ref{eq:hamiltonian_canonical_coupling_model}) is the circuit-level description for matter interacting with an external (classical) drive. The following exercise generalizes the idea further by modeling the external drive as a second oscillator, which allows us to also quantize the source as another QHO, just like how we quantized one mode of an electromagnetic field. When the coupling is weak, i.e., $C_g \ll C_1$ and $C_g \ll C_2$, we arrive at Eq.(\ref{eq:circuit_level_dipole_interaction}), the circuit-level analogy of the dipole interaction.

\begin{exercise} \label{ex:res_res_coupling}
    Suppose instead of driving an oscillator with an ideal voltage source, we couple two oscillators together. Use $\Phi_1$ and $\Phi_2$ annotated in Figure \ref{fig:coupled_LC_circuits} and the Lagrangian formalism to show that 
    \begin{multline}
        \mathcal{H}(\Phi_1, Q_1, \Phi_2, Q_2)
        = \frac{C_1}{2} 
                \left[ 
                    \frac{C_g Q_2 + (C_2 + C_g) Q_1}{C_1 C_2 + (C_1 + C_2) C_g} 
                \right]^2
            + \frac{C_2}{2} 
                \left[ 
                    \frac{C_g Q_1 + (C_1 + C_g) Q_2}{C_1 C_2 + (C_1 + C_2) C_g} 
                \right]^2
        \\
            + \frac{C_g}{2} 
                \left[ 
                    \frac{C_2 Q_1 - C_1 Q_2}{C_1 C_2 + (C_1 + C_2) C_g} 
                \right]^2
            + \frac{1}{2L_1} \Phi_1^2
            + \frac{1}{2L_2} \Phi_2^2,
    \end{multline}
    where the conjugate momenta are $Q_1 = C_1 \dot{\Phi}_1 + C_g \Bigsl( \dot{\Phi}_1 - \dot{\Phi}_2 \Bigsr)$ and $Q_2 = C_2 \dot{\Phi}_2 + C_g \Bigsl( \dot{\Phi}_2 - \dot{\Phi}_1 \Bigsr)$. Moreover, under the assumption that $C_g \ll C_1$ and $C_g \ll C_2$, show that
    \begin{align}
        \mathcal{H}(\Phi_1, Q_1, \Phi_2, Q_2)
        &\approx 
            \frac{1}{2C_1} Q_1^2
            + \frac{1}{2C_2} Q_2^2
            + \frac{1}{2L_1} \Phi_1^2
            + \frac{1}{2L_2} \Phi_2^2
            + \frac{C_g}{C_1 C_2} Q_1 Q_2
    \\ \label{eq:circuit_level_dipole_interaction_derivation}
        &\approx 
            \underbrace{
                \frac{1}{2C_1} \left( Q_1 + \frac{C_g}{C_2} Q_2\right)^2 
                + \frac{1}{2L_1} \Phi_1^2
            }_{
                \substack{
                    \text{linearized oscillator} \\ 
                    \text{with dipole interaction}
                }
            }
            + \underbrace{
                \frac{1}{2C_2} Q_2^2 
                + \frac{1}{2L_2} \Phi_2^2
            }_{
                \substack{
                    \text{drive field modeled} \\ 
                    \text{as an oscillator}
                }
            }
    \end{align}
    and, with $n_1 = Q_1 / 2e, n_2 = Q_2 / 2e, \varphi_1 = \Phi_1 / \phi_0$, and $\varphi_2 = \Phi_2 / \phi_0$,
    \begin{equation} \label{eq:circuit_level_dipole_interaction}
        \mathcal{H}(\varphi_1, n_1, \varphi_2, n_2)
        \approx 
            4 E_{C_1} n_1^2
            + 4 E_{C_2} n_2^2
            + \frac{1}{2} E_{L_1} \varphi_1^2
            + \frac{1}{2} E_{L_2} \varphi_2^2
            + 4e^2 \frac{C_g}{C_1 C_2} n_1 n_2,
    \end{equation}
    where $E_{C_1},E_{C_2}, E_{L_1}$, and $E_{L_2}$ are defined in the same way as before.

    Furthermore, we can define
    \begin{equation}
        V_{\text{d}} = \frac{Q_2}{C_2}
        = \dot{\Phi}_2 + \frac{C_g}{C_2} \Bigsl( \dot{\Phi}_2 - \dot{\Phi}_1 \Bigsr)
        \approx \dot{\Phi}_2,
    \end{equation}
    which is approximately the voltage developed on the second resonator, then
    \begin{equation} \label{eq:hamiltonian_res_driven_LC}
        \mathcal{H}(\Phi_1, Q_1, \Phi_2, Q_2)
        = \frac{1}{2C_1} 
                \left( 
                    Q_1 + C_g V_{\text{d}} 
                \right)^2 
            + \frac{1}{2L_1} \Phi_1^2
            + \frac{1}{2C_2} Q_2^2 
            + \frac{1}{2L_2} \Phi_2^2
    \end{equation}
    takes the same form as Eq.(\ref{eq:hamiltonian_driven_LC}). However, once we quantized the system, $V_{\text{d}}$ in Eq.(\ref{eq:hamiltonian_res_driven_LC}) will become an operator whereas $V_{\text{d}}$ in Eq.(\ref{eq:hamiltonian_driven_LC}) stay as a classical drive.
\end{exercise}
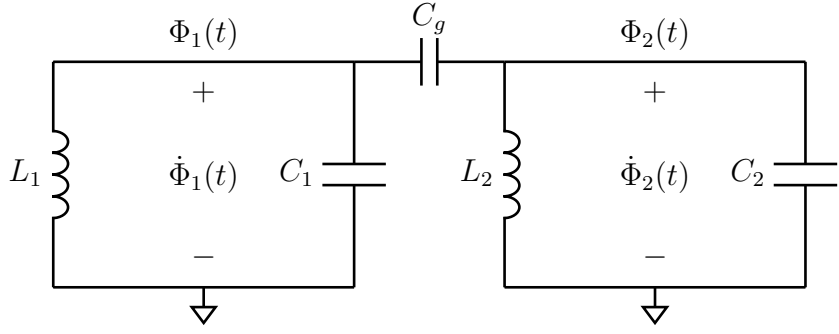
\begin{figure}
    \centering
    \begin{tikzpicture}[scale=1, transform shape]
    \ctikzset{tripoles/mos style/arrows};
        \draw 
            (-6,3) to (-2,3) 
            (-6,0) to (-2,0) 
            (-4,0) node[/tikz/circuitikz/bipoles/length=30pt,sground]{}
            (-6,3) to [/tikz/circuitikz/bipoles/length=40pt, L, l_=$L_1$] (-6,0)
            (-2,3) to (-2,2) to [/tikz/circuitikz/bipoles/length=40pt, C, l_=$C_1$] (-2,1) to (-2,0);
        \draw
            (0,3) to (4,3) 
            (4,0) to (0,0) 
            (2,0) node[/tikz/circuitikz/bipoles/length=30pt,sground]{}
            (0,3) to [/tikz/circuitikz/bipoles/length=40pt, L, l_=$L_2$] (0,0)
            (4,3) to (4,2) to [/tikz/circuitikz/bipoles/length=40pt, C, l_=$C_2$] (4,1) to (4,0);
        \draw 
            (0,3) to [/tikz/circuitikz/bipoles/length=30pt, C, l_=$C_{g}$] (-2,3);
        \draw
            (-4,2.6) node[]{$+$} 
            (-4,1.5) node[]{${\dot{\Phi}_1(t)}$}
            (-4,0.4) node[]{$-$}
            (2,2.6) node[]{$+$} 
            (2,1.5) node[]{${\dot{\Phi}_2(t)}$}
            (2,0.4) node[]{$-$}
            (-4,3) node[above]{$\Phi_1(t)$}
            (2,3) node[above]{$\Phi_2(t)$};
    \end{tikzpicture}
    \caption{Two coupled LC oscillators. This model can be applied to 1) a weakly anharmonic qubit coupled to a linear resonator or 2) two qubits coupled directly.}
    \label{fig:coupled_LC_circuits}
\end{figure}

\subsection{Circuit-Level Realization of Two-Qubit Gates}
In the last chapter, we introduced several two-qubit gates based on the dipole-interaction formalism. We now briefly discuss how one can realize these gates using lumped circuit elements at a low temperature by extending the idea used in the previous subsection. Again, we will use LC oscillators to model the linearized anharmonic oscillators as well as the resonator used to couple the anharmonic oscillators \cite{doi:10.1063/1.5089550}.

The simplest case where two qubits coupled directly with a capacitor was already solved in Exercise \ref{ex:res_res_coupling} if we treat the second oscillator also as a linearized anharmonic oscillator. Since the quantum operator of $n_1$ and $n_2$ can be expressed as (see Table \ref{table:normal_mode_expansion})
\begin{equation}
    \hat{n}_j
    = - \ci \left(\frac{E_{L_j}}{8E_{C_j}}\right)^{\!\! 1/4} \frac{\hat{a}_j - \hat{a}_j^{\dagger}}{\sqrt{2}}
    \ \ \text{ for } \ \ 
    j = 1,2,
\end{equation}
the coupling term in Eq.(\ref{eq:circuit_level_dipole_interaction}) becomes
\begin{equation}
    \hat{H}_{\text{int}}
    \doteq 4e^2 \frac{C_g}{C_1 C_2} \hat{n}_1 \hat{n}_2
    = - \frac{\hbar \sqrt{\omega_1 \omega_2} C_g}{2 \sqrt{C_1 C_2}} 
        \Big( \hat{a}_1 - \hat{a}_1^{\dagger} \Big)
        \Big( \hat{a}_2 - \hat{a}_2^{\dagger} \Big)
\end{equation}
after quantization, where $\omega_{1,2}$ are the resonant frequency of the two LC oscillators. By applying the RWA and restricting the two QHOs (with nonlinearity added) to the lowest two energy levels, we obtain the transverse coupling
\begin{equation}
    \hat{H}_{\text{int}}
    \approx 
    \frac{\hbar \sqrt{\omega_1 \omega_2} C_g}{2 \sqrt{C_1 C_2}} 
        \Big( 
            \hat{a}_1^{\dagger} \hat{a}_2 
            + \hat{a}_1 \hat{a}_2^{\dagger} 
        \Big)
    \approx 
        - \hbar g_{12} 
        \Big( 
            \hat{\sigma}_{+,1} \hat{\sigma}_{-,2} 
            + \hat{\sigma}_{-,1}  \hat{\sigma}_{+,2}
        \Big),
\end{equation}
where the coupling coefficient is defined to be
\begin{equation} \label{eq:coupling_coef_12_circuit}
    g_{12} 
    = - \frac{\sqrt{\omega_1 \omega_2} C_g}{2 \sqrt{C_1 C_2}} < 0.
\end{equation}

\begin{figure}[ht]
    \centering
    \begin{tikzpicture}[scale=0.9]
    \ctikzset{tripoles/mos style/arrows};
        \draw 
            (-6,3) to (-2,3) 
            (-6,0) to (-2,0) 
            (-4,0) node[/tikz/circuitikz/bipoles/length=30pt,sground]{}
            (-6,3) to [/tikz/circuitikz/bipoles/length=40pt, L, l_=$L_1$] (-6,0)
            (-2,3) to (-2,2) to [/tikz/circuitikz/bipoles/length=40pt, C, l_=$C_1$] (-2,1) to (-2,0);
        \draw
            (0,3) to (4,3) 
            (4,0) to (0,0) 
            (2,0) node[/tikz/circuitikz/bipoles/length=30pt,sground]{}
            (0,3) to [/tikz/circuitikz/bipoles/length=40pt, L, l_=$L_r$] (0,0)
            (4,3) to (4,2) to [/tikz/circuitikz/bipoles/length=40pt, C, l_=$C_r$] (4,1) to (4,0);
        \draw
            (6,3) to (10,3) 
            (10,0) to (6,0) 
            (8,0) node[/tikz/circuitikz/bipoles/length=30pt,sground]{}
            (6,3) to [/tikz/circuitikz/bipoles/length=40pt, L, l_=$L_2$] (6,0)
            (10,3) to (10,2) to [/tikz/circuitikz/bipoles/length=40pt, C, l_=$C_2$] (10,1) to (10,0);
        \draw 
            (0,3) to [/tikz/circuitikz/bipoles/length=30pt, C, l_=$C_{g1}$] (-2,3);
        \draw 
            (6,3) to [/tikz/circuitikz/bipoles/length=30pt, C, l_=$C_{g2}$] (4,3);
        \draw 
            (6,3) to (6,4.5) to [/tikz/circuitikz/bipoles/length=30pt, C, l_=${C_{g0}}$] (-2,4.5) to (-2,3);
        \draw
            (-4,2.6) node[]{$+$} 
            (-4,1.5) node[]{${\dot{\Phi}_1(t)}$}
            (-4,0.4) node[]{$-$}
            (2,2.6) node[]{$+$} 
            (2,1.5) node[]{${\dot{\Phi}_r(t)}$}
            (2,0.4) node[]{$-$}
            (8,2.6) node[]{$+$} 
            (8,1.5) node[]{${\dot{\Phi}_2(t)}$}
            (8,0.4) node[]{$-$}
            (-4,3) node[above]{$\Phi_1(t)$}
            (2,3) node[above]{$\Phi_r(t)$}
            (8,3) node[above]{$\Phi_2(t)$};
    \end{tikzpicture}
    \caption{Two LC oscillators interact via a common resonator and a direct coupling capacitance $C_{g0}$.}
    \label{fig:coupled_LC_circuits_via_resonator}
\end{figure}
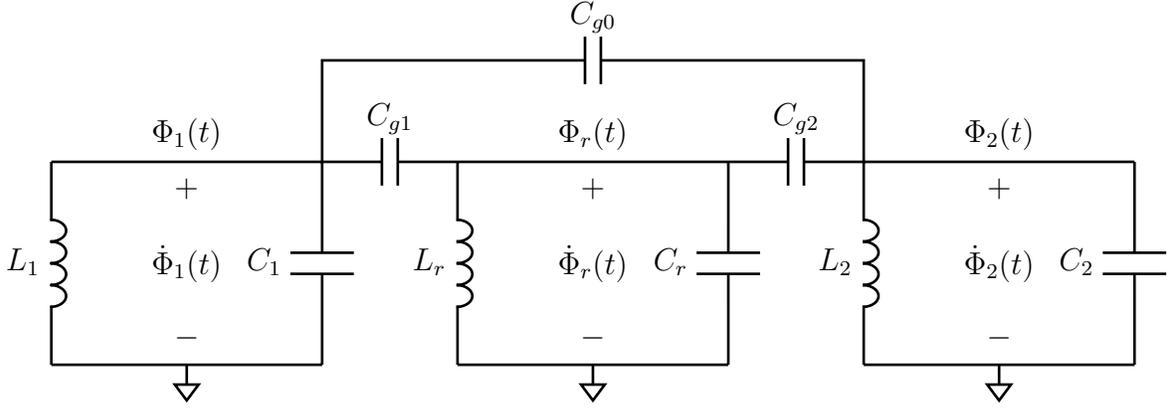

Besides the direct coupling, we can add the resonator-mediated coupling by inserting a linear oscillator in between the two anharmonic oscillators, as shown in Figure \ref{fig:coupled_LC_circuits_via_resonator}. For simplicity, consider $C_{g0} = 0$ first (i.e., no direct coupling between the two qubits). For a complex circuit, it's useful to use a matrix representation of the network; in particular, we define a vector of node flux quantities 
\begin{equation}
    \mathbf{\Phi} 
    = \begin{pmatrix}
            \Phi_1 & \Phi_2 & \Phi_r
        \end{pmatrix}^{\top},
\end{equation}
from which the Lagrangian of the system can be expressed as
\begin{equation}
    \mathcal{L}(\mathbf{\Phi}, \dot{\mathbf{\Phi}})
    = \frac{1}{2} \dot{\mathbf{\Phi}}^{\top} \mathsf{C} \dot{\mathbf{\Phi}}
        - \frac{1}{2L_1} \Phi_1^2
        - \frac{1}{2L_2} \Phi_2^2
        - \frac{1}{2L_r} \Phi_r^2,
\end{equation}
where the positive-definite (symmetric) matrix
\begin{equation}
    \mathsf{C} 
    = \begin{pmatrix}
            C_1 + C_{g1} & 0 & -C_{g1} \\ 
            0 & C_2 + C_{g2} & -C_{g2} \\ 
            -C_{g1} & -C_{g2} & C_{r} + C_{g1} + C_{g2} 
        \end{pmatrix} 
\end{equation}
is the Maxwell capacitance matrix of the network. We will come back to this formalism in the next section to study a general Jospheson network known as a multimon; for now, the introduction of the capacitance matrix merely simplifies the notation since the conjugate momentum can be compactly written as
\begin{equation}
    \mathbf{Q} 
    = (Q_1, Q_2, Q_r)
    \doteq \nabla_{\dot{\mathbf{\Phi}}} \mathcal{L} 
    = \mathsf{C} \dot{\mathbf{\Phi}}.
\end{equation}
Consequently, the Hamiltonian can also be computed with simple matrix multiplications
\begin{equation}
    \mathcal{H}(\mathbf{\Phi}, \mathbf{Q})
    = \mathbf{Q}^{\top} \dot{\mathbf{\Phi}} - \mathcal{L}
    = \frac{1}{2} 
            \mathbf{Q}^{\top} 
            \mathsf{C}^{-1} 
            \mathbf{Q}
        + \frac{1}{2L_1} \Phi_1^2
        + \frac{1}{2L_2} \Phi_2^2
        + \frac{1}{2L_r} \Phi_r^2.
\end{equation}
Specifically, our model gives
\begin{equation}
    \mathsf{C}^{-1}
    = \frac{1}{\bar{C}^3} 
        \begin{pmatrix}
            \substack{\displaystyle C_2 (C_{g1} + C_{g2} + C_r) \\ \displaystyle + C_{g2} (C_{g1} + C_r)} & C_{g1} C_{g2} & C_{g1} (C_2 + C_{g2}) \\[3mm] 
            C_{g1} C_{g2} & \substack{\displaystyle C_1 (C_{g1} + C_{g2} + C_r) \\ \displaystyle + C_{g1} (C_{g2} + C_r)} & C_{g2} (C_1 + C_{g1}) \\[3mm] 
            C_{g1} (C_2 + C_{g2}) & C_{g2} (C_1 + C_{g1}) & \substack{\displaystyle C_1 C_2 + C_1 C_{g2} \\ \displaystyle + C_2 C_{g1} + C_{g1} C_{g2}}
        \end{pmatrix},
\end{equation}
where 
\begin{equation}
    \bar{C}^3 
    = C_1 C_2 (C_{g1} + C_{g2} + C_r) 
        + C_1 C_{g2} (C_{g1} + C_r) 
        + C_2 C_{g1} (C_{g2} + C_r)
        + C_{g1} C_{g2} C_r
\end{equation}

For a real design, we usually have $C_{g1}, C_{g2} \ll C_1, C_2, C_r$, i.e., the coupling is weak. Under this assumption, we have
\begin{equation}
    \mathsf{C}^{-1}
    \approx 
        \begin{pmatrix}
            1/C_1 & 0 & C_{g1}/C_1 C_r \\[3mm] 
            0 & 1/C_2 & C_{g2}/C_2 C_r \\[3mm] 
            C_{g1}/C_1 C_r & C_{g2}/C_2 C_r & 1/C_r
        \end{pmatrix}
\end{equation}
and
\begin{align}
    \mathcal{H}(\mathbf{\Phi}, \mathbf{Q})
    &\approx \frac{1}{2 C_1} Q_1^2 
        + \frac{1}{2 C_2} Q_2^2 
        + \frac{1}{2 C_r} Q_r^2
        + \frac{1}{2L_1} \Phi_1^2
        + \frac{1}{2L_2} \Phi_2^2
        + \frac{1}{2L_r} \Phi_r^2
\nonumber \\
    & \ \ \ \  \ \ \ \  
        + \frac{C_{g1}}{C_1 C_r} Q_1 Q_r
        + \frac{C_{g2}}{C_2 C_r} Q_2 Q_r,
\end{align}
where the last two terms clearly give rise to the coupling of the individual qubits to the resonator. Then, we can apply unitary transformations to reveal the resonator-mediated coupling between the two qubits.

Moreover, if we include $C_{g0}$ in the calculation, we would have obtain
\begin{align}
    \mathcal{H}(\mathbf{\Phi}, \mathbf{Q})
    &\approx \frac{1}{2 C_1} Q_1^2 
        + \frac{1}{2 C_2} Q_2^2 
        + \frac{1}{2 C_r} Q_r^2
        + \frac{1}{2L_1} \Phi_1^2
        + \frac{1}{2L_2} \Phi_2^2
        + \frac{1}{2L_r} \Phi_r^2
\nonumber \\
    & \ \ \ \  \ \ \ \  
        + \frac{C_{g0}}{C_1 C_2} Q_1 Q_2
        + \frac{C_{g1}}{C_1 C_r} Q_1 Q_r
        + \frac{C_{g2}}{C_2 C_r} Q_2 Q_r,
\end{align}
which gives the circuit-level description of a tunable coupler after quantization, i.e.,
\begin{align}
    \hat{H}
    &= - \frac{1}{2} 
            \hbar \omega_1  \hat{\sigma}_{z,1}
        - \frac{1}{2} 
            \hbar \omega_2  \hat{\sigma}_{z,2}
        + \hbar \omega_r 
            \left( 
                \hat{a}_1^{\dagger} \hat{a}_1 
                + \frac{1}{2}
            \right)
\nonumber \\
    & \ \ \   \ \ \ \  
        - \hbar g_{12} 
        \Big( 
            \hat{\sigma}_{+,1} \hat{\sigma}_{-,2} 
            + \hat{\sigma}_{-,1}  \hat{\sigma}_{+,2}
        \Big)
        - \hbar g_1
        \Big( 
            \hat{a}_{r} \hat{\sigma}_{+,1}
            + \hat{a}_{r}^{\dagger} \hat{\sigma}_{-,1} 
        \Big)
        - \hbar g_2 
        \Big( 
            \hat{a}_{r} \hat{\sigma}_{+,2}
            + \hat{a}_{r}^{\dagger} \hat{\sigma}_{-,2}
        \Big),
\end{align}
where have approximated the two anharmonic oscillators as two-level systems and applied the RWA. The coupling coefficients takes the same form as that of Eq.(\ref{eq:coupling_coef_12_circuit}), i.e.,
\begin{equation}
    g_{12} 
    = - \frac{\sqrt{\omega_1 \omega_2} C_{g0}}{2 \sqrt{C_1 C_2}},
    \ \ \ \ 
    g_{2} 
    = - \frac{\sqrt{\omega_1 \omega_r} C_{g1}}{2 \sqrt{C_1 C_r}},
    \ \ \ \ 
    g_{1} 
    = - \frac{\sqrt{\omega_2 \omega_r} C_{g2}}{2 \sqrt{C_2 C_r}},
\end{equation}
and it's not hard to see that the effective resonator-mediated coupling strength
\begin{equation}
    J_{12} = - \frac{g_1 g_2 (\Delta_1 + \Delta_2)}{2 \Delta_1 \Delta_2} > 0
\end{equation}
if $\Delta_j = \omega_j - \omega_r < 0$ for $j = 1,2$ (i.e., the resonator frequency is above the two qubit frequencies). The sign difference between $J_{12}$ and $g_{12}$ is what enables the construction of a tunable coupler as mentioned in the last chapter.

\subsection{Nonlinearity and Canonical Quantization of the Cooper-Pair Box}
Now, we bring the nonlinearity back to the picture. The Lagrangian of the Cooper-pair box is still the difference between the capacitive and inductive energy, but we replace the inductive energy with the Josephson energy, which is non-quadratic:
\begin{equation}
    \mathcal{L}_{\text{CPB}}
        \Bigsl( 
            \Phi, \dot{\Phi}
        \Bigsr)
    = \frac{1}{2} (C_0 + C_g) \dot{\Phi}^2
        - C_{g} \dot{\Phi} V_{\text{d}}
        - E_{\text{J}} 
            \left[ 
                1 - \cos(\frac{\Phi}{\phi_0})
            \right].
\end{equation}
Since the conjugate momentum is the derivative of the Lagrangian with respect to $\dot{\Phi}$, modifying the inductive energy does not change the form of the conjugate momentum. Consequently, the Hamiltonian becomes
\begin{equation}
    \mathcal{H}_{\text{CPB}}(\delta, n) 
    = 4 E_C (n - n_{\text{g}})^2
        + E_{\text{J}} \Big( 1 - \cos \delta \Big),
\end{equation}
provided we identify $\delta = \varphi = \Phi/\phi_0$ as the reduced flux. (Recall that $C = C_0 + C_g$.)

With all the preparations, the canonical quantization simply involves the promotion of the classical conjugate variables\footnote{One might ask why the phase of the order parameter is classical since we know that superconductivity is a quantum effect. However, as discussed before, a superconducting state is well-modeled by a coherent state (i.e., the solution to the Ginzburg–Landau equations). Similarly, we can define a quantum operator (with some special care of the singularity) for the phase of a QHO; thus, by this analogy, it's unsurprising that we can quantize the phase of the order parameter. Experimentally, one can verify the quantumness of the phase by detecting quantum tunneling of the phase variable across the cosine potential at a temperature where the thermal activation can be ignored \cite{10.1143/PTP.69.80, PhysRevB.35.4682}.} $n$ and $\delta$ to the quantum operators $\hat{n}$ and $\hat{\delta}$ along with the canonical commutation relation
\begin{equation}
    \Big[
        \hat{\delta}, 
        \hat{n}
    \Big] = \ci 
    \quad
    \left(
    \text{more rigorously, }
    \left[\hat{n}, \widehat{e^{i \delta}}\right]=\widehat{e^{i \delta}},
    \text{see \cite{PhysicsPhysiqueFizika.1.49}}
    \right).
\end{equation}
Like any other conjugate observables, Fourier theorem relates the eigenstates of the flux and charge operators. However, due to the discrete change of Cooper-pair in the absence of a linear shunt inductor (equivalently, the periodicity of the phase/flux) \cite{PhysRevLett.103.217004, cQED_lectures_Girvin}, the Fourier transform is modified to a Fourier series\footnote{Recall that a periodic function should be expanded using a Fourier series instead of the Fourier transform to avoid the appearance of delta functions due to the non-normalizability of a periodic function.}; thus, 
\begin{gather}
    \ket{n}
    = \frac{1}{\sqrt{2\pi}}
        \int_{0}^{2\pi} 
            \mathrm{d} \delta 
            e^{\ci n \delta}
            \ket{\delta} ,
\\
    \ket{\delta}
    = \frac{1}{\sqrt{2\pi}} \sum_{n=-\infty}^{\infty}  e^{-\ci n \delta} \ket{n}.
\end{gather}
In addition, the orthogonality of the eigenstates is given by 
\begin{equation}
    \bra{n}\ket{n'} = \delta_{n,n'}
\ \ \text{ and } \ \ 
    \bra{\delta}\ket{\delta'} = \delta(\delta - \delta')
\end{equation}
and the completeness of the Hilbert space can be expressed as
\begin{equation}
    \hat{\delta}
    = \int_{0}^{2\pi} 
        \mathrm{d} \delta \, \delta\ket{\delta} \!
        \bra{\delta}
\ \ \text{ or } \ \ 
    \hat{n}
    = \sum_{n=-\infty}^{\infty}  n\ket{n} \!
        \bra{n}.
\end{equation}

Given the Fourier relation between charge and flux, the quantum Hamiltonian of the Cooper-pair box
\begin{equation}
    \hat{H}_{\text{CPB}}
    = 4E_C
            \left( 
                \hat{n} - n_{\text{g}} 
            \right)^2 
        + E_{\text{J}} 
            \left( 
                1 - \cos \hat{\delta} 
            \right)
\end{equation}
can be expressed in the charge basis as
\begin{equation}
    \hat{H}_{\text{CPB}}
    = 4E_C
            \left( 
                \hat{n} - n_{\text{g}} 
            \right)^2 
        - \frac{E_{\text{J}}}{2} 
            \sum_{n= - \infty}^{\infty} 
            \Big( 
                \ket{n} \! \bra{n+1} 
                + \ket{n+1} \! \bra{n} 
            \Big)
\end{equation}
or in the flux basis as
\begin{equation}
    \left[ 
            4E_C
                \left( 
                    - \ci \frac{\mathrm{d}}{\mathrm{d} \delta} 
                    - n_{\text{g}} 
                \right)^{\! 2} 
            + E_{\text{J}} 
                \left( 
                    1 - \cos \delta
                \right)
        \right] 
        \psi(\delta)
    = E \psi(\delta),
\end{equation}
where $\psi(\delta) = \bra{\delta}\ket{\psi}$ is the flux-basis wavefunction with eigenenergy $E$, satsifying $\psi(\delta + 2\pi) = \psi(\delta)$. To pick one representation against the other, we need to understand the energy landscape in the Hamiltonian; that is, whether the capacitive or the Josephson energy (i.e., $E_{C}$ or $E_{\text{J}}$) dominates the Hamiltonian. To see why the ratio $E_{\text{J}} / E_{C}$ plays a critical role in the qubit analysis, we examine the two limits of the ratio:

\begin{enumerate}
    \item[(i)] $E_{\text{J}} \ll E_{C}$ (charge qubits): The Josephson energy can be treated as a perturbation added to the ``free-particle'' Hamiltonian
    \begin{equation}
        \hat{H}_{\text{free}} 
        = 4E_C
            \left( 
                \hat{n} - n_{\text{g}} 
            \right)^2.
    \end{equation}
    Here, the free particle is an imaginary object that moves along the flux axis, just like how a real particle moves in the position space. Before adding the perturbation, the free particle can have any (discrete) charge values for a fixed gate charge $n_{\text{g}}$. (Note that we are treating $n_g$ as a classical drive; later we will also quantize the gate charge so that we can talk about coupling to a quantized cavity to which the qubit is coupled.) For the special case where $n_{\text{g}} = n + 1/2$ for some integer $n$, we observe a degeneracy between the energy states (i.e., the charge eigenstates) $\ket{n}$ and $\ket{n+1}$ since
    \begin{equation}
        \bra{n}\hat{H}\ket{n} = \bra{n+1}\hat{H}\ket{n+1} = E_{C}.
    \end{equation}
    Now, from basic solid-state physics, we know that the periodic potential will open an avoiding crossing to lift the degeneracy. Historically, the \textbf{charge qubit} (another name of the Cooper-pair box) \cite{V_Bouchiat_1998,nature19718, nature02851} is operating at the ``sweet spots'' $n_{\text{g}} = n + 1/2$ such that the two degenerate states are mixed equally by the perturbation to form the new computational basis
    \begin{equation}
        \ket{g} = \frac{\ket{n} + \ket{n+1}}{\sqrt{2}}
        \ \ \text{ and } \ \ 
        \ket{e} = \frac{\ket{n} - \ket{n+1}}{\sqrt{2}}.
    \end{equation}

    \hspace{\parindent} \hspace{\parindent} \hspace{\parindent} \hspace{\parindent} \hspace{\parindent} Since the community has moved away from the charge qubit, we will not give a full analysis of its operation. Nevertheless, it's straightforward to compute the energy difference opened by the perturbation by diagonalizing $\hat{H}_{\text{CPB}}$ restricted to the two-dimensional subspace spanned by unperturbed basis vectors $\ket{n}$ and $\ket{n+1}$:
    \begin{align}
        \hat{H}_{CPB} 
        &\rightarrow 
        \begin{pmatrix}
            \bra{n}\hat{H}\ket{n} & \bra{n}\hat{H}\ket{n+1}\\
            \bra{n+1}\hat{H}\ket{n} & \bra{n+1}\hat{H}\ket{n+1}
        \end{pmatrix}
    \nonumber \\
        &= \begin{pmatrix}
            4E_{\text{C}}[n-n_{\text{g}}]^2 & -E_{\text{J}}/2\\
            -E_{\text{J}}/2 & 4E_{\text{C}}[(n+1)-n_{\text{g}}]^2
        \end{pmatrix}
    \end{align}
    The charge qubit is not as practical as the transmon qubit to be discussed below because its energy levels (i.e., $\ket{g}$ and $\ket{e}$) are extremely sensitive to the noise of the gate charge. By solving $\ket{g}$ and $\ket{e}$ as a function of $n_{\text{g}}$, we can compute the first and second derivatives of the energy difference $E_{e} - E_{g}$ with respect to $n_{\text{g}}$. Even if we set $n_g = n + 1/2$ where the first derivative vanishes, the second derivative, given by $64E_C^2 / E_{\text{J}}$, is still considerably large if $E_C \gg E_{\text{J}}$. To improve the charge noise, we move to the other limit where the capacitive energy is much smaller than the Josephson energy.

    \item[(ii)] $E_{\text{J}} \gg E_{C}$ (transmon qubits): First, we go back to the particle analogy in the flux space and treat the Josephson energy as the potential energy experienced by the particle; this can be most easily understood in the flux-representation
    \begin{equation} \label{eq:TISE_CPB}
        \left[ 
                \underbrace{4E_C
                    \left( 
                        - \ci \frac{\mathrm{d}}{\mathrm{d} \delta} 
                        - n_{\text{g}} 
                    \right)^{\! 2} }_{\text{kinetic energy}}
                + \underbrace{E_{\text{J}} 
                    \left( 
                        1 - \cos \delta 
                    \right)}_{\text{potential energy}}
            \right] 
            \psi(\delta)
        = E \psi(\delta).
    \end{equation}
    From this perspective, $1/E_C$ is proportional to the ``mass'' of this imaginary particle and $E_{\text{J}}$ controls the depth of the periodic potential. The gate charge can be thought of as a constant kick to the particle, providing momentum for the particle to escape from one local minimum of the potential. 
    
    \hspace{\parindent} \hspace{\parindent} \hspace{\parindent} \hspace{\parindent} \hspace{\parindent} Then, from elementary quantum mechanics, we know that quantum tunneling can always happen since the potential is not infinitely high. Even from the classical point of view, if the particle has a small mass (i.e., $E_C$ is small), it has a better chance to swing from one well to the other. Combining these intuitions, we conclude that a large $E_{\text{J}}/E_C$ will result in a heavy particle trapped in a deep well, leading to a small displacement in the flux coordinate. Consequently, a particle localized at the minimum of a well will have less flux uncertainty compared to the charge uncertainty, thus, justifying the usage of the flux representation. Since the trapping is robust against a small kick due to $n_g$, we also expect the energy eigenstates in this high-$E_{\text{J}}/E_C$ regime to be insensitive to the charge noise. Indeed, one can use the WKB theory to show that the energy dispersion due to the gate charge is given by \cite{PhysRevA.76.042319, PhysRevLett.114.010501}
    \begin{equation} \label{eq:charge_noise_WKB}
        E_m (n_\text{g})
        = E_{m}(n_\text{g}=1/4)
            + \frac{(-1)^m E_C}{\sqrt{2\pi}} \frac{2^{4m + 5}}{m!} \left(\frac{E_{\text{J}}}{2E_C}\right)^{\frac{m}{2}+\frac{3}{4}} e^{-\sqrt{8E_{\text{J}}/ E_C}}
            \cos(2 \pi n_{\text{g}})
    \end{equation}
    for the $m$th eigenstate in the high-$E_{\text{J}}/E_C$ limit. The exponential suppression as a function of $8E_{\text{J}}/E_C$ proves quantitatively the advantage of going into the high-$E_{\text{J}}/E_C$ regime for gate-charge protection.
\end{enumerate}

\begin{figure}[t]
    \centering
    \includegraphics[scale=0.32]{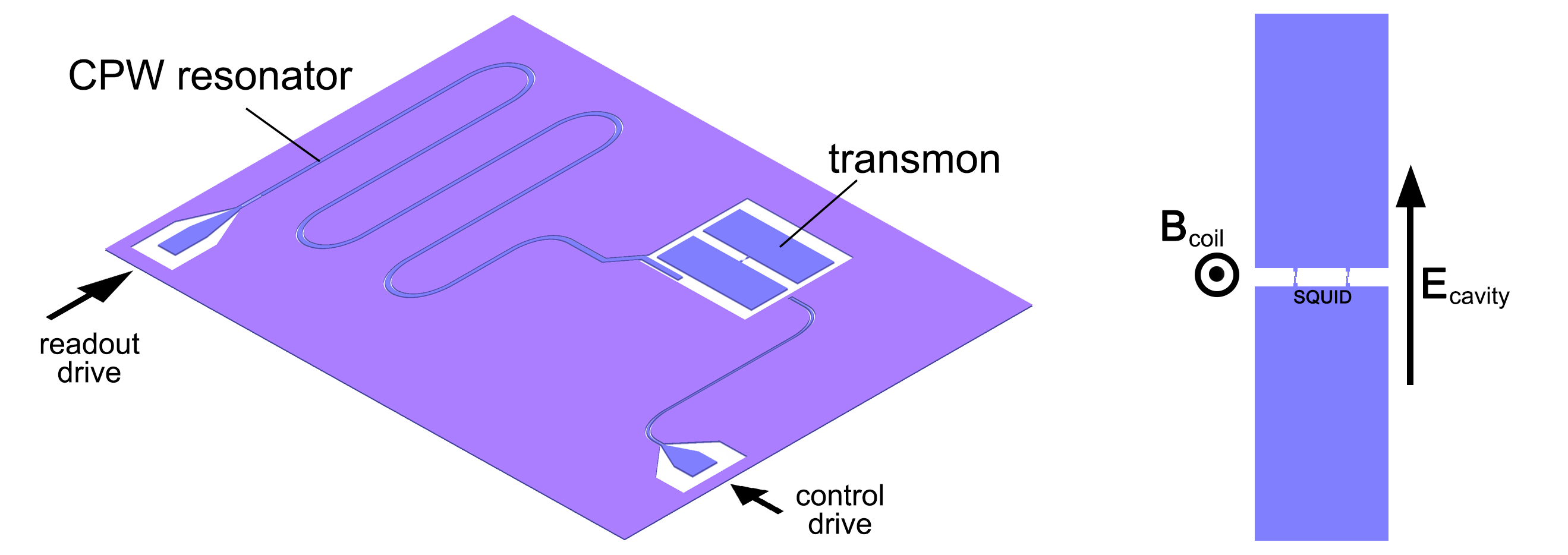}
    \caption{Examples of transmon designs. Left: A planar transmon surrounded by a huge metal ground plane. The transmon is coupled to a readout resonator (i.e., a CPW section with length $\lambda/2$) and a control line. Right: A 3D transmon before placing into the cavity. The two JJs between the capacitor pads form a SQUID, making the qubit frequency tunable.}
    \label{fig:transmon_picture}
\end{figure}

In an actual qubit design, achieving a high $E_{\text{J}}/E_C$ boils down to decreasing $E_C$ by shunting the Josephson junction with a large capacitor; such topology is known as a \textbf{transmission line shunted plasma oscillation qubit} (in short, a \textbf{transmon}) \cite{PhysRevA.76.042319}. Figure \ref{fig:transmon_picture} shows the layout of a planar transmon and a 3D transmon. 

\section{Analysis of Superconducting Qubits/Qudits}\label{section:qubit_analysis}
In this section, we combine the abstract ideas from the last two chapters with the anharmonic oscillator discussed in the current chapter to realize physical qubits/qudits. In addition, we will also examine another type of JJ-based qubit design that allows us to expand the computational space for large-scale superconducting quantum computing.

\begin{figure}[t]
    \centering
    \includegraphics[scale=0.33]{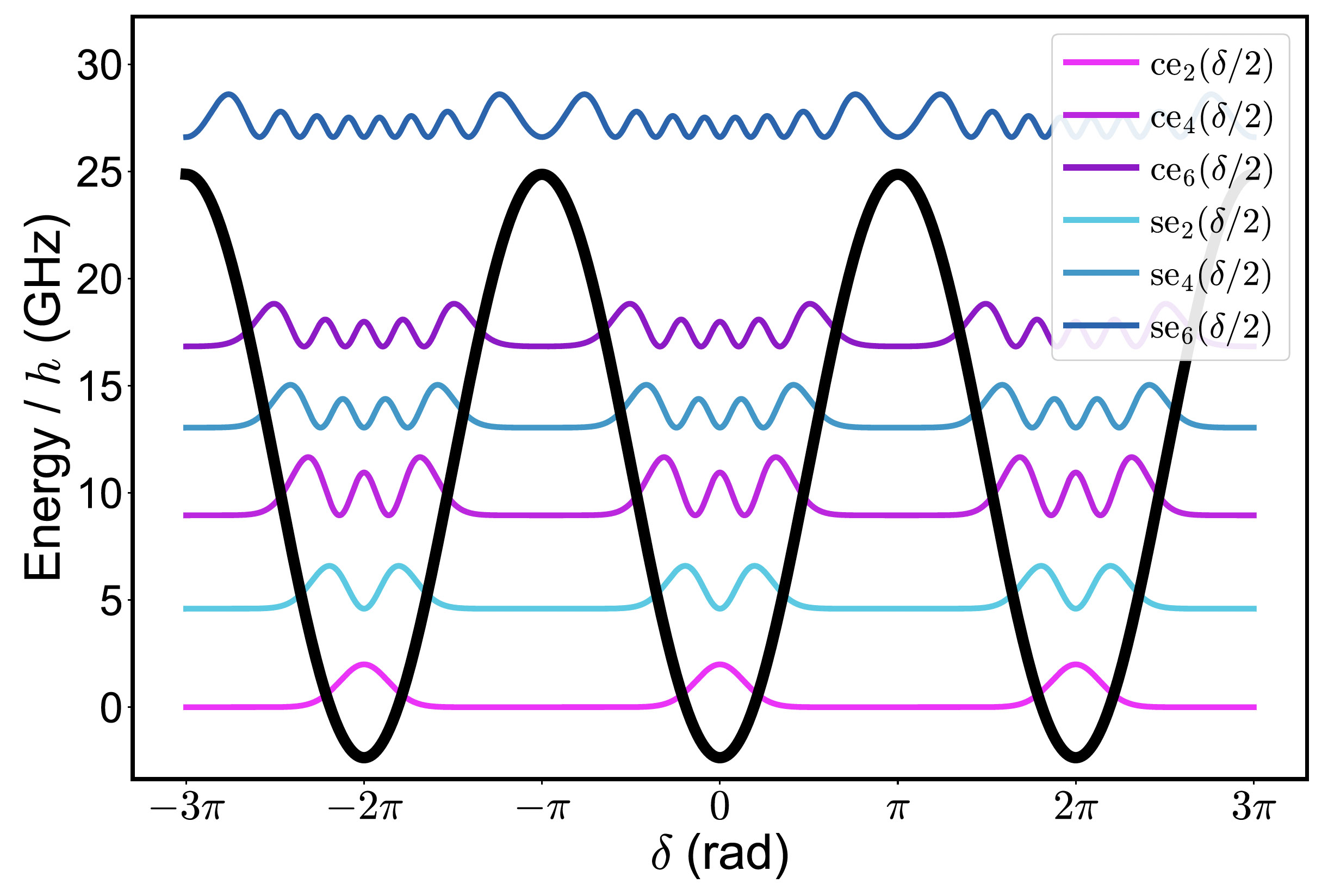}
    \caption{Magntidue of the solutions of the Mathieu differential equations ($E_{\mathrm{J}}/h = 13.61$ GHz, $E_C/h = 213.46$ MHz and $n_{\mathrm{g}} = 0$) superimposed onto the Josephson energy. The flux particle has a finite number of bound states. The lowest few wave functions resemble the Hermite-Gaussian functions of the QHO within one period of the flux. The probability is suppressed exponentially in the classically forbidden region such that the flux particle only sees a single well. ($\mathrm{se}$ and $\mathrm{ce}$ are the sine- and cosine-elliptic functions.)}
    \label{fig:mathieu_solutions}
\end{figure}

\subsection{An Isolated Transmon Qubit}

The Schr\"{o}dinger equation for a transmon (i.e., Eq.(\ref{eq:TISE_CPB})) can be solved analytically since it is of the form of a Mathieu differential equation. Figure \ref{fig:mathieu_solutions} shows an example of the transmon eigenstates solved analytically. However, instead of going over the Floquet theory for solving the Mathieu differential equation, it's more insightful to understand the energy level from the point of view of perturbation theory since we are already in the transmon limit where $E_{\text{J}} / E_C \gg 1$. In particular, we will ignore the charge noise from $n_{\text{g}}$ since the dependence of the energy levels on $n_{\text{g}}$ is exponentially suppressed in the transmon regime. However, do note that $n_{\text{g}}$ contains two components: One is the slow-varying classical charge noise (which can be any real number, without any quantization). The other one is the coherent drive either coming from the classical source or a quantized resonator (see Eq.(\ref{eq:hamiltonian_res_driven_LC})). We will assume that the charge noise has a much smaller magnitude compared to a coherent drive; in addition, since we will first study the isolated transmon, coupling to the drive is also turned off.

To solve the energy levels of a transmon in the perturbative limit, we first Taylor expand the Josephson energy to the fourth order. In the flux particle picture, what we are assuming is that the particle has a very large mass (because of the smallness of $E_C$) so that its flux coordinate is near one minimum of the cosine potential; thus, the potential can be expanded to the fourth order to good accuracy. As shown before, the second-order term acts as the quadratic inductive energy of a linear oscillator; hence,
\begin{equation}
    \hat{H}_{\text{T}} 
    \approx \left(4E_C \hat{n}^2 
        + \frac{1}{2} E_{\text{J}} \hat{\delta}^2\right)
        - \frac{1}{4!} E_{\text{J}} \hat{\delta}^4.
\end{equation}
Next, recall that the reduced flux operator can be expressed as (see Table \ref{table:normal_mode_expansion})
\begin{equation}
     \hat{\delta} 
     = \left(
            \frac{2E_C}{E_{\text{J}} }
        \right)^{\! 1/4} 
        \Big(
            \hat{a}_{\text{q}}^{\dagger} 
            + \hat{a}_{\text{q}}
        \Big),
\end{equation}
where we have used $\hat{a}_{\text{q}}$ to denote the annihilation operator of the harmonic part of the transmon. Thus, the Hamiltonian, to the fourth order, can be rewritten as
\begin{equation}
    \hat{H}_{\text{T}}  
    \approx 
    \sqrt{8E_C E_{\text{J}}} 
        \hat{a}_{\text{q}}^{\dagger} \hat{a}_{\text{q}}
        - \frac{E_C}{12}
        \Big(
            \hat{a}_{\text{q}}^{\dagger} 
            + \hat{a}_{\text{q}}
        \Big)^4.
\end{equation}
Next, we expand $\Bigsl(\hat{a}_{\text{q}}^{\dagger} + \hat{a}_{\text{q}} \Bigsr)^4$ with the help of the canonical commutation relation $\Bigsl[\hat{a}_{\text{q}}, \hat{a}_{\text{q}}^{\dagger} \Bigsr] = 1$:
\begin{align}
    \Big(
        \hat{a}_{\text{q}}^{\dagger} 
            + \hat{a}_{\text{q}}
        \Big)^4
    &= \Big(
            \hat{a}_{\text{q}}^{\dagger} \hat{a}_{\text{q}}^{\dagger} 
            + 2 \hat{a}_{\text{q}}^{\dagger} \hat{a}_{\text{q}}
            + \hat{a}_{\text{q}} \hat{a}_{\text{q}}
            + 1
        \Big)^2
\nonumber \\ \label{eq:RWA_in_transmon_hamiltonian}
    &\approx 
        4 \hat{a}_{\text{q}}^{\dagger} \hat{a}_{\text{q}}
        + \hat{a}_{\text{q}}^{\dagger}\hat{a}_{\text{q}}^{\dagger} 
            \hat{a}_{\text{q}} \hat{a}_{\text{q}}
        + 4 \hat{a}_{\text{q}}^{\dagger} \hat{a}_{\text{q}} 
            \hat{a}_{\text{q}}^{\dagger} \hat{a}_{\text{q}}
        + \hat{a}_{\text{q}} \hat{a}_{\text{q}} 
            \hat{a}_{\text{q}}^{\dagger}\hat{a}_{\text{q}}^{\dagger}
        + 1
\\ \label{eq:fourth_order_transmon_hamiltonian}
    &= 6 \hat{a}_{\text{q}}^{\dagger} \hat{a}_{\text{q}}^{\dagger} 
            \hat{a}_{\text{q}} \hat{a}_{\text{q}}
        + 12 \hat{a}_{\text{q}}^{\dagger} \hat{a}_{\text{q}}
        + 3,
\end{align}
where we have ignored all terms of the form $\hat{a}_{\text{q}}^{\dagger m} \hat{a}_{\text{q}}^n$ with $m \neq n$ in Eq.(\ref{eq:RWA_in_transmon_hamiltonian}). Such RWA is valid because in the interaction picture of the transmon, 
\begin{equation}
    \hat{a}_{\text{q}}^{\dagger m} \hat{a}_{\text{q}}^n
    \ \ \longrightarrow \ \ 
    \hat{a}_{\text{q}}^{\dagger m} \hat{a}_{\text{q}}^n e^{\ci (m-n) \omega_\text{q} t}
\end{equation}
will oscillate very fast unless $m = n$. Finally, substituting Eq.(\ref{eq:fourth_order_transmon_hamiltonian}) into the Hamiltonian yields
\begin{align}    
    \hat{H}_{\text{T}}  
    &= \sqrt{8E_C E_{\text{J}}} 
        \hat{a}_{\text{q}}^{\dagger} \hat{a}_{\text{q}}
        - \frac{E_C}{12}
            \Big( 
                6 \hat{a}_{\text{q}}^{\dagger} \hat{a}_{\text{q}}^{\dagger} 
                    \hat{a}_{\text{q}} \hat{a}_{\text{q}}
                + 12 \hat{a}_{\text{q}}^{\dagger} \hat{a}_{\text{q}}
            \Big)
\nonumber \\
    &= \left(
            \hbar \omega_{\text{J}} 
            - E_C
        \right) 
        \hat{a}_{\text{q}}^{\dagger} \hat{a}_{\text{q}}
        - \frac{E_C}{2}
            \hat{a}_{\text{q}}^{\dagger} \hat{a}_{\text{q}}^{\dagger} 
            \hat{a}_{\text{q}} \hat{a}_{\text{q}}
\end{align}
with the constant energy shift omitted. Since we are doing perturbation theory, the energy eigenstates to the first order are still given by those of the QHO, i.e., the Fock states (do not confuse the number states of the QHO with the charge number states which are the eigenstates of the charge operator only).

Finally, we define 
\begin{equation}
    \omega_{\text{q}} = \omega_{\text{J}} - E_C / \hbar
\end{equation}
as the transmon qubit frequency and 
\begin{equation} \label{eq:anharmonicity_transmon}
    \alpha_{\text{q}}  = - E_C / \hbar
\end{equation}
as the anharmonicity of the nonlinear oscillator. On the one hand, $\hbar \omega_{\text{q}}$ represents the transition energy between the ground state and the first excited state,
\begin{equation}
    \hbar \omega_{\text{q}} 
    = \bra{1} \hat{H}_{\text{T}} \ket{1} 
        -\bra{0} \hat{H}_{\text{T}} \ket{0}.
\end{equation}
On the other hand, $\hbar \alpha_{\text{q}} $ measures how much the nonlinearity alters the first two energy spacings, i.e.,
\begin{equation} \label{eq:anharmonicity_definition}
    \hbar \alpha_{\text{q}}  
    = \Big(
            \bra{2} \hat{H}_{\text{T}} \ket{2} 
            - \bra{1} \hat{H}_{\text{T}} \ket{1}
        \Big)
        - \Big(
            \bra{1} \hat{H}_{\text{T}} \ket{1} 
            - \bra{0} \hat{H}_{\text{T}} \ket{0}
            \Big).
\end{equation}
In fact, Eq.(\ref{eq:anharmonicity_definition}) is the definition of anharmonicity in general whereas Eq.(\ref{eq:anharmonicity_transmon}) is true only in the transmon limit (i.e., $E_{\text{J}}/E_C \gg 1$). By solving the transmon Hamiltonian analytically, one often finds a $10 \sim 20$-MHz difference between $E_C/\hbar$ and the actual $|\alpha|$. In summary, the transmon Hamiltonian, to the fourth order, is given by\footnote{Note that the perturbation theory ignores the fact that the potential has a finite depth $E_{\text{J}}$; hence, at high enough energy, the eigenstates are no longer bound states. For a full solution of the bound and scattering states, we would need to solve the Mathieu equation. See Figure \ref{fig:mathieu_solutions}.}
\begin{equation}
    \hat{H}_{\text{T}} 
    = \hbar \omega_{\text{q}} \hat{a}_{\text{q}}^{\dagger} \hat{a}_{\text{q}}
        + \frac{\hbar \alpha}{2}    
            \hat{a}_{\text{q}}^{\dagger} 
            \hat{a}_{\text{q}}^{\dagger} 
            \hat{a}_{\text{q}}
            \hat{a}_{\text{q}}.
\end{equation}

We, therefore, observe the tradeoff in a transmon design: To suppress the charge noise, we would like $E_C$ to be as small as possible. However, the transition frequencies between neighboring energy levels should be sufficiently different to avoid undesired excitations, thus, limiting how small $|\alpha_{\text{q}}|$ can be. In practice, $|\alpha_{\text{q}}/ 2\pi|$ is usually between 150 and 300 MHz and $E_{\text{J}} / h$ is between 10 and 20 GHz.

\subsection{A Transmon Qubit Coupled to a Resonator}
To couple a transmon to a resonator, we directly invoke Eq.(\ref{eq:circuit_level_dipole_interaction_derivation}), which is the Hamiltonian of two coupled linear oscillators. The procedure for adding the nonlinearity to one of the oscillators is unchanged; hence, the classical Hamiltonian now takes the form
\begin{align}
    \mathcal{H}(\Phi, Q, \Phi_{\text{r}}, Q_{\text{r}})
    &= \frac{1}{2C} 
            \left( 
                Q + \frac{C_g}{C_{\text{r}}} Q_{\text{r}}
            \right)^2 
        + E_{\text{J}} 
            \left[ 
                1 - \cos(\frac{\Phi}{\phi_0})
            \right]
        + \frac{1}{2C_{\text{r}}} Q_{\text{r}}^2 
        + \frac{1}{2L_{\text{r}}} \Phi_{\text{r}}^2
\nonumber \\
    &= 4 E_{C} n^2
        - E_{\text{J}}
            \Big(
                1 - \cos \delta
            \Big)
        + \frac{1}{2}
            \hbar \omega_{\text{r}} 
            \Big( 
                a_{\text{r}}^{*} a_{\text{r}} 
                + a_{\text{r}} a_{\text{r}}^{*}
            \Big)
        + 4e^2 \frac{C_g}{C_1 C_2} n n_{\text{r}},
\end{align}
where $\omega_{\text{r}}$ is the resonator frequency, $a_{\text{r}}$ the resonator mode amplitude, and $n_{\text{r}}$ the resonator reduced charge as defined in Chapter 2. 

Canonical quantization also follows from the isolated case, but with a coupling between the charge operators of the transmon and the resonator added,
\begin{align}
    \hat{H} 
    &= 4 E_{C} \hat{n}^2
        - E_{\text{J}}
            \Big(
                1 - \cos \hat{\delta}
            \Big)
        + \hbar \omega_{\text{r}} 
            \left(
                \hat{a}_{\text{r}}^{\dagger} \hat{a}_{\text{r}} 
                + \frac{1}{2}
            \right)
        + 4e^2 \frac{C_g}{C C_{\text{r}}} 
            \hat{n} \hat{n}_{\text{r}}
\nonumber \\
    &= 4 E_{C} \hat{n}^2
        - E_{\text{J}}
            \Big(
                1 - \cos \hat{\delta}
            \Big)
        + \hbar \omega_{\text{r}} 
            \left(
                \hat{a}_{\text{r}}^{\dagger} \hat{a}_{\text{r}} 
                + \frac{1}{2}
            \right)
        + \hbar g_{10} 
            \Big(
                \hat{a}_{\text{q}} 
                - \hat{a}_{\text{q}}^{\dagger}
            \Big)
            \Big(
                \hat{a}_{\text{r}} 
                - \hat{a}_{\text{r}}^{\dagger}
            \Big).
\end{align}
The coupling coefficient is given by \cite{RevModPhys.93.025005}
\begin{equation}
    g_{10} = - \omega_{\text{r}} 
        \left( 
            \frac{C_g}{C} 
        \right)
        \left(
            \frac{E_{\text{J}}}{2E_C}
        \right)^{\!\! 1/4}
        \sqrt{\frac{\pi Z_{\text{r}0}}{R_{\text{K}}}},
\end{equation}
where $Z_{\text{r}0} = \sqrt{L_{\text{r}} / C_{\text{r}}}$ is the resonator characteristic impedance and $R_{\text{K}} = h/e^2$ is the resistance quantum (von Klitzing constant). At this point, all the formalism introduced in Chapter 3 can be applied. For example, in the weakly anharmonic limit and under the RWA, the Hamiltonian simplifies to
\begin{align}
    \hat{H} 
    &= \hbar \omega_{\text{q}} 
            \hat{a}_{\text{q}}^{\dagger} 
            \hat{a}_{\text{q}}
        + \frac{\hbar \alpha}{2}    
            \hat{a}_{\text{q}}^{\dagger} 
            \hat{a}_{\text{q}}^{\dagger} 
            \hat{a}_{\text{q}}
            \hat{a}_{\text{q}}
        + \hbar \omega_{\text{r}} 
            \left(
                \hat{a}_{\text{r}}^{\dagger} \hat{a}_{\text{r}} 
                + \frac{1}{2}
            \right)
        - \hbar g_{10} 
            \Big(
                \hat{a}_{\text{r}}^{\dagger}
                    \hat{a}_{\text{q}}^{\dagger}
                - \hat{a}_{\text{r}}^{\dagger}
                    \hat{a}_{\text{q}}
            \Big)
\nonumber \\
    &= \sum_{j = 1}^{D-1}
            \hbar 
            \left[
                j \omega_{\text{q}} 
                + \frac{j (j-1)\alpha_{\text{q}}}{2} 
            \right]
            \ket{j} \! \bra{j}
        + \hbar \omega_{\text{r}} 
            \left(
                \hat{a}_{\text{r}}^{\dagger} \hat{a}_{\text{r}} 
                + \frac{1}{2}
            \right)
\nonumber \\[-2mm]
    & \ \ \ \ \ \ \ \ \ \ \ \ \ \ \ \ \ \ \ \ \ \ \ \ 
        - \sum_{j=0}^{D-2} 
            \hbar g
            \sqrt{j+1}
            \Big( 
                \hat{a}_{\text{r}} \ket{j+1} \! \bra{j}
                + \hat{a}_{\text{r}}^{\dagger} \ket{j} \! \bra{j+1}
            \Big).
\end{align}
Subsequently, one can show, based on Eq.(\ref{eq:general_chi_shifts}) and (\ref{eq:general_lambda_shifts}), that the Lamb and dispersive shifts on the resonator are given by 
\begin{equation}
    \Lambda_j 
    = \sum_{k=0}^{D-1} \chi_{jk}
    \approx \chi_{j,j-1}
    = \frac{jg_{10}^2}{\Delta_{\text{qr}} + (j-1) \alpha_{\text{q}}}
    \ \text{ for } \ j=1,2,...
\end{equation}
with $\Lambda_0 \approx 0$ and 
\begin{align}
    \chi_j
    \doteq 
        \left[
            \sum_{k=0}^{D-1} 
                \hbar (\chi_{jk} - \chi_{kj}) 
        \right]
    \approx 
        \left[
                \hbar (\chi_{j,j-1} - \chi_{j+1,j}) 
        \right]
    = \frac{jg_{10}^2}{\Delta_{\text{qr}} (j-1) \alpha_{\text{q}}} - \frac{(j+1)g_{10}^2}{\Delta_{\text{qr}} + j\alpha_{\text{q}}}
    \ \text{ for } \ j=1,2,...
\end{align}
with $\chi_0 = - g_{10}^2/\Delta_{\text{qr}}$. In other words, the qubit-resonator coupling in the dispersive regime is
\begin{equation}\label{eq:transmon_resonator_dispersive_hamiltonian}
    \hat{H}^{\text{disp}} 
    = \sum_{j = 1}^{D-1}
            \hbar 
            \left[
                j \omega_{\text{q}} 
                + \frac{j (j-1) \alpha_{\text{q}}}{2} 
                + \Lambda_j 
                + \chi_j 
                    \hat{a}_{\text{r}}^{\dagger} 
                    \hat{a}_{\text{r}} 
            \right] 
            \ket{j} \! \bra{j}
        + \hbar \omega_{\text{r}} 
                \hat{a}_{\text{r}}^{\dagger} \hat{a}_{\text{r}} .
\end{equation}
Note that the dispersive shifts are also commonly called the cross-Kerr coefficients between the qubit and resonator modes. Related to the cross-Kerr is the self-Kerr coefficient of each mode, defined to be twice the anharmonicity of the mode. For the composite system of qubit plus resonator, the self-Kerr is $2 \alpha_{\text{q}}$ for the qubit mode while approximately zero for the resonator. In general, self-Kerr represents the location of nonlinearity in a system. The difference between the self-Kerr coefficients of the qubit and resonator modes implies that the nonlinearity of the JJ is mainly concentrated in the qubit mode. Of course, Eq.(\ref{eq:transmon_resonator_dispersive_hamiltonian}) is derived in the dispersive regime where the qubit and resonator are only weakly coupled. In the case of strong and resonant coupling, the resonator will also share some of the nonlinearity of the JJ. In fact, building upon this idea of shared nonlinearity is the method of \textbf{energy participation ratio (EPR)} \cite{s41534_021_00461_8}, which finds the qubit-resonator Hamiltonian by computing how the total Josephson energy is distributed among the modes of the system.

\begin{figure}
    \centering
    \includegraphics[scale=0.31]{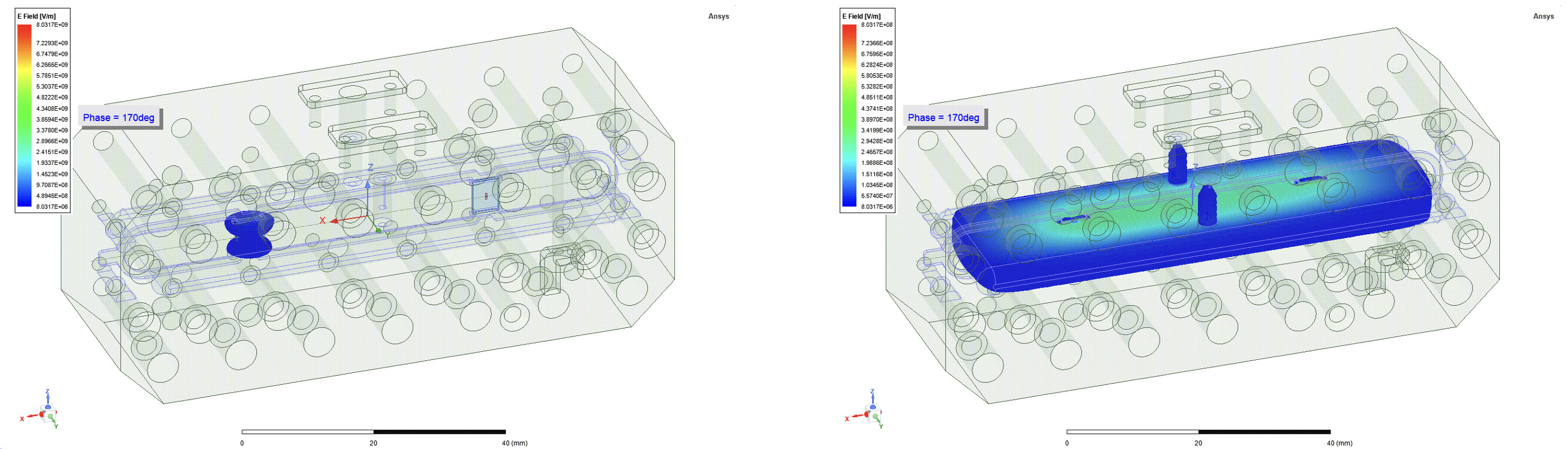}
    \caption{Eigenmode simulation of a linearized qubit hosted in a readout cavity. Left: the mode of a linearized transmon with the JJ replaced by a linear lumped inductor. Right: the fundamental mode of the cavity.}
    \label{fig:transmon_cavity_coupling}
\end{figure}

One can also ask to what extent is the circuit-level dipole interaction equivalent to the physical dipole interaction discussed in Chapter 2 and 3. To find the effective dipole and electric field operators in the cQED formalism, we use the basic relations among the charge, voltage, and electric field:
\begin{equation}
    \hat{H}_{\text{int}}
    = 4e^2 \frac{C_g}{C C_{\text{r}}} 
            \hat{n} \hat{n}_{\text{r}}
    = \frac{C_g}{C C_{\text{r}}} \hat{Q} \hat{Q}_{\text{r}}
    = - \underbrace{\Bigsl( \hat{Q} \ell \Bigsr) }_{\hat{d}_{\text{eff}}}
        \underbrace{\left( \frac{C_g}{C} \right) \!
        \left( - \frac{\hat{V}_{\text{r}}}{\ell} \right)}_{\hat{E}_{\text{eff}}}.
\end{equation}
The length $\ell$ represents the size of the dipole antenna formed by the two pads of the transmon. The charge operator $\hat{Q}$ of the qubit and the voltage operator $\hat{V}_{\text{r}}$ of the resonator are defined by Eq.(\ref{eq:res_charge_operator}) and (\ref{eq:res_voltage_operator}), respectively, which leads to the effective dipole and electric field operators
\begin{equation}
    \hat{d}_{\text{eff}} 
    = - \ci d_{0,\text{eff}}
        \Big(
            \hat{a}_{\text{q}} - \hat{a}_{\text{q}}^{\dagger}
        \Big)
    \ \ \text{ with } \ \ 
     d_{0,\text{eff}} 
     = 2e \ell 
        \left(
            \frac{E_{\text{J}}}{32E_C}
        \right)^{\!\! 1/4}
\end{equation}
and
\begin{equation}
    \hat{E}_{\text{eff}}
    = \ci \mathscr{E}_{0, \text{eff}}
        \Big(
            \hat{a}_{\text{r}} - \hat{a}_{\text{r}}^{\dagger}
        \Big)
    \ \ \text{ with } \ \ 
    \mathscr{E}_{0, \text{eff}}
    = \frac{C_g}{C} \frac{\sqrt{\hbar \omega_{\text{r}} / 2C_{\text{r}}}}{\ell},
\end{equation}
respectively. Figure \ref{fig:transmon_cavity_coupling} illustrates how a qubit is coupled to a 3D cavity. In practice, for distributed structures (e.g., a 3D cavity), it's better to model the qubit-resonator interaction using numerical solvers. For example, an eigenmode solver combined with the method of EPR allows one to first solve for the linearized mode in the classical simulation (e.g., Ansys HFSS) and then compute the Lamb-shifted qubit and resonator frequencies along with the Kerr coefficients.

\subsection{Multimon}
\begin{figure}
    \centering
    \includegraphics[scale=0.32]{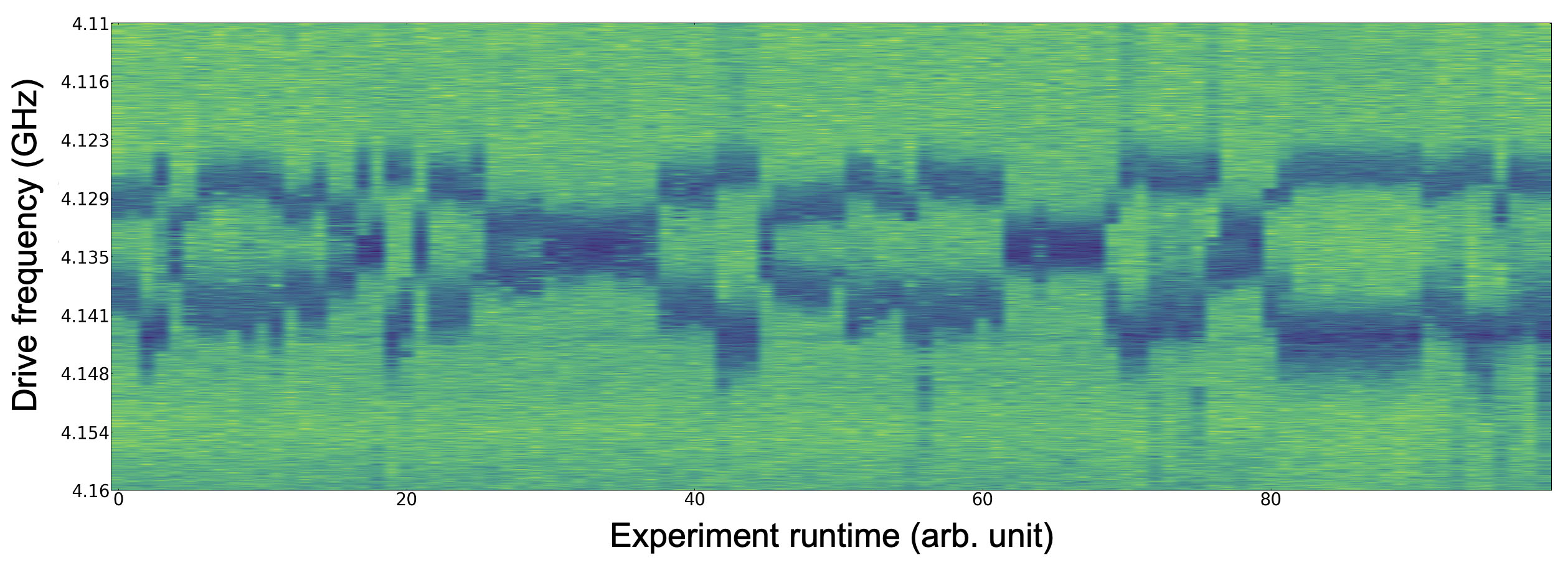}
    \caption{Monitoring of charge noise and parity switching in an hour. The transition frequency between $\ket{1} = \ket{e}$ and $\ket{2}= \ket{f}$ is probed for a long time to observe the effect of charge noise (the continuous change of the transition frequency) and quasiparticle tunneling (the discrete jumps due to parity switching).}
    \label{fig:charge_noise_monitoring}
\end{figure}

Although transmon acts as an anharmonic oscillator, it has a finite number of bound states since the Josephson energy is finite. More importantly, as implied by Eq.(\ref{eq:charge_noise_WKB}), the energy dispersion increases as we move to higher energy eigenstates, so $n_{\text{g}}$ can no longer be ignored. In reality, the charge noise, resulting from the break of Cooper pairs or some other quasiparticle tunnelings, can be observed experimentally \cite{PhysRevLett.114.010501, PRXQuantum.3.030307} by observing a beat note in the Ramsey interference or by monitoring the change of transition frequency directly over a long time as shown in Figure \ref{fig:charge_noise_monitoring}. The alternative approach to expand the computational basis relies on increasing the number of anharmonic oscillators within a small footprint \cite{PhysRevApplied.7.054025}. Instead of having one anharmonic oscillator with a single JJ, we can construct a more complicated capacitor network with multiple nonlinear inductors added so that the linearized network supports more than one mode. Then, we can restrict the computational space to be the tensor product of several two-dimensional manifolds each derived from an anharmonic oscillator in the network.

In order to design a \textbf{multimon}, i.e., a multimode quantum network, we start with a generic $N$-node network of capacitors, linear inductors, and JJs. We will use $\Phi_{i}$ and $\dot{\Phi}_{i}$ to denote the flux and voltage, respectively, at the $i$th node (for $i = 1 , ... , N$). Assume that there is no inductive element connected to the ground directly, but there is always a finite self-capacitance between any node and the ground. First, we consider only the linear part of the circuit (i.e., keep only the quadratic energy in $U_{\text{J}}(\delta)$) and set the total series inductance between node $i$ and $j$ to be
\begin{equation}
    L_{ij} = L_{0,ij} + L_{\text{J},ij},
\end{equation}
where $L_{0,ij}$ and $L_{\text{J},ij}$ are the linear and Josephson inductance, respectively. Similar to the transmon case, we define the inductive energy between node $i$ and $j$ (for $i \neq j$) as
\begin{equation}
    E_{L_{ij}} 
    = \frac{\phi_0^2}{L_{ij}}
    = \frac{\phi_0^2}{L_{0,ij} + L_{\text{J},ij}}
    = \left(\frac{1}{E_{L_{0,ij}}} + \frac{1}{E_{\text{J},ij}}\right)^{-1}.
\end{equation}
In addition, we set $E_{L_{ii}} = 0$ (since there is no inductance to the ground).

Restricted to the linear part of the circuit, we can describe its capacitive/kinetic energy as
\begin{equation}
        \displaystyle T_{\text{cap}} 
        = \frac{1}{2} 
            \begin{pmatrix}
                \dot{\Phi}_1 & \cdots & \dot{\Phi}_N
            \end{pmatrix}
            \begin{pmatrix}
                \sum_j C_{1j} & -C_{12} & \cdots & -C_{1N} \\[-1mm]
                -C_{21} & \sum_j C_{2j} & \cdots & -C_{2N} \\[-1mm]
                \vdots & \vdots & \ddots & \vdots \\[-1mm]
                -C_{N1} & -C_{N2} & \cdots & \sum_j C_{Nj} 
            \end{pmatrix}
            \begin{pmatrix}
                \dot{\Phi}_1 \\ 
                \vdots \\ 
                \dot{\Phi}_N
            \end{pmatrix}
        = \frac{1}{2} \dot{\boldsymbol{\Phi}}^{\top} \mathsf{C} \dot{\boldsymbol{\Phi}}
    \end{equation}
and its inductive/potential energy as
    \begin{align}
        U_{\text{ind}} 
        &= \frac{1}{2} 
            \sum_{i} 
            \sum_{j<i} 
                E_{L_{ij}} 
                \left(\frac{\Phi_i - \Phi_j}{\phi_0}\right)^2
        = \frac{1}{4} 
            \sum_{i,j} 
                E_{L_{ij}} 
                \left(\frac{\Phi_i - \Phi_j}{\phi_0}\right)^2
    \nonumber \\[-1mm]
        &= \frac{1}{2 \phi_0^2} 
            \begin{pmatrix}
                \Phi_1 & \cdots & \Phi_N
            \end{pmatrix}
            \begin{pmatrix}
                \sum_j E_{L_{1j}} & -E_{L_{12}} & \cdots & -E_{L_{1N}} \\[-1mm]
                -E_{L_{21}} & \sum_j E_{L_{2j}} & \cdots & -E_{L_{2N}} \\[-1mm]
                \vdots & \vdots & \ddots & \vdots \\[-1mm]
                -E_{L_{N1}} & -E_{L_{N2}} & \cdots & \sum_j E_{L_{Nj}}
            \end{pmatrix}
            \begin{pmatrix}
                \Phi_1 \\ 
                \vdots \\ 
                \Phi_N
            \end{pmatrix}
        = \frac{1}{2 \phi_0^2} \boldsymbol{\Phi}^{\top} 
                \mathsf{E}_{\mathsf{L}} 
                \boldsymbol{\Phi},
    \end{align}
where 
\begin{equation}
    \mathsf{C} 
    = \begin{pmatrix}
                \sum_j C_{1j} & -C_{12} & \cdots & -C_{1N} \\[-1mm]
                -C_{21} & \sum_j C_{2j} & \cdots & -C_{2N} \\[-1mm]
                \vdots & \vdots & \ddots & \vdots \\[-1mm]
                -C_{N1} & -C_{N2} & \cdots & \sum_j C_{Nj} 
            \end{pmatrix}
\end{equation}
is the Maxwell capacitance matrix and 
\begin{equation}
    \mathsf{E}_{\mathsf{L}} 
    = \begin{pmatrix}
                \sum_j E_{L_{1j}} & -E_{L_{12}} & \cdots & -E_{L_{1N}} \\[-1mm]
                -E_{L_{21}} & \sum_j E_{L_{2j}} & \cdots & -E_{L_{2N}} \\[-1mm]
                \vdots & \vdots & \ddots & \vdots \\[-1mm]
                -E_{L_{N1}} & -E_{L_{N2}} & \cdots & \sum_j E_{L_{Nj}}
            \end{pmatrix}
\end{equation}
the inductive energy matrix. Moreover, by introducing the inductance matrix $\mathsf{L}$ with
\begin{equation}
    \mathsf{L}^{-1} = \frac{1}{\phi_0^2} \mathsf{E}_{\mathsf{L}}
    = \begin{pmatrix}
                \sum_j L_{1j}^{-1} & -L_{12}^{-1} & \cdots & -L_{1N}^{-1} \\
                -L_{21}^{-1} & \sum_j L_{2j}^{-1} & \cdots & -L_{2N}^{-1} \\
                \vdots & \vdots & \ddots & \vdots \\
                -L_{N1}^{-1} & -L_{N2}^{-1} & \cdots & \sum_j L_{Nj}^{-1} 
            \end{pmatrix},
\end{equation}
the linear part of the Lagrangian can be cast in the more familiar form
\begin{equation}
    \mathcal{L}_{\text{lin}} 
            (\boldsymbol{\Phi}, \dot{\boldsymbol{\Phi}})
    = T_{\text{cap}}(\dot{\boldsymbol{\Phi}}) 
        - U_{\text{ind}}(\boldsymbol{\Phi})
    = \frac{1}{2} 
            \dot{\boldsymbol{\Phi}}^{\top} 
            \mathsf{C} 
            \dot{\boldsymbol{\Phi}}
        - \frac{1}{2} 
            \boldsymbol{\Phi}^{\top} 
            \mathsf{L}^{-1} 
            \boldsymbol{\Phi}.        
\end{equation}
We explicitly show the units of the capacitance and inductance matrices by introducing two identity matrices, $\mathsf{1}_{\text{cap}}$ and $\mathsf{1}_{\text{ind}}$, (or simply two scalars $1_{\text{cap}}$ and $1_{\text{ind}}$) with the unit of a capacitor (e.g., Farad) and of an inductor (e.g., Henry), respectively. Consequently, we have $\mathsf{C} = \mathsf{1}_{\text{cap}} \Bar{\mathsf{C}}$ and $\mathsf{L} = \mathsf{1}_{\text{ind}} \Bar{\mathsf{L}}$ for some unitless matrices $\Bar{\mathsf{C}}$ and $\Bar{\mathsf{L}}$; thus, 
\begin{equation}
    \mathcal{L}_{\text{lin}} 
            (\boldsymbol{\Phi}, \dot{\boldsymbol{\Phi}})
    = \frac{\mathsf{1}_{\text{cap}}}{2} 
            \dot{\boldsymbol{\Phi}}^{\top} 
            \Bar{\mathsf{C}} 
            \dot{\boldsymbol{\Phi}}
        - \frac{\mathsf{1}_{\text{ind}}^{-1}}{2} 
            \boldsymbol{\Phi}^{\top} 
            \Bar{\mathsf{L}}^{-1} 
            \boldsymbol{\Phi},   
\end{equation}
To find the linearized modes of the circuit, we need to diagonalize the Lagrangian:
\begin{align}
    &\mathcal{L}_{\text{lin}} 
        (\boldsymbol{\Phi}, \dot{\boldsymbol{\Phi}})
    = \frac{\mathsf{1}_{\text{cap}}}{2} 
            \dot{\boldsymbol{\Phi}}^{\top} 
            \Bar{\mathsf{C}} 
            \dot{\boldsymbol{\Phi}}
        - \frac{\mathsf{1}_{\text{ind}}^{-1}}{2} 
            \boldsymbol{\Phi}^{\top} 
            \Bar{\mathsf{L}}^{-1} 
            \boldsymbol{\Phi} 
\nonumber \\
    &= \frac{\mathsf{1}_{\text{cap}}}{2} 
        \dot{\boldsymbol{\Phi}}^{\top} 
        \mathsf{B} \mathsf{B}^{\top} 
        \dot{\boldsymbol{\Phi}} 
    - \frac{\mathsf{1}_{\text{ind}}^{-1}}{2} 
        \boldsymbol{\Phi}^{\top}
        \mathsf{B} 
        \mathsf{B}^{-1} 
        \bar{\mathsf{L}}^{-1} 
        \big(\mathsf{B}^{\top}\big)^{-1} 
        \mathsf{B}^{\top} 
        \boldsymbol{\Phi}
    &
    (\text{\scriptsize{Let $\bar{\mathsf{C}} = \mathsf{V} \mathsf{\Lambda} \mathsf{V}^{\top}$ and define $\mathsf{B} = \mathsf{V} \mathsf{\Lambda}^{1/2}$.}})
\nonumber \\
    &= \frac{\mathsf{1}_{\text{cap}}}{2} 
        \dot{\mathbf{X}}^{\top} 
        \dot{\mathbf{X}} 
    - \frac{\mathsf{1}_{\text{ind}}^{-1}}{2} 
        \mathbf{X} 
        \mathsf{J}^{-1} 
        \mathbf{X}
    &
    (\text{\scriptsize{Define $\mathbf{X} = \mathsf{B}^{\top} \boldsymbol{\Phi}$, $\Tilde{\mathsf{J}} = \mathsf{B}^{-1} \Bar{\mathsf{L}} \big(\mathsf{B}^{\top}\big)^{-1}$.}})
\nonumber \\
    &= \frac{\mathsf{1}_{\text{cap}}}{2} 
            \dot{\mathbf{X}}^{\top} 
            \tilde{\mathsf{V}} 
            \tilde{\mathsf{V}}^{\top} 
            \dot{\mathbf{X}} 
        - \frac{\mathsf{1}_{\text{ind}}^{-1}}{2} 
            \mathbf{X} 
            \tilde{\mathsf{V}} 
            \Tilde{\mathsf{L}}
            \tilde{\mathsf{V}}^{\top} 
            \mathbf{X}
    &
    (\text{\scriptsize{Let $\mathsf{J} = \tilde{\mathsf{V}} \tilde{\mathsf{L}} \tilde{\mathsf{V}}^{\top}$ and use $\tilde{\mathsf{V}} \tilde{\mathsf{V}}^{\top} = 1$.}})
\nonumber \\
    &= \frac{\mathsf{1}_{\text{cap}}}{2} 
        \dot{\tilde{\boldsymbol{\Phi}}}^{\top} 
        \dot{\tilde{\boldsymbol{\Phi}}}
    - \frac{\mathsf{1}_{\text{ind}}^{-1}}{2} 
        \tilde{\boldsymbol{\Phi}}
        \Bar{\mathsf{\Omega}}^2
        \tilde{\boldsymbol{\Phi}}
    &
    (\text{\scriptsize{Define $\tilde{\boldsymbol{\Phi}} = \tilde{\mathsf{V}}^{\top} \mathbf{X}$ and $\Bar{\mathsf{\Omega}} = \Tilde{\mathsf{L}}^{1/2}$.}})
\end{align}
Hence, by defining the new coordinates\footnote{We are numbering the new coordinates from 0 to signify the fact that the lowest mode is always the static mode in most multimon designs (e.g., a ring multimon). Hence, effectively there are only $N-1$ modes.} $\Tilde{\boldsymbol{\Phi}} = (\Tilde{\Phi}_{0}, \cdots \Tilde{\Phi}_{N-1})$, we have converted an LC network into $N$ independent normal modes (also known as the eigenmodes).

If we want to write a script to diagonalize the Lagrangian numerically, it's more straightforward to absorb the unit into the definition of the flux vector, i.e., defining
\begin{equation}
    \tilde{\mathbf{X}} 
    = \mathsf{1}_{\text{cap}}^{1/2} \tilde{\boldsymbol{\Phi}}
    \ \ \Big[\text{Time}\sqrt{\text{Energy}}\Big],
\end{equation}
so that the decoupled Lagrangian is of the form
\begin{equation}
    \mathcal{L}_{\text{lin}} 
    = \frac{1}{2} 
            \dot{\tilde{\mathbf{X}}}^{\top} 
            \dot{\tilde{\mathbf{X}}}
        - \frac{1}{2} 
            \tilde{\mathbf{X}}
            \Big(
                \mathsf{1}_{\text{ind}}^{-1} 
                \mathsf{1}_{\text{cap}}^{-1} 
                \Bar{\mathsf{\Omega}}^2 
            \Big)
            \tilde{\mathbf{X}}
    = \frac{1}{2} 
            \dot{\tilde{\mathbf{X}}}^{\top} 
            \dot{\tilde{\mathbf{X}}}
        - \frac{1}{2} 
            \tilde{\mathbf{X}}
            \mathsf{\Omega}^2 
            \tilde{\mathbf{X}}
    = \sum_{\mu=0}^{N-1} 
        \frac{1}{2} \left(
            \dot{\tilde{X}}_{\mu}^2 
            - \omega_{\mu}^2 
                \tilde{X}_{\mu}^2
        \right),
\end{equation}
where 
\begin{equation} \label{eq:multimon_angular_frequency}
    \mathsf{\Omega}
    = \mathsf{1}_{\text{ind}}^{-1/2} 
        \mathsf{1}_{\text{cap}}^{-1/2} 
        \Bar{\mathsf{\Omega}}
    = \text{diag}(\omega_0, ... , \omega_{N-1})
\end{equation}
is a diagonal matrix of angular frequencies, revealing the eigenfrequencies of the linearized circuit. 

\begin{figure}
    \centering
    \includegraphics[scale=0.35]{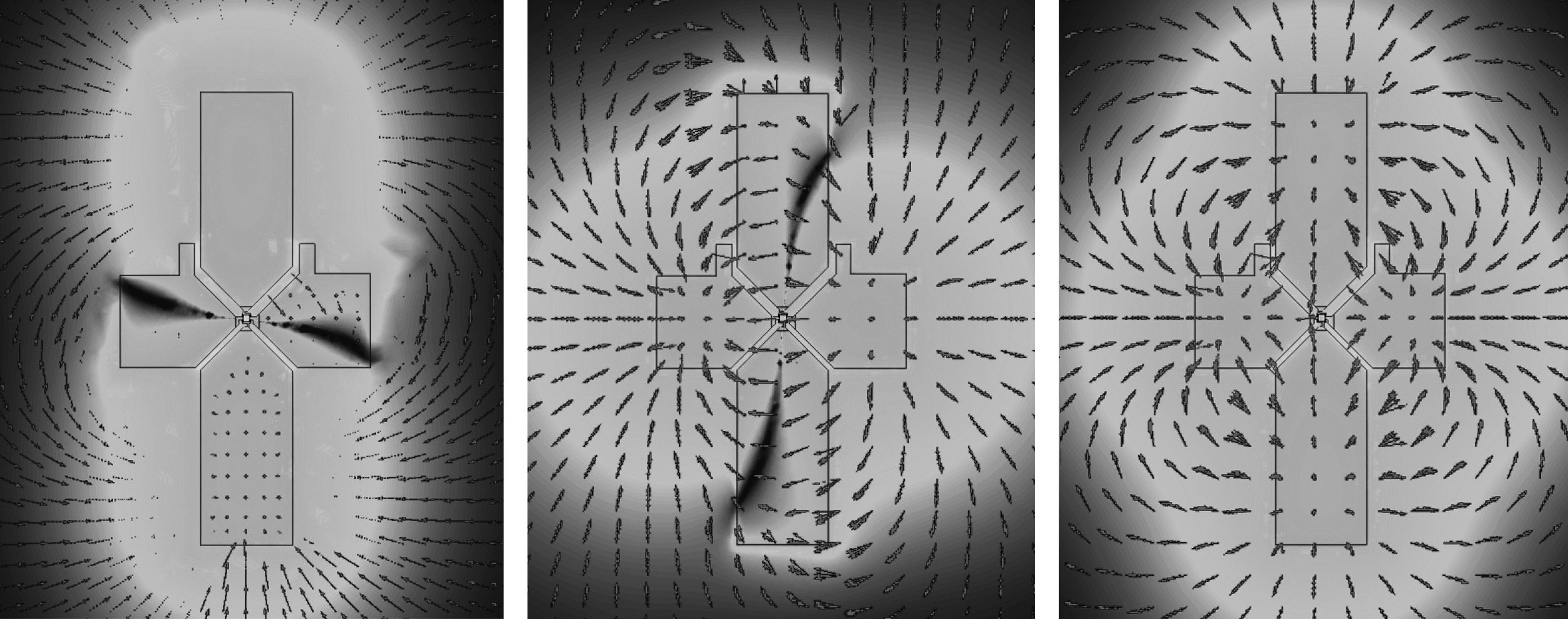}
    \caption{Eigenmode simulation of the trimon modes. Left: a vertical dipole mode. Middle: a horizontal dipole mode. Right: a quadrupole mode. Due to the small asymmetry of the four Josephson inductances placed in the design, the two dipole modes exhibit tilted lines of vanishing electric fields.}
    \label{fig:trimon_modes}
\end{figure}

In practice, we will first use electromagnetic solvers (e.g., Ansys Q3D) to find the capacitance matrix of the network without the JJ and then create the inductance matrix manually (which ignores the physical size of the JJs). Subsequently, the two matrices are fed into the algorithm described above, which is nothing more than a sequence of matrix multiplications and diagonalizations. Alternatively, one can also directly use the eigenmode solvers to find the linear modes in a three-dimensional structure, which, in principle, should give more accurate results compared to the lumped circuit analysis in exchange for a longer runtime. Figure \ref{fig:trimon_modes} shows the field profile of three linearized modes produced by a trimon -- a multimon with three modes. A trimon is a special case of a ring multimon, where the capacitor pads and JJs are placed in a ring fashion. There are four pads in a trimon with one JJ connecting a pair of neighboring pads. Due to the symmetry of the pads, there will be two dipole modes and one quadrupole mode as illustrated in Figure \ref{fig:trimon_modes}.

With the linearized circuit solved, we now add the nonlinearity of the JJ. Note that the nonlinearity is defined with respect to the old coordinates, so we will need the transformation that goes back to the old coordinates from the eigenmode coordinates, i.e., 
\begin{equation}
    \boldsymbol{\Phi}
    = \mathsf{V} \tilde{\mathsf{L}}^{-1/2} \tilde{\mathsf{V}} \tilde{\mathbf{X}}
    \doteq \mathsf{T} \tilde{\mathbf{X}}.
\end{equation}
Subsequently, we can express the nonlinear part of the Lagrangian in terms of the new coordinates $\tilde{\mathbf{X}}$. For simplicity, let us first start with the fourth-order expansion of the energy of each JJ:
\begin{align}
    \mathcal{L} 
    &= \mathcal{L}_{\text{lin}} 
            (\boldsymbol{\Phi}, 
            \dot{\boldsymbol{\Phi}})
        + \mathcal{L}_{\text{non}} (\boldsymbol{\Phi})
\nonumber \\
    &= \frac{1}{2} 
            \dot{\tilde{\mathbf{X}}}^{\top} 
            \dot{\tilde{\mathbf{X}}}
        - \frac{1}{2} 
            \tilde{\mathbf{X}} 
            \mathsf{\Omega}^2 
            \tilde{\mathbf{X}}
        - \sum_{(i,j)\in J} 
            \left[ 
                - E_{\text{J},ij} \cos \left( \frac{\Phi_i - \Phi_j}{\phi_0} \right)
                - \frac{1}{2} 
                    \tilde{\mathbf{X}} 
                    \mathsf{\Omega}^2 
                    \tilde{\mathbf{X}}
            \right]
\nonumber \\
    &\approx
        \frac{1}{2} 
            \dot{\tilde{\mathbf{X}}}^{\top} 
            \dot{\tilde{\mathbf{X}}}
        - \frac{1}{2} 
            \tilde{\mathbf{X}} 
            \mathsf{\Omega}^2 
            \tilde{\mathbf{X}}
        + \sum_{(i,j)\in J} 
            \frac{E_{\text{J},ij}}{4!} \left( \frac{\Phi_i - \Phi_j}{\phi_0} \right)^4
\nonumber \\
    &= \frac{1}{2} 
            \dot{\tilde{\mathbf{X}}}^{\top} 
            \dot{\tilde{\mathbf{X}}}
        - \frac{1}{2} 
            \tilde{\mathbf{X}} 
            \mathsf{\Omega}^2 
            \tilde{\mathbf{X}}
        + \sum_{(i,j)\in J} 
            \frac{E_{\text{J},ij}}{4!} 
            \left[ 
                \frac{
                    (\mathsf{T} 
                        \tilde{\mathbf{X}})_i 
                    - (\mathsf{T}   
                        \tilde{\mathbf{X}})_j
                }{\phi_0} 
            \right]^4,
\end{align}
where have defined the set $J = \{(i,j) | E_{\text{J},ij} \neq 0\}$. With the nonlinear terms included, the conjugate momenta with respect to the generalized coordinates $\tilde{\mathbf{X}}$ are given by
\begin{equation}
    \tilde{P}_{\mu}
    = \frac{\partial \mathcal{L}}{\partial \dot{\tilde{X}}_{\mu}}
    = \dot{\tilde{X}}_{\mu}.
\end{equation}
Note that since the nonlinear term is only a function of $\tilde{\mathbf{X}}$, the order of our approximation does not affect the expression of the conjugate momenta. Next, as usual, we define the normal mode amplitudes to be
\begin{equation}
    a_{\mu} 
    = \sqrt{\frac{\omega_{\mu}}{2\hbar}} 
        \left(
            \tilde{X}_{\mu} 
            + \frac{\ci}{\omega_{\mu}}
                \tilde{P}_{\mu}
        \right).
\end{equation}
The Hamiltonian can thus be rewritten in terms of $\alpha_{\mu}$ as
\begin{align}
    \mathcal{H}
    &= \tilde{\mathbf{P}}^{\top} \dot{\tilde{\mathbf{X}}}  - \mathcal{L}
\nonumber\\[-1mm]
    &= \sum_{\mu} 
            \hbar \omega_{\mu}
            \Big(   
                a^*_{\mu} a_{\mu}
                + a_{\mu} a^*_{\mu}
            \Big)
        - \sum_{(i,j)\in J} 
            \frac{E_{\text{J},ij}}{4! \phi_0^4} 
            \left[
                \sum_{\mu} 
                    (
                        \mathsf{T}_{i\mu} - \mathsf{T}_{j\mu}
                    )
                    \sqrt{\frac{\hbar}{2\omega_{\mu}}}
                    \Big(
                        a_{\mu} + a^*_{\mu}
                    \Big)
            \right]^4.
\end{align}

Now, following the procedure of canonical quantization, we promote the normal mode amplitudes to the annihilation operators and impose the bosonic commutation relations
\begin{equation}
    \Big[
        \hat{a}_{\mu}, \hat{a}_{\nu}^{\dagger} 
    \Big]
    = \delta_{\mu \nu} \hat{1}.
\end{equation}
The quadrature operators, $\hat{\tilde{X}}$ and $\hat{\tilde{P}}$, of each mode can be related to the mode annihilation operator in the familiar way:
\begin{equation}
    \hat{a}_{\mu} 
    = \sqrt{\frac{\omega_{\mu}}{2\hbar}} 
        \left(
            \hat{\tilde{X}}_{\mu} 
            + \frac{\ci}{\omega_{\mu}} \hat{\tilde{P}}_{\mu}
        \right)
    \ \ \text{ and } \ \ 
    \hat{\tilde{X}}_{\mu}
    = \sqrt{\frac{\hbar}{2\omega_{\mu}}} 
        \left(
            \hat{a}_{\mu} 
            + \hat{a}_{\mu}^{\dagger}
        \right).
\end{equation}
To finalize the quantum Hamiltonian, we use the bosonic commutation relations to reorder the appearance of the annihilation and creation operators. In addition, by assuming that the mode frequencies are sufficiently distinct, we can apply the RWA to remove terms with unequal numbers of creation and annihilation operators in the same mode (e.g., $\hat{a}_{\mu}^{\dagger} \hat{a}_{\mu}^{\dagger}\hat{a}_{\nu}^{\dagger} \hat{a}_{\nu}$). This results in
\begin{align}
    \hat{H}
    &= \sum_{\mu} 
        \hbar \omega_{\mu}
        \left(   
            \hat{a}^{\dagger}_{\mu}\hat{a}_{\mu}
            + \frac{1}{2}
        \right)
    - \sum_{(i,j)\in J} 
        \frac{E_{\text{J},ij}}{4! \phi_0^4} 
        \left[
            \sum_{\mu} 
                (
                    \mathsf{T}_{i\mu} - \mathsf{T}_{j\mu}
                )
                \sqrt{\frac{\hbar}{2\omega_{\mu}}}
                \left(\hat{a}_{\mu} + \hat{a}_{\mu}^{\dagger}\right)
        \right]^4
\nonumber\\[-1mm]
    &= \sum_{\mu} 
        \hbar \omega_{\mu}
        \left(   
            \hat{a}^{\dagger}_{\mu}\hat{a}_{\mu}
            + \frac{1}{2}
        \right)
        + f(x_0,...,x_{N-1}) \Big|_{x_{\mu} = \hat{a}_{\mu} + \hat{a}_{\mu}^{\dagger}}
\nonumber\\[-1mm]
    &= \sum_{\mu} \hbar 
        \left[ 
            (\omega_{\mu} + \Lambda_{\mu})
                \hat{a}_{\mu}^{\dagger} 
                \hat{a}_{\mu} 
            + \frac{\chi_{\mu \mu}}{4} \hat{a}_{\mu}^{\dagger 2} \hat{a}_{\mu}^2
        \right]
        + \sum_{\mu} \sum_{\nu < \mu} 
            \hbar \chi_{\mu \nu}
            \hat{a}_{\mu}^{\dagger} \hat{a}_{\mu}
            \hat{a}_{\nu}^{\dagger} \hat{a}_{\nu},
\end{align}
where 
\begin{equation}
    f(x_0,...,x_{N-1})
    = - \sum_{(i,j)\in J} 
        \frac{E_{\text{J},ij}}{4! \phi_0^4} 
        \left[
            \sum_{\mu} 
                (
                    \mathsf{T}_{i\mu} - \mathsf{T}_{j\mu}
                )
                \sqrt{\frac{\hbar}{2\omega_{\mu}}}
                \, x_{\mu}
        \right]^4
\end{equation}
has been defined for computing the entries of the (fourth-order) Kerr matrix
\begin{equation}
    \hbar \chi_{\mu \nu}
    = \frac{\partial^4 f}{\partial x_{\mu}^2 \partial x_{\nu}^2} \bigg|_{\mathbf{x} = \mathbf{0}}
    = - \frac{\hbar^2}{4 \phi_0^4}
        \sum_{(i,j)\in J} 
            \frac{E_{\text{J},ij}}{\omega_{\mu} \omega_{\nu}} 
            (
                \mathsf{T}_{i\mu} 
                - \mathsf{T}_{j\mu} 
            )^2 
            (
                \mathsf{T}_{i\nu} 
                - \mathsf{T}_{j\nu} 
            )^2
\end{equation}
and the Lamb shifts
\begin{equation}
    \Lambda_{\mu} 
    = \frac{1}{2} 
        \sum_{\nu} \chi_{\mu \nu} .
\end{equation}
As mentioned before, the anharmonicity, $\alpha_{\mu}$, is related to the self-Kerr coefficients $\chi_{\mu\mu}$ by
\begin{equation}
    \alpha_{\mu} = \frac{\chi_{\mu\mu}}{2};
\end{equation}
thus, the Hamiltonian can be rewritten as \cite{s41534_021_00461_8, PhysRevApplied.7.054025}
\begin{align} \label{eq:multimon_hamiltonian_using_alpha_mu}
    \hat{H}
    = \sum_{\mu} \hbar 
        \left(
            \tilde{\omega}_{\mu}\hat{a}_{\mu}^{\dagger} \hat{a}_{\mu} 
            + \frac{\alpha_{\mu}}{2} \hat{a}_{\mu}^{\dagger 2} \hat{a}_{\mu}^2
        \right)
        + \sum_{\mu} \sum_{\nu < \mu} 
            \hbar \chi_{\mu \nu}
            \hat{a}_{\mu}^{\dagger} \hat{a}_{\mu}
            \hat{a}_{\nu}^{\dagger} \hat{a}_{\nu}.
\end{align}
Note that we have also lumped the Lamb shifts into the frequency of the linearized mode so that the fundamental transition of each nonlinear mode has a frequency $\tilde{\omega}_{\mu} = \omega_{\mu} + \Lambda_{\mu}$. 

The (isolated) multimon Hamiltonian in Eq.(\ref{eq:multimon_hamiltonian_using_alpha_mu}) differs from the (isolated) transmon Hamiltonian by the longitudinal coupling term $\hat{a}_{\mu}^{\dagger} \hat{a}_{\mu} \hat{a}_{\nu}^{\dagger} \hat{a}_{\nu}$. However, one also realizes the similarity between the longitudinal coupling and the qubit-resonator dispersive interaction. Hence, the effect of the longitudinal coupling is to generate a frequency shift in one mode based on the state of the other modes. Since a frequency shift leads to the accumulation of an extra phase, we immediately obtain a phase gate.

A more pragmatic method for solving the energy levels of the multimon would be numerical diagonalization. In this case, there is no need to keep the Josephson energy to the fourth order; separating the Hamiltonian into the harmonic and anharmonic parts gives
\begin{align}
    \hat{H}
    &= \sum_{\mu} 
        \hbar \omega_{\mu}
        \left(   
            \hat{a}^{\dagger}_{\mu}\hat{a}_{\mu}
            + \frac{1}{2}
        \right)
\nonumber \\[-1mm]
    & \ \ \ \ \ \ \ \ 
        - \sum_{(i,j)\in J} 
            E_{\text{J},ij} 
            \Bigg\{ 
            \cos 
            \Bigg[
                \frac{1}{\phi_0}
                \sum_{\mu} 
                    (
                        \mathsf{T}_{i\mu} 
                        - \mathsf{T}_{j\mu}
                    )
                    \sqrt{\frac{\hbar}{2\omega_{\mu}}}
                    \left(\hat{a}_{\mu} + \hat{a}_{\mu}^{\dagger}\right)
            \Bigg]
\nonumber\\
    & \ \ \ \ \ \ \ \ \ \ \ \ \ \ \ \ \ \ \ \ \ \ \ \ \ \ \ \ 
        + \frac{1}{2}\left[
            \frac{1}{\phi_0}
            \sum_{\mu} 
                (
                    \mathsf{T}_{i\mu} 
                    - \mathsf{T}_{j\mu}
                )
                \sqrt{\frac{\hbar}{2\omega_{\mu}}}
                \left(\hat{a}_{\mu} + \hat{a}_{\mu}^{\dagger}\right)
        \right]^2 \Bigg\}.
\end{align}
However, one has to truncate the Hilbert space (i.e., the size of the matrix to be diagonalized). Typically, to diagonalize the Hamiltonian of a trimon, we need about ten Fock basis vectors per mode, which results in a thousand basis vectors after taking the tensor product between the three modes. The operators are also built from the tensor products; for example, the annihilation operator of mode $\mu$, with $d$ basis vectors in each mode, is given by
\begin{equation}
    \mathsf{a}_{\mu}
    = \mathsf{1}_{d}^{(0)} \otimes \mathsf{1}_{d}^{(1)} \otimes \cdots \otimes \begin{pmatrix}
        0 & \sqrt{1} & 0 & \cdots & 0 \\[-1mm]
        0 & 0 & \sqrt{2} & \cdots & 0 \\[-1mm]
        \vdots & \vdots & \vdots & \ddots & \vdots \\[-1mm]
        0 & 0 & 0 & \cdots & \sqrt{d-1} \\[-1mm]
        0 & 0 & 0 & \cdots & 0
    \end{pmatrix}
    \otimes \cdots \otimes \mathsf{1}_{d}^{(N-1)}
    \in \mathbf{C}^{d^{N} \times d^{N}}.
\end{equation}
In most numerical packages, the cosine of an operator, defined to be
\begin{equation}
    \cos{\mathsf{A}}
    = \mathsf{1}_{d^N} - \frac{1}{2!} \mathsf{A}^2 + \frac{1}{4!} \mathsf{A}^4 + \cdots,
\end{equation}
can also be computed easily (i.e., taking the real part of the matrix exponential). 

Moreover, the analysis of a multimon can be applied to that of a transmon. In practice, we do not just have a single capacitance between two conductors; for example, if a transmon is surrounded by a large ground plane and is coupled to the control and readout lines as shown in Figure \ref{fig:transmon_picture}, the effective capacitance between the two pads of the transmon must be calculated from the capacitance matrix. The associated inductance matrix will have four nonzero entries since there is only one JJ in the transmon.

A real circuit always has a finite size, so the lumped circuit model of the multimon is only an approximation. In addition, one needs to bring the resonator, let it be a 3D cavity or a planar CPW resonator, into the picture to examine the control and readout of the multimon. In the actual multimon design, one can also use the EPR method, which applies to a general electromagnetic structure, allowing us to include both the capacitor pads from the multimon and the distributed resonator in a single analysis. Nevertheless, the Hamiltonian computed by the EPR method will still take the form of Eq.(\ref{eq:multimon_hamiltonian_using_alpha_mu}) to the fourth order.

\subsection{A Multimon Coupled to a Resonator}
We can read out the state of a multimon by placing it near or inside a resonator. In general, each anharmonic mode of the multimon will be coupled to the resonator mode. Take a trimon as an example, the two dipole modes give rise to the usual dipole interaction if the electric field of the resonator has nonzero components along the two orthogonal directions of the dipole modes. If, in addition, the electric field has a gradient near the center of the trimon, the quadrupole mode can be excited through the interaction
\begin{equation}
    - \frac{1}{6}
            \sum_{i,j} 
                \hat{Q}_{ij} 
                \frac{\partial \hat{E}_{j}}{\partial r_i}(\mathbf{0}),
\end{equation}
where $\hat{Q}_{ij}$ is the quadrupole operator associated with the trimon mode and $\hat{E}_{j}$ is the $j$th component of the resonator field. Nevertheless, under the RWA, any multipole interaction can be represented abstractly as
\begin{equation}
    \hat{H}_{\text{int}, \mu} 
    = - \hbar \Big(
        g_{\mu} \hat{a}_{\mu}^{\dagger}   \hat{a}_{\text{r}}
        + g_{\mu}^* \hat{a}_{\mu}  \hat{a}_{\text{r}}^{\dagger}
    \Big),
\end{equation}
where $\hat{a}_{\text{r}}$ is the annihilation operator of the resonator and $g_{\mu}$ is coupling coefficient between the $\mu$th mode of the multimon and the resonator. In particular, for the multipoles other than the dipole, $g_{\mu}$ are the ZPF of the spatial derivatives of the electric field multiplied by the ZPF of the multipoles. Consequently, the Hamiltonian of the composite system is given by
\begin{align} 
    \hat{H}
    &= \sum_{\mu=0}^{N-1} \hbar 
            \left(
                \tilde{\omega}_{\mu}
                    \hat{a}_{\mu}^{\dagger} 
                    \hat{a}_{\mu} 
                + \frac{\alpha_{\mu}}{2} 
                    \hat{a}_{\mu}^{\dagger 2} 
                    \hat{a}_{\mu}^2
            \right)
        + \sum_{\mu=0}^{N-1} 
            \sum_{\nu < \mu} 
                \hbar \chi_{\mu \nu}
                \hat{a}_{\mu}^{\dagger} \hat{a}_{\mu}
                \hat{a}_{\nu}^{\dagger} \hat{a}_{\nu}
\nonumber\\ \label{eq:multimon_coupled_resonator}
    & \ \ \ \ \ \ \ \ \ \ \ \ 
        + \hbar\omega_{\text{r}}
            \left(
                \hat{a}_{\text{r}}^{\dagger} \hat{a}_{\text{r}} 
                + \frac{1}{2}
            \right)
        - \sum_{\mu=0}^{N-1}
            \hbar 
            \Big(
                g_{\mu} \hat{a}_{\mu}^{\dagger}   \hat{a}_{\text{r}}
                + g_{\mu}^* \hat{a}_{\mu}  \hat{a}_{\text{r}}^{\dagger} 
            \Big),
\end{align}
where $\omega_{\text{r}}$ is the frequency of the resonator.

We have already studied the case of a generic qudit coupled to a resonator. The interaction between a multimon and a resonator is slightly more complex since there are $N$ paths of coupling to the resonator; in addition, there are longitudinal couplings among the modes within the multimon. However, a similar analysis can still be executed if we examine the Hamiltonian in the number basis. To set up the notation, we first consider a single multimon mode without any longitudinal coupling to the other multimon modes or transverse coupling to the resonator; in other words, we look at mode $\mu$ with the Hamiltonian 
\begin{equation}
    \hat{H}_{\mu} 
    = \hbar \tilde{\omega}_{\mu}
                \hat{a}_{\mu}^{\dagger} 
                \hat{a}_{\mu} 
            + \frac{\hbar \alpha_{\mu}}{2} 
                \hat{a}_{\mu}^{\dagger 2} 
                \hat{a}_{\mu}^2
    = \hbar \tilde{\omega}_{\mu} \hat{N}_{\mu} 
        + \frac{\hbar \alpha_{\mu}}{2} 
            \Big( \hat{N}_{\mu} ^2 - \hat{N}_{\mu}  \Big),
\end{equation}
where $\hat{N}_{\mu} = \hat{a}_{\mu}^{\dagger} \hat{a}_{\mu}$ is the number operator of mode $\mu$.
Moreover, we denote the energy eigenstates of $\hat{H}_{\mu}$ by $\ket{j_{\mu}}$ for $j_{\mu} = 0,1,...,D-1$ (i.e., we consider the first $D$ energy levels of the anharmonic oscillator). In addition, let
\begin{equation}
    \omega_{\mu, j_\mu}
    = \bra{j_\mu} 
        \left(
            \tilde{\omega}_{\mu}
                \hat{a}_{\mu}^{\dagger} 
                \hat{a}_{\mu} 
            + \frac{\alpha_{\mu}}{2} 
                \hat{a}_{\mu}^{\dagger 2} 
                \hat{a}_{\mu}^2
        \right)
        \ket{j_\mu}
\end{equation}
for $j_{\mu} = 0,1,...,D-1$ be the transition frequencies within mode $\mu$ when the longitudinal coupling is turned off.

The resonator coupling in Eq.(\ref{eq:multimon_coupled_resonator}) assumes a weakly anharmonic model for each multimon mode so that the annihilation and creation operators of the modes are used directly; in addition, the interaction is under the RWA, i.e., $\hat{a}_{\mu} \hat{a}_{\text{r}}$ and $\hat{a}_{\mu}^{\dagger} \hat{a}_{\text{r}}^{\dagger}$ are omitted. For a more general consideration, we can define a coupling coefficient $g_{\mu, j_{\mu}, k_{\mu}}$ for the transition between any pair of eigenstates, $(\ket{j_{\mu}}, \ket{k_{\mu}})$, within a mode. The weakly anharmonic model corresponds to the case where
\begin{equation} \label{eq:weakly_anharmonic_coupling_coefficient_multimon}
    g_{\mu, j_{\mu}, k_{\mu}} 
    \approx 
    \begin{cases}
        \sqrt{j_{\mu}+1} g_{\mu} & \text{if } \ \  j_{\mu} - k_{\mu} = 1,  \\
        0 & \text{otherwise} 
    \end{cases}
\end{equation}
for all $j_{\mu}, k_{\mu}$, and $\mu$.

In the dispersive limit where $|\omega_{\mu} - \omega_{\text{r}}| \gg g_{\mu}$ for all $\mu$, the Schrieffer–Wolff transformation can be used to reveal the dispersive shift associated with the multimon-resonator coupling. By choosing 
\begin{align} 
    \hat{H}_0
    = \sum_{\mu=0}^{N-1} \hbar 
            \left(
                \tilde{\omega}_{\mu}
                    \hat{a}_{\mu}^{\dagger} 
                    \hat{a}_{\mu} 
                + \frac{\alpha_{\mu}}{2} 
                    \hat{a}_{\mu}^{\dagger 2} 
                    \hat{a}_{\mu}^2
            \right)
        + \sum_{\mu=0}^{N-1} 
            \sum_{\nu < \mu} 
                \hbar \chi_{\mu \nu}
                \hat{a}_{\mu}^{\dagger} \hat{a}_{\mu}
                \hat{a}_{\nu}^{\dagger} \hat{a}_{\nu}
        + \hbar\omega_{\text{r}}
            \left(
                \hat{a}_{\text{r}}^{\dagger} \hat{a}_{\text{r}} 
                + \frac{1}{2}
            \right),
\end{align}
\begin{equation}
    \hat{V} 
    = - \sum_{\mu=0}^{N-1}
            \hbar 
            \Big(
                g_{\mu} \hat{a}_{\mu}^{\dagger}   \hat{a}_{\text{r}}
                + g_{\mu}^* \hat{a}_{\mu}  \hat{a}_{\text{r}}^{\dagger} 
            \Big),
\end{equation}
and 
\begin{align}
    \hat{S} 
    &= \sum_{\mu = 0}^{N-1}
        \sum_{j_{0}= 0}^{D-1} 
        \cdots 
        \sum_{j_{N-1}= 0}^{D-1}
        \sum_{k_{\mu} = 0}^{D-1}
        \frac{1}{
            \left(
                \omega_{\mu, j_{\mu}} 
                + \sum_{\nu \neq \mu}
                    \chi_{\mu \nu} 
                    j_{\mu} 
                    j_{\nu}
            \right) 
            - \left(
                \omega_{\mu,k_{\mu}} 
                + \sum_{\nu \neq \mu} 
                    \chi_{\mu\nu} 
                    k_{\mu} 
                    j_{\nu}
            \right) 
            - \omega_{\text{r}}
        }
\nonumber \\ \label{eq:multimon_res_SW_transform}
    & \ \ \ \ \ \ \ \ \ \ \ \ \ \ \ \ \ \ \ \ 
        \times 
        \Big(
            g_{\mu, j_{\mu} k_{\mu}} 
                \ket{j_{\mu}} \! 
                \bra{k_{\mu}} 
                \hat{a}_{\text{r}}
            - g_{\mu, j_{\mu} k_{\mu}}^* 
                \ket{k_{\mu}} \! 
                \bra{j_{\mu}} 
                \hat{a}_{\text{r}}^{\dagger}
        \Big)
        \bigotimes_{\nu \neq \mu} 
            \ket{j_{\nu}} \! \bra{j_{\nu}}, 
\end{align}
one finds the Hamiltonian in the dispersive limit to be\footnote{Note that the Lamb shifts on the multimon from the resonator shows up as the ``$+1$'' in $\hat{N}_{\mu} \Bigsl( \hat{N}_{\text{r}} + 1 \Bigsr)$, which comes from the bosonic commutation relation of the resonator.}
\begin{align} 
    \hat{H}^{\text{disp}}
    &= \sum_{\mu=0}^{N-1} \hbar 
            \left[
                \tilde{\omega}_{\mu}
                    \hat{N}_{\mu}
                + \frac{\alpha_{\mu}}{2}
                    \Big( \hat{N}_{\mu} ^2 - \hat{N}_{\mu}  \Big)
            \right]
        + \sum_{\mu=0}^{N-1} 
            \sum_{\nu < \mu} 
                \hbar \chi_{\mu \nu}
                \hat{N}_{\mu}
                \hat{N}_{\nu}
        + \hbar\omega_{\text{r}}
            \left(
                \hat{N}_{\text{r}}
                + \frac{1}{2}
            \right)
\nonumber\\ \label{eq:dispersive_hamiltonian_multimon_res}
    & \ \ \
        + \sum_{\mu=0}^{N-1}
            \left[
                \frac{\hbar |g_{\mu}|^2 \hat{N}_{\mu} \Bigsl( \hat{N}_{\text{r}} + 1 \Bigsr)}{\Tilde{\omega}_{\mu} + \alpha_{\mu} \Bigsl(\hat{N}_{\mu} - 1\Bigsr) + \sum_{\nu \neq \mu} \chi_{\mu \nu} \hat{N}_{\nu} - \omega_{\text{r}}}
                - \frac{\hbar |g_{\mu}|^2 \Bigsl(\hat{N}_{\mu} + 1 \Bigsr) \hat{N}_{\text{r}}}{\Tilde{\omega}_{\mu} + \alpha_{\mu} \hat{N}_{\mu} + \sum_{\nu \neq \mu} \chi_{\mu \nu} \hat{N}_{\nu} - \omega_{\text{r}}}
            \right],
\end{align}
where Eq.(\ref{eq:weakly_anharmonic_coupling_coefficient_multimon}) is assumed\footnote{In particular, in the weakly anharmonic limit, $\hat{S}$ reduces to
\begin{align}
    \hat{S} 
    &= \sum_{\mu = 0}^{N-1}
        \Bigg(
            \hat{a}_{\mu}^{\dagger} 
            \frac{g_{\mu}}{\tilde{\omega}_{\mu} + \alpha_{\mu} \hat{N}_{\mu} + \sum_{\nu \neq \mu} \chi_{\mu\nu} \hat{N}_{\nu} - \omega_{\text{r}}}\hat{a}_{\text{r}}
            - \frac{g_{\mu}^*}{\tilde{\omega}_{\mu} + \alpha_{\mu} \hat{N}_{\mu} + \sum_{\nu \neq \mu} \chi_{\mu\nu} \hat{N}_{\nu} - \omega_{\text{r}}}\hat{a}_{\mu} \hat{a}_{\text{r}}^{\dagger}
        \Bigg).
\end{align}
}. Note that $\hat{S}$ defined in Eq.(\ref{eq:multimon_res_SW_transform}) is essentially the same as the one used to reveal the resonator-mediated two-qubit gate (see Eq.(\ref{eq:S_for_two_qubit_res_SW_transform})) since there are multiple coupling paths to the resonator in both cases. The only complication in Eq.(\ref{eq:multimon_res_SW_transform}) comes from the fact that the transition frequencies within a mode can be modified by the state of the other modes, resulting in the appearance of $\chi_{\mu\nu} j_{\mu} j_{\nu}$ and $\ket{j_{\nu}} \! \bra{j_{\nu}}$. Moreover, unlike the analysis of the two-qubit gate where the detuning between the qubit frequencies is assumed to be small for an ideal Rabi flopping, the derivation of Eq.(\ref{eq:dispersive_hamiltonian_multimon_res}) relies on the assumption that the mode frequencies of the multimon are sufficiently detuned from one another so that the transverse coupling (see Eq.(\ref{eq:two_qubit_gate_J_12})) between the multimon modes in $\hat{H}^{\text{disp}}$ has been omitted. In practice, we usually design the coupling such that $g_{\mu}$ are near zero for all but one multimon mode; thus, the coupling strength defined in Eq.(\ref{eq:two_qubit_gate_J_12}) can be ignored even if the detuning between two multimon modes is not sufficiently large.






%% file: include_chapters/chapter_open_systems.tex
\chapter{Open Quantum Systems}

To control a qubit or to excite the readout resonator, the quantum system must also contain the drive signals, usually coming from room-temperature electronics. On the one hand, a quantum system, as far as the axioms of quantum mechanics are concerned, is a closed and unitary system. On the other hand, however, it's infeasible to include all the degrees of freedom even for the simplest experiment, forcing us to introduce a new formalism -- the theory of open quantum systems. The chapter extends the ideal model discussed in the previous chapters but explains the realistic modeling of qubit control and readout signals. The methodology used in open quantum systems also allows us to discuss quantum measurement in terms of the ensemble average or the stochastic trajectories. In particular, we will end the chapter with the stochastic master equation for modeling the dispersive measurement of a qudit, which is the main result of the thesis. 
\section{Canonical Model of Open Quantum Systems}
\subsection{Resonator-Environment Coupling}
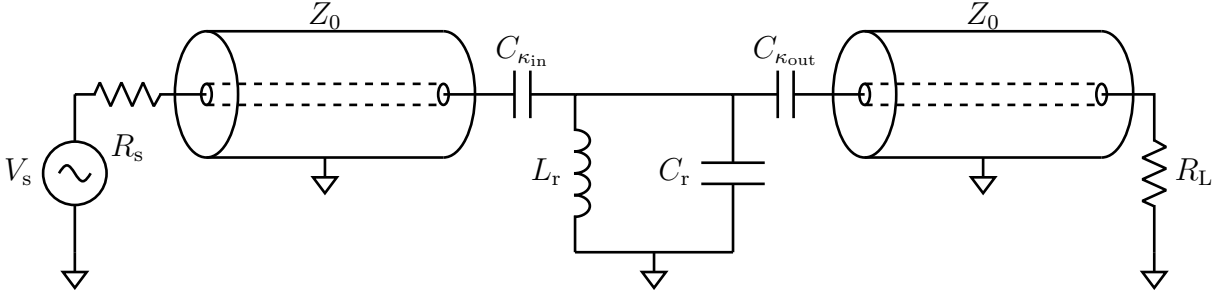
\begin{figure}[t]
    \centering
    \begin{tikzpicture}[scale=0.7]
    \ctikzset{tripoles/mos style/arrows};
    \begin{scope}[shift={(-7,3)}]
        \draw 
            (0,0) ellipse (0.6 and 1.2);
        \draw 
            (4.5,-1.2) arc(-90:90:0.6 and 1.2);
        \draw
            (0,1.2) to (4.5,1.2)
            (0,-1.2) to (4.5,-1.2)
            (2.25, 1.5) node[]{$Z_0$};
        \draw[dashed]
            (0,0.2) to (4.5,0.2)
            (0,-0.2) to (4.5,-0.2);
        \draw 
            (0,0) ellipse (0.1 and 0.2)
            (4.5,0) ellipse (0.1 and 0.2);
        \draw 
            (2.25,-1.2) node[/tikz/circuitikz/bipoles/length=30pt,sground]{};
        \draw 
            (0,0) to (-0.5,0) to [/tikz/circuitikz/bipoles/length=30pt, R, l^=$R_{\text{s}}$] (-2.5,0) 
            (-2.5,0) to [/tikz/circuitikz/bipoles/length=40pt, sV, l_=$V_{\text{s}}$](-2.5,-3) node[/tikz/circuitikz/bipoles/length=30pt,sground]{};
    \end{scope}
        \draw
            (0,3) to (3,3) 
            (3,0) to (0,0) 
            (0,3) to [/tikz/circuitikz/bipoles/length=40pt, L, l_=$L_{\text{r}}$] (0,0)
            (3,3) to (3,2) to [/tikz/circuitikz/bipoles/length=40pt, C, l_=$C_{\text{r}}$] (3,1) to (3,0)
            (1.5, 0) node[/tikz/circuitikz/bipoles/length=30pt,sground]{};
        \draw
            (-2.5,3) to (-2,3) to [/tikz/circuitikz/bipoles/length=30pt, C, l^=$C_{\kappa_{\text{in}}}$] (0,3);
        \draw
            (3,3) to [/tikz/circuitikz/bipoles/length=30pt, C, l^=$C_{\kappa_{\text{out}}}$] (5,3) to (5.5,3);
    \begin{scope}[shift={(5.5,3)}]
        \draw 
            (0,0) ellipse (0.6 and 1.2);
        \draw 
            (4.5,-1.2) arc(-90:90:0.6 and 1.2);
        \draw
            (0,1.2) to (4.5,1.2)
            (0,-1.2) to (4.5,-1.2)
            (2.25, 1.5) node[]{$Z_0$};
        \draw[dashed]
            (0,0.2) to (4.5,0.2)
            (0,-0.2) to (4.5,-0.2);
        \draw 
            (0,0) ellipse (0.1 and 0.2)
            (4.5,0) ellipse (0.1 and 0.2);
        \draw 
            (2.25,-1.2) node[/tikz/circuitikz/bipoles/length=30pt,sground]{};
        \draw 
            (4.5,0) to (5.5,0) to [/tikz/circuitikz/bipoles/length=30pt, R, l^=$R_{\text{L}}$] (5.5,-3)  node[/tikz/circuitikz/bipoles/length=30pt,sground]{};
    \end{scope}
    \end{tikzpicture}
    \caption{Coupling of a two-port resonator to the readout signal generator ($V_{\mathrm{s}}$ and $R_{\mathrm{s}}$) and the readout ADC ($R_{\mathrm{L}}$). The transmission lines in this model represent the coaxial cables used to connect the readout cavity inside the dilution fridge to the room-temperature electronics. The coupling capacitors $C_{\kappa_{\mathrm{in}}}$ and $C_{\kappa_{\mathrm{out}}}$ are controlled, for example, by how deep the core conductors of the SMA ports are inserted into the readout cavity (see Figure \ref{fig:cavity_example} for more details).}
    \label{fig:tline_resonator_coupling_two_sides}
\end{figure}

The theory of open quantum systems is a rich topic and the results generally rely on the specific model of the environment coupled to the system of interest. For superconducting quantum computation, a simplified schematics of the experiment is shown in Figure \ref{fig:tline_resonator_coupling_two_sides}. The voltage source represents the signal generator at room temperature, which is connected to the readout resonator via a long transmission line. The transmitted signal then leaves the resonator and reaches the ADC card via another transmission line. Alternatively, one can also measure the reflection by inserting a circulator to decouple the input and output signals. The qubit coupled to the readout resonator is not shown explicitly in Figure \ref{fig:tline_resonator_coupling_two_sides}. To drive a qubit inside a 3D readout cavity, the signal must first interact with the cavity before reaching the qubit; hence, it is easier to deal with the cavity alone first.

\begin{figure}
    \centering
    \begin{tikzpicture}[scale=0.7]
        \draw
            (0,3) to (3,3) 
            (3,0) to (0,0) 
            (0,3) to [/tikz/circuitikz/bipoles/length=40pt, L, l_=$L_{\text{r}}$] (0,0)
            (3,3) to (3,2) to [/tikz/circuitikz/bipoles/length=40pt, C, l_=$C_{\text{r}}$] (3,1) to (3,0)
            (1.5, 0) node[/tikz/circuitikz/bipoles/length=30pt,sground]{};
        \draw
            (0,3) to [/tikz/circuitikz/bipoles/length=30pt, C, l_=$C_{\kappa_{\text{in}}}$] (-2,3) to [/tikz/circuitikz/bipoles/length=30pt, R, l_=${Z_0}$] (-2,0)  node[/tikz/circuitikz/bipoles/length=30pt,sground]{};
        \draw
            (3,3) to [/tikz/circuitikz/bipoles/length=30pt, C, l^=$C_{\kappa_{\text{out}}}$] (5,3) to [/tikz/circuitikz/bipoles/length=30pt, R, l^=${Z_0}$] (5,0)  node[/tikz/circuitikz/bipoles/length=30pt,sground]{};
    \end{tikzpicture}
    \caption{Equivalent circuit when the source and load impedances are matched to the characteristic impedance of the coaxial cables.}
    \label{fig:equivalent_coupling_circuit}
\end{figure}
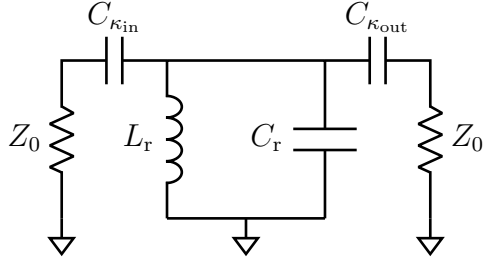

From the perspective of the resonator, it sees an effective impedance presented by the transmission lines. Under the assumption that the source and load are matched to the characteristic impedance of the cables (usually $50 \, \Omega$), power leaving the system is fully absorbed by the load and source impedance and the transmission lines only add a phase decay. In this case, we can remove the transmission and reduce the distributed circuit to a simple RLC network shown in Figure. \ref{fig:equivalent_coupling_circuit}.

Coupling to the readout resonator is modelled by two capacitors $C_{\kappa_{\text{in}}}$ and $C_{\kappa_{\text{out}}}$. Physically, they can be SMA ports on a 3D cavity or a real interdigitated capacitor between CPW lines. Since it's much easier to study the parallel or series RLC circuits, we shall transform the series-connected coupling capacitor into a parallel-connected one \cite{doi:10.1063/1.3010859, 2007PhDT.......167S}. The technique used here is the same as the $L$-network used in lumped-element impedance matching \cite{pozar1990microwave}: Given an impedance $Z = R + \cj X$ (we use the convention $\cj = -\ci$ in electrical engineering), we define the quality factor $Q = |X|/R$. Then, the impedance can be represented as a resistance $R_{\text{p}} = (1+Q^2) R$ in shunt with a reactance $X_{\text{p}} = (1+1/Q^{2})X$. Apply this transformation to both the input and output RC network yield Figure \ref{fig:equivalent_coupling_circuit_transformed}, where we have fixed the drive frequency $\omega_{\mathrm{d}}$ and defined
\begin{equation}
    Q_{\mathrm{in}} 
    = \frac{1}{\omega_{\mathrm{d}} C_{\kappa_{\mathrm{in}}}Z_0} 
    \ \ \text{ and } \ \ 
    Q_{\mathrm{out}} 
    = \frac{1}{\omega_{\mathrm{d}} C_{\kappa_{\mathrm{out}}}Z_0}.
\end{equation}
Note that this impedance transformation assumes a fixed frequency of operation since the impedance will change with the frequency. For large $Q_{\text{in,out}}$, i.e., weakly coupled resonator, the transformed reactance is approximately the same as the untransformed one, i.e., $C_{\kappa_{\text{in}}, \text{p}} \approx C_{\kappa_{\text{in}}}$ and $C_{\kappa_{\text{out}}, \text{p}} \approx C_{\kappa_{\text{out}}}$.

\begin{figure}[t]
    \centering
    \begin{tikzpicture}[scale=0.7]
        \draw
            (0,3) to (3,3) 
            (3,0) to (0,0) 
            (0,3) to [/tikz/circuitikz/bipoles/length=40pt, L, l_=$L_{\text{r}}$] (0,0)
            (3,3) to (3,2) to [/tikz/circuitikz/bipoles/length=40pt, C, l_=$C_{\text{r}}$] (3,1) to (3,0)
            (1.5, 0) node[/tikz/circuitikz/bipoles/length=30pt,sground]{};
        \draw
            (0,3) to (-2,3) to [/tikz/circuitikz/bipoles/length=30pt, C, l_={\Large${\substack{C_{\kappa_{\text{in}},\text{p}} \\[0.5mm] \approx C_{\kappa_{\text{in}}}}}$}] (-2,0) 
            (-2,0)  node[/tikz/circuitikz/bipoles/length=30pt,sground]{}
            (-2,3) to (-5,3) to [/tikz/circuitikz/bipoles/length=30pt, R, l_={\Large${\substack{R_{{\text{in,p}}} \\[0.5mm] \approx Q_{\text{in}}^2 Z_0}}$}] (-5,0) 
            (-5,0) node[/tikz/circuitikz/bipoles/length=30pt, sground]{};
        \draw
            (3,3) to (5,3) to [/tikz/circuitikz/bipoles/length=30pt, C, l^={\Large${\substack{C_{\kappa_{\text{out}},\text{p}} \\[0.5mm] \approx C_{\kappa_{\text{out}}}}}$}] (5,0) node[/tikz/circuitikz/bipoles/length=30pt,sground]{}
            (5,3) to (8,3) to [/tikz/circuitikz/bipoles/length=30pt, R, l^={\Large${\substack{R_{{\text{out,p}}} \\[0.5mm] \approx Q_{\text{out}}^2 Z_0}}$}] (8,0)
            (8,0) node[/tikz/circuitikz/bipoles/length=30pt, sground]{};
    \end{tikzpicture}
    \caption{Equivalent $L$-network at a given frequency. The series-connected coupling capacitor is transformed into a parallel-connected capacitor to the ground. As a result of this transformation, the LC circuit sees a source impedance boosted by the quality factor $Q_{\mathrm{in}} = 1/\omega_{\mathrm{d}}C_{\kappa_{\mathrm{in}}}Z_0$ of the RC input network. The same thing happens for the effective load impedance.}
    \label{fig:equivalent_coupling_circuit_transformed}
\end{figure}
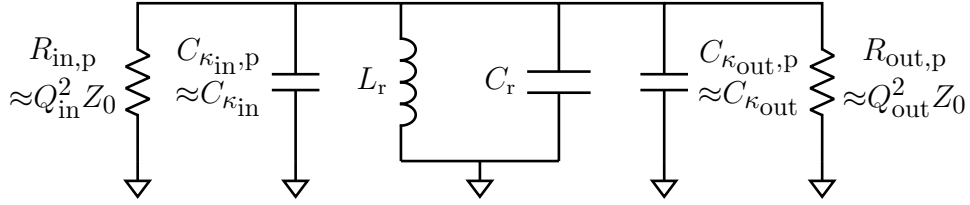

\begin{figure}
    \centering
    \begin{tikzpicture}[scale=0.7]
    \ctikzset{tripoles/mos style/arrows};
        \draw
            (-2,3) to (3,3) 
            (3,0) to (-2,0) 
            (-2,3) to [/tikz/circuitikz/bipoles/length=40pt, L, l_=$L_{\text{r}}$] (-2,0)
            (1,3) to (1,2) to [/tikz/circuitikz/bipoles/length=40pt, C, l_={$\substack{\displaystyle C_{\text{r}}+\\\displaystyle C_{\kappa, \text{p}}}$}] (1,1) to (1,0);
        \draw 
            (-2,3) node[above]{${\Phi_{\text{s}}}$}
            (1,2) node[right]{${+Q_{\text{s}}}$}
            (1,1) node[right]{${-Q_{\text{s}}}$}
            (3.5,2.6) node[]{$+$} 
            (3.5,1.5) node[]{${\substack{\displaystyle V(0,t) \\ \displaystyle \ \ =\dot{\Phi}(0,t)}}$}
            (3.5,0.4) node[]{$-$};
        \draw
            (3,3) to [short, i>^={${I(0,t) = \dot{Q}(0,t)}$}](5,3) to (5, 1.2) to (6, 1.2) 
            (6,0) to (5,0) to [short, i>^={${I(0,t)}$}] (3,0); 
    \begin{scope}[shift={(6,1.2)}]
        \draw 
            (0,0) ellipse (0.6 and 1.2);
        \draw 
            (6,-1.2) arc(-90:90:0.6 and 1.2);
        \draw
            (0,1.2) to (6,1.2)
            (0,-1.2) to (6,-1.2)
            (3, 1.2) node[above]{${R_{\text{p}}}$};
        \draw[dashed]
            (0,0.2) to (6,0.2)
            (0,-0.2) to (6,-0.2);
        \draw 
            (0,0) ellipse (0.1 and 0.2)
            (6,0) ellipse (0.1 and 0.2);
        \draw 
            (3,-1.2) node[/tikz/circuitikz/bipoles/length=30pt,sground]{};
        \draw
            (0,3) to (0,2.6)
            (0,3) node[above]{$x=0$}
            (6,3) node[above]{$x\rightarrow +\infty$};
        \draw[-{Latex[length=2mm]}]
            (0,2.8) to (7,2.8);
    \end{scope}
    \end{tikzpicture}
    \caption{A parallel resonator coupled to a transmission line. $\Phi_{\text{s}}$ and $Q_{\text{s}}$ are the conjugate variables of the LC oscillator, which is treated as the ``system''. The environment is described by a transmission, i.e., a one-dimensional electromagnetic free space. As shown in Chapter 2, the transmission line is described by $\Phi(x,t)$ and $Q(x,t)$.}
    \label{fig:tline_resonator_coupling_analysis}
\end{figure}
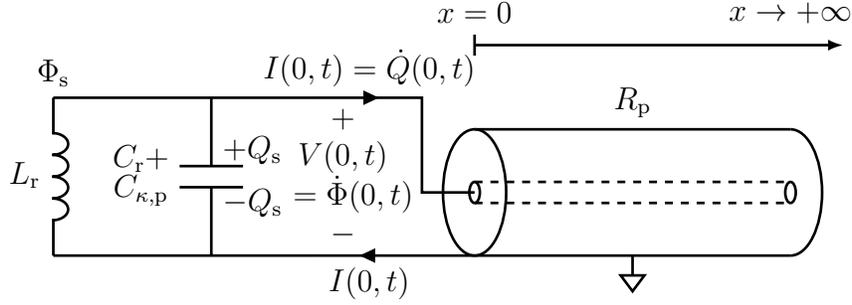

\subsection{Modeling Dissipation with a Unitary System}\label{section:unitary_model_of_dissipation}
The circuit in Figure \ref{fig:equivalent_coupling_circuit_transformed} is not ready for quantization since resistance, which represents losses in the system, is not describable by unitary time evolution. In addition, on the microscopic level, dissipation should also be accompanied by fluctuation from the environment, as required by thermodynamics; however, we clearly do not have a means to add quantum noise using the simple RLC model. 

To avoid the non-unitary nature of the dissipation, we bring the transmission line back to the model. Specifically, we replace the resistors (which are different from the original $Z_0$ due to impedance transformation) with semi-infinite and lossless transmission lines of the same impedance. For simplicity, let us first consider a one-port resonator with transformed impedance $R_{\text{p}}$ (say we measure the reflection only so that the input and output share the same coupling). The resulting circuit is shown in Figure \ref{fig:tline_resonator_coupling_analysis}. From the perspective of the LC oscillator, the resistor and the transmission line present the same impedance, so the energy initially stored in the LC circuit will be dissipated in exactly the same way. Moreover, since the transmission is extended to infinity, the energy left from the LC circuit is still kept somewhere on the line, making the composite system of the resonator plus the transmission line a \textit{conservative system}. Furthermore, the transmission line can support both the outgoing and incoming signals, providing a path for the noise to enter the resonator. Since Figure \ref{fig:tline_resonator_coupling_analysis} includes both fluctuation and dissipation, it will be our canonical model for understanding the control and readout of the quantum system.

\section{Classical vs. Quantum Probability}
This section serves as a review of quantum probability. In particular, we emphasize the similarities and also differences between the classical and quantum probability theories from various aspects (See Table \ref{tab:classical_quantum_probability_theory} for a summary).
\subsection{Classical and Quantum Density Functions}
In classical mechanics, a deterministic system can be described by a phase-space density $\rho$ using the Liouville equation
\begin{equation}
    \frac{\partial \rho}{\partial t} 
    = \{\mathcal{H}, \rho\}_{\text{PB}},
\end{equation}
where $\{A,B\}_{\text{PB}} = \partial_q A \partial_p B - \partial_q B \partial_p A$ is the Poisson bracket.
Once we enter the regime of classical statistical mechanics, we can replace the Liouville equation with equations in the form of a Fokker-Planck equation. Moving to the quantum regime, for a closed system (i.e., without any loss to the environment and any measurement), we have a similar equation called the \textbf{quantum Liouville equation} or \textbf{Liouville–von Neumann equation}
\begin{equation}
    \frac{\mathrm{d} \hat{\rho}}{\mathrm{d} t} 
    = - \frac{\ci}{\hbar}
        \Bigsl[ 
            \hat{H}, \hat{\rho} \,
        \Bigsr].
\end{equation}
The quantum Liouville equation is just another way to write the Schr\"odinger equation, so the time evolution is still unitary/reversible. (It is assumed that the reader has basic exposure to the density matrices and how to take the partial trace of a density matrix. For a review of the related concepts, see \cite{breuer2002theory}.)

However, once the system is subject to the measurement from the environment (also referred to as the reservoir or bath), it is no longer closed, i.e., an open quantum system, requiring us to modify the quantum Liouville equation to something like
\begin{align}
    \frac{\mathrm{d} \hat{\rho}}{\mathrm{d} t} 
    &= - \frac{\ci}{\hbar}
        \Bigsl[ 
            \hat{H}_{\text{eff}}, \hat{\rho} \,
        \Bigsr]
\nonumber \\
        & \ \ \ \ \ \
        + \text{backaction due to the measurement done by the environment}
\nonumber \\
        & \ \ \ \ \ \ 
        + \text{backaction due to the measurement done by us},
\end{align}
where $\hat{\rho}$ is the \textit{conditional} state of the system given that we and/or the environment have partially measured the state to possess some characteristics $J(t)$ (e.g., $J$ can be a continuous monitoring of the position of a molecule for time $s<t$ by us in the laboratory). The probabilistic nature of the measurement outcome $J$ is captured by some random process $W_t$ (e.g., a Wiener process for the homodyne detection or a counting process for a photodetector), and the history of $J$ clearly depends on the sample paths of the random processes. Hence, mathematically, we might look for a stochastic differential equation of the form\footnote{I will write any random process $X$ of time as $X_t$ instead of as $X(t)$. Expressions like $X(t)$ will be reserved for deterministic functions of time.}
\begin{equation}
    \mathrm{d} \hat{\rho}_t
    =  - \frac{\ci}{\hbar}
            \Bigsl[ 
                \hat{H}_{\text{eff}}, \hat{\rho}_t \,
            \Bigsr] 
            \, \mathrm{d} t
        + f_{\text{D}} \! \left( \hat{\rho}_t, t, W_t^{(\text{D})} \right)
        + f_{\text{M}} \! \left( \hat{\rho}_t, t, W_t^{(\text{M})} \right),
\end{equation}
where $W_t^{(\text{D})}$ and $W_t^{(\text{M})}$ represent the randomness in the measurement done by the environment and our measurement device, respectively. Note that we have separated the backaction $f_{\text{D}}$ due to the environment's monitoring of the system from the backaction $f_{\text{M}}$ of our active measurement because we can't keep track of zillions of particles in the environment. In fact, it's necessary to average the stochastic differential equation over all possible measurements from the environment (i.e., in the language of probability theory, we need to marginalize the uncertainty due to the environment) so the time evolution becomes
\begin{equation} \label{eq:SME_proposed_averaged}
    \mathrm{d} \! \left \langle \hat{\rho}_t\right \rangle_{\mathcal{E}}
    =  - \frac{\ci}{\hbar}
            \Bigsl[ 
                \hat{H}_{\text{eff}}, 
                \left \langle 
                    \hat{\rho}_t 
                \right \rangle_{\mathcal{E}} 
            \Bigsr] 
            \, \mathrm{d} t
        + \left \langle 
                f_{\text{D}} \left( 
                        \left 
                            \langle \hat{\rho}_t
                        \right \rangle_{\mathcal{E}}, 
                        W_t^{(\text{D})} 
                    \right) 
            \right \rangle_{\mathcal{E}} 
        + f_{\text{M}} \left( 
                \left \langle 
                    \hat{\rho}_t 
                \right \rangle_{\mathcal{E}}, 
                W_t^{(\text{M})} 
            \right),
\end{equation}
where $\langle\cdot \rangle_{\mathcal{E}}$ represents the ensemble average over the system-environment interaction.
In other words, stochasticity is explicitly written out only for our active measurement.

Intuitively, the random process will break the reversibility of the quantum Liouville equation (analogous to the heat equation). To go back to a unitary evolution, the only thing we can do is to include the entire universe (i.e., keeping track of all the particles, even those that are light years away) as the system such that no external measurement is possible (by definition the ``environment'' does not exist anymore). Since monitoring the entire universe is impossible, we can say that quantum mechanics is effectively irreversible in a small region of interest, leading to the classical phenomena observed.

\subsection{Classical to Quantum Dynamical Systems}\label{section:classical_to_quantum_dynamical_systems}
From the point of view of linear stochastic control, we can formulate a general control problem in terms of two equations:
\begin{equation} \label{eq:classical_state_transition}
    \mathrm{d} x_t 
    = A(t) x_t \, \mathrm{d} t 
        + B(t) u_t \, \mathrm{d} t 
        + G(t) \, \mathrm{d} W_t,
\end{equation}
\begin{equation} \label{eq:classical_observation}
    \mathrm{d} y_t 
    = C(t) x_t \, \mathrm{d} t 
        + H(t) \, \mathrm{d}  V_t.
\end{equation}
The first equation allows us to calculate the state evolution under some control $u_t$ but subject to a noise $W_t$, and the second equation quantifies the observability of the system with noise $V_t$ added to the detection. In addition, it's common to perform state estimation and then define feedback controls based on the estimation; one example of such a closed-loop estimation/control is the Kalman filter.

As mentioned before, a quantum measurement is always accompanied by a backaction (i.e., projection and re-normalization of the post-measurement state), so we can interpret the measurement process as some state estimation followed by a fixed form of feedback enforced by nature. As we will see later, the net effect is compactly rewritten as two equations:
\begin{align}
    \mathrm{d} \hat{\rho}_t
    &= - \frac{\ci}{\hbar} 
            \Bigsl[ 
                \hat{H}_{\text{eff}}, 
                \hat{\rho}_t
            \Bigsr] 
            \, \mathrm{d} t
        + \sum_{\mu \in \mathcal{D} \cup \mathcal{M}}
            k_{\mu} 
            \bigg[
                \hat{L}_{\mu} \hat{\rho}_t \hat{L}_{\mu}^{\dagger}
                - \frac{1}{2}
                    \Big( 
                        \hat{L}_{\mu}^{\dagger} \hat{L}_{\mu} \hat{\rho}_t
                        + \hat{\rho}_t \hat{L}_{\mu}^{\dagger} \hat{L}_{\mu}
                    \Big)
            \bigg] \, \mathrm{d} t
    \nonumber \\[0mm] \label{eq:SME_general}
        &\ \ \ \ \ \ \ \ \ \ \ \ \ \ \ \ \ \ \ \ \ \ \ \ \ \ \ \ \ \
            +  \sum_{\mu \in \mathcal{M}}
                \sqrt{\eta_{\mu} k_{\mu}} 
                \bigg[
                        \hat{L}_{\mu} \hat{\rho}_t
                        + \hat{\rho}_t \hat{L}_{\mu}^{\dagger}
                        - \Tr(
                            \hat{L}_{\mu} 
                                \hat{\rho}_t
                            + \hat{\rho}_t 
                                \hat{L}_{\mu}^{\dagger}
                            ) 
                            \hat{\rho}_t
                \bigg] \, \mathrm{d} W_t^{(\mu)},
\end{align}
\begin{equation} \label{eq:SME_y_general}
    \mathrm{d} y^{(\mu)}_t 
    = \sqrt{\eta_{\mu} k_{\mu}} 
        \Tr(
            \hat{L}_{\mu} \hat{\rho}_t
            + \hat{\rho}_t \hat{L}_{\mu}^{\dagger}
        ) \, \mathrm{d} t 
        + \mathrm{d} W^{(\mu)}_t
    \ \ \text{ for } \ \ \mu \in \mathcal{M}.
\end{equation}
Analogous to the classical case, the first equation, known as the \textbf{quantum stochastic master equation}, describes the time evolution of the quantum state under random measurement ``force'' $W^{(\mu)}_t$, which is related to the \textit{classical} measurement outcome $y_t^{(\mu)}$ in the second equation. In general, the trajectory of $\hat{\rho}_t$ depends on the measurement history $J_t = \Bigsl\{ y_s^{(\mu)} \ | \ \mu \in \mathcal{M} \text{ and } 0 \leq s < t \Bigsr\}$.

Although Eq.(\ref{eq:SME_general}) seems not to have a control term, we can imagine that the measurement $y_t$ is always automatically fed back to the system due to the axiom of quantum mechanics; hence, intuitively, we can say that Eq.(\ref{eq:SME_general}) has already been put into a closed-loop form without referring to $y^{(\mu)}_t$ explicitly. Theoretically, we can solve Eq.(\ref{eq:SME_general}) by drawing a random sample path of $W^{(\mu)}_t$; however, in practice, $y^{(\mu)}_t$ is what we measured, which helps us pick out one realization of $W^{(\mu)}_t$ and hence $\hat{\rho}_t$ (if we know the initial condition $\hat{\rho}_0$).

Moreover, we see that $W^{(\mu)}_t$ is multiplied by some operator $\hat{L}_{\mu}$ related to the set $\mathcal{M}$, but not to the other set $\mathcal{D}$; this is because we have marginalized the effect of any implicit measurement done by the environment, which on average causes the state $\hat{\rho}_t$ to decay smoothly to some fixed point in the Hilbert space. Moreover, we see that Eq.(\ref{eq:SME_general}) matches Eq.(\ref{eq:SME_proposed_averaged}) if we group terms related to $\mathcal{D}$ and $\mathcal{M}$ into $f_{\text{D}}$ and $f_{\text{M}}$, respectively, and treat $\hat{\rho}_t$ in Eq.(\ref{eq:SME_general}) as $\langle \hat{\rho}_t \rangle$ in Eq.(\ref{eq:SME_proposed_averaged}).

In general, $\Bigsl \{\hat{L}_{\mu} \Bigsr \}$, are known as the \textbf{Lindblad operators}; if $\mu \in \mathcal{D}$ (or $\mathcal{M}$), then $\hat{L}_{\mu}$ is called the Lindblad operator of the decay (or measurement) channel. Nevertheless, quantum mechanics forces both channels to leave a trace on the state via the term
\begin{equation}
    \mathcalboondox{D}\Bigsl[\hat{L}_{\mu} \Bigsr] \hat{\rho}_t
    = \hat{L}_{\mu} \hat{\rho}_t \hat{L}_{\mu}^{\dagger}
                - \frac{1}{2}
                    \left( 
                        \hat{L}_{\mu}^{\dagger} \hat{L}_{\mu} \hat{\rho}_t
                        + \hat{\rho}_t \hat{L}_{\mu}^{\dagger} \hat{L}_{\mu}
                    \right).
\end{equation}
The superoperator $\mathcalboondox{D}\Bigsl[\hat{L}_{\mu} \Bigsr]$ acting on a density operator is sometimes called the \textbf{dissipator} since it is the term responsible for the decoherence. 

As a final remark, let's look at the ensemble-averaged equation of the quantum stochastic master equation: If we assume $\mathrm{d} W^{(\mu)}_t$ has a zero mean and is independent of $\hat{\rho}_t$, all the terms with $\mathrm{d} W^{(\mu)}_t$ vanish when taking the expectation; thus, we obtain a deterministic differential equation
\begin{align} \label{eq:Lindblad_master_eqn}
    \frac{\mathrm{d}}{\mathrm{d} t}
        \mathbb{E}(\hat{\rho}_t)
    &= - \frac{\ci}{\hbar} 
        \left[ 
            \hat{H}_{\text{eff}}, \mathbb{E}(\hat{\rho}_t)
        \right]
        + \sum_{\mu \in \mathcal{D} \cup \mathcal{M}}
            k_{\mu}
            \mathcalboondox{D}\Bigsl[\hat{L}_{\mu} \Bigsr] 
            \mathbb{E}(\hat{\rho}_t),
\end{align}
which is the well-known \textbf{Lindblad master equation}. In other words, if we do not care about all the sample paths that the quantum system can go through, the quantum stochastic master equation reduces to the usual master equation that describes any completely positive trace-preserving map (subject to some mild assumption from operator algebra) \cite{watrous2018theory}.

\subsection{Stochasticity in the Evolution of the Density}
Despite the various analogies discussed so far, there is one decisive difference between the classical and quantum density functions. Classically, a Fokker-Plank equation is generally a deterministic PDE of the density ${\rho}$, and it does not contain information about each realization of the random process. In contrast, the equation for the quantum density operator ${\hat{\rho}}$, i.e., the quantum stochastic master equation, is itself stochastic as shown by Eq.($\ref{eq:SME_general}$). This is because quantum mechanics allows the existence of two layers of randomness. The first layer of randomness is unique to quantum mechanics because our measurement will still be probabilistic even if we know the state of a quantum system. On top of this intrinsic randomness, there is the ``classical'' randomness, which describes our ignorance of a system. Once we average out the stochasticity in the measurement, i.e., Eq.($\ref{eq:Lindblad_master_eqn}$), we indeed obtain a deterministic differential equation.

\subsection{Heisenberg picture}
It should be noted that we have been looking at the problem from the perspective of the quantum state, also known as the Schr\"odinger picture. There is an equivalent way of formulating quantum mechanics, known as the Heisenberg picture, which looks at the time evolution of observables directly without thinking about the quantum state. This equivalence is similar to the relationship between the stochastic differential equation and its Fokker-Planck equation for a classical random process. The Schr\"odinger picture deals with the probability density while the Heisenberg picture looks at the actual random variables/processes. Classically, the Langevin equation, a stochastic differential equation, describes the random evolution of the momentum of a Brownian particle. In the quantum setting, we will see that the quantum Langevin equation can be considered a prescription of open quantum systems in the Heisenberg picture.

\begin{table}
    \centering
    \scriptsize
    \setlength\tabcolsep{6pt}
    \renewcommand{\arraystretch}{0.5}
\begin{tabular}{c | c  c }
\toprule
     {} &  Classical  &  Quantum 
\\  \midrule \midrule \\
    Sample space 
    & $\substack{
            \displaystyle \text{a set $\Omega$ consisted of all possible} 
        \\[1mm]
            \displaystyle \text{real-valued measurement outcomes}
        \\[1mm]
            \displaystyle \text{of a particular experiment}
    }$
    & $\substack{
            \displaystyle \text{a set $\Omega$ consisted of all possible} 
        \\[1mm]
            \displaystyle \text{real-valued measurement outcomes}
        \\[1mm]
            \displaystyle \text{of a particular experiment}
    }$
\\  \\ \hline \\
    Event space & a $\sigma$-algebra $\mathcal{F} = \{F_i\}_i$ on $\Omega$ 
    & $\substack{
            \displaystyle \text{a $\sigma$-algebra $\mathcal{F} = \{F_i\}_i$ on $\Omega$,}
        \\[1mm]
            \displaystyle \text{e.g., the Borel set $\mathcal{B}$ of $\mathbf{R}$}
    }$
\\ \\ \hline \\
    $\substack{
            \displaystyle \text{Probability} 
        \\[1mm]
            \displaystyle \text{measure}
    }$
    & $\mathbb{P}(F_i) = \mu(F_i)$
    & $\substack{
            \displaystyle \text{a POVM $F(\mathcal{F}) = \Bigsl \{ \hat{F}_i \Bigsr \}_{i}$ such that}
        \\[1mm] 
            \displaystyle \text{$\mathbb{P}(F_i) = \bra{\Psi} \hat{F}_i \ket{\Psi}
            = \Tr \Bigsl( \hat{\rho} \hat{F} \Bigsr)$ for $\hat{\rho} \in \mathscr{H}$}
    }$
\\ \\ \hline \\
    Completeness 
    & $\mu (\Omega) = 1$
    & $F(\Omega) = \hat{1}$
\\ \\ \hline \\
    Observable 
    & a real-valued random variable $X:\Omega \rightarrow \mathcal{X}$ 
    & $\text{the spectrum of some Hermitian } \hat{A}$
\\ \\ \hline \\
    Density 
    & $\displaystyle f_{X} (x) \text{ or } \rho(q,p,t)$ 
    & $\displaystyle \hat{\rho} \in \text{a projective Hilbert space } \mathscr{H}$
\\ \\ \hline \\
    Normalization
    & $\displaystyle \int_{\mathcal{X}} f_{X}(x) \, \mathrm{d} x = 1$ 
    & $\displaystyle \Tr ( \hat{\rho} ) = 1$
\\ \\ \hline \\
    Joint Density 
    & $\displaystyle f_{XY} (\cdot, \cdot)$ 
    & $\displaystyle \hat{\rho}_{\mathcal{SE}} \in \mathscr{H}_{\mathcal{S}} \otimes \mathscr{H}_{\mathcal{E}}$ 
\\ \\ \hline \\
    $\substack{
            \displaystyle \text{Marginal} 
        \\[1.5mm]
            \displaystyle \text{density}
    }$
    & $\displaystyle f_X(x) = \int_{\mathcal{Y}} f_{XY}(x,y) \, \mathrm{d} y $ 
    & $\displaystyle \hat{\rho}_{\mathcal{S}} = \Tr_{\mathcal{E}} ( \hat{\rho}_{\mathcal{SE}} )$
\\ \\ \hline \\
    Expectation 
    & $\substack{
            \text{\scriptsize $\displaystyle \mathbb{E}[g(X,Y)] = \iint_{\mathcal{X} \times \mathcal{Y}} g(x,y) f_{XY}(x,y) \, \mathrm{d} x \mathrm{d} y$}
        \\[0.5mm]
            \text{\scriptsize $\displaystyle \mathbb{E}[g(X)] = \int_{\mathcal{X}} g(x) f_{X}(x) \, \mathrm{d} x $}
    }$ 
    & $\substack{
            \displaystyle \Bigsl \langle \hat{A}_{\mathcal{SE}} \Bigsr \rangle = \Tr \Bigsl(\hat{\rho}_{\mathcal{SE}} \hat{A}_{\mathcal{SE}} \Bigsr)
        \\[0.5mm]
            \displaystyle \Bigsl \langle \hat{A}_{\mathcal{S}} \Bigsr \rangle = \Tr_{\mathcal{S}} \Bigsl(\hat{\rho}_{\mathcal{S}} \hat{A}_{\mathcal{S}} \Bigsr)
        }$
\\ \\ \hline \\
    $\substack{
            \displaystyle \text{Cauchy-} 
        \\[1mm]
            \displaystyle \text{Schwarz}
    }$
    & $\displaystyle \sigma_X \sigma_Y \geq \mathrm{Cov}(X,Y)$
    & $\! \! \text{\scriptsize $
        \Delta \hat{A} \Delta \hat{B} 
        \! \geq \! \sqrt{ \left( 
            \frac{1}{2\ci}
                \Bigsl \langle 
                    \Bigsl[ \hat{A}, \hat{B} \Bigsr]
                \Bigsr \rangle
        \right)^{\! 2}
        \! + \! \left(
                \frac{1}{2}
                    \Bigsl \langle 
                        \Bigsl\{ 
                            \hat{A}, \hat{B} 
                        \Bigsr \}
                    \Bigsr \rangle
                \! - \! \Bigsl\langle 
                        \hat{A} 
                    \Bigsr\rangle  
                    \Bigsl\langle 
                        \hat{B} 
                    \Bigsr\rangle 
            \right)^{\! 2}}
    $} \! \!$
\\ \\ \hline \\
    $\substack{
            \displaystyle \text{Liouville} 
        \\[1mm]
            \displaystyle \text{equation}
    }$  & $\displaystyle \frac{\partial \rho}{\partial t} = \{H, \rho\}$ & $\displaystyle \frac{\mathrm{d} \hat{\rho}}{\mathrm{d} t} 
    = - \frac{\ci}{\hbar}
        \Bigsl[ 
            \hat{H}, \hat{\rho} \,
        \Bigsr]$
\\ \\ \hline \\
    $\substack{
            \displaystyle \text{Langevin} 
        \\[1mm]
            \displaystyle \text{equation}
        \\[1mm]
            \displaystyle \text{(white noise)}
    }$
    & $\substack{
            \displaystyle \frac{\mathrm{d} p}{\mathrm{d}t} = - \gamma p_t + \xi_t
        \\[1mm]
            \displaystyle \text{with } \mathbb{E}(\xi_t \xi_{s}) = 2 m \gamma k_{\text{B}} T \delta(t-s)
    }$
    & $\substack{ 
            \displaystyle 
            \frac{\mathrm{d} \hat{a}}{\mathrm{d}t}
            = - \ci \omega_{\text{r}} \hat{a}
                - \frac{\kappa}{2} \hat{a} 
                + \sqrt{\kappa} \hat{a}_{\text{in}}
    \\[1mm]
            \displaystyle \text{with } 
            \left \langle 
                \frac{1}{2} 
                \left \{
                    \hat{a}_{\text{in}}(t),
                    \hat{a}_{\text{in}}^{\dagger}(s)
                \right \}
            \right \rangle 
            = \left(\bar{n} + \frac{1}{2} \right)\delta(t-s)
    }$
\\ \\ \bottomrule
    \end{tabular}
    \caption{A summary of various probabilistic concepts in classical and quantum mechanics. The mapping between classical and quantum-mechanical quantities is not meant to be rigorous. For example, one can show that a classical joint probability distribution does not exist for non-commuting random variables, so the joint density $\hat{\rho}_{\mathcal{SE}}$ is more than a generalization of $f_{XY}$.}
    \label{tab:classical_quantum_probability_theory}
\end{table}


\section{Quantum Channels}
\subsection{Positive Operator Valued Measures}
Similar to the classical control theory, the time evolution of a quantum state in a closed system is determined by a state transition map known as the quantum channel $\mathcalboondox{Q}$\footnote{A quantum channel, in a mathematically precise way, is defined to be a linear completely-positive trace-preserving map. The input and output of the map do not need to have the same dimension. But, of course, for a closed system, the input and output live in the same Hilbert space, so they are of the same dimension.}. If the system is closed, then the channel simply conjugates the initial state $\hat{\rho}_{\mathcal{S}\mathcal{E}}(0)$ by the unitary time-evolution operator $\hat{U}(t)$, i.e., 
\begin{equation}
        \hat{\rho}_{\mathcal{S}\mathcal{E}}(t) 
        = \mathcalboondox{Q}(\hat{\rho}_{\mathcal{S}\mathcal{E}}(0))
        = \hat{U}(t) 
            \hat{\rho}_{\mathcal{S}\mathcal{E}}(0)
            \hat{U}^{\dagger}(t) 
        = \hat{U}(t) 
            \Big[
                \hat{\rho}_{\mathcal{S}}(0)
                \otimes 
                \hat{\rho}_{\mathcal{E}}(0)
            \Big]
            \hat{U}^{\dagger}(t) .
\end{equation}
We have assumed that the state is prepared in a product state (i.e., $\hat{\rho}_{\mathcal{S}\mathcal{E}}(0) = \hat{\rho}_{\mathcal{S}}(0) \otimes \hat{\rho}_{\mathcal{E}}(0)$) such that there is no entanglement between the system and the environment initially.

As motivated before, since we cannot keep track of all events correlated to the environment, the best we can do is to marginalize the environment from the joint probability density. In quantum mechanics, marginalization is executed as taking the partial trace of the composite state over the Hilbert space of the environment. To derive a closed-form formula of the marginal density, suppose $\{\ket{\nu}\}_{\nu}$ diagonalize $\hat{\rho}_{\mathcal{E}}(0)$, i.e.,
\begin{equation}
    \hat{\rho}_{\mathcal{E}}(0) 
    = \sum_{\nu} 
        \lambda_{\nu}
        \ket{\nu} \! \bra{\nu},
\end{equation}
and form an orthonormal basis of $\mathscr{H}_{\mathcal{E}}$. Then, the marginal density, also known as the \textbf{reduced density operator}, is given by
\begin{align}
    \hat{\rho}_{\mathcal{S}}(t)
    &= \Tr_{\mathcal{E}}
        \! \Big[
            \hat{\rho}_{\mathcal{S}\mathcal{E}} (t)
        \Big]
    = \Tr_{\mathcal{E}}
        \! \bigg \{
        \hat{U}(t) 
        \Big[
            \hat{\rho}_{\mathcal{S}}(0)
            \otimes 
            \hat{\rho}_{\mathcal{E}}(0)
        \Big]
        \hat{U}^{\dagger}(t) 
        \bigg \}
\nonumber \\
    &= \sum_{\nu}
        \bra{\nu} 
        \hat{U}(t) 
            \!\left(
                \hat{\rho}_{\mathcal{S}}(0)
                \otimes 
                \sum_{\mu} 
                    \lambda_{\mu}
                    \ket{\mu} \! \bra{\mu}
            \right) \!
        \hat{U}^{\dagger}(t) 
        \ket{\nu}
\nonumber \\
    &= \sum_{\mu, \nu}
        \underbrace{\sqrt{\lambda_{\mu}}
        \bra{\nu} 
            \hat{U}(t) 
        \ket{\mu}}_{\displaystyle \hat{K}_{\nu \mu} \in L(\mathscr{H}_{\mathcal{S}})}
        \hat{\rho}_{\mathcal{S}}(0)
        \underbrace{\sqrt{\lambda_{\mu}}
        \bra{\mu} 
            \hat{U}(t) 
        \ket{\nu}}_{\displaystyle \hat{K}_{\nu \mu}^{\dagger}}.
\end{align}
Hence, we have shown that a quantum channel restricted/marginalized to the system can be represented in the so-called \textbf{Kraus representation} \cite{breuer2002theory, nielsen_chuang_2010, watrous2018theory}
\begin{equation}
    \mathcalboondox{Q}
        (\hat{\rho}_{\mathcal{S}}(0))
    = \sum_{\mu, \nu} 
        \hat{K}_{\nu \mu} \hat{\rho}_{\mathcal{S}}(0) \hat{K}^{\dagger}_{\nu \mu},\\[-3mm]
\end{equation}
where $\{\hat{K}_{\nu \mu}\}$ are the \textbf{Kraus operators}, usually non-Hermitian, satisfying the completeness relation
\begin{equation} \label{eq:kraus_op_completeness}
    \sum_{\mu, \nu}
        \hat{K}_{\nu \mu}^{\dagger}
        \hat{K}_{\nu \mu} 
    = \hat{1}.
\end{equation}
The derivation above assumes a specific initial state (i.e., the environment is uncorrelated with the system at $t=0$); nevertheless, it turns out that any quantum channel has at least one Kraus representation \cite{watrous2018theory}.

In addition, we call
\begin{equation}
    \hat{F}_{\nu \mu} = \hat{K}_{\nu \mu}^{\dagger}
        \hat{K}_{\nu \mu} 
\end{equation}
the \textbf{effect}. According to Eq.(\ref{eq:kraus_op_completeness}), the effects sum to unity and thus are called the \textbf{positive operator valued measures (POVM)} since they generalize the concept of probability measures (see Table \ref{tab:classical_quantum_probability_theory} for more details).

\subsection{Example: Qubit Decay Channel}
As an example, let's consider a simple model of spontaneous emission, which is described by the two Kraus operators
\begin{equation}
    \hat{K}_0 
    = \begin{pmatrix}
        1 & 0 \\
        0 & \mu
    \end{pmatrix}
    \ \ \text{ and } \ \ 
    \hat{K}_1 
    = \begin{pmatrix}
        0 & \lambda \\
        0 & 0
    \end{pmatrix}
\end{equation}
with $|\mu|^2 + |\lambda|^2 = 1$ for conservation of probability. We are using the convention that $\ket{g} = (1 \ 0)^{\top}$ and $\ket{e} = (0 \ 1)^{\top}$ represent ground and excited state, respectively. In the Kraus operator $\hat{\Phi}_1$, $|\lambda|^2$ is the probability that the atom decays to the ground state by emitting a photon; hence, if the atom is initially excited, the state will go from $\ket{e}$ to $\ket{g}$ with \textit{probability amplitude} $\lambda$. However, there is also a probability $1 - |\lambda|^2 = |\mu|^2$ that the atom stays excited, which is why $\mu$ appears in $\hat{\Phi}_0$. If the atom is initially in the ground state, then it will remain unexcited forever as implied by $\hat{\Phi}_0$. The effects corresponding to the Kraus operators are
\begin{equation}
    \hat{F}_0 
    = \begin{pmatrix}
        1 & 0 \\
        0 & |\mu|^2
    \end{pmatrix}
    \ \ \text{ and } \ \ 
    \hat{F}_1 
    = \begin{pmatrix}
        0 & 0 \\
        0 & |\lambda|^2
    \end{pmatrix},
\end{equation}
which clearly satisfy the completeness relation.

\subsection{The Measurement Channel}
The ideal quantum measurement can be formulated under the Kraus formalism if we identify all the orthogonal projection operators as the Kraus operators. To see this, recall that each post-measurement state takes the form
\begin{equation}
    \hat{\rho}_{\mathcal{S}, a}
    =  \mathcalboondox{Q}_{a}(\hat{\rho}_{\mathcal{S}})
    = \frac{\hat{\Pi}_a 
        \hat{\rho}_{\mathcal{S}} 
        \hat{\Pi}_a }{\Tr(\hat{\Pi}_a \hat{\rho}_{\mathcal{S}} \hat{\Pi}_a^{\dagger})}
    \ \ \text{ with probability } \ \ 
    \Tr(\hat{\Pi}_a \hat{\rho}_{\mathcal{S}} \hat{\Pi}_a^{\dagger}).
\end{equation}
In addition, the projection operators clearly satisfy the completeness relation. Summing over all possible measurement outcomes (and weighted by the corresponding probability) yields the Kraus representation of the measurement channel
\begin{equation}
    \hat{\rho}_{\mathcal{S}}'
    = \mathbb{E}[ \mathcalboondox{Q}_{a}(\hat{\rho}_{\mathcal{S}})]
    = \sum_{a} \Tr(\hat{\Pi}_a \hat{\rho}_{\mathcal{S}} \hat{\Pi}_a^{\dagger})  \mathcalboondox{Q}_a(\hat{\rho}_{\mathcal{S}})
    = \sum_{a} \hat{\Pi}_a \hat{\rho}_{\mathcal{S}} \hat{\Pi}^{\dagger}
    \doteq  \mathcalboondox{Q}(\hat{\rho}_{\mathcal{S}}).
\end{equation}
    
The Kraus representation is derived by discarding all information gained from the system-environment interaction; however, when doing the active measurement, we shouldn't simply average over all the measurement outcomes. In fact, we know that the state will be projected onto one of $ \mathcalboondox{Q}_{a}(\hat{\rho}_{\mathcal{S}})$, conditioned on the measurement outcome. From this perspective, we see that the role of the Kraus representation is very much like that of the Fokker-Planck equation, whereas tracking one sample path requires us to model the sequence of outcomes as a random process. 

\subsection{The Generalized Measurement Channel}
As a generalization to the projective measurement, the Kraus operators can be used to define a non-projective measurement of some physical observable $Y$:
\begin{equation}\label{eq:POVM_with_prob}
    \hat{\rho}_{\mathcal{S},y}
    = \mathcalboondox{Q}_{y}(\hat{\rho})
    = \frac{\hat{K}_{y} \hat{\rho}_{\mathcal{S}} \hat{K}_{y}^{\dagger} }{\Tr(\hat{K}_{y} \hat{\rho}_{\mathcal{S}} \hat{K}_{y}^{\dagger} )}
    \ \ \text{ with probability } \ \
    \Tr(\hat{K}_{y} \hat{\rho}_{\mathcal{S}} \hat{K}_{y}^{\dagger} ),
\end{equation}
where $Y=y$ represents some partial information gathered about the system by passing the state through a non-unitary quantum channel and $\hat{K}_{y}$ is the non-projective backaction associated with the partial information $y$. For example, imagine we have a photodetector used to measure the emission of atoms from a system. Because the detector is imperfect, it may only capture $50 \%$ of the photons coming out from the system. Since we only gain partial information about the system, the state conditioned on our imperfect measurement will be determined by a non-unitary quantum channel whose job is to average over all the possible ways that the other $50\%$ of the photons can leak into the environment. Eq.(\ref{eq:POVM_with_prob}) thus successfully combines two approaches in dealing with an open quantum system:
\begin{enumerate}
    \item[(i)]
        For the environment or anything we don't have access to, we will average over all the conditional states to obtain a deterministic expression for the (unconditional) density.
    \item[(ii)]
        For information that we can and want to gather, we will simulate the sample path conditioned on the past measurement outcomes.
\end{enumerate}

\subsection{The Discrete-Time Stochastic Equation}
In deriving the Kraus operators, we have ignored the details that happen within $[0,t]$, which can be thought of as a discretization of the time axis. Suppose we want to know the density at $T$, we can always divide the interval into $(0,T/N), (T/N, 2T/N), ..., ((N-1)T/N, T)$ and apply the time evolution in a discretized fashion:
\begin{gather} \label{eq:discrete_stochastic_equation}
    \hat{\rho}_{\mathcal{S},k+1}
    =  \mathcalboondox{Q}^{(k)}(\hat{\rho}_{\mathcal{S},k})
    = \sum_{\mu, \nu} 
        \hat{K}^{(k)}_{\nu \mu} \hat{\rho}_{\mathcal{S}, k} \hat{K}^{(k)\dagger}_{\nu \mu},
        \ \ \ \ 
        k = 0,1,..., N-1,
\\ 
    \text{with initial state} \ \ 
    \hat{\rho}_{\mathcal{S}, 0}.
\nonumber
\end{gather}

By thinking about time evolution in this way, we can add measurement as a part of the chain of quantum channels even though the backaction happens instantaneously. In fact, it is the active measurement that adds randomness to the time evolution\footnote{Again, the monitoring from the environment also adds randomness to $\hat{\rho}_{\mathcal{S}}$; however, the partial trace has averaged out the effect, so we don't need any random process to describe this part. In other words, everything being averaged over becomes ``classical'', and things not being averaged over are ``quantum''. What should be considered as the ``environment'' inspires lots of discussions in the field; this problem of assigning the boundary among the system, detector, and environment is known as the \textbf{Heisenberg cut}.}, so the state transition should be defined in terms of the conditional probability, which has the same spirit as the transition matrix for describing a time-inhomogeneous Markov chain: For example, if we made a non-projective measurement at time step $k$, the conditional state at $k+1$ would be
\begin{equation} \label{eq:discrete_time_SME}
    \hat{\rho}_{\mathcal{S},k+1}
    = \mathcalboondox{Q}^{(k)}_{y}(\hat{\rho}_{k})
    \doteq \frac{\hat{K}_{y}^{(k)} \hat{\rho}_{\mathcal{S},k} \hat{K}_{y}^{(k)\dagger} }{\Tr(\hat{K}_{y}^{(k)} \hat{\rho}_{\mathcal{S},k} \hat{K}_{y}^{(k)\dagger} )}
    \ \ \text{if $Y=y$ is measured}
\end{equation}
and the transition probability is
\begin{equation}
    \mathbb{P}\left[\hat{\rho}_{\mathcal{S},k+1}
    =  \mathcalboondox{Q}^{(k)}_{y}(\hat{\rho}_{\mathcal{S},k}) \, | \, \hat{\rho}_{\mathcal{S},k}\right]
    = \mathbb{P}(Y=y \, | \,  \hat{\rho}_{\mathcal{S},k})
    = \Tr(\hat{K}_{y}^{(k)} \hat{\rho}_{\mathcal{S},k} \hat{K}_{y}^{(k)\dagger}).
\end{equation}
Naturally, our next step is to take the limit of Eq.(\ref{eq:discrete_time_SME}) as $T/N \rightarrow 0$ and model the uncertainty in the measurement as a continuous-time random process.
    
\subsection{Example: Measurement via Dispersive Coupling}\label{section:measurement_via_dispersive_coupling_kraus}
\begin{figure}[t]
    \centering
    \includegraphics[scale=0.28]{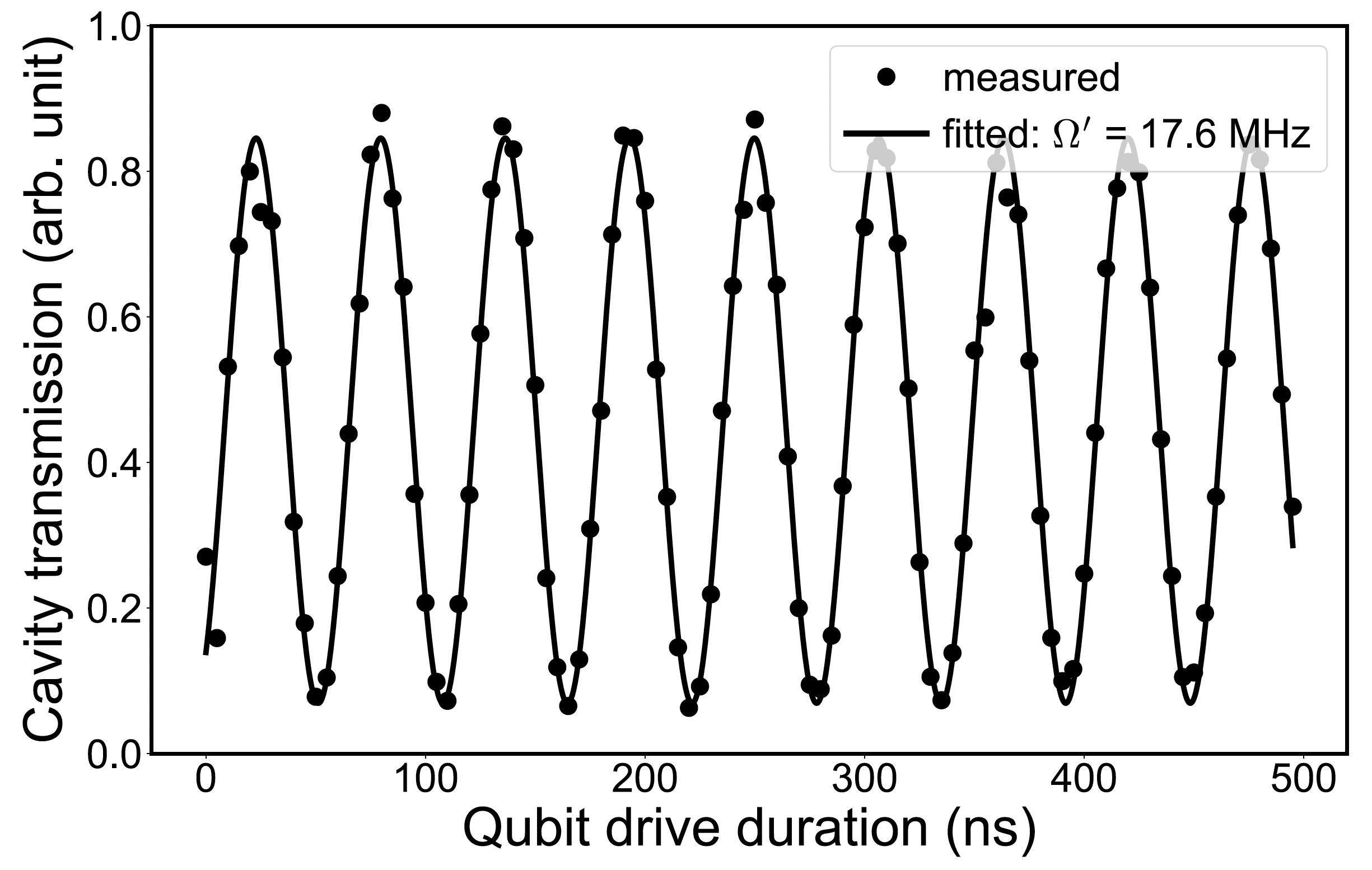}
    \caption{Experimentally measured Rabi oscillations inferred from the averaged cavity transmission. The readout cavity is in resonance with the readout drive when the qubit is in the ground state. When the qubit is excited by a control drive, the cavity frequency shifts slightly and causes the transmission to drop.}
    \label{fig:rabi_amplitude}
\end{figure}

A slightly more involved example is the measurement of coherent states from a resonator coupled to a qubit dispersively \cite{PhysRevA.74.042318, PhysRevA.76.042319, cQED_lectures_Girvin, doi:10.1063/1.5089550}. In a typical experiment, one quadrature of the electromagnetic field leaking out of the readout resonator will be measured and the state of the qubit will be inferred by looking at the change of transmission as shown in Figure \ref{fig:rabi_amplitude}. However, since the coherent state has an uncertainty along both quadrature axes, the outcome of the measurement will be random. More importantly, since the qubit is coupled to the resonator dispersively, a projective measurement of one quadrature operator of the resonator will also project the qubit to some new state. It is this measurement backaction that we are interested in eventually since we do not want the qubit state to vary while making a continuous measurement.

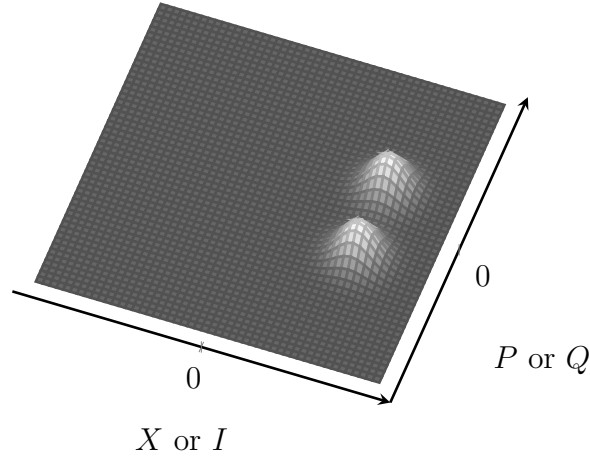
\begin{figure}[t]
    \centering
    \begin{tikzpicture}
    \begin{axis}[
        colormap={bw}{
        gray(0cm)=(0.4);
        gray(1cm)=(1);
        },
        hide z axis,
        view={20}{88},
        xlabel=${X \text{ or } I}$,
        ylabel=${P \text{ or } Q}$,
        axis x line=bottom,
        axis y line=left,
        xmin=-5.5,
        xmax=5.5,
        ymin=-5.5,
        ymax=5.5,
        zmax=0.3,
        xtick={0},
        ytick={0},
        ]
    \addplot3[
        surf,
        samples={51},
        samples y={51},
        domain=-5:5,
        domain y=-5:5,
        ] 
        {e^(-(2*(x-3)^2 + 2*(y-1.2)^2)) + e^(-(2*(x-3)^2 + 2*(y+1.2)^2))};
    \end{axis}
    \end{tikzpicture}
    \caption{Probability distribution associated with the heterodyne detection when the qubit is in an equal superposition state. For a homodyne detection, which measures only one of the quadratures, simply take the projection of the joint probability onto the $Q$-axis.}
    \label{fig:dispersive_shift_3D}
\end{figure}

Note that Figure \ref{fig:rabi_amplitude} measures the change of amplitude of the cavity transmission by intentionally setting the readout frequency to be $\omega_{\text{r}} + \chi$. Hence, a drop in the transmission appears when the cavity frequency shifts to $\omega_{\text{r}} - \chi$. Alternatively, we can also set the readout frequency at $\omega_{\text{r}}$. Although the amplitude of the transmission will be the same when the qubit is in $\ket{g}$ or $\ket{e}$, there will be a phase difference in the transmitted signal in the two cases. In this case, for each possible measurement outcome $q \in \mathbf{R}$ of the quadrature observable $Q_k$ ($k$ being the time index), the Kraus operator for the qubit (i.e., the ``system'' in this example) is given by
\begin{equation}\label{eq:dispersive_measurement_channel}
    \hat{K}_{q} 
    = \left(\frac{2}{\pi} \right)^{\frac{1}{4}} 
            e^{-(\Bar{q} - q)^2}
            \ket{g} \! \bra{g}
        + \left(\frac{2}{\pi} \right)^{\frac{1}{4}} 
            e^{-(\Bar{q} + q)^2}
            \ket{e} \! \bra{e},
\end{equation}
and the probability density\footnote{Notice that each $Q_k$ is a continuous random variable, so we work with probability density instead and define the channel to be
\begin{equation}
     \mathcalboondox{Q}(\hat{\rho})
    = \int_{-\infty}^{\infty}
            \hat{K}_{q} \hat{\rho} 
            \hat{K}_{q}^{\dagger}
            dq.
\end{equation}} of measuring $Q_k = q$ is 
\begin{align}
    f(Q_k = q \, | \, \hat{\rho}_{\mathcal{S},k} = \hat{\rho})
    &= \Tr(
            \hat{K}_{q}
            \hat{\rho}
            \hat{K}_{q}^{\dagger}
        )
\nonumber \\ \label{eq:prob_measuring_q}
    &= \left(\frac{2}{\pi} \right)^{\frac{1}{2}} 
            e^{-2(\Bar{q} - q)^2}
            \bra{g} \hat{\rho} \ket{g}
        + \left(\frac{2}{\pi} \right)^{\frac{1}{2}} 
            e^{-2(\Bar{q} + q)^2}
            \bra{e} \hat{\rho} \ket{e}.
\end{align}
The number $\pm \Bar{q}$, to be discussed in detail in the later sections, is the expected quadrature field gathered from the cavity within $\mathrm{d}t$ depending on the state of the qubit. It is a function of the dispersive shift $\chi$, the cavity decay rate, and the number of photons used to perform the readout. Moreover, it's not hard to see that Eq.(\ref{eq:prob_measuring_q}) is essentially a superposition of two Gaussian distributions as shown in Figure \ref{fig:dispersive_shift_3D}, hence:
\begin{enumerate}
    \item[(i)] If the qubit is mostly in the ground state, i.e., $\bra{g} \hat{\rho} \ket{g} \gg \bra{e} \hat{\rho}\ket{e}$, then our measurement $Q_k$ is drawn from a Gaussian centered at $\Bar{q}$ with a variance of $1/4$ as expected from measuring a single coherent state. In addition, it's assumed that the measurement is ideal at this point; later, we will include measurement efficiency so that the variance will be larger than $1/4$.
    \item[(ii)] If the qubit is mostly in the excited state, i.e., $\bra{g} \hat{\rho} \ket{g} \ll \bra{e} \hat{\rho} \ket{e}$, then $Q_k$ is drawn from a Gaussian centered at $-\Bar{q}$ with the same variance $1/4$.
    \item[(iii)] If the qubit is in an equal superposition of $\ket{g}$ and $\ket{e}$, it's hard to infer useful information from the measurement besides the ratio of population in the $z$-basis. This is because the dispersive coupling is designed to gain information along the $z$-axis in the Bloch sphere. To retrieve the relative phase in the superposition state, we need to first rotate the qubit so that the dispersive measurement can give out information along another axis. 
\end{enumerate}

In general, the measurement channel would be different at each time step; however, the channel defined in Eq.(\ref{eq:dispersive_measurement_channel}) implies a \textit{time-homogeneous Markov process} since it's identical for all time steps.

\section{Quantum Langevin Equations}
The discussions in the last two sections were considerably abstract. To quantify some of the parameters (e.g., the strength $k_{\mu}$ of decay and measurement, the Kraus operator $\hat{K}_{\mu \nu}$, the dispersive shift $\bar{q}$, etc.), we have to start with a physical model and derive the time evolution of the observables or the associated density. The starting point will be the equivalent circuit introduced in Section \ref{section:unitary_model_of_dissipation} and the circuit schematic is redrawn in Figure \ref{fig:tline_resonator_coupling_analysis_QLE}. Keep in mind that the capacitor $C$ and transmission line impedance $Z_0$ in Figure \ref{fig:tline_resonator_coupling_analysis_QLE} are the transformed capacitance and resistance.
\subsection{Quantum Analog of Kirchhoff's Current/Voltage Law} 
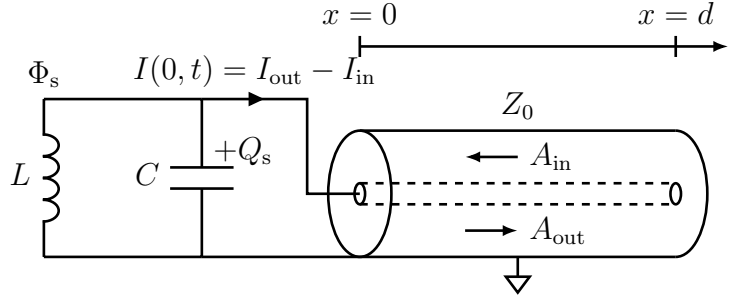
\begin{figure}
    \centering
    \begin{tikzpicture}[scale=0.7]
    \ctikzset{tripoles/mos style/arrows};
        \draw
            (0,3) to (3,3) 
            (3,0) to (0,0) 
            (0,3) to [/tikz/circuitikz/bipoles/length=40pt, L, l_=$L$] (0,0)
            (3,3) to (3,2) to [/tikz/circuitikz/bipoles/length=40pt, C, l_=$C$] (3,1) to (3,0);
        \draw 
            (0,3) node[above]{${\Phi_{\text{s}}}$}
            (3,2) node[right]{${+Q_{\text{s}}}$};
        \draw
            (3,3) to [short, i>^={${I(0,t)= I_{\text{out}} - I_{\text{in}}}$}](5,3) to (5, 1.2) to (6, 1.2) 
            (6,0) to (5,0) to (3,0); 
    \begin{scope}[shift={(6,1.2)}]
        \draw 
            (0,0) ellipse (0.6 and 1.2);
        \draw 
            (6,-1.2) arc(-90:90:0.6 and 1.2);
        \draw
            (0,1.2) to (6,1.2)
            (0,-1.2) to (6,-1.2)
            (3, 1.2) node[above]{$Z_0$};
        \draw[dashed]
            (0,0.2) to (6,0.2)
            (0,-0.2) to (6,-0.2);
        \draw 
            (0,0) ellipse (0.1 and 0.2)
            (6,0) ellipse (0.1 and 0.2);
        \draw 
            (3,-1.2) node[/tikz/circuitikz/bipoles/length=30pt,sground]{};
        \draw
            (0,3) to (0,2.6)
            (0,3) node[above]{$x=0$}
            (6,3) to (6,2.6)
            (6,3) node[above]{$x=d$};
        \draw[-{Latex[length=2mm]}]
            (0,2.8) to (7,2.8);
        \draw[-{Latex[length=2mm]}]
            (2,-0.7) to (3,-0.7);
        \draw[-{Latex[length=2mm]}]
            (3,0.7) to (2,0.7);
        \draw 
            (3, 0.7) node[right]{$A_{\text{in}}$}
            (3, -0.7) node[right]{$A_{\text{out}}$};
    \end{scope}
    \end{tikzpicture}
    \caption{A damped parallel LC circuit used for the derivation of the QLE.}
    \label{fig:tline_resonator_coupling_analysis_QLE}
\end{figure}

The general derivation of a \textbf{quantum Langevin equation (QLE)} can be found in standard textbooks \cite{gardiner2004quantum}. Instead of deriving the QLE for an arbitrary system, we focus on the example of Figure \ref{fig:tline_resonator_coupling_analysis_QLE} and show a heuristic construction of the QLE from classical equations of circuit analysis \cite{Vool_2017_into_cQED}. 

We are interested in the time evolution of the system observables (i.e., charge on $C$ and node flux of $L$). For instance, if we want to know the rate of change of the charge $Q_{\text{s}}$ on $C$, we essentially need to calculate the net current flowing towards the positive plate of $C$. Let the current flowing into the transmission line be $I(t) = I(x=0,t)$ and the voltage at $x=0$ be $V(t) = V(x=0,t)$, then
\begin{equation*}
    I = \frac{V}{Z_0} - \frac{V}{Z_0} + I
    = \frac{V}{Z_0} 
        - \frac{2}{\sqrt{Z_0}} \frac{1}{2\sqrt{Z_0}} 
            \Big(
                V - Z_0 I
            \Big)
    = \frac{V}{Z_0} - \frac{2 A_{\text{in}}}{\sqrt{Z_0}},
\end{equation*}
where the input and output power wave amplitudes are defined to be
\begin{equation} \label{eq:power_wave_amplitude}
    A_{\text{in}/\text{out}}(t)
    = A_{\text{in}/\text{out}}(x=0, t) 
    = \frac{1}{2\sqrt{Z_0}}
        \Big[ 
            V(x=0, t)
            \mp Z_0 I(x=0, t)
        \Big]
\end{equation}
based on classical microwave theory \cite{pozar1990microwave}. In addition, recall that we can decompose the net voltage/current on the transmission into the input and output voltages/currents:
\begin{equation}
    V(x,t) = V_{\text{out}} (x,t) + V_{\text{in}} (x,t),
\end{equation}
\begin{equation}
    I(x,t) = I_{\text{out}} (x,t) - I_{\text{in}} (x,t).
\end{equation}
These quantities are further related to $A_{\text{in}}$ and $A_{\text{out}}$ by
\begin{equation}
    A_{\text{in}/\text{out}} (x,t)
    = \frac{1}{\sqrt{Z_0}} V_{\text{in}/\text{out}}  (x,t)
    = \sqrt{Z_0} \, I_{\text{in}/\text{out}}  (x,t).
\end{equation}

Next, recall that the Heisenberg equation for the charge operator is given by
\begin{equation}
    \frac{\mathrm{d}}{\mathrm{d} t} \hat{Q}_{\text{s}, \text{H}}
    = - \frac{\ci}{\hbar} 
        \Big[
            \hat{Q}_{\text{s}, \text{H}},
            \hat{H}_{\text{LC}, \text{H}}
        \Big]
        + \left(
                \frac{\partial \hat{Q}_{\text{s}, \text{S}}}{\partial t}
            \right)_{\text{H}},
\end{equation}
where we have explicitly shown the time-dependence of the operator in the Schr\"{o}dinger picture (with the subscript $\text{S}$) which is usually ignored. The partial time derivative allows us to add input and output to the LC oscillator; in other words, we identify the time-dependence of the charge operator in the Schr\"{o}dinger picture as the current leaving the load (i.e., $\hat{I}$). This results in 
\begin{equation} \label{eq:QLE_for_LC}
    \frac{\mathrm{d}}{\mathrm{d} t} \hat{Q}_{\text{s}, \text{H}}
    = - \frac{\hat{\Phi}_{\text{s}, \text{H}}}{L} - \hat{I}
    = - \frac{\hat{\Phi}_{\text{s}, \text{H}}}{L} 
        - \frac{1}{Z_0} \frac{\mathrm{d}}{\mathrm{d} t} \hat{\Phi}_{\text{s}, \text{H}} 
        + \frac{2 \hat{A}_{\text{in}, \text{H}}}{\sqrt{Z_0}},
\end{equation}
where we have used the commutation relation $\Bigsl[\hat{\Phi}, \hat{Q}\Bigsr] = \ci \hbar$ and have promoted all the classical variables to the corresponding operators.

Eq.(\ref{eq:QLE_for_LC}) is the QLE for a parallel LC oscillator coupled to a one-dimensional free space, which is nothing more than a quantum-mechanical restatement of the classical Kirchhoff's current law. Nevertheless, the quantum equation allows one to discuss the fluctuation and dissipation of the system in the quantum language. 

\subsection{The General QLE} 
For a wide range of problems, the environment $\mathcal{E}$ defined in an open quantum system can be modeled as a collection of harmonic oscillators with degrees of freedoms $\{(\hat{q}_m, \hat{p}_m)\}_m$ (e.g., the electromagnetic field is a continuum of harmonic oscillators):
\begin{equation}
    \hat{H}_{\mathcal{E}} 
    = \sum_n 
        \frac{1}{2}
        \Big( 
            \hat{p}_n^2 
            + \omega_n^2 \hat{q}_n^2
        \Big),
\end{equation}
where $\omega_n$ is the frequency of the $n$th oscillator. As usual, we define the annihilation operators of the harmonic oscillators in the environment to be
\begin{equation}
    \hat{a}_n = \frac{\omega_n \hat{q}_n + \ci \hat{p}_n}{\sqrt{2\hbar \omega_n}}.
\end{equation}
Next, given the system of interests $\mathcal{S}$, e.g., a transmon qubit or a cavity, we denote the system Hamiltonian by some generic function $\hat{H}_{\mathcal{S}}(\{\hat{Z}_{\alpha}\})$ of the system operators $\{\hat{Z}_{\alpha}\}_{\alpha}$. The composite system of $\mathcal{S}$ and $\mathcal{E}$ is described by the tensor product of the two corresponding Hilbert spaces, i.e., $\mathscr{H}_{\mathcal{SE}} = \mathscr{H}_{\mathcal{S}} \otimes \mathscr{H}_{\mathcal{E}}$ with the usual commutation relations
\begin{gather}
    \Big[\hat{Z}_{\alpha} , \hat{q}_n\Big]
    = \Big[\hat{Z}_{\alpha} , \hat{p}_n\Big]
    = 0, \ \ \ \ 
    \forall \, \alpha, n,
\\
    \Big[ \hat{p}_n, \hat{p}_m \Big]
    = \Big[ \hat{q}_n, \hat{q}_m \Big]
    = 0, \ \ \ \ 
    \forall \, m, n,
\\
    \Big[ \hat{q}_n, \hat{p}_m \Big]
    = \ci \hbar \delta_{nm},
    \ \ \ \ 
    \forall \, m, n.
\end{gather}

In the weak-coupling limit, we can always express the interaction between the system and environment as
\begin{equation} \label{eq:general_QLE_hamiltonian_for_derivation}
    \hat{H} 
    = \hat{H}_{\mathcal{S}}(\{\hat{Z}_{\alpha}\}) + \frac{1}{2} 
    \sum_n \left[ \left(\hat{p}_n - \kappa_n \hat{X}\right)^{\!2} + \omega_n^2 \hat{q}_n^2\right],
\end{equation}
where $\hat{X} \in \{\hat{Z}_{\alpha}\}$ is some system operator that is responsible for the coupling and $\kappa_n$ is the frequency-dependent coupling strength. Consequently, we can show that any system observables $\hat{Y} \in \{\hat{Z}_{\alpha}\}$, starting from time $t_0$, follows the Heisenberg equation \cite{gardiner2004quantum}
\begin{equation} \label{eq:general_QLE_no_approximation}
    \dot{\hat{Y}}
    = - \frac{\ci}{\hbar} 
            \Bigsl[
                \hat{Y}, 
                \hat{H}_{\mathcal{S}} 
            \Bigsr]
    - \frac{\ci}{2\hbar}
        \bigg\{
            \Bigsl[
                \hat{X}, \hat{Y}
            \Bigsr], 
            \hat{\xi}(t)
            - f(t - t_0)\hat{X}(t_0)
            - \int_{t_0}^{t} 
                \mathrm{d} \tau
                f(t-\tau) \dot{\hat{X}}(\tau)
        \bigg\},
\end{equation}
where
\begin{equation}
    \hat{\xi}(t)
    = \ci \sum_n \kappa_n \sqrt{\frac{\hbar \omega_n}{2}}
        \Big[
            - e^{-\ci\omega_n (t-t_0)}\hat{a}_n(t_0)
            + e^{\ci\omega_n (t-t_0)}\hat{a}_n^{\dagger}(t_0)
        \Big]
\end{equation}
can be thought of as the information entering the system and
\begin{equation} \label{eq:memory_function_definition}
    f(t) 
    = \sum_n \kappa_n^2
        \cos(\omega_n t)
\end{equation}
is a memory function describing how long the correlation between the system and the environment would last. Eq.(\ref{eq:general_QLE_no_approximation}) is the general QLE for any arbitrary system, and we emphasize that everything at this point follows from the Hamiltonian in Eq.(\ref{eq:general_QLE_hamiltonian_for_derivation}) without any approximation.

\subsection{A Series LC Oscillator Coupled to a Transmission Line}
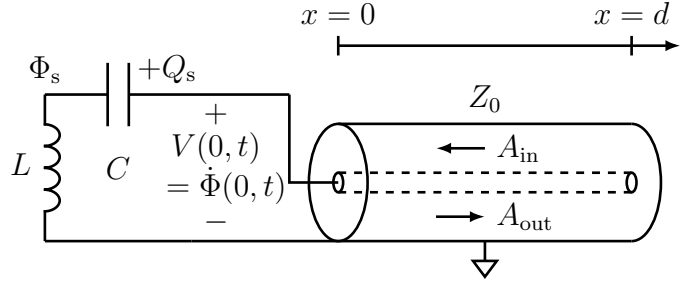
\begin{figure}[t]
    \centering
    \begin{tikzpicture}[scale=0.65]
    \ctikzset{tripoles/mos style/arrows};
        \draw
            (3,0) to (0,0) 
            (0,3) to [/tikz/circuitikz/bipoles/length=40pt, L, l_=$L$] (0,0)
            (0,3) to [/tikz/circuitikz/bipoles/length=40pt, C, l_=$C$] (3,3) to (5,3);
        \draw 
            (0,3) node[above]{${\Phi_{\text{s}}}$}
            (2.5,3) node[above]{${+Q_{\text{s}}}$};
        \draw 
            (3.5,2.6) node[]{$+$} 
            (3.5,1.5) node[]{${\substack{\displaystyle V(0,t) \\ \displaystyle \ \ =\dot{\Phi}(0,t)}}$}
            (3.5,0.4) node[]{$-$};
        \draw
            (5,3) to (5, 1.2) to (6, 1.2) 
            (6,0) to (5,0) to (3,0); 
    \begin{scope}[shift={(6,1.2)}]
        \draw 
            (0,0) ellipse (0.6 and 1.2);
        \draw 
            (6,-1.2) arc(-90:90:0.6 and 1.2);
        \draw
            (0,1.2) to (6,1.2)
            (0,-1.2) to (6,-1.2)
            (3, 1.2) node[above]{$Z_0$};
        \draw[dashed]
            (0,0.2) to (6,0.2)
            (0,-0.2) to (6,-0.2);
        \draw 
            (0,0) ellipse (0.1 and 0.2)
            (6,0) ellipse (0.1 and 0.2);
        \draw 
            (3,-1.2) node[/tikz/circuitikz/bipoles/length=30pt,sground]{};
        \draw
            (0,3) to (0,2.6)
            (0,3) node[above]{$x=0$}
            (6,3) to (6,2.6)
            (6,3) node[above]{$x=d$};
        \draw[-{Latex[length=2mm]}]
            (0,2.8) to (7,2.8);
        \draw[-{Latex[length=2mm]}]
            (2,-0.7) to (3,-0.7);
        \draw[-{Latex[length=2mm]}]
            (3,0.7) to (2,0.7);
        \draw 
            (3, 0.7) node[right]{$A_{\text{in}}$}
            (3, -0.7) node[right]{$A_{\text{out}}$};
    \end{scope}
    \end{tikzpicture}
    \caption{A damped series LC circuit used for the derivation of the QLE.}
    \label{fig:tline_series_resonator_coupling_QLE}
\end{figure}
Since we have already solved the circuit in Figure \ref{fig:tline_resonator_coupling_analysis_QLE}, which is a parallel LC circuit coupled to a transmission line, let us use the general QLE to solve for a series LC circuit coupled to a transmission line (see Figure \ref{fig:tline_series_resonator_coupling_QLE}) instead. To do so, we need a full Hamiltonian describing the series LC, the semi-infinite transmission line, and as well as their coupling. By classical circuit analysis \cite{PhysRevA.29.1419}, one finds the Lagrangian and Hamiltonian to be
\begin{align}
    &\mathcal{L}(\Phi_{\text{s}}(t), \dot{\Phi}_{\text{s}}(t), \Phi(x,t), \dot{\Phi}(x,t))
\nonumber \\[-1mm]
    &= \frac{C}{2} [\dot{\Phi}_{\text{s}}(t) - \dot{\Phi}(0,t)]^2 
    - \frac{1}{2L}\Phi_{\text{s}}(t)^2
    + \int_{0}^{\infty}
        \mathrm{d} x \left[\frac{C_l}{2} \left(\frac{\partial \Phi }{\partial t}\right)^2
        - \frac{1}{2L_l} \left(\frac{\partial \Phi }{\partial x}\right)^2\right]
\nonumber \\[-1mm] 
    &= \frac{C}{2} \dot{\Phi}_{\text{s}}(t)^2 
        - \frac{1}{2L}\Phi_{\text{s}}(t)^2
\nonumber \\
    & \ \ \ \ 
    + \underbrace{
        \int_{0}^{\infty}
            \mathrm{d} x \left[\frac{C_l + 2 \delta(x) C}{2} \left(\frac{\partial \Phi }{\partial t}\right)^2
            - \frac{1}{2L_l} \left(\frac{\partial \Phi }{\partial x}\right)^2\right]
        - \int_{0}^{\infty}
            \mathrm{d} x \, 2 \delta(x) C \dot{\Phi}_{\text{s}}(t) \dot{\Phi}(x,t)}_{
        \text{(We can identify the integrand as a Lagrangian density $\mathscr{L}$.)}
        }
\end{align}
and\footnote{Note that 
\begin{equation}
    \int_{0}^{\infty} \delta(x) \mathrm{d}x = \frac{1}{2}. 
\end{equation}
} 
\begin{align}
    &\mathcal{H}(\Phi_{\text{s}}(t), Q_{\text{s}}(t), \Phi(x,t), \Pi(x,t))
\nonumber \\[-1mm] \label{eq:series_LC_t_line_hamiltonian}
    &= \frac{1}{2C} Q_{\text{s}}(t)^2 
    + \frac{1}{2L}\Phi_{\text{s}}(t)^2
    + \int_{0}^{\infty}
        \mathrm{d} x \left\{\frac{1}{2C_l} \big[\Pi(x,t) - 2\delta(x)Q_{\text{s}}(t)\big]^2
        + \frac{1}{2L_l} \left(\frac{\partial \Phi }{\partial x}\right)^2\right\}.
\end{align}
Due to the coupling, the conjugate variables $Q_{\text{s}}(t)$ of the LC oscillator and $\Pi(x,t)$ of the transmission line are no longer given by Eq.(\ref{eq:LC_conjugate_momentum}) and (\ref{eq:conjugate_variable_t_line}), respectively; instead, we have\footnote{To match the definition of $Q_{\text{s}}$ in Figure \ref{fig:tline_series_resonator_coupling_QLE}, we need an extra minus sign. This changes the sign of the Poisson bracket between the conjugate variables and thus the commutation relation becomes $\Bigsl[\hat{Q}_{\text{s}}, \Phi_{\text{s}}\Bigsr] = \ci \hbar$.}
\begin{equation}
    - Q_{\text{s}}(t) 
    = \frac{\partial \mathcal{L}}{\partial \dot{\Phi}_{\text{s}}} \Big|_{\dot{\Phi}_{\text{s}} 
                    = \dot{\Phi}_{\text{s}}(t)} 
    = C [\dot{\Phi}_{\text{s}}(t)- \dot{\Phi}(0,t)]
\end{equation}
and 
\begin{align}
    \Pi(x,t) 
    = \frac{\partial \mathscr{L}}{\partial \dot{\Phi}} \Big|_{\dot{\Phi} 
    = \dot{\Phi}(x,t)} 
    &= [C_l + 2 \delta(x) C] 
            \dot{\Phi}(x,t) 
        - 2\delta(x) C \dot{\Phi}_{\text{s}}(t) 
\nonumber \\
    &= C_l \dot{\Phi}(x,t) 
        + 2 \delta(x) Q_{\text{s}}(t).
\end{align}

Notice the appearance of the delta function when describing a continuous set of degrees of freedom coupled with a discrete degree of freedom. This singularity results from the fact that a lumped circuit does not have a physical size. In reality, however, the resonator also has a finite footprint and thus is distributive. For example, a planar CPW resonator is itself a section of a transmission line, so the coupling will not be a delta function. Mathematically, a point-coupling in real space implies a flat coupling spectrum in the frequency domain; consequently, the memory function in Eq.(\ref{eq:memory_function_definition}) is also a delta function, which is equivalent to assuming the system is Markovian. In the subsequent calculations, we will replace $2\delta(x)$ in the above expressions with a generic coupling function $\kappa(x)$ so that our derivation can account for non-Markovian coupling as well. For a short correlation time, we can always move back to the Markovian regime.

Now, we can quantize the composite system with the usual rule introduced in Chapter 2. However, compared with the general Hamiltonian shown in Eq.(\ref{eq:general_QLE_hamiltonian_for_derivation}), Eq.(\ref{eq:series_LC_t_line_hamiltonian}) is still not suitable for computing the QLE since the environment (i.e., the transmission line) is not written in terms of a sum of harmonic oscillators. As shown in Chapter 2, we can decompose the Hamiltonian of the transmission line by going into the spatial Fourier domain. But, because the line is semi-infinite, we will use the Fourier cosine transform
\begin{align}
    \hat{\tilde{\Phi}}(k,t) 
    = \sqrt{\frac{2}{\pi}}
        \int_{0}^{\infty} 
                \mathrm{d}x \, 
                \hat{\Phi}(x,t) 
                \cos(k x)
    & \ \ \longleftrightarrow \ \ 
    \hat{\Phi}(x,t) 
    = \sqrt{\frac{2}{\pi}}
        \int_{0}^{\infty} 
                \mathrm{d}k \, 
                \hat{\tilde{\Phi}}(k,t) 
                \cos(k x),
\\
    \hat{\tilde{\Pi}}(k,t) 
    = \sqrt{\frac{2}{\pi}}
        \int_{0}^{\infty} 
                \mathrm{d}x \, 
                \hat{\Pi}(x,t) 
                \cos(k x)
    & \ \ \longleftrightarrow \ \ 
    \hat{\Pi}(x,t) 
    = \sqrt{\frac{2}{\pi}}
        \int_{0}^{\infty} 
                \mathrm{d}k \, 
                \hat{\tilde{\Pi}}(x,t) 
                \cos(k x),
\\
    \tilde{\kappa}(k) 
    = \sqrt{\frac{2}{\pi}}
        \int_{0}^{\infty} 
                \mathrm{d}x \, 
                \kappa(x) 
                \cos(k x)
    & \ \ \longleftrightarrow \ \
    \kappa(x) 
    = \sqrt{\frac{2}{\pi}}
        \int_{0}^{\infty} 
                \mathrm{d}k \, 
                \tilde{\kappa}(x) 
                \cos(k x).
\end{align}
In other words, imagine the semi-infinite line is even-extended to form an infinitely long line. We do not consider the Fourier sine transform since the excitation at $x=0$ should not be zero\footnote{In other words, all the modes with a node at $x=0$ will not interact with the LC oscillator and they live in a separable Hilbert space which is omitted in our description.}. With some mathematical manipulation, we arrive at the Hamiltonian
\begin{align}
    \hat{H}(t)
    &= \frac{1}{2C} \hat{Q}_{\text{s}}(t)^2 
    + \frac{1}{2L} \hat{\Phi}_{\text{s}}(t)^2
\nonumber \\
    & \ \ \ \ + \int_{0}^{\infty}
        dx \, \Bigg\{\frac{1}{2C_l} \!
        \left[
            \sqrt{\frac{2}{\pi}} \! 
            \int_{0}^{\infty} \!
                    \mathrm{d} k \, 
                    \hat{\tilde{\Pi}}(k,t) 
                    \cos(k x) - \hat{Q}_{\text{s}}(t) \sqrt{\frac{2}{\pi}} \! 
            \int_{0}^{\infty} \!
                    \mathrm{d}k \, 
                    \tilde{\kappa}(k) 
                    \cos(k x)
        \right]^2
\nonumber \\
    & \ \ \ \ \ \ \ \ \ \ \ \ \ \ \ \ 
        + \frac{1}{2L_l} \!  \left[-\sqrt{\frac{2}{\pi}} k \! 
        \int_{0}^{\infty} \!
                \mathrm{d}k \, 
                \hat{\tilde{\Phi}}(k,t) 
                \sin(k x)\right]^2\Bigg\} \ \ \ \ \ \ \ \
\nonumber \\ \label{eq:Hamiltonian_of_QHO_coupled_to_QHOs}
    &= \frac{1}{2C} \hat{Q}_{\text{s}}(t)^2 
    + \frac{1}{2L} \hat{\Phi}_{\text{s}}(t)^2
    + \int_{0}^{\infty}
        \mathrm{d}k  
        \left\{         
            \frac{1}{2C_l}
            \left[
                \hat{\tilde{\Pi}}(k,t) 
                - \hat{Q}_{\text{s}}(t) \tilde{\kappa}(k)
            \right]^2 
            + \frac{\omega_k^2 C_l}{2} \hat{\tilde{\Phi}}(k,t)^2
        \right\}.
\end{align}
At this point, we make the mapping
\begin{equation}
    \hat{X} \sim \hat{Q}_{\text{s}}(t)
    \ \ \ \ 
    \hat{q}_n \sim 
    \sqrt{C_l} \hat{\tilde{\Phi}}(k,t),
    \ \ \ \ 
    \hat{p}_n \sim 
    \frac{\hat{\tilde{\Pi}}(k,t) }{\sqrt{C_l}},
    \ \ \ \ 
    \kappa_n \sim \frac{\tilde{\kappa}(k)}{\sqrt{C_l}},
\end{equation}
\begin{equation}
    \hat{a}_n \sim \hat{a}(k,t) 
    = \sqrt{\frac{\omega_k C_l}{2\hbar}} \, \hat{\tilde{\Phi}}(k,t)
        + \frac{\ci}{\sqrt{2\hbar \omega_k C_l}} \, \hat{\tilde{\Pi}}(k,t),
\end{equation}
so that the Hamiltonian can be put into the form of Eq.(\ref{eq:general_QLE_hamiltonian_for_derivation}):
\begin{equation}
    \hat{H}(t)
    = \frac{\hat{Q}_{\text{s}}(t)^2}{2C}
    + \frac{\hat{\Phi}_{\text{s}}(t)^2}{2L}
    + \frac{1}{2} 
        \int_{0}^{\infty}
        \! \mathrm{d}k
        \left\{
            \left[\frac{\hat{\tilde{\Pi}}(k,t) }{\sqrt{C_l}} - \frac{\tilde{\kappa}(k)}{\sqrt{C_l}} \hat{X}\right]^{\!2} + \omega_k^2 \left[\sqrt{C_l} \hat{\tilde{\Phi}}(k,t)\right]^2
        \right\}.
\end{equation}

With all the effort follows the QLE of the system operator
\begin{align}
    \dot{\hat{\Phi}}_{\text{s}}
    &= - \frac{\ci}{\hbar} \left[\hat{\Phi}_{\text{s}}, \hat{H}_{\mathcal{S}} \right]
    - \frac{\ci}{2\hbar} 
        \left\{
            \left[
                \hat{Q}_{\text{s}}, \hat{\Phi}_{\text{s}}
            \right], 
            \hat{\xi}(t)
            - f(t - t_0)\hat{Q}_{\text{s}}(t_0)
            - \int_{t_0}^{t} \mathrm{d} \tau
                f(t-\tau) \dot{\hat{Q}}_{\text{s}}(\tau)
        \right\}
\nonumber\\ \label{eq:QLE_series_LC_t_line_derivation}
    &= - \frac{\hat{Q}_{\text{s}}}{C}
        + \hat{\xi}(t)
        - f(t - t_0)\hat{Q}_{\text{s}}(t_0)
        - \int_{t_0}^{t} \mathrm{d} \tau
                f(t-\tau) \dot{\hat{Q}}_{\text{s}}(\tau)
\end{align}
along with
\begin{gather}
    \hat{\xi}(t)
    = \ci \int_0^{\infty} dk \, \frac{\tilde{\kappa}(k)}{\sqrt{C_l}} \sqrt{\frac{\hbar \omega_k}{2}}
        \left[
            - e^{-\ci \omega_k (t-t_0)} 
                \hat{a}(k,t_0)
            + e^{\ci \omega_k (t-t_0)}
                \hat{a}^{\dagger}(k,t_0)
        \right],
\\[-1mm]
    f(t) 
    = \int_0^{\infty} 
        \mathrm{d}k \, 
        \frac{\tilde{\kappa}(k)^2}{C_l}
        \cos(\omega_k t).
\end{gather}
In addition, since $\hat{X} = \hat{Q}_{\text{s}}$ is the operator responsible for the coupling, the QLE for $\hat{Q}_{\text{s}}$ is trivially given by 
\begin{equation} \label{eq:QLE_for_Q_s}
    \dot{\hat{Q}}_s = - \frac{\ci}{\hbar} \left[\hat{Q}_{\text{s}}, \hat{H}_{\mathcal{S}} \right]
    = \frac{\hat{\Phi}}{L}.
\end{equation}
We can interpret the QLE of a series LC resonator, again, as a circuit law: On the left-hand side of Eq.(\ref{eq:QLE_series_LC_t_line_derivation}), we have the voltage $\partial_t \hat{\Phi}_{\text{s}}$ developed across the inductor. By Kirchhoff's voltage law, $\partial_t \hat{\Phi}_{\text{s}}$ must be equal to the voltage $-\hat{Q}_{\text{s}} / C$ across the capacitor plus the net voltage on the transmission line at $x=0$. Hence, the expression 
\begin{equation*} 
    \hat{\xi}(t)
        - f(t - t_0)\hat{Q}_{\text{s}}(t_0)
        - \int_{t_0}^{t} \mathrm{d} \tau
                f(t-\tau) \dot{\hat{Q}}_{\text{s}}(\tau)
\end{equation*}
on the right-hand side of Eq.(\ref{eq:QLE_series_LC_t_line_derivation}) is precisely the sum of the incoming and outgoing voltages; in other words, we can define, in the quantum-mechanical sense,
\begin{equation} \label{eq:sum_of_incoming_and_outgoing_waves}
    \hat{V}_{\text{out}}(x=0, t) + \hat{V}_{\text{in}}(x=0, t)
    = \hat{\xi}(t)
        - f(t - t_0)\hat{Q}_{\text{s}}(t_0)
        - \int_{t_0}^{t} \mathrm{d} \tau
                f(t-\tau) \dot{\hat{Q}}_{\text{s}}(\tau).
\end{equation}
For an infinitely-long line, the left- and right-traveling waves can exist independently; however, for any line with a boundary condition, the two waves must be related by the load impedance, which is why $\hat{Q}_{\text{s}}$ appears on the right-hand side of Eq.(\ref{eq:sum_of_incoming_and_outgoing_waves}).

\textbf{Markov Assumption}: If we set $\kappa(x) = 2 \delta(x)$ (and thus $\Tilde{\kappa}(k) = \sqrt{2/\pi}$) for the point-coupling, the memory function reduces to $f(t) = 2 Z_0 \delta(t) = 2\sqrt{L_{l} / C_{l}} \, \delta(t)$; hence, for $t>t_0$,
\begin{align}
    \dot{\hat{\Phi}}_{\text{s}}
    &= - \frac{\hat{Q}_s}{C}
        + 2 \sqrt{Z_0} \, \hat{A}_{\text{in}}(t)
        - 2 Z_0 \delta(t-t_0)\hat{Q}_{\text{s}}(t_0)
        - \int_{t_0}^{t} 
                \mathrm{d} \tau \,
                2 Z_0 \delta(t - \tau) 
                \dot{\hat{Q}}_{\text{s}}(\tau)
\nonumber \\[-1mm] \label{eq:QLE_series_LC_t_line_markov}
    &= - \frac{\hat{Q}_{\text{s}}}{C} 
        + 2 \sqrt{Z_0} \, 
            \hat{A}_{\text{in}}(t) 
        - Z_0 \dot{\hat{Q}}_{\text{s}}(t),
\end{align}
where we have redefined $\hat{\xi}/2\sqrt{Z_0}$ as $\hat{A}^{\text{in}}(t) = \hat{A}^{\text{in}}(x=0,t)$ with
\begin{align} \label{eq:quantum_input_wave_amplitude_definition}
    \hat{A}^{\text{in}}(x,t)
    = \frac{-\ci}{\sqrt{2\pi}} \int_{0}^{\infty}
        dk \sqrt{\frac{\hbar \omega_k v_{\text{p}}}{2}}
        \left[ 
            e^{- \ci kx - \ci\omega_k(t-t_0)} 
                \hat{a}(k,t_0)
            - e^{\ci kx + \ci\omega_k(t-t_0)}
                \hat{a}^{\dagger}(k,t_0) 
        \right],
\end{align}
where $v_{\text{p}} = 1/\sqrt{L_{l} C_{l}}$ is the phase velocity. One can show (see Appendix \ref{appendix:LC_to_TL}\footnote{In the appendix, $\hat{a}(k,t_0)$ and $\hat{a}^{\dagger}(k,t_0)$ of Eq.(\ref{eq:quantum_input_wave_amplitude_definition}) are replaced with $\hat{a}(-k,t_0)$ and $\hat{a}^{\dagger}(-k,t_0)$. In general, the latter is the correct definition. However, recall that we have even-extended all the Hermitian observables on the transmission line to define the Fourier cosine transform; this means the transformed operators are also Hermitian and even in $k$. Hence, only for the semi-infinite line, we have $\hat{a}(-k,t_0) = \hat{a}(k,t_0)$.}) that 
\begin{equation}
    \hat{A}^{\text{in}}(x,t) 
    = \frac{1}{2} 
        \left[
            \frac{1}{\sqrt{Z_0}}\hat{V}(x,t)
            - \sqrt{Z_0} \hat{I}(x,t)
        \right]
    = \frac{1}{\sqrt{Z_0}} \hat{V}_{\text{in}}(x,t)
    = \sqrt{Z_0} \hat{I}_{\text{in}}(x,t)
\end{equation}
is the quantum operator for the input power wave amplitude defined by Eq.(\ref{eq:power_wave_amplitude}). Now, Eq.(\ref{eq:QLE_series_LC_t_line_markov}) is identical to Kirchhoff's voltage law for ideal lumped-element coupling.

Furthermore, the input-output relation shown in Eq.(\ref{eq:sum_of_incoming_and_outgoing_waves}) can also be simplified under the Markov assumption. Given that we have defined the incoming wave $\hat{A}_{\text{in}}$, the rest of Eq.(\ref{eq:sum_of_incoming_and_outgoing_waves}) must be the outgoing wave, i.e.,
\begin{align}
    \hat{A}_{\text{out}} (x=0, t)
    &= \frac{1}{\sqrt{Z_0}} \hat{V}_{\text{out}} (x=0, t)
\nonumber \\
    &= \frac{1}{\sqrt{Z_0}} 
        \left[ 
            \hat{\xi}(t) -\int_{t_0}^{t} 
            \mathrm{d} \tau \,
                2 Z_0 \delta(t-\tau) \dot{\hat{Q}}_{\text{s}}(\tau)
            - \hat{V}_{\text{in}} (x=0, t)
        \right]
\nonumber \\
    &= \frac{1}{\sqrt{Z_0}} 
        \left[ 
            2\sqrt{Z_0} \hat{A}_{\text{in}}(x=0, t)
            - Z_0 \dot{\hat{Q}}_{\text{s}}(t)
            - \sqrt{Z_0} \hat{A}_{\text{in}}(x=0, t)
        \right]
\nonumber \\ \label{eq:input_output_relation_power_amp}
    &= \hat{A}_{\text{in}}(x=0,t) 
        - \frac{\sqrt{Z_0}}{L} \hat{\Phi}_{\text{s}}(t),
\end{align}
where we have used Eq.(\ref{eq:QLE_for_Q_s}) to replace $\partial_t \hat{Q}_{\text{s}}$ with  $\hat{\Phi}_{\text{s}}$.

\subsection{The Rotating Wave Approximation}
Since the LC resonator is also a QHO, it's customary to write the QLE in terms of the annihilation operator of the system. Using the definition given in Chapter 2 (see Eq.(\ref{eq:annihilation_op_LC_definition})), we obtain
\begin{align}
    \dot{\hat{a}}_{\text{s}} (t) 
    &= \sqrt{\frac{Z_{0,\text{r}}}{2 \hbar} } 
        \dot{\hat{Q}}_{\text{s}} (t) 
        + \ci \sqrt{\frac{1}{2 \hbar Z_{0,\text{r}}}} \dot{\hat{\Phi}}_{\text{s}} (t) 
\nonumber \\
    &= \sqrt{\frac{Z_{0,\text{r}}}{2 \hbar} } \frac{\hat{\Phi}_{\text{s}}(t)}{L}
        + \ci \sqrt{\frac{1}{2 \hbar Z_{0,\text{r}}}} 
            \left[ 
                - \frac{\hat{Q}_{\text{s}}(t)}{C}
                + \hat{\xi}(t)
                - \int_{-\infty}^{t} 
                        \mathrm{d} \tau
                        f(t-\tau) \dot{\hat{Q}}_{\text{s}}(\tau)
            \right]
\nonumber \\
    &= -\ci \omega_{\text{r}} \hat{a}_{\text{s}}(t)   
        + \frac{\ci}{\sqrt{2\hbar Z_{0,\text{r}}}} \hat{\xi}(t) 
        - \frac{1}{2L} 
            \int_{-\infty}^{t} 
                \mathrm{d} \tau
                f(t-\tau)
                    \left[
                        \hat{a}_{\text{s}}(\tau)
                        - \hat{a}_{\text{s}}^{\dagger}(\tau)
                    \right],
\end{align}
where $\omega_{\text{r}}$ is the resonant frequency of the LC oscillator. Also note that $Z_{0,\text{r}} = \sqrt{L/C}$ is the characteristic impedance of the LC oscillator while $Z_0$ is the characteristic impedance of the transmission line. Furthermore, under the Markov assumption, $f(t) \approx 2 Z_0 \delta(t)$ and 
\begin{equation}
    \dot{\hat{a}}_{\text{s}}(t)  
    = -\ci \omega_{\text{r}} \hat{a}_{\text{s}}(t)   
        + \frac{\ci}{\sqrt{2 \hbar Z_{0,\text{r}}}} \hat{\xi}(t) 
        - \frac{Z_0}{2 L} 
            \left[
                \hat{a}_{\text{s}}(t)
                - \hat{a}_{\text{s}}^{\dagger}(t)
            \right].
\end{equation}

Now, we argue that we can make another RWA \cite{gardiner2004quantum} if the LC resonator has a narrow bandwidth and the memory function is also sharply peaked at zero. As always, we first move to the interaction picture of the LC oscillator to identify the fast-rotating terms:
\begin{align}
    \dot{\hat{\tilde{a}}}_{\text{s}}(t)  
    &= \frac{\ci}{\sqrt{2 \hbar Z_{0,r}}}
        \, e^{\ci \omega_{\text{r}} t}
        \, \hat{\xi}(t) 
        - \frac{Z_0}{2L} 
            \Big[
                \hat{\tilde{a}}_{\text{s}}(t)
                - e^{\ci 2 \omega_{\text{r}} t} \, \hat{\tilde{a}}_{\text{s}}^{\dagger}(t)
            \Big]
\nonumber \\
     &\approx
     \frac{\ci}{\sqrt{2\hbar Z_{0,r}}}
        \, e^{\ci \omega_{\text{r}} t}
        \, \hat{\xi}(t) 
        - \frac{Z_0}{2L} 
                \hat{\tilde{a}}_{\text{s}}(t).
\end{align}
We have omitted the creation operator since the oscillation in the interaction picture is twice the resonant frequency. Moreover, we can expand $\hat{\xi}$ using the annihilation and creation operators of the transmission line; combine with the factor $e^{\ci \omega_{\text{r}} t}$, we have
\begin{equation}
    e^{\ci \omega_{\text{r}} t} 
        \hat{\xi}(t) 
    = \ci 
        \int_0^{\infty} 
            \mathrm{d} k \, 
            \frac{\tilde{\kappa}(k)}{\sqrt{C_l}} \sqrt{\frac{\hbar \omega_k}{2}}
            \left[
                - e^{\ci (\omega_{\text{r}}-\omega_k)t + \ci \omega_k t_0} 
                    \hat{a}(k,t_0)
                + e^{\ci (\omega_{\text{r}} + \omega_k)t - \ci \omega_k t_0}
                    \hat{a}^{\dagger}(k,t_0)
            \right].
\end{equation}
Since $\omega_{\text{r}}$ and $\omega_{k}$ are both positive, $e^{\ci (\omega_{\text{r}} + \omega_k)t}$ in the integrand oscillates fast and can be ignored. It's possible that $e^{\ci (\omega_{\text{r}}-\omega_k)t} \hat{a}(k,t_0)$ also oscillates fast, but at least it allows the possibility that $\omega_{\text{r}} = \omega_{k}$. Therefore, in the Markovian regime and under the RWA,
\begin{align}
    \dot{\hat{\tilde{a}}}_{\text{s}}(t)  
    \approx
    e^{\ci \omega_{\text{r}} t}
        \, \hat{a}_{\text{in}}(t) 
        - \frac{Z_0}{2L} 
                \hat{\tilde{a}}_{\text{s}}(t),
\end{align}
where we have normalized $\hat{\xi}$ with the introduction of the \textbf{input traveling-wave annihilation operator} (under the RWA)
\begin{align} \label{eq:input_traveling_wave_annihilation_op_def_main}
    \hat{a}_{\text{in}}(t)
    = \frac{\ci}{\sqrt{2\hbar \omega_{\text{r}} Z_0}} \hat{\xi}(t)
    &\approx \frac{\sqrt{v_{\text{p}}}}{2}
            \int_{0}^{\infty}
                \mathrm{d} k \,
                \tilde{\kappa}(k)
                e^{-\ci \omega_k (t-t_0)} 
                \hat{a}(k,t_0)
    && \text{(RWA)}
\nonumber \\
    &\approx \sqrt{\frac{v_p}{2\pi}} 
            \int_{0}^{\infty}
                \mathrm{d} k \,
                e^{-\ci \omega_k (t-t_0)} 
                \hat{a}(k,t_0).
    && \text{(Markov)}
\end{align}
Alternatively, the input traveling-wave annihilation operator can be defined as
\begin{equation} \label{eq:input_traveling_wave_annihilation_op_def_main_2}
    \hat{a}_{\text{in}}(t) 
    = \sqrt{\frac{2}{\hbar \omega_{\text{r}}}} \hat{A}_{\text{in}}(t)
\end{equation} 
before dropping the creation operators in $\hat{A}_{\text{in}}$ (see Appendix \ref{appendix:LC_to_TL}) since $\hat{A}_{\text{in}}$ is proportional to $\hat{\xi}$\footnote{The two definitions given by Eq.(\ref{eq:input_traveling_wave_annihilation_op_def_main}) and (\ref{eq:input_traveling_wave_annihilation_op_def_main_2}) actually differ by a phase due to imaginary unit $\ci$ in Eq.(\ref{eq:input_traveling_wave_annihilation_op_def_main}). This difference does not cause any problem since we can always redefine the phase of the input field.}. In addition, due to the RWA, $\omega_{k}$ under the square root in Eq.(\ref{eq:quantum_input_wave_amplitude_definition}) is replaced with $\omega_{\text{r}}$ so that Eq.(\ref{eq:input_traveling_wave_annihilation_op_def_main}) and (\ref{eq:input_traveling_wave_annihilation_op_def_main_2}) are indeed equivalent. Similarly, we can also define an \textbf{output traveling-wave annihilation operator} by normalizing $\hat{A}_{\text{out}}$. From Eq.(\ref{eq:input_output_relation_power_amp}), we immediately obtain the \text{input-output} relation
\begin{equation}
    \hat{a}_{\text{out}}(t) 
    = \hat{a}_{\text{in}}(t) - \sqrt{\frac{Z_0}{L}} \hat{a}_{\text{s}}(t),
\end{equation}
where $\hat{a}_{\text{s}}^{\dagger}$ in $\hat{\Phi}_{\text{s}}$ is omitted due to the RWA. Physically, $\hat{a}^{\dagger}_{\text{in}} \hat{a}_{\text{in}}$ and $\hat{a}^{\dagger}_{\text{out}} \hat{a}_{\text{out}}$ give the photon flux traveling on the transmission line. Moreover, under the RWA where $\omega$ is replaced by $\omega_{\text{r}}$ and the counter-rotating terms are dropped, $\hat{a}_{\text{in}}$ and $\hat{a}_{\text{out}}$ satisfy the bosonic commutation relations
\begin{equation}
    \Bigsl[\hat{a}_{\text{in}}(t), \hat{a}_{\text{in}}^{\dagger}(t') \Bigsr]
    = \Bigsl[\hat{a}_{\text{out}}(t), \hat{a}_{\text{out}}^{\dagger}(t') \Bigsr]
    = \delta(t-t'),
\end{equation}
just like the normal-mode annihilation operators. To avoid the singularity at $t=t'$, we can define the annihilation and creation operators for photon wavepackets propagating along the transmission line \cite{loudon2000quantum}; subsequently, the bosonic commutation relation reduces to
\begin{equation}
    \Bigsl[\hat{a}_{\text{in}}^{\zeta}, \hat{a}_{\text{in}}^{\zeta\dagger} \Bigsr]
    = \Bigsl[\hat{a}_{\text{out}}^{\zeta}, \hat{a}_{\text{out}}^{\zeta\dagger} \Bigsr]
    = 1
\end{equation}
for each wavelet $\zeta(t)$, which was used to derive the quantum amplification in Chapter 2.



Finally, going back to the lab frame, the QLE takes the simple form
\begin{equation}
    \dot{\hat{a}}_{\text{s}}(t) 
    = - \ci \omega_{\text{r}} \hat{a}_{\text{s}}(t) 
        - \frac{Z_0}{2L} \,
            \hat{a}_{\text{s}}(t)
        + \sqrt{\frac{Z_0}{L}} \, 
        \hat{a}_{\text{in}}(t).
\end{equation}
From classical microwave theory, if a series LC resonator is loaded by an impedance $Z_0$, the external quality factor is given by $Q_{\text{ext}} = \omega_{\text{r}}L/Z_0$, implying a 3-dB bandwidth
\begin{equation}
    \kappa = \frac{\omega_{\text{r}}}{Q_{\text{ext}}} = \frac{Z_0}{L}.
\end{equation}
For a resonating structure, the bandwidth of the response function is mathematically equivalent to the decay rate of the transient response; for this reason, $\kappa$ is also known as the \textbf{decay rate} of the system. Using the expression for $\kappa$, the QLE and the associated input-output relation become
\begin{equation}
    \dot{\hat{a}}_{\text{s}}(t) 
    = - \ci \omega_{\text{r}} \hat{a}_{\text{s}}(t) 
        - \frac{\kappa}{2} \,
            \hat{a}_{\text{s}}(t)
        + \sqrt{\kappa} \, 
        \hat{a}_{\text{in}}(t)
\end{equation}
and 
\begin{equation}
    \hat{a}_{\text{out}}(t) 
    = \hat{a}_{\text{in}}(t)
        - \sqrt{\kappa} \hat{a}_{\text{s}}(t),
\end{equation}
respectively. One can go through the dual calculation for a parallel LC oscillator coupled to the transmission line; the results for the series and parallel LC circuits are summarized in Table \ref{tab:QLE_summary}. For both cases, the QLE is a manifestation of the fluctuation-dissipation theorem since $- (\kappa/2) \hat{a}_{\text{s}}(t)$ represents the dissipation of the system to the environment while $\sqrt{\kappa} \hat{a}_{\text{in}}(t)$ characterizes the noise entering into the system from the environment. In particular, quantum fluctuation from the transmission line enters into the LC resonator even when the line is in a vacuum state. 

\begin{table}
    \centering
    \footnotesize
    \setlength\tabcolsep{6pt}
    \renewcommand{\arraystretch}{0.75}
\begin{tabular}{c | c  c }
\toprule
     {} &  Series LC  &  Parallel LC 
\\  \midrule \midrule \\
    QLE 
    & $ \displaystyle \dot{\hat{a}}_{\text{s}}(t) 
    = - \ci \omega_{\text{r}} \hat{a}_{\text{s}}(t) 
        - \frac{\kappa}{2} \,
            \hat{a}_{\text{s}}(t)
        + \sqrt{\kappa} \, 
        \hat{a}_{\text{in}}(t)$
    & $ \displaystyle \dot{\hat{a}}_{\text{s}}(t) 
    = - \ci \omega_{\text{r}} \hat{a}_{\text{s}}(t) 
        - \frac{\kappa}{2} \,
            \hat{a}_{\text{s}}(t)
        + \sqrt{\kappa} \, 
        \hat{a}_{\text{in}}(t)$
\\ \\ \hline \\
    $\substack{
        \displaystyle\text{Input-output} \\[1.5mm] \displaystyle\text{relation}
    }$
    & $\hat{a}_{\text{out}}(t) 
    = \hat{a}_{\text{in}}(t)
        - \sqrt{\kappa} \hat{a}_{\text{s}}(t)$
    & $\hat{a}_{\text{out}}(t) 
    = - \hat{a}_{\text{in}}(t)
        + \sqrt{\kappa} \hat{a}_{\text{s}}(t)$
\\ \\ \hline \\
    Decay rate $\kappa$
    & $\displaystyle\frac{Z_0}{L}$ 
    & $\displaystyle\frac{1}{Z_0 C}$
\\ \\ \bottomrule
    \end{tabular}
    \caption{A summary of the important expressions related to the QLEs for a series or parallel LC circuit coupled to a semi-infinite transmission line.}
    \label{tab:QLE_summary}
\end{table}

By revisiting the effective parallel RLC circuit shown in Figure \ref{fig:equivalent_coupling_circuit_transformed}, one can now compute the decay rate of each port as $\kappa_{i} = 1 / \{(1+Q_{i}^2)Z_0 [C + C_{\kappa_{i}}(1+Q_{i}^{-2})]\} \approx \omega_{\text{r}}^2 Z_0 C_{\kappa_{i}}^2 / C$ for $i = 1$ (input port) or $2$ (output port). The QLE can be extended to describe multiple decay channels; for a two-port resonator modeled as Figure \ref{fig:equivalent_coupling_circuit_transformed}, the QLE changes to
\begin{equation}
    \dot{\hat{a}}_{\text{s}}(t) 
    = - \ci \omega_{\text{r}} \hat{a}_{\text{s}}(t) 
        - \frac{\kappa_{\text{1}} + \kappa_{\text{2}}}{2} \,
            \hat{a}_{\text{s}}(t)
        + \sqrt{\kappa_{\text{1}}} \, 
        \hat{a}_{\text{in,1}}(t)
        + \sqrt{\kappa_{\text{2}}} \, 
        \hat{a}_{\text{in,2}}(t).
\end{equation}
In addition, each port is constrained by an input-output relation. The application of QLE is itself a rich topic. As a generalization of the classical Langevin equation, one can introduce quantum white noise to the QLE and define quantum stochastic differential equations for the observables. However, we will not pursue this path in the following sections. Instead, we will construct a stochastic differential equation for the quantum state of a qubit or qudit.

\subsection{Stiff-Pump Limit}
The QLE also provides us a way to add input to a quantum system via $\hat{a}_{\text{in}}$. As mentioned in Chapter 3, the readout drive is usually weak but the control pulse can be treated classically. Following the same argument, for a single-frequency operation, we can demote the input traveling-wave operator to a classical traveling wave, i.e.,
\begin{equation}
    \hat{a}_{\text{in}} \ \ \longrightarrow \ \ 
    \ci \bar{a}_{\text{in}} e^{- \ci \omega_{\text{d}} t},
\end{equation}
where $\bar{a}_{\text{in}}$ is the mean (complex) amplitude of the drive. Scaling $\bar{a}_{\text{in}}$ by $\sqrt{\hbar \omega_{\text{d}}/2}$ gives precisely the classical power wave amplitude $A_{\text{in}}$. The QLE now reads
\begin{equation} \label{eq:QLE_stiff_pump}
    \dot{\hat{a}}_{\text{s}}(t) 
    = - \ci \omega_{\text{r}} \hat{a}_{\text{s}}(t) 
        - \frac{\kappa}{2} \,
            \hat{a}_{\text{s}}(t)
        + \ci \sqrt{\kappa} \, 
        \bar{a}_{\text{in}} e^{- \ci \omega_{\text{d}} t}.
\end{equation}

Eq.(\ref{eq:QLE_stiff_pump}) is known as the \textbf{stiff-pump limit} \cite{PRXQuantum.2.040202} since quantum fluctuation in the traveling wave is ignored. This also means that the transmission line is completely eliminated from the Hamiltonian. However, one can reverse-engineer the interaction Hamiltonian and show that the Heisenberg equation associated with the following Hamilontian is precisely Eq.(\ref{eq:QLE_stiff_pump}):
\begin{equation}
    \hat{H} 
    = \hbar \left( \omega_{\text{r}} - \frac{\ci \kappa}{2} \right) \!
        \left(\hat{a}_{\text{s}}^{\dagger} \hat{a}_{\text{s}}
        + \frac{1}{2}
        \right)
        - \hbar
            \Big( 
                \sqrt{\kappa} \, 
                    \bar{a}_{\text{in}} e^{-\ci\omega_{\text{d}} t} \, \hat{a}_{\text{r}}^{\dagger} 
                + \sqrt{\kappa} \, 
                    \bar{a}_{\text{in}}^* e^{\ci\omega_{\text{d}} t} \, \hat{a}_{\text{r}}
            \Big) .
\end{equation}
If the decay rate $\kappa$ is much smaller than $\omega_{\text{r}}$ and $\sqrt{\kappa} \, |\bar{a}_{\text{in}}|$, the imaginary part of the oscillator frequency can be ignored. In fact, we have already studied this Hamiltonian (see Eq.(\ref{eq:semiclassically_drive_QHO})) whose time evolution operator in the interaction picture is the displacement operator. Hence, if we send a classical drive to a QHO initially in a vacuum state, we can excite the QHO to a coherent state.

\section{An Artificial Atom in an Open System}\label{section:artificial_atom_in_an_open_system}
\subsection{Qubit Control Revisited in the Displaced Frame}
Before introducing the QLE, we avoided talking about the drive strength quantitatively in the semiclassical analysis of Rabi flopping. Obviously, the so-called classical drive must enter the system through some coupling mechanism. For a qubit placed in a 3D cavity, it does not see the drive directly because the drive field must first enter the cavity and then be coupled to the qubit via the qubit-resonator interaction. 

The effective Hamiltonian for driving a cavity mode, based on the QLE, is given by
\begin{equation} \label{eq:stiff_pump_for_qubit_control}
    \hat{H}_{\text{d}}/\hbar
    = - \varepsilon_{\text{d}}(t) 
            \hat{a}_{\text{r}}^{\dagger} 
        - \varepsilon_{\text{d}}^*(t) 
            \hat{a}_{\text{r}}
    = - \sqrt{\kappa} \, 
            \bar{a}_{\text{in}} e^{-\ci\omega_{\text{d}} t} \, \hat{a}_{\text{r}}^{\dagger} 
        - \sqrt{\kappa} \, 
            \bar{a}_{\text{in}}^* e^{\ci\omega_{\text{d}} t} \, \hat{a}_{\text{r}},
\end{equation}
where we adopt the classical limit by setting the input traveling wave to be some large, classical amplitude $\bar{a}_{\text{in}} = \abs{\bar{a}_{\text{in}}} e^{\ci \phi_{\text{d}}}$ and ignoring any quantum fluctuation around the mean amplitude. We also assume that $\omega_{\text{d}} \approx \omega_{\text{q}}$ and $|\Delta_{\text{rd}}| = |\omega_{\text{r}} - \omega_{\text{d}}| \gg \kappa$, i.e., we are in the dispersive coupling regime. The total Hamiltonian now consists of the qubit-resonator and resonator-environment coupling:
\begin{align}
    \hat{H} 
    &= \hat{H}_{\text{JC}}
        + \hat{H}_{\text{d}}
\nonumber \\
    &= - \frac{1}{2} 
        \hbar 
        \omega_{\text{q}} 
        \hat{\sigma}_z
    + \hbar 
        \omega_{\text{r}} 
        \bigg(
            \hat{a}_{\text{r}}^{\dagger}
                \hat{a}_{\text{r}} 
            + \frac{1}{2}
        \bigg)
    - \hbar
        \Big(
            g \hat{a}_{\text{r}} 
                \hat{\sigma}_{+}
            + g^* \hat{a}_{\text{r}}^{\dagger}     
                \hat{\sigma}_{-}
        \Big) 
    - \hbar 
        \Big[
            \varepsilon_{\text{d}}(t) 
                \hat{a}_{\text{r}} ^{\dagger} 
            +  \varepsilon_{\text{d}}^*(t) 
                \hat{a}_{\text{r}}
        \Big]
\end{align}
Ignore the qubit part for a moment and consider the steady state behavior of the resonator under the drive. The QLE for the resonator\footnote{We have assumed that the qubit-resonator coupling is weak enough (i.e., $|g| \ll |\Delta_{\text{rd}} - \ci \kappa / 2|$) to be ignored in the QLE; otherwise, the QLE of the resonator should be changed to
\begin{equation}
    \dot{\hat{a}}_{\text{r}}(t) 
    = - \ci \omega_{\text{r}} \hat{a}_{\text{r}}(t) 
        - \frac{\kappa}{2} \hat{a}_{\text{r}}(t)
        + \ci \varepsilon_{\text{d}}(t) 
        + \ci g^* \hat{\sigma}_{-}(t)
\end{equation}
where $\ci g^* \hat{\sigma}_{-}$ comes from the Heisenberg equation of $\hat{a}_{\text{r}}$.} can be treated as an eigenvalue equation if $\hat{a}_{\text{r}}$ is replaced by its eigenstate in the classical limit (i.e., a coherent state $\ket{\alpha}$)
\begin{equation} \label{eq:alpha_dispaced_QLE}
    \dot{\alpha}(t) 
    = - \ci \omega_{\text{r}} \alpha(t) 
        - \frac{\kappa}{2} \alpha(t)
        + \ci \varepsilon_{\text{d}}(t) 
    \approx - \ci \omega_{\text{r}} \alpha(t) 
        + \ci \varepsilon_{\text{d}}(t),
\end{equation}
resulting in the steady-state\footnote{The ``steady-state'' behavior means any transient response due to the initial condition of $\alpha$ dies out. It does NOT mean $\alpha$ is a constant. Taking the Fourier transform of the QLE gives the steady-state solution oscillating at the drive frequency.} amplitude 
\begin{equation}\label{eq:steady_state_alpha_dispaced}
    \alpha(t)
    = \frac{
            \sqrt{\kappa} \, 
            \bar{a}_{\text{in}}
        }{
            (\omega_{\text{r}} 
                - \omega_{\text{d}})
            - \ci \kappa / 2
        } 
        e^{-\ci \omega_{\text{d}} t}
    \approx \frac{
            \sqrt{\kappa} \, 
            \bar{a}_{\text{in}}
        }{
            \Delta_{\text{rd}}
        } 
        e^{-\ci \omega_{\text{d}} t},
\end{equation}
which means the average photon number in the resonator is about $\bar{n} = \abs{\alpha(t)}^2 = \kappa |\bar{a}_{\text{in}}|^2/\Delta_{\text{dr}}^2$. Intuitively, we expect the qubit to extract energy based on this steady-state photon number, so the Rabi frequency should be about $2 \abs{g} \sqrt{\bar{n}}$. 

To mathematically reveal the qubit-environment coupling, we would like to adiabatically eliminate the resonator-environment coupling by evoking the displacement operator \cite{PhysRevA.75.032329}
\begin{equation}
    \hat{D}_{\text{r}}(\alpha(t)) 
    = \exp[\alpha(t)\, \hat{a}_{\text{r}}^{\dagger} - \alpha^*(t)\, \hat{a}_{\text{r}}].
\end{equation}
In particular, we are using the steady-state amplitude $\alpha(t)$ so that we can effectively remove all the steady-state photons in the resonator due to the drive and transfer the corresponding energy to the qubit. Recall that given any unitary operator $\hat{U}(t)$, the transformed Hamiltonian is given by $\hat{H}'(t) = \hat{U}^{\dagger}(t) \hat{H} \hat{U}(t) - \ci \hbar \hat{U}^{\dagger}(t) \dot{\hat{U}}(t)$. Letting $\hat{U} = \hat{D}_{\text{r}}$ and using Eq.(\ref{eq:alpha_dispaced_QLE}) and (\ref{eq:steady_state_alpha_dispaced}) yields
\begin{align}
    \hat{D}_{\text{r}}^{\dagger}(\alpha(t)) 
        \hat{H} 
        \hat{D}_{\text{r}}(\alpha(t)) 
    &= - \frac{1}{2} 
            \hbar 
            \omega_{\text{q}} 
            \hat{\sigma}_z
        + \hbar \omega_{\text{r}}
            \bigg(
                \hat{a}_{\text{r}}^{\dagger}
                    \hat{a}_{\text{r}} 
                + \frac{1}{2}
            \bigg)
        + \hbar \omega_{\text{r}} 
            \Big(
                \alpha^* \hat{a}_{\text{r}} 
                + \alpha\hat{a}_{\text{r}}^{\dagger}
            \Big)
        + \hbar \omega_{\text{r}} |\alpha|^2
\nonumber\\
    & \ \ \ \ \ \ \ 
    - \hbar
        \Big(
            g \hat{a}_{\text{r}} 
                \hat{\sigma}_{+}
            + g^* \hat{a}_{\text{r}}^{\dagger} 
                \hat{\sigma}_{-}
        \Big) 
    - \hbar
        \Big(
            g \alpha \hat{\sigma}_{+}
            + g^* \alpha^* \hat{\sigma}_{-} 
        \Big)
\nonumber \\
& \ \ \ \ \ \ \ 
    - \hbar
        \Big(
            \varepsilon_{\text{d}}
                \hat{a}_{\text{r}}^{\dagger} 
            + \varepsilon_{\text{d}}^*
                \hat{a}_{\text{r}}
        \Big)
    - \hbar
        \Big(
            \varepsilon_{\text{d}}
                \alpha^*
            + \varepsilon_{\text{d}}^*
                \alpha
        \Big)
\nonumber\\
        &= - \frac{1}{2} 
            \hbar 
            \omega_{\text{q}} 
            \hat{\sigma}_z
        + \hbar \omega_{\text{r}}
            \bigg(
                \hat{a}_{\text{r}}^{\dagger}
                    \hat{a}_{\text{r}} 
                + \frac{1}{2}
            \bigg)
        + \hbar \omega_{\text{r}} 
            \Big(
                \alpha^* \hat{a}_{\text{r}} 
                + \alpha\hat{a}_{\text{r}}^{\dagger}
            \Big)
        + \hbar (\omega_{\text{d}} - \Delta_{\text{rd}}) |\alpha|^2
\nonumber\\
    & \ \ \ \ \ \ \ 
    - \hbar
        \Big(
            g \hat{a}_{\text{r}} 
                \hat{\sigma}_{+}
            + g^* \hat{a}_{\text{r}}^{\dagger} 
                \hat{\sigma}_{-}
        \Big) 
    - \hbar
        \Big(
            g \alpha \hat{\sigma}_{+}
            + g^* \alpha^* \hat{\sigma}_{-} 
        \Big)
    - \hbar
        \Big(
            \varepsilon_{\text{d}}
                \hat{a}_{\text{r}}^{\dagger} 
            + \varepsilon_{\text{d}}^*
                \hat{a}_{\text{r}}
        \Big)
\end{align}
and
\begin{align}
    \ci \hbar 
        \hat{D}_{\text{r}}^{\dagger}(\alpha(t))
        \frac{\mathrm{d}}{\mathrm{d} t} 
            \hat{D}_{\text{r}}(\alpha(t))
    &= \ci \hbar 
        \hat{D}_{\text{r}}^{\dagger}(\alpha(t))
        \frac{\mathrm{d}}{\mathrm{d} t}
        \Big[
            e^{-\abs{\alpha(t)}^2/2}
            \,
            e^{\alpha(t) \hat{a}_{\text{r}}^{\dagger}}
            \, 
            e^{-\alpha^*(t) \hat{a}_{\text{r}}}
        \Big]
\nonumber \\
    & = \ci \hbar 
        \hat{D}_{\text{r}}^{\dagger}(\alpha)
        \Big[
            \dot{\alpha} 
                \hat{a}_{\text{r}}^{\dagger} 
                \hat{D}_{\text{r}}(\alpha)
            - \hat{D}_{\text{r}}(\alpha) 
                \dot{\alpha}^* 
                \hat{a}_{\text{r}}
        \Big]
\nonumber \\
    &= \ci \hbar 
        \Big[
            \dot{\alpha}
                \Big(
                    \hat{a}_{\text{r}}^{\dagger} 
                    + \alpha^*
                \Big) 
            - \dot{\alpha}^* 
                \hat{a}_{\text{r}}
        \Big] 
\nonumber \\
    &= \hbar \omega_{\text{r}} 
        \Big(
            \alpha^*
                \hat{a}_{\text{r}}
            + \alpha
                \hat{a}_{\text{r}}^{\dagger}  
        \Big)
        - \hbar 
            \Big(
                \varepsilon_{\text{d}}
                    \hat{a}_{\text{r}}^{\dagger} 
                +
                \varepsilon_{\text{d}}^*
                    \hat{a}_{\text{r}}
            \Big)
        + \hbar \omega_{\text{d}} \abs{\alpha}^2.
\end{align}
Combining the two calculations above and ignoring the constant $\hbar \Delta_{\text{rd}}|\alpha|^2$ gives
\begin{align}
    \hat{H}'(t)
    &= \hat{D}_{\text{r}}^{\dagger}(\alpha(t))
            \hat{H}
            \hat{D}_{\text{r}}(\alpha(t)) 
        - \ci \hbar     
            \hat{D}_{\text{r}}^{\dagger}(\alpha(t))
            \frac{\mathrm{d}}{\mathrm{d} t} 
            \hat{D}_{\text{r}}(\alpha(t))
\nonumber \\ \label{eq:displaced_H}
    &= - \frac{1}{2} \hbar \omega_{\text{q}} \hat{\sigma}_z
    + \hbar \omega_{\text{r}}  
        \bigg(
            \hat{a}_{\text{r}}^{\dagger}
                \hat{a}_{\text{r}} 
            + \frac{1}{2}
        \bigg)
    - \hbar
        \Big(
            g \hat{a}_{\text{r}} 
                \hat{\sigma}_{+}
            + g^* \hat{a}_{\text{r}}^{\dagger} 
                \hat{\sigma}_{-}
        \Big) 
    - \hbar
        \Big[
            g \alpha(t) \hat{\sigma}_{+}
            + g^* \alpha^*(t) \hat{\sigma}_{-}
        \Big].
\end{align}
Note that the constant energy $\hbar \Delta_{\text{rd}}|\alpha|^2$ omitted is the energy required to displace the resonator by $\alpha$, one consequence of going into the displaced frame.

The last term in Eq.(\ref{eq:displaced_H}) is the same as the semiclassical interaction introduced before; however, the coupling coefficient is not $g$ anymore because $\alpha$ also contributes a factor $\abs{\alpha} = \sqrt{\bar{n}}$ to the coupling:
\begin{align}
    \hbar
        \Big[
            g \alpha(t) \hat{\sigma}_{+}
            + g^* \alpha^*(t) \hat{\sigma}_{-}
        \Big]
    &= \hbar
        \Big( 
            g \sqrt{\bar{n}} 
                e^{-\ci\omega_{\text{d}}t + \ci \phi_{\text{d}}} 
                \hat{\sigma}_{+}
            + g^* 
                \sqrt{\bar{n}} 
                e^{\ci\omega_{\text{d}}t - \ci \phi_{\text{d}}} 
                \hat{\sigma}_{-}
        \Big)
\nonumber \\ \label{eq:effective_qubit_drive_in_displaced_frame}
    &= \hbar 
        \Big(
            g_{\text{qd}} 
                e^{-\ci\omega_{\text{d}} t} 
                \hat{\sigma}_{+}
            + g_{\text{qd}}^* 
                e^{\ci\omega_{\text{d}} t} 
                \hat{\sigma}_{-}
        \Big),
\end{align}
where $g_{\text{qd}} = \abs{g_{\text{qd}}} e^{\ci \phi_{\text{qd}}} = g \sqrt{\bar{n}} e^{\ci \phi_{\text{d}}}$. Thus, indeed, the effective coupling strength is $\abs{g_{\text{qd}}} = \abs{g} \sqrt{\bar{n}}$ and the Rabi frequency is given by $\Omega_{\text{qd}} = 2\abs{g}\sqrt{\bar{n}}$. As a common simplification, we can move to a frame rotating at $\omega_{\text{d}}$ by using $\hat{R}(t) = \exp \Bigsl[ \ci \omega_{\text{d}} t \Bigsl(\hat{\sigma}_z / 2 - \hat{a}^{\dagger} \hat{a}\Bigsr) \Bigsr]$, producing
\begin{align}
    \hat{H}'_{\text{rot}} 
    = \hat{R}^{\dagger} \hat{H}' \hat{R}
        - \ci \hbar 
            \hat{R}^{\dagger} \frac{\mathrm{d}}{\mathrm{d}t} \hat{R}
    &= - \frac{1}{2} \hbar 
        \Big[
            \Delta_{\text{qd}} \hat{\sigma}_z
            + \Omega_{\text{qd}} 
                \cos(\phi_{\text{qd}})     
                \hat{\sigma}_x
            + \Omega_{\text{qd}} 
                \sin(\phi_{\text{qd}})     
                \hat{\sigma}_y
        \Big]
\nonumber \\ \label{eq:driven_qubit_displace_frame}
    & \ \ \ \ \ \ \ \ \ \ \ \ \ \ \ \ \ \ \ \ \ \ \ \ \ 
        + \hbar \Delta_{\text{rd}} \hat{a}_{\text{r}}^{\dagger}\hat{a}_{\text{r}}
    - \hbar
        \Big(
            g \hat{a}_{\text{r}} 
                \hat{\sigma}_{+}
            + g^* \hat{a}_{\text{r}}^{\dagger}     
                \hat{\sigma}_{-}
        \Big) 
\end{align}
that is time-independent\footnote{The reader can verify that $\hat{D}(\alpha(t)) \hat{R} = \hat{R} \hat{D}(\alpha(0))$ given that $\alpha(t) \propto e^{-\ci \omega_{\text{d}} t}$. In other words, we can also first go to the rotating frame and then apply the field displacement, but the displacement in the displaced frame will not be oscillating anymore.} (a constant $\hbar \omega_{\mathrm{r}} /2$ is omitted). Compared to Eq.(\ref{eq:JC_Hamiltonian_general}), we still have the quantized resonator and the qubit-resonator interaction. In other words, $\hat{H}'_{\text{rot}}$ can describe both the semi-classical qubit control and a fully-quantized dispersive coupling. Specifically, we can move to the dispersive regime by applying the Schrieffer–Wolff transformation in Eq.(\ref{eq:SW_transform_qubit_resonator_dispersive_coupling}), resulting in
\begin{align}
    \hat{H}_{\text{rot}}^{'\text{disp}}
    &= \hat{T}^{\dagger} 
        \hat{H}'_{\text{rot}} 
        \hat{T}
\nonumber \\
    &\approx - \frac{1}{2} \hbar
        \Big(
            \Delta_{\text{qd}} + \chi
            + 2\chi \hat{a}_{\text{r}}^{\dagger}\hat{a}_{\text{r}}
        \Big)
        \hat{\sigma}_z
    - \frac{1}{2} 
        \hbar \Omega_{\text{qd}} 
        \hat{\sigma}_{\phi_{\text{qd}}}
    + \hbar 
        \Delta_{\text{rd}} 
        \hat{a}_{\text{r}}^{\dagger}
        \hat{a}_{\text{r}}
\nonumber \\ \label{eq:driven_qubit_in_dispersive_regime}
    & \ \ \    \ \ \ \  \ \ \ \  \ \ \ \  \ \ \ \  \ \ \ \  \ \ \ \  \ \ \ \  \ \ \ \  \ \ \ \  \ \ \ \ 
    - \hbar 
        \frac{\Omega_{\text{qd}} |g|}{4 \Delta_{\text{qr}}} 
        \Big(
            e^{-\ci \phi_{\text{qd}}} 
                \hat{a}_{\text{r}} 
            + e^{\ci \phi_{\text{qd}}} 
                \hat{a}_{\text{r}}^{\dagger}
        \Big) 
        \hat{\sigma}_z,
\end{align}
where we have adopted the notation 
\begin{equation}
    \hat{\sigma}_{\phi} 
    = \cos \phi \hat{\sigma}_x
        + \sin \phi  \hat{\sigma}_y,
\end{equation}
i.e., $\hat{\sigma}_{\phi} $ is the projection of the Pauli vector along the unit vector $(1,\pi/2, \phi)$ on the Bloch sphere. Finally, the last term in Eq.(\ref{eq:driven_qubit_in_dispersive_regime}) can be dropped in comparison to the Rabi flopping term $-\hbar \Omega_{\text{qd}} \hat{\sigma}_{\phi_{\text{qd}}} /2$ since $|g| \ll |\Delta_{\text{qr}}|$; therefore, we obtain
\begin{equation}
    \hat{H}_{\text{rot}}^{'\text{disp}}
    \approx 
    - \frac{1}{2} \hbar
            \Big(
                \Delta'_{\text{qd}}
                    \hat{\sigma}_z
                + \Omega_{\text{qd}} 
                    \hat{\sigma}_x
            \Big)
    + \hbar 
        (
            \Delta_{\text{rd}} - \chi \hat{\sigma}_z 
        )
        \hat{a}_{\text{r}}^{\dagger} 
        \hat{a}_{\text{r}},
\end{equation}
where $\Delta'_{\text{qd}} = \Delta_{\text{qd}} + \chi$ is the Lamb-shifted detuning.


\subsection{Purcell Effect}
Given the similarity between Eq.(\ref{eq:stiff_pump_for_qubit_control}) and (\ref{eq:effective_qubit_drive_in_displaced_frame}), one might argue that the qubit acquires an effective decay rate due to the coupling to the resonator. Consider the following mapping:
\begin{align*}
    \text{cavity drive strength: } \sqrt{\kappa} \hat{a}_{\text{in}} 
    \ \ &\rightarrow \ \ \text{cavity decay rate: } \kappa
\\
    \text{effective qubit drive strength: } g \sqrt{\frac{\kappa}{\Delta_{\text{rd}}^2}} \bar{a}_{\text{in}} 
    \ \ &\rightarrow \ \ \text{effective qubit decay rate: } 
        \left(
            g \sqrt{\frac{\kappa}{\Delta_{\text{rd}}^2}} 
        \right)^2
\end{align*}
We can thus conclude that the qubit has an effective decay rate of 
\begin{equation}
    \gamma_{\kappa} = \frac{g^2}{\Delta_{\text{rd}}^2} \kappa,
\end{equation}
known as the \textbf{Purcell rate} ($|g|^2$ should be used if $g$ is complex). Due to the stiff-pump assumption, Eq.(\ref{eq:driven_qubit_in_dispersive_regime}) does not contain any cavity dissipation; hence, the induced qubit decay is not obvious. Nevertheless, the validity of the mapping above actually relies on the duality between fluctuation and dissipation. In other words, we are arguing for a dissipation based on the extent a fluctuation could affect the system.

One can also characterize the Purcell rate more rigorously. According to the second-order time-dependent perturbation theory and Fermi's golden rule (assume the cavity has no photon initially), the decay rate of a qubit inside a cavity is given by \cite{PhysRevB.60.13276, cohen1992atom}
\begin{align}
    \Gamma 
    &= \frac{2}{\hbar} 
        \Im \Bigg(
            \lim_{\epsilon \rightarrow 0^{+}} \sum_n \frac{\Bigsl| \bra{g, 1_n} \hat{\mathbf{d}} \cdot \hat{\mathbf{E}}(\mathbf{r}_0) \ket{e,0}\Bigsr|^2}{\hbar \Tilde{\omega}_n - \hbar \omega_{\text{q}} - \ci \epsilon}
        \Bigg)
\nonumber \\ \label{eq:Purcell_effect_general}
    &\approx
        \frac{2\pi}{\hbar^2}
        \sum_{n}
            \underbrace{\left[
                \sqrt{\frac{\hbar \omega_n}{2\epsilon_0 V_{\text{eff},n}}}
                \hat{\mathbf{e}}_{n}
                \cdot 
                \mathbf{d}_{ge}
            \right]^2}_{\substack{\text{matrix elements of} \\ \text{dipole interaction}}}
            \underbrace{\frac{\kappa_n / 2\pi}{(\omega_n - \omega_{\text{q}})^2 + (\kappa_n / 2)^2}}_{\text{generalization of $\delta(\omega_{\text{q}} - \omega_n)$}},
\end{align}
where $\mathbf{d}_{ge} = \bra{g} \hat{\mathbf{d}} \ket{e}$ is the dipole matrix element of the qubit located at $\mathbf{r}_0$ and $\Tilde{\omega}_{n} = \omega_n - \ci \kappa_n /2$ are all the normal mode frequencies of the cavity with the decay rates $\kappa_n$ included. 
The effective mode volume $V_{\text{eff},n}$ is defined\footnote{If the mode profile is uniform over the physical volume $V_{\text{cav}}$ of the cavity, then $|\mathbf{f}_n(\mathbf{r}_0)|^2 = 1/V_{\text{cav}}$ due to normalization. Hence, the effective mode volume is a generalization of the physical volume of the cavity when the mode profile is nonuniform.} such that
\begin{equation}
    \mathbf{f}_n(\mathbf{r}_0) \mathbf{f}^{\dagger}_n(\mathbf{r}_0)
    = \frac{\hat{\mathbf{e}}_n \hat{\mathbf{e}}_n^{\dagger}}{V_{\text{eff},n}},
\end{equation}
where $\mathbf{f}_n$ is the spatial profile of the $n$th normal mode as defined in Section \ref{section:microwave_cavity} and $\hat{\mathbf{e}}_n$ is the direction of the electric field at $\mathbf{r}_0$. If a qubit lives in a free space without any confinement, the available electromagnetic modes would form a continuum, resulting in an integration of delta functions in Fermi's golden rule. However, the presence of a cavity modifies the decay rate of the qubit by reshaping the density of states into a set of discrete Lorentzian densities; this phenomenon is known as the Purcell effect. Clearly, the enhancement (or suppression) of the decay rate depends on the detuning $\omega_n - \omega_{\text{q}}$, the effective mode volume $V_{\text{eff},n}$, and the quality factor $Q_n = \omega_n / \kappa_n$. We now discuss two cases implied by Eq.(\ref{eq:Purcell_effect_general}) which are qualitatively different.

\textbf{Near-Resonance Coupling}: Near the resonance of the $n$th eigenmode, i.e., $\omega_{\text{q}} \approx \omega_n$, we can compare the decay rate of the qubit inside a cavity to that inside free space. Recall that the free-space density of states (i.e., the number of states per volume and per angular frequency) in a periodic box of size $V$ and the spontaneous decay rate are given, respectively, by
\begin{equation}
    \rho_{\text{free}}(\omega_{\text{q}}) 
    = \frac{\omega_{\text{q}}^2 }{\pi^2 c^3}
\ \ \text{ and } \ \ 
    \Gamma_{\text{free}}
    = \frac{\omega_{\text{q}}^3 |\mathbf{d}_{ge}|^2}{3\pi \epsilon_0 \hbar c^3}.
\end{equation}
Hence, the enhancement is given by \cite{PhysRevA.40.5516,Gerard_99}
\begin{equation}
    \frac{\Gamma }{\Gamma_{\text{free}}}
    = \frac{3\lambda_{\text{q}}^3}{4\pi^2} \frac{Q_n}{V_{\text{eff},n}} \frac{\kappa_n^2}{4(\omega_n - \omega_{\text{q}})^2 + \kappa_n^2},
\end{equation}
where $\lambda_{\text{q}} = 2\pi c /\omega_{\text{q}}$ is the free-space wavelength of the qubit transition. We have also assumed that the dipole is aligned with the electric field at $\mathbf{r}_0$; otherwise, the ratio should be modulated by $\eta_n^2 = (\hat{\mathbf{e}}_{\text{di}} \cdot \hat{\mathbf{e}}_n)^2$, where $\hat{\mathbf{e}}_{\text{di}} = \mathbf{d}_{ge} / |\mathbf{d}_{ge}|$ is the direction of the dipole. If the space is filled with a dielectric material with a refractive index $n_{\text{r}}$ homogeneously, we can replace $\lambda_{\text{q}}$ by $\lambda_{\text{q}} / n_{\text{r}}$. At resonance, we arrive at the Purcell factor
\begin{equation}
    F_{\text{P}} 
    = \frac{
            \Gamma(\omega_{\text{q}} 
            = \omega_n)
        }{
            \Gamma_{\text{free}}
        }
    = \frac{3\lambda_{\text{q}}^3}{4\pi^2} 
        \frac{Q_n}{V_{\text{eff},n}}.
\end{equation}
Clearly, a larger $Q_n$ and a more confined mode give a stronger decay. However, be cautious that the calculation based on the perturbation theory only holds for weak qubit-resonator coupling, which aligns with our previous assumption that $|g| \ll |\Delta_{\text{rd}} - \ci \kappa / 2| \approx |\kappa/2|$. 

\textbf{Dispersive Coupling}: As usual, we define the coupling strength $g_n$ associated with the dipole interaction as
\begin{equation}
    g_n 
    = \frac{1}{\hbar}  
        \underbrace{\sqrt{\frac{\hbar \omega_n}{2\epsilon_n V_{\text{eff},n}}}}_{\text{ZPF of  $\hat{\mathbf{E}}$}}
        \hat{\mathbf{e}}_n \cdot \mathbf{d}_{ge}.
\end{equation}
Then, when $\Delta_n = |\omega_{\text{q}} - \omega_n| \gg \kappa_n$ for all $n$, the decay rate is approximately \cite{PhysRevLett.101.080502}
\begin{equation}\label{eq:purcell_de_enhancement}
    \Gamma 
    \approx 
    \sum_{n} 
        \frac{\kappa_n g_n^2}{\Delta_n^2},
\end{equation}
which is exactly the same as the expression argued heuristically.
Compared to the free-space decay rate, we obtain the suppression 
\begin{equation}
    \frac{\Gamma }{\Gamma_{\text{free}}}
    \approx \sum_{n} 
        \frac{3\lambda_{\text{q}}^3}{16 \pi^2} \frac{1}{Q_n V_{\text{eff},n}} \frac{\omega_n^2}{\Delta_n^2}.
\end{equation}
To suppress the qubit decay via the cavity normal modes, we need a high-$Q$ cavity with a large effective volume. In practice, one often builds filters \cite{PhysRevApplied.10.034040, PhysRevA.92.012325} near the readout resonator to actively suppress the density of states away from the resonator frequency.

\subsection{Qudit Control}
The displacement of the cavity amplitude introduced for the two-level case did not affect the qubit and qubit-cavity parts of the Hamiltonian, so the previous derivation goes through for the qudit control as well (as long as $|g_{ij}| \ll |\Delta_{\text{rd}} - \ci \kappa / 2|$). If we adopt the usual selection rule of a weakly anharmonic qudit (i.e., $g_{ij} \approx 0$ if $|i-j| \neq 1$) and assume the coupling coefficients $g_{j,j+1}$ are real, we can immediately write down a generalization of Eq.(\ref{eq:displaced_H}) for a qudit
\begin{align}
    \hat{H}'
    &= \sum_{j=0}^{D-1}
        \hbar \omega_j 
        \ket{j}\!\bra{j} 
    + \hbar \omega_{\text{r}}
        \bigg(
            \hat{a}_{\text{r}}^{\dagger}
                \hat{a}_{\text{r}} 
            + \frac{1}{2}
        \bigg)
    - \sum_{j=0}^{D-2} 
        \hbar g_{j,j+1}
        \Big(
            \hat{a}_{\text{r}} \ket{j+1}\!\bra{j} 
            + \hat{a}_{\text{r}}^{\dagger} \ket{j}\!\bra{j+1}
        \Big)
\nonumber \\
    & \ \ \ \ \ \ \ \ \ \ \ \ \ \ \ \ \  \ \ \ \ \ \ \ \ \ \ \ \ \ \ \ \ \ 
    - \sum_{j=0}^{D-2} 
        \hbar g_{j,j+1} \sqrt{\bar{n}}
        \Big[
            e^{-\ci\omega_{\text{d}}t}  \ket{j+1}\!\bra{j} 
            + e^{\ci\omega_{\text{d}}t}  \ket{j}\!\bra{j+1}
        \Big],
\end{align}
where $\bar{n} = \kappa |\bar{a}_{\text{in}}|^2/\Delta_{\text{dr}}^2$ remains the same.
    
\section{Introduction to Master Equations}
As pointed out in the previous sections, a QLE describes the time evolution of the system observables in the Heisenberg picture. The state of the system, i.e., the density operator, is static in this picture. An alternative approach is to understand the time evolution of the density in the Schr\"{o}dinger picture, which leads to the master equation description of the open quantum system. For this introductory section, we will focus on the ensemble-averaged state where information leaked from the system is ignored by taking the partial trace of the composite density with respect to the environment (including our measurement devices). In other words, we look for differential equations of the form of Eq.(\ref{eq:Lindblad_master_eqn}).

\subsection{The Born-Markov Master Equation}
Derivation of a master equation under the Born and Markov assumption can be found in any textbook on open quantum systems. Here we simply restate some of the important assumptions and skip the proof. However, before discussing anything, it should be clear that a master equation cannot be a full description of a subsystem. By the axioms of quantum mechanics, the quantum Liouville equation is a general differential equation of the composite system, which satisfies the semigroup axiom\footnote{The semigroup axiom simply states that one can concatenate two state-transition maps for time intervals $[t_0,t_1]$ and $[t_1, t_2]$ to build a new state-transition map for the interval $[t_0,t_2]$. Since the state-transition maps do not need to be invertible, they form a semigroup instead of a group.} of dynamical systems. In general, there is no reason we should expect that a subsystem can also be described by a simple differential equation in isolation since information is exchanged between different subsystems. Hence, what a master equation assumes is the fact that we can approximate the time evolution of a subsystem also by a semigroup; this assumption is the so-called quantum semigroup axiom.

As usual, the starting point is always to introduce the Hamiltonian of the composite system, which can always be written as 
\begin{equation}
    \hat{H}
    = \hat{H}_{\mathcal{S}} 
        +\hat{H}_{\mathcal{E}} 
        +\hat{H}_{\text{int}},
\end{equation}
where $\hat{H}_{\mathcal{S}}$ and $\hat{H}_{\mathcal{E}}$ are the Hamiltonian of the system $\mathcal{S}$ and environment $\mathcal{E}$ in isolation, respectively, and 
\begin{equation}
    \hat{H}_{\text{int}}
    = \sum_{\alpha} 
        \hat{S}_{\alpha}
        \otimes 
        \hat{R}_{\alpha}
\end{equation}
is the interaction Hamiltonian. Like the derivation of the QLE, we assume that the interaction is weak so that to the first order any coupling can be expressed as the product of a system operator  $\hat{S}_{\alpha}$ and an environment operator $\hat{R}_{\alpha}$. In general, there can be multiple paths of coupling, so we have a sum of weak interactions indexed by $\alpha$.

Given the general form of the Hamiltonian, it's, in general, impossible to find a master equation as explained at the beginning of the subsection. Thus, we adopt two assumptions to make a master equation possible:
\begin{enumerate}
    \item[(i)] \textbf{Born assumption}: The environment has a huge number of degrees of freedom and is only weakly coupled to the system. In addition, the environment is always at thermal equilibrium. Hence, the density operator of the environment (i.e., with the system traced out) will be approximately constant 
    \begin{equation}
        \hat{\rho}_{\mathcal{E}}(t)
        = \Tr_{\mathcal{S}}
            \! \big[
                \hat{\rho}_{\mathcal{SE}}(t)
            \big]
        \approx \hat{\rho}_{\mathcal{E}}(0)
        \doteq \hat{\rho}_{\mathcal{E}}
    \end{equation}
    since it is much larger than the system and can be hardly affected by the weak interactions. In other words, if we write composite density operator as
    \begin{equation}
        \hat{\rho}_{\mathcal{SE}}(t)
        = \Tr_{\mathcal{E}}
            \! \big[
                \hat{\rho}_{\mathcal{SE}}(t)
            \big] 
            \otimes 
            \Tr_{\mathcal{S}}
            \! \big[
                \hat{\rho}_{\mathcal{SE}}(t)
            \big]
            + \hat{\rho}_{\text{corr}}(t)
    \end{equation}
    with some arbitrary correlation part $\hat{\rho}_{\text{corr}}(t)$, the Born assumption allows us to ignore $\hat{\rho}_{\text{corr}}(t)$ and write
    \begin{equation}
        \hat{\rho}_{\mathcal{SE}}(t)
        \approx
            \hat{\rho}_{\mathcal{S}}(t) \otimes
            \hat{\rho}_{\mathcal{E}}.
    \end{equation}

    \item[(ii)] \textbf{Markov assumption}: In the derivation of the master equation, one encounters the so-called environment correlation function
    \begin{equation}
        G_{\alpha \beta} (\tau)
        = \Tr 
            \!\Bigsl[ 
                \hat{\rho}_{\mathcal{E}} 
                \hat{\Tilde{R}}_{\alpha} (\tau)
                \hat{R}_{\beta}
            \Bigsr],
    \end{equation}
    where $\hat{\tilde{R}}(\tau)$ is the environment operator $\hat{R}$ in the interaction picture. The Markov assumption states that $G_{\alpha \beta} (\tau)$ is sharply peaked at $\tau = 0$, which implies that the self-correlation of the environment between time $t=0$ and $t=\tau$ is negligible for $\tau$ greater than some short characteristic time $\tau_{\text{corr}}$. This assumption is consistent with the fact that the environment is a large reservoir with infinitely many degrees of freedom. A typical example associated with this argument is the picture that a photon emitted by an atom into a mode of free space will never come back and can hence be ignored after $\tau_{\text{corr}}$. Clearly, this argument does not hold for an atom in a cavity since the photon in a cavity mode can be re-absorbed by the atom and cause Rabi flopping; thus, the large size and continuous nature of the environments are important for the argument to stand.
\end{enumerate}

The derivation now becomes straightforward. Starting from the quantum Liouville equation of the composite system in the interaction picture,
\begin{equation}
    \ci \hbar \frac{\mathrm{d}}{\mathrm{d} t} \hat{\Tilde{\rho}}(t)
    = \left[
            \hat{\Tilde{H}}_{\text{int}}(t),
            \hat{\Tilde{\rho}}(t)
        \right],
\end{equation}
we trace out the environment's degrees of freedom to obtain an equation for the reduced density operator $\hat{\rho}_{\mathcal{S}}$. With the Born and Markov assumptions applied, we would arrive at the following differential equation of $\hat{\rho}_{\mathcal{S}}$ in the Schr\"{o}dinger picture \cite{schlosshauer2007decoherence}:
\begin{align} \label{eq:born_markov_master_equation_derivation}
    \frac{\mathrm{d}}{\mathrm{d} t} \hat{\rho}_{\mathcal{S}}(t)
    &= -\frac{\ci}{\hbar} 
        \left[
            \hat{H}_{\mathcal{S}},
            \hat{\rho}_{\mathcal{S}}(t)
        \right]
\nonumber \\
    & \ \ \ 
        - \frac{1}{\hbar^2} \!
        \int_{0}^{\infty} 
            \! \mathrm{d} \tau
            \sum_{\alpha,\beta}
            \! \bigg\{
                G_{\alpha \beta}(\tau) \!
                \left[
                \hat{S}_{\alpha} (t)
                \hat{\Tilde{S}}_{\beta} (-\tau)
                \hat{\rho}_{\mathcal{S}}(t)
                - \hat{\Tilde{S}}_{\beta} (-\tau)
                \hat{\rho}_{\mathcal{S}}(t)
                \hat{S}_{\alpha} (t)
                \right]
\nonumber \\[-1mm]
        & \ \ \ \ \ \ \ \ \ \ \ 
        \ \ \ \ \ \ \ \ \ \ \ \ \ \ 
            + G_{\beta\alpha}(-\tau) \!
                \left[
                \hat{\rho}_{\mathcal{S}}(t)
                \hat{\Tilde{S}}_{\beta} (-\tau)
                \hat{S}_{\alpha} (t)
                - \hat{S}_{\alpha} (t) \hat{\rho}_{\mathcal{S}}(t)
                \hat{\Tilde{S}}_{\beta} (-\tau)
                \right]
            \! \bigg\},
\end{align}
where, as always, operators with tildes are in the interaction picture. 

To summarize, Eq.(\ref{eq:born_markov_master_equation_derivation}) is the \textbf{Born-Markov master equation} in its most general form. One can then apply the master equation to specific interaction Hamiltonian and make further assumptions such as the RWA (also known as the secular approximation in the literature) \cite{cohen1992atom}. Furthermore, by using the formalism of eigenoperators, one can also put Eq.(\ref{eq:born_markov_master_equation_derivation}) in the well-celebrated Lindblad form:
\begin{align}
    \frac{\mathrm{d}}{\mathrm{d} t}
        \hat{\rho}_{\mathcal{S}}
    &= - \frac{\ci}{\hbar} 
        \left[ 
            \hat{H}_{\text{eff}}, \hat{\rho}_{\mathcal{S}}
        \right]
        + \sum_{\mu=1}^{N^2-1}
            k_{\mu}
            \mathcalboondox{D}\Bigsl[\hat{L}_{\mu} \Bigsr] 
            \hat{\rho}_{\mathcal{S}},
\nonumber \\ \label{eq:Lindblad_master_eqn_derived}
    &= - \frac{\ci}{\hbar} 
        \left[ 
            \hat{H}_{\text{eff}}, \hat{\rho}_{\mathcal{S}}
        \right] 
        + \sum_{\mu=1}^{N^2-1}
            k_{\mu} \left( 
                \hat{L}_{\mu} 
                \hat{\rho}_{\mathcal{S}} 
                \hat{L}_{\mu}^{\dagger}
                - \frac{1}{2}
                    \hat{L}_{\mu}^{\dagger} 
                    \hat{L}_{\mu} 
                    \hat{\rho}_{\mathcal{S}}
                - \frac{1}{2}
                    \hat{\rho}_{\mathcal{S}} 
                    \hat{L}_{\mu}^{\dagger} 
                    \hat{L}_{\mu}
            \right),
\end{align}
which is the same as Eq.(\ref{eq:Lindblad_master_eqn}) but the averaging over the measurement results is formalized by the reduced density operator $\hat{\rho}_{\mathcal{S}}$. Note that the Hamiltonian in the commutator is no longer $\hat{H}_{\mathcal{S}}$ due to the Lamb shifts hidden inside the integral in Eq.(\ref{eq:born_markov_master_equation_derivation}). Most of the time, the Lamb shifts can be absorbed into the original system Hamiltonian; nevertheless, one should be aware of this subtlety. The Lindblad operators $\hat{L}_{\mu}$ and the associated strength $k_{\mu}$ can be related to the system operators $\hat{S}_{\alpha}$ directly; however, in practice, people usually start with Eq.(\ref{eq:born_markov_master_equation_derivation}) for a specific microscopic description of the system-environment coupling (e.g., the dipole interaction for spontaneous emission) and try to cast the entire master equation to the Lindblad form directly. Alternatively, one can also show, from the perspective of the quantum channels, that a Lindblad master equation is the limit of Eq.(\ref{eq:discrete_stochastic_equation}) as $T/N \rightarrow 0$. In fact, the Lindblad operators are the infinitesimal versions of the Kraus operators. Of course, any information gained from the measurement is ignored since we are tracing out the environment.




\subsection{Example: Qubit and Qudit Decay (at Zero Temperature)}
It's well known that we can use a master equation to model the spontaneous emission of an atom when coupled to free space via a dipole interaction. A two-level system coupled to free space that has no thermal excitation, of course, falls into the general description as well. One can show from the first principle or phenomenologically that the decay of a two-level system is associated with the Lindblad operator $\hat{L}_1 = \hat{\sigma}_{-}$ with some decay rate $k_1 = \gamma_{1}$. 
Substitute the Lindblad operator into Eq.(\ref{eq:Lindblad_master_eqn}) yields the master equation \cite{cohen1992atom}
\begin{align}
    \dot{\hat{\rho}}_{\mathcal{S}}
    &= -\frac{\ci}{\hbar} 
        \left[ 
            - \frac{1}{2} \Tilde{\omega}_{\text{q}} \hat{\sigma}_z, \hat{\rho}_{\mathcal{S}}
        \right]
        +  \gamma_1 
            \hat{\sigma}_{-} 
            \hat{\rho}_{\mathcal{S}}
            \hat{\sigma}_{+}
        - \frac{\gamma_1}{2}
            \hat{\sigma}_{+} 
            \hat{\sigma}_{-} 
            \hat{\rho}_{\mathcal{S}}
        - \frac{\gamma_1}{2}
            \hat{\rho}_{\mathcal{S}} 
            \hat{\sigma}_{+}
            \hat{\sigma}_{-} 
\nonumber \\ \label{eq:qubit_decay_master_equation}
    &= \begin{pmatrix}
                0 & \ci \Tilde{\omega}_{\text{q}} \rho_{ge} \\
                -\ci \Tilde{\omega}_{\text{q}} \rho_{eg} & 0
            \end{pmatrix}
        + \begin{pmatrix}
                \gamma_1 \rho_{ee} & 0 \\
                0 & 0 
            \end{pmatrix}
        + \begin{pmatrix}
                0 & 0 \\
                \displaystyle - \gamma_1 \rho_{eg} /2 & \displaystyle - \gamma_1 \rho_{ee} /2
            \end{pmatrix}
        + \begin{pmatrix}
                0 & \displaystyle -\gamma_1\rho_{ge}/2 \\
                0 & \displaystyle -\gamma_1 \rho_{ee}/2 
            \end{pmatrix},
\end{align}
where $\rho_{ab} = \bra{a}\hat{\rho}_{\mathcal{S}} \ket{b}$ for $a,b \in \{g, e\}$ and $\Tilde{\omega}_{\text{q}}$ is the Lamb-shifted qubit frequency. 

Before summing the matrices, let us examine each term in the master equation: The first matrix on the right-hand side of Eq.(\ref{eq:qubit_decay_master_equation}) is the free evolution of the qubit. The second matrix is the main term that describes the decay of the qubit into the ground state with a rate $\gamma_1$. However, in order to conserve the total probability (i.e., the sum of the diagonal terms of the density operator), the excited state must show a decay at the same rate, which is exactly demonstrated by the last two terms. As a consequence of being trace-preserving, we naturally obtain a dephasing of the off-diagonal terms (i.e., the coherence between the $z$-basis vectors when a qubit is in a superposition state expressed in the $z$-basis) at a rate $\gamma_1 / 2$. The decay-induced dephasing is unavoidable since it's mathematically enforced by probability conservation; thus, knowing the decay time $T_1 = 1/\gamma_1$ gives us an upper bound on the total dephasing time $T_2^*$ (i.e., the decay rate of the off-diagonal terms). In practice, the qubit will also have a \text{pure dephasing} contribution in the master equation which will be discussed below. Let the pure dephasing rate be $\gamma_{\phi}$, then the total dephasing time is $T_2^* = 1/(\gamma_{\phi} + \gamma_1/2) \leq 2T_1$.

Finally, combining the matrices gives the matrix differential equation
\begin{equation}
    \dot{\hat{\rho}}_{\mathcal{S}}
    = \begin{pmatrix}
            \gamma_1 \rho_{ee} & \ci \Tilde{\omega}_{\text{q}} \rho_{ge} \displaystyle -\gamma_1\rho_{ge}/2 \\
            -\ci \Tilde{\omega}_{\text{q}} \rho_{eg} - \gamma_1 \rho_{eg} /2 & -\gamma_1 \rho_{ee}
        \end{pmatrix},
\end{equation}
which is trivial to solve since the diagonal and the off-diagonal terms are decoupled given an initial quantum state. At a finite temperature, one can also add an equilibrium population to the master equation so that the steady-state solution of the master equation is not $\ket{g}\!\bra{g}$. Nevertheless, at the superconducting temperature where $\hbar \tilde{\omega}_{\text{q}} \gg k_{\text{B}} T$, the thermal excitation can usually be ignored.

The decay of a qudit can be modeled in a similar way. One can easily verify that $\hat{L}_{ij} = \ket{i} \! \bra{j}$ is the Lindblad operator responsible for the decay from $\ket{j}$ to $\ket{i}$, assuming $\ket{j}$ is more energetic than $\ket{i}$. Thus, for a qutrit with energy levels $ \{\ket{g}, \ket{e}, \ket{f} \}$ (ordered with increasing energy), we can phenomenologically construct a master equation of the form
\begin{equation} 
    \dot{\hat{\rho}}_{\mathcal{S}}
    = -\frac{\ci}{\hbar} 
        \left[
            \hat{H}_{\text{eff}}, \hat{\rho}_{\mathcal{S}}
        \right]
    + \gamma_{1, ge}
        \mathcalboondox{D}\big[\ket{g} \! \bra{e}\big]
        \hat{\rho}_{\mathcal{S}}
    + \gamma_{1, gf} 
        \mathcalboondox{D}\big[\ket{g} \! \bra{f}\big]
        \hat{\rho}_{\mathcal{S}}
    + \gamma_{1, ef}
        \mathcalboondox{D}\big[\ket{e} \! \bra{f}\big]
        \hat{\rho}_{\mathcal{S}},
\end{equation}
where $\hat{H}_{\text{eff}}$ is the Lamb-shifted Hamiltonian of the qutrit and $\gamma_{1, ab}$ is the decay rate from $\ket{b}$ to $\ket{a}$.

\subsection{Example: Resonator Decay (at a Finite Temperature)}
A similar decay model can be created for a resonator coupled to the free space. Instead of assuming the free space is held at zero temperature, we consider an environment with a finite temperature $T$ such that the mean photon number in each mode ($\omega$) of the environment is given by
\begin{equation}
    \Bar{N}(\omega) = \frac{1}{e^{\hbar \omega/k_{\text{B}}T} - 1}.
\end{equation}
As for the Hamiltonian of the composite system, Eq.(\ref{eq:Hamiltonian_of_QHO_coupled_to_QHOs}) used in the derivation of the QLE provides the appropriate Hamiltonian for an oscillator coupled to infinitely many oscillators in the environment. By substituting Eq.(\ref{eq:Hamiltonian_of_QHO_coupled_to_QHOs}) into the Born-Markov master equation, one finds \cite{breuer2002theory}
\begin{equation} \label{eq:resonator_decay_master_equation}
    \dot{\hat{\rho}}_{\mathcal{S}}
    = -\frac{\ci}{\hbar} 
        \left[ 
            \hat{H}_{\text{eff}}, \hat{\rho}_{\mathcal{S}}
        \right]
        + \kappa \Bigsl[\Bar{N}(\Tilde{\omega}) + 1 \Bigsr]
            \mathcalboondox{D}\Bigsl[\hat{a}_{\text{r}} \Bigsr] 
            \hat{\rho}_{\mathcal{S}}
        + \kappa \Bar{N}(\Tilde{\omega})
            \mathcalboondox{D}\Bigsl[\hat{a}_{\text{r}}^{\dagger} \Bigsr] 
            \hat{\rho}_{\mathcal{S}},
\end{equation}
where $\kappa$ is the decay rate defined from before and $\hat{H}_{\text{eff}} = \hbar \tilde{\omega}_{\text{r}} \hat{a}_{\text{r}}^{\dagger} \hat{a}_{\text{r}}$ is the Lamb-shifted Hamiltonian with the zero-point energy omitted. If the environment is at zero temperature, then the last term on the right-hand side of Eq.(\ref{eq:resonator_decay_master_equation}) disappears, and the resulting master equation
\begin{equation} \label{eq:resonator_decay_master_equation_no_thermal_photon}
    \dot{\hat{\rho}}_{\mathcal{S}}
    = -\frac{\ci}{\hbar} 
        \left[ 
            \hat{H}_{\text{eff}}, \hat{\rho}_{\mathcal{S}}
        \right]
        + \kappa \mathcalboondox{D}\Bigsl[\hat{a}_{\text{r}} \Bigsr] 
            \hat{\rho}_{\mathcal{S}}
\end{equation}
is essentially the same as that for the two-level system but with $\hat{\sigma}$ and $\gamma_1$ replaced with $\hat{a}_{\text{r}}$ and $\kappa$, respectively, as one might expect.

More assuring is the fact that we can establish the equivalence between the master equation and QLE. Recall that the QLE written in terms of the annihilation operator of the resonator is given by
\begin{equation}
    \Bigsl\langle \dot{\hat{a}}_{\text{r}} \Bigsr\rangle
    = - \ci \omega_{\text{r}}
            \langle \hat{a}_{\text{r}} \rangle
        - \frac{\kappa}{2} \,
            \langle \hat{a}_{\text{r}} \rangle
        + \sqrt{\kappa} \, 
            \langle \hat{a}_{\text{in}} \rangle
\end{equation}
when taking the expectation on both sides. For a thermal bath, the averaged input field $\langle \hat{a}_{\text{in}} \rangle$ is zero; thus, the time evolution of the average amplitude of the resonator is given by
\begin{equation} \label{eq:time_evolution_dampled_amplitude}
    \langle \hat{a}_{\text{r}} (t) \rangle
    = \langle \hat{a}_{\text{r}} (0) \rangle e^{(-\ci \omega_{\text{r}} - \kappa/2) t},
\end{equation}
which is what one would expect for a damped resonator. Similarly, one can also find the time evolution of $\langle \hat{a}_{\text{r}} \rangle$ in the Schr\"{o}dinger picture from the master equation\footnote{In general, one can construct a differential equation for any system operator from the master equation. Such an equation is known as the \textbf{adjoint equation}.}. The resulting expression is exactly the same as Eq.(\ref{eq:time_evolution_dampled_amplitude}) (up to a negligible Lamb shift ignored in the QLE). 

One might wonder what is the effect of a finite temperature because Eq.(\ref{eq:time_evolution_dampled_amplitude}) seems to be independent of $\bar{N}$. The influence of the thermal photon is hidden in the variance of the amplitude. One can show that the mean photon number follows the transient behavior
\begin{equation}\label{eq:time_evolution_resonator_mean_photon}
    \Bigsl \langle 
            \hat{a}_{\text{r}}^{\dagger}(t) \hat{a}_{\text{r}}(t)
        \Bigsr\rangle
    = e^{-\kappa t} 
            \hat{a}_{\text{r}}^{\dagger}(0) 
            \hat{a}_{\text{r}}(0)
        + \bar{N}(\Tilde{\omega}_{\text{r}}) 
            \big(
                1 - e^{-\kappa t}
            \big).
\end{equation}
Thus, when the system equilibrates with the thermal bath (i.e., as $t \rightarrow \infty$), the noise added to the resonator is exactly the mean photon number $\bar{N}$ of the thermal bath at the resonator frequency. This is again a manifestation of the fluctuation-dissipation theorem\footnote{In fact, based on Eq.(\ref{eq:time_evolution_dampled_amplitude}) and (\ref{eq:time_evolution_resonator_mean_photon}), we can argue that the amplitude $\alpha_t = \langle \hat{a}_{\text{r}} (t) \rangle$ should satisfy the stochastic differential equation 
\begin{equation} \label{eq:QLE_as_SDE}
    \mathrm{d} \alpha_t = \left(-\ci \omega_{\text{r}} - \frac{\kappa}{2} \right) \alpha_t \mathrm{d}t + \sqrt{\kappa \bar{N}(\omega_{\text{r}})} \, \mathrm{d}W_t,
\end{equation}
where $W_t$ is the complex Wiener process with $\mathbb{E}(\mathrm{d}W_t) = 0$, $\mathbb{E}(\mathrm{d}W_t \mathrm{d}W_t) = 0$, and $\mathbb{E}(\mathrm{d}W_t^* \mathrm{d}W_t) = \mathrm{d}t$. In other words, the complex Wiener process can be defined as $W_t = \Bigsl(W^{(1)}_t + \ci W^{(2)}_t\Bigsr) / \sqrt{2}$, where $W^{(1)}_t$ and $W^{(2)}_t$ are real Wiener processes satisfying $\smash{\mathbb{E}\Bigsl(\mathrm{d} W^{(1)}_t \Bigsr) = \mathbb{E}\Bigsl(\mathrm{d} W^{(2)}_t \Bigsr) = \mathbb{E}\Bigsl(\mathrm{d} W^{(1)}_t \mathrm{d} W^{(2)}_t\Bigsr) = 0}$ and $\mathbb{E}\Big[\Bigsl(\mathrm{d} W^{(1) }_t\Bigsr)^2\Big] = \mathbb{E}\Big[\Bigsl(\mathrm{d} W^{(2)}_t\Bigsr)^2\Big]= \mathrm{d} t$. Eq.(\ref{eq:QLE_as_SDE}) can be treated as two classical Langevin equations.
}. 

\subsection{Bloch Equations}
For describing the decoherence of a qubit in practice, an effective master equation is usually proposed in a phenomenological way. We have already seen how to describe the qubit decay. To add the pure dephasing due to quasi-static fluctuations from the environment, we introduce the Lindblad operator $\hat{L}_2 = \hat{\sigma}_z$ with a rate $\gamma_{\phi}/2$, resulting in
\begin{equation} \label{eq:qubit_T1_T2_master equation}
    \dot{\hat{\rho}}_{\mathcal{S}}
    = -\frac{\ci}{\hbar} 
        \left[
            \hat{H}_{\text{eff}}, \hat{\rho}_{\mathcal{S}}
        \right]
        + \gamma_1 
        \mathcalboondox{D}[\hat{\sigma}_{-}] 
        \hat{\rho}_{\mathcal{S}}
    + \frac{\gamma_{\phi}}{2} 
        \mathcalboondox{D}[\hat{\sigma}_{z}]
        \hat{\rho}_{\mathcal{S}}.
\end{equation}
The master equation, written entry-wisely, is given by
\begin{align}
    \dot{\rho}_{gg} 
    & = \gamma_1 \rho_{ee} ,
\\ 
    \dot{\rho}_{ee} 
    &= - \gamma_1 \rho_{ee} ,
\\
    \dot{\rho}_{eg} 
    &= - \ci \tilde{\omega}_{\text{q}} \rho_{eg}
        - \left(\frac{\gamma_1}{2} + \gamma_{\phi} \right) \rho_{eg}. 
\end{align}
Indeed, the correct pure dephasing rate is added to the off-diagonal term. In addition, it's customary to define the total dephasing rate  
\begin{equation}
    \gamma_2 = \frac{\gamma_1}{2} + \gamma_{\phi},
\end{equation}
which can be measured in the experiment by using the Ramsey interference.

Suppose the qubit is initially in the pure state $\ket{\Psi(0)} = \alpha \ket{g} + \beta \ket{e}$, then the time evolution of its density matrix is given by
\begin{equation}\label{eq:general_solution_T1_T2_decay}
    \hat{\rho}_{\mathcal{S}}(t)
    = \begin{pmatrix}
        \displaystyle 1 + (|\alpha|^2-1) e^{-\gamma_1 t} 
        & \displaystyle \alpha \beta^* e^{\ci \Tilde{\omega}_{\text{q}} t} e^{-\gamma_2 t} \\[4mm]
        \displaystyle \alpha^* \beta e^{-\ci \Tilde{\omega}_{\text{q}} t} e^{-\gamma_2 t}  
        & \displaystyle |\beta|^2 e^{-\gamma_1 t} 
    \end{pmatrix}.
\end{equation}
At $t \rightarrow \infty$, the qubit ends up in the pure state $\ket{g}$. However, it's not true that the qubit stays pure while decaying toward the ground state from an initially pure state. For instance, consider the case where $\ket{\Psi(0)} = \ket{e}$. Then, the coherence $\rho_{eg}$ of the qubit stays at zero in the $z$-basis as implied by Eq.(\ref{eq:general_solution_T1_T2_decay}). Recall that any pure state of a two-level system can be plotted as a unit vector on the Bloch sphere; hence, to go continuously from the south pole ($\ket{e}$) to the north pole ($\ket{g}$) of the sphere, one must visit states with a nonzero value $\alpha\beta^*$ to stay on the surface of the sphere. In contrast, Eq.(\ref{eq:general_solution_T1_T2_decay}) implies that the quantum state starting from $\ket{e}$ will go into the unit sphere, passing through the origin, and move straight towards $\ket{g}$. 

A state that does not live on the surface of the Bloch sphere is known as a mixed state since it can be thought of as an ensemble of pure quantum states. Physically, each spontaneous emission of the qubit will result in one-half of a Rabi oscillation from $\ket{e}$ to $\ket{g}$; thus a single trajectory of the experiment will move on the surface of the Bloch sphere. However, since there are infinitely many possible trajectories going from $\ket{e}$ to $\ket{g}$ (due to the arbitrariness of $\phi$-coordinate), we have no way of retrieving the phases in an ensemble-averaged experiment. Therefore, the average trajectory will give rise to a set of Bloch vectors that are aligned with the $z$-axis by symmetry. Mathematically, non-unit Bloch vectors are also allowed by the definition of the density matrix. One can verify that any $2 \times 2$ positive Hermitian operator with a unit trace can be expressed as
\begin{equation}
    \hat{\rho} 
    = \frac{1}{2} 
        \left(
            \hat{1} 
            + \langle \hat{\boldsymbol{\sigma}} \rangle_{\hat{\rho} } \cdot \hat{\boldsymbol{\sigma}}
        \right),
\end{equation}
where the expectation value
\begin{equation}
    \langle \hat{\boldsymbol{\sigma}} \rangle_{\hat{\rho} }
    = \Tr(\hat{\rho}\hat{\boldsymbol{\sigma}})
\end{equation}
is the Bloch-vector representation of the state. In particular, $\langle \hat{\boldsymbol{\sigma}} \rangle_{\hat{\rho} }$ is a unit vector for a pure state and a vector with a length strictly less than unity for any mixed state. As we will see in the more complicated master equations, the system's evolution obtained by tracing out a part of the composite system will often result in a decay of the coherence, thus making the quantum state impure.

\subsection{Ramsey Interference}
To experimentally characterize the loss of coherence of a qubit, one often deploys the so-called Ramsey sequence. In a standard Ramsey sequence, the qubit is first excited to an equal superposition state by a $\pi/2$-pulse. The qubit is then allowed to evolve freely based on Eq.(\ref{eq:general_solution_T1_T2_decay}). After some time $T_{\text{free}}$, another $\pi/2$-pulse is sent to the qubit, and the state of the qubit is measured. For simplicity, we will consider the idealized case where $T_1 \gg T_2^*$ so that the decoherence is due to pure dephasing. We will also assume that the duration of the $\pi/2$-pulses, $T_{\pi/2}$, are much shorter than $T_2^*$ so that the decay within the pulse time can be ignored to a good approximation. In practice, the second condition should always be satisfied for a well-designed experiment. 

We now show a quantitative analysis of the result of the experiment. As emphasized before, the axis of the Rabi oscillation is an important physical parameter in the experiment. Suppose all of our control pulses are generated by a single local oscillator (LO) operating at $\omega_{\text{d}}$. Now, imagine at time $t=0$, the phase of the local oscillator is set in such as way that the phase of the Rabi flopping is exactly $\phi_{g} = 0$. Then, the first $\pi/2$-pulse is described by the unitary matrix
\begin{equation}
    \hat{U}_{\pi/2, 1} 
    = \frac{1}{\sqrt{2}}
        \begin{pmatrix}
            e^{\ci \omega_{\text{d}} T_{\pi/2}/2} & 
            \ci e^{\ci \omega_{\text{d}} T_{\pi/2}/2} \\ 
            \ci e^{-\ci \omega_{\text{d}} T_{\pi/2}/2}  & 
            e^{-\ci \omega_{\text{d}} T_{\pi/2}/2}
        \end{pmatrix}
\end{equation}
as computed from Eq.(\ref{eq:semiclassical_single_qubit_gate}). We have assumed that $\Tilde{\Delta}_{\text{qd}} = \Tilde{\omega}_{\text{q}} - \omega_{\text{d}} \ll \Omega$ so that the generalized Rabi frequency is $\Omega' \approx \Omega$. Applying the gate to the initial state $\ket{\Psi_0} = \ket{g}$ gives the new state
\begin{equation} \label{eq:ramsey_state_1}
    \ket{\Psi_1} 
    = \hat{U}_{\pi/2, 1} \ket{\Psi_0} 
    = \frac{e^{\ci \omega_{\text{d}} T_{\pi/2}/2}}{\sqrt{2}} \ket{g}
        + \frac{\ci e^{-\ci \omega_{\text{d}} T_{\pi/2}/2}}{\sqrt{2}} \ket{e}.
\end{equation}
Next, in the limit $T_1 \gg T_2^*$, the qubit ends up in the mixed state
\begin{equation} \label{eq:ramsey_state_2}
    \hat{\rho}_2 
    = \frac{1}{2} 
        \begin{pmatrix} 
        \displaystyle 1
        & \displaystyle - \ci e^{\ci \omega_{\text{d}} T_{\pi/2}} e^{\ci \Tilde{\omega}_{\text{q}} T_{\text{free}}} e^{-\gamma_2 T_{\text{free}}} \\[4mm]
        \displaystyle \ci e^{-\ci \omega_{\text{d}} T_{\pi/2}}  e^{-\ci \Tilde{\omega}_{\text{q}} T_{\text{free}}} e^{-\gamma_2 T_{\text{free}}}  
        & \displaystyle 1
    \end{pmatrix}
\end{equation}
by substituting Eq.(\ref{eq:ramsey_state_2}) as the initial condition into Eq.(\ref{eq:general_solution_T1_T2_decay}). In the meanwhile, the LO is not turned off even though we are not sending any signals to the qubit; therefore, after $T_{\text{free}}$, the LO has accumulated a phase of $- \omega_{\text{d}} T_{\text{free}}$. This extra phase will now change the axis of rotation for the second $\pi/2$-pulse, i.e., the unitary matrix describing the second pulse is given by
\begin{equation}
    \hat{U}_{\pi/2, 2} 
    = \frac{1}{\sqrt{2}}
        \begin{pmatrix}
            e^{\ci \omega_{\text{d}} T_{\pi/2}/2} & 
            \ci e^{\ci \omega_{\text{d}} T_{\pi/2}/2 + \ci \omega_{\text{d}} T_{\text{free}}} 
        \\ 
            \ci e^{-\ci \omega_{\text{d}} T_{\pi/2}/2 - \ci \omega_{\text{d}} T_{\text{free}}}  
            & 
            e^{-\ci \omega_{\text{d}} T_{\pi/2}/2}
        \end{pmatrix}.
\end{equation}
Applying $\hat{U}_{\pi/2, 2} $ to the state $\hat{\rho}_2 $ gives the final state
\begin{align}
    &\hat{\rho}_3 
    = \hat{U}_{\pi/2, 2} 
        \hat{\rho}_2 
        \hat{U}_{\pi/2, 2}^{\dagger} 
\nonumber \\
    &= \frac{1}{4}\!\!
        \begin{pmatrix}
            2 - 2\cos\Bigsl(\omega_{\text{d}} T_{\pi/2} + \Tilde{\Delta}_{\text{qd}} T_{\text{free}}\Bigsr) e^{-\gamma_2 T_{\text{free}}}
            & \!\!\!\!\!\! \ci e^{(-\ci \tilde{\omega}_{\text{q}} + 2\ci \tilde{\omega}_{\text{d}} - \gamma_2)T_{\text{free}}}
            - \ci e^{2\ci \omega_{\text{d}} T_{\pi/2} + (\ci \tilde{\omega}_{\text{q}}
            -\gamma_2)T_{\text{free}}}\!
        \\
            - \ci e^{(\ci \tilde{\omega}_{\text{q}} - 2\ci \tilde{\omega}_{\text{d}} - \gamma_2)T_{\text{free}}} 
            + \ci e^{-2\ci \omega_{\text{d}} T_{\pi/2} + (-\ci \tilde{\omega}_{\text{q}}
            -\gamma_2)T_{\text{free}}}
            &
            \!\!\!\!\!\! 2 + 2\cos\Bigsl(\omega_{\text{d}} T_{\pi/2} + \Tilde{\Delta}_{\text{qd}} T_{\text{free}}\Bigsr) e^{-\gamma_2 T_{\text{free}}}
        \end{pmatrix}\!\! .
\end{align}
Finally, by performing the projective measurement in the $z$-basis, we obtain
\begin{gather} \label{eq:ramsey_p_g}
    p_{g}(T_{\text{free}}) 
    = \rho_{3,gg}(T_{\text{free}}) 
    = \frac{1}{2} 
        - \frac{1}{2} 
            \cos 
                \Bigsl(
                    \omega_{\text{d}} T_{\pi/2} + \Tilde{\Delta}_{\text{qd}} T_{\text{free}} 
                \Bigsr) 
            e^{-\gamma_2 T_{\text{free}}},
\\ \label{eq:ramsey_p_e}
    p_{e}(T_{\text{free}}) 
    = \rho_{3,ee}(T_{\text{free}}) 
    = \frac{1}{2} 
        + \frac{1}{2} 
            \cos 
                \Bigsl(
                    \omega_{\text{d}} T_{\pi/2} + \Tilde{\Delta}_{\text{qd}} T_{\text{free}} 
                \Bigsr) 
            e^{-\gamma_2 T_{\text{free}}}.
\end{gather}

If the qubit has no dephasing, Eq.(\ref{eq:ramsey_p_g}) and (\ref{eq:ramsey_p_e}) indicate an oscillation, known as the Ramsey fringes, between the ground and excited states at the beat frequency $\Tilde{\Delta}_{\text{qd}}$. This interference can be explained nicely using the Bloch sphere: When a qubit in the equal superposition state undergoes an ideal free evolution (i.e., $\gamma_1 = \gamma_2 = 0$), its Bloch vector simply rotates in the $xy$-plane with a precession frequency $\Tilde{\omega}_{\text{q}}$ ($\Tilde{\omega}_{\text{q}} = \omega_{\text{q}}$ ideally). In the meanwhile, the LO oscillates with a frequency $\omega_{\text{d}}$; hence, in the frame rotating at $\omega_{\text{d}}$ (i.e., the point of view of the LO), the qubit precesses at $\tilde{\Delta}_{\text{qr}} = \Tilde{\omega}_{\text{q}} - \omega_{\text{d}}$. But the phase of the drive is always zero in the rotating frame, so the axes of rotation for the first and second $\pi/2$-pulses stay static while the qubit has rotated by $\tilde{\Delta}_{\text{qr}} T_{\text{free}}$. Hence, depending on $T_{\text{free}}$, the Bloch vector of the qubit will make different angles with the axis of rotation. On the one hand, if the Bloch vector is aligned with the axis of rotation, then the qubit stays in the $xy$-plane after the second $\pi/2$-pulse; on the other hand, when the Bloch vector is perpendicular to the rotation axis, the state will be brought to either $\ket{g}$ or $\ket{e}$.

Consequently, adding a nonzero dephasing creates a decay of the interference. In particular, in an actual experiment, one finds $\gamma_2$ by fitting the envelope of the decaying oscillation with an exponential function of $T_{\text{free}}$. Recall that our derivation assumes that $T_2^* \ll T_1$, i.e., the system is subject to strong pure dephasing, so the fitted decay rate is $\gamma_{\phi}$ in essence. If we add the decay-induced dephasing back, we will conclude that the oscillation decays with $\gamma_2 = \gamma_{\phi} + \gamma_1/2$. Thus, in reality, one finds $T_2^*$ with two contributions.

\begin{figure}[t]
    \centering
    \includegraphics[scale=0.3]{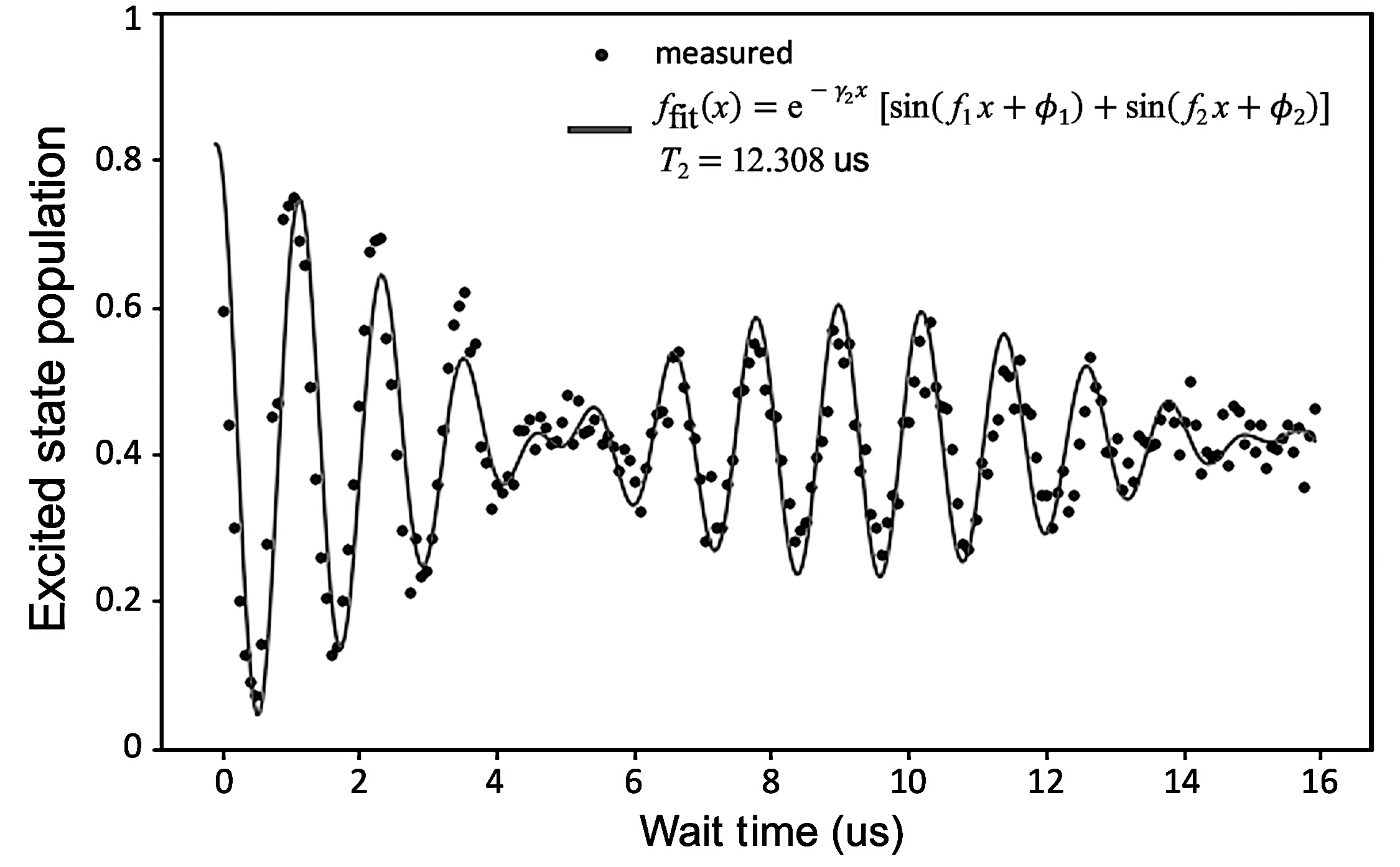}
    \caption{Effect of charge noise and parity switching in Ramsey measurements.}
    \label{fig:ramsey_charge_noise}
\end{figure}

When measuring the Ramsey fringes associated with an anharmonic oscillation, we need to the effect of charge noise and party switching. As mentioned before, a large $E_{\text{J}} / E_C$ makes a transmon less sensitive to external changes. However, for 
an intermediate value of $E_{\text{J}} / E_C$ (i.e., around $20 \sim 40$), the Cooper pair box is not in the deep transmon regime and thus is still affected by charge noise and party switching. For example, the transition frequency between $\ket{g}$ and $\ket{e}$ could drift between two frequencies\footnote{Recall that in the charge-basis picture, the Josephson energy opens bandgaps in the energy spectrum. For each energy band, there will be a minimum and maximum energy; the charge noise can steer the state between the two limits continuously. If $E_{\text{J}} \gg E_C$, the energy difference between the minimum and maximum points of the band will be small; thus, we say that the transmon is less sensitive to the charge noise. However, for a transmon with $E_{\text{J}}/E_C \sim 30$, we can still observe this frequency difference sometimes.}, $f_{\text{min}}$ and $f_{\text{max}}$, in a long experiment. By repeating the Ramsey measurement multiple times and averaging the fringes, one observes a beating at the frequency difference $f_{\text{max}} - f_{\text{min}}$, as shown in Figure \ref{fig:ramsey_charge_noise}.

\subsection{Dephasing Modeling and Dynamical Decoupling}
The dephasing term is added phenomenologically in Eq.(\ref{eq:qubit_T1_T2_master equation}). In reality, they are resulting from various low-frequency noises in the system. For example, the charge and flux noises mentioned before can cause the qubit frequency to jump stochastically \cite{s41467_021_21098_3} (see Figure \ref{fig:charge_noise_monitoring}). To model the random shift of the qubit frequency, we can modify the nominal qubit frequency by some random processes attributed to the different noise sources. As shown in Appendix \ref{eq:appendix_dephasing}, under certain approximations, a noise source can cause the coherence term of the density operator to decay:
\begin{equation}\label{eq:dephasing_noise_derivation}
    \rho_{ge} (t) 
    = e^{\ci \omega_{\text{q},0} t} \,
        \exp
        \left[ 
            - \frac{t^2}{2}
            \int_{-\infty}^{\infty}
                \frac{\mathrm{d} \omega }{2\pi} \, 
                \frac{\sin^2(\omega t /2)}{(\omega t /2)^2} 
                S_{v}(\omega)
        \right] 
        \rho_{ge}(0),
\end{equation}
where $S_{v}(\omega)$ is the power spectral density of the corresponding noise source. Hence, the noise in the environment is coupled to the off-diagonal terms of the qubit density operator by an overlapping integral between $S_{v}(\omega)$ and a filter function
\begin{equation}
    g_0 (\omega)
    = \frac{\sin^2(\omega t /2)}{(\omega t /2)^2} 
                S_{v}(\omega),
\end{equation}
which is the square of a sinc function.

To understand the decay shown in Eq.(\ref{eq:dephasing_noise_derivation}), recall that a sinc function in the frequency domain corresponds to a rectangular function in the time domain. Hence, the filtering of the noise amplitude (i.e., the square root of the noise power) in the frequency domain can be thought of as sampling the noise within a window of length $t$ in the time domain. As a result, any noise component with a frequency higher than $1/t$ will be averaged out within the observation window while the low-frequency noise is integrated with $g_0 (\omega)$. In fact, in the Ramsey experiment discussed above, we wait for $T_{\text{free}}$ in between two $\pi/2$-pulses, which effectively creates a window of size $T_{\text{free}}$. 

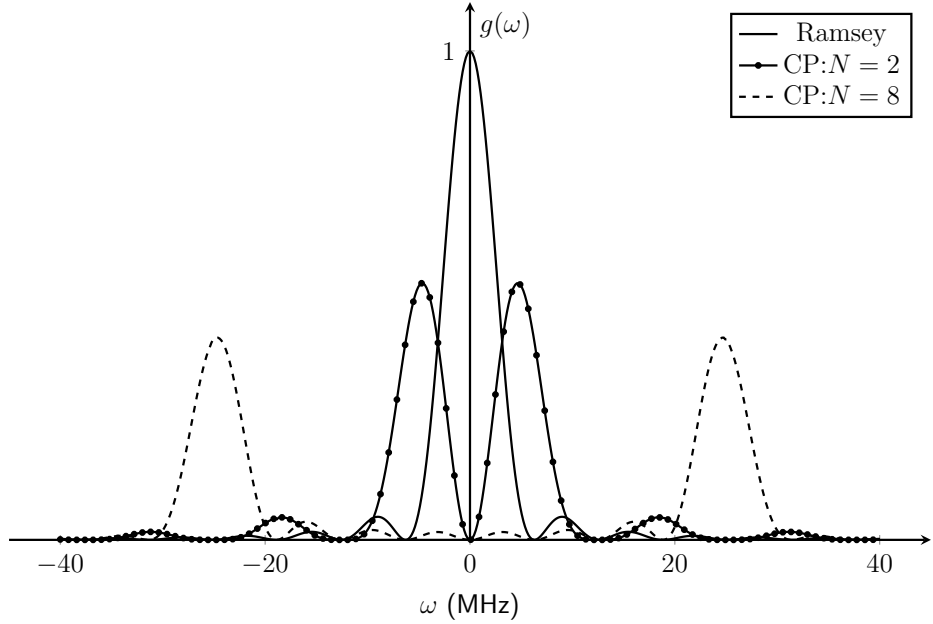
\begin{figure}[t]
    \centering
    \begin{tikzpicture}[scale=0.85]
    \draw
        ;
    \begin{axis}[
        width=16cm, height=10cm,
        axis x line=bottom,
        axis y line=center,
        xlabel={$\omega$ \textsf{(MHz)}},
        ylabel={${g(\omega)}$},
        xtick={-40,-20,0,20,40},
        ytick={0, 1},
        samples=500,
        domain=-40:40,
        xmin=-45,
        xmax=45,
        ymin=0,
        ymax=1.1
        ]
        \addplot[]
            {( sin(deg(\x / 2)) / (\x / 2) )^2};
        \addplot[mark=*, 
                mark size=1,
                mark repeat=5]
            {( tan(deg(\x / 4)) * sin(deg(\x / 2)) / (\x / 2) )^2};
        \addplot[dashed]
            {( tan(deg(\x / 16)) * sin(deg(\x / 2)) / (\x / 2) )^2};
        \legend{${\text{Ramsey}}$, ${\text{CP:} N=2}$, ${\text{CP:} N=8}$}
    \end{axis}
    \end{tikzpicture}
    \caption{The effect of CP sequences.}
    \label{fig:dynamical_decoupling}
\end{figure}

The fact that the wait time can affect the amount of noise observed also suggests that we should design a pulse sequence whose Fourier transform filters out the low-frequency noise. For example, we can shift the center of the sinc function from the origin to some high frequency if we can somehow modulate the rectangular wait window. This is achieved by the so-called CP sequences, in which an even number of $\pi$-pulses are inserted in between the two $\pi/2$-pulses to flip the qubit back and forth between opposite states in the $xy$-plane of the Bloch sphere. Based on Fourier analysis, the more $\pi$-pulses we add, the faster the modulation in the time domain and thus the larger the shift of the center of the sinc function. In general, the filter function of a CP sequence is given by
\begin{equation}
    g_{\text{CP}}^{(N)} (\omega)
    = \tan^2 \!
        \left( 
            \frac{\omega t}{2 N}
        \right) 
        \frac{\sin^2(\omega t / 2)}{(\omega t / 2)^2} 
    = \tan^2 \!
        \left( 
            \frac{\omega t}{2 N}
        \right) 
        g_{0} (\omega)
\end{equation}
if $N$ $\pi$-pulses are inserted at $t/2N$, $3t/2N$, ..., $(2N-1)t/2N$ within a total wait time $t$. Figure \ref{fig:dynamical_decoupling} plots the filter functions corresponding to the Ramsey and the CP sequences. As we increase the number of $\pi$-pulses from $N=2$ to 8, we indeed observe a large shift of the peak of the filter function.

Techniques that decouple the quasi-static noise of the environment from the qubit are known as dynamical decoupling (DD) \cite{PhysRevA.87.042309, nphys1994, PhysRevLett.106.240501}. One can imagine more complicated DD sequences (e.g. CPMG, XY4, Knill, etc.) so that
\begin{equation}
    \rho_{ge} (t) 
    = e^{\ci \omega_{\text{q},0} t} \,
        \exp
        \left[ 
            - \frac{t^2}{2}
            \int_{-\infty}^{\infty}
                \frac{\mathrm{d} \omega }{2\pi} \, 
                g_{DD}(\omega)
                S_{v}(\omega)
        \right] 
        \rho_{ge}(0)
\end{equation}
can be reduced further. Moreover, often time these sequences also mitigate systematic errors in the experiment. In practice, the LO frequency, pulse time, and pulse amplitudes are subject to noise and discretization errors in the instruments. A universally robust (UR) \cite{PhysRevLett.118.133202} sequence, for example, can be used to compensate for these errors at the cost of making the pulse sequence more complicated.

\section{Decoherence Analysis: Dispersive Readout of a Qutrit} \label{section:composite_system_master_equations}
With the QLE and the decay model of the qudit and resonator introduced, we are now ready to examine the dispersive readout with full rigor. Prior to sending the classical readout pulse into the resonator, it is assumed that the qubit is in its final state to be measured. As mentioned before, a driven resonator should have its state evolves from a vacuum state to some coherent state. The dispersive coupling between the qudit and the resonator changes the resonator frequency, thus modifying the amplitude and phase of the resonator coherent state. In the meanwhile, photons inside the resonator leak out of the resonator either via reflection (for a one-port resonator) or transmission (for a two-port resonator). The reflected or transmitted signal is then amplified and filtered before being discretized by the ADC card at room temperature. Our main objective is to understand 1) the amount of information we can infer as a function of the dispersive shift, the readout frequency and duration, and the decay rate of the cavity, and 2) the backaction of the continuous measurement on the qudit.

The dispersive shift for a qubit has been studied extensively \cite{PhysRevA.74.042318}. Here we generalize the conclusions to an arbitrary qudit in the dispersive-coupling regime. To avoid writing too many equations, we will show the derivation for a qutrit measured dispersively; nevertheless, the results can be easily extended to higher dimensional systems.

\subsection{Zero Temperature} 
To set up the problem, let the composite system be a qutrit (labeled as $\mathcal{S}$) coupled to a resonator (labeled as $\mathcal{R}$) dispersively. Note that the environment is not a part of the composite system, i.e., we have already traced out the environment to write down a master equation. The state of the composite system, denoted by $\hat{\rho}_{\mathcal{SR}}$, lives in the Hilbert space $\mathscr{H}_{\mathcal{S}} \otimes \mathscr{H}_{\mathcal{R}}$. We study the time evolution of the composite state under the usual Born and Markov approximations in which the state transition maps form a quantum dynamical semigroup and are described by the Lindblad master equation.

Typically, the resonator is a 3D cavity or a planar CPW section; it is coupled to the environment with a total decay rate of $\kappa$. For a microwave cavity, $\kappa$ is the sum of the input decay rate $\kappa_{\text{in}}$, output decay rate $\kappa_{\text{out}}$, and the internal decay rates $\kappa_{\text{int}}$ due to material losses. If the resonator is configured in the reflection mode, then $\kappa_{\text{in}} = \kappa_{\text{out}} \doteq \kappa_{\text{ext}}$ and the total decay rate is $\kappa = \kappa_{\text{int}} + \kappa_{\text{ext}}$. In reality, a resonator can support infinitely many modes; we focus on only one mode (usually the fundamental mode) of the resonator with frequency $\omega_{\text{r}}$. The resonator-environment interaction is modeled as a harmonic oscillator coupled to a continuum of bath oscillators. In addition, at superconducting temperature, we assume that the bath is in the vacuum state (i.e., the mean photon number at the resonator frequency is $\bar{N}(\omega_{\text{r}}) = 0$) so that the usual terms in the master equation, i.e., 
\begin{align}
    &\kappa 
        \big[ 
            \bar{N}(\omega_{\text{r}}) + 1 
        \big]
        \bigg[ 
            \hat{a} \hat{\rho}_{\mathcal{SR}}(t) \hat{a}^{\dagger}
            - \frac{1}{2} 
            \hat{\rho}_{\mathcal{SR}}(t) \hat{a}^{\dagger}\hat{a} 
            - \frac{1}{2} 
           \hat{a}^{\dagger} \hat{a} \hat{\rho}_{\mathcal{SR}}(t) 
        \bigg]
\nonumber \\
    & \ \ \ \ 
        + \kappa 
            \bar{N}(\omega_{\text{r}})
        \bigg[ 
            \hat{a}^{\dagger} \hat{\rho}_{\mathcal{SR}}(t) \hat{a}
            - \frac{1}{2} 
            \hat{\rho}_{\mathcal{SR}}(t) \hat{a} \hat{a}^{\dagger} 
            - \frac{1}{2} 
           \hat{a}\hat{a}^{\dagger} \hat{\rho}_{\mathcal{SR}}(t) 
        \bigg],
\end{align}
reduces to 
\begin{equation}
    \kappa 
        \bigg[ 
            \hat{a} \hat{\rho}_{\mathcal{R}}(t) \hat{a}^{\dagger}
            - \frac{1}{2} 
            \hat{\rho}_{\mathcal{R}}(t) \hat{a}^{\dagger}\hat{a} 
            - \frac{1}{2} 
           \hat{a}^{\dagger} \hat{a} \hat{\rho}_{\mathcal{R}}(t) 
        \bigg].
\end{equation}

For the qutrit, we study both spontaneous decay and pure dephasing. Without imposing any selection rule, we assume the qutrit can decay from $\ket{f}$ to $\ket{e}$, from $\ket{f}$ to $\ket{g}$, and from $\ket{e}$ to $\ket{g}$ with decay rates $\gamma_{1,ef}$, $\gamma_{1,gf}$, and $\gamma_{1,ge}$, respectively. We also include the pairwise pure dephasing with rates $\gamma_{\phi,ge}, \gamma_{\phi,gf},$ and $\gamma_{\phi,ef}$ to study the coherence time of superposition states.

By including the decoherence channels mentioned above, we can write down the Lindblad master equation of the composite system 
\begin{align}
    \dot{\hat{\rho}}_{\mathcal{SR}}(t)
    &= - \frac{\ci}{\hbar}
            \left[ 
                \hat{H}_{\text{eff}}(t), 
                \hat{\rho}_{\mathcal{SR}}(t)
            \right]
        + \kappa 
            \mathcalboondox{D}[\hat{a}] 
            \hat{\rho}_{\mathcal{SR}}(t)
\nonumber \\
        &\ \ \    \ \ \ \ 
        + \gamma_{1,ge}
            \mathcalboondox{D}\big[\ket{g}\!\bra{e}\big]
            \hat{\rho}_{\mathcal{SR}}(t)
        + \gamma_{1,gf}
            \mathcalboondox{D}\big[\ket{g}\!\bra{f}\big]
            \hat{\rho}_{\mathcal{SR}}(t)
\nonumber \\
        &\ \ \    \ \ \ \  \ \ \ \ 
        + \gamma_{1,ef}
            \mathcalboondox{D}\big[\ket{e}\!\bra{f}\big]
            \hat{\rho}_{\mathcal{SR}}(t)
        + \frac{\gamma_{\phi,ge}}{2} 
            \mathcalboondox{D}\big[\ket{g}\!\bra{g} - \ket{e}\!\bra{e}\big]
            \hat{\rho}_{\mathcal{SR}}(t)
\nonumber \\
        &\ \ \    \ \ \ \ \ \ \ \  \ \ \ \ 
        + \frac{\gamma_{\phi,gf}}{2} 
            \mathcalboondox{D}\big[\ket{g}\!\bra{g} - \ket{f}\!\bra{f}\big]
            \hat{\rho}_{\mathcal{SR}}(t)
        + \frac{\gamma_{\phi,ef}}{2} 
            \mathcalboondox{D}\big[\ket{e}\!\bra{e} - \ket{f}\!\bra{f}\big]
            \hat{\rho}_{\mathcal{SR}}(t)
\nonumber \\
    &\approx - \frac{\ci}{\hbar}
            \left[ 
                \hat{H}_{\text{eff}}(t), 
                \hat{\rho}_{\mathcal{SR}}(t)
            \right]
        + \kappa 
            \mathcalboondox{D}[\hat{a}] 
            \hat{\rho}_{\mathcal{SR}}(t)
        + \frac{\gamma_{2,ge}}{2} 
            \mathcalboondox{D}\big[\ket{g}\!\bra{g} - \ket{e}\!\bra{e}\big]
            \hat{\rho}_{\mathcal{SR}}(t)
\nonumber \\ \label{eq:general_qutrit_cavity_master_equation}
        &\ \ \    \ \ \ \ 
        + \frac{\gamma_{2,gf}}{2} 
            \mathcalboondox{D}\big[\ket{g}\!\bra{g} - \ket{f}\!\bra{f}\big]
            \hat{\rho}_{\mathcal{SR}}(t)
        + \frac{\gamma_{2,ef}}{2} 
            \mathcalboondox{D}\big[\ket{e}\!\bra{e} - \ket{f}\!\bra{f}\big]
            \hat{\rho}_{\mathcal{SR}}(t).
\end{align}
In addition, by assuming that $T_{1,ab} = 1/\gamma_{1,ab}$ is much longer than other decoherence timescales, we have removed the qutrit decay terms and have lumped the extra dephasing rates $\gamma_{1,ab}/2$ with the pure dephasing rates $\gamma_{\phi,ab}$ to define $\gamma_{2,ab} = \gamma_{\phi,ab} +\gamma_{1,ab}/2$ for $(a,b) =(g,e),(g,f)$, $(e,f)$. Later, we will add $\gamma_{1,ab}$ back, but the result stays qualitatively the same. 

The Hamiltonian of a qutrit coupled with a resonator in the dispersive regime subject to a classical drive $\varepsilon_{\text{d}}(t)$ (under the RWA) is given by
\begin{align}
    \frac{\hat{H}^{\text{disp}}_{\mathcal{SR}}(t)}{\hbar}
    &= \omega_{\text{q}} \ket{e}\!\bra{e}
        + ( 2\omega_{\text{q}} + \alpha_{\text{q}} ) 
            \ket{f}\!\bra{f}
        + \omega_{\text{r}} \hat{a}^{\dagger} \hat{a} 
\nonumber \\
    & \ \ \ \ \ \ \ \ \ \ \
        + \chi_{\text{qr}} 
            (\ket{e}\!\bra{e} + 2 \ket{f}\!\bra{f})\hat{a}^{\dagger} \hat{a} 
        - \left[ 
            \varepsilon_{\text{d}}(t) \hat{a}^{\dagger}
            + \varepsilon_{\text{d}}^{*}(t) \hat{a} 
        \right]
\nonumber \\
    &= \omega_{\text{q}} \ket{e}\!\bra{e}
        + (2\omega_{\text{q}} + \alpha_{\text{q}}) \ket{f}\!\bra{f}
        + \omega_{\text{r}} \hat{a}^{\dagger} \hat{a} 
\nonumber \\
    & \ \ \ \ \ \ \ \ \ \ \
        + \chi_{\text{qr}} (\ket{e}\!\bra{e} + 2 \ket{f}\!\bra{f})\hat{a}^{\dagger} \hat{a} 
        - \left(
            \sqrt{\kappa_{\text{in}}} \bar{a}_{\text{in}} 
                e^{-\ci\omega_{\text{d}} t} 
                \,\hat{a}^{\dagger} 
            + \sqrt{\kappa_{\text{in}}} \bar{a}_{\text{in}}^{*} 
                e^{\ci\omega_{\text{d}} t} 
                \, \hat{a} 
        \right),
\end{align}
where we have set the zero-energy reference to be the ground-state energy of the dressed system and used $\omega_{\text{q}}$ to denote the qubit frequency with the Lamb shift included. To address the state $\ket{f}$, we also introduce the anharmonicity $\alpha_{\text{q}} = \omega_{fg} - 2 \omega_{\text{q}}$, which is negative for a transmon. Moreover, for a weakly anharmonic qudit, we use the fact that the dispersive shift (to the fourth order in the reduced flux variable) is a linear function of the number of excitations in the qudit, i.e., the cavity frequency shifts by $\chi_{\text{qr}}$ when exited from $\ket{g}$ to $\ket{e}$ and shifts by $2\chi_{\text{qr}}$ when exited from $\ket{g}$ to $\ket{f}$. One can also verify this linear relation from the dispersive coupling terms in the multi-mode Hamiltonian (see Eq.(\ref{eq:multimon_hamiltonian_using_alpha_mu})). It should be mentioned that we do not need to assume a specific value for the dispersive shifts. The derivation below applied to any dispersive shifts, i.e., we could have used $(\chi_{e\text{r}} \ket{e}\!\bra{e} + \chi_{e\text{r}} \ket{f}\!\bra{f})\hat{a}^{\dagger} \hat{a}$.

The subsequent calculation can be simplified if we move the cavity part of the Hamiltonian to a frame that rotates at the drive frequency $\omega_{\text{d}}$. Then, the time-varying drive $\varepsilon_{\text{d}}(t)$ reduces to a complex scalar $\epsilon = \sqrt{\kappa_{\text{in}}} \bar{a}_{\text{in}}$ and the Hamiltonian in this rotating frame
\begin{align}
    \frac{\hat{H}_{\mathcal{SR},\text{rot}}^{\text{disp}}}{\hbar}
    = \omega_{\text{q}} 
            \ket{e}\!\bra{e}
        + (
                2\omega_{\text{q}} 
                &+ \alpha_{\text{q}}
            ) 
            \ket{f}\!\bra{f}
        + \Delta_{\text{rd}} 
            \hat{a}^{\dagger} \hat{a} 
\nonumber \\
        &+ \chi_{\text{qr}} 
            (
                \ket{e}\!\bra{e} + 2 \ket{f}\!\bra{f}
            )
            \hat{a}^{\dagger} \hat{a} 
        - \left( 
            \epsilon \hat{a}^{\dagger}
            + \epsilon^* \hat{a}
        \right)
\end{align}
is now time-independent. Then, the master equation of the composite system in the rotating frame is obtained by making the substitution $\hat{H}_{\text{eff}} = \hat{H}_{\mathcal{SR},\text{rot}}^{\text{disp}}$ in Eq.(\ref{eq:general_qutrit_cavity_master_equation}). 

To solve the master equation, we project the density operator of the composite system onto the energy eigenbasis of the qutrit and thus introduce the operators
\begin{equation}
    \hat{\rho}_{ab}(t) 
    = \bra{a}
            \hat{\rho}_{\mathcal{SR}}(t)
        \ket{b}
        \in L(\mathscr{H}_{\mathcal{R}})
\end{equation}
for $a, b \in \{ g, e, f\}$; in other words, the reduced density operator can be decomposed into
\begin{align}
    \hat{\rho}_{\mathcal{SR}}(t)
    &= \hat{\rho}_{gg}
        \ket{g} \! \bra{g} 
        + \hat{\rho}_{ge}
        \ket{g} \! \bra{e} 
        + \hat{\rho}_{gf}
        \ket{g} \! \bra{f} 
\nonumber \\
    & \ \ \ \ 
        + \hat{\rho}_{eg}
        \ket{e} \! \bra{g} 
        + \hat{\rho}_{ee}
        \ket{e} \! \bra{e} 
        + \hat{\rho}_{ef}
        \ket{e} \! \bra{f} 
\nonumber \\
    & \ \ \ \ \ \ \ \ 
        + \hat{\rho}_{fg}
        \ket{f} \! \bra{g} 
        + \hat{\rho}_{fe}
        \ket{f} \! \bra{e} 
        + \hat{\rho}_{ff}
        \ket{f} \! \bra{f},
\end{align}
and, after expanding the master equation using the nine operators, we obtain nine coupled \textit{operator} differential equations
\begin{align} \label{eq:rho_qutrit_and_cavity_gg}
    \dot{\hat{\rho}}_{gg}
    &= - \ci \Delta_{\text{rd}} \left[ 
            \hat{a}^{\dagger}\hat{a}, \hat{\rho}_{gg}
        \right]
        + \ci 
            \left[ 
                \epsilon \hat{a}^{\dagger}
                    + \epsilon^* \hat{a},
                \hat{\rho}_{gg} 
            \right]
            + \kappa \mathcalboondox{D}[\hat{a}]
                \hat{\rho}_{gg},
\\[1.5mm] \label{eq:rho_qutrit_and_cavity_ee}
    \dot{\hat{\rho}}_{ee}
    &= - \ci (\chi_{\text{qr}} + \Delta_{\text{rd}}) \left[ 
            \hat{a}^{\dagger}\hat{a}, \hat{\rho}_{ee}
        \right]
        + \ci \left[ 
                \epsilon \hat{a}^{\dagger}
                    + \epsilon^* \hat{a},
                \hat{\rho}_{ee} 
            \right]
            + \kappa \mathcalboondox{D}[\hat{a}]
                \hat{\rho}_{ee},
\\[1.5mm] \label{eq:rho_qutrit_and_cavity_ff}
    \dot{\hat{\rho}}_{ff}
    &= - \ci (2\chi_{\text{qr}} + \Delta_{\text{rd}}) \left[ 
            \hat{a}^{\dagger}\hat{a}, \hat{\rho}_{ff}
        \right]
        + \ci \left[ 
                \epsilon \hat{a}^{\dagger}
                    + \epsilon^* \hat{a},
                \hat{\rho}_{ff} 
            \right]
            + \kappa \mathcalboondox{D}[\hat{a}]
                \hat{\rho}_{ff},
\end{align}
\begin{align}
    \dot{\hat{\rho}}_{ge}
    &= \ci \omega_{\text{q}} \hat{\rho}_{ge} 
        - \ci \Delta_{\text{rd}} \left[ 
            \hat{a}^{\dagger}\hat{a}, \hat{\rho}_{ge}
        \right] \!
        + \ci\chi_{\text{qr}} \hat{\rho}_{ge}
            \hat{a}^{\dagger}\hat{a}
\nonumber \\ \label{eq:rho_qutrit_and_cavity_ge}
    & \ \ \ \ \ \ \ \ \ \ \ \ \ \ \ \ \ \ \ \ \ \ \ \ 
    \ \ \ \ \ \ \ \ \ \ \ \ \ \ \ \ \ \ \ \ \ \
        + \ci \left[ 
                \epsilon \hat{a}^{\dagger}
                    + \epsilon^* \hat{a},
                \hat{\rho}_{ge} 
            \right] \!
        + \kappa \mathcalboondox{D}[\hat{a}]
                \hat{\rho}_{ge}
        - \gamma_{2,ge} \hat{\rho}_{ge},
\\
    \dot{\hat{\rho}}_{eg}
    &= - \ci \omega_{\text{q}} \hat{\rho}_{eg} 
        + \ci \Delta_{\text{rd}} \left[ 
            \hat{a}^{\dagger}\hat{a}, \hat{\rho}_{eg}
        \right] \!
        - \ci\chi_{\text{qr}} \hat{\rho}_{eg}
            \hat{a}^{\dagger}\hat{a}
\nonumber \\ \label{eq:rho_qutrit_and_cavity_eg}
    & \ \ \ \ \ \ \ \ \ \ \ \ \ \ \ \ \ \ \ \ \ \ \ \ 
    \ \ \ \ \ \ \ \ \ \ \ \ \ \ \ \ \ \ \ \ \ \
        - \ci \left[ 
                \epsilon \hat{a}^{\dagger}
                    + \epsilon^* \hat{a},
                \hat{\rho}_{eg} 
            \right] \!
        + \kappa \mathcalboondox{D}[\hat{a}]
                \hat{\rho}_{eg}
        - \gamma_{2,eg} \hat{\rho}_{eg},
\\
    \dot{\hat{\rho}}_{gf}
    &= \ci (2\omega_{\text{q}} + \alpha_{\text{q}})\hat{\rho}_{gf} 
        - \ci \Delta_{\text{rd}} \left[ 
            \hat{a}^{\dagger}\hat{a}, \hat{\rho}_{gf}
        \right] \!
        + \ci 2 \chi_{\text{qr}} \hat{\rho}_{gf}
            \hat{a}^{\dagger}\hat{a}
\nonumber \\ \label{eq:rho_qutrit_and_cavity_gf}
    & \ \ \ \ \ \ \ \ \ \ \ \ \ \ \ \ \ \ \ \ \ \ \ \ 
    \ \ \ \ \ \ \ \ \ \ \ \ \ \ \ \ \ \ \ \ \ \
        + \ci \left[ 
                \epsilon \hat{a}^{\dagger}
                    + \epsilon^* \hat{a},
                \hat{\rho}_{gf} 
            \right] \!
        + \kappa \mathcalboondox{D}[\hat{a}]
                \hat{\rho}_{gf}
        - \gamma_{2,gf} \hat{\rho}_{gf},
\\ 
    \dot{\hat{\rho}}_{fg}
    &= -\ci (2\omega_{\text{q}} + \alpha_{\text{q}})\hat{\rho}_{fg} 
        + \ci \Delta_{\text{rd}} \left[ 
            \hat{a}^{\dagger}\hat{a}, \hat{\rho}_{fg}
        \right] \!
        - \ci 2 \chi_{\text{qr}} \hat{\rho}_{fg}
            \hat{a}^{\dagger}\hat{a}
\nonumber \\ \label{eq:rho_qutrit_and_cavity_fg}
    & \ \ \ \ \ \ \ \ \ \ \ \ \ \ \ \ \ \ \ \ \ \ \ \ 
    \ \ \ \ \ \ \ \ \ \ \ \ \ \ \ \ \ \ \ \ \ \
        - \ci \left[ 
                \epsilon \hat{a}^{\dagger}
                    + \epsilon^* \hat{a},
                \hat{\rho}_{fg} 
            \right] \!
        + \kappa \mathcalboondox{D}[\hat{a}]
                \hat{\rho}_{fg}
        - \gamma_{2,fg} \hat{\rho}_{fg},
\\ 
    \dot{\hat{\rho}}_{ef}
    &= \ci (\omega_{\text{q}} + \alpha_{\text{q}}) \hat{\rho}_{ef} 
        - \ci \Delta_{\text{rd}} \left[ 
            \hat{a}^{\dagger}\hat{a}, \hat{\rho}_{ef}
        \right] \!
        + \ci \chi_{\text{qr}} (2\hat{\rho}_{ef}
            \hat{a}^{\dagger}\hat{a}
            - \hat{a}^{\dagger}\hat{a} \hat{\rho}_{ef})
\nonumber \\ \label{eq:rho_qutrit_and_cavity_ef}
    & \ \ \ \ \ \ \ \ \ \ \ \ \ \ \ \ \ \ \ \ \ \ \ \
    \ \ \ \ \ \ \ \ \ \ \ \ \ \ \ \ \ \ \ \ \ \
        + \ci \left[ 
                \epsilon \hat{a}^{\dagger}
                    + \epsilon^* \hat{a},
                \hat{\rho}_{ef} 
            \right] \!
        + \kappa \mathcalboondox{D}[\hat{a}]
                \hat{\rho}_{ef}
        - \gamma_{2,ef} \hat{\rho}_{ef},
\\ 
    \dot{\hat{\rho}}_{fe}
    &= -\ci (\omega_{\text{q}} + \alpha_{\text{q}}) \hat{\rho}_{fe} 
        + \ci \Delta_{\text{rd}} \left[ 
            \hat{a}^{\dagger}\hat{a}, \hat{\rho}_{fe}
        \right] \!
        - \ci \chi_{\text{qr}} (2\hat{\rho}_{fe}
            \hat{a}^{\dagger}\hat{a}
            - \hat{a}^{\dagger}\hat{a} \hat{\rho}_{fe})
\nonumber \\ \label{eq:rho_qutrit_and_cavity_fe}
    & \ \ \ \ \ \ \ \ \ \ \ \ \ \ \ \ \ \ \ \ \ \ \ \ 
    \ \ \ \ \ \ \ \ \ \ \ \ \ \ \ \ \ \ \ \ \ \
        - \ci \left[ 
                \epsilon \hat{a}^{\dagger}
                    + \epsilon^* \hat{a},
                \hat{\rho}_{fe} 
            \right] \!
        + \kappa \mathcalboondox{D}[\hat{a}]
                \hat{\rho}_{fe}
        - \gamma_{2,fe} \hat{\rho}_{fe}.
\end{align}

Note that each operator $\hat{\rho}_{ab}$ lives in an infinite dimensional space since $\mathscr{H}_{\mathcal{R}}$ is a Fock space. Nevertheless, it's possible to find a closed-form solution by invoking the positive $P$-representation
\begin{equation}
    \hat{\rho}_{ab}(t) 
    = \int \mathrm{d}^2 \alpha
        \int \mathrm{d}^2 \beta
            \frac{\ket{\alpha}\!\bra{\beta^*}}{\braket{\beta^*}{\alpha}} P_{ab}(\alpha, \beta, t),
\end{equation}
commonly used when a QHO is in a coherent state (since coherent states are represented by delta functions in the $P$-representation). More importantly, one can verify that the action of the creation and annihilation operators in the operator space can be translated to some simple operations in the positive $P$-representation \cite{P_D_Drummond_1980}:
\begin{align} \label{eq:operator_correspondance_1}
    \hat{a} \hat{\rho} (t)  
    &\ \ \longrightarrow \ \ 
    \alpha 
    P(\alpha, \beta, t),
\\ \label{eq:operator_correspondance_2}
    \hat{a}^{\dagger} \hat{\rho} (t)  
    &\ \ \longrightarrow \ \ 
    \left(
        \beta - \frac{\partial}{\partial \alpha} 
    \right)
    P(\alpha, \beta, t),
\\ \label{eq:operator_correspondance_3}
    \hat{\rho} (t) \hat{a}^{\dagger}
    &\ \ \longrightarrow \ \ 
    \beta P(\alpha, \beta, t),
\\ \label{eq:operator_correspondance_4}
    \hat{\rho} (t) \hat{a}
    &\ \ \longrightarrow \ \ 
    \left(
        \alpha - \frac{\partial}{\partial \beta} 
    \right) P(\alpha, \beta, t).
\end{align}
This ``operator correspondence'' allows us to deal with scalars instead of operators. As an example, let us use Eq.(\ref{eq:operator_correspondance_1})-(\ref{eq:operator_correspondance_4}) to transform Eq.(\ref{eq:rho_qutrit_and_cavity_gg}) into a scalar equation:
\begin{align}
    \dot{P}_{gg}
    &= - \ci \Delta_{\text{rd}}
            \left [
                \left(
                    \beta - \frac{\partial}{\partial \alpha} 
                \right)
                (\alpha P_{gg})
                - \left(
                    \alpha - \frac{\partial}{\partial \beta} 
                \right)
                (\beta P_{gg})
            \right]
\nonumber \\
    & \ \ \ \ \ \ \ \ \ \ \ \ 
        + \ci 
            \left[
                \epsilon 
                    \left(
                        \beta 
                        - \frac{\partial}{\partial \alpha} 
                    \right) P_{gg}
                + \epsilon^* \alpha P_{gg}
                - \epsilon \beta P_{gg}
                - \epsilon^* 
                    \left(
                        \alpha 
                        - \frac{\partial}{\partial \beta} 
                    \right) P_{gg}
            \right]
\nonumber \\
    & \ \ \ \ \ \ \ \ \ \ \ \
        + \kappa 
            \left[
                \alpha \beta P_{gg}
                - \frac{1}{2} 
                \left(
                    \beta - \frac{\partial}{\partial \alpha} 
                \right)
                (\alpha P_{gg})
                - \frac{1}{2} \left(
                    \alpha - \frac{\partial}{\partial \beta} 
                \right)
                (\beta P_{gg})
            \right]
\nonumber \\
    &= \frac{\partial}{\partial \alpha}
        \left[
            \left(
                -\ci \epsilon 
                + \ci \Delta_{\text{rd}} \alpha 
                + \frac{\kappa \alpha}{2} 
            \right) 
            P_{gg}
        \right]
        + \frac{\partial}{\partial \beta}
        \left[
            \left(
                \ci \epsilon^{*} 
                - \ci \Delta_{\text{rd}} \beta
                + \frac{\kappa \beta}{2}
            \right) 
            P_{gg}
        \right],
\end{align}
where $\partial_t P_{gg} (\alpha, \beta, t)$ is shorthanded as $\dot{P}_{gg}$. By applying the same procedure to the other eight operators, we get, in total, nine coupled \textit{scalar} differential equations
\begin{align} \label{eq:complex_P_gg}
    \dot{P}_{gg}
    &= \frac{\partial}{\partial \alpha}
        \left[
            (-\ci \epsilon 
            + \ci \Delta_{\text{rd}} \alpha 
            + \kappa \alpha/2) 
            P_{gg}
        \right]
        + \frac{\partial}{\partial \beta}
        \left[
            (\ci \epsilon^{*}  
            - \ci \Delta_{\text{rd}} \beta
            + \kappa \beta/2) 
            P_{gg}
        \right],
\\[2mm]  \label{eq:complex_P_ee}
    \dot{P}_{ee}
    &= \frac{\partial}{\partial \alpha}
        \left[
            (-\ci \epsilon 
            + \ci \chi_{\text{qr}} \alpha 
            + \ci \Delta_{\text{rd}} \alpha 
            + \kappa \alpha/2) 
            P_{ee}
        \right]
        + \frac{\partial}{\partial \beta}
        \left[
            (\ci \epsilon^{*} 
            - \ci \chi_{\text{qr}} \beta
            - \ci \Delta_{\text{rd}} \beta
            + \kappa \beta/2) 
            P_{ee}
        \right],
\\[2mm]  \label{eq:complex_P_ff}
    \dot{P}_{ff}
    &= \frac{\partial}{\partial \alpha}
        \left[
            (-\ci \epsilon 
            + \ci 2\chi_{\text{qr}} \alpha 
            + \ci \Delta_{\text{rd}} \alpha 
            + \kappa \alpha/2) 
            P_{ff}
        \right]
        + \frac{\partial}{\partial \beta}
        \left[
            (\ci \epsilon^{*} 
            - \ci 2\chi_{\text{qr}} \beta
            - \ci \Delta_{\text{rd}} \beta
            + \kappa \beta/2) 
            P_{ff}
        \right],
\end{align}
\begin{align}
    \dot{P}_{ge}
    &= \frac{\partial}{\partial \alpha}
        \left[
            (- \ci \epsilon 
            + \ci \Delta_{\text{rd}} \alpha 
            + \kappa \alpha/2) 
            P_{ge}
        \right]
        + \frac{\partial}{\partial \beta}
        \left[
            (\ci \epsilon^{*}  
            - \ci \chi_{\text{qr}} \beta
            - \ci \Delta_{\text{rd}} \beta
            + \kappa \beta/2) 
            P_{ge}
        \right]
\nonumber \\ \label{eq:complex_P_ge}
    & \ \ \ \ \ \ \ \ \ \ \ \ \ \ \ \ \ \ \ \ \ \ \ \ \ \ \ \ \ \ \ \ \ \ \ \ \ \ \ \ \ \ \ \ \ \ \ \ \ \ \ \ \ \ \ \ 
        + \ci \chi_{\text{qr}} \alpha \beta P_{ge}
        + \ci \omega_{\text{q}} P_{ge} - \gamma_{2,ge} P_{ge},
\\[2mm]
    \dot{P}_{eg}
    &= \frac{\partial}{\partial \alpha}
        \left[
            (- \ci \epsilon 
            + \ci \chi_{\text{qr}} \alpha
            + \ci \Delta_{\text{rd}} \alpha 
            + \kappa \alpha/2) 
            P_{eg}
        \right]
        + \frac{\partial}{\partial \beta}
        \left[
            (\ci \epsilon^{*}  
            - \ci \Delta_{\text{rd}} \beta
            + \kappa \beta/2) 
            P_{eg}
        \right]
\nonumber \\ \label{eq:complex_P_eg}
    & \ \ \ \ \ \ \ \ \ \ \ \ \ \ \ \ \ \ \ \ \ \ \ \ \ \ \ \ \ \ \ \ \ \ \ \ \ \ \ \ \ \ \ \ \ \ \ \ \ \ \ \ \ \ \ \ 
        - \ci \chi_{\text{qr}} \alpha \beta P_{eg}
        - \ci \omega_{\text{q}} P_{eg} - \gamma_{2,ge} P_{eg},
\\[2mm]
    \dot{P}_{gf}
    &= \frac{\partial}{\partial \alpha}
        \left[
            (- \ci \epsilon 
            + \ci \Delta_{\text{rd}} \alpha 
            + \kappa \alpha/2) 
            P_{gf}
        \right]
        + \frac{\partial}{\partial \beta}
        \left[
            (\ci \epsilon^{*}  
            - \ci 2\chi_{\text{qr}} \beta
            - \ci \Delta_{\text{rd}} \beta
            + \kappa \beta/2) 
            P_{gf}
        \right]
\nonumber \\ \label{eq:complex_P_gf}
    & \ \ \ \ \ \ \ \ \ \ \ \ \ \ \ \ \ \ \ \ \ \ \ \ \ \ \ \ \ \ \ \ \ \ \ \ \ \ \ \ \ \ \ 
        + \ci 2\chi_{\text{qr}} \alpha \beta P_{gf}
        + \ci (2\omega_{\text{q}} + \alpha_{\text{q}}) P_{gf} - \gamma_{2,gf} P_{gf},
\\[2mm]
    \dot{P}_{fg}
    &= \frac{\partial}{\partial \alpha}
        \left[
            (- \ci \epsilon 
            + \ci 2\chi_{\text{qr}} \alpha 
            + \ci \Delta_{\text{rd}} \alpha 
            + \kappa \alpha/2) 
            P_{fg}
        \right]
        + \frac{\partial}{\partial \beta}
        \left[
            (\ci \epsilon^{*}  
            - \ci \Delta_{\text{rd}} \beta
            + \kappa \beta/2) 
            P_{fg}
        \right]
\nonumber \\ \label{eq:complex_P_fg}
    & \ \ \ \ \ \ \ \ \ \ \ \ \ \ \ \ \ \ \ \ \ \ \ \ \ \ \ \ \ \ \ \ \ \ \ \ \ \ \ \ \ \ \ 
        - \ci 2\chi_{\text{qr}} \alpha \beta P_{fg}
        - \ci (2\omega_{\text{q}} + \alpha_{\text{q}}) P_{fg} - \gamma_{2,gf} P_{fg},
\end{align}
\begin{align}
    \dot{P}_{ef}
    &= \frac{\partial}{\partial \alpha}
        \left[
            (- \ci \epsilon 
            + \ci \chi_{\text{qr}} \alpha 
            + \ci \Delta_{\text{rd}} \alpha 
            + \kappa \alpha/2) 
            P_{ef}
        \right]
        + \frac{\partial}{\partial \beta}
        \left[
            (\ci \epsilon^{*}  
            - \ci 2\chi_{\text{qr}} \beta
            - \ci \Delta_{\text{rd}} \beta
            + \kappa \beta/2) 
            P_{ef}
        \right]
\nonumber \\ \label{eq:complex_P_ef}
    & \ \ \ \ \ \ \ \ \ \ \ \ \ \ \ \ \ \ \ \ \ \ \ \ \ \ \ \ \ \ \ \ \ \ \ \ \ \ \ \ \ \ \ \ \ \ \ \ \ 
        + \ci \chi_{\text{qr}} \alpha \beta P_{ef}
        + \ci (\omega_{\text{q}} + \alpha_{\text{q}}) P_{ef} - \gamma_{2,ef} P_{ef},
\\[2mm] 
    \dot{P}_{fe}
    &= \frac{\partial}{\partial \alpha}
        \left[
            (- \ci \epsilon 
            + \ci 2\chi_{\text{qr}} \alpha 
            + \ci \Delta_{\text{rd}} \alpha 
            + \kappa \alpha/2) 
            P_{fe}
        \right]
        + \frac{\partial}{\partial \beta}
        \left[
            (\ci \epsilon^{*}  
            - \ci \chi_{\text{qr}} \beta
            - \ci \Delta_{\text{rd}} \beta
            + \kappa \beta/2) 
            P_{fe}
        \right]
\nonumber \\ \label{eq:complex_P_fe}
    & \ \ \ \ \ \ \ \ \ \ \ \ \ \ \ \ \ \ \ \ \ \ \ \ \ \ \ \ \ \ \ \ \ \ \ \ \ \ \ \ \ \ \ \ \ \ \ \ \
        - \ci \chi_{\text{qr}} \alpha \beta P_{fe}
        - \ci (\omega_{\text{q}} + \alpha_{\text{q}}) P_{fe} - \gamma_{2,ef} P_{fe}.
\end{align}
It should be noted that the differential equations of $P_{ab}$ are usually of the type of Fokker-Planck equations, which also include the diffusive terms (i.e., the second partial derivatives with respect to $\alpha$ and $\beta$). However, since we have assumed that $\Bar{N}(\omega_{\text{r}}) = 0$, there is no terms of the form
\begin{equation}
    \hat{a}^{\dagger} \hat{\rho}_{ab} \hat{a} 
    \ \longrightarrow \ \ 
    \left(
        \beta - \frac{\partial}{\partial \alpha} 
    \right)
    \left(
        \alpha - \frac{\partial}{\partial \beta} 
    \right)
    P(\alpha, \beta, t).
\end{equation}
Even in the case where $\Bar{N} > 0$, the method of the positive $P$-representation will still work but a sharp coherent state (see below) inside the cavity will broaden itself diffusively in the phase plane.

\begin{figure}[h]
    \centering
    \includegraphics[scale=0.31]{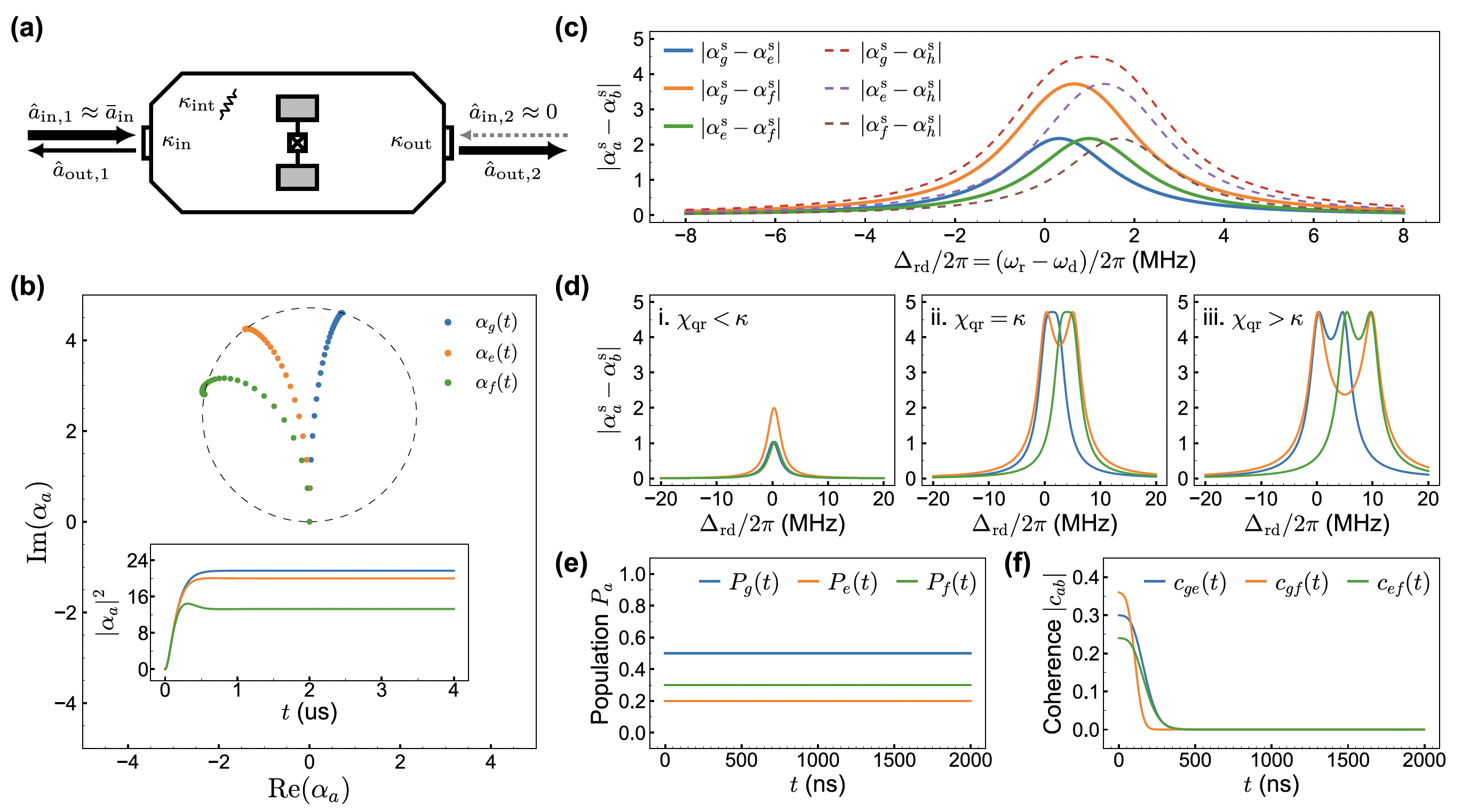}
    \caption{Schematic of qutrit readout and solution of the composite-system master equation. \textbf{a}. The input-output perspective of the transmission-mode dispersive measurement. The readout signal $\hat{a}_{\text{in}}$ entering the cavity from the left port (i.e., port 1) is approximated by a classical drive with complex amplitude $\Bar{a}_{\text{in}}$ while the transmitted signal at the right port (i.e., port 2) described by the traveling-wave annihilation operator $\hat{a}_{\text{out},2}$ in order to capture the quadrature uncertainty. \textbf{b}. The transient complex amplitude of the three coherent states $\ket{\alpha_a}$ of the resonator associated with the $\ket{a}$ for $a=g,e,f$. The steady state of each coherent state amplitude lies on a circle going through the origin of the phase plane. Inset: The build-up of mean photon number of $\ket{\alpha}_a$ as a function of time. \textbf{c}. Distance between two coherent state amplitudes as a function of the readout frequency. To illustrate a more general trend, we also include the fourth energy level $\ket{h}$ of the transmon. \textbf{d}. Same as \textbf{c}, but plotted with $\chi_{\text{qr}}$ smaller, equal, or larger than a fixed $\kappa$. \textbf{e}/\textbf{f}. Time evolution of the composite state as solved from the composite-system master equation in the long-$T_1$ limit.}
    \label{fig:cavity_amplitude_plot_simulations}
\end{figure}

Although looking complicated, the nine coupled equations admit simple trajectories in the complex planes of $\alpha$ and $\beta$. We use the ansatze 
\begin{align}
    P_{gg}(\alpha, \beta, t)
    &= \delta^{(2)}(\alpha - \alpha_g(t)) 
        \delta^{(2)}(\beta - \alpha_g^*(t)),
\\
    P_{ee}(\alpha, \beta, t)
    &= \delta^{(2)}(\alpha - \alpha_e(t)) 
        \delta^{(2)}(\beta - \alpha_e^*(t)),
\\ 
    P_{ff}(\alpha, \beta, t)
    &= \delta^{(2)}(\alpha - \alpha_f(t)) 
        \delta^{(2)}(\beta - \alpha_f^*(t))
\end{align}
for the diagonal terms and 
\begin{align}
    P_{ge}(\alpha, \beta, t)
    &= c_{ge}(t) \delta^{(2)}(\alpha - \alpha_g(t)) 
        \delta^{(2)}(\beta - \alpha_e^*(t)),
\\
    P_{eg}(\alpha, \beta, t)
    &= c_{eg}(t) \delta^{(2)}(\alpha - \alpha_e(t)) 
        \delta^{(2)}(\beta - \alpha_g^*(t)),
\\ 
    P_{gf}(\alpha, \beta, t)
    &= c_{gf}(t) \delta^{(2)}(\alpha - \alpha_g(t)) 
        \delta^{(2)}(\beta - \alpha_f^*(t)),
\\ 
    P_{fg}(\alpha, \beta, t)
    &= c_{fg}(t) \delta^{(2)}(\alpha - \alpha_f(t)) 
        \delta^{(2)}(\beta - \alpha_g^*(t)),
\\ 
    P_{ef}(\alpha, \beta, t)
    &= c_{ef}(t) \delta^{(2)}(\alpha - \alpha_e(t)) 
        \delta^{(2)}(\beta - \alpha_f^*(t)),
\\
    P_{fe}(\alpha, \beta, t)
    &= c_{fe}(t)\delta^{(2)}(\alpha - \alpha_f(t)) 
        \delta^{(2)}(\beta - \alpha_e^*(t))
\end{align}
for the off-diagonal terms. Each diagonal term $P_{aa}$ represents a single coherent state whose amplitude is specified by the two delta functions. Plugging the ansatze into Eq.(\ref{eq:complex_P_gg})-(\ref{eq:complex_P_ff}), we obtain the time evolution of the coherent states 
\begin{align} \label{eq:differential_eqn_alpha_g}
    \dot{\alpha}_g
    &= - \ci (\Delta_{\text{rd}} - \ci \kappa/2) \alpha_g
        + \ci \epsilon,
\\ \label{eq:differential_eqn_alpha_e}
    \dot{\alpha}_e
    &=  - \ci (\Delta_{\text{rd}} + \chi_{\text{qr}} - \ci \kappa/2) \alpha_e
        + \ci \epsilon,
\\ \label{eq:differential_eqn_alpha_f}
    \dot{\alpha}_f
    &= - \ci (\Delta_{\text{rd}} + 2 \chi_{\text{qr}} - \ci \kappa/2) \alpha_f
        + \ci \epsilon.
\end{align}
Besides the time evolution brought by the coherent states, each off-diagonal term $P_{ab}$ ($a \neq b$) is modulated by an envelope function $c_{ab}$. By substituting the ansatze together with Eq.(\ref{eq:differential_eqn_alpha_g})-(\ref{eq:differential_eqn_alpha_f}) into Eq.(\ref{eq:complex_P_ge})-(\ref{eq:complex_P_fe}), we deduce that
\begin{align} \label{eq:c_ge_diff_eqn}
    \dot{c}_{ge}
    &= \ci(\omega_{\text{q}} + \ci \gamma_{2,ge}) c_{ge} 
        + \ci \chi_{\text{qr}} \alpha_g \alpha_e^* c_{ge},
\\ \label{eq:c_eg_diff_eqn}
    c_{eg}
    &= c_{ge}^{*},
\\ \label{eq:c_gf_diff_eqn}
    \dot{c}_{gf}
    &= \ci(2\omega_{\text{q}} + \alpha_{\text{q}} + \ci \gamma_{2,gf}) c_{gf} 
        + \ci 2\chi_{\text{qr}} \alpha_g \alpha_f^* c_{gf},
\\ \label{eq:c_fg_diff_eqn}
    c_{fg}
    &= c_{gf}^{*},
\\ \label{eq:c_ef_diff_eqn}
    \dot{c}_{ef}
    &= \ci(\omega_{\text{q}} + \alpha_{\text{q}} + \ci \gamma_{2,ef}) c_{ef} 
        + \ci \chi_{\text{qr}} \alpha_e \alpha_f^* c_{ef},
\\ \label{eq:c_fe_diff_eqn}
    c_{fe}
    &= c_{ef}^{*}.
\end{align}
Hence, besides the dephasing rate $\gamma_{2,ab}$, the time evolution of each off-diagonal term is also modified by a term related to the dispersive shift $\chi_{\text{qr}}$ and the amplitudes of the resonator coherent states. Clearly, the real part of this extra term contributes to an additional dephasing while the imaginary part generates a frequency shift. Figure \ref{fig:cavity_amplitude_plot_simulations}(b) and (f) record a simulation of $\alpha_a$ and $c_{ab}$.

Given the arbitrary initial conditions, the detailed time evolution of $\alpha_{a}$ and $c_{ab}$ can be solved numerically. After integrating the delta functions in the complex plane of $\alpha$ and $\beta$, we arrive at the general solution (in the rotating frame) of the density operator
\begin{align}
    \hat{\rho}_{\mathcal{SR}}(t) 
    &= p_g(0) \ket{g} \! \bra{g} \otimes  \ket{\alpha_g(t)} \! \bra{\alpha_g(t)}
\nonumber \\
    &\ \ \ \ 
        + p_e(0) \ket{e} \! \bra{e} \otimes  \ket{\alpha_e(t)} \! \bra{\alpha_e(t)}
\nonumber \\
    &\ \ \ \ 
        + p_f(0) \ket{f} \! \bra{f} \otimes  \ket{\alpha_f(t)} \! \bra{\alpha_f(t)}
\nonumber \\
    &\ \ \ \ 
        + \frac{c_{ge}(t)}{\bra{\alpha_{e}(t)}\ket{\alpha_g(t)}} \ket{g} \! \bra{e} \otimes  \ket{\alpha_g(t)} \! \bra{\alpha_e(t)}
        + \frac{c_{eg}(t)}{\bra{\alpha_{g}(t)}\ket{\alpha_e(t)}} \ket{e} \! \bra{g} \otimes  \ket{\alpha_e(t)} \! \bra{\alpha_g(t)}
\nonumber \\
    &\ \ \ \ 
        + \frac{c_{gf}(t)}{\bra{\alpha_{f}(t)}\ket{\alpha_g(t)}} \ket{g} \! \bra{f} \otimes  \ket{\alpha_g(t)} \! \bra{\alpha_f(t)}
        + \frac{c_{fg}(t)}{\bra{\alpha_{g}(t)}\ket{\alpha_f(t)}} \ket{f} \! \bra{g} \otimes  \ket{\alpha_f(t)} \! \bra{\alpha_g(t)}
\nonumber \\ \label{eq:general_qutrit_resonator_solution}
    &\ \ \ \ 
        + \frac{c_{ef}(t)}{\bra{\alpha_{f}(t)}\ket{\alpha_e(t)}} \ket{e} \! \bra{f} \otimes  \ket{\alpha_e(t)} \! \bra{\alpha_f(t)}
        + \frac{c_{fe}(t)}{\bra{\alpha_{e}(t)}\ket{\alpha_f(t)}} \ket{f} \! \bra{e} \otimes  \ket{\alpha_f(t)} \! \bra{\alpha_e(t)},
\end{align}
where $p_{g,e,f}(0)$ are the initial populations in $\ket{g}$, $\ket{e}$, and $\ket{f}$, respectively. Since we have ignored $\gamma_{1,ab}$, we observe that the populations do not change over time, a critical feature of the \textbf{quantum non-demolition measurement}. However, the coherence term will decay to zero and the exact dephasing rate will be discussed in the following sections. The time evolution of the matrix elements of $\hat{\rho}_{\mathcal{SR}}$ is shown in Figure \ref{fig:cavity_amplitude_plot_simulations}(e) and (f).

Given the general solution, of particular interest are the steady-state amplitudes of the cavity coherent states
\begin{align} \label{eq:steady_state_cavity_amplitude_alpha_g}
    \alpha_g (+\infty)
    &= \frac{\sqrt{\kappa_{\text{in}}} \bar{a}_{\text{in}}}{\Delta_{\text{rd}} - \ci \kappa/2},
\\ \label{eq:steady_state_cavity_amplitude_alpha_e}
    \alpha_e (+\infty)
    &= \frac{\sqrt{\kappa_{\text{in}}} \bar{a}_{\text{in}}}{\Delta_{\text{rd}} + \chi_{\text{qr}} - \ci \kappa/2},
\\ \label{eq:steady_state_cavity_amplitude_alpha_f}
    \alpha_f (+\infty)
    &= \frac{\sqrt{\kappa_{\text{in}}} \bar{a}_{\text{in}}}{\Delta_{\text{rd}} + 2\chi_{\text{qr}} - \ci \kappa/2},
\end{align}
which are the kind of results we would expect from the QLE when the resonator is driven by a classical source\footnote{To go back to the rest frame, we just need to restore the phase $e^{- \ci \omega_{\text{d}} t}$ in each coherent state.}, i.e.,
\begin{equation}
    \dot{\hat{a}}(t)
    = - \ci \Big[
            \Delta_{\text{rd}} 
            + (\chi_{\text{qr}} \ket{e} \! \bra{e}
                + 2 \chi_{\text{qr}} \ket{f} \! \bra{f})
            - \ci \kappa/2
        \Big] \hat{a}(t)
        + \ci \sqrt{\kappa_{\text{in}}} \bar{a}_{\text{in}}.
\end{equation}
What is not obvious by looking at the QLE is the dephasing rate captured in Eq.(\ref{eq:c_ge_diff_eqn})-(\ref{eq:c_fe_diff_eqn}).

We can go one step further by tracing out the resonator part; in other words, the reduced density operator for the qutrit is given by
\begin{align}
    \hat{\rho}_{\mathcal{S}}(t) 
    &= \Tr_{\mathcal{R}}
        \Bigsl[
            \hat{\rho}_{\mathcal{SR}}(t) 
        \Bigsr]
\nonumber \\
    &= p_g(0) \ket{g} \! \bra{g} 
        + p_e(0) \ket{e} \! \bra{e} 
        + p_f(0) \ket{f} \! \bra{f}
        + c_{ge}(t) \ket{g} \! \bra{e}
        + c_{eg}(t) \ket{e} \! \bra{g}
\nonumber \\ \label{eq:composite_state_long_T1_limit}
    &\ \ \ \ 
        + c_{gf}(t) \ket{g} \! \bra{f}
        + c_{fg}(t) \ket{f} \! \bra{g}
        + c_{ef}(t) \ket{e} \! \bra{f}
        + c_{fe}(t) \ket{f} \! \bra{e};
\end{align}
hence, $c_{ab}$ are simply the coherence of the qutrit and can be solved from Eq.(\ref{eq:differential_eqn_alpha_g})-(\ref{eq:c_fe_diff_eqn}). Later, we will approach the same problem with a different technique; nevertheless, the result will be identical, except that we will include $\gamma_{1,ab}$ for completeness.

\subsection{Nonzero Temperature}
Before discussing the other approach, we briefly mention the case when the thermal bath is equilibrated at a nonzero temperature. Since $\Bar{N} > 0$, the master equation takes the form
\begin{align}
    \dot{\hat{\rho}}_{\mathcal{SR}}(t)
    &= - \frac{\ci}{\hbar}
            \left[ 
                \hat{H}_{\text{eff}}(t), 
                \hat{\rho}_{\mathcal{SR}}(t)
            \right]
        + \kappa \big(\bar{N} + 1 \big)
            \mathcalboondox{D}[\hat{a}] 
            \hat{\rho}_{\mathcal{SR}}(t)
        + \kappa \bar{N} 
            \mathcalboondox{D}\Bigsl[\hat{a}^{\dagger}\Bigsr] 
            \hat{\rho}_{\mathcal{SR}}(t)
\nonumber \\ 
        &\ \ \    \ \ \ \ 
        + \frac{\gamma_{2,ge}}{2} 
            \mathcalboondox{D}\big[\ket{g}\!\bra{g} - \ket{e}\!\bra{e}\big]
            \hat{\rho}_{\mathcal{SR}}(t)
        + \frac{\gamma_{2,gf}}{2} 
            \mathcalboondox{D}\big[\ket{g}\!\bra{g} - \ket{f}\!\bra{f}\big]
            \hat{\rho}_{\mathcal{SR}}(t)
\nonumber \\ \label{eq:general_qutrit_cavity_master_equation_finite_temp}
        &\ \ \    \ \ \ \ \ \ \ \ 
        + \frac{\gamma_{2,ef}}{2} 
            \mathcalboondox{D}\big[\ket{e}\!\bra{e} - \ket{f}\!\bra{f}\big]
            \hat{\rho}_{\mathcal{SR}}(t),
\end{align}
in the long-$T_1$ limit. The operator differential equations of $\hat{\rho}_{ab}$ are almost the same as before except that we replace $\kappa \mathcalboondox{D}[\hat{a}] \hat{\rho}_{ab}$ with $\kappa \big(\bar{N} + 1 \big) \mathcalboondox{D}[\hat{a}] \hat{\rho}_{ab} + \kappa \bar{N} \mathcalboondox{D}\Bigsl[\hat{a}^{\dagger}\Bigsr] \hat{\rho}_{ab}$. Consequently, the scalar differential equations for $P_{ab}$ acquire the second partial derivatives mention before, i.e.,
\begin{equation} \label{eq:fokker_planck_P_ab}
    \dot{P}_{ab}
    = \big( \text{terms from the case $\bar{N} = 0$} \big) 
        + \kappa \bar{N} 
            \frac{\partial^2}{\partial \alpha \partial \beta} P_{ab}.
\end{equation}

Since Eq.(\ref{eq:fokker_planck_P_ab}) with $a=b$ has the same form as the classical Fokker-Planck equation, we use Gaussian distributions now as the new ansatze
\begin{align}
    P_{gg}(\alpha, \beta, t)
    &= \frac{1}{\pi N(t)} 
        \exp
            \left\{
                - \frac{1}{N(t)} 
                \big[ \alpha - \alpha_g(t) \big]
                \big[ \beta - \alpha_g^*(t) \big]
            \right\},
\\
    P_{ee}(\alpha, \beta, t)
    &= \frac{1}{\pi N(t)} 
        \exp
            \left\{
                - \frac{1}{N(t)} 
                \big[ \alpha - \alpha_e(t) \big]
                \big[ \beta - \alpha_e^*(t) \big]
            \right\},
\\ 
    P_{ff}(\alpha, \beta, t)
    &= \frac{1}{\pi N(t)} 
        \exp
            \left\{
                - \frac{1}{N(t)} 
                \big[ \alpha - \alpha_f(t) \big]
                \big[ \beta - \alpha_f^*(t) \big]
            \right\},
\end{align}
where we require that $\alpha_g$, $\alpha_e$, and $\alpha_f$ still satisfy Eq.(\ref{eq:differential_eqn_alpha_g}), (\ref{eq:differential_eqn_alpha_e}), and (\ref{eq:differential_eqn_alpha_f}), respectively. Substituting the Gaussian distributions into the Fokker-Planck equations, we obtain a differential equation for the variance $N/2$ of each $P_{aa}$:
\begin{equation}
    \dot{N}(t) = - \kappa [N(t) - \bar{N}]
\end{equation}

Suppose the composite system was in thermal equilibrium with the bath before receiving the drive $\varepsilon_{\text{d}}(t)$, then we will simply use $N(+\infty) = \Bar{N}$ in $P_{gg}$, $P_{ee}$, and $P_{ff}$. In other words, instead of building up a coherent state in the resonator, the external drive will excite a Gaussian state with a quadrature uncertainty broadened by the thermal bath. This also means that the resonator state is a linear combination of a continuum of coherent states with amplitudes near $\alpha_g$, $\alpha_e$, or $\alpha_f$. In contrast, if the bath is in the vacuum state, i.e., $N(+\infty) = 0$, a coherent state excited in the resonator will remain coherent forever.

Unlike the case where $\bar{N} = 0$, the coherence $c_{ab}$ also depends on $\alpha$ and $\beta$ now. This is because $c_{ab}$ are affected by an infinite collection of coherent states around $\alpha_g$, $\alpha_e$, and $\alpha_f$. Thus, it becomes much more cumbersome to write down the exact expressions of $c_{ab}$. Nevertheless, we expect $c_{ab}$ to vanish on similar timescales set by Eq.(\ref{eq:c_ge_diff_eqn})-(\ref{eq:c_fe_diff_eqn}).

\subsection{Master Equation of the Composite System in the Displaced Frame}
Eq.(\ref{eq:composite_state_long_T1_limit}) gives the time evolution of the qutrit state in the long-$T_1$ limit. Now, we want to be more general and find the time evolution with $\gamma_{1,ab}$. In fact, we will go one step further and try to formulate an effective master equation for the qutrit alone since the resonator is only an auxiliary part of the superconducting quantum computation. One way of finding a master equation of the qutrit is to first move the master equation of the composite system into the so-called displaced frame and subsequently use the fact that the displaced resonator has no photons to help trace out the resonator part of the density operator \cite{PhysRevA.77.012112}.

To begin with, due to dispersive coupling, each eigenstate of the qutrit is entangled with a coherent state of the resonator. In the last two subsections, we have found three differential equations for the complex amplitudes $\alpha_g$, $\alpha_e$, and $\alpha_f$ of the coherent states. To make this entanglement explicit, we can define a unitary operator
\begin{equation} \label{eq:dispaced_frame_transformation}
    \hat{\mathsf{P}}(t)
    = \hat{\Pi}_g \hat{D}(\alpha_g(t))
        + \hat{\Pi}_e \hat{D}(\alpha_e(t))
        + \hat{\Pi}_f \hat{D}(\alpha_f(t)),
\end{equation}
where $\hat{\Pi}_a = \ket{a} \! \bra{a}$ for $a \in \{g,e,f\}$ are the projection operator onto the energy eigenstate $\ket{a}$ of the qutrit. $\hat{\mathsf{P}}$ entangles each projection $\hat{\Pi}_a = \ket{a} \! \bra{a}$ (for $a \in \{g,e,f\}$) of the qutrit with a displacement operator of the resonator such that if the qutrit is in an energy eigenstate, the resonator coherent state will be displaced to the vacuum state. For the subsequent derivation, we follow the notation in \cite{PhysRevA.77.012112} and use
\begin{equation}
    \hat{O}^{\mathsf{P}}
    = \mathsf{P}^{\dagger} \hat{O}^{\mathsf{P}} \mathsf{P}
\end{equation}
to denote any operator $\hat{O}$ in the displaced frame. 

In the new frame, the density operator of the composite system is given by
\begin{equation}
    \hat{\rho}^{\mathsf{P}}(t) 
    = \mathsf{P}^{\dagger} \hat{\rho}(t) \mathsf{P},
\end{equation}
where, to simplify the notation, to will start to use $\hat{\rho} = \hat{\rho}_{\mathcal{SR}}$.
In addition, if we define
\begin{equation}
    \hat{\rho}^{\mathsf{P}}_{nmab} (t)
    = \bra{n, a} 
        \hat{\rho}^{\mathsf{P}} (t)
            \ket{m, b} 
\end{equation}
to be the matrix element of $\hat{\rho}^{\mathsf{P}}$ in the energy basis of the qutrit and the number basis of the resonator, then
\begin{equation}
    \hat{\rho}^{\mathsf{P}}
    = \sum_{n,m = 0}^{\infty}
        \sum_{a,b \in \{g,e,f\}}
            \hat{\rho}^{\mathsf{P}}_{nmab} 
                \ket{n, a} \! \bra{m, b} .
\end{equation}
Our goal is to find the time evolution of the qutrit reduced density operator, i.e.,
\begin{equation}
    \hat{\rho}_{\mathcal{S}}(t)
    = \Tr_{\mathcal{R}}
        \Bigsl[
            \hat{\rho}(t) 
        \Bigsr]
    = \Tr_{\mathcal{R}}
        \Bigsl[
            \mathsf{P} \hat{\rho}^{\mathsf{P}}(t) \mathsf{P}^{\dagger}
        \Bigsr].
\end{equation}
By using Eq.(\ref{eq:dispaced_frame_transformation}), we obtain
\begin{align}
    \hat{\rho}_{\mathcal{S}}(t)
    &= \sum_{n} 
            \Big(
                \rho^{\mathsf{P}}_{nngg}
                    \ket{g} \! \bra{g}
                + \rho^{\mathsf{P}}_{nnee}
                    \ket{e} \! \bra{e}
                + \rho^{\mathsf{P}}_{nnff}
                    \ket{f} \! \bra{f}
            \Big)
\nonumber \\
    & \ \ \ \ 
        + \sum_{n,m} 
            \Big(
                \lambda_{nmmn}^{ge}
                    \ket{g} \! \bra{e}
                + \lambda_{mnnm}^{ge *}
                    \ket{e} \! \bra{g}
            \Big)
\nonumber \\
    & \ \ \ \ \ \ \ \ 
        + \sum_{n,m} 
            \Big(
                \lambda_{nmmn}^{gf}
                    \ket{g} \! \bra{f}
                + \lambda_{mnnm}^{gf *}
                    \ket{f} \! \bra{g}
            \Big)
\nonumber \\ \label{eq:reduced_rho_in_terms_of_displaced_rho}
    & \ \ \ \ \ \ \ \ \ \ \ \ 
        + \sum_{n,m} 
            \Big(
                \lambda_{nmmn}^{ef}
                    \ket{e} \! \bra{f}
                + \lambda_{mnnm}^{ef *}
                    \ket{f} \! \bra{e}
            \Big),
\end{align}
where
\begin{align} \label{eq:lambda_nmpq_ge_def}
    \lambda_{nmpq}^{ge}(t)
    &= \rho^{\mathsf{P}}_{nmge} e^{-\ci \Im(\alpha_e \alpha_g^*)} d_{pq},
\\  \label{eq:lambda_nmpq_gf_def}
    \lambda_{nmpq}^{gf}(t)
    &= \rho^{\mathsf{P}}_{nmgf} e^{-\ci \Im(\alpha_f \alpha_g^*)} d_{pq},
\\  \label{eq:lambda_nmpq_ef_def}
    \lambda_{nmpq}^{ef}(t)
    &= \rho^{\mathsf{P}}_{nmef} e^{-\ci \Im(\alpha_f \alpha_e^*)} d_{pq},
\end{align}
with
\begin{equation}
    d_{pq}(t) = \bra{p} \hat{D}(\beta_{ge}) \ket{q}.
\end{equation}
To arrive at Eq.(\ref{eq:reduced_rho_in_terms_of_displaced_rho}), we have used the fact that
\begin{equation}
    \sum_{p} 
        \bra{m} \hat{D}^{\dagger}(\alpha) \ket{p}
        \bra{p} \hat{D}(\alpha) \ket{n}
    = \delta_{mn}
\end{equation}
since $\hat{D}(\alpha)$ is unitary.
Once the matrix elements of $\rho^{\mathsf{P}}$ (and thus $\lambda_{nmpq}^{ab}$) are known, the matrix elements of the qutrit reduced density operator can be computed trivially. For example, 
\begin{equation}
    \hat{\rho}_{\mathcal{S},gg}
    = \bra{g} \hat{\rho}_{\mathcal{S}} \ket{g}
    = \sum_{n} 
            \rho^{\mathsf{P}}_{nngg}
\end{equation}
and 
\begin{equation} \label{eq:rho_s_ge_and_lambda_nmmn_ge}
    \hat{\rho}_{\mathcal{S},ge}
    = \bra{g} \hat{\rho}_{\mathcal{S}} \ket{e}
    = \sum_{n,m} 
        \lambda_{nmmn}^{ge}.
\end{equation}

The density operator in the displaced frame satisfies the master equation
\begin{align} 
    \dot{\hat{\rho}}^{\mathsf{P}}
    &= - \frac{\ci}{\hbar}
            \Big[ 
                \hat{H}_{\text{eff}}^{\mathsf{P}}, 
                \hat{\rho}^{\mathsf{P}}
            \Big]
        - \hat{\mathsf{P}}^{\dagger} 
            \dot{\hat{\mathsf{P}}}
            \hat{\rho}^{\mathsf{P}} 
        - \hat{\rho}^{\mathsf{P}}   
            \dot{\hat{\mathsf{P}}}^{\dagger}
            \hat{\mathsf{P}}
        + \kappa 
            \mathcalboondox{D}
                \Bigsl[
                    \hat{a}^{\mathsf{P}}
                \Bigsr] 
            \hat{\rho}^{\mathsf{P}}
\nonumber \\[2mm]
    & \ \ \   \ \ \ \ 
        + \gamma_{1,ge} 
            \mathcalboondox{D}
                \Bigsl[
                    \hat{\sigma}_{ge}^{\mathsf{P}}
                \Bigsr] 
            \hat{\rho}^{\mathsf{P}}
        + \gamma_{1,gf} 
            \mathcalboondox{D}
                \Bigsl[
                    \hat{\sigma}_{gf}^{\mathsf{P}}
                \Bigsr] 
            \hat{\rho}^{\mathsf{P}}
        + \gamma_{1,ef} 
            \mathcalboondox{D}
                \Bigsl[
                    \hat{\sigma}_{ef}^{\mathsf{P}}
                \Bigsr] 
            \hat{\rho}^{\mathsf{P}}
\nonumber \\[2mm] \label{eq:displaced_frame_composite_master_equation_before_simplification}
    & \ \ \   \ \ \ \ \ \ \ \ 
        + \frac{\gamma_{\phi,ge}}{2} 
            \mathcalboondox{D}
                \Bigsl[
                    \hat{\sigma}_{z,ge}^{\mathsf{P}}
                \Bigsr]
            \hat{\rho}^{\mathsf{P}}
        + \frac{\gamma_{\phi,gf}}{2} 
            \mathcalboondox{D}
                \Bigsl[
                    \hat{\sigma}_{z,gf}^{\mathsf{P}}
                \Bigsr]
            \hat{\rho}^{\mathsf{P}}
        + \frac{\gamma_{\phi,ef}}{2} 
            \mathcalboondox{D}
                \Bigsl[
                    \hat{\sigma}_{z,ef}^{\mathsf{P}}
                \Bigsr]
            \hat{\rho}^{\mathsf{P}},
\end{align}
where, for simplicity, we adopt the notations
\begin{equation}
\ \ \ \ \ \ \ \
    \hat{\sigma}_{z,ge}
    = \ket{g}\!\bra{g} - \ket{e}\!\bra{e},
\ \ \ \ 
    \hat{\sigma}_{z,gf}
    = \ket{g}\!\bra{g} - \ket{f}\!\bra{f},
\ \ \ \ 
    \hat{\sigma}_{z,ef}
    = \ket{e}\!\bra{e} - \ket{f}\!\bra{f},
\end{equation}
\begin{equation}
    \hat{\sigma}_{ge}
    = \ket{g}\!\bra{e},
\ \ \ \ 
    \hat{\sigma}_{gf}
    = \ket{g}\!\bra{f},
\ \ \ \ 
    \hat{\sigma}_{ef}
    = \ket{e}\!\bra{f}.
\end{equation}
As for any time-dependent unitary transformation, the extra terms $- \hat{\mathsf{P}}^{\dagger} \dot{\hat{\mathsf{P}}} \hat{\rho}^{\mathsf{P}} - \hat{\rho}^{\mathsf{P}} \dot{\hat{\mathsf{P}}}^{\dagger} \hat{\mathsf{P}}$ appear in the new master equation to eliminate the readout drive terms in the original master equation. Moreover, the Hamiltonian of the composite system (qutrit + cavity) still takes the form
\begin{equation}
    \frac{\hat{H}_{\text{eff}}}{\hbar}
    = \frac{\hat{H}_{\mathcal{SR},\text{rot}}^{\text{disp}}}{\hbar}
    = \omega_{\text{q}} 
            \hat{\Pi}_e
        + (
                2 \omega_{\text{q}} 
                + \alpha_{\text{q}}
            ) 
            \hat{\Pi}_f
        + \Delta_{\text{rd}} 
            \hat{a}^{\dagger} \hat{a} 
        + \chi_{\text{qr}} 
                \Bigsl(
                    \hat{\Pi}_e 
                    + 2 \hat{\Pi}_f
                \Bigsr)
            \hat{a}^{\dagger} \hat{a} 
        - \left( 
            \epsilon \hat{a}^{\dagger}
            + \epsilon^* \hat{a}
        \right).
\end{equation}
Note, however, we have kept the qutrit decay (i.e., $\gamma_{1,ab}$) in the master equation for full generality. The subsequent calculations are all about the simplification of Eq.(\ref{eq:displaced_frame_composite_master_equation_before_simplification}); readers who are not interested in the detailed mathematics can move to Eq.(\ref{eq:displaced_frame_composite_state_master_equation}) directly.

To begin simplifying each term, we will need various operators rewritten in the displaced frame. For the cavity operators, we have
\begin{align} \label{eq:displaced_a_op}
    \hat{a}^{\mathsf{P}}
    &= \hat{a}
        + \left(
            \alpha_g \hat{\Pi}_g 
            + \alpha_g \hat{\Pi}_g
            + \alpha_g \hat{\Pi}_g
            \right)
    = \hat{a} + \hat{\Pi}_{\alpha},
\\
    \Bigsl( \hat{a}^{\dagger}\hat{a} \Bigsr)^{\mathsf{P}}
    &= \hat{a}^{\dagger}\hat{a} 
        + \hat{a}^{\dagger} \hat{\Pi}_{\alpha}
        + \hat{a} \hat{\Pi}_{\alpha}^{\dagger}
        + \hat{\Pi}_{\alpha}^{\dagger} 
            \hat{\Pi}_{\alpha}
\nonumber \\
    &= \hat{a}^{\dagger}\hat{a} 
        + \hat{a}^{\dagger} \hat{\Pi}_{\alpha}
        + \hat{a} \hat{\Pi}_{\alpha}^{\dagger}
        + |\alpha_g|^2 \hat{\Pi}_{g}
        + |\alpha_e|^2 \hat{\Pi}_{e}
        + |\alpha_f|^2 \hat{\Pi}_{f},
\end{align}
where we denote
\begin{equation}
    \hat{\Pi}_{\alpha}(t) 
    = \alpha_g(t) \hat{\Pi}_g 
        + \alpha_g(t) \hat{\Pi}_g
        + \alpha_g(t) \hat{\Pi}_g.
\end{equation}
Similarly, the operators associated with the qutrit subspace in the displaced frame are given by
\begin{equation} \label{eq:displaced_sigma_z_op}
    \hat{\sigma}_{z,ge}^{\mathsf{P}}
    = \hat{\sigma}_{z,ge},
\ \ \ \ 
    \hat{\sigma}_{z,gf}^{\mathsf{P}}
    = \hat{\sigma}_{z,gf},
\ \ \ \ 
    \hat{\sigma}_{z,ef}^{\mathsf{P}}
    = \hat{\sigma}_{z,ef},
\end{equation}
\begin{equation} \label{eq:displaced_sigma_minus_op}
\ \ \ \ \ \ \ \
    \hat{\sigma}_{ge}^{\mathsf{P}}
    = \hat{\sigma}_{ge} 
        \hat{D}^{\dagger}(\alpha_g)
        \hat{D}(\alpha_e),
\ \ \ \ 
    \hat{\sigma}_{gf}^{\mathsf{P}}
    = \hat{\sigma}_{gf} 
        \hat{D}^{\dagger}(\alpha_g)
        \hat{D}(\alpha_f),
\ \ \ \ 
    \hat{\sigma}_{ef}^{\mathsf{P}}
    = \hat{\sigma}_{ef} 
        \hat{D}^{\dagger}(\alpha_e)
        \hat{D}(\alpha_f).
\end{equation}

First, we start with the transformed Hamiltonian. By using Eq.(\ref{eq:displaced_a_op})-(\ref{eq:displaced_sigma_minus_op}), we obtain
\begin{align}
    \frac{\hat{H}_{\text{eff}}^{\mathsf{P}}}{\hbar}
    &= \omega_{\text{q}} 
            \ket{e}\!\bra{e}
        + (
                2\omega_{\text{q}} 
                + \alpha_{\text{q}}
            ) 
            \ket{f}\!\bra{f}
        - \left[ 
                \epsilon    
                    \Big(
                        \hat{a}^{\dagger} + \hat{\Pi}_{\alpha}^{\dagger} 
                    \Big)
                + \epsilon^*
                    \Big(
                        \hat{a} + \hat{\Pi}_{\alpha}
                    \Big)
        \right]
\nonumber \\
    & \ \ \ \
        + \Big[ 
                \Delta_{\text{rd}} 
                + \chi_{\text{qr}} 
                    \Bigsl(
                        \hat{\Pi}_e 
                        + 2 \hat{\Pi}_f
                    \Bigsr)
            \Big]
            \Big(
                \hat{a}^{\dagger}\hat{a} 
                + \hat{a}^{\dagger} \hat{\Pi}_{\alpha}
                + \hat{a} \hat{\Pi}_{\alpha}^{\dagger}
                + |\alpha_g|^2 \hat{\Pi}_{g}
                + |\alpha_e|^2 \hat{\Pi}_{e}
                + |\alpha_f|^2 \hat{\Pi}_{f}
            \Big)
\nonumber \\
    &= \omega_{\text{q}} 
            \ket{e}\!\bra{e}
        + (
                2\omega_{\text{q}} 
                + \alpha_{\text{q}}
            ) 
            \ket{f}\!\bra{f}
        - \Big(
            \epsilon \hat{a}^{\dagger}
            + \epsilon^{*} \hat{a}
            \Big)
        - \Big(
            \epsilon \hat{\Pi}_{\alpha}^{\dagger}
            + \epsilon^{*} \hat{\Pi}_{\alpha}
            \Big)
\nonumber \\
    & \ \ \ \
        + \Big[ 
                \Delta_{\text{rd}} 
                + \chi_{\text{qr}} 
                    \Bigsl(
                        \hat{\Pi}_e 
                        + 2 \hat{\Pi}_f
                    \Bigsr)
            \Big]
            \hat{a}^{\dagger}\hat{a} 
        + \Big[ 
                \Delta_{\text{rd}} 
                + \chi_{\text{qr}} 
                (
                \ket{e}\!\bra{e} + 2 \ket{f}\!\bra{f}
                )
            \Big]
            \Big( 
                \hat{a}^{\dagger} \hat{\Pi}_{\alpha}
                + \hat{a} \hat{\Pi}_{\alpha}^{\dagger}
            \Big)
\nonumber \\
    & \ \ \ \ \ \ \ \ 
        + \Big[ 
                \Delta_{\text{rd}} 
                + \chi_{\text{qr}} 
                    \Bigsl(
                        \hat{\Pi}_e 
                        + 2 \hat{\Pi}_f
                    \Bigsr)
            \Big]
            \Big(
                |\alpha_g|^2 \hat{\Pi}_{g}
                + |\alpha_e|^2 \hat{\Pi}_{e}
                + |\alpha_f|^2 \hat{\Pi}_{f}
            \Big).
\end{align}
Next, to simplify $- \hat{\mathsf{P}}^{\dagger} \dot{\hat{\mathsf{P}}} \hat{\rho}^{\mathsf{P}} - \hat{\rho}^{\mathsf{P}} \dot{\hat{\mathsf{P}}}^{\dagger} \hat{\mathsf{P}}$, we invoke the identity
\begin{align} \label{eq:time_derivative_displacement_op}
    \frac{\mathrm{d}}{\mathrm{d} t} 
            \hat{D}(\alpha(t))
    &= \dot{\alpha} \hat{a}^{\dagger} \hat{D}(\alpha(t))
        - \hat{D}(\alpha(t)) \dot{\alpha}^* \hat{a}
        - \frac{\alpha^* \dot{\alpha} + \dot{\alpha}^* \alpha}{2} \hat{D}(\alpha(t))
\nonumber\\
    &= \left[ 
            \dot{\alpha} \hat{a}^{\dagger}
            - \dot{\alpha}^* \hat{a}
            - \frac{\alpha^* \dot{\alpha} - \dot{\alpha}^* \alpha}{2}
        \right]
        \hat{D}(\alpha(t))
\end{align}
for the displacement operator, resulting in
\begin{align}
    \dot{\hat{\mathsf{P}}}
    &= \hat{\Pi}_g
            \left[ 
                \left(
                    \dot{\alpha}_g \hat{a}^{\dagger} 
                    - \dot{\alpha}_g^* \hat{a}
                \right) 
                + \left(
                    \dot{\alpha}_g^* \alpha_g
                    - \alpha_g^* \dot{\alpha}_g
                \right)/2 
            \right] \hat{D}(\alpha_g)
\nonumber \\
    & \ \ \ \ 
        + \hat{\Pi}_e
            \left[
                \left(
                    \dot{\alpha}_e \hat{a}^{\dagger} 
                    - \dot{\alpha}_e^* \hat{a}
                \right) 
                + \left(
                    \dot{\alpha}_e^* \alpha_e
                    - \alpha_e^* \dot{\alpha}_e
                \right) /2
            \right] \hat{D}(\alpha_e)
\nonumber \\
    & \ \ \ \ \ \ \ \ 
        + \hat{\Pi}_f 
            \left[
                \left(
                    \dot{\alpha}_f \hat{a}^{\dagger} 
                    - \dot{\alpha}_f^* \hat{a}
                \right) 
                + \left(
                    \dot{\alpha}_f^* \alpha_f
                    - \alpha_f^* \dot{\alpha}_f
                \right) /2
            \right] \hat{D}(\alpha_f)
\end{align}
and
\begin{align}
    \hat{\mathsf{P}}^{\dagger} \dot{\hat{\mathsf{P}}}
    &= \hat{\Pi}_g 
            \hat{D}^{\dagger}(\alpha_g)
            \left[ 
                \left(
                    \dot{\alpha}_g \hat{a}^{\dagger} 
                    - \dot{\alpha}_g^* \hat{a}
                \right) 
                + \left(
                    \dot{\alpha}_g^* \alpha_g
                    - \alpha_g^* \dot{\alpha}_g
                \right) /2
            \right] 
            \hat{D}(\alpha_g)
\nonumber \\
    & \ \ \ \ 
        + \hat{\Pi}_e 
            \hat{D}^{\dagger}(\alpha_e)
            \left[
                \left(
                    \dot{\alpha}_e \hat{a}^{\dagger} 
                    - \dot{\alpha}_e^* \hat{a}
                \right) 
                + \left(
                    \dot{\alpha}_e^* \alpha_e
                    - \alpha_e^* \dot{\alpha}_e
                \right) /2
            \right] 
            \hat{D}(\alpha_e)
\nonumber \\
    & \ \ \ \ \ \ \ \ 
        + \hat{\Pi}_f 
            \hat{D}^{\dagger} (\alpha_f)
            \left[
                \left(
                    \dot{\alpha}_f \hat{a}^{\dagger} 
                    - \dot{\alpha}_f^* \hat{a}
                \right) 
                + \left(
                    \dot{\alpha}_f^* \alpha_f
                    - \alpha_f^* \dot{\alpha}_f
                \right) /2
            \right] 
            \hat{D}(\alpha_f)
\nonumber \\
    &= \hat{\Pi}_g 
            \left[ 
                \left(
                    \dot{\alpha}_g \hat{a}^{\dagger} 
                    - \dot{\alpha}_g^* \hat{a}
                \right) 
                - \left(
                    \dot{\alpha}_g^* \alpha_g
                    - \alpha_g^* \dot{\alpha}_g
                \right) /2
            \right]
\nonumber \\
    & \ \ \ \ 
        + \hat{\Pi}_e 
            \left[
                \left(
                    \dot{\alpha}_e \hat{a}^{\dagger} 
                    - \dot{\alpha}_e^* \hat{a}
                \right) 
                - \left(
                    \dot{\alpha}_e^* \alpha_e
                    - \alpha_e^* \dot{\alpha}_e
                \right) /2
            \right] 
\nonumber \\
    & \ \ \ \ \ \ \ \ 
        + \hat{\Pi}_f 
            \left[
                \left(
                    \dot{\alpha}_f \hat{a}^{\dagger} 
                    - \dot{\alpha}_f^* \hat{a}
                \right) 
                - \left(
                    \dot{\alpha}_f^* \alpha_f
                    - \alpha_f^* \dot{\alpha}_f
                \right) /2
            \right] 
\nonumber \\
    &= \dot{\hat{\Pi}}_{\alpha} \hat{a}^{\dagger} 
        - \dot{\hat{\Pi}}_{\alpha}^{\dagger}  \hat{a}
        + \ci \Im(\alpha_g^* \dot{\alpha}_g) \hat{\Pi}_g
        + \ci \Im(\alpha_e^* \dot{\alpha}_e) \hat{\Pi}_e
        + \ci \Im(\alpha_f^* \dot{\alpha}_f) \hat{\Pi}_f.
\end{align}
Since $\dot{\hat{\mathsf{P}}}^{\dagger} \hat{\mathsf{P}} = - \hat{\mathsf{P}}^{\dagger} \dot{\hat{\mathsf{P}}}$, we have
\begin{align}
    &- \hat{\mathsf{P}}^{\dagger} 
            \dot{\hat{\mathsf{P}}}
            \hat{\rho}^{\mathsf{P}} 
        - \hat{\rho}^{\mathsf{P}}   
            \dot{\hat{\mathsf{P}}}^{\dagger}
            \hat{\mathsf{P}}
\nonumber \\ \label{eq:displaced_frame_time_derivative_terms}
    &= - \left[
                \dot{\hat{\Pi}}_{\alpha} 
                    \hat{a}^{\dagger}
                - \dot{\hat{\Pi}}_{\alpha}^{\dagger} 
                    \hat{a},
                \hat{\rho}^{\mathsf{P}} 
            \right]
        - \ci 
            \left[
                \Im(\alpha_g^* \dot{\alpha}_g) \hat{\Pi}_g
                + \Im(\alpha_e^* \dot{\alpha}_e) \hat{\Pi}_e
                + \Im(\alpha_f^* \dot{\alpha}_f) \hat{\Pi}_f,
                \hat{\rho}^{\mathsf{P}} 
            \right].
\end{align}
To proceed further, we substitute Eq.(\ref{eq:differential_eqn_alpha_g})-(\ref{eq:differential_eqn_alpha_f}), i.e., 
\begin{align} 
    \dot{\alpha}_g
    &= - \ci (\Delta_{\text{rd}} - \ci \kappa/2) \alpha_g
        + \ci \epsilon,
\\ 
    \dot{\alpha}_e
    &=  - \ci (\Delta_{\text{rd}} + \chi_{\text{qr}} - \ci \kappa/2) \alpha_e
        + \ci \epsilon,
\\ 
    \dot{\alpha}_f
    &= - \ci (\Delta_{\text{rd}} + 2 \chi_{\text{qr}} - \ci \kappa/2) \alpha_f
        + \ci \epsilon,
\end{align}
found for the combined system into Eq.(\ref{eq:displaced_frame_time_derivative_terms}). In particular, 
\begin{align}
    \dot{\hat{\Pi}}_{\alpha} 
    &= \dot{\alpha}_g \hat{\Pi}_g
        + \dot{\alpha}_e \hat{\Pi}_e
        + \dot{\alpha}_f \hat{\Pi}_f
\nonumber \\
    &= \Big[
             - \ci (\Delta_{\text{rd}} - \ci \kappa/2) \alpha_g
            + \ci \epsilon
        \Big] \hat{\Pi}_g
\nonumber \\
    & \ \ \ \ 
        + \Big[
             - \ci (\Delta_{\text{rd}} + \chi_{\text{qr}} - \ci \kappa/2) \alpha_e
            + \ci \epsilon
        \Big] \hat{\Pi}_e
\nonumber \\
    & \ \ \ \ \ \ \ \ 
        + \Big[
             - \ci (\Delta_{\text{rd}} + 2 \chi_{\text{qr}} - \ci \kappa/2) \alpha_f
             + \ci \epsilon
        \Big] \hat{\Pi}_f
\nonumber \\
    &= \ci \epsilon
        - \ci 
            \Big[ 
                \Delta_{\text{rd}} 
                + \chi_{\text{qr}} 
                    \Bigsl(
                        \hat{\Pi}_e 
                        + 2 \hat{\Pi}_f
                    \Bigsr)
            \Big]
            \hat{\Pi}_{\alpha}
        - \frac{\kappa}{2} \hat{\Pi}_{\alpha}
\end{align}
and the first term in Eq.(\ref{eq:displaced_frame_time_derivative_terms}) becomes
\begin{align}
    \dot{\hat{\Pi}}_{\alpha} \hat{a}^{\dagger}
        - \dot{\hat{\Pi}}_{\alpha}^{\dagger} \hat{a}
    &= \ci \epsilon \hat{a}^{\dagger}
        - \ci 
            \Big[ 
                \Delta_{\text{rd}} 
                + \chi_{\text{qr}} 
                    \Bigsl(
                        \hat{\Pi}_e 
                        + 2 \hat{\Pi}_f
                    \Bigsr)
            \Big]
            \hat{\Pi}_{\alpha} 
            \hat{a}^{\dagger}
        - \frac{\kappa}{2} \hat{\Pi}_{\alpha}
            \hat{a}^{\dagger}
\nonumber \\
    & \ \ \ \ 
        + \ci \epsilon^* \hat{a}
        - \ci 
            \Big[ 
                \Delta_{\text{rd}} 
                + \chi_{\text{qr}} 
                    \Bigsl(
                        \hat{\Pi}_e 
                        + 2 \hat{\Pi}_f
                    \Bigsr)
            \Big]
            \hat{\Pi}_{\alpha}^{\dagger}
            \hat{a}
        + \frac{\kappa}{2} \hat{\Pi}_{\alpha}^{\dagger}
            \hat{a}
\nonumber \\
    &= \ci \Big(
            \epsilon \hat{a}^{\dagger}
            + \epsilon^* \hat{a}
        \Big)
        - \ci 
            \Big[ 
                \Delta_{\text{rd}} 
                + \chi_{\text{qr}} 
                    \Bigsl(
                        \hat{\Pi}_e 
                        + 2 \hat{\Pi}_f
                    \Bigsr)
            \Big]
            \Big(
                \hat{\Pi}_{\alpha} 
                    \hat{a}^{\dagger}
                + \hat{\Pi}_{\alpha}^{\dagger}
                    \hat{a}
            \Big)
        - \frac{\kappa}{2} 
            \Big(
                \hat{\Pi}_{\alpha} 
                    \hat{a}^{\dagger}
                - \hat{\Pi}_{\alpha}^{\dagger}
                    \hat{a}
            \Big).
\end{align}
With a similar manipulation, the second term in Eq.(\ref{eq:displaced_frame_time_derivative_terms}) reduces to 
\begin{align}
    &
    - \ci \Im(\alpha_g^* \dot{\alpha}_g) \hat{\Pi}_g
        - \ci \Im(\alpha_e^* \dot{\alpha}_e) \hat{\Pi}_e
        - \ci \Im(\alpha_f^* \dot{\alpha}_f) \hat{\Pi}_f
\nonumber \\
    &= -\ci \Big[
                \Im(\ci \alpha_g^* \epsilon ) 
                - \Delta_{\text{rd}} |\alpha_g|^2
            \Big]
            \hat{\Pi}_g
\nonumber \\
    & \ \ \   \ \ \ \
        -\ci \Big[ 
                \Im( \ci \alpha_e^* \epsilon ) 
                - (\Delta_{\text{rd}} + \chi_{\text{qr}}) |\alpha_e|^2
            \Big]
            \hat{\Pi}_e
\nonumber \\
    & \ \ \   \ \ \ \ \ \ \ \ 
        -\ci \Big[ 
                \Im( \ci \alpha_f^* \epsilon )
                - (\Delta_{\text{rd}} + 2\chi_{\text{qr}}) |\alpha_e|^2
            \Big] \hat{\Pi}_f
\nonumber \\
    &= - \ci 
            \Big[ 
                \Im(\ci \alpha_g^* \epsilon ) 
                    \hat{\Pi}_g
                + \Im(\ci \alpha_e^* \epsilon ) 
                    \hat{\Pi}_e
                + \Im(\ci \alpha_f^* \epsilon ) 
                    \hat{\Pi}_f
            \Big]
\nonumber \\ \label{eq:expression_with_identity_in_it}
    & \ \ \   \ \ \ \
        + \ci \Big[ 
                \Delta_{\text{rd}} 
                + \chi_{\text{qr}} 
                (
                \ket{e}\!\bra{e} + 2 \ket{f}\!\bra{f}
                )
            \Big]
            \Big( 
                |\alpha_g|^2 \hat{\Pi}_g
                + |\alpha_e|^2 \hat{\Pi}_e
                + |\alpha_f|^2 \hat{\Pi}_f
            \Big).
\end{align}
Since $\Bigsl[\hat{1}, \hat{\rho}^{\mathsf{P}}\Bigsr] = 0$, we can remove a multiple of the identity operator from the first term of Eq.(\ref{eq:expression_with_identity_in_it}); specifically, we can write
\begin{align}
    &- \ci 
        \Big[ 
            \Im(\ci \alpha_g^* \epsilon ) 
                \hat{\Pi}_g
            + \Im(\ci \alpha_e^* \epsilon ) 
                \hat{\Pi}_e
            + \Im(\ci \alpha_f^* \epsilon ) 
                \hat{\Pi}_f
        \Big]
\nonumber \\
    &= - \ci \frac{\alpha_g^* \epsilon + \alpha_g \epsilon^*}{2} 
        \hat{\Pi}_g
        - \ci \frac{\alpha_e^* \epsilon + \alpha_e \epsilon^*}{2} 
        \hat{\Pi}_e
        - \ci \frac{\alpha_f^* \epsilon + \alpha_f \epsilon^*}{2} 
        \hat{\Pi}_f
\nonumber \\
    &= - \frac{\ci}{2} \frac{\epsilon^*(\alpha_g+\alpha_e+\alpha_f)+\epsilon(\alpha_g+\alpha_e+\alpha_f)^*}{3} \hat{1} 
\nonumber \\
    & \ \ \   \ \ \ \
        - \frac{\ci}{2} \frac{\epsilon^* \beta_{ge} + \epsilon \beta_{ge}^*}{3} 
        \hat{\sigma}_{z,ge}
        - \frac{\ci}{2} \frac{\epsilon^* \beta_{gf} + \epsilon \beta_{gf}^*}{3} 
        \hat{\sigma}_{z,gf}
        - \frac{\ci}{2} \frac{\epsilon^* \beta_{ef} + \epsilon \beta_{ef}^*}{3} 
        \hat{\sigma}_{z,ef}
\\[2mm]
    &= - \ci C_1 \hat{1} 
        + \ci \Delta_{g,1}
            \hat{\Pi}_{g}
        + \ci \Delta_{e,1}
            \hat{\Pi}_{e}
        + \ci \Delta_{f,1}
            \hat{\Pi}_{f},
\end{align}
where we have defined $C_1 = [\epsilon^*(\alpha_g+\alpha_e+\alpha_f)+\epsilon(\alpha_g+\alpha_e+\alpha_f)^*]/6$, 
\begin{equation}
    \beta_{ge}(t) = \alpha_g(t) - \alpha_e(t),
\ \ \ \ 
    \beta_{gf}(t) = \alpha_g(t) - \alpha_f(t),
\ \ \ \ 
    \beta_{ef}(t) = \alpha_e(t) - \alpha_f(t),
\end{equation}
and 
\begin{align}
    \Delta_{g,1}(t) 
    &= \frac{1}{6} \Big[
            - \Big(
                \epsilon^* \beta_{ge} 
                +\epsilon \beta_{ge}^*
            \Big)
            - \Big(
                \epsilon^* \beta_{gf} 
                +\epsilon \beta_{gf}^*
            \Big)
        \Big],
\\
    \Delta_{e,1}(t)
    &= \frac{1}{6} \Big[
            + \Big(
                \epsilon^* \beta_{ge} 
                +\epsilon \beta_{ge}^*
            \Big)
            - \Big(
                \epsilon^* \beta_{ef} 
                +\epsilon \beta_{ef}^*
            \Big)
        \Big],
\\
    \Delta_{f,1}(t)
    &= \frac{1}{6} \Big[
            + \Big(
                \epsilon^* \beta_{gf} 
                +\epsilon \beta_{gf}^*
            \Big)
            + \Big(
                \epsilon^* \beta_{ef} 
                +\epsilon \beta_{ef}^*
            \Big)
        \Big].
\end{align}
In addition, the same argument can be applied to terms in $\hat{H}_{\text{eff}}^{\mathsf{P}}$, i.e.,
\begin{equation}
    \epsilon \hat{\Pi}_{\alpha}^{\dagger}
        + \epsilon^{*} \hat{\Pi}_{\alpha}
    = 2C_1 \hat{1} 
        - 2 \Delta_{g,1}
            \hat{\Pi}_{g}
        - 2 \Delta_{e,1}
            \hat{\Pi}_{e}
        - 2 \Delta_{f,1}
            \hat{\Pi}_{f},
\end{equation}
and the net effect is that
\begin{align}
    & -\frac{\ci}{\hbar}    
            \Bigsl[
                \hat{H}_{\text{eff}}^{\mathsf{P}},
                \hat{\rho}^{\mathsf{P}}   
            \Bigsr]
        - \hat{\mathsf{P}}^{\dagger} 
            \dot{\hat{\mathsf{P}}}
            \hat{\rho}^{\mathsf{P}} 
        - \hat{\rho}^{\mathsf{P}}   
            \dot{\hat{\mathsf{P}}}^{\dagger}
            \hat{\mathsf{P}}
\nonumber\\
    &= - \ci
            \Big[ 
                    \Delta_{g,1} 
                    \hat{\Pi}_g
                + (
                    \omega_{\text{q}} 
                    + \Delta_{e,1}
                    )
                    \hat{\Pi}_e
                + (
                    2\omega_{\text{q}} 
                    + \Delta_{f,1})
                    + \alpha_{\text{q}}
                    ) 
                    \hat{\Pi}_f,
                \hat{\rho}^{\mathsf{P}}   
            \Big]
\nonumber \\ \label{eq:H_eff_and_displaced_frame_derivative_terms}
    & \ \ \   \ \ \ \
        - \ci 
            \Big[ 
                \Bigsl[ 
                    \Delta_{\text{rd}} 
                    + \chi_{\text{qr}} 
                    \Bigsl(
                        \hat{\Pi}_e 
                        + 2 \hat{\Pi}_f
                    \Bigsr)
                \Bigsr]
                \hat{a}^{\dagger}\hat{a}, 
                \hat{\rho}^{\mathsf{P}}   
            \Big]
        + \frac{\kappa}{2} 
            \Big[
                \hat{\Pi}_{\alpha} 
                    \hat{a}^{\dagger}
                - \hat{\Pi}_{\alpha}^{\dagger}
                    \hat{a},
                \hat{\rho}^{\mathsf{P}}   
            \Big].
\end{align}

Now, we focus our attention on the cavity decay term
\begin{align}
    \mathcalboondox{D}
        \Bigsl[
            \hat{a}^{\mathsf{P}}
        \Bigsr] 
        \hat{\rho}^{\mathsf{P}}
    &= \Big( 
                \hat{a} + \hat{\Pi}_{\alpha}
            \Big)
            \hat{\rho}^{\mathsf{P}}
            \Big( 
                \hat{a}^{\dagger} + \hat{\Pi}_{\alpha}^{\dagger}
            \Big)
\nonumber \\
    & \ \ \ \ 
        - \frac{1}{2}
            \hat{\rho}^{\mathsf{P}}
            \Big(
                \hat{a}^{\dagger}\hat{a} 
                + \hat{a}^{\dagger} 
                    \hat{\Pi}_{\alpha}
                + \hat{a} 
                    \hat{\Pi}_{\alpha}^{\dagger}
                + \hat{\Pi}_{\alpha}^{\dagger}
                    \hat{\Pi}_{\alpha}
            \Big)
        - \frac{1}{2}
            \Big(
                \hat{a}^{\dagger}\hat{a} 
                + \hat{a}^{\dagger} 
                    \hat{\Pi}_{\alpha}
                + \hat{a} 
                    \hat{\Pi}_{\alpha}^{\dagger}
                + \hat{\Pi}_{\alpha}^{\dagger}
                    \hat{\Pi}_{\alpha}
            \Big)
            \hat{\rho}^{\mathsf{P}}
\nonumber \\
    &= \mathcalboondox{D}
            \Bigsl[
                \hat{a}
            \Bigsr] 
            \hat{\rho}^{\mathsf{P}}
        + \mathcalboondox{D}
            \Bigsl[
                \hat{\Pi}_{\alpha} 
            \Bigsr] 
            \hat{\rho}^{\mathsf{P}}
        + \hat{a} 
            \hat{\rho}^{\mathsf{P}}  
            \hat{\Pi}^{\dagger}
        + \hat{a}^{\dagger}
            \hat{\rho}^{\mathsf{P}}  
            \hat{\Pi}
\nonumber \\
    & \ \ \ \ 
        - \frac{1}{2}
            \hat{\rho}^{\mathsf{P}}
            \hat{a}^{\dagger} 
            \hat{\Pi}_{\alpha}
        - \frac{1}{2}
            \hat{\rho}^{\mathsf{P}}
            \hat{a} 
            \hat{\Pi}_{\alpha}^{\dagger}
        - \frac{1}{2}
            \hat{a}^{\dagger} 
            \hat{\Pi}_{\alpha}
            \hat{\rho}^{\mathsf{P}}
        - \frac{1}{2} \hat{a} 
            \hat{\Pi}_{\alpha}^{\dagger}
            \hat{\rho}^{\mathsf{P}}.
\end{align}
The second term $\mathcalboondox{D} \Bigsl[\hat{\Pi}_{\alpha} \Bigsr] \hat{\rho}^{\mathsf{P}}$ contains both frequency shifts and dephasing. To separate the two effects, we can simply expand the expression in the energy eigenbasis of the qutrit. For example,
\begin{align}
    \bra{g} \mathcalboondox{D} \Bigsl[\hat{\Pi}_{\alpha} \Bigsr] \hat{\rho}^{\mathsf{P}} \ket{e}
    &= \Big( 
            \alpha_g \alpha_e^* 
            - \frac{1}{2} |\alpha_g|^2 
            - \frac{1}{2} |\alpha_e|^2
        \Big)
        \bra{g} \hat{\rho}^{\mathsf{P}} \ket{e}
\nonumber \\
    &= \Big( 
            - \frac{1}{2} |\beta_{ge}|^2 
            -\ci \Im(\alpha_e \alpha_g^*)
        \Big) \bra{g} \hat{\rho}^{\mathsf{P}} \ket{e}.
\end{align}
By applying the same calculation to the other off-diagonal terms and noting that the diagonal terms vanish in the chosen basis, we find that
\begin{align}
    \mathcalboondox{D} \Bigsl[\hat{\Pi}_{\alpha} \Bigsr] \hat{\rho}^{\mathsf{P}}
    &= \frac{\Gamma_{\text{m},ge}}{4\kappa} 
        \mathcalboondox{D} 
            \Bigsl[
                \hat{\sigma}_{z,ge}
            \Bigsr] 
            \hat{\rho}^{\mathsf{P}}
        + \frac{\Gamma_{\text{m},gf}}{4\kappa} |\beta_{gf}|^2 
        \mathcalboondox{D} 
            \Bigsl[
                \hat{\sigma}_{z,gf}
            \Bigsr] 
            \hat{\rho}^{\mathsf{P}}
        + \frac{\Gamma_{\text{m},ef}}{4\kappa} |\beta_{ef}|^2 
        \mathcalboondox{D} 
            \Bigsl[
                \hat{\sigma}_{z,ef}
            \Bigsr] 
            \hat{\rho}^{\mathsf{P}}
\nonumber \\[1.5mm] \label{eq:dephasing_terms_in_displaced_frame}
    & \ \ \ \ 
        - \frac{\ci}{2}
            \Bigsl[
                \Im(\alpha_e \alpha_g^*) 
                    \hat{\sigma}_{z,ge}
                + \Im(\alpha_f \alpha_g^*) 
                    \hat{\sigma}_{z,gf}
                + \Im(\alpha_f \alpha_e^*) 
                    \hat{\sigma}_{z,ef},
                \hat{\rho}^{\mathsf{P}}   
            \Bigsr],
\end{align}
where we have introduced three dephasing rates
\begin{equation}
    \Gamma_{\text{m},ge} = \kappa |\beta_{ge}|^2,
\ \ \ \ 
    \Gamma_{\text{m},gf} = \kappa |\beta_{gf}|^2,
\ \ \ \ 
    \Gamma_{\text{m},ef} = \kappa |\beta_{ef}|^2.
\end{equation}
With the help of Eq.(\ref{eq:dephasing_terms_in_displaced_frame}) and the observation that
\begin{equation}
    \hat{\Pi}_{\alpha}
    = (\alpha_g + \alpha_e + \alpha_f) \hat{1}
        + \frac{\beta_{ge}}{3} \hat{\sigma}_{z,ge}
        + \frac{\beta_{gf}}{3} \hat{\sigma}_{z,gf}
        + \frac{\beta_{ef}}{3} \hat{\sigma}_{z,ef},
\end{equation}
we find
\begin{align}
    \mathcalboondox{D}
        \Bigsl[
            \hat{a}^{\mathsf{P}}
        \Bigsr] 
        \hat{\rho}^{\mathsf{P}}
    &= \Big( 
                \hat{a} + \hat{\Pi}_{\alpha}
            \Big)
            \hat{\rho}^{\mathsf{P}}
            \Big( 
                \hat{a}^{\dagger} + \hat{\Pi}_{\alpha}^{\dagger}
            \Big)
\nonumber \\
    & \ \ \ \ 
        - \frac{1}{2}
            \hat{\rho}^{\mathsf{P}}
            \Big(
                \hat{a}^{\dagger}\hat{a} 
                + \hat{a}^{\dagger} 
                    \hat{\Pi}_{\alpha}
                + \hat{a} 
                    \hat{\Pi}_{\alpha}^{\dagger}
                + \hat{\Pi}_{\alpha}^{\dagger}
                    \hat{\Pi}_{\alpha}
            \Big)
        - \frac{1}{2}
            \Big(
                \hat{a}^{\dagger}\hat{a} 
                + \hat{a}^{\dagger} 
                    \hat{\Pi}_{\alpha}
                + \hat{a} 
                    \hat{\Pi}_{\alpha}^{\dagger}
                + \hat{\Pi}_{\alpha}^{\dagger}
                    \hat{\Pi}_{\alpha}
            \Big)
            \hat{\rho}^{\mathsf{P}}
\nonumber \\
    &= \left(
            \hat{a} 
                \hat{\rho}^{\mathsf{P}} 
                \hat{a}^{\dagger} 
            - \frac{1}{2} 
                \hat{\rho}^{\mathsf{P}} 
                \hat{a}^{\dagger}
                \hat{a}
            - \frac{1}{2} 
                \hat{a}^{\dagger}
                \hat{a}
                \hat{\rho}^{\mathsf{P}} 
        \right)
        + \left(
            \hat{\Pi}_{\alpha} 
                \hat{\rho}^{\mathsf{P}} 
                \hat{\Pi}_{\alpha} ^{\dagger} 
            - \frac{1}{2} 
                \hat{\rho}^{\mathsf{P}} 
                \hat{\Pi}_{\alpha} ^{\dagger}
                \hat{\Pi}_{\alpha} 
            - \frac{1}{2} 
                \hat{\Pi}_{\alpha} ^{\dagger}
                \hat{\Pi}_{\alpha} 
                \hat{\rho}^{\mathsf{P}} 
        \right)
\nonumber \\
    & \ \ \ \ 
        + \hat{a} 
            \hat{\rho}^{\mathsf{P}}  
            \hat{\Pi}^{\dagger}
        + \hat{a}^{\dagger}
            \hat{\rho}^{\mathsf{P}}  
            \hat{\Pi}
        - \frac{1}{2}
            \hat{\rho}^{\mathsf{P}}
            \hat{a}^{\dagger} 
            \hat{\Pi}_{\alpha}
        - \frac{1}{2}
            \hat{\rho}^{\mathsf{P}}
            \hat{a} 
            \hat{\Pi}_{\alpha}^{\dagger}
        - \frac{1}{2}
            \hat{a}^{\dagger} 
            \hat{\Pi}_{\alpha}
            \hat{\rho}^{\mathsf{P}}
        - \frac{1}{2} \hat{a} 
            \hat{\Pi}_{\alpha}^{\dagger}
            \hat{\rho}^{\mathsf{P}}
\nonumber \\
    &= \mathcalboondox{D}
            \Bigsl[
                \hat{a}
            \Bigsr] 
            \hat{\rho}^{\mathsf{P}}
        - \frac{1}{2}
            \Big[
                \hat{\Pi}_{\alpha} 
                    \hat{a}^{\dagger}
                - \hat{\Pi}_{\alpha}^{\dagger}
                    \hat{a},
                \hat{\rho}^{\mathsf{P}}
            \Big]
        + \frac{\beta_{ge}^*}{3} 
            \hat{a}
            \Big[
                \hat{\rho}^{\mathsf{P}},
                \hat{\sigma}_{z,ge}
            \Big]
        + \frac{\beta_{ge}}{3} 
            \Big[
                \hat{\sigma}_{z,ge}, 
                \hat{\rho}^{\mathsf{P}}
            \Big]
            \hat{a}^{\dagger}
\nonumber \\
    & \ \ \ \ 
        + \frac{\beta_{gf}^*}{3} 
            \hat{a}
            \Big[
                \hat{\rho}^{\mathsf{P}},
                \hat{\sigma}_{z,gf}
            \Big]
        + \frac{\beta_{gf}}{3} 
            \Big[
                \hat{\sigma}_{z,gf}, 
                \hat{\rho}^{\mathsf{P}}
            \Big]
            \hat{a}^{\dagger}
        + \frac{\beta_{ef}^*}{3} 
            \hat{a}
            \Big[
                \hat{\rho}^{\mathsf{P}},
                \hat{\sigma}_{z,ef}
            \Big]
        + \frac{\beta_{ef}}{3} 
            \Big[
                \hat{\sigma}_{z,ef}, 
                \hat{\rho}^{\mathsf{P}}
            \Big]
            \hat{a}^{\dagger}
\nonumber \\   
    & \ \ \ \ \ \ \ \  
        +  \frac{\Gamma_{\text{m},ge}}{4\kappa} 
        \mathcalboondox{D} 
            \Bigsl[
                \hat{\sigma}_{z,ge}
            \Bigsr] 
            \hat{\rho}^{\mathsf{P}}
        + \frac{\Gamma_{\text{m},gf}}{4\kappa} |\beta_{gf}|^2 
        \mathcalboondox{D} 
            \Bigsl[
                \hat{\sigma}_{z,gf}
            \Bigsr] 
            \hat{\rho}^{\mathsf{P}}
        + \frac{\Gamma_{\text{m},ef}}{4\kappa} |\beta_{ef}|^2 
        \mathcalboondox{D} 
            \Bigsl[
                \hat{\sigma}_{z,ef}
            \Bigsr] 
            \hat{\rho}^{\mathsf{P}}
\nonumber \\ \label{eq:cavity_decay_terms_displaced_frame}
    & \ \ \ \ \ \ \ \ \ \ \ \  
        - \frac{\ci}{2}
            \Bigsl[
                \Im(\alpha_e \alpha_g^*) 
                    \hat{\sigma}_{z,ge}
                + \Im(\alpha_f \alpha_g^*) 
                    \hat{\sigma}_{z,gf}
                + \Im(\alpha_f \alpha_e^*) 
                    \hat{\sigma}_{z,ef},
                \hat{\rho}^{\mathsf{P}}   
    \Bigsr].
\end{align}

Finally, by combining Eq.(\ref{eq:displaced_sigma_z_op}), (\ref{eq:displaced_sigma_minus_op}), (\ref{eq:H_eff_and_displaced_frame_derivative_terms}), and (\ref{eq:cavity_decay_terms_displaced_frame}), we arrive at the master equation of the composite system in the displaced frame
\begin{align} 
    \dot{\rho}^{\mathsf{P}}
    &= - \frac{\ci}{\hbar} \Big[ 
                \hat{H}'_{\text{q,eff}},
                \hat{\rho}^{\mathsf{P}}
            \Big]
        - \ci \Big[  
                \Bigsl[ 
                    \Delta_{\text{rd}} 
                    + \chi_{\text{qr}} 
                    \Bigsl(
                        \hat{\Pi}_e 
                        + 2 \hat{\Pi}_f
                    \Bigsr)
                \Bigsr]
                \hat{a}^{\dagger}\hat{a},
                \hat{\rho}^{\mathsf{P}}
            \Big]
        + \kappa \mathcalboondox{D}
            \Bigsl[
                \hat{a}
            \Bigsr]
            \hat{\rho}^{\mathsf{P}}
\nonumber \\[1.5mm]
    & \ \ \   \ \ \ \
        + \frac{\kappa \beta_{ge}^*}{3} 
            \hat{a}
            \Big[
                \hat{\rho}^{\mathsf{P}},
                \hat{\sigma}_{z,ge}
            \Big]
        + \frac{\kappa \beta_{gf}^*}{3} 
            \hat{a}
            \Big[
                \hat{\rho}^{\mathsf{P}},
                \hat{\sigma}_{z,gf}
            \Big]
        + \frac{\kappa \beta_{ef}^*}{3} 
            \hat{a}
            \Big[
                \hat{\rho}^{\mathsf{P}},
                \hat{\sigma}_{z,ef}
            \Big]
        + \frac{\kappa \beta_{ge}}{3} 
            \Big[
                \hat{\sigma}_{z,ge}, 
                \hat{\rho}^{\mathsf{P}}
            \Big]
            \hat{a}^{\dagger}
\nonumber \\[1.5mm]
    & \ \ \   \ \ \ \ \ \ \ \ 
        + \frac{\kappa \beta_{gf}}{3} 
            \Big[
                \hat{\sigma}_{z,gf}, 
                \hat{\rho}^{\mathsf{P}}
            \Big]
            \hat{a}^{\dagger}
        + \frac{\kappa \beta_{ef}}{3} 
            \Big[
                \hat{\sigma}_{z,ef}, 
                \hat{\rho}^{\mathsf{P}}
            \Big]
            \hat{a}^{\dagger}
        + \gamma_{1,ge} 
            \mathcalboondox{D}
                \Bigsl[
                    \hat{\sigma}_{ge} 
                    \hat{D}^{\dagger}(\alpha_g)
                    \hat{D}(\alpha_e)
                \Bigsr] 
            \hat{\rho}^{\mathsf{P}}
\nonumber \\[1.5mm]
    & \ \ \   \ \ \ \ \ \ \ \ \ \ \ \
        + \gamma_{1,gf} 
            \mathcalboondox{D}
                \Bigsl[
                    \hat{\sigma}_{gf} 
                    \hat{D}^{\dagger}(\alpha_g)
                    \hat{D}(\alpha_f)
                \Bigsr] 
            \hat{\rho}^{\mathsf{P}}
        + \gamma_{1,ef} 
            \mathcalboondox{D}
                \Bigsl[
                    \hat{\sigma}_{ef} 
                    \hat{D}^{\dagger}(\alpha_e)
                    \hat{D}(\alpha_f)
                \Bigsr] 
            \hat{\rho}^{\mathsf{P}}
\nonumber \\[1.5mm]
    & \ \ \   \ \ \ \ \ \ \ \ \ \ \ \ \ \ \ \ 
        + \frac{\gamma_{\phi,ge}}{2} 
            \mathcalboondox{D}
                \Bigsl[
                    \hat{\sigma}_{z,ge}
                \Bigsr]
            \hat{\rho}^{\mathsf{P}}
        + \frac{\gamma_{\phi,gf}}{2} 
            \mathcalboondox{D}
                \Bigsl[
                    \hat{\sigma}_{z,gf}
                \Bigsr]
            \hat{\rho}^{\mathsf{P}}
        + \frac{\gamma_{\phi,ef}}{2} 
            \mathcalboondox{D}
                \Bigsl[
                    \hat{\sigma}_{z,ef}
                \Bigsr]
            \hat{\rho}^{\mathsf{P}}
\nonumber \\  \label{eq:displaced_frame_composite_state_master_equation}
    & \ \ \   \ \ \ \ \ \ \ \ \ \ \ \ \ \ \ \ \ \ \ \ 
        + \frac{\Gamma_{\text{m},ge}}{4} 
            \mathcalboondox{D} 
            \Bigsl[
                \hat{\sigma}_{z,ge}
            \Bigsr] 
            \hat{\rho}^{\mathsf{P}}
        + \frac{\Gamma_{\text{m},gf}}{4} |\beta_{gf}|^2 
        \mathcalboondox{D} 
            \Bigsl[
                \hat{\sigma}_{z,gf}
            \Bigsr] 
            \hat{\rho}^{\mathsf{P}}
        + \frac{\Gamma_{\text{m},ef}}{4} |\beta_{ef}|^2 
        \mathcalboondox{D} 
            \Bigsl[
                \hat{\sigma}_{z,ef}
            \Bigsr] 
            \hat{\rho}^{\mathsf{P}},
\end{align}
where we have defined
\begin{align}
    \hat{H}'_{\text{q,eff}}
    &= \hbar 
            \Delta_{g,1}
            \hat{\Pi}_g
        + \hbar 
            (
                \omega_{\text{q}} 
                + \Delta_{e,1} 
            )
            \hat{\Pi}_e
        + \hbar 
            (
                2\omega_{\text{q}} 
                + \Delta_{f,1}
                + \alpha_{\text{q}}
            ) 
            \hat{\Pi}_f
\nonumber \\
    & \ \ \ \ 
        + \frac{\hbar \kappa }{2}
            \Bigsl[
                \Im(\alpha_e \alpha_g^*) 
                    \hat{\sigma}_{z,ge}
                + \Im(\alpha_f \alpha_g^*) 
                    \hat{\sigma}_{z,gf}
                + \Im(\alpha_f \alpha_e^*) 
                    \hat{\sigma}_{z,ef}
            \Bigsr]
\nonumber \\
    &\doteq 
        \hbar \Tilde{\omega}_g \hat{\Pi}_g
        + \hbar \Tilde{\omega}_e \hat{\Pi}_e
        + \hbar \Tilde{\omega}_f \hat{\Pi}_f
\end{align}
to be the effective qubit Hamiltonian in the displaced frame. Note that $\hat{H}'_{\text{q,eff}}$ is not the final effective Hamiltonian of the qutrit since we are still in the displaced frame; transforming back to the laboratory frame will cancel some of the shifts seen in the displaced frame.

\subsection{Effective Master Equation of a Qutrit}
In the last subsection, we have established the connection between the matrix elements of the density operator in the displaced frame to the qutrit density operator in the laboratory frame. To find the effective master equation of the qutrit, we first rewrite the master equation in the displaced frame in terms of the matrix elements of $\hat{\rho}^{\mathsf{P}}$:
\begin{align} 
    \dot{\rho}^{\mathsf{P}}_{nmgg}
    &= \left[
            -\ci \Delta_{\text{rd}} (n-m) 
            - \kappa (n + m) / 2
        \right]
        \rho^{\mathsf{P}}_{nmgg}
\nonumber \\[1mm]
    & \ \ \ \ 
        + \gamma_{1, ge} 
            \sum_{p,q} 
                d_{pn}^{*} 
                d_{qm}
                \rho^{\mathsf{P}}_{pqee} 
        + \gamma_{1, gf} 
            \sum_{p,q} 
                d_{pn}^{*} 
                d_{qm}
                \rho^{\mathsf{P}}_{pqff}  
\nonumber \\  \label{eq:dispaced_frame_rho_nmgg}
    & \ \ \ \ \ \ \ \ 
        + \kappa \sqrt{(n+1)(m+1)} \rho^{\mathsf{P}}_{(n+1)(m+1)gg},
\\
    \dot{\rho}^{\mathsf{P}}_{nmee}
    &= \left[
            -\ci (\Delta_{\text{rd}} + \chi_{\text{qr}}) (n-m) 
            - \gamma_{1, ge}
            - \kappa (n + m) / 2
        \right]
        \rho^{\mathsf{P}}_{nmee}
\nonumber \\[1mm] \label{eq:dispaced_frame_rho_nmee}
    & \ \ \ \ 
        + \gamma_{1, ef} 
            \sum_{p,q} 
                d_{pn}^{*} 
                d_{qm}
                \rho^{\mathsf{P}}_{pqff}
        + \kappa \sqrt{(n+1)(m+1)} \rho^{\mathsf{P}}_{(n+1)(m+1)ee},
\\
    \dot{\rho}^{\mathsf{P}}_{nmff}
    &= \left[
            -\ci (\Delta_{\text{rd}} + 2\chi_{\text{qr}}) (n-m) 
            - (\gamma_{1, gf} + \gamma_{1, ef})
            - \kappa (n + m) / 2
        \right]
        \rho^{\mathsf{P}}_{nmff}
\nonumber \\[1mm] \label{eq:dispaced_frame_rho_nmff}
    & \ \ \ \
        + \kappa \sqrt{(n+1)(m+1)} \rho^{\mathsf{P}}_{(n+1)(m+1)ff},
\end{align}
\begin{align} \label{eq:dispaced_frame_rho_nmge}
    \dot{\rho}^{\mathsf{P}}_{nmge}
    &= \left[
            \ci \Tilde{\omega}_{eg}
            -\ci \Delta_{\text{rd}} (n-m) 
            + \ci \chi_{\text{qr}} m
            - \gamma_{1,ge} / 2 - \gamma_{\phi,ge} 
            - \kappa (n + m) / 2
            - \Gamma_{m,ge}
        \right]
        \rho^{\mathsf{P}}_{nmge}
\nonumber \\[1.5mm]
    & \ \ \ \ 
        + \kappa \sqrt{(n+1)(m+1)}
            \rho^{\mathsf{P}}_{(n+1)(m+1)ge}
\nonumber \\
    & \ \ \ \ \ \ \ \ 
        - \frac{2\kappa \beta_{ge}^*}{3} 
            \sqrt{n+1}
            \rho^{\mathsf{P}}_{(n+1)mge}
        + \frac{2\kappa \beta_{ge}}{3} 
            \sqrt{m+1}
            \rho^{\mathsf{P}}_{n(m+1)ge},
\\ \label{eq:dispaced_frame_rho_nmgf}
    \dot{\rho}^{\mathsf{P}}_{nmgf}
    &= \left[
            \ci \Tilde{\omega}_{fg}
            -\ci \Delta_{\text{rd}} (n-m) 
            + \ci \chi_{\text{qr}} 2 m
            - \gamma_{1,gf} / 2 - \gamma_{\phi,gf} 
            - \kappa (n + m) / 2
            - \Gamma_{m,gf}
        \right]
        \rho^{\mathsf{P}}_{nmgf}
\nonumber \\[1.5mm]
    & \ \ \ \ 
        + \kappa \sqrt{(n+1)(m+1)}
            \rho^{\mathsf{P}}_{(n+1)(m+1)gf}
\nonumber \\
    & \ \ \ \ \ \ \ \ 
        - \frac{2\kappa \beta_{gf}^*}{3} 
            \sqrt{n+1}
            \rho^{\mathsf{P}}_{(n+1)mgf}
        + \frac{2\kappa \beta_{gf}}{3} 
            \sqrt{m+1}
            \rho^{\mathsf{P}}_{n(m+1)gf},
\\ \label{eq:dispaced_frame_rho_nmef}
    \dot{\rho}^{\mathsf{P}}_{nmef}
    &= \left[
            \ci \Tilde{\omega}_{fe}
            -\ci \Delta_{\text{rd}} (n-m) 
            + \ci \chi_{\text{qr}} (2m-n)
            - \gamma_{1,ef} / 2 - \gamma_{\phi,ef} 
            - \kappa (n + m) / 2
            - \Gamma_{m,ef}
        \right]
        \rho^{\mathsf{P}}_{nmef}
\nonumber \\[1.5mm]
    & \ \ \ \ 
        + \kappa \sqrt{(n+1)(m+1)}
            \rho^{\mathsf{P}}_{(n+1)(m+1)ef}
\nonumber \\
    & \ \ \ \ \ \ \ \ 
        - \frac{2\kappa \beta_{ef}^*}{3} 
            \sqrt{n+1}
            \rho^{\mathsf{P}}_{(n+1)mef}
        + \frac{2\kappa \beta_{ef}}{3} 
            \sqrt{m+1}
            \rho^{\mathsf{P}}_{n(m+1)ef}.
\end{align}
There are three other differential equations but they are simply the complex conjugates of Eq.(\ref{eq:dispaced_frame_rho_nmge})-(\ref{eq:dispaced_frame_rho_nmef}). 
Now, given Eq.(\ref{eq:dispaced_frame_rho_nmgg})-(\ref{eq:dispaced_frame_rho_nmff}), the differential equations governing the time evolution of the diagonal matrix elements of $\hat{\rho}_{\mathcal{S}}$ (i.e., the populations of the qutrit eigenstates) are found to be
\begin{align}
    \dot{\rho}_{\mathcal{S},gg}
    &= \sum_{n}
        \dot{\rho}^{\mathsf{P}}_{nngg}
    = \gamma_{1, ge} 
            \sum_{p,q} 
                \rho^{\mathsf{P}}_{pqee} 
                \sum_{n}
                    d_{pn}^{*} 
                    d_{qn}
        + \gamma_{1, gf} 
            \sum_{p,q} 
                \rho^{\mathsf{P}}_{pqff}  
                \sum_{n} 
                    d_{pn}^{*} 
                    d_{qn}
\nonumber \\ \label{eq:qutrit_population_gg}
    &= \gamma_{1, ge} \rho_{\mathcal{S},ee} 
        + \gamma_{1, gf} \rho_{\mathcal{S},ff} ,
\\[2mm]
    \dot{\rho}_{\mathcal{S},ee}
    &= \sum_{n}
        \dot{\rho}^{\mathsf{P}}_{nnee}
    =  - \gamma_{1, ge} \sum_{n} \rho^{\mathsf{P}}_{nnee}
        + \gamma_{1, ef} 
            \sum_{p,q}
                \rho^{\mathsf{P}}_{pqff} 
                \sum_{n}
                    d_{pn}^{*} 
                    d_{qn}
\nonumber \\ \label{eq:qutrit_population_ee}
    &= - \gamma_{1, ge} \rho_{\mathcal{S},ee} 
        + \gamma_{1, ef} \rho_{\mathcal{S},ff} ,
\\[2mm] \label{eq:qutrit_population_ff}
    \dot{\rho}_{\mathcal{S},ff}
    &= \sum_{n}
        \dot{\rho}^{\mathsf{P}}_{nnff}
    = - (\gamma_{1, gf} + \gamma_{1, ef})
        \sum_{n} \rho^{\mathsf{P}}_{nnff}
    = - (\gamma_{1, gf} + \gamma_{1, ef}) \rho_{\mathcal{S},ff}.
\end{align}
These are nothing else but the rate equations one could have hoped for from a semi-classical treatment. Due to the conservation of probability, the population leaving from a higher energy eigenstate must be accepted by the lower energy eigenstate with the same rate. Thus, the resonator has made no modification to the qutrit decay rate. However, we know from the previous discussion in the long-$T_1$ limit that the coherence terms of the qutrit are clearly affected by the dispersive coupling.

Next, according to Eq.(\ref{eq:reduced_rho_in_terms_of_displaced_rho}), computing the off-diagonal terms of $\hat{\rho}_{\mathcal{S}}$ also requires us to know all $\lambda_{nmpq}$, whose time derivatives follow (see Eq.(\ref{eq:lambda_nmpq_ge_def})-(\ref{eq:lambda_nmpq_ef_def}))
\begin{align}  \label{eq:lambda_nmpq_ge}
    \dot{\lambda}_{nmpq}^{ge}
    &= \dot{\rho}_{nmge}
            d_{p,q} 
            e^{-\ci \Im(\alpha_e \alpha_g^*)}
        - \ci \frac{\mathrm{d} \Im(\alpha_e \alpha_g^*)}{\mathrm{d} t} 
            \lambda_{nmpq}^{ge}
\nonumber \\
    & \ \ \ \ 
        + \sqrt{p} \dot{\beta}_{ge} \lambda_{nm(p-1)q}^{ge}
        - \sqrt{q} \dot{\beta}_{ge}^* \lambda_{nmp(q-1)}^{ge}
        - \frac{\dot{\beta}_{ge}^* \beta_{ge} + \dot{\beta}_{ge} \beta_{ge}^*}{2} \lambda_{nmpq}^{ge},
\\
    \dot{\lambda}_{nmpq}^{gf}
    &= \dot{\rho}_{nmgf}
            d_{p,q} 
            e^{-\ci \Im(\alpha_f \alpha_g^*)}
        - \ci \frac{\mathrm{d} \Im(\alpha_f \alpha_g^*)}{\mathrm{d} t} 
            \lambda_{nmpq}^{gf}
\nonumber \\
    & \ \ \ \ 
        + \sqrt{p} \dot{\beta}_{gf} \lambda_{nm(p-1)q}^{gf}
        - \sqrt{q} \dot{\beta}_{gf}^* \lambda_{nmp(q-1)}^{gf}
        - \frac{\dot{\beta}_{gf}^* \beta_{gf} + \dot{\beta}_{gf} \beta_{gf}^*}{2} \lambda_{nmpq}^{gf},
\\
    \dot{\lambda}_{nmpq}^{ef}
    &= \dot{\rho}_{nmef}
            d_{p,q} 
            e^{-\ci \Im(\alpha_f \alpha_e^*)}
        - \ci \frac{\mathrm{d} \Im(\alpha_f \alpha_e^*)}{\mathrm{d} t} 
            \lambda_{nmpq}^{ef}
\nonumber \\
    & \ \ \ \ 
        + \sqrt{p} \dot{\beta}_{ef} \lambda_{nm(p-1)q}^{ef}
        - \sqrt{q} \dot{\beta}_{ef}^* \lambda_{nmp(q-1)}^{ef}
        - \frac{\dot{\beta}_{ef}^* \beta_{ef} + \dot{\beta}_{ef} \beta_{ef}^*}{2} \lambda_{nmpq}^{ef}.
\end{align}
The three equations take the same forms; we show the simplification of Eq.(\ref{eq:lambda_nmpq_ge}) as an example. Since we have a differential equation for $\dot{\rho}_{nmge}$, the first term on the RHS of Eq.(\ref{eq:lambda_nmpq_ge}) reduces to
\begin{align}
    &\dot{\rho}_{nmge}
            d_{p,q} 
            e^{-\ci \Im(\alpha_e \alpha_g^*)}
\nonumber\\[1.5mm]
    &= \Big[
            \ci \Tilde{\omega}_{eg}
            -\ci \Delta_{\text{rd}} (n-m) 
            + \ci \chi_{\text{qr}} m
            - \gamma_{1,ge} / 2 - \gamma_{\phi,ge} 
            - \kappa (n + m) / 2
            - \Gamma_{\text{m},ge} / 2
        \Big]
        \lambda_{nmpq}^{ge}
\nonumber \\[1.5mm]
    & \ \ \ \ 
        + \kappa \sqrt{(n+1)(m+1)}
            \lambda_{(n+1)(m+1)pq}^{ge}
\nonumber \\
    & \ \ \ \ \ \ \ \ 
        - \frac{2\kappa \beta_{ge}^*}{3} 
            \sqrt{n+1}
            \lambda_{(n+1)mpq}^{ge}
        + \frac{2\kappa \beta_{ge}}{3} 
            \sqrt{m+1}
            \lambda_{n(m+1)pq}^{ge}.
\end{align}
To simplify the second term on the RHS of Eq.(\ref{eq:lambda_nmpq_ge}), we first compute
\begin{align}
    \dot{\alpha}_e \alpha_g^* + \alpha_e \dot{\alpha_g}^*
    &= \Big[
            - \ci (\Delta_{\text{rd}} + \chi_{\text{qr}} - \ci \kappa/2) \alpha_e
            + \ci \epsilon
        \Big] \alpha_g^*
        + \alpha_e
        \Big[
            - \ci (\Delta_{\text{rd}} - \ci \kappa/2) \alpha_g
            + \ci \epsilon
        \Big]^*
\nonumber \\
    &= - \ci \chi_{\text{qr}} \alpha_e\alpha_g^*
        - \kappa \alpha_e \alpha_g^*
        + \ci ( \epsilon \alpha_g^* - \epsilon^* \alpha_e).
\end{align}
Then,
\begin{align}
    - \ci \frac{\mathrm{d} \Im(\alpha_e \alpha_g^*)}{\mathrm{d} t} 
        \lambda_{nmpq}^{ge}
    &= - \ci \Im \!
        \Big[
            - \ci \chi_{\text{qr}} \alpha_e\alpha_g^*
            - \kappa \alpha_e \alpha_g^*
            + \ci ( \epsilon \alpha_g^* - \epsilon^* \alpha_e)
        \Big]
        \lambda_{nmpq}^{ge}
\nonumber \\
    &= \ci 
        \Big[
            \chi_{\text{qr}} \Re(\alpha_e \alpha_g^*)
            + \kappa \Im(\alpha_e \alpha_g^*)
            - \frac{\epsilon^* \beta_{ge} + \epsilon \beta_{ge}^*}{2}
        \Big]
        \lambda_{nmpq}^{ge}.
\end{align}
We keep the terms involving $\lambda_{nm(p-1)q}^{ge}$ and $\lambda_{nmp(q-1)}^{ge}$ untouched and directly go to the last term on the RHS of Eq.(\ref{eq:lambda_nmpq_ge}). Using
\begin{equation}
    \dot{\beta}_{ge} \beta_{ge}^*
    = (\dot{\alpha}_g - \dot{\alpha}_e)\beta_{ge}^*
    = - \ci \Delta_{\text{rd}} |\beta_{ge}|^2 - \frac{\kappa |\beta_{ge}|^2}{2} + \ci \chi_{\text{qr}} \alpha_{e}\beta_{ge}^*,
\end{equation}
we obtain
\begin{equation}
    - \frac{\dot{\beta}_{ge}^* \beta_{ge} + \dot{\beta}_{ge} \beta_{ge}^*}{2} \lambda_{nmpq}^{ge}
    = \left[ 
            \frac{\kappa |\beta_{ge}|^2}{2}
            + \chi_{\text{qr}} \Im(\alpha_e \alpha_g^*)
        \right] 
        \lambda_{nmpq}^{ge}.
\end{equation}
Finally, combining all the pieces yields
\begin{align}
    \dot{\lambda}_{nmpq}^{ge}
    &= \Big[
            \ci \Bar{\omega}_{eg}
            - \gamma_{1,ge} / 2 
            - \gamma_{\phi,ge} 
            + \chi_{\text{qr}} \Im(\alpha_e \alpha_g^*)
\nonumber \\
    & \ \ \ \ \ \ \ \ 
            -\ci \Delta_{\text{rd}} (n-m) 
            + \ci \chi_{\text{qr}} m
            - \kappa (n + m) / 2
        \Big]
        \lambda_{nmpq}^{ge}
\nonumber \\[1.5mm]
    & \ \ \ \ 
        + \kappa \sqrt{(n+1)(m+1)}
            \lambda_{(n+1)(m+1)pq}^{ge}
\nonumber \\
    & \ \ \ \ \ \ \ \ 
        - \frac{2\kappa \beta_{ge}^*}{3} 
            \sqrt{n+1}
            \lambda_{(n+1)mpq}^{ge}
        + \frac{2\kappa \beta_{ge}}{3} 
            \sqrt{m+1}
            \lambda_{n(m+1)pq}^{ge},
\end{align}
where the net frequency difference between $\ket{e}$ and $\ket{g}$ is found to be
\begin{align}
    \bar{\omega}_{eg}
    &= (\Tilde{\omega}_{e} - \Tilde{\omega}_{g} )
            + \chi_{\text{qr}} \Re(\alpha_e \alpha_g^*)
            + \kappa \Im(\alpha_e \alpha_g^*)
            - \Big(
                    \epsilon^* \beta_{ge} 
                    + \epsilon \beta_{ge}^* 
                \Big) / 2
\nonumber \\
    &= \Big[ \omega_{\text{q}} 
            + \Delta_{e,1}
            - \kappa \Im(\alpha_e \alpha_g^*)
            - \Delta_{g,1} 
        \Big]
\nonumber \\
    &\ \ \ \ 
        + \chi_{\text{qr}} \Re(\alpha_e \alpha_g^*)
        + \kappa \Im(\alpha_e \alpha_g^*)
        - \Big(
                \epsilon^* \beta_{ge} 
                + \epsilon \beta_{ge}^* 
            \Big) / 2
\nonumber \\
    &= \omega_{\text{q}}
        + \chi_{\text{qr}} \Re(\alpha_e \alpha_g^*).
\end{align}
Applying the same procedure to the other two equations, we find
\begin{align}
    \dot{\lambda}_{nmpq}^{gf}
    &= \Big[
            \ci \Bar{\omega}_{fg}
            - \gamma_{1,gf} / 2 
            - \gamma_{\phi,gf} 
            + 2 \chi_{\text{qr}} \Im(\alpha_f \alpha_g^*)
\nonumber \\
    & \ \ \ \ \ \ \ \ 
            -\ci \Delta_{\text{rd}} (n-m) 
            + \ci 2 \chi_{\text{qr}} m
            - \kappa (n + m) / 2
        \Big]
        \lambda_{nmpq}^{gf}
\nonumber \\[1.5mm]
    & \ \ \ \ 
        + \kappa \sqrt{(n+1)(m+1)}
            \lambda_{(n+1)(m+1)pq}^{gf}
\nonumber \\
    & \ \ \ \ \ \ \ \ 
        - \frac{2\kappa \beta_{gf}^*}{3} 
            \sqrt{n+1}
            \lambda_{(n+1)mpq}^{gf}
        + \frac{2\kappa \beta_{gf}}{3} 
            \sqrt{m+1}
            \lambda_{n(m+1)pq}^{gf}
\end{align}
and
\begin{align}
    \dot{\lambda}_{nmpq}^{ef}
    &= \Big[
            \ci \Bar{\omega}_{fe}
            - \gamma_{1,ef} / 2 
            - \gamma_{\phi,ef} 
            + 2 \chi_{\text{qr}} \Im(\alpha_f \alpha_e^*)
\nonumber \\
    & \ \ \ \ \ \ \ \ 
            -\ci \Delta_{\text{rd}} (n - m) 
            + \ci \chi_{\text{qr}} (2m - n)
            - \kappa (n + m) / 2
        \Big]
        \lambda_{nmpq}^{ef}
\nonumber \\[1.5mm]
    & \ \ \ \ 
        + \kappa \sqrt{(n+1)(m+1)}
            \lambda_{(n+1)(m+1)pq}^{ef}
\nonumber \\
    & \ \ \ \ \ \ \ \ 
        - \frac{2\kappa \beta_{ef}^*}{3} 
            \sqrt{n+1}
            \lambda_{(n+1)mpq}^{ef}
        + \frac{2\kappa \beta_{ef}}{3} 
            \sqrt{m+1}
            \lambda_{n(m+1)pq}^{ef}
\end{align}
with the net frequency differences
\begin{align}
    \Bar{\omega}_{fg}
    &= 2 \omega_{\text{q}} + \alpha_{\text{q}}
        + 2\chi_{\text{qr}} \Re(\alpha_f \alpha_g^*),
\\
    \Bar{\omega}_{fe}
    &= \omega_{\text{q}} + \alpha_{\text{q}}
        + \chi_{\text{qr}} \Re(\alpha_f \alpha_e^*).
\end{align}

Since we are in the transformed frame, the photon population is initially displaced to the vacuum state, i.e., ${\lambda}_{nmpq}^{ab} \propto \rho^{\mathsf{P}}_{nm ab}= 0$. In addition, there is no mechanism to excite $\lambda_{nmpq}^{ab}$ with $n,m,p,q >0$ because the three displacement operators are designed to keep the photon number zero in the displaced frame; hence, $\rho_{\mathcal{S}, ab} = \lambda_{0000}^{ab}$ (see Eq.(\ref{eq:rho_s_ge_and_lambda_nmmn_ge})) and 
\begin{align} \label{eq:lambda_0000_ge}
    \dot{\rho}_{\mathcal{S}, ge}
    &= \dot{\lambda}_{0000}^{ge}
    = \Big[
            \ci \Bar{\omega}_{eg}
            - \gamma_{1,ge} / 2 
            - \gamma_{\phi,ge} 
            + \chi_{\text{qr}} \Im(\alpha_e \alpha_g^*)
        \Big]
        \lambda_{0000}^{ge},
\\ \label{eq:lambda_0000_gf}
    \dot{\rho}_{\mathcal{S}, gf}
    &= \dot{\lambda}_{0000}^{gf}
    = \Big[
            \ci \Bar{\omega}_{gf}
            - \gamma_{1,ge} / 2 
            - \gamma_{\phi,gf} 
            + 2 \chi_{\text{qr}} \Im(\alpha_f \alpha_g^*)
        \Big]
        \lambda_{0000}^{gf},
\\ \label{eq:lambda_0000_ef}
    \dot{\rho}_{\mathcal{S}, ef}
    &= \dot{\lambda}_{0000}^{ef}
    = \Big[
            \ci \Bar{\omega}_{ef}
            - \gamma_{1,ef} / 2 
            - \gamma_{\phi,ef} 
            + \chi_{\text{qr}} \Im(\alpha_f \alpha_e^*)
        \Big]
        \lambda_{0000}^{ef}.
\end{align}
At this point, one might naively write down an effective master equation for the qutrit based on Eq.(\ref{eq:qutrit_population_gg})-(\ref{eq:qutrit_population_ff}) and (\ref{eq:lambda_0000_ge})-(\ref{eq:lambda_0000_ef}); however, the net frequency differences calculated above, in general, do not satisfy the relation
\begin{equation}
    \Bar{\omega}_{fg} - \Bar{\omega}_{fe} 
    = \Bar{\omega}_{ge}
\end{equation}
for a three-level system, so we cannot write down an exact master equation of the Lindblad form. Such a problem does not appear in the qubit case since the single transition energy is not subject to any constraint. Nevertheless, if the frequency shifts are much smaller than other rate parameters, we can still approximate the qutrit as a simple Markovian system, thus writing down an effective master equation
\begin{align}
    \dot{\hat{\rho}}_{\mathcal{S}}
    &= - \frac{\ci}{\hbar}
            \Big[ 
                \hat{H}_{\text{q,eff}}, 
                \hat{\rho}_{\mathcal{S}}
            \Big]
        + \gamma_{1,ge} 
            \mathcalboondox{D}
                \Bigsl[
                    \hat{\sigma}_{ge}
                \Bigsr] 
            \hat{\rho}_{\mathcal{S}}
        + \gamma_{1,gf} 
            \mathcalboondox{D}
                \Bigsl[
                    \hat{\sigma}_{gf}
                \Bigsr] 
            \hat{\rho}_{\mathcal{S}}
        + \gamma_{1,ef} 
            \mathcalboondox{D}
                \Bigsl[
                    \hat{\sigma}_{ef}
                \Bigsr] 
            \hat{\rho}_{\mathcal{S}}
\nonumber \\[2mm]
    & \ \ \ \
        + \frac{\gamma_{\phi,ge} + \Gamma_{\text{d},ge}}{2} 
            \mathcalboondox{D}
                \Bigsl[
                    \hat{\sigma}_{z,ge}
                \Bigsr]
            \hat{\rho}_{\mathcal{S}}
        + \frac{\gamma_{\phi,gf}+ \Gamma_{\text{d},gf}}{2} 
            \mathcalboondox{D}
                \Bigsl[
                    \hat{\sigma}_{z,gf}
                \Bigsr]
            \hat{\rho}_{\mathcal{S}}
        + \frac{\gamma_{\phi,ef}+ \Gamma_{\text{d},ef}}{2} 
            \mathcalboondox{D}
                \Bigsl[
                    \hat{\sigma}_{z,ef}
                \Bigsr]
            \hat{\rho}_{\mathcal{S}},
\end{align}
where the effective Hamiltonian is assumed to describe a self-consistent set of energy levels and the \textbf{measurement-induced dephasing rates} are given by
\begin{equation} \label{eq:measurement_induced_dephasing_definition}
    \Gamma_{\text{d},ge}(t)
    = \chi_{\text{qr}} \Im(\alpha_g \alpha_e^*),
\ \ \ \
    \Gamma_{\text{d},gf}(t)
    = 2 \chi_{\text{qr}} \Im(\alpha_g \alpha_f^*),
\ \ \ \
    \Gamma_{\text{d},ef}(t)
    = \chi_{\text{qr}} \Im(\alpha_e \alpha_f^*).
\end{equation}
Furthermore, note that, by substituting the expression of $\Bar{\omega}_{ab}$ into Eq.(\ref{eq:lambda_0000_ge})-(\ref{eq:lambda_0000_ef}), we obtain
\begin{align} 
    \dot{\rho}_{\mathcal{S}, ge}
    &= \Big[
            \ci \omega_{\text{q}}
            - \gamma_{1,ge} / 2 
            - \gamma_{\phi,ge} 
            + \ci \chi_{\text{qr}} \alpha_g \alpha_e^*
        \Big]
        {\rho}_{\mathcal{S}, ge},
\\ 
    \dot{\rho}_{\mathcal{S}, gf}
    &= \Big[
            \ci (2\omega_{\text{q}} + \alpha_{\text{q}})
            - \gamma_{1,ge} / 2 
            - \gamma_{\phi,gf} 
            + \ci 2 \chi_{\text{qr}} \alpha_g \alpha_f^*
        \Big]
        {\rho}_{\mathcal{S}, gf},
\\
    \dot{\rho}_{\mathcal{S}, ef}
    &= \Big[
            \ci (\omega_{\text{q}} + \alpha_{\text{q}})
            - \gamma_{1,ef} / 2 
            - \gamma_{\phi,ef} 
            + \ci \chi_{\text{qr}} \alpha_e \alpha_f^*
        \Big]
        {\rho}_{\mathcal{S}, ef},
\end{align}
matching Eq.(\ref{eq:c_ge_diff_eqn})-(\ref{eq:c_fe_diff_eqn}).

\begin{figure}
    \centering
    \begin{tikzpicture}[scale=0.75]
    \begin{axis}[
        width=10cm,
        height=10cm,
        colormap={bw}{
        gray(0cm)=(1);
        gray(1cm)=(0);
        },
        domain=-8:8,
        domain y=-8:8,
        view={0}{90},
        xlabel={$I$},
        ylabel={$Q$},
        xtick={0},
        ytick={0},
        ]
    \addplot3[
        contour filled={
        number=50,
        },
        samples={40}
    ]
        {exp((-(x-4)^2-(y-2.5)^2)) + exp((-(x-6)^2-y^2)) + exp((-(x-5)^2-(y+2)^2))};
    \end{axis}
    \begin{axis}[
        width=10cm,
        height=10cm,
        xmin=-8,
        xmax=8,
        ymin=-8,
        ymax=8,
        domain y=-8:8,
        hide x axis,
        hide y axis,
        nodes near coords,
        ]
    \addplot+[scatter, 
            only marks, 
            point meta=explicit symbolic,
            color=white,
            mark size=0.001
    ]
        coordinates{
            ({4},{2.5}) [${\ket{f}}$]
            ({6},{0}) [${\ket{e}}$]
            ({5},{-2}) [${\ket{g}}$]
        };
    \end{axis}
    \end{tikzpicture}
    \caption{The expected scattering plot of the dispersive measurement for a qutrit prepared in an equal superposition state. A single measurement of the state leads to a dot in the phase plane. When we perform the same experiment repeatedly, we would obtain the scattering plot with three Gaussian blobs centered at $\alpha_g$, $\alpha_e$, and $\alpha_f$ (up to an overall scaling due to the amplification along the output chain). One can thus interpret the plot as the probability distribution associated with the quantum measurement, similar to the qubit case shown in Figure \ref{fig:dispersive_shift_3D}.}
    \label{fig:phase_plane_coherent_states_for_dispersive_measurement}
\end{figure}
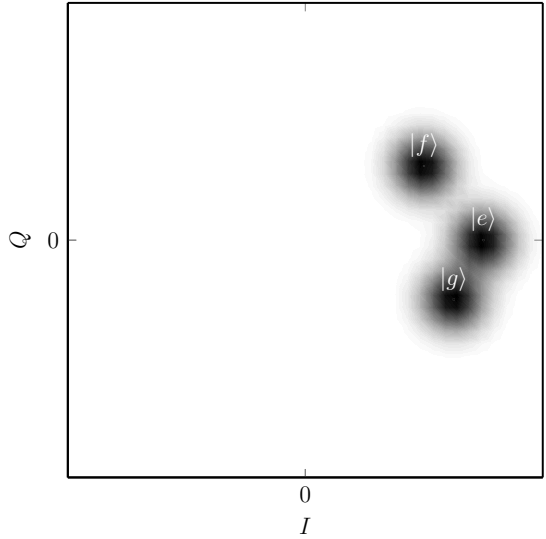

\subsection{Measurement-Induced Dephasing}
There has been an overwhelmingly long calculation in this section. Hence, we should pause for a second and extract the physical interpretation of the time evolution of the qutrit density operator. We have already pointed out the connection between the diagonal terms of the effective master equation and the semi-classical rate equation. The more interesting phenomenon lies in the time evolution of the coherence terms. In particular, we see that the product of the dispersive shift and the imaginary part of $\alpha_a \alpha_b^*$ induces a dephasing for each energy level of the qutrit. There are three factors that affect the measurement-induced dephasing rates: 
\begin{enumerate}
    \item [(i)] A large readout drive leads to large coherent state amplitudes and thus a stronger dephasing. From the point of view of information theory, since the field leaks out from the resonator will have a larger amplitude as well, we are more likely to gain useful information from the measurement since the signal-to-noise ratio increases as the power of the readout signal goes up\footnote{However, remember that the dispersive coupling is valid only in the low-photon case, so the readout power is usually not a good parameter to adjust.}. Nevertheless, our measurement of the coherent states leaked out of the resonator is subject to the quadrature uncertainty, thus, the measurement result will be distributed as Gaussians centered at $\alpha_{g}$, $\alpha_{e}$, or $\alpha_{f}$ as shown in Figure \ref{fig:phase_plane_coherent_states_for_dispersive_measurement}. The uncertainty in the measurement will then lead to a random backaction on the qutrit conditioned on the measurement results. It is this random backaction that leads to the dephasing of the qutrit. For a qubit, we have already introduced the Kraus operator description in Section \ref{section:measurement_via_dispersive_coupling_kraus} without giving a rigorous derivation. In the next section, we will finally formalize this idea in the limit of continuous measurement.

    \item[(ii)] Related to the coherent state amplitudes is the readout frequency. As shown in Eq.(\ref{eq:steady_state_cavity_amplitude_alpha_g})-(\ref{eq:steady_state_cavity_amplitude_alpha_f}), the field amplitudes are Lorentzian functions of the detuning. Hence, for the same drive strength $\bar{a}_{\text{in}}$, the amplitude built up inside the resonator will be the highest when the detuning is zero. However, we cannot drive all three dressed frequencies with zero detuning simultaneously, which means that we have to play with the readout frequency so that the separations among the three coherent states are maximized for state classification. Figure \ref{fig:cavity_amplitude_plot_simulations}(b) shows the steady-state amplitudes of the resonator for some arbitrary readout frequency near the bare resonator frequency; in general, the complex steady-state amplitudes lie on a circle that goes through the origin of the phase plane. In addition, Figure \ref{fig:cavity_amplitude_plot_simulations}(c) plots the distance between two coherent state amplitudes as a function of the readout frequency.

    \item [(iii)] A larger dispersive shift $\chi_{\text{qr}}$ will also lead to a faster decoherence time. This, again, can be argued from an information-theory point of view. The dispersive shift determines how well we can separate the three qutrit-dependent resonator frequencies; hence the larger the dispersive shifts, the easier the state classification. However, as we have seen in the analysis of the qubit-resonator coupling, $\chi_{\text{qr}}$ is proportional to the square of the coupling coefficient and is inversely proportional to the detuning. For the dispersive coupling to be valid, we cannot make $\sqrt{n} g/\Delta$ too large, thus limiting the amount of $\chi_{\text{qr}}$ realizable in practice. Furthermore, there is another ratio we can design to improve the state classification -- the ratio between the dispersive shift $\chi_{\text{qr}}$ and cavity decay rate $\kappa$. The effect of $\kappa$ is hidden in the expression of the steady-state amplitudes. As shown in Figure \ref{fig:cavity_amplitude_plot_simulations}(d), the distance between the coherent states can be improved by making $\chi_{\text{qr}} > \kappa$.
\end{enumerate}

\section{Quantum Stochastic Master Equations for Qubit and Qutrit Measurement}
As discussed in Section \ref{section:classical_to_quantum_dynamical_systems}, an unconditioned master equation can be interpreted as the stochastic trajectories ensemble-averaged over all the possible measurement outcomes. The combined system of the qudit and the resonator can be measured either actively by us or implicitly by the environment. In the previous section, we ignored the information coming out of the resonator, which is equivalent to assuming that all measurements are implicitly made by the environment. To describe an active dispersive measurement by us on a qudit, we thus need to retrieve the information that has been so far averaged out. Since measurements are probabilistic in quantum mechanics, we need to introduce a \textit{stochastic} master equation \cite{PhysRevA.47.642, WISEMAN200191, gardiner2004quantum, PhysRevA.77.012112} to model the random measurement outcomes.

In this section, we first revisit the quantum channel for the dispersive measurement of a qubit introduced in Section \ref{section:measurement_via_dispersive_coupling_kraus}. In particular, the field coming out from the resonator is detected using a homodyne scheme where only one quadrature of information is measured. Since a typical superconducting qubit is a weakly-anharmonic oscillator, the multi-level model should produce a more realistic result. Nevertheless, we can already gather tremendous intuition through the qubit example. After solving the stochastic master equation for a qubit, we move to the dispersive measurement of a qutrit using the heterodyne scheme where both quadratures of the outgoing field are measured. Finally, the theory is compared with the experiment on a transmon qutrit.

\subsection{Dispersive Measurement of a Qubit Using the Homodyne Detection} \label{section:dispersive_measurement_of_a_qubit}
To derive a stochastic master equation, we need to quantify the leakage rate of information from the readout cavity. (To be concrete, Figure \ref{fig:cavity_amplitude_plot_simulations} (a) shows a schematic of the readout cavity with a transmon placed inside.) As usual, assume that the cavity has a leakage rate of $\kappa_{\text{out}}$ at the output port; that is, there are $\kappa_{\text{out}} \Bar{n} \Delta t$ photons coming out from the cavity during a short time $\Delta t$ if the cavity stores $\Bar{n}$ photons on average. For simplicity, let us assume $\kappa_{\text{in}} \ll \kappa_{\text{out}}$ so that the photon leaking from the input port can be ignored; we will relax this condition later for the qutrit case. To follow the assumption made in Section \ref{section:measurement_via_dispersive_coupling_kraus}, the readout frequency will be set to the bare cavity frequency $\omega_{\text{r}}$ so that the readout signal is detuned by $\pm \chi$ when the qubit is in the ground and excited states, respectively. This assumption will also be relaxed when discussing the qutrit measurement. For a qubit, however, placing the readout frequency exactly in the middle will make the problem symmetric and simplify the algebra.

The quadrature field coming out from the cavity should have an average field amplitude $\sqrt{\kappa_{\text{out}} \Bar{n} \Delta t}$, where $\Bar{n}(t) = |\alpha(t)|^2$ depends on the coherent state amplitude of the cavity. For example, the resonator should be in $\ket{\alpha_g(t)}$ (or $\ket{\alpha_e(t)}$) if the qubit is in the ground state (or excited state) as solved in the last section. For simplicity, we work in the steady-state condition where the coherent state amplitudes are given by
\begin{equation} \label{eq:steady_state_amp_cavity_qubit}
    \alpha_{g,e}(+\infty) 
    = \frac{\sqrt{\kappa_{\text{in}}} \bar{a}_{\text{in}}}{\pm \chi - \ci \kappa/2}.
\end{equation}
Then, one can plot the two steady-state amplitudes on the phase plane, which are already shown in Figure \ref{fig:dispersive_shift_3D} with the uncertainty of the coherent states taken into account. Note that we have chosen the phase of $\bar{a}_{\text{in}}$ so that the two coherent states are symmetrically placed about the $I$-axis as shown in \ref{fig:dispersive_shift_3D} so that all the information is encoded along the $Q$-axis. Consequently, by using Eq.(\ref{eq:steady_state_amp_cavity_qubit}), one can express the $Q$-coordinate of the coherent states by $\mp \sqrt{\Bar{n}} \sin\phi_{\chi}$, where $\bar{n} = \abs{\alpha_{g,e}(+\infty)}$ and $\phi_{\chi} = \tan^{-1}(\kappa/2\chi)$. In other words, the quadrature signal coming out of the cavity is given by
\begin{equation} \label{eq:Q_amp}
    \mp \Bar{q} 
    = \mp\sqrt{\kappa_{\text{out}} \Bar{n} \Delta t} \sin \phi_{\chi}.
\end{equation}
(Of course, the other coordinate we ignore has an amplitude $\sqrt{\kappa_{\text{out}} \Bar{n} \Delta t} \cos \phi_{\chi}$.) Intuitively speaking, a bigger $\sin \phi_{\chi}$ would give a larger separation between the two Gaussian distributions, leading to faster state estimation.

Looking back to Section \ref{section:measurement_via_dispersive_coupling_kraus}, we now have quantified the value of $\bar{q}$ for specific qubit and cavity parameters. As a reminder, the Kraus operator defined in Eq.(\ref{eq:dispersive_measurement_channel}) assumes a discretized time axis where each measurement is performed with an integration time $\Delta t$; however, in reality, the measurement is ``almost'' continuous because the readout signal will interact with the cavity for a duration defined by the pulse length\footnote{We use the word ``almost'' to indicate that the signal traveling along the cable will eventually be discretized by the ADC whose sampling frequency is finite.}. Hence, we look for a continuum limit of the quantum channel described by Eq.(\ref{eq:dispersive_measurement_channel}). 

By using Eq.(\ref{eq:prob_measuring_q}) and (\ref{eq:Q_amp}), we can calculate the first and second moments of the measurement $Q_k$ conditioned on the qubit state $\hat{\rho}_{k} = \hat{\rho}$:
\begin{align}
    \mathbb{E}
        \Bigsl(
            Q_k \, | \, \hat{\rho}_{k} = \hat{\rho}
        \Bigsr) 
    &= \Bar{q} \bra{g} \hat{\rho} \ket{g} 
        - \Bar{q} \bra{e} \hat{\rho} \ket{e}
    = \Bar{q} 
        \Tr(\hat{\sigma}_z 
        \hat{\rho})
\nonumber\\
    &= \sqrt{\kappa \Bar{n} \Delta t } 
        \sin \phi 
            \Tr(\hat{\sigma}_z 
            \hat{\rho})
    = \sqrt{k_m \Delta t} 
        \Tr(\hat{\sigma}_z 
        \hat{\rho}),
\\
    \mathbb{E}
        \Bigsl(
            Q_k^2 \, | \, \hat{\rho}_{k} = \hat{\rho}
        \Bigsr)
    &= \int_{- \infty}^{\infty} 
        \mathrm{d} q
        \left[
            \left(\frac{2}{\pi} \right)^{\frac{1}{2}} 
                e^{-2(\Bar{q} - q)^2}
                \bra{g} \hat{\rho} \ket{g}
            + \left(\frac{2}{\pi} \right)^{\frac{1}{2}} 
                e^{-2(\Bar{q} + q)^2}
                \bra{e} \hat{\rho} \ket{e}
        \right] q^2
\nonumber \\
    &= \int_{- \infty}^{\infty} 
            \mathrm{d} q
            \left(\frac{2}{\pi} \right)^{\frac{1}{2}} 
            e^{-2(\Bar{q} - q)^2} q^2
    = \frac{1}{4} 
        + \Bar{q}^2
    = \frac{1}{4} 
        + k_m \Delta t,
\end{align}
where we lump all the constants into a parameter
\begin{equation}
    k_m = \kappa \Bar{n} \sin^2 \phi_{\chi}.
\end{equation}
Next, by assuming $\Delta t = \Bar{q}^2/k_m \ll 1$, we compute the updated qubit state $\hat{\rho}_{k+1}$ conditioned on the previous state $\hat{\rho}_{k} = \hat{\rho}$ and measurement $Q_k = q$ up to the second order in $q\Bar{q} = q\sqrt{k_m \Delta t}$ \cite{arxiv.2208.07416}:
\begin{align}
    \! \! \! \! 
    \hat{\rho}_{k+1}
    &= \frac{\hat{K}_{q}
            \hat{\rho}
            \hat{K}_{q}^{\dagger}}{\Tr(\hat{K}_{q}
            \hat{\rho}
            \hat{K}_{q}^{\dagger})}
\nonumber \\
    &= \frac{
        \Big[
            e^{-(\Bar{q} - q)^2}
            \ket{g} \! \bra{g}
            + e^{-(\Bar{q} + q)^2}
            \ket{e} \! \bra{e}  
        \Big]
        \hat{\rho}    
        \Big[
            e^{-(\Bar{q} - q)^2}
            \ket{g} \! \bra{g}
            + e^{-(\Bar{q} + q)^2}
            \ket{e} \! \bra{e} 
        \Big]
        }{ 
            e^{-2(\Bar{q} - q)^2}
            \bra{g} \hat{\rho} \ket{g}
        + e^{-2(\Bar{q} + q)^2}
            \bra{e} \hat{\rho} \ket{e}}
\nonumber \\
    &= \hat{\rho} 
        + 4q^2 \Bar{q}^2 \left( 
                \hat{\sigma}_z \hat{\rho} \hat{\sigma}_z 
                - \hat{\rho} 
            \right) 
\nonumber \\
        & \ \ \ \ \ \ \ \ \ \ \ \ \ \ \ \ \ \ \ \ \ \ \ \ \
        + \Big[
                \hat{\sigma}_z \hat{\rho}_t
                + \hat{\rho}_t \hat{\sigma}_z^{\dagger}
                - 2\Tr(\hat{\sigma}_z \hat{\rho})
                \hat{\rho}
            \Big] 
            \Big[
                2 q \Bar{q} 
                - 8 q^2 \Bar{q}^2  
                    \Tr(\hat{\sigma}_z \hat{\rho}) 
            \Big]
        + \mathbf{o}(q^2 \Delta t)
\nonumber \\
    &= \hat{\rho} 
        + 4 k_m q^2 \Delta t \left( 
                \hat{\sigma}_z \hat{\rho} \hat{\sigma}_z 
                - \hat{\rho} 
            \right) 
\nonumber \\ \label{eq:updated_rho_small_dt}
        & \ \ \ \ \ \
        + \sqrt{k_m} \Big[
                \hat{\sigma}_z \hat{\rho}_t
                + \hat{\rho}_t \hat{\sigma}_z^{\dagger}
                - 2\Tr(\hat{\sigma}_z \hat{\rho})
                \hat{\rho}
            \Big] 
            \Big[
                2 q \sqrt{\Delta t}
                - 8 \sqrt{k_m} q^2
                    \Tr(\hat{\sigma}_z \hat{\rho}) \Delta t
            \Big]
        + \mathbf{o}(q^2 \Delta t).
\end{align}

To proceed further, let us examine the expression in the last square brackets, i.e., (I change $q$ and $\hat{\rho}$ back to their corresponding random variables)
\begin{equation} \label{eq:defining_wiener}
    \Delta W_k 
    \doteq 2 Q_k \sqrt{\Delta t}
        - 8 \sqrt{k_m} Q_k^2
            \Tr(\hat{\sigma}_z \hat{\rho}_{k}) \Delta t.
\end{equation}
In particular, the first and second moments of $\Delta W_k$ to first order in $\Delta t$ are given by
\begin{align}
    \mathbb{E}\! 
        \left(
            \Delta W_k \, | \, \hat{\rho}_{k} = \hat{\rho}
        \right) 
    &= 2\sqrt{\Delta t} 
            \, 
            \mathbb{E}\!
                \left(
                    Q_k \, | \, \hat{\rho}_{k} = \hat{\rho}
                \right) 
        - 8 \sqrt{k_m} \, \mathbb{E}\!
                \left(
                    Q_k^2 \, | \, \hat{\rho}_{k} = \hat{\rho}
                \right) 
            \Tr(\hat{\sigma}_z \hat{\rho}) \Delta t
\nonumber \\
    &= 2 \sqrt{k_m} 
        \Tr(
            \hat{\sigma}_z
            \hat{\rho}
            ) 
            \Delta t 
        - 8 \sqrt{k_m} 
            \left(\frac{1}{4} + k_m \Delta t \right)
            \Tr(\hat{\sigma}_z
            \hat{\rho}) \Delta t
    = \mathbf{o}(\Delta t),
\end{align}
\begin{equation}
    \mathbb{E}\!
        \left(
            \Delta W_k^2 \, | \, \hat{\rho}_{k} 
            = \hat{\rho}
        \right) 
    = 4 \, 
        \mathbb{E}\!
            \left(
                Q_k^2 \, | \, \hat{\rho}_{k} = \hat{\rho}
            \right) 
            dt 
    = \Delta t + \mathbf{o}(\Delta t).
\end{equation}
Since the expectation and variance are independent of $\hat{\rho}_{k}$ to first order in $\Delta t$, we can drop the conditioning on $\hat{\rho}_{k}$; in other words, $W_k$ satisfies the Markov property. Using the fact that $\Delta W_k$ has a variance $\Delta t$, we conclude that $W_k$ can be approximated by a classical Wiener process $W_t$ (i.e., $\mathbb{E}(\mathrm{d} W_t) = 0$ and $\mathbb{E}(\mathrm{d} W_t^2) = \mathrm{d}t$) as $\Delta t \rightarrow 0$. In addition, from Eq.(\ref{eq:defining_wiener}), we also have
\begin{equation}
    Q^2_k \Delta t
    = \frac{1}{4}
        \Big[
            \Delta W_k
            + 8 \sqrt{k_m} 
                Q_k^2
                \Tr(\hat{\sigma}_z \hat{\rho}_{k})
                \Delta t
        \Big]^2
    =  \frac{1}{4} \Delta W_k^2 + \mathrm{o}(\Delta t),
\end{equation}
which can now be replaced (to the first order in $\Delta t$) by
\begin{equation}
    Q^2_t \mathrm{d} t
    = \frac{1}{4} \mathrm{d} t
\end{equation}
with the help of It\^{o}'s formula $\mathrm{d} W_t^2 = \mathrm{d} t$ (almost surely).

Finally, by taking $\Delta t \rightarrow 0$ and substituting $\mathrm{d} W_t$ and $Q_t^2 \mathrm{d} t$ into Eq.(\ref{eq:updated_rho_small_dt}), we obtain the \textbf{quantum stochastic master equation in the diffusive limit} \cite{arxiv.2208.07416, naghiloo2019introduction, wiseman2010quantum}
\begin{equation} \label{eq:SME_for_dispersive_coupling}
    \mathrm{d} \hat{\rho}_{t}
    = k_m
            \Big(
                \hat{\sigma}_z 
                \hat{\rho}_t
                \hat{\sigma}_z^{\dagger}
                - \hat{\rho}_t
            \Big) 
            \, \mathrm{d} t 
        + \sqrt{k_m} 
            \Big[
                \hat{\sigma}_z \hat{\rho}_t
                + \hat{\rho}_t \hat{\sigma}_z^{\dagger}
                - 2\Tr(\hat{\sigma}_z \hat{\rho}_t)
                \hat{\rho}_t
            \Big] 
            \, \mathrm{d} W_t.
\end{equation}
Besides the missing decay channel, Eq.($\ref{eq:SME_for_dispersive_coupling}$) follows the general form of the stochastic master equation introduced in Section \ref{section:classical_to_quantum_dynamical_systems}. The measurement channel has the Lindblad operator $\hat{L}_m = \hat{L}_m^{\dagger} = \hat{\sigma}_z$. The measurement strength is characterized by $k_m$, which is proportional to the cavity leakage rate $\kappa_{\text{out}}$, the number of photons in the cavity $\Bar{n}$, and the dispersive coupling strength $\chi$; from the perspective of information theory, the measurement strength represents the rate of the information leakage via the measurement channel. 

In practice, it's custom to define a time-integrated quantity $y_k$, whose change at step $k$ (i.e., $\Delta y_k = y_k - y_{k-1}$) is defined to be $Q_k$ multiplied by the square root of the integration time $\Delta t$:
\begin{equation}
    \Delta y_k
    = 2 Q_k \sqrt{\Delta t} 
    = 8 \sqrt{k_m} 
            Q_k^2
            \Tr(\hat{\sigma}_z \hat{\rho}_{k}) 
            \Delta t 
        + \Delta W_t.
\end{equation}
In the diffusive limit, we obtain the second stochastic differential equation\footnote{Sometimes it is also helpful to define a quantity called the \textbf{classical white noise} as the time derivative of the Wiener process
\begin{equation}
    \xi_t = \dot{W}_t
\end{equation}
such that the stochastic differential equation of $y_t$ can be rewritten as
\begin{equation} \label{eq:observability_SME_white_noise}
    \dot{y}_t = 2\sqrt{k_m} \Tr(\hat{\sigma}_z \hat{\rho}_t) + \xi_t.
\end{equation}
The noise $\xi_t$ is said to be white since it can be shown that the autocorrelation function of $\xi_t$ is a delta function, resulting in a flat power spectral density.}
\begin{equation} \label{eq:observability_SME}
    \mathrm{d} y_t 
    = 2\sqrt{k_m} 
            \Tr(\hat{\sigma}_z \hat{\rho}_t) 
            \, \mathrm{d} t 
        + \mathrm{d} W_t,
\end{equation}
analogous to Eq.(\ref{eq:classical_observation}) in the classical control theory. $y_t$ is useful because the expectation of its time derivative is an estimator of the qubit state in the $z$-basis:
\begin{equation} 
    \mathbb{E}[\Tr(\hat{\sigma}_z \hat{\rho}_t)]
    = \frac{1}{2 \sqrt{k_m}}
        \frac{\mathrm{d}}{\mathrm{d} t} 
        \mathbb{E}(y_t) .
\end{equation}
Such a quantity can be numerically computed in an actual experiment since $\Delta t$ is not infinitesimally small due to the finite sampling rate of the analog-to-digital conversion.

\subsection{Understanding Decoherence}
Let us now use the quantum stochastic master equation for dispersive coupling to understand the cause of decoherence. Recall that any qubit state (pure or mixed) has a Bloch-vector representation with the coordinates given by
\begin{gather}
    x = \langle \hat{\sigma}_x \rangle
    = \Tr(\hat{\sigma}_x \hat{\rho}_t),
\ \ \ \
    y = \langle \hat{\sigma}_y \rangle 
    = \Tr(\hat{\sigma}_y \hat{\rho}_t),
\ \ \ \ 
    z = \langle \hat{\sigma}_z \rangle 
    = \Tr(\hat{\sigma}_z \hat{\rho}_t).
\end{gather}
In other words, the state of the qubit is completely specified if we know the expectation of each Pauli matrix. Given any stochastic master equation
\begin{equation}
    \mathrm{d} \hat{\rho}_t 
    = F(\hat{\rho}_t, t) 
            \, \mathrm{d} t
        + G(\hat{\rho}_t, t) 
            \, \mathrm{d} W_t,
\end{equation}
we can use It\^{o}'s rule to derive a stochastic differential equation for each component of the Bloch vector. Since the second derivative of $\Tr\!\big(A \hat{\rho}_t \big)$ with respect to $\hat{\rho}_t$ vanishes for any constant matrix $A$, the It\^{o}'s rule for $\Tr \!\big(A\hat{\rho}_t \big)$ reduces to
\\[-4mm]\begin{align}
    \mathrm{d} \Tr \!\big( A \hat{\rho}_t \big) 
    &= \left\langle 
                \frac{
                    \mathrm{d}
                        \Tr \!
                        \big(   
                            A
                            \hat{\rho}_t 
                        \big) 
                }{
                    \mathrm{d} \hat{\rho}_t
                }, 
                F
            \right\rangle 
            \mathrm{d} t
        + \left\langle 
                \frac{
                    \mathrm{d} 
                        \Tr \!
                        \big(
                            A
                            \hat{\rho}_t 
                        \big) 
                }{
                    \mathrm{d} 
                    \hat{\rho}_t
                },
                G
            \right\rangle \mathrm{d} W_t
\nonumber \\
    &= \Tr\!\big(A F \big) \, \mathrm{d} t + \Tr\!\big(A G \big) \, \mathrm{d} W_t,
\end{align}
where $\mathrm{d}/\mathrm{d}\hat{\rho}_t$ represents the elementwise derivative\footnote{To be precise, the $(i,j)$-entry of the derivative is given by
\begin{equation}
    \left[
            \frac{
                \mathrm{d} 
                    \Tr \!
                    \big(
                        A
                        \hat{\rho}_t 
                    \big) 
            }{
                \mathrm{d} 
                \hat{\rho}_t
            }
        \right]_{ij}
    = \frac{
        \partial 
            \Tr \!
            \big(
                A
                \hat{\rho}_t 
            \big) 
        }{
            \partial (\hat{\rho}_t)_{ij}
        }.
\end{equation}} with respect to the matrix elements of $\hat{\rho}_t$ and $\langle \cdot,\cdot \rangle$ is the Frobenius inner product\footnote{Recall that $\langle A, B \rangle = A^{\dagger} B$ for any matrices $A$ and $B$ of the same size.} of two matrices. Apply this to $A = \hat{\sigma}_x,  \hat{\sigma}_y,  \hat{\sigma}_z$, we obtain
\begin{align} \label{eq:SME_for_x}
    \mathrm{d} \langle \hat{\sigma}_x \rangle 
    &= -2 k_m 
            \langle \hat{\sigma}_x \rangle  
            \, \mathrm{d} t
        - 2 \sqrt{k_m} 
            \langle \hat{\sigma}_x \rangle 
            \langle \hat{\sigma}_z \rangle 
            \, \mathrm{d} W_t,
\\ \label{eq:SME_for_y}
    \mathrm{d} \langle \hat{\sigma}_y \rangle 
    &= -2 k_m 
            \langle \hat{\sigma}_y \rangle  
            \, \mathrm{d} t
        - 2 \sqrt{k_m} 
            \langle \hat{\sigma}_y \rangle 
            \langle \hat{\sigma}_z \rangle 
            \, \mathrm{d} W_t,
\\ \label{eq:SME_for_z}
    \mathrm{d} \langle \hat{\sigma}_z \rangle 
    &= 2\sqrt{k_m} 
        \Big(
            1
            - \langle \hat{\sigma}_z \rangle^2
        \Big) 
        \, \mathrm{d} W_t.
\end{align}

To gain some intuition about the behavior of each component, we take the expectation of each stochastic differential equation:
\begin{align}
    \mathrm{d} \mathbb{E}( \langle \hat{\sigma}_x \rangle )
    &= -2 k_m 
        \mathbb{E}( \langle \hat{\sigma}_x \rangle )  
        \, \mathrm{d} t,
\\
    \mathrm{d} 
        \mathbb{E}( \langle \hat{\sigma}_y \rangle )
    &= -2 k_m 
        \mathbb{E}( \langle \hat{\sigma}_y \rangle ) 
        \, \mathrm{d} t,
\\
    \mathrm{d} 
        \mathbb{E}( \langle \hat{\sigma}_z \rangle )
    &= 0.
\end{align}
Clearly, on average, the vector shrinks to the $z$-axis. However, despite the fact that measurement creates backaction, the $z$-component is not affected in expectation, which allows us to gather information about the z-component of the qubit. Such an invariance against the measurement backaction is called the quantum non-demolition (QND) measurement.

Nevertheless, the measurement does destroy information in the transverse direction, leading to measurement-induced decoherence. Let us go back to the stochastic differential equations before taking the expectation. Suppose we have a qubit, which is initialized in a superposition state $(\ket{g} + \ket{e})/\sqrt{2}$, represented by the Bloch vector $\mathbf{r} = (1, 0, 0)$. In quantum mechanics, this state is as quantum as all the other pure states because the laws of quantum mechanics do not have a preferred basis. However, a real measurement will happen along one specific axis (in our case, it is the $z$-axis). According to Eq.(\ref{eq:SME_for_z}) the state will be pushed towards either $z = 1$ or $z = -1$ due to the term $(1- \langle \hat{\sigma}_z \rangle ^2)$. Before the state is steered to $z = \pm 1$, Eq.(\ref{eq:SME_for_x}) will also push the value of $x$ towards the $z$-axis. Hence, the net effect is that $\mathbf{r}$ will converge to $\mathbf{r} = (0,0,\pm 1)$, i.e., $\ket{g}$ or $\ket{e}$\footnote{In fact, it can be shown that the solution to the stochastic master equation converges to $\ket{g}$ or $\ket{e}$ almost surely by using the Lyapunov function
\begin{equation}
    V(\hat{\rho}_t) 
    = \sqrt{\bra{g} \hat{\rho}_t \ket{g} \bra{e} \hat{\rho}_t \ket{e}}\\[-2mm]
\end{equation}
and (at least in the discrete-time case)
\begin{equation}
    \mathbb{E}
        \big[
            V(\hat{\rho}_{k+1}) 
            \, | \, 
            \hat{\rho}_k = \hat{\rho}
        \big]
    = e^{-k_m \Delta t} V(\hat{\rho}).
\end{equation}}.
In other words, a qubit subject to a continuous-time measurement along the $z$-basis will lose its quantumness regardless of its initial state and behave just like a classical coin with only two possible states. In fact, we do not need to be the witnesses of any measurement; any real macroscopic system will inevitably interact with the environment and thus lose its coherence quickly due to a similar convergence. The set of states that a quantum system can converge to in the classical limit is known as the \textbf{pointer states}. The measurement strength $k_m$ will be huge for a macroscopic object, so the decoherence process would appear to be instantaneous.

\subsection{Dispersive Measurement of a Qutrit Using the Heterodyne Detection}
Now, we consider a qutrit coupled to the cavity dispersively. To read out the resonator state, we perform a heterodyne detection where the readout signal coming out of the resonator is mixed with a strong (i.e., the annihilation operator can be replaced by its eigenvalues) local oscillator (LO) signal 
\begin{equation}
    \hat{V}_{\text{LO}} (t)
    = \frac{\hat{a}_{\text{LO}}(t) e^{-\ci \phi_{\text{LO}}} + \hat{a}^{\dagger}_{\text{LO}}(t) e^{\ci \phi_{\text{LO}}}}{2}
    \approx 
    V_{\text{LO}} \cos(\omega_{\text{LO}} t - \phi_{\text{LO}}),
\end{equation}
whose frequency $\omega_{\text{LO}}$ is different from the readout signal $\omega_{\text{d}}$ by the intermediate frequency (IF) $\omega_{\text{IF}}$, e.g., $\omega_{\text{IF}} = \omega_{\text{LO}} - \omega_{\text{d}}$. In fact, in a typical IQ demodulation stage, the amplified readout signal 
is first divided equally in power and then mixed separately by two LO signals whose phases are $90^{\circ}$ out of phase. Subsequently, the analog IF signals, passing through an analog-to-digital converter (ADC), are processed digitally and demodulated finally to DC (zero frequency). Unlike a homodyne detection where $\omega_{\text{LO}} = \omega_{\text{d}}$, the heterodyne scheme allows us to measure both quadratures of the field at the same time (but still constrained by the uncertainty principle). In addition, by first moving to an IF frequency (usually $\omega_{\text{IF}} \sim 100 \text{ MHz}$), the signal experiences less $1/f$ noise.

We will, however, not attempt to model the analog or digital demodulation part of the heterodyne detection using quantum mechanics. Instead, we will work directly with the coherent state coming out of the resonator and assume that we can process the signal in the way described above and retrieve information about the coherent states subject to quantum-mechanical noise and imperfect measurement efficiency. For a fully quantum-mechanical description of the output chain, including filtering and amplification, see \cite{RevModPhys.93.025005}.

Unlike the qubit measurement discussed in Section \ref{section:dispersive_measurement_of_a_qubit}, there isn't any symmetry we can utilize to describe a general qutrit measured using the heterodyne scheme. To analyze the information encoded in the complex amplitude of a coherent state, we first define
\begin{equation}
    \hat{I}_{\phi} = \frac{\hat{a} e^{-\ci \phi} + \hat{a}^{\dagger} e^{\ci \phi}}{2}
\ \ \text{ and } \ \ 
    \hat{Q}_{\phi} = \frac{\hat{a} e^{-\ci \phi} - \hat{a}^{\dagger} e^{\ci \phi}}{2\ci}
\end{equation}
to be the two quadrature operators of the resonator field, where $\phi$ models the net phase of the LO, cable delay, and any rotation applied during data processing. Similarly, for the coherent states introduced in Section \ref{section:composite_system_master_equations}, we define
\begin{align} 
    \bar{I}_{g}(t) &= \Re(\alpha_g e^{-\ci \phi}),
    \ \ \ \ 
    \bar{Q}_{g}(t) = \Im(\alpha_g e^{-\ci \phi}),
\\
    \bar{I}_{e}(t) &= \Re(\alpha_e e^{-\ci \phi}),
    \ \ \ \ 
    \bar{Q}_{e}(t) = \Im(\alpha_e e^{-\ci \phi}),
\\
    \bar{I}_{f}(t) &= \Re(\alpha_f e^{-\ci \phi}),
    \ \ \ \ 
    \bar{Q}_{f}(t) = \Im(\alpha_f e^{-\ci \phi}).
\end{align}

Note, however, that the quadrature fields defined above live in the resonator and what we observe is only the leakage of the resonator. Recall that associated with the QLE for a resonator configured in the transmission mode is the boundary condition
\begin{equation} \label{eq:input_output_at_output_port}
    \hat{a}_{\text{out}}(t)
    =  - \hat{a}_{\text{in}}(t) 
        + \sqrt{\kappa_{\text{out}}} \hat{a}(t)
    \approx \sqrt{\kappa_\text{out}} \hat{a}(t),
\end{equation}
where we have dropped $\hat{a}_{\text{in}}$ by assuming that the incoming signal at the output port is isolated by a well-designed circulator stage and is not amplified at the output stage (i.e., the parametric amplifier and HEMT are approximately unilateral). Since in the transmission mode, the resonator is a two-port device, there should be another boundary condition for the input port (see Figure \ref{fig:cavity_amplitude_plot_simulations}(a)); in fact, we have been implicitly using it to define the drive $\epsilon = \sqrt{\kappa_{\text{in}}} \bar{a}_{\text{in}}$, where $\bar{a}_{\text{in}}$ is a classical (i.e., deterministic) signal entering at the input port of the resonator.

In the Schr\"{o}dinger picture, this boundary condition at the output port is equivalent to saying that the outgoing traveling signal is in a coherent state whose amplitude is given by
\begin{equation}
    \alpha_{\text{out}}(t) = \sqrt{\kappa_{\text{out}}} \alpha_{a}(t)
\end{equation}
if the resonator is in the coherent state $\ket{\alpha_a}(t)$ for $a \in \{e,g,f\}$. Moreover, recall that $\hat{a}_{\text{out}}^{\dagger} \hat{a}_{\text{out}}$ is the outgoing photon flux, so the mean photon number leaving the resonator from the output port within an infinitesimally short time $\Delta t$ is given by
\begin{equation}
    \bar{n}(t) 
    = \kappa_{\text{out}} \Delta t |\alpha_{a}(t)|^2,
\end{equation}
the same as the expression argued heuristically in Section \ref{section:dispersive_measurement_of_a_qubit}. In terms of the outgoing quadrature fields on the transmission line, we have 
\begin{equation}
    I_{\text{out}}(t) 
    = \sqrt{\kappa_{\text{out}} \Delta t} \bar{I}_a(t)
\ \ \text{ and } \ \ 
    Q_{\text{out}}(t) 
    = \sqrt{\kappa_{\text{out}} \Delta t} \bar{Q}_a(t),
\end{equation}
where $\bar{I}_a$ and $\bar{Q}_a$ are approximately constant over short $\Delta t$. 

In reality, however, measurements are not perfect and not all the information encoded in the photon flux can be captured; thus, the effective photon number we can measure is only 
\begin{equation}
    \bar{n}_{\text{eff}}(t) 
    = \eta \kappa \Delta t |\alpha_{a}(t)|^2,
\end{equation}
where $\eta \in [0,1]$ is the measurement efficiency. Since $\eta_{\text{r}} = \kappa_{\text{out}} / \kappa < 1$, the efficiency $\eta$ is naturally lowered by $\eta_{\text{r}}$. Note that $\eta_{\text{r}}$ also contains the effect of $\kappa_{\text{int}}$ since photons lost internally are inaccessible to the detector. In addition, by performing a heterodyne detection, we automatically half the efficiency due to the power division in an IQ mixer. Moreover, note that when $\eta = 0$, we would gain zero information about the system and can only talk about the behavior of the qutrit in an averaged sense; thus, the stochastic differential equation to be constructed should reduce to the unconditioned master equation and we will verify this point later.

Unlike the mean amplitude, the uncertainty/variance associated with a coherent state traveling on the transmission line is fixed (each quadrature has a variance of $1/4$), so the signal-to-noise ratio is proportional to $\Delta t$; in other words, the photon shot noise can be effectively reduced by increasing the measurement time. Concretely, suppose the resonator is in one of the coherent states $\ket{\alpha_a}$ associated with an energy eigenstate $\ket{a}$ of the qutrit. Then, the \textit{conditional} probability density of measuring a particular point $(I,Q)$ in the phase plane with an integration time of $\Delta t$ \textit{given the qutrit state $\ket{a}$} is a two-dimensional Gaussian
\begin{equation}
    f (I,Q | \hat{\rho}_{\mathcal{S}}(t) = \ket{a}\!\bra{a}) 
    \propto \exp[- \frac{1}{2} \frac{(I - \sqrt{\eta \kappa \Delta t} \, \bar{I}_a)^2 + (Q - \sqrt{\eta \kappa \Delta t} \, \bar{Q}_a)^2}{1/4}]
\end{equation}
with a variance $1/4$ in each quadrature. We have also assumed that the two arms of the mixer output have the same conversion loss to use a single $\eta$; in other words, the measurement is balanced.

More generally, if the qutrit is in a superposition state, then the cavity state, after tracing out the qutrit state, is given by
\begin{equation}
    \hat{\rho}_{\mathcal{R}}(t) 
    = \Tr_{\mathcal{S}}[\hat{\rho}_{\mathcal{SR}}(t)]
    = \rho_{\mathcal{S}, gg}(t) 
            \ket{\alpha_{g}}\!\bra{\alpha_{g}}
        + \rho_{\mathcal{S}, ee} (t)
            \ket{\alpha_{e}}\!\bra{\alpha_{e}}
        + \rho_{\mathcal{S}, ff} (t)
            \ket{\alpha_{f}}\!\bra{\alpha_{f}},
\end{equation}
as suggested by Eq.(\ref{eq:general_qutrit_resonator_solution}). Hence, the total conditional probability density of measuring $(I,Q)$ in the phase plane is now given by
\begin{equation} \label{eq:prob_IQ_general_state}
    f(I,Q | \hat{\rho}_{\mathcal{SR}}(t)) 
    \propto 
    \sum_{a \in \{g,e,f\}}
        \rho_{\mathcal{S}, aa}(t)
        \exp[- \frac{1}{2} \frac{(I - \sqrt{\eta \kappa \Delta t} \, \bar{I}_a)^2 + (Q - \sqrt{\eta \kappa \Delta t} \, \bar{Q}_a)^2}{1/4}].
\end{equation}
Consequently, the entanglement generated by the dispersive coupling will project the qutrit state to a new state (possibly mixed) based on the measurement outcome $(I,Q)$ after $\Delta t$. 

By introducing the integration time $\Delta t$, we have discretized the continuous measurement into time steps $t_0, t_1, t_2,...$ with a step size of $\Delta t$. We formalize measurement and the backaction induced by the measurement output by introducing, at each time step $t_k$, a continuum of POVM 
\begin{equation}
    \Big\{ 
        \hat{E}_{IQ}(t_k) = \hat{K}_{IQ}^{\dagger}(t_k) \hat{K}_{IQ}(t_k)
        \ | \ 
        (I,Q) \in \mathbf{R}^2 
    \Big \}
\end{equation}
for the qutrit (i.e., the resonator is traced out in this description) with the Krauss operators
\begin{align}
    \hat{K}_{IQ}(t_k)
    = \mathscr{N}_{k} 
        \sum_{a \in \{g,e,f\}} 
            \exp{- \Big[ I - \sqrt{\eta \kappa \Delta t} \, \bar{I}_a(t_k) \Big]^2 - \Big[ Q - \sqrt{\eta \kappa \Delta t} \, \bar{Q}_a(t_k) \Big]^2}
            \hat{\Pi}_a
\end{align}
for any point $(I,Q)$ in the phase plane. The normalization constant $\mathscr{N}_k$ can be found by imposing
\begin{equation}
    \int_{-\infty}^{\infty} 
        \int_{-\infty}^{\infty} 
            \mathrm{d} I \mathrm{d} Q \,
            \hat{E}_{IQ}(t_k)
    = 1,
\end{equation}
as required by the completeness of the POVM. Using the Kraus operators, the probability density of measuring $(I,Q)$ is
\begin{equation}
    \Tr[\hat{\rho}_{\mathcal{S}}(t_k) \hat{E}_{IQ}(t_k)] 
    = \Tr[\hat{K}_{IQ}(t_k) \hat{\rho}_{\mathcal{S}}(t_k) \hat{K}_{IQ}^{\dagger}(t_k)] ,
\end{equation}
which, of course, must agree with Eq.(\ref{eq:prob_IQ_general_state}). Furthermore, the post-measurement state \textit{conditioned} on the outcome $(I,Q)$ is 
\begin{equation} \label{eq:qutrit_state_update_rule}
    \hat{\rho}_{\mathcal{S}}(t_{k+1})
    = \hat{\rho}_{\mathcal{S}}(t_k + \Delta t)
    = \frac{\hat{K}_{IQ}(t_k) \hat{\rho}_{\mathcal{S}}(t_k) \hat{K}_{IQ}^{\dagger}(t_k)}{\Tr[\hat{K}_{IQ} (t_k)\hat{\rho}_{\mathcal{S}}(t_k) \hat{K}_{IQ}^{\dagger}(t_k)] }.
\end{equation}
We emphasize that $\hat{\rho}_{\mathcal{S}}(t_{k+1})$ is the conditional reduced density operator and thus is \textit{not} the same as the reduced density operator used in the effective qutrit master equation before. Nevertheless, we should reproduce the unconditional density operator once averaged over all the possible measurement history. Furthermore, it should be clear from Eq.(\ref{eq:qutrit_state_update_rule}) that the series of quantum channels form a Markov chain, making the entire mathematical formalism easier to deal with.

Based on Eq.(\ref{eq:qutrit_state_update_rule}), we look for a stochastic differential equation in the diffusive limit as $\Delta t \rightarrow 0$. We start with $\eta = 1$ so that we do not need to worry about averaging over the unobserved information; of course, such an assumption is unphysical, so we will need to relax it later. To introduce random processes that capture the measurement noise, we first examine various moments of the measurement outcomes $I_k = I(t_{k+1})$ and $Q_k = Q(t_{k+1})$ within $[t_k, t_k + \Delta t)$ conditioned on the qutrit state at $t_k$. To begin with, the conditional expectation of $I_k$ and $Q_k$ are
\begin{align}
    \mathbb{E}
        \Bigsl[
            I_k \, | \, 
            \hat{\rho}_{\mathcal{S}}(t_{k})
        \Bigsr] 
    &= \iint 
            \mathrm{d} I' 
            \mathrm{d} Q' \, 
            I' 
            f(I',Q'|\hat{\rho}_{\mathcal{S}}(t_k))
    = \sqrt{\eta \kappa \Delta t} 
        \Tr[\hat{\rho}_{\mathcal{S}}(t_k) \hat{L}_{I}(t_k)]
    = \mathcal{O}(\sqrt{\Delta t}),
\\
    \mathbb{E}
        \Bigsl[
            Q_k \, | \, 
            \hat{\rho}_{\mathcal{S}}(t_{k})
        \Bigsr] 
    &= \iint 
            \mathrm{d} I' 
            \mathrm{d} Q' \, 
            Q' 
            f(I',Q'|\hat{\rho}_{\mathcal{S}}(t_k))
    = \sqrt{\eta \kappa \Delta t} 
        \Tr[\hat{\rho}_{\mathcal{S}}(t_k) \hat{L}_{Q}(t_k)]
    = \mathcal{O}(\sqrt{\Delta t}),
\end{align}
where we have defined
\begin{gather}
    \hat{L}_I (t)
    = \bar{I}_g(t) \hat{\Pi}_g
        + \bar{I}_e(t) \hat{\Pi}_e
        + \bar{I}_f(t) \hat{\Pi}_f,
\\
    \hat{L}_Q(t) 
    = \bar{Q}_g(t) \hat{\Pi}_g
        + \bar{Q}_e(t) \hat{\Pi}_e
        + \bar{Q}_f(t) \hat{\Pi}_f.
\end{gather}
More importantly, we also have
\begin{align}
    \mathbb{E}
        \Bigsl[
            I_k^2 \, | \, 
            \hat{\rho}_{\mathcal{S}}(t_{k})
        \Bigsr]
    &= \sum_{a} \rho_{\mathcal{S},aa}
            \left(
                \eta \kappa \Delta t \bar{I}_a^2 
                + \frac{1}{4}
            \right)
    = \eta \kappa \Delta t \Tr(\hat{\rho}_{\mathcal{S}} \hat{L}_{I}^{\dagger} \hat{L}_{I})
        + \frac{1}{4}
    = \mathcal{O}(1),
\\
    \mathbb{E}
        \Bigsl[
            Q_k^2 \, | \, 
            \hat{\rho}_{\mathcal{S}}(t_{k})
        \Bigsr]
    &= \sum_{a} \rho_{\mathcal{S},aa}
            \left(
                \eta \kappa \Delta t \bar{Q}_a^2 
                + \frac{1}{4}
            \right)
    = \eta \kappa \Delta t \Tr(\hat{\rho}_{\mathcal{S}} \hat{L}_{Q}^{\dagger} \hat{L}_{Q})
        + \frac{1}{4}
    = \mathcal{O}(1),
\\
    \mathbb{E}\Bigsl[
            I_k Q_k \, | \, 
            \hat{\rho}_{\mathcal{S}}(t_{k})
        \Bigsr]
    &= \eta \kappa \Delta t
            \Tr(\hat{\rho}_{\mathcal{S}} \hat{L}_{I} \hat{L}_{Q})
    = \mathcal{O}(\Delta t).
\end{align}
We observe that the second moments of $I_k$ and $Q_k$ are nonvanishing as $\Delta t \rightarrow 0$, but the correlation between $I_k$ and $Q_k$ would vanish for small $\Delta t$, which implies that measurements of $I_k$ and $Q_k$ are related to two independent random processes. We can keep computing higher moments, such as
\begin{align}
    \mathbb{E}\Bigsl[
            I_k^2 Q_k \, | \, 
            \hat{\rho}_{\mathcal{S}}(t_{k})
        \Bigsr]
    &= \frac{1}{4} \sqrt{\eta \kappa \Delta t}
            \Tr(\hat{\rho}_{\mathcal{S}} \hat{L}_{Q})
        + \eta \kappa \Delta t
            \Tr(\hat{\rho}_{\mathcal{S}} \hat{L}_{I}^2 \hat{L}_{Q})
    = \mathcal{O}(\sqrt{\Delta t}),
\\
    \mathbb{E}\Bigsl[
            I_k Q_k^2 \, | \,
            \hat{\rho}_{\mathcal{S}}(t_{k})
        \Bigsr]
    &= \frac{1}{4} \sqrt{\eta \kappa \Delta t}
            \Tr(\hat{\rho}_{\mathcal{S}} \hat{L}_{I})
        + \eta \kappa \Delta t
            \Tr(\hat{\rho}_{\mathcal{S}} \hat{L}_{I} \hat{L}_{Q}^2)
    = \mathcal{O}(\sqrt{\Delta t}),
\end{align}
but it's clear that terms containing $I_k^2$ and $Q_k^2$ are not negligible and should be examined carefully as $\Delta t \rightarrow 0$. 

In the diffusive limit, we introduce two random processes, $W_I(t)$ and $W_Q(t)$, related to $I_k$ and $Q_k$ by
\begin{gather} \label{eq:Wiener_I_def}
    I_k 
    = \sqrt{\eta \kappa \Delta t} 
            \Tr[ 
                \hat{\rho}_{\mathcal{S}}(t_k)
                \hat{L}_{I} (t_k)
            ]
        + \frac{\Delta W_{I}(t_k)}{2\sqrt{\Delta t}},
\\ \label{eq:Wiener_Q_def}
    Q_k 
    = \sqrt{\eta \kappa \Delta t} 
            \Tr[ 
                \hat{\rho}_{\mathcal{S}}(t_k)
                \hat{L}_{Q} (t_k)
            ]
        + \frac{\Delta W_{Q}(t_k)}{2\sqrt{\Delta t}},
\end{gather}
where $\Delta W_{I}(t_k) =  W_{I}(t_{k+1}) -  W_{I}(t_k)$ and $\Delta W_{Q}(t_k) =  W_{Q}(t_{k+1}) -  W_{Q}(t_k)$. From the moments of $I_k$ and $Q_k$, one can easily verify that, up to the first order in $\Delta t$, 
\begin{gather}
    \mathbb{E}\Bigsl[
            \Delta W_{I}(t_k) \, | \, 
            \hat{\rho}_{\mathcal{S}}(t_{k})
        \Bigsr]
    = \mathbb{E}\Bigsl[
            \Delta W_{Q}(t_k) \, | \, 
            \hat{\rho}_{\mathcal{S}}(t_{k})
        \Bigsr]
    = 0,
\\
    \mathbb{E}\Bigsl[
            (\Delta W_{I}(t_k))^2 \, | \, 
            \hat{\rho}_{\mathcal{S}}(t_{k})
        \Bigsr]
    = \mathbb{E}\Bigsl[
            (\Delta W_{Q}(t_k))^2 \, | \, 
            \hat{\rho}_{\mathcal{S}}(t_{k})
        \Bigsr]
    = \Delta t,
\\
    \mathbb{E}\Bigsl[
            \Delta W_{I}(t_k) \Delta W_{Q}(t_k) \, | \, 
            \hat{\rho}_{\mathcal{S}}(t_{k})
        \Bigsr]
    = 0.
\end{gather}
Hence, $W_{I}$ and $W_{Q}$ can be treated as two independent Wiener processes. Moreover, since the moments of the Wiener increments are independent of the qutrit state, we can drop the conditioning above. However, note that the qutrit state depends on the past trajectory of $W_{I}$ and $W_{Q}$, which is why $\hat{\rho}_{\mathcal{S}}$ should always be interpreted as the conditional states. Putting it differently, according to Eq.(\ref{eq:Wiener_I_def}) and (\ref{eq:Wiener_Q_def}), we notice that the history of $I$ and $Q$ are determined once we have specified a realization of $W_I$ and $W_Q$; therefore, we can generate the quantum trajectories of the qutrit during the dispersive measurement by simulating all possible realizations of $W_I$ and $W_Q$.

\begin{figure}[t]
    \centering
    \includegraphics[scale=0.3]{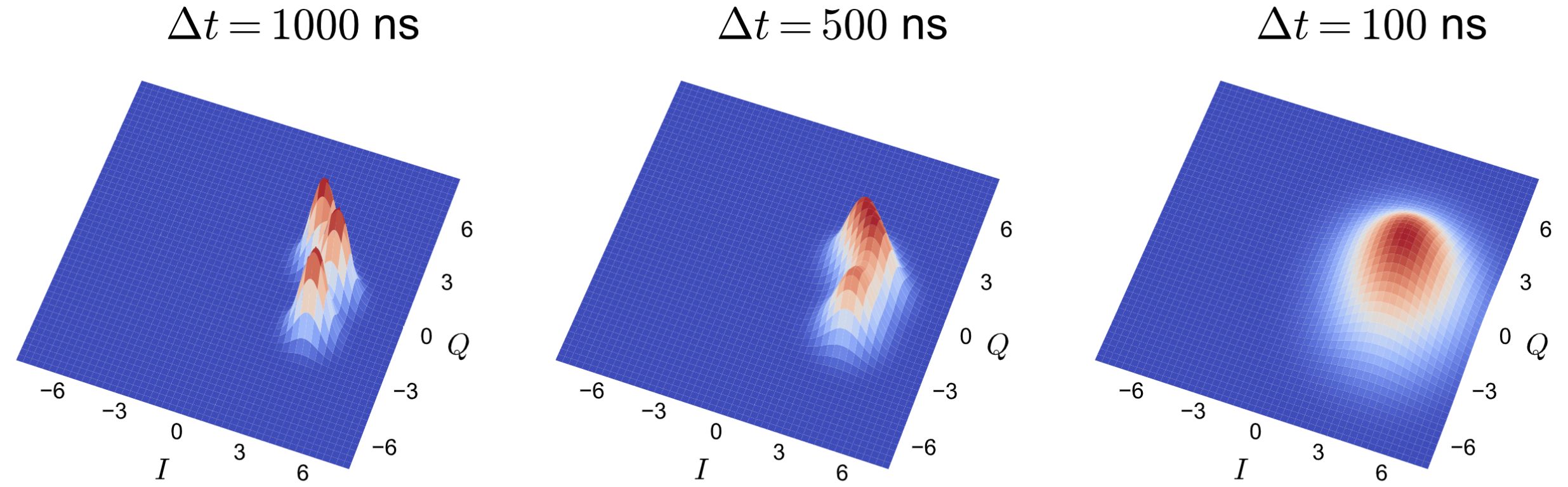}
    \caption{Illustration of a weak measurement in the diffusive limit. As the integration time of each sample reduces, the three Gaussian distributions merge and can be approximated by a single Gaussian. The mean vector of the approximated distribution is given by the centroid of the three original mean vectors (i.e., $(\Bar{I}_a, \Bar{Q}_a)$) weighted by the probability of the corresponding qutrit state $\ket{a}$.}
    \label{fig:diffusive_limit_measurement}
\end{figure}

With the preparation above, we are ready to derive the stochastic master equation in the diffusive limit from Eq.(\ref{eq:qutrit_state_update_rule}). In the limit as $\Delta t \rightarrow 0$, we can expand Eq.(\ref{eq:qutrit_state_update_rule}) to first order in $\Delta t$ with the caution that the Wiener processes follow It\^{o}'s rule, i.e., $(\Delta W_{I})^2 = (\Delta W_{Q})^2 = \Delta t$ and $\Delta W_{I} \Delta W_{Q} = 0$. In addition, the calculation is considerably simplified with the following observation: The Kraus operator
\begin{align}
    \hat{K}_{IQ}(t_k)
    &= \mathscr{N}_{k} 
        \sum_{a \in \{g,e,f\}} 
            \exp{- \Big[ I - \sqrt{\eta \kappa \Delta t} \, \bar{I}_a(t_k) \Big]^2 - \Big[ Q - \sqrt{\eta \kappa \Delta t} \, \bar{Q}_a(t_k) \Big]^2}
            \hat{\Pi}_a
\nonumber \\
    &\approx 
        \Tilde{\mathscr{N}}_{k} 
        \exp{
            - \left[
                    I - \sqrt{\eta \kappa \Delta t} 
                    \sum_{a}      
                        \bar{I}_a(t_k) 
                        \hat{\Pi}_a
                \right]^2 
            - \left[ 
                    Q - \sqrt{\eta \kappa \Delta t} 
                    \sum_{a} 
                        \bar{Q}_a(t_k)
                        \hat{\Pi}_a
                \right]^2
        }
\nonumber \\
    &=
        \Tilde{\mathscr{N}}_{k} 
        \exp{
            - \Big[
                    I - \sqrt{\eta \kappa \Delta t} 
                    \hat{L}_I(t_k)
                \Big]^2 
            - \Big[ 
                    Q - \sqrt{\eta \kappa \Delta t} 
                    \hat{L}_Q(t_k)
                \Big]^2
        }
\end{align}
to the first order in $\Delta t$. A graphical justification is provided in Figure \ref{fig:diffusive_limit_measurement}. Since $\Bigsl[\hat{L}_I, \hat{L}_Q\Bigsr] = 0$, we immediately have
\begin{equation}
    \hat{K}_{IQ}(t_k)
    \approx 
    \Tilde{\mathscr{N}}_{k} 
        \exp{
            - \Big[
                    I - \sqrt{\eta \kappa \Delta t} 
                    \hat{L}_I(t_k)
                \Big]^2 
        }
        \exp{
            - \Big[
                    Q - \sqrt{\eta \kappa \Delta t} 
                    \hat{L}_Q(t_k)
                \Big]^2
        },
\end{equation}
which is a great simplification and a verification that the values of $I$ and $Q$ are uncorrelated in the diffusive limit. Next, we replace $I$ and $Q$ using Eq.(\ref{eq:Wiener_I_def}) and (\ref{eq:Wiener_Q_def}) so that the Kraus operator at $t_k$ is implicitly fixed by a realization of $W_I$ and $W_Q$:
\begin{align}
    \hat{K}_{I_k Q_k}(t_k)
    &= \Tilde{\mathscr{N}}_{k} 
        \exp \Big[
                \sqrt{\eta \kappa} \hat{L}_I \Delta W_{I}(t_k)
                + 2 \eta \kappa \Delta t 
                    \Bigsl \langle
                        \hat{L}_{I}(t_k)
                    \Bigsr \rangle
                    \hat{L}_{I}(t_k)
                - \eta \kappa \Delta t \hat{L}_{I}^2(t_k)
\nonumber \\
    & \ \ \ \ \ \ \ \ \ \ \ \ \ \ \ \ \ \ \ \ \ \ \ \ 
                - \eta \kappa \Delta t \Bigsl \langle
                        \hat{L}_{I}(t_k)
                    \Bigsr \rangle^2
                - \sqrt{\Delta t} /2
            \Big]
\nonumber \\
    & \ \ \ \ \ \ \ \ \ \ \ \ 
        \exp \Big[
                \sqrt{\eta \kappa} \hat{L}_Q \Delta W_{Q}(t_k)
                + 2 \eta \kappa \Delta t 
                    \Bigsl \langle
                        \hat{L}_{Q}(t_k)
                    \Bigsr \rangle
                    \hat{L}_{Q}(t_k)
                - \eta \kappa \Delta t \hat{L}_{Q}^2(t_k)
\nonumber \\
    & \ \ \ \ \ \ \ \ \ \ \ \ \ \ \ \ \ \ \ \ \ \ \ \ \ \ \ \ 
                - \eta \kappa \Delta t \Bigsl \langle
                        \hat{L}_{Q}(t_k)
                    \Bigsr \rangle^2
                - \sqrt{\Delta t} /2
            \Big]
\nonumber \\
    &= \Tilde{N}_{k}
        \exp \Big[
                \sqrt{\eta \kappa} \hat{L}_I \Delta W_{I}(t_k)
                + 2 \eta \kappa \Delta t 
                    \Bigsl \langle
                        \hat{L}_{I}(t_k)
                    \Bigsr \rangle
                    \hat{L}_{I}(t_k)
                - \eta \kappa \Delta t \hat{L}_{I}^2(t_k)
        \Big]
\nonumber \\
    & \ \ \ \ \ \ \ \ \ \ \ \
        \exp \Big[
                \sqrt{\eta \kappa} \hat{L}_Q \Delta W_{Q}(t_k)
                + 2 \eta \kappa \Delta t 
                    \Bigsl \langle
                        \hat{L}_{Q}(t_k)
                    \Bigsr \rangle
                    \hat{L}_{Q}(t_k)
                - \eta \kappa \Delta t \hat{L}_{Q}^2(t_k)
        \Big],
\end{align}
where we have used It\^{o}'s rule and lumped all scalar terms into $\Tilde{N}_k$. We have also adopted the notation $\Bigsl \langle \hat{A} \Bigsr \rangle = \Tr\Bigsl[\hat{\rho}_{\mathcal{S}}(t) \hat{A}\Bigsr]$. In addition, note that we do not need to know the value of $\Tilde{N}_k$ because it appears in both the numerator and denominator of Eq.(\ref{eq:qutrit_state_update_rule}) and will thus be canceled.

At this point, there is not much simplification possible, but the math is also quite straightforward. After some algebra and several applications of the It\^{o}'s rule, we arrive at 
\begin{align}
    \hat{\rho}_{\mathcal{S}}(t_{k+1})
    &= \frac{\hat{K}_{I_k Q_k}(t_k) \hat{\rho}_{\mathcal{S}}(t_k) \hat{K}_{I_k Q_k}^{\dagger}(t_k)}{\Tr[\hat{K}_{I_k Q_k} (t_k)\hat{\rho}_{\mathcal{S}}(t_k) \hat{K}_{I_k Q_k}^{\dagger}(t_k)] }
\nonumber \\
    &= \hat{\rho}_{\mathcal{S}}(t_{k})
        + \eta \kappa
            \left[ 
                \hat{L}_{I}(t_k) 
                    \hat{\rho}_{\mathcal{S}}(t_k)  
                    \hat{L}_{I}(t_k)
                - \frac{1}{2} 
                    \hat{L}_{I}^2(t_k) 
                    \hat{\rho}_{\mathcal{S}}(t_k)
                - \frac{1}{2} 
                    \hat{\rho}_{\mathcal{S}}(t_k)
                    \hat{L}_{I}^2(t_k) 
            \right]
            \Delta t
\nonumber \\
    & \ \ \ \
        + \sqrt{\eta \kappa} 
            \left[ 
                \hat{L}_{I}(t_k) 
                    \hat{\rho}_{\mathcal{S}}(t_k) 
                + \hat{\rho}_{\mathcal{S}}(t_k)
                    \hat{L}_{I}(t_k) 
                - 2 \Bigsl \langle
                        \hat{L}_{I}(t_k)
                    \Bigsr \rangle 
                    \hat{\rho}_{\mathcal{S}}(t_k)
            \right]
            \Delta W_{I}(t_k)
\nonumber \\
    & \ \ \ \ \ \ \ \ 
        + \eta \kappa
            \left[ 
                \hat{L}_{Q}(t_k) \hat{\rho}_{\mathcal{S}}(t_k)
                    \hat{L}_{Q}(t_k)
                - \frac{1}{2} 
                    \hat{L}_{Q}^2(t_k) 
                    \hat{\rho}_{\mathcal{S}}(t_k)
                - \frac{1}{2} 
                    \hat{\rho}_{\mathcal{S}}(t_k)
                    \hat{L}_{Q}^2(t_k) 
            \right]
            \Delta t
\nonumber \\
    & \ \ \ \ \ \ \ \ \ \ \ \
        + \sqrt{\eta \kappa} 
            \left[ 
                \hat{L}_{Q}(t_k) 
                    \hat{\rho}_{\mathcal{S}}(t_k) 
                + \hat{\rho}_{\mathcal{S}}(t_k)
                    \hat{L}_{Q}(t_k) 
                - 2 \Bigsl \langle
                        \hat{L}_{Q}(t_k)
                    \Bigsr \rangle 
                    \hat{\rho}_{\mathcal{S}}(t_k)
            \right]
            \Delta W_{Q}(t_k)
\end{align}
to the first order in $\Delta t$. By replacing $\Delta t$ and $\Delta W(t)$ by differentials $\mathrm{d} t$ and $\mathrm{d} W(t)$, respectively, and let $\mathrm{d} \hat{\rho}_{\mathcal{S}}(t) = \hat{\rho}_{\mathcal{S}}(t+\Delta t) - \hat{\rho}_{\mathcal{S}}(t)$ as $\Delta t \rightarrow 0$, we finally obtain the stochastic master equation\footnote{The dependence of the stochastic quantities on time is shown in the parentheses since there are too many subscripts.}
\begin{align}
    \mathrm{d} \hat{\rho}_{\mathcal{S}}(t)
    &= \eta \kappa 
        \mathcalboondox{D}
                \Bigsl[
                    \hat{L}_{I}(t)
                \Bigsr]
            \hat{\rho}_{\mathcal{S}} (t) \mathrm{d} t
        + \eta \kappa 
        \mathcalboondox{D}
                \Bigsl[
                    \hat{L}_{Q}(t)
                \Bigsr]
            \hat{\rho}_{\mathcal{S}} (t) \mathrm{d} t
\nonumber \\
    &\ \ \ \ 
        + \sqrt{\eta \kappa} 
            \Big[
                    \hat{L}_{I}(t)
                        \hat{\rho}_{\mathcal{S}}(t)
                    + \hat{\rho}_{\mathcal{S}}(t)
                        \hat{L}_{I}(t)
                    - 2 \Bigsl \langle
                            \hat{L}_{I} (t)
                        \Bigsr \rangle
                        \hat{\rho}_{\mathcal{S}}(t)
                \Big]
                \mathrm{d} W_{I}(t)
\nonumber \\
    &\ \ \ \ \ \ \ \ 
        + \sqrt{\eta \kappa} 
            \Big[
                    \hat{L}_{Q}(t)
                        \hat{\rho}_{\mathcal{S}}(t)
                    + \hat{\rho}_{\mathcal{S}}(t)
                        \hat{L}_{Q}(t)
                    - 2 \Bigsl \langle
                            \hat{L}_{Q} (t)
                        \Bigsr \rangle
                        \hat{\rho}_{\mathcal{S}}(t)
                \Big]
                \mathrm{d} W_{Q}(t).
\end{align}
In practice, we do not have control over $W_I$ and $W_Q$; instead, we observe current/voltage-like quantities of the form
\begin{gather} 
    V_{I,k}
    \doteq \frac{2 I_k}{\sqrt{\Delta t}}
    = \sqrt{\eta \kappa} 
            \Bigsl \langle
                2 \hat{L}_{I} (t_k)
            \Bigsr \rangle
        + \frac{\Delta W_{I}(t_k)}{\Delta t},
\\ 
    V_{Q,k}
    \doteq \frac{2 Q_k}{\sqrt{\Delta t}}
    = \sqrt{\eta \kappa} 
            \Bigsl \langle
                2 \hat{L}_{Q} (t_k)
            \Bigsr \rangle
        + \frac{\Delta W_{Q}(t_k)}{\Delta t},
\end{gather}
which, in the continuous limit, become
\begin{gather} \label{eq:voltage_I_def}
    V_{I}(t)
    = \sqrt{\eta \kappa} 
            \Bigsl \langle
                2 \hat{L}_{I} (t)
            \Bigsr \rangle
        + \xi_{I}(t),
\\ \label{eq:voltage_Q_def}
    V_{Q}(t)
    = \sqrt{\eta \kappa} 
            \Bigsl \langle
                2 \hat{L}_{Q} (t)
            \Bigsr \rangle
        + \xi_{Q}(t),
\end{gather}
where $\xi_{I}(t) = \dot{W}_I(t)$ and $\xi_{Q}(t) = \dot{W}_Q(t)$ are classical white-noise signals defined by
\begin{gather}
    \mathbb{E}[\xi_{I}(t)] 
    = \mathbb{E}[\xi_{Q}(t)] 
    = \mathbb{E}[\xi_{I}(t)\xi_{Q}(t')] 
    = 0,
\\
    \mathbb{E}[\xi_{I}(t)\xi_{I}(t')]
    = \mathbb{E}[\xi_{Q}(t)\xi_{Q}(t')]
    = \delta(t-t').
\end{gather}
Since the ensemble average of the white noise is zero, Eq.(\ref{eq:voltage_I_def}) and (\ref{eq:voltage_Q_def}) explain why we can determine the qutrit state, which is encoded in $\Bigsl \langle \hat{L}_{I} (t) \Bigsr \rangle$ and $\Bigsl \langle \hat{L}_{Q} (t) \Bigsr \rangle$, by making measurement of $V_I$ and $V_Q$.

There is still one detail to be corrected, which is the fact that we have set $\eta = 1$ in the above derivation. Following the standard treatment of detection inefficiency (i.e., $0 \leq \eta < 1$), we introduce two new Wiener processes $W_{I}'(t)$ and $W_{Q}'(t)$ such that the efficiencies associated with them are both set to be $(1-\eta)$. In addition, the four Wiener processes should be independent (think of $(W_I, W_I')$ and $(W_Q, W_Q')$ as two Poisson branching processes but in the diffusive limit). Consequently, the stochastic master equation has four stochastic terms
\begin{align}
    \mathrm{d} \hat{\rho}_{\mathcal{S}}(t)
    &= \eta \kappa 
        \mathcalboondox{D}
                \Bigsl[
                    \hat{L}_{I}(t)
                \Bigsr]
            \hat{\rho}_{\mathcal{S}} (t) \mathrm{d} t
        + (1-\eta) \kappa 
        \mathcalboondox{D}
                \Bigsl[
                    \hat{L}_{I}(t)
                \Bigsr]
            \hat{\rho}_{\mathcal{S}} (t) \mathrm{d} t
\nonumber \\
    & \ \ \ \ 
        + \eta \kappa 
        \mathcalboondox{D}
                \Bigsl[
                    \hat{L}_{Q}(t)
                \Bigsr]
            \hat{\rho}_{\mathcal{S}} (t) \mathrm{d} t
        + (1-\eta) \kappa 
        \mathcalboondox{D}
                \Bigsl[
                    \hat{L}_{Q}(t)
                \Bigsr]
            \hat{\rho}_{\mathcal{S}} (t) \mathrm{d} t
\nonumber \\
    &\ \ \ \ \ \ \ \ 
        + \sqrt{\eta \kappa} 
            \Big[
                    \hat{L}_{I}(t)
                        \hat{\rho}_{\mathcal{S}}(t)
                    + \hat{\rho}_{\mathcal{S}}(t)
                        \hat{L}_{I}(t)
                    - 2 \Bigsl \langle
                            \hat{L}_{I} (t)
                        \Bigsr \rangle
                        \hat{\rho}_{\mathcal{S}}(t)
                \Big]
                \mathrm{d} W_{I}(t)
\nonumber \\
    &\ \ \ \ \ \ \ \ \ \ \ \ 
        + \sqrt{\eta \kappa} 
            \Big[
                    \hat{L}_{Q}(t)
                        \hat{\rho}_{\mathcal{S}}(t)
                    + \hat{\rho}_{\mathcal{S}}(t)
                        \hat{L}_{Q}(t)
                    - 2 \Bigsl \langle
                            \hat{L}_{Q} (t)
                        \Bigsr \rangle
                        \hat{\rho}_{\mathcal{S}}(t)
                \Big]
                \mathrm{d} W_{Q}(t)
\nonumber \\
    &\ \ \ \ \ \ \ \ \ \ \ \ \ \ \ \ 
        + \sqrt{(1-\eta) \kappa} 
            \Big[
                    \hat{L}_{I}(t)
                        \hat{\rho}_{\mathcal{S}}(t)
                    + \hat{\rho}_{\mathcal{S}}(t)
                        \hat{L}_{I}(t)
                    - 2 \Bigsl \langle
                            \hat{L}_{I} (t)
                        \Bigsr \rangle
                        \hat{\rho}_{\mathcal{S}}(t)
                \Big]
                \mathrm{d} W_{I}(t)
\nonumber \\
    &\ \ \ \ \ \ \ \ \ \ \ \ \ \ \ \ \ \ \ \ 
        + \sqrt{(1-\eta) \kappa} 
            \Big[
                    \hat{L}_{Q}(t)
                        \hat{\rho}_{\mathcal{S}}(t)
                    + \hat{\rho}_{\mathcal{S}}(t)
                        \hat{L}_{Q}(t)
                    - 2 \Bigsl \langle
                            \hat{L}_{Q} (t)
                        \Bigsr \rangle
                        \hat{\rho}_{\mathcal{S}}(t)
                \Big]
                \mathrm{d} W_{Q}'(t).
\end{align}
where the information encoded in the last two terms is assumed to be lost in the measurement; thus, marginalizing $W_{I}'(t)$ and $W_{Q}'(t)$ yields
\begin{align}
    \mathrm{d} \hat{\rho}_{\mathcal{S}}
    &= \kappa 
        \mathcalboondox{D}
                \Bigsl[
                    \hat{L}_{I}
                \Bigsr]
            \hat{\rho}_{\mathcal{S}} \mathrm{d} t
        + \kappa 
        \mathcalboondox{D}
                \Bigsl[
                    \hat{L}_{Q}
                \Bigsr]
            \hat{\rho}_{\mathcal{S}} \mathrm{d} t
\nonumber \\
        & \ \ \ \ 
        + \sqrt{\eta \kappa} 
            \Big[
                    \hat{L}_{I}
                        \hat{\rho}_{\mathcal{S}}
                    + \hat{\rho}_{\mathcal{S}}
                        \hat{L}_{I}
                    - 2 \Bigsl \langle
                            \hat{L}_{I}
                        \Bigsr \rangle
                        \hat{\rho}_{\mathcal{S}}
                \Big]
                \mathrm{d} W_{I}
        + \sqrt{\eta \kappa} 
            \Big[
                    \hat{L}_{Q}
                        \hat{\rho}_{\mathcal{S}}
                    + \hat{\rho}_{\mathcal{S}}
                        \hat{L}_{Q}
                    - 2 \Bigsl \langle
                            \hat{L}_{Q}
                        \Bigsr \rangle
                        \hat{\rho}_{\mathcal{S}}
                \Big]
                \mathrm{d} W_{Q}.
\end{align}
Note that despite reducing the amount of information gained in an imperfect measurement, the qubit still dephases as if the efficiency is 1.

So far, we have been ignoring all the other decoherence channels of the qutrit. We can generalize the stochastic master equation by simply adding the qutrit decay and pure dephasing terms; however, at first glance, the measurement-induced dephasing rates defined in Eq.(\ref{eq:measurement_induced_dephasing_definition}) seem to be missed by the stochastic master equation. In fact, $\kappa \mathcalboondox{D} \Bigsl[ \hat{L}_{I} \Bigsr] \hat{\rho}_{\mathcal{S}} + \kappa \mathcalboondox{D} \Bigsl[ \hat{L}_{Q} \Bigsr] \hat{\rho}_{\mathcal{S}}$ did not show up in the any of the master equations introduced before, at least not obvious in its current form. Nevertheless, one can show that
\begin{align}
    & \kappa 
        \mathcalboondox{D}
                \Bigsl[
                    \hat{L}_{I}
                \Bigsr]
            \hat{\rho}_{\mathcal{S}} 
        + \kappa 
        \mathcalboondox{D}
                \Bigsl[
                    \hat{L}_{Q}
                \Bigsr]
            \hat{\rho}_{\mathcal{S}}
\nonumber\\
    &= \frac{\kappa |\beta_{ge}|^2}{4} 
            \mathcalboondox{D}
                \Bigsl[
                    \hat{\sigma}_{z,ge}
                \Bigsr]
            \hat{\rho}_{\mathcal{S}} 
        + \frac{\kappa |\beta_{gf}|^2}{4} 
            \mathcalboondox{D}
                \Bigsl[
                    \hat{\sigma}_{z,gf}
                \Bigsr]
            \hat{\rho}_{\mathcal{S}} 
        + \frac{\kappa |\beta_{ef}|^2}{4} 
            \mathcalboondox{D}
                \Bigsl[
                    \hat{\sigma}_{z,ef}
                \Bigsr]
            \hat{\rho}_{\mathcal{S}} 
\nonumber\\
    &= \frac{\Gamma_{\text{m}, ge}}{4} 
                \mathcalboondox{D}
                    \Bigsl[
                        \hat{\sigma}_{z,ge}
                    \Bigsr]
                \hat{\rho}_{\mathcal{S}} 
            + \frac{\Gamma_{\text{m}, gf}}{4} 
                \mathcalboondox{D}
                    \Bigsl[
                        \hat{\sigma}_{z,gf}
                    \Bigsr]
                \hat{\rho}_{\mathcal{S}} 
            + \frac{\Gamma_{\text{m}, ef}}{4} 
                \mathcalboondox{D}
                    \Bigsl[
                        \hat{\sigma}_{z,ef}
                    \Bigsr]
                \hat{\rho}_{\mathcal{S}} 
\end{align}
by a simple rearrangement of the dissipation superoperators. Thus, by including the self-evolution of the qutrit and all the other Lindbald operators, we obtain
\begin{align}
    \mathrm{d} \hat{\rho}
    &= \bigg(
        - \frac{\ci}{\hbar}
            \Big[ 
                \hat{H}_{\text{q,eff}}, 
                \hat{\rho}
            \Big] 
        + \gamma_{1,ge} 
            \mathcalboondox{D}
                \Bigsl[
                    \hat{\sigma}_{ge}
                \Bigsr] 
            \hat{\rho} 
        + \gamma_{1,gf} 
            \mathcalboondox{D}
                \Bigsl[
                    \hat{\sigma}_{gf}
                \Bigsr] 
            \hat{\rho} 
        + \gamma_{1,ef} 
            \mathcalboondox{D}
                \Bigsl[
                    \hat{\sigma}_{ef}
                \Bigsr] 
            \hat{\rho} 
\nonumber \\
    & \ \ \ \ \ \ \ \ 
        + \frac{\gamma_{\phi,ge}}{2} 
            \mathcalboondox{D}
                \Bigsl[
                    \hat{\sigma}_{z,ge}
                \Bigsr]
            \hat{\rho}
        + \frac{\gamma_{\phi,gf}}{2} 
            \mathcalboondox{D}
                \Bigsl[
                    \hat{\sigma}_{z,gf}
                \Bigsr]
            \hat{\rho}
        + \frac{\gamma_{\phi,ef}}{2} 
            \mathcalboondox{D}
                \Bigsl[
                    \hat{\sigma}_{z,ef}
                \Bigsr]
            \hat{\rho}
        \bigg)
        \mathrm{d} t
\nonumber \\
    &  \ \ \ \ 
        + \left(
            \frac{\Gamma_{\text{m}, ge}}{4} 
                \mathcalboondox{D}
                    \Bigsl[
                        \hat{\sigma}_{z,ge}
                    \Bigsr]
                \hat{\rho} 
            + \frac{\Gamma_{\text{m}, gf}}{4} 
                \mathcalboondox{D}
                    \Bigsl[
                        \hat{\sigma}_{z,gf}
                    \Bigsr]
                \hat{\rho} 
            + \frac{\Gamma_{\text{m}, ef}}{4} 
                \mathcalboondox{D}
                    \Bigsl[
                        \hat{\sigma}_{z,ef}
                    \Bigsr]
                \hat{\rho} 
        \right)
        \mathrm{d} t
\nonumber\\
        &  \ \ \ \ \ \ \ \ 
            + \sqrt{\eta \kappa} 
            \mathcalboondox{M}
                    \Bigsl[
                        \hat{L}_{I}
                    \Bigsr]
                \hat{\rho} \,
                \mathrm{d} W_{I}
        + \sqrt{\eta \kappa} 
            \mathcalboondox{M}
                    \Bigsl[
                        \hat{L}_{Q}
                    \Bigsr]
                \hat{\rho} \,
                \mathrm{d} W_{Q},
\end{align}
where the measurement superoperator $\mathcalboondox{M}\Bigsl[\hat{L}\Bigsr]$ associated with an operator $\hat{L}$ is defined via
\begin{equation}
    \mathcalboondox{M}
        \Bigsl[
            \hat{L}
        \Bigsr]
    \hat{\rho}
    = \hat{L}
            \hat{\rho}
        + \hat{\rho}
            \hat{L}^{\dagger}
        - \Tr(
                \hat{L}
                \hat{\rho}
                + \hat{\rho}
                \hat{L}^{\dagger}
            )
            \hat{\rho}
\end{equation}
and the subscript $\mathcal{S}$ is dropped with the understanding that $\hat{\rho}$ represents the conditional density operator of the qutrit only.

\begin{figure}
    \centering
    \includegraphics[scale=0.28]{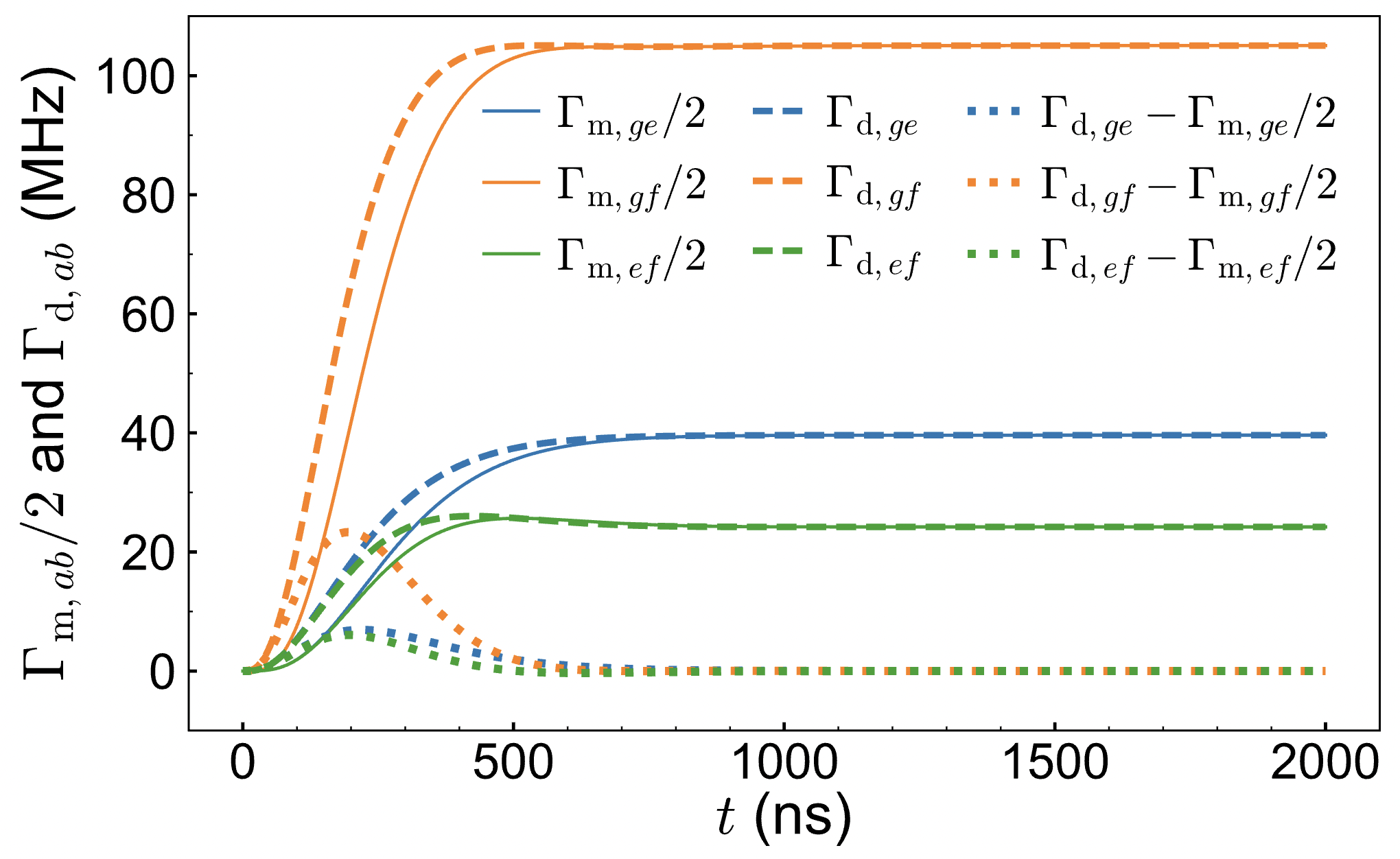}
    \caption{A comparison of the measurement dephasing rates derived in the laboratory frame and the measurement rates found in the displaced frame. Since we have assumed a Markovian system in the derivation of the SME, the measurement rate and measurement-induced dephasing rates are not exactly the same in the transient parts. However, when the system reaches its steady state, the two rates are equivalent.}
    \label{fig:dephasing_rate_comparison_plot}
\end{figure}

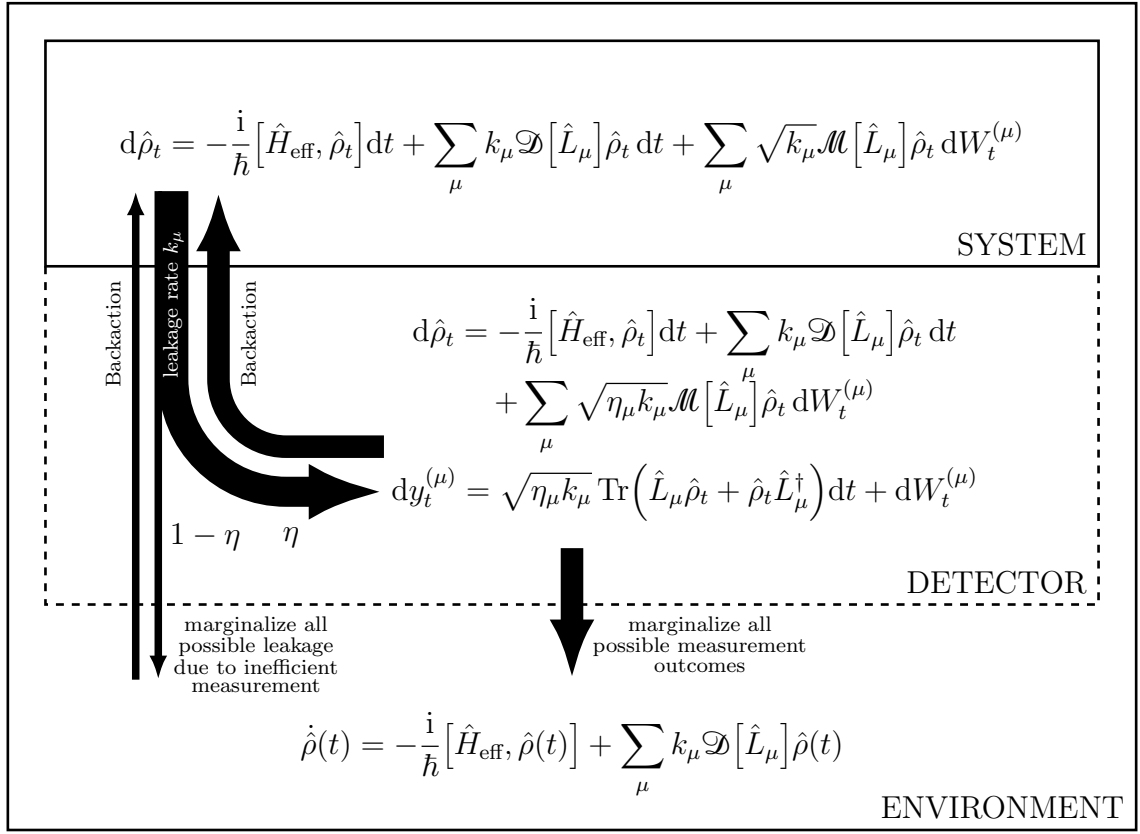
\begin{figure}[t]
    \centering
    \begin{tikzpicture}
    \draw 
        (0,3) to (0,-1) to (15,-1) to (15,10) to (0,10) to (0,3);
    \draw 
        (0.5,8) to (0.5,6.5) to (14.5,6.5) to (14.5,9.5) to (0.5,9.5) to (0.5,8);
    \draw[dashed]
        (0.5,6.5) to (0.5,2) to (14.5,2) to (14.5,6.5);
    \draw 
        (14.5, 6.8) node[left]{${\text{SYSTEM}}$}
        (14.5, 2.3) node[left]{${\text{DETECTOR}}$}
        (15, -0.7) node[left]{${\text{ENVIRONMENT}}$};
    \draw 
        (3.5, 2.9) node[right]{${\displaystyle \eta}$}
        (2, 2.9) node[right]{${\displaystyle 1 - \eta}$}
        (1.4, 5.6) node[]{\rotatebox{90}{${\scriptstyle \text{Backaction}}$}}
        (3.2, 5.6) node[]{\rotatebox{90}{${\scriptstyle \text{Backaction}}$}};
    \draw 
        (9.2, 2) node[below]{${\substack{\text{marginalize all} \\ \text{possible measurement} \\ \text{outcomes}}}$}
        (3.3, 2) node[below]{${\substack{\text{marginalize all} \\ \text{possible leakage} \\ \text{due to  inefficient} \\ \text{measurement}}}$};
    \draw 
        (7.5,8) node[]{${ \displaystyle
            \mathrm{d} \hat{\rho}_t
            = - \frac{\ci}{\hbar} 
                \Bigsl[ 
                    \hat{H}_{\text{eff}}, \hat{\rho}_t
                \Bigsr] \mathrm{d} t
                + \sum_{\mu}
                    k_{\mu} 
                    \mathcalboondox{D}
                        \Bigsl[ 
                            \hat{L}_{\mu} 
                        \Bigsr] 
                    \hat{\rho}_t 
                    \, \mathrm{d} t
                + \sum_{\mu} \sqrt{k_{\mu}} 
                    \mathcalboondox{M}
                        \Bigsl[ 
                            \hat{L}_{\mu} 
                        \Bigsr] 
                    \hat{\rho}_t
                    \, \mathrm{d} W^{(\mu)}_t
        }$}
        (9,5.5) node[]{${ \displaystyle
            \mathrm{d} \hat{\rho}_t
            = - \frac{\ci}{\hbar} 
                    \Bigsl[ 
                        \hat{H}_{\text{eff}}, \hat{\rho}_t
                    \Bigsr] 
                    \mathrm{d} t
                + \sum_{\mu}
                    k_{\mu} 
                    \mathcalboondox{D} 
                        \Bigsl[ 
                            \hat{L}_{\mu} 
                        \Bigsr] 
                    \hat{\rho}_t 
                    \, \mathrm{d} t
        }$}
        (9,4.5) node[]{${ \displaystyle
                + \sum_{\mu} \sqrt{\eta_{\mu} k_{\mu}} 
                    \mathcalboondox{M}
                        \Bigsl[ 
                            \hat{L}_{\mu} 
                        \Bigsr] 
                    \hat{\rho}_t
                    \, \mathrm{d} W^{(\mu)}_t
        }$}
        (9,3.5) node[]{${
            \mathrm{d} y^{(\mu)}_{t} 
            = \sqrt{\eta_{\mu} k_{\mu}} 
                \Tr(
                    \hat{L}_{\mu} \hat{\rho}_t
                    + \hat{\rho}_t \hat{L}_{\mu}^{\dagger}
                ) \mathrm{d} t 
                + \mathrm{d} W^{(\mu)}_t
        }$}
        (7.5,0) node[]{${ \displaystyle
            \dot{\hat{\rho}}(t)
            = - \frac{\ci}{\hbar} 
                \Bigsl[ 
                    \hat{H}_{\text{eff}}, \hat{\rho}(t)
                \Bigsr]
                + \sum_{\mu}
                    k_{\mu} 
                    \mathcalboondox{D} 
                        \Bigsl[ 
                            \hat{L}_{\mu} 
                        \Bigsr] 
                    \hat{\rho}(t)
        }$};
    \draw[line width=1mm, -{Latex[length=3.2mm]}] 
        (2, 7.5) to (2, 1);
    \draw[line width=1mm, -{Latex[length=3.2mm]}] 
        (1.7, 1) to (1.7, 7.5);
    \draw[line width=4mm, -{Latex[length=10mm]}] 
        (2.2, 7.5) to (2.2, 5) arc(180:270:1.5 and 1.5) to (5,3.5);
    \draw[color=white] (2.2, 6) node[]{\rotatebox{90}{${\scriptstyle \text{leakage rate } k_{\mu}}$}};
    \draw[line width=3mm, {Latex[length=8mm]}-] 
        (2.8, 7.5) to (2.8, 5) arc(180:270:0.9 and 0.9) to (5,4.1);
    \draw[line width=3mm, -{Latex[length=8mm]}] 
        (7.5, 2.75) to (7.5, 1);
    \end{tikzpicture}
    \caption{Open quantum system modeled with a diffusive random process. By ignoring a portion of the information lost in the environment, we can reduce a fully stochastic time evolution to a partially stochastic one by introducing the detection efficiency $\eta$. Further marginalization of the measured information results in the unconditioned master equation.}
    \label{fig:open_quantum_system_summary}
\end{figure}

Even though we have reproduced something that appeared before in the displaced frame, i.e., $\Gamma_{\text{m}, ab}$, we have still not derived the exact measurement-induced dephasing rates $\Gamma_{\text{d}, ab}$ that appear in the unconditioned master equation of the qutrit. However, if we use the expression for $\alpha_g$, $\alpha_e$, and $\alpha_f$ at steady state, i.e.,
\begin{align}
    \alpha_g (+\infty)
    &= \frac{\epsilon}{\Delta_{\text{rd}} - \ci \kappa/2},
\ \ 
    \alpha_e (+\infty)
    &= \frac{\epsilon}{\Delta_{\text{rd}} + \chi_{\text{qr}} - \ci \kappa/2},
\ \ 
    \alpha_f (+\infty)
    &= \frac{\epsilon}{\Delta_{\text{rd}} + 2\chi_{\text{qr}} - \ci \kappa/2},
\end{align}
we can show that
\begin{align}
    \Gamma_{\text{m}, ge} (+\infty)
    &= 2 \Gamma_{\text{d}, ge}(+\infty)
    = \frac{\kappa |\epsilon|^2 \chi_{\text{qr}}^2}{[\Delta_{\text{rd}}^2 + (\kappa/2)^2][(\Delta_{\text{rd}} + \chi_{\text{qr}})^2 + (\kappa/2)^2]},
\\
    \Gamma_{\text{m}, gf} (+\infty)
    &= 2 \Gamma_{\text{d}, gf}(+\infty)
    = \frac{\kappa |\epsilon|^2 \chi_{\text{qr}}^2}{[\Delta_{\text{rd}}^2 + (\kappa/2)^2][(\Delta_{\text{rd}} + 2\chi_{\text{qr}})^2 + (\kappa/2)^2]},
\\
    \Gamma_{\text{m}, ef} (+\infty)
    &= 2 \Gamma_{\text{d}, ef}(+\infty)
    = \frac{\kappa |\epsilon|^2 \chi_{\text{qr}}^2}{[(\Delta_{\text{rd}} + \chi_{\text{qr}})^2 + (\kappa/2)^2][(\Delta_{\text{rd}} + 2\chi_{\text{qr}})^2 + (\kappa/2)^2]},
\end{align}
and, thus, at steady state, the heterodyne measurement indeed induces a dephasing at rate $\Gamma_{\text{d}, ab}$ between two energy levels of the qutrit. Figure \ref{fig:dephasing_rate_comparison_plot} provides a detailed comparison between $\Gamma_{\text{m}, ge}$ and $2\Gamma_{\text{d}, ge}$ before reaching the steady state. Such a discrepancy can exist since we have been assuming a Markovian system; from the derivation of the effective qutrit master equation, however, we know that the qutrit alone is not really Markovian. Nevertheless, by replacing $\Gamma_{\text{m}, ab}(t)$ by $2\Gamma_{\text{d}, ab}(+\infty)$ at steady state, we arrive at an effective qutrit stochastic master equation
\begin{align}
    \mathrm{d} \hat{\rho}
    &= \bigg(
        - \frac{\ci}{\hbar}
            \Big[ 
                \hat{H}_{\text{q,eff}}, 
                \hat{\rho}
            \Big] 
        + \gamma_{1,ge} 
            \mathcalboondox{D}
                \Bigsl[
                    \hat{\sigma}_{ge}
                \Bigsr] 
            \hat{\rho} 
        + \gamma_{1,gf} 
            \mathcalboondox{D}
                \Bigsl[
                    \hat{\sigma}_{gf}
                \Bigsr] 
            \hat{\rho} 
        + \gamma_{1,ef} 
            \mathcalboondox{D}
                \Bigsl[
                    \hat{\sigma}_{ef}
                \Bigsr] 
            \hat{\rho} 
\nonumber \\
    &  \ \ \ \ \ \ \ \ 
        + \frac{\gamma_{\phi,ge} + \Gamma_{\text{d},ge}}{2} 
            \mathcalboondox{D}
                \Bigsl[
                    \hat{\sigma}_{z,ge}
                \Bigsr]
            \hat{\rho}
        + \frac{\gamma_{\phi,gf}+ \Gamma_{\text{d},gf}}{2} 
            \mathcalboondox{D}
                \Bigsl[
                    \hat{\sigma}_{z,gf}
                \Bigsr]
            \hat{\rho}
        + \frac{\gamma_{\phi,ef}+ \Gamma_{\text{d},ef}}{2} 
            \mathcalboondox{D}
                \Bigsl[
                    \hat{\sigma}_{z,ef}
                \Bigsr]
            \hat{\rho}
        \bigg)
        \mathrm{d} t
\nonumber\\ \label{eq:sme_final_form}
    &  \ \ \ \ 
        + \sqrt{\eta \kappa} 
            \mathcalboondox{M}
                    \Bigsl[
                        \hat{L}_{I}
                    \Bigsr]
                \hat{\rho} \,
                \mathrm{d} W_{I}
        + \sqrt{\eta \kappa} 
            \mathcalboondox{M}
                    \Bigsl[
                        \hat{L}_{Q}
                    \Bigsr]
                \hat{\rho} \,
                \mathrm{d} W_{Q}.
\end{align}
By taking the ensemble average of the stochastic master equation, we indeed reproduce the effective qutrit master equation. As a summary, Figure \ref{fig:open_quantum_system_summary} illustrates the relationship between all the master equations we have encountered.

\subsection{Simulation and Experimental Verification}
Eq.(\ref{eq:sme_final_form}) can be simulated using finite difference just like a normal differential equation. However, since the measurement is random, each finite-difference simulation of Eq.(\ref{eq:sme_final_form}) must be accompanied by a specific realization of $W_{I}$ and $W_{Q}$. In other words, Eq.(\ref{eq:sme_final_form}) can only be simulated in the Monte-Carlo sense. In addition, since a finite-difference simulation discretizes the time into steps of size $\Delta t$, we need to draw $\Delta W_I$ and $\Delta W_Q$ from two independent Gaussian distributions both with mean zero and variance $\Delta t$\footnote{Recall that for a standard Wiener process $W_t$, $W_{t+\tau} - W_t \sim \mathcal{N}(0,\tau)$.}.

\begin{figure}[t]
    \centering
    \includegraphics[scale=0.33]{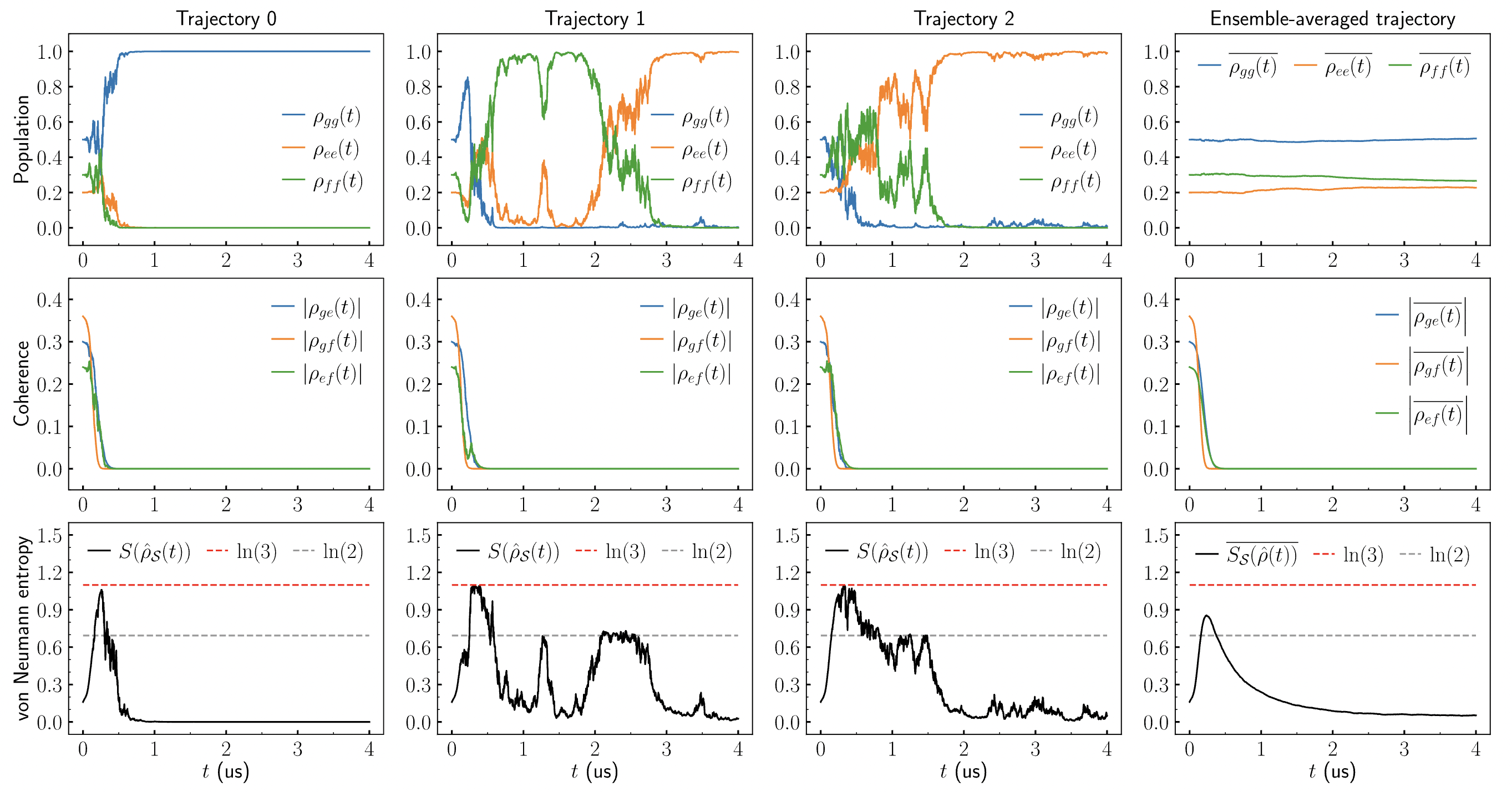}
    \caption{Monte Carlo simulation of the qutrit stochastic master equation with an initial state given by Eq.(\ref{eq:sme_sim_initial_state}). Three sample trajectories are shown in the first three columns. Then, a thousand trajectories are averaged to produce the last column. The first two rows plot the matrix elements of the qutrit density operator. The last row shows the von Neumann entropy $S(\hat{\rho}) = - \Tr(\hat{\rho} \ln \hat{\rho})$, which peaks when the coherence drops to zero. This is because the measurement-induced dephasing will first erase the phase information in a superposition state; however, since the state will be steered towards one of the energy eigenstates, the uncertainty drops to zero eventually.}
    \label{fig:sme_MC_sim_plot}
\end{figure}

For example, consider an arbitrary qutrit state  
\begin{equation} \label{eq:sme_sim_initial_state}
    \hat{\rho}(0)
    = \begin{pmatrix}
        0.5  & 0.3 & 0.36 \\[1mm] 
        0.3  & 0.2 & 0.24 \\[1mm] 
        0.36 &0.24 & 0.3
    \end{pmatrix}.
\end{equation}
If we repeat the simulation with this initial state a thousand times, we will obtain a thousand distinct quantum trajectories of the qutrit. Three sample trajectories are shown in Figure \ref{fig:sme_MC_sim_plot}, along with the corresponding von Neumann entropy. The last column of the figure gives the sample-averaged state and entropy. Indeed, the sample-averaged population and coherence agree with the ensemble-averaged ones shown in Figure \ref{fig:cavity_amplitude_plot_simulations}, verifying that the unconditioned master equation is the expectation of the stochastic master equation.

\begin{figure}[t!]
    \centering
    \includegraphics[scale=0.26]{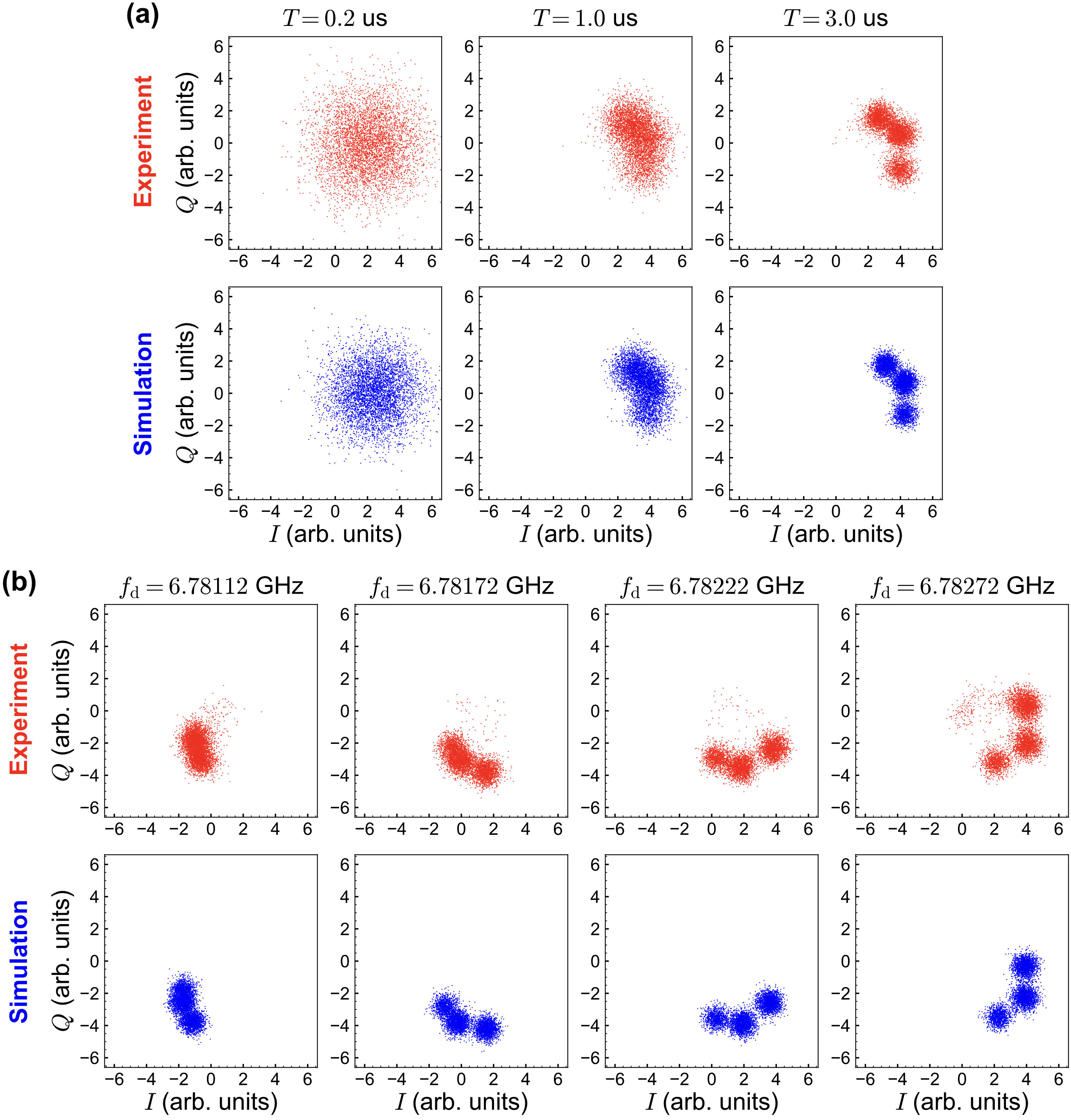}
    \caption{Comparison between the simulation and experiment. The total cavity decay rate is measured to be $\kappa/2\pi = 2.7$ MHz and the dispersive shift is $\chi_{\text{qr}}/2\pi = 0.6$ MHz. The same parameters are used in the simulation with the measurement efficiency set to $\eta = 0.04$ (the experiment is performed without a quantum-limited amplifier). The quadrature signals are collected by AlazarTech digitizer with a sampling frequency of $1$ GHz, which can be treated at $1/\Delta t$ in the derivation of the diffusive SME. \textbf{a.} Changing the measurement time $T$. \textbf{b.} Sweeping the readout frequency $f_{\text{d}}$ (i.e., the cavity drive frequency).}
    \label{fig:sme_sim_exp_comparison}
\end{figure}

What is not obvious from the unconditioned master equation is the convergence of the qutrit state to one of the energy eigenstates, as shown in the first row of Figure \ref{fig:sme_MC_sim_plot}. Although the qutrit is measured continuously and weakly, each infinitesimal measurement can push the qutrit to a new state; consequently, due to the nature of the dispersive coupling, the qutrit will slowly converge to the pointer states $\ket{g}$, $\ket{e}$, and $\ket{f}$.

\begin{figure}
    \centering
    \includegraphics[scale=0.28]{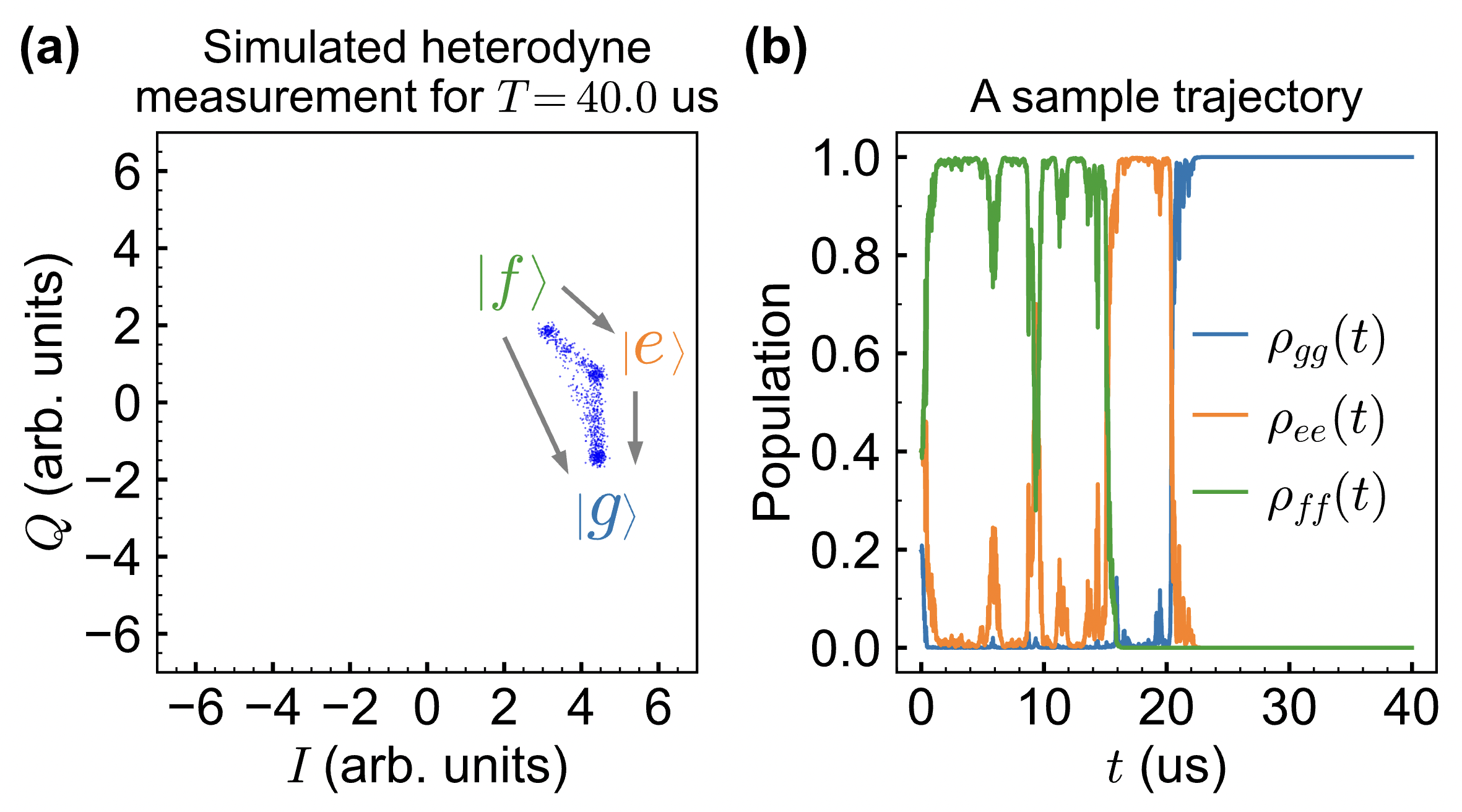}
    \caption{Effect of taking a long measurement. A thousand random quantum trajectories have been simulated with a measurement time $T=40 \text{ us}$ and with the qutrit parameters $1/\gamma_{1,ge} =1/\gamma_{1,ef} = 35 \text{ us}$ and $1/\gamma_{1,gf} = 1 \text{ ms}$. \textbf{a}. For each quantum trajectory, the heterodyne measurement outcomes $V_I(t)$ and $V_Q(t)$ are averaged over the measurement time $T$ and are plotted as a point in the $I-Q$ plane. Besides the three Gaussian clusters expected from a QND measurement, we also observe streams of points connecting the three clusters, representing the population's leakage from the excited states to the ground states. \textbf{b}. When the measurement time is on the order of the decay time, the random process exhibits a jumping characteristic on the large time scale.}
    \label{fig:long_measurement_time_effect}
\end{figure}

In a real dispersive measurement, what we have access to are the transmitted signals $V_I(t)$ and $V_Q(t)$. By calculating the time averages of $V_I(t)$ and $V_Q(t)$, denoted by $\bar{V}_I$ and $\bar{V}_Q$, we obtain a complex number $\bar{V}_I + \ci \bar{V}_Q$, which can be plotted on a phase plane. Repeating the experiment
a large number of times generates a scattering plot similar to Figure \ref{fig:phase_plane_coherent_states_for_dispersive_measurement}. One can then compare the simulated $\bar{V}_I + \ci \bar{V}_Q$ with the one measured in experiments to verify the correctness of the stochastic master equation. In Figure \ref{fig:sme_sim_exp_comparison}, two sets of comparisons are made: (a) We vary the measurement time to see the effect of the shot noise. As mentioned before, increasing the measurement time allows one to effectively obtain a larger signal-to-noise ratio and thus suppress the classification error. (b) We sweep the readout frequency with a fixed readout time and observe the change in the amplitude and phase of the (time-averaged) transmitted signal. As illustrated in Figure \ref{fig:cavity_amplitude_plot_simulations}(b), the center of each Gaussian cluster lies on a circle that goes through the center. More critically, we can now find an optimal readout frequency corresponding to a large separation between the clusters, thus giving the best state classification.

We can gain insight into other aspects of the measurement process from the stochastic master equation. As one final example, recall that the qutrit can also decay due to the Lindblad operators $\hat{\sigma}_{ab}$ while the measurement is happening. As shown in Figure \ref{fig:long_measurement_time_effect}, if we set the measurement time to be longer than $1/\gamma_{1,ab}$, we would observe the shift of population in the phase plane. In other words, the sample-averaged populations, i.e., $\rho_{gg}$, $\rho_{ee}$, and $\rho_{ff}$, are no longer a constant as shown in Figure \ref{fig:sme_MC_sim_plot}. Hence, in practice, we cannot simply improve the signal-to-noise ratio by increasing the measurement time arbitrarily. 

In conclusion, the qutrit unconditioned and stochastic master equations allow one to understand quantitatively the dispersive measurement used in superconducting quantum computation. We can simulate the expected measurement outcome to guide our design of the qudit, the cavity, and their coupling strength. Moreover, the results demonstrated for a qutrit can be easily generalized to a qudit with an arbitrary number of energy levels. To further enhance our understanding, it is worthwhile to investigate the non-Markovian nature of the system and detail the model of the noise around the qudit, thus enabling more accurate prediction of quantum measurement in realistic scenarios.

\newpage

%% file: include_appendices/appendix_Fourier.tex
\chapter{Useful Formulae and Fourier-Transform Pairs}\label{appendix:fourier_properties}

\begin{enumerate}
    \item Parseval's identity
    \begin{equation}
        \int 
            \mathrm{d}^3 r \, 
            F^{*}(\mathbf{r}) G(\mathbf{r})
        = \int 
            \mathrm{d}^3 k \, 
            \mathscr{F}^{*}(\mathbf{k}) 
            \mathscr{G}(\mathbf{k})
    \end{equation}
    
    \item Conjugate-symmetry of real functions: If $\mathbf{E}(\mathbf{r},t)$ is real, then
    \begin{equation}
        \mathscr{E}(-\mathbf{k},t) 
        = \mathscr{E}^{*}(\mathbf{k},t)
    \end{equation}
    
    \item Fourier-transform pairs
    \begin{align} \label{eq: S1_Appendix_1A_FT_1}
        \delta(\mathbf{r} - \mathbf{r}_0)
        & \longleftrightarrow
        \frac{1}{(2 \pi)^{3/2}} e^{-\ci\mathbf{k} \cdot \mathbf{r}_0}
    \\ \label{eq: S1_Appendix_1A_FT_2}
        \frac{1}{4 \pi r}
        & \longleftrightarrow
        \frac{1}{(2 \pi)^{3/2}} \frac{1}{k^2}
    \\ \label{eq: S1_Appendix_1A_FT_3}
        \frac{\hat{\mathbf{r}}}{4 \pi r^2}
        = \frac{\mathbf{r}}{4 \pi r^3}
        & \longleftrightarrow
        \frac{1}{(2 \pi)^{3/2}} \frac{-\ci\mathbf{k}}{k^2}
    \end{align}
    \item The convolution theorem
    \begin{equation} \label{eq: S1_Appendix_1A_FT_4}
        (F * G)(\mathbf{r}) 
        =  \int 
            \mathrm{d}^3 r' \,
            F(\mathbf{r}') 
            G(\mathbf{r} - \mathbf{r}')
        \longleftrightarrow
        (2 \pi)^{3/2} 
            \mathscr{F}(\mathbf{k})
            \mathscr{G}(\mathbf{k})
    \end{equation}
\end{enumerate}

%% file: include_appendices/appendix_QHO_Field.tex
\chapter{Classical Harmonic Oscillator}
    It is a well-known fact that the eigen-solution to a harmonic oscillator with the Hamiltonian (or simply the total energy)
    \begin{equation}
        \mathcal{H}(t) 
        = \frac{p(t)^2}{2m} + \frac{1}{2}m \omega^2 q(t)^2
    \end{equation}
    is given by a rotation in the phase plane with a natural frequency $\omega$. Since the phase plane is isomorphic to a complex plane, we can represent the pair $(q,p)$ as a complex number; in particular, we can define a dimensionless complex quantity, called the normal mode, as
    \begin{equation}
        a(t) 
        = \sqrt{\frac{m\omega}{2\hbar}} \, q(t) 
        + \ci \sqrt{\frac{1}{2\hbar m \omega}} \, p(t) .
    \end{equation}
    By design, $q$ and $p$ are (up to some normalization) the real and imaginary parts of the normal mode:
    \begin{align}
        q(t) 
        &= \sqrt{\frac{\hbar}{2m\omega}}
            \left[
                a(t) + a^*(t)
            \right]
    \\
        p(t) 
        &= -\ci \sqrt{\frac{m \hbar \omega}{2}}
            \left[
                a(t) - a^*(t)
            \right]
    \end{align}
    Then, the Hamiltonian can be rewritten as
    \begin{align}
        \mathcal{H}(t) 
        &= \frac{p^2}{2m} + \frac{1}{2}m \omega^2 q^2
    \nonumber \\
        &= - \frac{\hbar \omega}{4} (a - a^*)^2 
            + \frac{\hbar \omega}{4} (a + a^*)^2
    \nonumber \\
        &= \frac{\hbar \omega}{2} (a^* a + a a^*).
    \end{align}
    At this point, we can promote the normal mode to the annihilation operator and obtain the familiar Hamiltonian operator for a quantum harmonic oscillator. However, we will stay in the classical regime for now and try to formulate the Lagrangian mechanics in terms of the normal mode.
    
    Given that we have an expression for the Hamiltonian, the Lagrangian can be computed by a Legendre transformation:
    \begin{align}
        \mathcal{L}
        &= p \dot{q} 
            - \mathcal{H}
        = - \ci \sqrt{\frac{m \hbar \omega}{2}}
            (a - a^*)
            \sqrt{\frac{\hbar}{2m\omega}}
            (\dot{a}+ \dot{a}^*)
            - \frac{\hbar \omega}{2} (a^* a + a a^*)
    \nonumber \\
        &= \frac{\ci \hbar}{2}
            (- a \dot{a} - a \dot{a}^* + a^* \dot{a} + a^* \dot{a}^*)
            - \frac{\hbar \omega}{2} (a^* a + a a^*)
    \nonumber \\
        &= \frac{\ci \hbar}{2}
            (a^* \dot{a} - a \dot{a}^* )
            - \frac{\hbar \omega}{2} (a^* a + a a^*)
            + \frac{\ci \hbar}{4} \frac{d}{dt} 
                \left[ a^2 + (a^*)^2\right].
    \end{align}
    Since the action functional is invariant up to the addition of a total time derivative, we can ignore the last term. Moreover, integrating the first term by parts changes $-a\dot{a}^*$ to $\dot{a} a^*$ (the total time derivative produced is also omitted). Hence, we obtain
    \begin{equation}
        \mathcal{L} (a, \dot{a}, a^*, \dot{a}^*) 
        = \frac{\ci \hbar}{2}
            (a^* \dot{a} + \dot{a} a^* )
            - \frac{\hbar \omega}{2} (a^* a + a a^*).
    \end{equation}
    For simplicity, we will ignore the zero-point energy which comes from the noncommutativity of $a$, $a^*$ and $\dot{a}$ after quantization; this allows us to exchange the order of multiplication and write the Lagrangian as
    \begin{equation} \label{eq:lagrangian_sho_new}
        \mathcal{L}(a, \dot{a}, a^*, \dot{a}^*) 
        = \ci \hbar a^* \dot{a}
            - \hbar \omega a^* a.
    \end{equation}
    
    It's straightforward to verify that the Euler Lagrange equation associated with Eq.(\ref{eq:lagrangian_sho_new})\footnote{There is actually another Euler-Lagrange equation for $a^*$, but the resulting differential equation is exactly the same.}
    \begin{equation}
        \frac{\mathrm{d}}{\mathrm{d} t} 
            \left[ 
                \frac{\partial\mathcal{L} }{\partial \dot{a}} 
                (a(t), \dot{a}(t), a^*(t), \dot{a}^*(t)) 
            \right]
        = \frac{\partial\mathcal{L} }{\partial a} (a(t), \dot{a}(t), a^*(t), \dot{a}^*(t)) 
    \end{equation}
    reduces to 
    \begin{equation}
        \dot{a}(t) = - \ci \omega a(t),
    \end{equation}
    which indeed describes a rotation in the complex plane.

%% file: include_appendices/appendix_LC_TL.tex
\chapter{Quantization of a Transmission Line}\label{appendix:LC_to_TL}
\section{Field Theory of an Infinite Transmission Line}
We start from a discretized model of the transmission line and subsequently take the limit of each physical quantity as $\Delta x \rightarrow 0$. To begin with, the capacitive energy, identified as the kinetic energy, is given by
\begin{align}
    T_{\text{cap}}(t) 
    &= \lim_{\Delta x \rightarrow 0} 
        \sum_{k} 
            \frac{1}{2} (C_l \Delta x) 
            V(x_k,t)^2
    = \int_{-\infty}^{\infty}
        \mathrm{d} x \, \frac{C_l}{2} \left(\frac{\partial \Phi }{\partial t}\right)^2.
\end{align}
Similarly, the inductive energy, identified as the potential energy, is
\begin{align}
    U_{\text{ind}}(t) 
    &= \lim_{\Delta x \rightarrow 0} 
        \sum_{k} 
            \frac{1}{2} (L_l \Delta x) 
            I(x_k,t)^2
\nonumber \\
    &= \lim_{\Delta x \rightarrow 0} 
        \sum_{k} 
            \frac{1}{2} (L_l \Delta x) 
            \left[
                \frac{\Phi(x_k,t) 
                - \Phi(x_k + \Delta x,t)}{L_l \Delta x}
            \right]^2
    = \int_{-\infty}^{\infty}
            \mathrm{d} x \, 
            \frac{1}{2 L_l} 
            \left(
                \frac{\partial \Phi }{\partial x}
            \right)^2.
\end{align}
Note that we have used the position-dependent flux $\Phi(x,t)$ as the generalized coordinates. It's clear that $\{ ... , \Phi(x_{k-1},t)$,  $\Phi(x_k,t)$,  $\Phi(x_{k+1},t), ... \}$ in the discretized case are all independent degrees of freedom; hence, when moving to the continuum, we should treat $\Phi(x,t)$ as infinitely many degrees of freedom labeled by a continuous index $x$ instead of a single degree of freedom as it might appear to be. In other words, we are dealing with a classical field. 

Just like the discrete case, we can write down the Lagrangian $\mathcal{L}_{\text{TL}}$ as the difference between the kinetic and potential energy
\begin{align}
    \mathcal{L}_{\text{TL}}(\Phi(x), \dot{\Phi}(x))
    &= \int_{-\infty}^{\infty}
        \mathrm{d} x 
            \left[
                \frac{C_l}{2} 
                    \dot{\Phi}(x)^2
                - \frac{1}{2 L_l}
                    (\partial_x \Phi(x))^2
            \right]
\nonumber \\
    &\doteq \int_{-\infty}^{\infty}
        \mathrm{d} x \, 
        \mathscr{L}_{\text{TL}} 
        (\Phi(x), \dot{\Phi}(x), \partial_x \Phi(x)),
\end{align}
where\footnote{Although the Lagrangian density is independent of the field $\Phi$ in our example, we keep the general dependence in the definition so that we can write down a general one-dimensional Euler-Lagrange equation.} 
\begin{equation}\label{eq:TL_Lagrangian_density}
    \mathscr{L}_{\text{TL}} (\Phi, \dot{\Phi}, \partial_x \Phi)
    = \frac{C_l}{2} 
                    \dot{\Phi}^2
                - \frac{1}{2 L_l}
                    (\partial_x \Phi)^2
\end{equation}
is the Lagrangian density, satisfying the Euler-Lagrange equation \cite{peskin1995introQFT}
\begin{equation}
        \frac{\partial}{\partial t} 
        \left[
            \frac{\partial \mathscr{L}_{\text{TL}}}{\partial \dot{\Phi}}
            \Bigg|_{\substack{\Phi(x,t) \\ \dot{\Phi}(x,t) \\ \partial_x \Phi(x,t)}}
        \right]
        = \frac{\partial \mathscr{L}_{\text{TL}}}{\partial \Phi}
                \Bigg|_{\substack{\Phi(x,t) \\ \dot{\Phi}(x,t) \\ \partial_x \Phi(x,t)}} 
            - \frac{\partial}{\partial x} 
                \left[
                    \frac{\partial \mathscr{L}_{\text{TL}}}{\partial (\partial_x\Phi)}
                    \Bigg|_{\substack{\Phi(x,t) \\ \dot{\Phi}(x,t) \\ \partial_x \Phi(x,t)}}
                \right].
    \end{equation}
Given the Lagrangian density defined in Eq.(\ref{eq:TL_Lagrangian_density}), it's straightforward to show that the Euler-Lagrange equation reduces to a one-dimensional wave equation
\begin{equation}
    \left(
        \frac{\partial^2}{\partial x^2} 
        - \frac{1}{v_{\text{p}}^2} \frac{\partial^2}{\partial t^2}
    \right)
    \Phi(x,t) = 0,
\end{equation}
where $v_{\text{p}} = 1/\sqrt{L_l C_l}$ is interpreted as the phase velocity of the waves traveling on the line.

To perform the field quantization, we rewrite the equations of motion in terms of the Hamiltonian. To do so, we first need to define the conjugate momenta of $\Phi(x)$
\begin{equation}
    \Pi(x) 
    \doteq \frac{\partial \mathscr{L}_{\text{TL}}}{\partial \dot{\Phi}} 
        \Bigg|_{\dot{\Phi} = \dot{\Phi}(x)} 
    = C_l \dot{\Phi}(x) 
    = C_l V(x),
\end{equation}
which can be interpreted as the charge stored on an infinitesimally small capacitor located at $x$.\footnote{Note that the unit of $\Pi$ is $[\mathrm{C}/\mathrm{m}]$, so it's the charge per unit length.} Consequently, the Hamiltonian is defined as the Legendre transform of $\mathcal{L}_{\text{TL}}$:
\begin{equation}
    \mathcal{H}_{\text{TL}}
    = \int_{-\infty}^{\infty}
        \mathrm{d} x 
        \left( 
            \Pi \dot{\Phi}
            - \mathscr{L}_{\text{TL}}
        \right)
    = \int_{-\infty}^{\infty}
        \mathrm{d} x \left[\frac{1}{2C_l} \Pi(x)^2
        + \frac{1}{2L_l} (\partial_x \Phi(x))^2\right]
\end{equation}
The Hamiltonian is the sum of the capacitive and inductive energy as expected.

Since we will be looking at the modes (let it be a standing-wave or a traveling-wave mode) on the line, it's useful to convert physical quantities to their counterparts in the frequency domain: Define the spatial Fourier transform of the flux and charge to be, respectively, 
\begin{equation}
    \tilde{\Phi}(k,t)
    = \frac{1}{\sqrt{2\pi}} \int_{- \infty}^{\infty} \mathrm{d} x \, \Phi(x,t) e^{- \ci kx},
\end{equation}
\begin{equation}
    \tilde{\Pi}(k,t)
    = \frac{1}{\sqrt{2\pi}} \int_{- \infty}^{\infty} \mathrm{d} x \, \Pi(x,t) e^{- \ci kx}.
\end{equation}
The dual variable $k$ is assigned the meaning of a wavevector, whose sign indicates the propagating direction of the traveling mode. For future reference, we will also define the frequency corresponding to a specific $k$ as
\begin{equation}
    \omega_k 
    = \omega_{-k}
    = v_{\text{p}} |k|. 
\end{equation}
Note that we have restricted the frequency to be nonnegative. In this one-dimensional problem, each frequency is two-fold degenerate since the wavevector has a sense of direction. Moreover, since $\Phi$ and $\Pi$ are real fields, their Fourier transform must satisfy 
\begin{align}
    \tilde{\Phi}^*(k,t) &= \tilde{\Phi}(-k,t),
\\
    \tilde{\Pi}^*(k,t) &= \tilde{\Pi}(-k,t).
\end{align}

Applying Parseval's theorem yields
\begin{equation}\label{eq:TL_Hamiltonian_k_space}
    \mathcal{H}_{\text{TL}}(t)
    = \int_{-\infty}^{\infty}
        \mathrm{d} k 
            \left[
                \frac{1}{2C_l} 
                    \Bigsl| \tilde{\Pi}(k, t) \Bigsr|^2
                + \frac{\omega_k^2 C_l}{2} 
                    \Bigsl| \tilde{\Phi}(k, t) \Bigsr|^2
            \right].
\end{equation}
Since Eq.(\ref{eq:TL_Hamiltonian_k_space}) represents an infinite collection of harmonic oscillators indexed by $k$, it's natural to define a normal mode for each $k$
\begin{equation}
    a(k,t) 
    = \sqrt{\frac{\omega_k C_l}{2\hbar}} \, \tilde{\Phi}(k,t)
        + \frac{\ci}{\sqrt{2\hbar \omega_k C_l}} \, \tilde{\Pi}(k,t),
\end{equation}
which will become the annihilation operator as a result of the field quantization. Due to the conjugate symmetry of $\tilde{\Phi}$ and $\tilde{\Pi}$, we have
\begin{equation}
    a^*(k,t) 
    = \sqrt{\frac{\omega_k C_l}{2\hbar}} \, \tilde{\Phi}(-k,t)
        - \frac{\ci}{\sqrt{2\hbar \omega_k C_l}} \, \tilde{\Pi}(-k,t)
    \neq a(-k,t) .
\end{equation}
As the final step, we write the Hamiltonian in terms of these normal modes:
\begin{equation}
    \mathcal{H}_{\text{TL}} (t)
    = \int_{-\infty}^{\infty} 
        \mathrm{d} k \, 
        \frac{1}{2} 
        \hbar \omega_k 
        \Big[
            a^*(k,t) a(k,t) 
            + a(k,t) a^*(k,t)
        \Big].
\end{equation}

\section{Quantization of an Infinite Transmission Line}
\subsection{The Schr\"{o}dinger Picture}
Since an infinite transmission line is mathematically equivalent to a one-dimensional free space, its quantization amounts to the introduction of the commutation relations
\begin{equation}
    \left[\hat{a}(k), \hat{a}^{\dagger}(k')\right]
    = \delta(k-k')
\end{equation}
for any pair of wavevectors $k$ and $k'$. In terms of the conjugate variables, we have
\begin{equation}
    \left[\hat{\tilde{\Phi}}(k), \hat{\tilde{\Pi}}(k')\right]
    =  \ci \hbar \delta(k-k').
\end{equation}
As expected, the quantum Hamiltonian reduces to
\begin{equation}
    \hat{H}_{\text{TL}}
    = \int_{-\infty}^{\infty} 
        \mathrm{d} k \, 
        \hbar \omega_k 
        \left[
            \hat{a}^{\dagger}(k) \hat{a}(k) 
            + \frac{1}{2}
        \right].
\end{equation}
\subsection{The Heisenberg Picture}
Next, we examine the time evolution of the observables in the Heisenberg picture; it turns out that the expressions obtained in this picture are very similar to the classical counterparts.

As usual, we start from the time evolution of the annihilation operator governed by the Heisenberg equation
\begin{equation}
    \dot{\hat{a}}(k,t)
    = - \frac{\ci}{\hbar} 
        \left[
            \hat{a}(k,t), 
            \hat{H}_{\text{TL}}
        \right]
    = - \ci \omega_k \hat{a}(k,t).
\end{equation}
Next, we use the solution
\begin{equation}
    \hat{a}(k,t) 
    = e^{-\ci \omega_k (t-t_0)} \hat{a}(k,t_0)
\end{equation}
to find the time evolution of $\hat{\tilde{\Phi}}$ and $\hat{\tilde{\Pi}}$:
\begin{align}
    \hat{\tilde{\Phi}}(k,t) 
    &= \sqrt{\frac{\hbar}{2\omega_k C_l}}   
        \left[
            \hat{a}(k,t) 
            + \hat{a}^{\dagger}(k,t)
        \right]
\nonumber \\
    &= \sqrt{\frac{\hbar}{2\omega_k C_l}} 
        \left[
            e^{-\ci \omega_k (t-t_0)} \hat{a}(k,t_0)
            + e^{\ci \omega_k (t-t_0)} \hat{a}^{\dagger}(k,t_0)
        \right],
\\
    \hat{\tilde{\Pi}}(k,t) 
    &= - \ci \sqrt{\frac{\hbar \omega_k C_l}{2}} 
        \left[
            \hat{a}(k,t) 
            - \hat{a}^{\dagger}(k,t)
        \right]
\nonumber \\
    &= - \ci \sqrt{\frac{\hbar \omega_k C_l}{2}} 
        \left[
            e^{-\ci \omega_k (t-t_0)} \hat{a}(k,t_0)
            - e^{\ci \omega_k (t-t_0)} \hat{a}^{\dagger}(k,t_0)
        \right].
\end{align}
To observe a wave motion in real space, we take the inverse Fourier transform of $\hat{\tilde{\Phi}}(k,t)$ and $\hat{\tilde{\Pi}}(k,t)$:
\begin{align}
    \hat{\Phi}(x,t) 
    &= \frac{1}{\sqrt{2\pi}}
        \int_{-\infty}^{\infty}
            \mathrm{d} k \,
            \hat{\tilde{\Phi}}(k,t) e^{\ci k x} 
\nonumber \\
    &= \int_{-\infty}^{\infty}  
        \mathrm{d} k  
        \sqrt{\frac{\hbar}{4\pi \omega_k C_l}}
        \left[
            e^{\ci [kx-\omega_k (t-t_0)]} \hat{a}(k,t_0)
            + e^{- \ci[kx-\omega_k (t-t_0)]} \hat{a}^{\dagger}(k,t_0)
        \right]
\nonumber \\
    &= \int_{0}^{\infty}  
        \mathrm{d} k  
        \sqrt{\frac{\hbar}{4\pi \omega_k C_l}}
        \Big[
            e^{\ci [kx-\omega_k (t-t_0)]} \hat{a}(k,t_0)
            + e^{\ci [-kx-\omega_k (t-t_0)]} \hat{a}(-k,t_0)
\nonumber \\
    &\ \ \ \ \ \ \ \ \ \ \ \ \ \ \ \ \ \ \ \ \ \ \ \ \ \ \ 
            + e^{- \ci[kx-\omega_k (t-t_0)]} \hat{a}^{\dagger}(k,t_0)
            + e^{- \ci[-kx-\omega_k (t-t_0)]} \hat{a}^{\dagger}(-k,t_0)
        \Big],
\\
    \hat{\Pi}(x,t) 
    &= \frac{1}{\sqrt{2\pi}}
        \int_{-\infty}^{\infty}
            \mathrm{d} k \,
            \hat{\tilde{\Pi}}(k,t) e^{\ci k x} 
\nonumber \\
    &= - \ci \int_{-\infty}^{\infty}  
        \mathrm{d} k  
        \sqrt{\frac{\hbar \omega_k C_l}{4\pi}}
        \left[
            e^{\ci [k x - \omega_k (t - t_0)]} \hat{a}(k, t_0)
            - e^{- \ci[k x - \omega_k (t - t_0)]} \hat{a}^{\dagger}(k, t_0)
        \right]
\nonumber \\
    &= - \ci \int_{0}^{\infty}  
        \mathrm{d} k  
        \sqrt{\frac{\hbar \omega_k C_l}{4 \pi}}
        \Big[
            e^{\ci [k x - \omega_k (t - t_0)]} \hat{a}(k, t_0)
            + e^{\ci [-k x - \omega_k (t - t_0)]} \hat{a}(-k, t_0)
\nonumber \\
    &\ \ \ \ \ \ \ \ \ \ \ \ \ \ \ \ \ \ \ \ \ \ \ \ \ \ \ \
            - e^{- \ci[k x - \omega_k (t - t_0)]} \hat{a}^{\dagger}(k, t_0)
            - e^{- \ci[- k x - \omega_k (t - t_0)]} \hat{a}^{\dagger}(-k, t_0)
        \Big].
\end{align}
Note that we have used the change of variables to combine integrals with positive and negative wavevectors. Alternatively, one can think of the new resulting expression as an integration of the frequency since $\omega_{k} = \omega_{-k}$ is always positive in our definition.

Moreover, like the classical case, we define the voltage and current operators from the flux operator
\begin{align}
    \hat{V}(x,t) 
    &= \frac{\partial}{\partial t} \hat{\Phi}(x, t) 
\nonumber \\
    &= -\ci \int_{0}^{\infty}  
        \mathrm{d} k  
        \sqrt{\frac{\hbar \omega_k}{4\pi C_l}}
        \Big[
            e^{\ci [kx-\omega_k (t-t_0)]} \hat{a}(k,t_0)
            + e^{\ci [-kx-\omega_k (t-t_0)]} \hat{a}(-k,t_0)
\nonumber \\ \label{eq:def_tline_voltage_operator}
    &\ \ \ \ \ \ \ \ \ \ \ \ \ \ \ \ \ \ \ \ \ \ \ \ \ \ \ \
            - e^{- \ci[kx-\omega_k (t-t_0)]} \hat{a}^{\dagger}(k,t_0)
            - e^{- \ci[-kx-\omega_k (t-t_0)]} \hat{a}^{\dagger}(-k,t_0)
        \Big],
\\
    \hat{I}(x,t) 
    &= -\frac{1}{L_l}\frac{\partial}{\partial x} \hat{\Phi}(x,t) 
\nonumber \\
    &= -\frac{\ci}{Z_0} \int_{0}^{\infty}  
        \mathrm{d} k  
        \sqrt{\frac{\hbar \omega_k}{4\pi C_l}}
        \Big[
            e^{\ci [kx-\omega_k (t-t_0)]} \hat{a}(k,t_0)
            - e^{\ci [-kx-\omega_k (t-t_0)]} \hat{a}(-k,t_0)
\nonumber \\ \label{eq:def_tline_current_operator}
    &\ \ \ \ \ \ \ \ \ \ \ \ \ \ \ \ \ \ \ \ \ \ \ \ \ \ \ \ 
            - e^{- \ci[kx-\omega_k (t-t_0)]} \hat{a}^{\dagger}(k,t_0)
            + e^{- \ci[-kx-\omega_k (t-t_0)]} \hat{a}^{\dagger}(-k,t_0)
        \Big],
\end{align}
respectively. At this point, there is not a clear connection between the classical and quantum picture. This is because we usually consider traveling waves on a transmission line in the classical microwave theory.

\subsection{Standing Waves v.s. Travelling Waves}

Classically, one can define the left- and right-traveling voltages and currents via
\begin{gather}
    V(x,t)
    = V^{\rightarrow}(x,t) + V^{\leftarrow}(x,t),
\\
    I(x,t)
    = I^{\rightarrow}(x,t) - I^{\leftarrow}(x,t)
\end{gather}
because a one-dimensional wave equation, in general, admits two independent solutions, i.e.,
\begin{equation}
    f(x - v_{\text{p}} t) + g(x + v_{\text{p}} t).
\end{equation}
The function $f$ represents a right-traveling signal while $g$ represents a left-traveling signal. Looking closely at the voltage and current operators defined in Eq.(\ref{eq:def_tline_voltage_operator}) and (\ref{eq:def_tline_current_operator}), we realize that the right- and left-traveling waves can be defined as
\begin{align}
    \hat{V}^{\rightleftarrows}(x,t)
    &= - \ci \int_{0}^{\infty}
        \mathrm{d} k \sqrt{\frac{\hbar \omega_k}{4\pi C_l}}
        \left[ 
            e^{\pm \ci kx - \ci\omega_k(t-t_0)} 
                \hat{a}(\pm k,t_0)
            - e^{\mp \ci kx + \ci\omega_k(t-t_0)}
                \hat{a}^{\dagger}(\pm k,t_0) 
        \right],
\\
    \hat{I}^{\rightleftarrows}(x,t)
    &= -\frac{\ci}{Z_0} \int_{0}^{\infty}
        \mathrm{d} k \sqrt{\frac{\hbar \omega_k}{4\pi C_l}}
        \left[ 
            e^{\pm \ci kx - \ci\omega_k(t-t_0)} 
                \hat{a}(\pm k,t_0)
            - e^{\mp \ci kx + \ci\omega_k(t-t_0)}
                \hat{a}^{\dagger}(\pm k,t_0) 
        \right]
\end{align}
such that 
\begin{gather}
    \hat{V}(x,t)
    = \hat{V}^{\rightarrow}(x,t) + \hat{V}^{\leftarrow}(x,t),
\\
    I(x,t)
    = \hat{V}^{\rightarrow}(x,t) - \hat{V}^{\leftarrow}(x,t).
\end{gather}
In addition, we naturally obtain the impedance relations
\begin{equation}
    \hat{V}^{\leftrightarrows}(x,t)
    = Z_0 \hat{I}^{\leftrightarrows}(x,t),
\end{equation}
which is the same as the classical results. Consequently, the quantum operators for power wave amplitudes can be defined as \cite{PhysRevB.98.045405, Vool_2017_into_cQED}
\begin{align}
    \hat{A}^{\rightleftarrows}(x,t)
    &= \frac{1}{2} 
        \left[
            \frac{1}{\sqrt{Z_0}}\hat{V}(x,t)
            \pm \sqrt{Z_0} \hat{I}(x,t)
        \right]
\\
    &= \frac{-\ci}{\sqrt{2\pi}} \int_{0}^{\infty}
        \mathrm{d} k \sqrt{\frac{\hbar \omega_k v_{\text{p}}}{2}}
        \left[ 
            e^{\pm \ci kx - \ci\omega_k(t-t_0)} 
                \hat{a}(\pm k,t_0)
            - e^{\mp \ci kx + \ci\omega_k(t-t_0)}
                \hat{a}^{\dagger}(\pm k,t_0) 
        \right].
\end{align}
Since $x$ and $t$ are always related by the velocity $v_{\text{p}}$, we have
\begin{equation}
    \hat{A}^{\rightleftarrows}(x,t) = \hat{A}^{\rightleftarrows}(t \mp x/v_{\text{p}}).
\end{equation}
In other words, the power wave amplitude can be thought of as a function of time only, i.e.,
\begin{align}
    \hat{A}^{\rightleftarrows}(t)
    \doteq \hat{A}^{\rightleftarrows}(0,t)
    = \frac{-\ci}{\sqrt{2\pi}} \int_{0}^{\infty}
        \mathrm{d} k \sqrt{\frac{\hbar \omega_k v_{\text{p}}}{2}}
        \left[ 
            e^{- \ci\omega_k(t-t_0)} 
                \hat{a}(\pm k,t_0)
            - e^{\ci\omega_k(t-t_0)}
                \hat{a}^{\dagger}(\pm k,t_0) 
        \right].
\end{align}

The quantity $\hat{A}^{\rightleftarrows \dagger}(t)\hat{A}^{\rightleftarrows}(t)$ represents the power traveling on the transmission line. In practice, we often are interested in the power within a narrow bandwidth; hence, let us examine the spectrum of $\hat{A}^{\rightleftarrows}(t)$ by taking their Fourier transform:
\begin{align}
    \hat{A}^{\rightleftarrows}(\omega)
    = \frac{1}{\sqrt{2\pi}}
        \int_{-\infty}^{\infty}
            \mathrm{d} t \,
            \hat{A}^{\rightleftarrows}(t)
            e^{\ci \omega t}
    = \begin{cases}\displaystyle
        - \ci \sqrt{\frac{\hbar \abs{\omega}}{2 v_{\text{p}}}}
        \, e^{\ci \abs{\omega} t_0} \hat{a}(\pm \abs{\omega}/v_{\text{p}},t_0),
        & \text{if } \omega > 0,
        \\[6mm] \displaystyle
        \ci \sqrt{\frac{\hbar \abs{\omega}}{2 v_{\text{p}}}}
        \, e^{-\ci \abs{\omega} t_0} \hat{a}^{\dagger}(\pm \abs{\omega}/v_{\text{p}},t_0),
        & \text{if } \omega < 0.
    \end{cases}
\end{align}
Intuitively, $\hat{A}^{\rightleftarrows \dagger}(\omega) \hat{A}^{\rightleftarrows}(\omega)$ represent the power traveling on the line with a frequency $\omega$. If we normalize $\hat{A}^{\rightleftarrows}(\omega)$ by the zero-point energy of the mode at frequency $\omega$, we would obtain the traveling-wave annihilation operators
\begin{equation} \label{eq:definition_traveling_wave_ops}
    \hat{a}^{\rightleftarrows}(\omega)
    = \sqrt{\frac{2}{\hbar \abs{\omega}}}
        \hat{A}^{\rightleftarrows}(\omega)
    = \begin{cases}\displaystyle
        - \frac{\ci}{\sqrt{v_{\text{p}}}}
        \, e^{\ci \abs{\omega} t_0} \hat{a}(\pm \abs{\omega}/v_{\text{p}},t_0),
        & \text{if } \omega > 0,
        \\[6mm] \displaystyle
        \frac{\ci}{\sqrt{v_{\text{p}}}}
        \, e^{-\ci \abs{\omega} t_0} \hat{a}^{\dagger}(\pm \abs{\omega}/v_{\text{p}},t_0),
        & \text{if } \omega < 0.
    \end{cases}
\end{equation}
Consequently, $\hat{a}^{\rightleftarrows \dagger}(\omega) \hat{a}^{\rightleftarrows}(\omega)$ give the photon fluxes going to the right and left of the line. Finally, we can move back to the time domain to find the traveling-wave annihilation operators used in the quantum Langevin equation:
\begin{align}
    \hat{a}^{\rightleftarrows}(t)
    &= \frac{1}{\sqrt{2\pi}} 
        \int_{-\infty}^{\infty}
            \mathrm{d} \omega \, 
            e^{-\ci \omega t}
            \hat{a}^{\rightleftarrows}(\omega)
\nonumber \\
    &= - \frac{\ci}{\sqrt{2\pi v_{\text{p}}}} 
        \int_{0}^{\infty}
            \mathrm{d} \omega 
            \Big[
                e^{-\ci \omega (t-t_0)} \hat{a}(\pm \omega/v_{\text{p}},t_0)
                - e^{-\ci \omega (t-t_0)} \hat{a}^{\dagger}(\pm \omega/v_{\text{p}},t_0)
            \Big]
\nonumber \\ \label{eq:traveling_wave_annihilation_operator_full_derivation}
    &= - \ci \sqrt{\frac{v_{\text{p}}}{2\pi}} 
        \int_{0}^{\infty}
            \mathrm{d} k
            \Big[
                e^{-\ci \omega_k (t-t_0)} \hat{a}(\pm k,t_0)
                - e^{\ci \omega_k (t-t_0)} \hat{a}^{\dagger}(\pm k,t_0)
            \Big]
\end{align}
To be more precise, Eq.(\ref{eq:traveling_wave_annihilation_operator_full_derivation}) defines the traveling-wave annihilation operators before making any rotating-wave approximation (RWA); clearly, $\hat{a}^{\rightleftarrows}(t)$ is Hermitian. Making the RWA amounts to ignoring the creation operator in Eq.(\ref{eq:traveling_wave_annihilation_operator_full_derivation}); thus, making $\hat{a}^{\rightleftarrows}(t)$ non-Hermitian. It should also be noted that $\hat{a}^{\rightleftarrows}(t)$ is not proportional to $\hat{A}^{\rightleftarrows}(t)$ before make the RWA because $\hat{a}^{\rightleftarrows}(\omega)$ and $\hat{A}^{\rightleftarrows}(\omega)$ are related by $|\omega|$ in the frequency domain. Technically, $\hat{A}^{\rightleftarrows}(t)$ should be used in the quantum Langevin equation; nevertheless, one consequence of the RWA is the replacement of $\omega$ with $\omega_{\text{r}}$, where $\omega_{\text{r}}$ is the center frequency of a narrow-band signal. Thus, under the RWA, $\hat{a}^{\rightleftarrows}(t)$ is directly proportional to $\hat{A}^{\rightleftarrows}(t)$. In addition, by dropping the counter-rotating terms (i.e., the negative frequencies), Eq.(\ref{eq:traveling_wave_annihilation_operator_full_derivation}) leads to the commutation relation
\begin{equation}
    \Big[ \hat{a}^{\rightarrow}(t), \hat{a}^{\rightarrow \dagger}(t')
    \Big]
    = \Big[ \hat{a}^{\leftarrow}(t), \hat{a}^{\leftarrow \dagger}(t')
    \Big]
    = \delta(t - t').
\end{equation}


%% file: include_appendices/appendix_Semiclassical_Rabi.tex
\chapter{Semi-Classical Approach to Matter-Light Interaction} \label{appendix:semiclassical_matter_light}

\section{Hamiltonian}
We start with the dipole interaction
\begin{align}
    \hat{H}
    = - \frac{1}{2} \hbar \omega_{\text{q}} \hat{\sigma}_z
        - \hat{\mathbf{d}} \cdot \mathbf{E}(\mathbf{0}, t),
\end{align}
where the dipole is a quantum operator
\begin{equation}
    \hat{\mathbf{d}}
    = \hat{\mathbf{e}}_x d_0 
        \Big(
           e^{\ci \phi_{\text{di}}} \hat{\sigma}_{+} 
           + e^{-\ci \phi_{\text{di}}} \hat{\sigma}_{-} 
        \Big)
\end{equation}
but the field is classical function of time
\begin{equation}
    \mathbf{E}(\mathbf{0}, t)
    = \mathbf{E}(\mathbf{0})
        \sin(\omega t - \phi)
    = \hat{\mathbf{e}}_x \frac{\ci E_0}{2}
        \left(
            e^{- \ci \omega t + \ci \phi} 
            - e^{\ci \omega t - \ci \phi} 
        \right).
\end{equation}
The Hamiltonian is thus given by
\begin{align}
    \hat{H}
    &= - \frac{1}{2} \hbar \omega_{\text{q}} \hat{\sigma}_z
        - \frac{\ci d_0 E_0}{2}  \left(
            e^{- \ci \omega t + \ci \phi} 
            - e^{\ci \omega t - \ci \phi} 
        \right) 
        \left(
            e^{\ci \phi_{\text{di}}} \hat{\sigma}_{+} 
            + e^{-\ci \phi_{\text{di}}} \hat{\sigma}_{-} 
        \right)
\nonumber \\
    &= - \frac{1}{2} \hbar \omega_{\text{q}} \hat{\sigma}_z
        - \frac{\ci}{2} \hbar \Omega 
            \left(
                e^{-\ci \omega t + \ci \phi}
                    e^{\ci \phi_{\text{di}}} 
                    \hat{\sigma}_{+}
                - e^{\ci \omega t - \ci \phi}
                    e^{-\ci \phi_{\text{di}}} 
                    \hat{\sigma}_{-} 
                + e^{-\ci \omega t + \ci \phi}
                    e^{-\ci \phi_{\text{di}}} 
                    \hat{\sigma}_{-} 
                - e^{\ci \omega t - \ci \phi}
                    e^{\ci \phi_{\text{di}}} 
                    \hat{\sigma}_{+}
            \right),
\end{align}
where $\Omega = d_0 E_0 / \hbar$ is the Rabi frequency.

\section{Rabi Oscillation}
Given the Hamiltonian above, our goal is to solve the time evolution of the two-level system 
\begin{equation} \label{eq:general_2level_state_appendix}
    \ket{\Psi(t)} = a(t) \ket{0} + b(t) \ket{1}
\end{equation}
subject to some arbitrary initial condition. Substitute Eq.(\ref{eq:general_2level_state_appendix}) into the Schr\"{o}dinger equation, we obtain two coupled differential equations:
\begin{equation}
    \ci \!
    \begin{pmatrix}
        \dot{a}(t) \\[2mm]
        \dot{b}(t)
    \end{pmatrix}\!
    = - \frac{1}{2} 
        \begin{pmatrix}
            \omega_{\text{q}} 
                & \!\!\! - \ci \Omega e^{\ci \omega t - \ci \phi} 
                        e^{-\ci \phi_{\text{di}}}
                    + \ci \Omega e^{-\ci \omega t + \ci \phi} 
                        e^{-\ci \phi_{\text{di}}} 
            \\[2mm]
                \ci \Omega e^{-\ci \omega t + \ci \phi} 
                        e^{\ci \phi_{\text{di}}} 
                    - \ci \Omega e^{\ci \omega t - \ci \phi} 
                        e^{\ci \phi_{\text{di}}}  
                & - \omega_{\text{q}} 
        \end{pmatrix}
        \!\!
        \begin{pmatrix}
            a(t) \\[2mm]
            b(t)
        \end{pmatrix}
        \! .
\end{equation}

We solve this equation by going into the interaction picture of $\hat{H}_{\text{q}} = - \hbar \omega_{\text{q}} \hat{\sigma}_{z} /2$, i.e., we define slow-varying coefficients $\Tilde{a}(t)$ and $\Tilde{b}(t)$ such that
\begin{gather}
    a(t) 
    = e^{\ci \omega_{\text{q}} t/2} \Tilde{a}(t),
    \\ 
    b(t) 
    = e^{-\ci \omega_{\text{q}} t/2} \Tilde{b} (t).
\end{gather}
Consequently, we obtain
\begin{gather}
    \ci \dot{\Tilde{a}} (t)
    = - \frac{\Omega}{2} 
        \left[
            - \ci e^{ - \ci (\omega_{\text{q}} - \omega) t} 
                e^{-\ci (\phi + \phi_{\text{di}})}
            + \ci e^{- \ci (\omega_{\text{q}} + \omega) t} 
                e^{\ci (\phi - \phi_{\text{di}})} 
        \right] \Tilde{b}(t) ,
\\
    \ci \dot{\Tilde{b}} (t)
    = - \frac{\Omega }{2}
        \left[
            \ci e^{\ci (\omega_{\text{q}} - \omega) t} 
                e^{\ci (\phi + \phi_{\text{di}})} 
            - \ci e^{\ci (\omega_{\text{q}} + \omega) t} 
                e^{- \ci (\phi - \phi_{\text{di}})} 
        \right] \Tilde{a}(t).
\end{gather}

Suppose both $\Omega$ and the detuning $\Delta = \omega_{\text{q}} - \omega$ are much smaller than the qubit transition frequency $\omega_{\text{q}}$, then terms that oscillate at the frequency $\omega_{\text{q}} + \omega \approx 2\omega_{\text{q}}$ will be integrated to 0 on the time scale $2\pi/2\omega_{\text{q}}$ before $\Tilde{a}$ and $\Tilde{b}$ vary appreciably. Hence, we can make the RWA by discarding the fast oscillating terms, i.e.,
\begin{gather} \label{eq:alpha_ode}
    \ci \dot{\Tilde{a}} (t)
    = - \frac{\Omega }{2}
            e^{ - \ci \Delta t - \ci \phi_g}
            \Tilde{b}(t) ,
\\ \label{eq:beta_ode}
    \ci \dot{\Tilde{b}}(t) 
    = - \frac{\Omega}{2}
            e^{\ci \Delta t + \ci \phi_g} 
            \Tilde{a}(t),
\end{gather}
where $\phi_g = \phi + \phi_{\text{di}} + \pi/2$. Eventually, we will show that $\Tilde{a}$ and $\Tilde{b}$ vary on the time scale $2\pi/\sqrt{\Omega^2 + \Delta^2} $; with the assumption that $\omega_{\text{q}} \gg \Omega$ and $\omega_{\text{q}} \gg \Delta$, the RWA is thus self-consistent.  

To solve the coupled equations, we eliminate $\Tilde{a}$ to obtain a single second-order equation of $\Tilde{b}$:
\begin{equation}
    \ddot{\Tilde{b}}(t) 
        - \ci \Delta \dot{\Tilde{b}}(t)
        + \left( \frac{\Omega}{2}\right)^{\! 2} \Tilde{b}(t) 
    = 0.
\end{equation}
The roots of the characteristic polynomial are
\begin{equation}
    \lambda_{1,2} 
    = \frac{\ci \Delta \pm \sqrt{(-\ci\Delta)^2 - 4 (\Omega/2)^2}}{2}
    = \frac{\ci}{2} \Delta 
        \pm \frac{\ci}{2}
            \sqrt{\Omega^2 + \Delta^2},
\end{equation}
leading to a general solution of the form
\begin{equation}\label{eq:general_soln_beta}
    \Tilde{b}(t) 
    = e^{\ci \Delta /2} 
        \left[ 
            A \exp\left(\frac{\ci}{2}\sqrt{\Omega^2 + \Delta^2} \,  t \right) 
            + B \exp\left(- \frac{\ci}{2}\sqrt{\Omega^2 + \Delta^2} \,  t \right) 
        \right].
\end{equation}

To determine the constant $A$ and $B$, we use the initial conditions
\begin{align}
    a(0) = 1 \ \ &\Longrightarrow \ \ \Tilde{a}(0) = 1,
\\
    b(0) = 0 \ \ &\Longrightarrow \ \ \Tilde{b}(0) = 0.
\end{align}
Then, from Eq.(\ref{eq:general_soln_beta}), we have
\begin{equation}
   0 =  \Tilde{b}(0) = A+B
   \Longrightarrow \ \ B = - A,
\end{equation}
which means
\begin{align}
    \Tilde{b}(t) 
    &= A e^{\ci \Delta /2} 
        \left[ 
            \exp\left(\frac{\ci}{2}\sqrt{\Omega^2 + \Delta^2} \,  t \right) 
            - \exp\left(- \frac{\ci}{2}\sqrt{\Omega^2 + \Delta^2} \,  t \right) 
        \right]
\nonumber \\ \label{eq:beta_intermediate}
    &= 2\ci A e^{\ci \Delta /2} 
        \sin\left( \frac{1}{2}\sqrt{\Omega^2 + \Delta^2} \,  t \right) .
\end{align}
To find $A$, we also need an expression for $\Tilde{a}(t)$. Substituting Eq.(\ref{eq:beta_intermediate}) into Eq.(\ref{eq:beta_ode}) yields
\begin{align}
    \Tilde{a}(t) 
    &= - \frac{2 \ci e^{-\ci \Delta t - \ci \phi_g}}{\Omega}
        \, \dot{\Tilde{b}} (t)
\nonumber \\
    &= 2A e^{-\ci \Delta t/2 - \ci \phi_g} 
        \left[
            \ci \frac{\Delta}{\Omega}  \sin\left( \frac{1}{2}\sqrt{\Omega^2 + \Delta^2} \,  t \right)
            + \frac{\sqrt{\Omega^2 + \Delta^2} \, }{\Omega}
            \cos\left( \frac{1}{2}\sqrt{\Omega^2 + \Delta^2} \,  t \right)
        \right].
\end{align}
By using the other initial condition $\Tilde{a}(0) = 1$, we obtain
\begin{equation}
    A = \frac{e^{\ci \phi_g} \Omega}{2 \sqrt{\Omega^2+\Delta^2}}
\end{equation}
and, hence,
\begin{equation}
    \Tilde{a}(t) 
    = e^{-\ci \Delta t /2} 
        \left[
            \ci \frac{\Delta}{\sqrt{\Omega^2 + \Delta^2} \, }  \sin\left( \frac{1}{2}\sqrt{\Omega^2 + \Delta^2} \,  t \right)
            +
            \cos\left( \frac{1}{2}\sqrt{\Omega^2 + \Delta^2} \,  t \right)
        \right] ,
\end{equation}
\begin{equation}
    \Tilde{b}(t)
    =  \ci e^{\ci \Delta t/2 + \ci \phi_g}  \frac{\Omega}{\sqrt{\Omega^2 + \Delta^2} \, } 
        \sin\left( \frac{1}{2}\sqrt{\Omega^2 + \Delta^2} \,  t \right).
\end{equation}
Or, in the Schr\"{o}dinger picture, 
\begin{equation}
    a(t) 
    = e^{\ci \omega t /2} 
        \left[
            \ci \frac{\Delta}{\sqrt{\Omega^2 + \Delta^2} \, }  \sin\left( \frac{1}{2}\sqrt{\Omega^2 + \Delta^2} \,  t \right)
            +
            \cos\left( \frac{1}{2}\sqrt{\Omega^2 + \Delta^2} \,  t \right)
        \right] ,
\end{equation}
\begin{equation}
    b(t)
    =  \ci e^{- \ci \omega t/2 + \ci \phi_g}  \frac{\Omega}{\sqrt{\Omega^2 + \Delta^2} \, } 
        \sin\left( \frac{1}{2}\sqrt{\Omega^2 + \Delta^2} \,  t \right).
\end{equation}

In conclusion, the probability of measuring the qubit in $\ket{0}$ and $\ket{1}$ are given, respectively, as
\begin{equation}
    p_0(t)
    = \abs{a(t)}^2 = \abs{\Tilde{a}(t)}^2
    = 1- \frac{\Omega^2}{\Omega^2 + \Delta^2}
        \sin^2 \left( \frac{1}{2}\sqrt{\Omega^2 + \Delta^2} \,  t \right) ,
\end{equation}
\begin{equation}
    p_1(t) =\abs{b(t)}^2 = \Bigsl| \Tilde{b}(t) \Bigsr|^2
    = \frac{\Omega^2}{\Omega^2 + \Delta^2}  
        \sin^2 \left( \frac{1}{2}\sqrt{\Omega^2 + \Delta^2} \,  t \right) .
\end{equation}
By defining $\Omega' = \sqrt{\Omega^2 + \Delta^2}$ to be generalized Rabi frequency, we can also write
\begin{equation}
    p_0(t)
    = 1- \frac{\Omega^2}{\Omega'^2}
        \sin^2 \left( \frac{1}{2} \Omega' t \right) ,
\end{equation}
\begin{equation}
    p_1(t)
    = \frac{\Omega^2}{\Omega'^2}  
        \sin^2 \left( \frac{1}{2}\Omega'  t \right) .
\end{equation}
\section{General Time Evolution}
For arbitrary initial conditions $a(0)$ and $b(0)$, we would have
\begin{equation}
    a(t) 
    = a(0) e^{\ci \omega t /2} 
            \left[
                \cos\left( \frac{1}{2} \Omega' t \right) 
                + \ci \frac{\Delta}{\Omega'} \sin \left( \frac{1}{2}\Omega' t \right) 
            \right]
        + \ci b(0) e^{\ci \omega t /2 -\ci \phi_{g}} 
                \frac{\Omega}{\Omega'} \sin\left( \frac{1}{2}\Omega' t \right),
\end{equation}
\begin{equation}
    b(t)
    = \ci a(0) e^{-\ci \omega t/2 + \ci \phi_{g}} 
            \frac{\Omega}{\Omega'} \sin\left( \frac{1}{2}\Omega' t \right) 
        + b(0) e^{-\ci \omega t/2} 
            \left[
                \cos \left( \frac{1}{2}\Omega' t \right) 
                - \ci \frac{\Delta}{\Omega'} \sin \left( \frac{1}{2}\Omega' t \right) 
            \right],
\end{equation}
which can also be described by the following time evolution operator
\begin{equation}
\text{$
    \hat{U}(t)
    = \begin{pmatrix} \displaystyle
        e^{\ci \omega t /2} \cos\left( \frac{1}{2} \Omega' t \right) 
        + \ci e^{\ci \omega t /2} \frac{\Delta}{\Omega'} \sin \left( \frac{1}{2}\Omega' t \right) 
        & \displaystyle \ci e^{\ci \omega t /2 -\ci \phi_{g}} \frac{\Omega}{\Omega'} \sin\left( \frac{1}{2}\Omega' t \right) \\[4mm]
        \displaystyle \ci e^{-\ci \omega t /2 + \ci \phi_{g}} \frac{\Omega}{\Omega'} \sin\left( \frac{1}{2}\Omega' t \right) 
        & \displaystyle e^{-\ci \omega t /2} \cos \left( \frac{1}{2}\Omega' t \right) 
        - \ci e^{-\ci \omega t /2} \frac{\Delta}{\Omega'} \sin \left( \frac{1}{2}\Omega' t \right) 
    \end{pmatrix}.
$}
\end{equation}
For completeness, in the interaction picture, the time evolution operator is given by
\begin{equation}
\text{$
    \hat{\Tilde{U}}(t)
    = \hat{U}_0^{\dagger}(t) \hat{U}(t)
    = \begin{pmatrix} \displaystyle
        \cos\left( \frac{1}{2} \Omega' t \right) 
        + \ci \frac{\Delta}{\Omega'} \sin \left( \frac{1}{2}\Omega' t \right) 
        & \displaystyle \ci e^{-\ci \phi_{g}} \frac{\Omega}{\Omega'} \sin\left( \frac{1}{2}\Omega' t \right) \\[4mm]
        \displaystyle \ci e^{\ci \phi_{g}} \frac{\Omega}{\Omega'} \sin\left( \frac{1}{2}\Omega' t \right) 
        & \displaystyle \cos \left( \frac{1}{2}\Omega' t \right) 
        - \ci \frac{\Delta}{\Omega'} \sin \left( \frac{1}{2}\Omega' t \right) 
    \end{pmatrix},
$}
\end{equation}
where
\begin{equation}
    \hat{U}_0(t) = \exp(-\frac{\ci \hat{H}_{\text{q}} t}{\hbar}) 
    = \begin{pmatrix} \displaystyle
        \displaystyle e^{\ci \omega_{\text{q}} t}
        & \displaystyle 0 \\[2mm]
        \displaystyle 0
        & \displaystyle e^{-\ci \omega_{\text{q}} t}
    \end{pmatrix}
    = \hat{R}_z(-\omega_{\text{q}} t).
\end{equation}

%% file: include_appendices/appendix_Displacement_op.tex
\chapter{Displacement Operators}
\section{Definition}
The displacement operator
\begin{equation}
    \hat{D}(\alpha)
    = e^{\alpha \hat{a}^{\dagger} - \alpha^* \hat{a}}
\end{equation}
is a unitary operator, i.e.,
\begin{equation}
    \hat{D}^{\dagger}(\alpha)
        \hat{D}(\alpha)
    = \hat{D}(\alpha)
        \hat{D}^{\dagger}(\alpha)
\end{equation}
with the property
\begin{equation}
    \hat{D}^{\dagger}(\alpha) 
    = \hat{D}(-\alpha).
\end{equation}
The complex amplitude $\alpha$ indicates the displacement when $\hat{D}(\alpha)$ acts on a coherent state.

One can also rewrite the displacement operator using the real and imaginary parts of $\alpha$ and $\hat{a}$:
Convention (i): ($\hbar = 1/2$)
\begin{equation}
    \hat{a} = \hat{X} + \ci \hat{P}
\end{equation}
\begin{equation}
    \hat{X} = \frac{\hat{a} + \hat{a}^{\dagger}}{2}
    \ \ \text{ and } \ \ 
    \hat{P} = -\ci \frac{\hat{a} - \hat{a}^{\dagger}}{2}
\end{equation}
\begin{equation}
    \hat{D}(\alpha) = e^{2 \ci (P\hat{X} - X\hat{P})}
    = \hat{D}(X,P)
\end{equation}
Convention (ii): ($\hbar = 1$)
\begin{equation}
    \hat{a} = \frac{\hat{X} + \ci \hat{P}}{\sqrt{2}}
\end{equation}
\begin{equation}
    \hat{X} = \frac{\hat{a} + \hat{a}^{\dagger}}{\sqrt{2}}
    \ \ \text{ and } \ \ 
    \hat{P} = -\ci \frac{\hat{a} - \hat{a}^{\dagger}}{\sqrt{2}}
\end{equation}
\begin{equation}
    \hat{D}(\alpha) = e^{\ci (P\hat{X} - X\hat{P})}
    = \hat{D}(X,P)
\end{equation}

\section{Equivalent Forms}
The following forms of the displacement operator might come in handy:
\begin{align}
    \hat{D}(\alpha)
    = e^{\alpha \hat{a}^{\dagger} - \alpha^* \hat{a}} 
    = e^{-|\alpha|^2 /2} e^{\alpha \hat{a}^{\dagger}} e^{-\alpha^* \hat{a}}
    = e^{|\alpha|^2 /2} e^{-\alpha^* \hat{a}} e^{\alpha \hat{a}^{\dagger}}
\end{align}

\section{Composition} 
Displacement operators usually do not commute with one another:
\begin{equation}
    \hat{D}(\alpha) \hat{D}(\beta) 
    = e^{(\alpha \beta^* - \alpha^* \beta)/2} \hat{D}(\alpha + \beta)
    = e^{\ci \Im(\alpha \beta^*)} \hat{D}(\alpha + \beta)
\end{equation}
\begin{equation}
    \hat{D}(\alpha) \hat{D}(\beta) 
    = e^{\alpha \beta^* - \alpha^* \beta}
        \hat{D}(\beta) \hat{D}(\alpha)
    = e^{\ci 2 \Im(\alpha \beta^*)}
        \hat{D}(\beta) \hat{D}(\alpha)
\end{equation}
This property is, for example, considered in the construction of the GKP states for bosonic qubits.

\section{Actions on the Annihilation and Creation Operators}
Displacement operators are often used to perform a change of frame. In the Schr\"{o}dinger picture, a displacement operator can move the amplitude of a coherent state arbitrarily in the phase space. In the Heisenberg picture, this is equivalent to adding some complex amplitude to the annihilation and creation operators:
\begin{equation}
    \hat{D}^{\dagger}(\alpha)
    \hat{a}
    \hat{D}(\alpha)
    = \hat{a} + \alpha,
\end{equation}
\begin{equation}
    \hat{D}^{\dagger}(\alpha)
    \hat{a}^{\dagger}
    \hat{D}(\alpha)
    = \hat{a} + \alpha^*.
\end{equation}
Taking the Hermitian conjugate of the two equations also gives
\begin{equation}
    \hat{D}(\alpha)
    \hat{a}
    \hat{D}^{\dagger}(\alpha)
    = \hat{a} - \alpha,
\end{equation}
\begin{equation}
    \hat{D}(\alpha)
    \hat{a}^{\dagger}
    \hat{D}^{\dagger}(\alpha)
    = \hat{a} - \alpha^*.
\end{equation}

\section{Time Dependence}
One often encounters the terms $\hat{D}^{\dagger}(\alpha) \dot{\hat{D}}(\alpha)$ or $\dot{\hat{D}}^{\dagger}(\alpha) \hat{D}(\alpha)$ when going into a displaced frame with a time-dependent $\alpha(t)$. This section is devoted to derivative a general expression for the derivatives of the displacement operator.

To begin with, note that the statement
\begin{equation}
     \frac{\mathrm{d}}{\mathrm{d} t} 
            e^{f(t) \hat{A}}
    = \dot{f}(t) e^{f(t) \hat{A}}
\end{equation}
is in general true for a scalar function $f(t)$ because $[f(t) \hat{A}, f(s) \hat{A}] = 0$ for any $t$ and $s$. However, it's \textit{incorrect} to make the generalization
\begin{equation}
    \frac{\mathrm{d}}{\mathrm{d} t} 
            e^{\hat{A}(t)}
    = \dot{\hat{A}} (t) e^{\hat{A}(t)}
\end{equation}
because $[\hat{A}(t), \hat{A}(s)] \neq 0$ in general. Hence, when taking the time derivative of the displacement operator, extra care must be taken:
\begin{align}
    \hat{D}^{\dagger}(\alpha(t))
        \frac{\mathrm{d}}{\mathrm{d} t} 
            \hat{D}(\alpha(t))
    &= \hat{D}^{\dagger}(\alpha(t))
        \frac{\mathrm{d}}{\mathrm{d} t}
        \Big[
            e^{-\abs{\alpha(t)}^2/2}
            \,
            e^{\alpha(t) \hat{a}^{\dagger}}
            \, 
            e^{-\alpha^*(t) \hat{a}}
        \Big]
\nonumber \\
    & = \hat{D}^{\dagger}(\alpha)
        \left[
            - \frac{\dot{\alpha}^* \alpha + \alpha^* \dot{\alpha}}{2} \hat{D}(\alpha)
            + \dot{\alpha} 
                \hat{a}^{\dagger} 
                \hat{D}(\alpha)
            - \hat{D}(\alpha) 
                \dot{\alpha}^* 
                \hat{a}
        \right]
\nonumber \\
    &= \dot{\alpha} \hat{a}^{\dagger}
        - \dot{\alpha}^* \hat{a}
        + \frac{\alpha^* \dot{\alpha} - \dot{\alpha}^* \alpha}{2}
\end{align}
and, using the fact that $\hat{D}^{\dagger}\hat{D} = \hat{1}$, 
\begin{align}
    \frac{\mathrm{d}}{\mathrm{d} t} \hat{D}^{\dagger}(\alpha(t))
            \hat{D}(\alpha(t))
    &= \frac{\mathrm{d}}{\mathrm{d} t} 
        \cancel{\left[
            \hat{D}^{\dagger}(\alpha(t))
            \hat{D}(\alpha(t))
        \right]}
        - \hat{D}^{\dagger}(\alpha(t))
        \frac{\mathrm{d}}{\mathrm{d} t} 
            \hat{D}(\alpha(t))
\nonumber \\
    &= - \dot{\alpha} \hat{a}^{\dagger}
        + \dot{\alpha}^* \hat{a}
        - \frac{\alpha^* \dot{\alpha} - \dot{\alpha}^* \alpha}{2}.
\end{align}
In addition, from the above derivation, it's also easy to show that 
\begin{equation}
    \frac{\mathrm{d}}{\mathrm{d} t} 
            \hat{D}(\alpha(t))
    = \left[ 
            \dot{\alpha} \hat{a}^{\dagger}
            - \dot{\alpha}^* \hat{a}
            - \frac{\alpha^* \dot{\alpha} - \dot{\alpha}^* \alpha}{2}
        \right]
        \hat{D}(\alpha(t)),
\end{equation}
before multiplied by the adjoint.

%% file: include_appendices/appendix_Dephasing.tex
\chapter{Modelling Dephasing}\label{eq:appendix_dephasing}
\section{Longitudinal Qubit-Environment Coupling}
Consider a generic noise source\footnote{The noise sources should really be represented by quantum operators \cite{schaller2014open}. Since we focus more on the ensemble-averaged behavior in the weak coupling regime, we will apply a semi-classical analysis here.} 
\begin{equation}
    \lambda_{t} = \Bar{\lambda} + \delta\lambda_{t} 
\end{equation}
specified by its mean $\Bar{\lambda}$ (i.e., the desired value) and a random process $\delta \lambda_{t}$. $\lambda_{t}$ can represent, for example, noise on the flux line when tuning the qubit or charge noise on the control line. If the noise source couples to the qubit weakly in the longitudinal direction (i.e., $\hat{H}_{\text{int}} \propto \delta \lambda_{t} \hat{\sigma}_z$), we can represent the perturbed Hamiltonian as
\begin{align}
    \hat{H}
    &= - \frac{1}{2} 
        \hbar 
        \omega_{\text{q}} (\lambda_{t}) \, 
        \hat{\sigma}_z
    = - \frac{1}{2} 
        \hbar 
        \omega_{\text{q}} (\Bar{\lambda}) \, 
        \hat{\sigma}_z
     - \frac{1}{2} 
        \hbar \hat{\sigma}_z 
        \frac{\partial \omega_{\text{q}}}{\partial \lambda}     
        \bigg|_{\Bar{\lambda}} 
        \delta \lambda_{t} 
    + \mathcal{O}(\delta \lambda_{t}^2)
\nonumber \\
    &\doteq - \frac{1}{2} \hbar \omega_{\text{q},0} \, \hat{\sigma}_z
        - \frac{1}{2} \hbar v_{t}\hat{\sigma}_z 
        + \mathcal{O}(\delta \lambda_{t}^2),
\end{align}
where $\omega_{\text{q},0}$ is the nominal qubit frequency and
\begin{equation}\label{eq:definition_noise_freq_v}
    v_{t}
    = \frac{\partial \omega_\text{q}}{\partial \lambda}     
        \bigg|_{\Bar{\lambda}} 
        \delta \lambda_{t} 
\end{equation}
is the noise in the frequency. For weak coupling, we only keep the first-order noise.

Given the Hamiltonian above, we can write down the Liouville equation for the qubit. In particular, we will examine the coherence $\rho_{eg,t} = \bra{e} \hat{\rho}_t \ket{g}$, which evolves according to
\begin{equation}
    \dot{\rho}_{eg,t}
    = \ci [\omega_{\text{q},0} + v_{t}] \rho_{eg, t}
\end{equation}
or
\begin{equation}
    \rho_{eg,t}
    = \exp
        \left[ 
            \ci \omega_{\text{q},0} t
            + \ci 
                \int_{0}^{t} 
                    \mathrm{d} \tau 
                    \, v_{\tau}
        \right] 
        \rho_{eg,0}.
\end{equation}
To understand the average dephasing time, we can take the expected value over all realizations of $\delta \lambda_{t}$. Denote the expected coherence by $\overline{\rho}_{eg} (t) = \mathbb{E}[\rho_{eg,t}]$, then
\begin{equation}\label{eq:average_rho_eg}
    \overline{\rho}_{eg} (t) 
    = e^{\ci \omega_{\text{q},0} t} \,
        \mathbb{E} 
            \left\{ 
                \exp
                    \left[ 
                        - \ci \int_{0}^{t} 
                            \mathrm{d} \tau 
                            v_{\tau}
                    \right]
            \right\} \overline{\rho}_{eg}(0).
\end{equation}
Moreover, we assume $\delta\lambda_{t}$ and thus the frequency uncertainty $v_t$ can be described by Gaussian random processes \cite{doi:10.1063/1.5089550}, so the extra phase accumulated is also Gaussian-distributed (at any fixed time $t$)\footnote{Recall that a linear combination of jointly Gaussian random variables is still Gaussian-distributed.}:
\begin{equation}
    \Phi(t)
    = \int_{0}^{t} 
        \mathrm{d} \tau 
        v_{\tau}
    = \frac{\partial \omega_{\text{q}}}{\partial \lambda}     
        \int_{0}^{t} 
            \mathrm{d} \tau 
            \delta \lambda_{\tau} 
    \sim 
    \mathcal{N} 
        \Big(
            0,\sigma^2_{\Phi}(t)
        \Big),
\end{equation}
where the variance $\sigma^2_{\Phi}(t)$ clearly depends on the details of $\delta \lambda_{t}$. Under this definition, the expected value in Eq.(\ref{eq:average_rho_eg}) becomes the characteristic function of $\Phi(t)$:
\begin{equation}
    \mathbb{E}
        \left\{ 
            \exp 
                \left[ 
                    - \ci \int_{0}^{t} 
                        \mathrm{d} \tau 
                        v_{\tau}
                \right]
        \right\}
    = \mathbb{E}
        \Big[ 
            e^{\ci \Phi(t) \theta} 
        \Big]\bigg|_{\theta = -1}
    = \exp\left[- \frac{1}{2} \sigma_{\Phi}^2(t)\right]
\end{equation}

To quantify $\sigma^2_{\Phi}(t)$, we need to characterize the random process $\delta \lambda_{t}$ first. To do so, we invoke its autocorrelation function
\begin{equation}
    R_{\delta\lambda} (\tau) 
    = \mathbb{E}
        \big[
            \delta\lambda_{t} 
            \delta\lambda_{t + \tau}
        \big],
\end{equation}
whose Fourier transform is the (double-sided) power spectral density \cite{Schoelkopf2003, Clerk_2010}
\begin{equation}
    S_{\delta \lambda}(\omega) 
    = \int_{-\infty}^{\infty} 
        \mathrm{d} \tau \, 
        R_{\delta \lambda} (\tau) 
        e^{-\ci \omega \tau}.
\end{equation}
This allows us to write
\begin{align}
    \sigma_{\Phi}^2(t) 
    = \mathbb{E} 
        \big[
            \Phi^2(t)
        \big] 
    &= \mathbb{E}
        \left[
            \left(
                \frac{\partial \omega_{\text{q}}}{\partial \lambda}     
            \right)^2
            \int_{0}^{t} 
                \mathrm{d} t_1
                \int_{0}^{t} 
                    \mathrm{d} t_2 \, 
                    \delta \lambda_{t_1}
                    \delta \lambda_{t_2}
        \right]
\nonumber \\ \label{eq:variance_phi}
    &= \left(
            \frac{\partial \omega_{\text{q}}}{\partial \lambda}     
        \right)^2
            \int_{0}^{t} 
            \mathrm{d} t_1
            \int_{0}^{t} 
            \mathrm{d} t_2 \, 
            R_{\delta\lambda} (t_1 - t_2).
\end{align}
Next, we can extend the double integral to the entire $t_1$-$t_2$ plane by defining a rectangular function $\Pi_t(\tau)$ which represents the window of integration (note that $t$ is just a constant in $\Pi_t$):
\begin{equation}
    \Pi_t(\tau) 
    = \begin{cases}
            1, & \text{ if } 0 < \tau < t, \\
            0, & \text{ otherwise}.
        \end{cases}
\end{equation}
More importantly, the Fourier transform of $\Pi_t(\tau)$, i.e.,
\begin{equation}
    \tilde{\Pi}_t(\omega) 
    = \int_{-\infty}^{\infty} 
        \mathrm{d} \tau 
        \Pi_t(\tau) e^{-\ci \omega \tau}
    = \int_{0}^{t} 
        \mathrm{d} \tau 
        e^{-\ci \omega \tau}
    = e^{-\ci \omega t/2} \, \frac{\sin(\omega t /2)}{\omega /2},
\end{equation}
acts as a filter, allowing the low-frequency noise to pass. To see this, notice that the inner integral in Eq.(\ref{eq:variance_phi}) can be treated as the convolution of $\Pi_t(t_1)$ and $R_{\delta \lambda}(t_1)$ and, thus, can be converted to the frequency domain by the convolution theorem:
\begin{align}
    \sigma_{\Phi}^2(t)
    &= \left(
            \frac{\partial \omega_{\text{q}}}{\partial \lambda}     
        \right)^2
        \int_{-\infty}^{\infty} 
            \mathrm{d} t_1
            \Pi_t(t_1)
            \int_{-\infty}^{\infty} 
                \mathrm{d} t_2 \,
                \Pi_t(t_2) 
                R_{\delta\lambda}(t_1 - t_2)
\nonumber \\
    &= \left(
            \frac{\partial \omega_{\text{q}}}{\partial \lambda}     
        \right)^2 
        \int_{-\infty}^{\infty} 
            \mathrm{d} t_1
            \Pi_t(t_1)
            \left[
                \frac{1}{2\pi}
                \int_{-\infty}^{\infty} 
                    \mathrm{d} \omega \, 
                    e^{\ci \omega t_1}
                    \tilde{\Pi}_t(\omega) 
                    S_{\delta \lambda}(\omega)
            \right]
\nonumber\\
    & = \left(
            \frac{\partial \omega_{\text{q}}}{\partial \lambda}     
        \right)^2 
        \int_{-\infty}^{\infty}
            \frac{\mathrm{d} \omega }{2\pi} \, 
            \tilde{\Pi}_t(\omega)
            S_{\delta \lambda}(\omega)
            \left[
                \int_{-\infty}^{\infty} 
                \mathrm{d} t_1 \, 
                e^{\ci \omega t_1}
                \Pi_t(t_1)
            \right]
\nonumber\\ \label{eq:sigma_phi_derivation}
    & = \left(
            \frac{\partial \omega_{\text{q}}}{\partial \lambda}     
        \right)^2 
        \int_{-\infty}^{\infty}
            \frac{\mathrm{d} \omega }{2\pi} \, 
            \abs{\tilde{\Pi}_t(\omega)}^2
                S_{\delta \lambda}(\omega)
    = \left(
            \frac{\partial \omega_{\text{q}}}{\partial \lambda}     
        \right)^2 
        \int_{-\infty}^{\infty}
            \frac{\mathrm{d} \omega }{2\pi} \, 
            \frac{\sin^2(\omega t /2)}{(\omega /2)^2} 
            S_{\delta \lambda}(\omega)
\end{align}
Note that the noise spectrum of $\delta \lambda$ is filtered by a $\mathrm{sinc}^2$ function, which unfortunately makes the measurement susceptible to the $1/f$-noise.

Usually, we use $S_{v}(\omega)$ instead of $S_{\delta \lambda}(\omega)$ to remove the partial derivative in Eq.(\ref{eq:sigma_phi_derivation}) based on Eq.(\ref{eq:definition_noise_freq_v}). Combining all the pieces, we obtain \cite{PhysRevLett.88.047901}
\begin{equation}
    \overline{\rho}_{eg} (t) 
    = e^{\ci \omega_{\text{q},0} t} \,
        \exp
        \left[ 
            - \frac{t^2}{2}
            \int_{-\infty}^{\infty}
                \frac{\mathrm{d} \omega }{2\pi} \, 
                \frac{\sin^2(\omega t /2)}{(\omega t /2)^2} 
                S_{v}(\omega)
        \right] 
        \overline{\rho}_{eg}(0).
\end{equation}